\pdfoutput=1
\RequirePackage[log]{snapshot}
\documentclass[12pt,twoside]{article}

\usepackage[utf8]{inputenc}
\usepackage[T1]{fontenc}
\usepackage{amsmath}
\usepackage{amssymb}
\usepackage[nottoc,notlot,notlof]{tocbibind}
\usepackage{calc}
\usepackage[pdftex]{graphicx} %
\usepackage[figuresright]{rotating}

\usepackage{authblk}
\usepackage{caption}
\usepackage{array}
\usepackage{longtable}
\usepackage{etoolbox}
\usepackage{float}

\usepackage{adjustbox}
\usepackage{tabularx}

\usepackage{multirow}
\usepackage{booktabs}
\usepackage{color, colortbl}
\usepackage[dvipsnames]{xcolor} %
\usepackage{footmisc}
\usepackage{afterpage}
\usepackage{relsize}                                      %
\usepackage{wasysym}
\usepackage[math]{cellspace}
\usepackage{bm}
\usepackage{lineno}  %

\definecolor{Gray}{gray}{0.9}
\definecolor{LightGray}{gray}{0.97}

\usepackage{silence}
\usepackage{hyperref}
\usepackage[all]{hypcap}
\usepackage{cite}    %
\usepackage{mciteplus}
\usepackage{ifthen} %
\usepackage{orcidlink}

\newcommand{\mysection}[1]{\section{\boldmath #1}}
\newcommand{\mysubsection}[1]{\subsection[#1]{\boldmath #1}}
\newcommand{\mysubsubsection}[1]{\subsubsection[#1]{\boldmath #1}}
\newcommand{\mysubsubsubsection}[1]{\subsubsubsection{\boldmath #1}}

\def\dof{{\rm dof}}

\renewcommand\Im{{\rm Im}} 
\renewcommand\Re{{\rm Re}}

\newcommand\Abar{\kern 0.18em\overline{\kern -0.18em A}{}}

\newcommand {\cbf}{\ensuremath{{\cal B}}}

\newcommand {\vcb}{\ensuremath{|V_{cb}|}}
\newcommand {\vub}{\ensuremath{|V_{ub}|}}

\def\Bp      {\ensuremath{B^{+}}}

\def\Bz      {\ensuremath{B^{0}}}
\def\Bs      {\ensuremath{B_{s}}}

\newcommand{\BzbDplnu}    {\ensuremath{\Bzb \to D^{+}\ell^{-}\nub_\ell}}
\newcommand{\BzbDstarlnu} {\ensuremath{\Bzb \to D^{*+}\ell^{-}\nub_\ell}}

\usepackage{relsize}

\def\beq{\begin{equation}}
\def\eeq#1{\label{#1}\end{equation}}
\def\eeqn{\end{equation}}

\def\beqa{\begin{eqnarray}}
\def\eeqa#1{\label{#1}\end{eqnarray}}
\def\eeqan{\end{eqnarray}}

\let\bar=\overbar

\def\ie{{\it i.e.}}
\def\eg{{\it e.g.}\xspace}

\def\W{{\cal W}}

\def\Dslash{\ensuremath{\not{\hbox{\kern-4pt $D$}}}\xspace}
\def\dslash{\not{\hbox{\kern-2pt $\del$}}}

\def\BR{\mbox{\rm BR}}
\def\ee{e^+e^-}

\def\alphas{\alpha_s}
\def\msb{{\bar{\ssstyle M \kern -1pt S}}}

\RequirePackage{xspace}

\usepackage{relsize}
\def\atlas{\mbox{\normalfont ATLAS}\xspace}
\def\babar{\mbox{\slshape B\kern-0.1em{\smaller A}\kern-0.1em
    B\kern-0.1em{\smaller A\kern-0.2em R}}\xspace}
\def\belle{\mbox{\normalfont Belle}\xspace}
\def\belletwo   {\mbox{\normalfont Belle II}\xspace}
\def\besthree{\mbox{\normalfont BES III}\xspace}
\def\cdf{\mbox{\normalfont CDF}\xspace}
\def\dzero{\mbox{\normalfont D0}\xspace}
\def\lhcb{\mbox{\normalfont LHCb}\xspace}

\def\ee         {\ensuremath{e^-e^-}\xspace}

\def\mup        {\ensuremath{\mu^+}\xspace}
\def\mun        {\ensuremath{\mu^-}\xspace}    %

\def\mtau       {\ensuremath{\tau}\xspace}

\def\ellp       {\ensuremath{\ell^+}\xspace}

\def\nub        {\ensuremath{\overline{\nu}}\xspace}

\def\nub        {\ensuremath{\overline{\nu}}\xspace}

\def\W      {\ensuremath{W}\xspace}

\def\Z      {\ensuremath{Z^0}\xspace}

\def\dbar  {\ensuremath{\overline d}\xspace}
\def\ddbar {\ensuremath{d\overline d}\xspace}

\def\b  {\ensuremath{b}\xspace}
\def\bbar  {\ensuremath{\overline b}\xspace}

\def\pip   {\ensuremath{\pi^+}\xspace}

\def\etapr {\ensuremath{\eta^{\prime}}\xspace}

\def\Kbar  {\kern 0.2em\overline{\kern -0.2em K}{}\xspace}

\def\Kp    {\ensuremath{K^+}\xspace}
\def\Km    {\ensuremath{K^-}\xspace}
\def\KS    {\ensuremath{K^0_{\scriptscriptstyle S}}\xspace} 
\def\KL    {\ensuremath{K^0_{\scriptscriptstyle L}}\xspace} 
\def\Kstarz  {\ensuremath{K^{*0}}\xspace}

\def\Kz   {\ensuremath{K^0}\xspace}
\def\Kzb   {\ensuremath{\Kbar^0}\xspace}
\def\KzKzb {\ensuremath{K^0 \kern -0.16em \Kzb}\xspace}

\def\Dz    {\ensuremath{D^0}\xspace}
\def\Dbar  {\kern 0.2em\overline{\kern -0.2em D}{}\xspace}

\def\Dzb   {\ensuremath{\Dbar^0}\xspace}
\def\DzDzb {\ensuremath{D^0 {\kern -0.16em \Dzb}}\xspace}
\def\Dp    {\ensuremath{D^+}\xspace}
\def\Dm    {\ensuremath{D^-}\xspace}

\def\Ds    {\ensuremath{D^+_s}\xspace}
\def\Dsp   {\ensuremath{D^+_s}\xspace}
\def\Dsm   {\ensuremath{D^-_s}\xspace}

\def\Bz    {\ensuremath{B^0}\xspace}
\def\B     {\ensuremath{B}\xspace}
\def\Bbar  {\kern 0.18em\overline{\kern -0.18em B}{}\xspace}

\def\Bzb   {\ensuremath{\Bbar^0}\xspace}
\def\Bu    {\ensuremath{B^+}\xspace}

\def\Bpm   {\ensuremath{B^\pm}\xspace}

\def\Bs    {\ensuremath{B_s}\xspace}
\def\Bsb   {\ensuremath{\Bbar_s^0}\xspace}
\def\BB    {\ensuremath{B\Bbar}\xspace} 
\def\BzBzb {\ensuremath{B^0 {\kern -0.16em \Bzb}}\xspace}

\def\jpsi  {\ensuremath{{J\mskip -3mu/\mskip -2mu\psi\mskip 2mu}}\xspace}

\mathchardef\Upsilon="7107
\def\Y#1S{\ensuremath{\Upsilon{(#1S)}}\xspace}%

\mathchardef\Deltares="7101
\mathchardef\Xi="7104
\mathchardef\Lambda="7103
\mathchardef\Sigma="7106
\mathchardef\Omega="710A
\def\Deltabar   {\kern 0.25em\overline{\kern -0.25em \Deltares}{}\xspace}
\def\Lbar {\kern 0.2em\overline{\kern -0.2em\Lambda\kern 0.05em}\kern-0.05em{}\xspace}
\def\Sigbar{\kern 0.2em\overline{\kern -0.2em \Sigma}{}\xspace}
\def\Xibar{\kern 0.2em\overline{\kern -0.2em \Xi}{}\xspace}
\def\Obar{\kern 0.2em\overline{\kern -0.2em \Omega}{}\xspace}
\def\Nbar{\kern 0.2em\overline{\kern -0.2em N}{}\xspace}
\def\Xb{\kern 0.2em\overline{\kern -0.2em X}{}}

\newcommand{\particle}[1]{\ensuremath{#1}\xspace}
\renewcommand{\ee}{\particle{e^+e^-}}
\newcommand{\Ups}{\particle{\Upsilon(4S)}}

\renewcommand{\b}{\particle{b}}
\renewcommand{\B}{\particle{B}}
\newcommand{\Bd}{\particle{B^0}}
\renewcommand{\Bs}{\particle{B^0_s}}
\renewcommand{\Bu}{\particle{B^+}}
\newcommand{\Bc}{\particle{B^+_c}}
\newcommand{\Bdbar}{\particle{\bar{B}^0}}
\newcommand{\Bsbar}{\particle{\bar{B}^0_s}}
\newcommand{\Lb}{\particle{\Lambda_b^0}}

\newcommand{\Xibd}{\particle{\Xi_b^-}}
\newcommand{\Xibu}{\particle{\Xi_b^0}}
\newcommand{\Omegab}{\particle{\Omega_b^-}}
\newcommand{\Lc}{\particle{\Lambda_c^+}}

\def\BR{{\ensuremath{\cal B}}}

\def\Btopilnu   {\ensuremath{B \to \pi\ell\nu}}

\newcommand{\tev}{\ensuremath{\mathrm{Te\kern -0.1em V}}\xspace}
\newcommand{\gev}{\ensuremath{\mathrm{Ge\kern -0.1em V}}\xspace}
\newcommand{\mev}{\ensuremath{\mathrm{Me\kern -0.1em V}}\xspace}
\newcommand{\kev}{\ensuremath{\mathrm{ke\kern -0.1em V}}\xspace}
\newcommand{\ev}{\ensuremath{\mathrm{e\kern -0.1em V}}\xspace}
\newcommand{\gevc}{\ensuremath{{\mathrm{Ge\kern -0.1em V\!/}c}}\xspace}
\newcommand{\mevc}{\ensuremath{{\mathrm{Me\kern -0.1em V\!/}c}}\xspace}
\newcommand{\gevcc}{\ensuremath{{\mathrm{Ge\kern -0.1em V\!/}c^2}}\xspace}
\newcommand{\gevgevcccc}{\ensuremath{{\mathrm{Ge\kern -0.1em V^2\!/}c^4}}\xspace}
\newcommand{\mevcc}{\ensuremath{{\mathrm{Me\kern -0.1em V\!/}c^2}}\xspace}

\def\pb {\ensuremath{\rm \,pb}\xspace}

\def\fb   {\ensuremath{\mbox{\,fb}}\xspace}
\def\invfb   {\ensuremath{\mbox{\,fb}^{-1}}\xspace}
\def\mus  {\ensuremath{\rm \,\mus}\xspace}

\def\ps   {\ensuremath{\rm \,ps}\xspace}

\def\mus        {\ensuremath{\,\mu{\rm s}}\xspace}    %
\def\ps         {\ensuremath{{\rm \,ps}}\xspace}  %

\def\gsim{{~\raise.15em\hbox{$>$}\kern-.85em
          \lower.35em\hbox{$\sim$}~}\xspace}
\def\lsim{{~\raise.15em\hbox{$<$}\kern-.85em
          \lower.35em\hbox{$\sim$}~}\xspace}
\def\qsq                {\ensuremath{q^2}\xspace}
\def\CP                 {\ensuremath{C\!P}\xspace}
\def\CPT                {\ensuremath{C\!PT}\xspace}
\def\ra                 {\ensuremath{\to}\xspace}
\def\pep2{PEP-II}

\def\Vub  {\ensuremath{|V_{ub}|}\xspace}
\def\Vcb  {\ensuremath{|V_{cb}|}\xspace}

\def\deltamd{\ensuremath{{\rm \Delta}m_d}\xspace}

\xspace
\newcommand{\fds}{\ensuremath{f_{D_s}}\xspace}

\newcommand{\tabref}[1]{Table~\ref{tab:#1}}

\def\jetset74   {\mbox{\tt Jetset \hspace{-0.5em}7.\hspace{-0.2em}4}}
\newcommand{\ks}    {\KS}

\xspace
\xspace

\newboolean{unpublished}\setboolean{unpublished}{false}

\newboolean{history}\setboolean{history}{false}

\newcommand{\unpublished}[2]{\ifthenelse{\boolean{unpublished}}{#2}{#1}}
\newcommand{\history}[2]{\ifthenelse{\boolean{history}}{#2}{#1}}
\newcommand{\citehistory}[2]{\ifthenelse{\boolean{history}}{\cite{#2}}{\cite{#1}}}

\newcommand{\definemath}[2]{\newcommand{#1}{\ensuremath{#2}\xspace}}

\definemath{\hflavCHIDWval}{0.1860}%
\definemath{\hflavCHIDWerr}{\pm0.0011}%
\definemath{\hflavCHIDWUval}{0.1860}%
\definemath{\hflavCHIDWUerr}{\pm0.0011}%
\definemath{\hflavXDWval}{0.7698}%
\definemath{\hflavXDWerr}{\pm0.0035}%
\definemath{\hflavXDWUval}{0.7697}%
\definemath{\hflavXDWUerr}{\pm0.0035}%
\definemath{\hflavCHIDUval}{0.182}%
\definemath{\hflavCHIDUerr}{\pm0.015}%
\definemath{\hflavDMDWval}{0.5069}%
\definemath{\hflavDMDWsta}{\pm0.0016}%
\definemath{\hflavDMDWsys}{\pm0.0011}%
\definemath{\hflavDMDWerr}{\pm0.0019}%
\definemath{\hflavDMDWUval}{0.5069}%
\definemath{\hflavDMDWUsta}{\pm0.0016}%
\definemath{\hflavDMDWUsys}{\pm0.0011}%
\definemath{\hflavDMDWUerr}{\pm0.0019}%
\definemath{\hflavDMDLHCbval}{0.5063}%
\definemath{\hflavDMDLHCbsta}{\pm0.0019}%
\definemath{\hflavDMDLHCbsys}{\pm0.0010}%
\definemath{\hflavDMDLHCberr}{\pm0.0022}%
\definemath{\hflavDMDLval}{0.495}%
\definemath{\hflavDMDLsta}{\pm0.010}%
\definemath{\hflavDMDLsys}{\pm0.010}%
\definemath{\hflavDMDLerr}{\pm0.014}%
\definemath{\hflavDMDTval}{0.509}%
\definemath{\hflavDMDTsta}{\pm0.017}%
\definemath{\hflavDMDTsys}{\pm0.013}%
\definemath{\hflavDMDTerr}{\pm0.022}%
\definemath{\hflavDMDBval}{0.5109}%
\definemath{\hflavDMDBsta}{\pm0.0030}%
\definemath{\hflavDMDBsys}{\pm0.0034}%
\definemath{\hflavDMDBerr}{\pm0.0045}%
\definemath{\hflavTAUBSJPSIKSHORTval}{1.63}%
\definemath{\hflavTAUBSJPSIKSHORTerr}{\pm0.07}%
\definemath{\hflavRTAUBSval}{1.0017}%
\definemath{\hflavRTAUBSerr}{\pm0.0034}%
\definemath{\hflavONEMINUSRTAUBSMEANCpercent}{(-0.2\pm0.3)\%}%
\definemath{\hflavRTAUBSMEANCval}{1.0017}%
\definemath{\hflavRTAUBSMEANCerr}{\pm0.0034}%
\definemath{\hflavTAUBSSLval}{1.527}%
\definemath{\hflavTAUBSSLerr}{\pm0.011}%
\definemath{\hflavRTAUBSSLval}{1.006}%
\definemath{\hflavRTAUBSSLerr}{\pm0.008}%
\definemath{\hflavTAUBSJFval}{1.480}%
\definemath{\hflavTAUBSJFerr}{\pm0.007}%
\definemath{\hflavTAUBSJPSIPIPIval}{1.646}%
\definemath{\hflavTAUBSJPSIPIPIerr}{\pm0.013}%
\definemath{\hflavTAUBSLONGval}{1.646}%
\definemath{\hflavTAUBSLONGerr}{\pm0.013}%
\definemath{\hflavTAUBSKKval}{1.408}%
\definemath{\hflavTAUBSKKerr}{\pm0.017}%
\definemath{\hflavTAUBSDSDSval}{1.379}%
\definemath{\hflavTAUBSDSDSerr}{\pm0.031}%
\definemath{\hflavTAUBSJPSIETAval}{1.452}%
\definemath{\hflavTAUBSJPSIETAerr}{\pm0.016}%
\definemath{\hflavTAUBSSHORTval}{1.437}%
\definemath{\hflavTAUBSSHORTerr}{\pm0.014}%
\definemath{\hflavGSval}{0.6632}%
\definemath{\hflavGSerr}{\pm0.0034}%
\definemath{\hflavDGSval}{+0.076}%
\definemath{\hflavDGSerr}{\pm0.004}%
\definemath{\hflavRHOGSDGS}{-0.14}%
\definemath{\hflavDGSGSval}{+0.114}%
\definemath{\hflavDGSGSerr}{\pm0.006}%
\definemath{\hflavTAUBSMEANval}{1.508}%
\definemath{\hflavTAUBSMEANerr}{\pm0.008}%
\definemath{\hflavTAUBSLval}{1.427}%
\definemath{\hflavTAUBSLerr}{\pm0.007}%
\definemath{\hflavTAUBSHval}{1.599}%
\definemath{\hflavTAUBSHerr}{\pm0.011}%
\definemath{\hflavGSCOval}{0.6595}%
\definemath{\hflavGSCOerr}{\pm0.0029}%
\definemath{\hflavDGSCOval}{+0.078}%
\definemath{\hflavDGSCOerr}{\pm0.004}%
\definemath{\hflavRHOGSDGSCO}{-0.16}%
\definemath{\hflavDGSGSCOval}{+0.118}%
\definemath{\hflavDGSGSCOerr}{\pm0.006}%
\definemath{\hflavTAUBSMEANCOval}{1.516}%
\definemath{\hflavTAUBSMEANCOerr}{\pm0.007}%
\definemath{\hflavTAUBSLCOval}{1.432}%
\definemath{\hflavTAUBSLCOerr}{\pm0.006}%
\definemath{\hflavTAUBSHCOval}{1.611}%
\definemath{\hflavTAUBSHCOerr}{\pm0.010}%
\definemath{\hflavGSCONval}{0.6595}%
\definemath{\hflavGSCONerr}{\pm0.0025}%
\definemath{\hflavDGSCONval}{+0.0781}%
\definemath{\hflavDGSCONerr}{\pm0.0035}%
\definemath{\hflavRHOGSDGSCON}{-0.09}%
\definemath{\hflavDGSGSCONval}{+0.119}%
\definemath{\hflavDGSGSCONerr}{\pm0.005}%
\definemath{\hflavTAUBSMEANCONval}{1.516}%
\definemath{\hflavTAUBSMEANCONerr}{\pm0.006}%
\definemath{\hflavTAUBSMEANCval}{1.516}%
\definemath{\hflavTAUBSMEANCerr}{\pm0.006}%
\definemath{\hflavTAUBSval}{1.516}%
\definemath{\hflavTAUBSerr}{\pm0.006}%
\definemath{\hflavTAUBSLCONval}{1.432}%
\definemath{\hflavTAUBSLCONerr}{\pm0.006}%
\definemath{\hflavTAUBSHCONval}{1.612}%
\definemath{\hflavTAUBSHCONerr}{\pm0.008}%
\definemath{\hflavDMSval}{17.766}%
\definemath{\hflavDMSerr}{\pm0.006}%
\definemath{\hflavDMSsta}{\pm0.004}%
\definemath{\hflavDMSsys}{\pm0.004}%
\definemath{\hflavXSval}{26.94}%
\definemath{\hflavXSerr}{\pm0.10}%
\definemath{\hflavCHISval}{0.499314}%
\definemath{\hflavCHISerr}{\pm0.000005}%
\definemath{\hflavRATIODGSDMSval}{0.0044}%
\definemath{\hflavRATIODGSDMSerr}{\pm0.0002}%
\definemath{\hflavRATIODMDDMSval}{0.02853}%
\definemath{\hflavRATIODMDDMSerr}{\pm0.00011}%
\definemath{\hflavXILQCDval}{1.206}%
\definemath{\hflavXILQCDerr}{\pm0.017}%
\definemath{\hflavVTDVTSLQCDval}{0.2054}%
\definemath{\hflavVTDVTSLQCDerr}{\pm0.0029}%
\definemath{\hflavVTDVTSLQCDexx}{\pm0.0004}%
\definemath{\hflavVTDVTSLQCDthe}{\pm0.0029}%
\definemath{\hflavXISRULval}{1.201}%
\definemath{\hflavXISRULerp}{^{+0.006}}%
\definemath{\hflavXISRULern}{_{-0.007}}%
\definemath{\hflavVTDVTSSRULval}{0.2045}%
\definemath{\hflavVTDVTSSRULerp}{^{+0.0011}}%
\definemath{\hflavVTDVTSSRULern}{_{-0.0013}}%
\definemath{\hflavVTDVTSSRULexx}{\pm0.0004}%
\definemath{\hflavVTDVTSSRULthp}{^{+0.0011}}%
\definemath{\hflavVTDVTSSRULthn}{_{-0.0012}}%
\definemath{\hflavTANPHIval}{-0.1}%
\definemath{\hflavTANPHIerr}{\pm0.6}%
\definemath{\hflavAZEROCCKKSF}{1.29}%
\definemath{\hflavAPERPCCKKSF}{3.64}%
\definemath{\hflavDPACCKKSF}{1.72}%
\definemath{\hflavDPECCKKSF}{2.30}%
\definemath{\hflavGSCCKKSF}{2.51}%
\definemath{\hflavDGSCCKKSF}{1.60}%
\definemath{\hflavPHISCCKKSF}{1.00}%
\definemath{\hflavAZEROSQBCCval}{0.5185}%
\definemath{\hflavAZEROSQBCCerr}{\pm0.0027}%
\definemath{\hflavAPERPSQBCCval}{0.259}%
\definemath{\hflavAPERPSQBCCerr}{\pm0.004}%
\definemath{\hflavASSQBCCval}{0.0069}%
\definemath{\hflavASSQBCCerr}{\pm0.0032}%
\definemath{\hflavASSQBCCLHCBval}{0.08}%
\definemath{\hflavASSQBCCLHCBerr}{\pm0.05}%
\definemath{\hflavASSQBCCATLASval}{0.035}%
\definemath{\hflavASSQBCCATLASerr}{\pm0.005}%
\definemath{\hflavDPABCCval}{3.21}%
\definemath{\hflavDPABCCerr}{\pm0.07}%
\definemath{\hflavDPEBCCval}{3.01}%
\definemath{\hflavDPEBCCerr}{\pm0.12}%
\definemath{\hflavSPERPBCCval}{0.45}%
\definemath{\hflavSPERPBCCerr}{\pm0.14}%
\definemath{\hflavGSBCCval}{0.6632}%
\definemath{\hflavGSBCCerr}{\pm0.0034}%
\definemath{\hflavDGSBCCval}{+0.076}%
\definemath{\hflavDGSBCCerr}{\pm0.004}%
\definemath{\hflavPHISBCCval}{-0.052}%
\definemath{\hflavPHISBCCerr}{\pm0.013}%
\definemath{\hflavLAMBBCCval}{0.996}%
\definemath{\hflavLAMBBCCerr}{\pm0.009}%
\definemath{\hflavDMSBCCval}{17.753}%
\definemath{\hflavDMSBCCerr}{\pm0.021}%
\definemath{\hflavAZEROBCCSF}{1.29}%
\definemath{\hflavAPERPBCCSF}{3.64}%
\definemath{\hflavDPABCCSF}{1.72}%
\definemath{\hflavDPEBCCSF}{2.30}%
\definemath{\hflavGSBCCSF}{2.51}%
\definemath{\hflavDGSBCCSF}{1.60}%
\definemath{\hflavPHISBCCSF}{1.00}%
\definemath{\hflavAZEROSQJPSIKKval}{0.5182}%
\definemath{\hflavAZEROSQJPSIKKerr}{\pm0.0026}%
\definemath{\hflavAPERPSQJPSIKKval}{0.259}%
\definemath{\hflavAPERPSQJPSIKKerr}{\pm0.004}%
\definemath{\hflavASSQJPSIKKval}{0.0069}%
\definemath{\hflavASSQJPSIKKerr}{\pm0.0032}%
\definemath{\hflavASSQJPSIKKLHCBval}{0.08}%
\definemath{\hflavASSQJPSIKKLHCBerr}{\pm0.05}%
\definemath{\hflavASSQJPSIKKATLASval}{0.035}%
\definemath{\hflavASSQJPSIKKATLASerr}{\pm0.005}%
\definemath{\hflavDPAJPSIKKval}{3.20}%
\definemath{\hflavDPAJPSIKKerr}{\pm0.07}%
\definemath{\hflavDPEJPSIKKval}{3.00}%
\definemath{\hflavDPEJPSIKKerr}{\pm0.12}%
\definemath{\hflavSPERPJPSIKKval}{0.45}%
\definemath{\hflavSPERPJPSIKKerr}{\pm0.14}%
\definemath{\hflavGSJPSIKKval}{0.6637}%
\definemath{\hflavGSJPSIKKerr}{\pm0.0033}%
\definemath{\hflavDGSJPSIKKval}{+0.074}%
\definemath{\hflavDGSJPSIKKerr}{\pm0.004}%
\definemath{\hflavPHISJPSIKKval}{-0.060}%
\definemath{\hflavPHISJPSIKKerr}{\pm0.014}%
\definemath{\hflavLAMBJPSIKKval}{0.996}%
\definemath{\hflavLAMBJPSIKKerr}{\pm0.009}%
\definemath{\hflavDMSJPSIKKval}{17.748}%
\definemath{\hflavDMSJPSIKKerr}{\pm0.022}%
\definemath{\hflavPHISLHCBval}{-0.033}%
\definemath{\hflavPHISLHCBerr}{\pm0.018}%
\definemath{\hflavDGSLHCBval}{+0.085}%
\definemath{\hflavDGSLHCBerr}{\pm0.004}%
\definemath{\hflavPHISCMSval}{-0.074}%
\definemath{\hflavPHISCMSerr}{\pm0.023}%
\definemath{\hflavDGSCMSval}{+0.078}%
\definemath{\hflavDGSCMSerr}{\pm0.004}%
\definemath{\hflavGSCMSval}{0.6640}%
\definemath{\hflavGSCMSerr}{\pm0.0028}%
\definemath{\hflavGSCCHHLHCBval}{0.6567}%
\definemath{\hflavGSCCHHLHCBerr}{\pm0.0019}%
\definemath{\hflavDGSCCHHLHCBval}{+0.085}%
\definemath{\hflavDGSCCHHLHCBerr}{\pm0.004}%
\definemath{\hflavAZEROJPSIKKSF}{1.23}%
\definemath{\hflavAPERPJPSIKKSF}{3.54}%
\definemath{\hflavDPAJPSIKKSF}{1.73}%
\definemath{\hflavDPEJPSIKKSF}{2.36}%
\definemath{\hflavGSJPSIKKSF}{2.42}%
\definemath{\hflavDGSJPSIKKSF}{1.51}%
\definemath{\hflavPHISJPSIKKSF}{1.00}%
\definemath{\hflavTAUBSJPSIPIPILHCBADJUSTEDval}{1.641}%
\definemath{\hflavTAUBSJPSIPIPILHCBADJUSTEDerr}{\pm0.015}%
\definemath{\hflavBETASBCCval}{+0.026}%
\definemath{\hflavBETASBCCerr}{\pm0.007}%
\definemath{\hflavPHISSMval}{-0.0376}%
\definemath{\hflavPHISSMerp}{^{+0.0006}}%
\definemath{\hflavPHISSMern}{_{-0.0005}}%
\definemath{\hflavPHISSMUTFITval}{-0.0367}%
\definemath{\hflavPHISSMUTFITerr}{\pm0.0010}%
\definemath{\hflavPHISTWELVESMval}{0.0046}%
\definemath{\hflavPHISTWELVESMerr}{\pm0.0012}%
\definemath{\hflavPHISTWELVEval}{-0.010}%
\definemath{\hflavPHISTWELVEerr}{\pm0.014}%
\definemath{\hflavTAUBSMMval}{1.79}%
\definemath{\hflavTAUBSMMerr}{\pm0.17}%
\definemath{\hflavTAULBval}{1.468}%
\definemath{\hflavTAULBsta}{\pm0.008}%
\definemath{\hflavTAULBsys}{\pm0.005}%
\definemath{\hflavTAULBerr}{\pm0.009}%
\definemath{\hflavTAUXBDval}{1.578}%
\definemath{\hflavTAUXBDsta}{\pm0.019}%
\definemath{\hflavTAUXBDsys}{\pm0.009}%
\definemath{\hflavTAUXBDerr}{\pm0.021}%
\definemath{\hflavTAUXBUval}{1.477}%
\definemath{\hflavTAUXBUsta}{\pm0.027}%
\definemath{\hflavTAUXBUsys}{\pm0.016}%
\definemath{\hflavTAUXBUerr}{\pm0.032}%
\definemath{\hflavRTAULBval}{0.969}%
\definemath{\hflavRTAULBerr}{\pm0.006}%
\definemath{\hflavRTAUXBUXBDval}{0.936}%
\definemath{\hflavRTAUXBUXBDerr}{\pm0.022}%
\definemath{\hflavTAUOBval}{1.64}%
\definemath{\hflavTAUOBsta}{\pm0.16}%
\definemath{\hflavTAUOBsys}{\pm0.03}%
\definemath{\hflavTAUOBerr}{\pm0.16}%
\definemath{\hflavRTAUOBval}{1.08}%
\definemath{\hflavRTAUOBerr}{\pm0.11}%
\definemath{\hflavTAUBDval}{1.517}%
\definemath{\hflavTAUBDerr}{\pm0.004}%
\definemath{\hflavCHIBARTEVval}{0.147}%
\definemath{\hflavCHIBARTEVerr}{\pm0.011}%
\definemath{\hflavCHIBARSFACTOR}{1.8}%
\definemath{\hflavCHIBARval}{0.1284}%
\definemath{\hflavCHIBARerr}{\pm0.0069}%
\definemath{\hflavSDGDGDval}{0.001}%
\definemath{\hflavSDGDGDerr}{\pm0.010}%
\definemath{\hflavCHIBARLEPval}{0.1259}%
\definemath{\hflavCHIBARLEPerr}{\pm0.0042}%
\definemath{\hflavNSIGMATAULBEXCLSEMI}{3.1}%
\definemath{\hflavTAUBUval}{1.638}%
\definemath{\hflavTAUBUerr}{\pm0.004}%
\definemath{\hflavRTAUBUval}{1.076}%
\definemath{\hflavRTAUBUerr}{\pm0.004}%
\definemath{\hflavTAULBSval}{1.247}%
\definemath{\hflavTAULBSerp}{^{+0.071}}%
\definemath{\hflavTAULBSern}{_{-0.069}}%
\definemath{\hflavTAULBEval}{1.471}%
\definemath{\hflavTAULBEerr}{\pm0.009}%
\definemath{\hflavTAUBCval}{0.510}%
\definemath{\hflavTAUBCerr}{\pm0.009}%
\definemath{\hflavNSIGMATAULBCDFTWO}{2.4}%
\definemath{\hflavQPDBval}{1.0009}%
\definemath{\hflavQPDBerr}{\pm0.0013}%
\definemath{\hflavQPDDval}{1.0000}%
\definemath{\hflavQPDDerr}{\pm0.0010}%
\definemath{\hflavQPDWval}{1.0005}%
\definemath{\hflavQPDWerr}{\pm0.0009}%
\definemath{\hflavQPDAval}{1.0005}%
\definemath{\hflavQPDAerr}{\pm0.0009}%
\definemath{\hflavASLDBval}{-0.0019}%
\definemath{\hflavASLDBerr}{\pm0.0027}%
\definemath{\hflavASLDDval}{+0.0001}%
\definemath{\hflavASLDDerr}{\pm0.0020}%
\definemath{\hflavASLDWval}{-0.0010}%
\definemath{\hflavASLDWerr}{\pm0.0018}%
\definemath{\hflavASLDAval}{-0.0010}%
\definemath{\hflavASLDAerr}{\pm0.0018}%
\definemath{\hflavREBDBval}{-0.0005}%
\definemath{\hflavREBDBerr}{\pm0.0007}%
\definemath{\hflavREBDDval}{+0.0000}%
\definemath{\hflavREBDDerr}{\pm0.0005}%
\definemath{\hflavREBDWval}{-0.0002}%
\definemath{\hflavREBDWerr}{\pm0.0004}%
\definemath{\hflavREBDAval}{-0.0002}%
\definemath{\hflavREBDAerr}{\pm0.0004}%
\definemath{\hflavASLLHCBDZERONSIGMA}{2.2}%
\definemath{\hflavASLLHCBDZEROPVALPERCENT}{3.1}%
\definemath{\hflavASLSval}{-0.0006}%
\definemath{\hflavASLSerr}{\pm0.0028}%
\definemath{\hflavQPSval}{1.0003}%
\definemath{\hflavQPSerr}{\pm0.0014}%
\definemath{\hflavASLDval}{-0.0021}%
\definemath{\hflavASLDerr}{\pm0.0017}%
\definemath{\hflavQPDval}{1.0010}%
\definemath{\hflavQPDerr}{\pm0.0008}%
\definemath{\hflavRHOASLSASLD}{-0.054}%
\definemath{\hflavCLPERCENTASLSASLD}{4.5}%
\definemath{\hflavREBDval}{-0.0005}%
\definemath{\hflavREBDerr}{\pm0.0004}%
\definemath{\hflavASLDNOMUval}{+0.0000}%
\definemath{\hflavASLDNOMUerr}{\pm0.0019}%
\definemath{\hflavASLSNOMUval}{+0.0016}%
\definemath{\hflavASLSNOMUerr}{\pm0.0030}%
\definemath{\hflavRHOASLSASLDNOMU}{+0.066}%
\definemath{\hflavASLDASLSNSIGMA}{0.5}%
\definemath{\hflavASLDASLSPVALPERCENT}{61.3}%

\newcommand{\unit}[1]{~\ensuremath{\rm #1}\xspace}
\renewcommand{\ps}{\unit{ps}}
\newcommand{\invps}{\unit{ps^{-1}}}

\newcommand{\MeVcc}{\unit{MeV/\mbox{$c$}^2}}

\definemath{\hflavCHIDW}{\hflavCHIDWval\hflavCHIDWerr}
\definemath{\hflavCHIDWU}{\hflavCHIDWUval\hflavCHIDWUerr}
\definemath{\hflavXDW}{\hflavXDWval\hflavXDWerr}
\definemath{\hflavXDWU}{\hflavXDWUval\hflavXDWUerr}
\definemath{\hflavCHIDU}{\hflavCHIDUval\hflavCHIDUerr}
\definemath{\hflavDMDW}{\hflavDMDWval\hflavDMDWerr\invps}
\definemath{\hflavDMDWnounit}{\hflavDMDWval\hflavDMDWerr}
\definemath{\hflavDMDWfull}{\hflavDMDWval\hflavDMDWsta\hflavDMDWsys\invps}
\definemath{\hflavDMDWnounitfull}{\hflavDMDWval\hflavDMDWsta\hflavDMDWsys}
\definemath{\hflavDMDWU}{\hflavDMDWUval\hflavDMDWUerr\invps}
\definemath{\hflavDMDWUnounit}{\hflavDMDWUval\hflavDMDWUerr}
\definemath{\hflavDMDWUfull}{\hflavDMDWUval\hflavDMDWUsta\hflavDMDWUsys\invps}
\definemath{\hflavDMDWUnounitfull}{\hflavDMDWUval\hflavDMDWUsta\hflavDMDWUsys}
\definemath{\hflavDMDLHCb}{\hflavDMDLHCbval\hflavDMDLHCberr\invps}
\definemath{\hflavDMDLHCbnounit}{\hflavDMDLHCbval\hflavDMDLHCberr}
\definemath{\hflavDMDLHCbfull}{\hflavDMDLHCbval\hflavDMDLHCbsta\hflavDMDLHCbsys\invps}
\definemath{\hflavDMDLHCbnounitfull}{\hflavDMDLHCbval\hflavDMDLHCbsta\hflavDMDLHCbsys}
\definemath{\hflavDMDL}{\hflavDMDLval\hflavDMDLerr\invps}
\definemath{\hflavDMDLnounit}{\hflavDMDLval\hflavDMDLerr}
\definemath{\hflavDMDLfull}{\hflavDMDLval\hflavDMDLsta\hflavDMDLsys\invps}
\definemath{\hflavDMDLnounitfull}{\hflavDMDLval\hflavDMDLsta\hflavDMDLsys}
\definemath{\hflavDMDT}{\hflavDMDTval\hflavDMDTerr\invps}
\definemath{\hflavDMDTnounit}{\hflavDMDTval\hflavDMDTerr}
\definemath{\hflavDMDTfull}{\hflavDMDTval\hflavDMDTsta\hflavDMDTsys\invps}
\definemath{\hflavDMDTnounitfull}{\hflavDMDTval\hflavDMDTsta\hflavDMDTsys}
\definemath{\hflavDMDB}{\hflavDMDBval\hflavDMDBerr\invps}
\definemath{\hflavDMDBnounit}{\hflavDMDBval\hflavDMDBerr}
\definemath{\hflavDMDBfull}{\hflavDMDBval\hflavDMDBsta\hflavDMDBsys\invps}
\definemath{\hflavDMDBnounitfull}{\hflavDMDBval\hflavDMDBsta\hflavDMDBsys}
\definemath{\hflavTAUBSJPSIKSHORT}{\hflavTAUBSJPSIKSHORTval\hflavTAUBSJPSIKSHORTerr\ps}
\definemath{\hflavTAUBSJPSIKSHORTnounit}{\hflavTAUBSJPSIKSHORTval\hflavTAUBSJPSIKSHORTerr}
\definemath{\hflavRTAUBS}{\hflavRTAUBSval\hflavRTAUBSerr}
\definemath{\hflavRTAUBSMEANC}{\hflavRTAUBSMEANCval\hflavRTAUBSMEANCerr}
\definemath{\hflavTAUBSSL}{\hflavTAUBSSLval\hflavTAUBSSLerr\ps}
\definemath{\hflavTAUBSSLnounit}{\hflavTAUBSSLval\hflavTAUBSSLerr}
\definemath{\hflavRTAUBSSL}{\hflavRTAUBSSLval\hflavRTAUBSSLerr}
\definemath{\hflavTAUBSJF}{\hflavTAUBSJFval\hflavTAUBSJFerr\ps}
\definemath{\hflavTAUBSJFnounit}{\hflavTAUBSJFval\hflavTAUBSJFerr}
\definemath{\hflavTAUBSJPSIPIPI}{\hflavTAUBSJPSIPIPIval\hflavTAUBSJPSIPIPIerr\ps}
\definemath{\hflavTAUBSJPSIPIPInounit}{\hflavTAUBSJPSIPIPIval\hflavTAUBSJPSIPIPIerr}
\definemath{\hflavTAUBSLONG}{\hflavTAUBSLONGval\hflavTAUBSLONGerr\ps}
\definemath{\hflavTAUBSLONGnounit}{\hflavTAUBSLONGval\hflavTAUBSLONGerr}
\definemath{\hflavTAUBSKK}{\hflavTAUBSKKval\hflavTAUBSKKerr\ps}
\definemath{\hflavTAUBSKKnounit}{\hflavTAUBSKKval\hflavTAUBSKKerr}
\definemath{\hflavTAUBSDSDS}{\hflavTAUBSDSDSval\hflavTAUBSDSDSerr\ps}
\definemath{\hflavTAUBSDSDSnounit}{\hflavTAUBSDSDSval\hflavTAUBSDSDSerr}
\definemath{\hflavTAUBSJPSIETA}{\hflavTAUBSJPSIETAval\hflavTAUBSJPSIETAerr\ps}
\definemath{\hflavTAUBSJPSIETAnounit}{\hflavTAUBSJPSIETAval\hflavTAUBSJPSIETAerr}
\definemath{\hflavTAUBSSHORT}{\hflavTAUBSSHORTval\hflavTAUBSSHORTerr\ps}
\definemath{\hflavTAUBSSHORTnounit}{\hflavTAUBSSHORTval\hflavTAUBSSHORTerr}
\definemath{\hflavGS}{\hflavGSval\hflavGSerr\invps}
\definemath{\hflavGSnounit}{\hflavGSval\hflavGSerr}
\definemath{\hflavDGS}{\hflavDGSval\hflavDGSerr\invps}
\definemath{\hflavDGSnounit}{\hflavDGSval\hflavDGSerr}
\definemath{\hflavDGSGS}{\hflavDGSGSval\hflavDGSGSerr}
\definemath{\hflavTAUBSMEAN}{\hflavTAUBSMEANval\hflavTAUBSMEANerr\ps}
\definemath{\hflavTAUBSMEANnounit}{\hflavTAUBSMEANval\hflavTAUBSMEANerr}
\definemath{\hflavTAUBSL}{\hflavTAUBSLval\hflavTAUBSLerr\ps}
\definemath{\hflavTAUBSLnounit}{\hflavTAUBSLval\hflavTAUBSLerr}
\definemath{\hflavTAUBSH}{\hflavTAUBSHval\hflavTAUBSHerr\ps}
\definemath{\hflavTAUBSHnounit}{\hflavTAUBSHval\hflavTAUBSHerr}
\definemath{\hflavGSCO}{\hflavGSCOval\hflavGSCOerr\invps}
\definemath{\hflavGSCOnounit}{\hflavGSCOval\hflavGSCOerr}
\definemath{\hflavDGSCO}{\hflavDGSCOval\hflavDGSCOerr\invps}
\definemath{\hflavDGSCOnounit}{\hflavDGSCOval\hflavDGSCOerr}
\definemath{\hflavDGSGSCO}{\hflavDGSGSCOval\hflavDGSGSCOerr}
\definemath{\hflavTAUBSMEANCO}{\hflavTAUBSMEANCOval\hflavTAUBSMEANCOerr\ps}
\definemath{\hflavTAUBSMEANCOnounit}{\hflavTAUBSMEANCOval\hflavTAUBSMEANCOerr}
\definemath{\hflavTAUBSLCO}{\hflavTAUBSLCOval\hflavTAUBSLCOerr\ps}
\definemath{\hflavTAUBSLCOnounit}{\hflavTAUBSLCOval\hflavTAUBSLCOerr}
\definemath{\hflavTAUBSHCO}{\hflavTAUBSHCOval\hflavTAUBSHCOerr\ps}
\definemath{\hflavTAUBSHCOnounit}{\hflavTAUBSHCOval\hflavTAUBSHCOerr}
\definemath{\hflavGSCON}{\hflavGSCONval\hflavGSCONerr\invps}
\definemath{\hflavGSCONnounit}{\hflavGSCONval\hflavGSCONerr}
\definemath{\hflavDGSCON}{\hflavDGSCONval\hflavDGSCONerr\invps}
\definemath{\hflavDGSCONnounit}{\hflavDGSCONval\hflavDGSCONerr}
\definemath{\hflavDGSGSCON}{\hflavDGSGSCONval\hflavDGSGSCONerr}
\definemath{\hflavTAUBSMEANCON}{\hflavTAUBSMEANCONval\hflavTAUBSMEANCONerr\ps}
\definemath{\hflavTAUBSMEANCONnounit}{\hflavTAUBSMEANCONval\hflavTAUBSMEANCONerr}
\definemath{\hflavTAUBSMEANC}{\hflavTAUBSMEANCval\hflavTAUBSMEANCerr\ps}
\definemath{\hflavTAUBSMEANCnounit}{\hflavTAUBSMEANCval\hflavTAUBSMEANCerr}
\definemath{\hflavTAUBS}{\hflavTAUBSval\hflavTAUBSerr\ps}
\definemath{\hflavTAUBSnounit}{\hflavTAUBSval\hflavTAUBSerr}
\definemath{\hflavTAUBSLCON}{\hflavTAUBSLCONval\hflavTAUBSLCONerr\ps}
\definemath{\hflavTAUBSLCONnounit}{\hflavTAUBSLCONval\hflavTAUBSLCONerr}
\definemath{\hflavTAUBSHCON}{\hflavTAUBSHCONval\hflavTAUBSHCONerr\ps}
\definemath{\hflavTAUBSHCONnounit}{\hflavTAUBSHCONval\hflavTAUBSHCONerr}
\definemath{\hflavDMS}{\hflavDMSval\hflavDMSerr\invps}
\definemath{\hflavDMSnounit}{\hflavDMSval\hflavDMSerr}
\definemath{\hflavDMSfull}{\hflavDMSval\hflavDMSsta\hflavDMSsys\invps}
\definemath{\hflavDMSnounitfull}{\hflavDMSval\hflavDMSsta\hflavDMSsys}
\definemath{\hflavXS}{\hflavXSval\hflavXSerr}
\definemath{\hflavCHIS}{\hflavCHISval\hflavCHISerr}
\definemath{\hflavRATIODGSDMS}{\hflavRATIODGSDMSval\hflavRATIODGSDMSerr}
\definemath{\hflavRATIODMDDMS}{\hflavRATIODMDDMSval\hflavRATIODMDDMSerr}
\definemath{\hflavXILQCD}{\hflavXILQCDval\hflavXILQCDerr}
\definemath{\hflavVTDVTSLQCD}{\hflavVTDVTSLQCDval\hflavVTDVTSLQCDerr}
\definemath{\hflavVTDVTSLQCDfull}{\hflavVTDVTSLQCDval\hflavVTDVTSLQCDexx\hflavVTDVTSLQCDthe}
\definemath{\hflavXISRULerr}{\hflavXISRULerp\hflavXISRULern}
\definemath{\hflavXISRUL}{\hflavXISRULval\hflavXISRULerr}
\definemath{\hflavVTDVTSSRULerr}{\hflavVTDVTSSRULerp\hflavVTDVTSSRULern}
\definemath{\hflavVTDVTSSRULthe}{\hflavVTDVTSSRULthp\hflavVTDVTSSRULthn}
\definemath{\hflavVTDVTSSRUL}{\hflavVTDVTSSRULval\hflavVTDVTSSRULerr}
\definemath{\hflavVTDVTSSRULfull}{\hflavVTDVTSSRULval\hflavVTDVTSSRULexx\hflavVTDVTSSRULthe}
\definemath{\hflavTANPHI}{\hflavTANPHIval\hflavTANPHIerr}
\definemath{\hflavAZEROSQBCC}{\hflavAZEROSQBCCval\hflavAZEROSQBCCerr}
\definemath{\hflavAPERPSQBCC}{\hflavAPERPSQBCCval\hflavAPERPSQBCCerr}
\definemath{\hflavASSQBCC}{\hflavASSQBCCval\hflavASSQBCCerr}
\definemath{\hflavASSQBCCLHCB}{\hflavASSQBCCLHCBval\hflavASSQBCCLHCBerr}
\definemath{\hflavASSQBCCATLAS}{\hflavASSQBCCATLASval\hflavASSQBCCATLASerr}
\definemath{\hflavDPABCC}{\hflavDPABCCval\hflavDPABCCerr}
\definemath{\hflavDPEBCC}{\hflavDPEBCCval\hflavDPEBCCerr}
\definemath{\hflavSPERPBCC}{\hflavSPERPBCCval\hflavSPERPBCCerr}
\definemath{\hflavGSBCC}{\hflavGSBCCval\hflavGSBCCerr\invps}
\definemath{\hflavGSBCCnounit}{\hflavGSBCCval\hflavGSBCCerr}
\definemath{\hflavDGSBCC}{\hflavDGSBCCval\hflavDGSBCCerr\invps}
\definemath{\hflavDGSBCCnounit}{\hflavDGSBCCval\hflavDGSBCCerr}
\definemath{\hflavPHISBCC}{\hflavPHISBCCval\hflavPHISBCCerr}
\definemath{\hflavLAMBBCC}{\hflavLAMBBCCval\hflavLAMBBCCerr}
\definemath{\hflavDMSBCC}{\hflavDMSBCCval\hflavDMSBCCerr\invps}
\definemath{\hflavDMSBCCnounit}{\hflavDMSBCCval\hflavDMSBCCerr}
\definemath{\hflavAZEROSQJPSIKK}{\hflavAZEROSQJPSIKKval\hflavAZEROSQJPSIKKerr}
\definemath{\hflavAPERPSQJPSIKK}{\hflavAPERPSQJPSIKKval\hflavAPERPSQJPSIKKerr}
\definemath{\hflavASSQJPSIKK}{\hflavASSQJPSIKKval\hflavASSQJPSIKKerr}
\definemath{\hflavASSQJPSIKKLHCB}{\hflavASSQJPSIKKLHCBval\hflavASSQJPSIKKLHCBerr}
\definemath{\hflavASSQJPSIKKATLAS}{\hflavASSQJPSIKKATLASval\hflavASSQJPSIKKATLASerr}
\definemath{\hflavDPAJPSIKK}{\hflavDPAJPSIKKval\hflavDPAJPSIKKerr}
\definemath{\hflavDPEJPSIKK}{\hflavDPEJPSIKKval\hflavDPEJPSIKKerr}
\definemath{\hflavSPERPJPSIKK}{\hflavSPERPJPSIKKval\hflavSPERPJPSIKKerr}
\definemath{\hflavGSJPSIKK}{\hflavGSJPSIKKval\hflavGSJPSIKKerr\invps}
\definemath{\hflavGSJPSIKKnounit}{\hflavGSJPSIKKval\hflavGSJPSIKKerr}
\definemath{\hflavDGSJPSIKK}{\hflavDGSJPSIKKval\hflavDGSJPSIKKerr\invps}
\definemath{\hflavDGSJPSIKKnounit}{\hflavDGSJPSIKKval\hflavDGSJPSIKKerr}
\definemath{\hflavPHISJPSIKK}{\hflavPHISJPSIKKval\hflavPHISJPSIKKerr}
\definemath{\hflavLAMBJPSIKK}{\hflavLAMBJPSIKKval\hflavLAMBJPSIKKerr}
\definemath{\hflavDMSJPSIKK}{\hflavDMSJPSIKKval\hflavDMSJPSIKKerr\invps}
\definemath{\hflavDMSJPSIKKnounit}{\hflavDMSJPSIKKval\hflavDMSJPSIKKerr}
\definemath{\hflavPHISLHCB}{\hflavPHISLHCBval\hflavPHISLHCBerr}
\definemath{\hflavDGSLHCB}{\hflavDGSLHCBval\hflavDGSLHCBerr\invps}
\definemath{\hflavDGSLHCBnounit}{\hflavDGSLHCBval\hflavDGSLHCBerr}
\definemath{\hflavPHISCMS}{\hflavPHISCMSval\hflavPHISCMSerr}
\definemath{\hflavDGSCMS}{\hflavDGSCMSval\hflavDGSCMSerr\invps}
\definemath{\hflavDGSCMSnounit}{\hflavDGSCMSval\hflavDGSCMSerr}
\definemath{\hflavGSCMS}{\hflavGSCMSval\hflavGSCMSerr\invps}
\definemath{\hflavGSCMSnounit}{\hflavGSCMSval\hflavGSCMSerr}
\definemath{\hflavGSCCHHLHCB}{\hflavGSCCHHLHCBval\hflavGSCCHHLHCBerr\invps}
\definemath{\hflavGSCCHHLHCBnounit}{\hflavGSCCHHLHCBval\hflavGSCCHHLHCBerr}
\definemath{\hflavDGSCCHHLHCB}{\hflavDGSCCHHLHCBval\hflavDGSCCHHLHCBerr\invps}
\definemath{\hflavDGSCCHHLHCBnounit}{\hflavDGSCCHHLHCBval\hflavDGSCCHHLHCBerr}
\definemath{\hflavTAUBSJPSIPIPILHCBADJUSTED}{\hflavTAUBSJPSIPIPILHCBADJUSTEDval\hflavTAUBSJPSIPIPILHCBADJUSTEDerr\ps}
\definemath{\hflavTAUBSJPSIPIPILHCBADJUSTEDnounit}{\hflavTAUBSJPSIPIPILHCBADJUSTEDval\hflavTAUBSJPSIPIPILHCBADJUSTEDerr}
\definemath{\hflavBETASBCC}{\hflavBETASBCCval\hflavBETASBCCerr}
\definemath{\hflavPHISSMerr}{\hflavPHISSMerp\hflavPHISSMern}
\definemath{\hflavPHISSM}{\hflavPHISSMval\hflavPHISSMerr}
\definemath{\hflavPHISSMUTFIT}{\hflavPHISSMUTFITval\hflavPHISSMUTFITerr}
\definemath{\hflavPHISTWELVESM}{\hflavPHISTWELVESMval\hflavPHISTWELVESMerr}
\definemath{\hflavPHISTWELVE}{\hflavPHISTWELVEval\hflavPHISTWELVEerr}
\definemath{\hflavTAUBSMM}{\hflavTAUBSMMval\hflavTAUBSMMerr\ps}
\definemath{\hflavTAUBSMMnounit}{\hflavTAUBSMMval\hflavTAUBSMMerr}
\definemath{\hflavTAULB}{\hflavTAULBval\hflavTAULBerr\ps}
\definemath{\hflavTAULBnounit}{\hflavTAULBval\hflavTAULBerr}
\definemath{\hflavTAULBfull}{\hflavTAULBval\hflavTAULBsta\hflavTAULBsys\ps}
\definemath{\hflavTAULBnounitfull}{\hflavTAULBval\hflavTAULBsta\hflavTAULBsys}
\definemath{\hflavTAUXBD}{\hflavTAUXBDval\hflavTAUXBDerr\ps}
\definemath{\hflavTAUXBDnounit}{\hflavTAUXBDval\hflavTAUXBDerr}
\definemath{\hflavTAUXBDfull}{\hflavTAUXBDval\hflavTAUXBDsta\hflavTAUXBDsys\ps}
\definemath{\hflavTAUXBDnounitfull}{\hflavTAUXBDval\hflavTAUXBDsta\hflavTAUXBDsys}
\definemath{\hflavTAUXBU}{\hflavTAUXBUval\hflavTAUXBUerr\ps}
\definemath{\hflavTAUXBUnounit}{\hflavTAUXBUval\hflavTAUXBUerr}
\definemath{\hflavTAUXBUfull}{\hflavTAUXBUval\hflavTAUXBUsta\hflavTAUXBUsys\ps}
\definemath{\hflavTAUXBUnounitfull}{\hflavTAUXBUval\hflavTAUXBUsta\hflavTAUXBUsys}
\definemath{\hflavRTAULB}{\hflavRTAULBval\hflavRTAULBerr}
\definemath{\hflavRTAUXBUXBD}{\hflavRTAUXBUXBDval\hflavRTAUXBUXBDerr}
\definemath{\hflavTAUOB}{\hflavTAUOBval\hflavTAUOBerr\ps}
\definemath{\hflavTAUOBnounit}{\hflavTAUOBval\hflavTAUOBerr}
\definemath{\hflavTAUOBfull}{\hflavTAUOBval\hflavTAUOBsta\hflavTAUOBsys\ps}
\definemath{\hflavTAUOBnounitfull}{\hflavTAUOBval\hflavTAUOBsta\hflavTAUOBsys}
\definemath{\hflavRTAUOB}{\hflavRTAUOBval\hflavRTAUOBerr}
\definemath{\hflavTAUBD}{\hflavTAUBDval\hflavTAUBDerr\ps}
\definemath{\hflavTAUBDnounit}{\hflavTAUBDval\hflavTAUBDerr}
\definemath{\hflavCHIBARTEV}{\hflavCHIBARTEVval\hflavCHIBARTEVerr}
\definemath{\hflavCHIBAR}{\hflavCHIBARval\hflavCHIBARerr}
\definemath{\hflavSDGDGD}{\hflavSDGDGDval\hflavSDGDGDerr}
\definemath{\hflavCHIBARLEP}{\hflavCHIBARLEPval\hflavCHIBARLEPerr}
\definemath{\hflavTAUBU}{\hflavTAUBUval\hflavTAUBUerr\ps}
\definemath{\hflavTAUBUnounit}{\hflavTAUBUval\hflavTAUBUerr}
\definemath{\hflavRTAUBU}{\hflavRTAUBUval\hflavRTAUBUerr}
\definemath{\hflavTAULBSerr}{\hflavTAULBSerp\hflavTAULBSern}
\definemath{\hflavTAULBS}{\hflavTAULBSval\hflavTAULBSerr\ps}
\definemath{\hflavTAULBSnounit}{\hflavTAULBSval\hflavTAULBSerr}
\definemath{\hflavTAULBE}{\hflavTAULBEval\hflavTAULBEerr\ps}
\definemath{\hflavTAULBEnounit}{\hflavTAULBEval\hflavTAULBEerr}
\definemath{\hflavTAUBC}{\hflavTAUBCval\hflavTAUBCerr\ps}
\definemath{\hflavTAUBCnounit}{\hflavTAUBCval\hflavTAUBCerr}
\definemath{\hflavQPDB}{\hflavQPDBval\hflavQPDBerr}
\definemath{\hflavQPDD}{\hflavQPDDval\hflavQPDDerr}
\definemath{\hflavQPDW}{\hflavQPDWval\hflavQPDWerr}
\definemath{\hflavQPDA}{\hflavQPDAval\hflavQPDAerr}
\definemath{\hflavASLDB}{\hflavASLDBval\hflavASLDBerr}
\definemath{\hflavASLDD}{\hflavASLDDval\hflavASLDDerr}
\definemath{\hflavASLDW}{\hflavASLDWval\hflavASLDWerr}
\definemath{\hflavASLDA}{\hflavASLDAval\hflavASLDAerr}
\definemath{\hflavREBDB}{\hflavREBDBval\hflavREBDBerr}
\definemath{\hflavREBDD}{\hflavREBDDval\hflavREBDDerr}
\definemath{\hflavREBDW}{\hflavREBDWval\hflavREBDWerr}
\definemath{\hflavREBDA}{\hflavREBDAval\hflavREBDAerr}
\definemath{\hflavASLS}{\hflavASLSval\hflavASLSerr}
\definemath{\hflavQPS}{\hflavQPSval\hflavQPSerr}
\definemath{\hflavASLD}{\hflavASLDval\hflavASLDerr}
\definemath{\hflavQPD}{\hflavQPDval\hflavQPDerr}
\definemath{\hflavREBD}{\hflavREBDval\hflavREBDerr}
\definemath{\hflavASLDNOMU}{\hflavASLDNOMUval\hflavASLDNOMUerr}
\definemath{\hflavASLSNOMU}{\hflavASLSNOMUval\hflavASLSNOMUerr}

\newcommand{\comment}[1]{}

\newcommand{\dmd}{\ensuremath{\Delta m_{\particle{d}}}\xspace}
\newcommand{\dms}{\ensuremath{\Delta m_{\particle{s}}}\xspace}
\newcommand{\xd}{\ensuremath{x_{\particle{d}}}\xspace}
\newcommand{\xs}{\ensuremath{x_{\particle{s}}}\xspace}
\newcommand{\yd}{\ensuremath{y_{\particle{d}}}\xspace}
\newcommand{\ys}{\ensuremath{y_{\particle{s}}}\xspace}

\newcommand{\chid}{\ensuremath{\chi_{\particle{d}}}\xspace}
\newcommand{\chis}{\ensuremath{\chi_{\particle{s}}}\xspace}
\newcommand{\Gd}{\ensuremath{\Gamma_{\particle{d}}}\xspace}
\newcommand{\DGd}{\ensuremath{\Delta\Gd}\xspace}
\newcommand{\DGGd}{\ensuremath{\DGd/\Gd}\xspace}
\newcommand{\Gs}{\ensuremath{\Gamma_{\particle{s}}}\xspace}
\newcommand{\DGs}{\ensuremath{\Delta\Gs}\xspace}

\newcommand{\ASLd}{\ensuremath{{\cal A}_{\rm SL}^\particle{d}}\xspace}
\newcommand{\ASLs}{\ensuremath{{\cal A}_{\rm SL}^\particle{s}}\xspace}
\newcommand{\ASLb}{\ensuremath{{\cal A}_{\rm SL}^\particle{b}}\xspace}

\newcommand{\DG}{\ensuremath{\Delta\Gamma}\xspace}
\newcommand{\phiccbars}{\ensuremath{\phi_s^{c\bar{c}s}}\xspace}

\newcommand{\CL}[1]{#1\%~\mbox{C.L.}}
\newcommand{\Qjet}{\ensuremath{Q_{\rm jet}}\xspace}

\newcommand{\labe}[1]{\label{equ:#1}}
\newcommand{\labs}[1]{\label{sec:#1}}
\newcommand{\labf}[1]{\label{fig:#1}}
\newcommand{\labt}[1]{\label{tab:#1}}
\newcommand{\refe}[1]{\ref{equ:#1}}
\newcommand{\refs}[1]{\ref{sec:#1}}
\newcommand{\reff}[1]{\ref{fig:#1}}
\newcommand{\reft}[1]{\ref{tab:#1}}
\newcommand{\Refx}[1]{Ref.~\cite{#1}}
\newcommand{\Refs}[1]{Refs.~\cite{#1}}

\newcommand{\Eq}[1]{Eq.~(\refe{#1})}

\newcommand{\Eqss}[2]{Eqs.~(\refe{#1}) and (\refe{#2})}

\newcommand{\Eqsss}[3]{Eqs.~(\refe{#1}), (\refe{#2}), and (\refe{#3})}
\newcommand{\Figure}[1]{Figure~\reff{#1}}

\newcommand{\Fig}[1]{Fig.~\reff{#1}}

\newcommand{\Sec}[1]{Sec.~\refs{#1}}

\newcommand{\Secss}[2]{Secs.~\refs{#1} and \refs{#2}}

\newcommand{\Table}[1]{Table~\reft{#1}}

\newcommand{\Tablesss}[3]{Tables~\reft{#1}, \reft{#2}, and \reft{#3}}

\newcommand{\subsubsubsection}[1]{\vspace{2ex}\par\noindent {\bf\boldmath\em #1} \vspace{2ex}\par}

\input{tau/tau-header} %

\renewcommand{\mysection}[1]{\section[#1]{#1}} %

\newif\ifref
\newif\ifhtml
\reftrue
\usepackage{threeparttable} %

\makeatletter
\patchcmd{\@sect}{#8}{\boldmath #8}{}{}
\let\ori@chapter\@chapter
\def\@chapter[#1]#2{\ori@chapter[\boldmath#1]{\boldmath#2}}
\makeatother

\newboolean{articletitles}
\setboolean{articletitles}{true} %

\newboolean{uprightparticles}
\setboolean{uprightparticles}{false} %

\begin{document}

\setcounter{page}{1}
\thispagestyle{empty}
\renewcommand\Affilfont{\itshape\small}

\title{
  Averages of $b$-hadron, $c$-hadron, and $\tau$-lepton properties
  as of 2023
\vskip0.20in
\large{\it Heavy Flavor Averaging Group (HFLAV):}
Sw.~Banerjee\,\orcidlink{0000-0001-8852-2409}, %
E.~Ben-Haim\,\orcidlink{0000-0002-9510-8414}, %
F.~Bernlochner\,\orcidlink{0000-0001-8153-2719}, %
E.~Bertholet\,\orcidlink{0000-0002-3792-2450},
M.~Bona\,\orcidlink{0000-0002-9660-580X},
A.~Bozek\,\orcidlink{0000-0002-5915-1319}, %
C.~Bozzi\,\orcidlink{0000-0001-6782-3982}, %
J.~Brodzicka\,\orcidlink{0000-0002-8556-0597}, %
V.~Chobanova\,\orcidlink{0000-0002-1353-6002}, %
M.~Chrzaszcz\,\orcidlink{0000-0001-7901-8710}, %
U.~Egede\,\orcidlink{0000-0001-5493-0762}, %
M.~Gersabeck\,\orcidlink{0000-0002-0075-8669}, %
P.~Goldenzweig\,\orcidlink{0000-0001-8785-847X},
N.~Gharbi\,\orcidlink{0009-0007-2038-4864}, %
L.~Grillo\,\orcidlink{0000-0001-5360-0091}, %
K.~Hayasaka\,\orcidlink{0000-0002-6347-433X}, %
T.~Humair\,\orcidlink{0000-0002-2922-9779},
D.~Johnson\,\orcidlink{0000-0003-3272-6001}, %
T.~Kuhr\,\orcidlink{0000-0001-6251-8049}, %
O.~Leroy\,\orcidlink{0000-0002-2589-240X},
A.~Lusiani\,\orcidlink{0000-0002-6876-3288}, %
H.-L.~Ma\,\orcidlink{0000-0001-9771-2802}, %
M.~Margoni\,\orcidlink{0000-0003-1797-4330},
R.~Mizuk\,\orcidlink{0000-0002-2209-6969}, %
P.~Naik\,\orcidlink{0000-0001-6977-2971}, %
T.~Nanut~Petri\v{c}\,\orcidlink{0000-0002-5728-9867}, %
A.~Pereiro Castro\,\orcidlink{0000-0001-9721-3325}, %
M.~Prim\,\orcidlink{0000-0002-1407-7450}, %
M.~Roney\,\orcidlink{0000-0001-7802-4617}, %
M.~Rotondo\,\orcidlink{0000-0001-5704-6163}, %
O.~Schneider\,\orcidlink{0000-0002-6014-7552},
C.~Schwanda\,\orcidlink{0000-0003-4844-5028}, %
A.~J.~Schwartz\,\orcidlink{0000-0002-7310-1983}, %
J.~Serrano\,\orcidlink{0000-0003-2489-7812},
B.~Shwartz\,\orcidlink{0000-0002-1456-1496}, %
A.~Soffer\,\orcidlink{0000-0002-0749-2146}, %
M.~Whitehead\,\orcidlink{0000-0002-2142-3673} %
and
J.~Yelton\,\orcidlink{0000-0001-8840-3346} %
}
\author[ ]{}
\date{\today} %
\maketitle

\begin{abstract}
\noindent
This paper reports world averages of measurements of $b$-hadron, $c$-hadron,
and $\tau$-lepton properties obtained by the Heavy Flavour Averaging Group using results available before October 2023. In rare cases, significant results obtained several months later are also used.
For the averaging, common input parameters used in the various analyses are adjusted (rescaled) to common values, and known correlations are taken into account.
The averages include branching fractions, lifetimes, neutral meson mixing
parameters, \CP~violation parameters, parameters of semileptonic decays, and
Cabibbo-Kobayashi-Maskawa matrix elements.
\end{abstract}

\newpage
\tableofcontents
\newpage

\section{Executive Summary}
\label{sec:summary}

This paper provides updated world averages of measurements of $b$-hadron, $c$-hadron, and $\tau$-lepton properties using results available by September 2023. 
In a few cases, important results that appeared later are included and are clearly labeled as such. 
In addition, older results, including those from previous generations of experiments, are still very important and contribute to the averages that we report. 
Significant additional results are expected in the near future.

Since our previous paper~\cite{HFLAV:2022esi}, 
the \b-hadron lifetime and mixing averages have been improved both in precision and in the averaging procedures.
In total, new \Bz results from one publication (from Belle II), new \Bs results from eight publications
(one from ATLAS, three from CMS, and four from LHCb) and new \Xibd results from one publication (from LHCb)
have been incorporated into these averages.

The lifetime hierarchy for the most abundant weakly-decaying \b-hadron species
is well established, with a precision below 10~fs
for the meson and \Lb-baryon lifetimes, 
and is compatible with the expectations from the Heavy Quark Expansion.
However, small sample sizes still limit the precision for the lifetimes of the heavier, observed weakly decaying \b baryons, namely, \Xibd, \Xibu, and $\Omega_b$.
A sizable value of the decay width difference in the $\Bs$--$\Bsb$ system
is measured with a relative precision of 4\% and is well predicted by the
Standard Model~(SM). In contrast,
the experimental results for the decay width difference in the
$\Bd$--$\Bzb$ system are not yet precise enough to distinguish
the small (expected) value from zero.
The mass differences in both the $\Bd$--$\Bzb$ and $\Bs$--$\Bsb$ systems are accurately known, to a few per mil or tenths of per mil, respectively. On the other hand, \CP violation in the mixing of either system has not been observed yet. 
Although measured with uncertainties at the per mil level, the \CP asymmetries are
still consistent with both zero and their SM predictions.
A similar conclusion holds for the \CP violation induced
by \Bs mixing in the $b\to c\bar{c}s$ transition.
Many measurements are still dominated by statistical uncertainties, which will improve once new results from the LHC become available, especially from Run~3, and from \belletwo.

In exclusive semileptonic $b$~hadron decays, determinations of the CKM elements \vcb\ and \vub\ are based on the decays $B\to D^{(*)}\ell\nu$, $B_s\to D^{(*)}_s\mu\nu$, $B\to\pi\ell\nu$, $B_s\to K\mu\nu$ and $\Lambda_b\to p\mu\nu$. 
A global fit to all exclusive results yields $\vcb=(39.77\pm 0.46)\times 10^{-3}$ and $\vub=(3.43\pm 0.12)\times 10^{-3}$. The tension with the determinations from inclusive $B$~meson decays is larger than $3\sigma$ for both \vcb\ and \vub. New measurements of $R(D^*)$ and $R(D)$, characterising semitauonic decays $B\to D^{(*)}\tau\nu_\tau$, have been added to the averages since our previous paper. With respect to the most recent theory calculations for $R(D^*)$ and $R(D)$, the combined tension with the SM expectation persists, at $3.14\sigma$.

We have compiled 717 measurements involving $b$-hadron decays into states with open or hidden charm from 305 papers by the \babar, Belle, CDF, D0, LHCb, CMS, and ATLAS collaborations into 503 averages.
Following the averaging procedures established in our previous paper, correlations among averages are taken into account.
The large samples of $b$ hadrons available in current experiments allow precise measurements.
In addition to improvements in precision for branching fractions, new decay modes of $B$ mesons and $b$ baryons have been discovered.

Among charmless $b$-hadron decays, an important update from LHCb is tests of lepton flavor universality in $b\to s \ell^+\ell^-$ transitions. 
Previously seen anomalies in the
measurements of the branching-fraction ratios $R_K = {\cal B}(\Bp \to K^+\mu^+\mu^-)/{\cal B}(\Bp \to K^+e^+e^-)$ and $R_{K^*}={\cal B}(\Bz \to \Kstarz\mu^+\mu^-)/(\Bz \to \Kstarz e^+e^-)$ have
vanished, as updated measurements based on 9~fb$^{-1}$ are compatible with SM prediction at the level of one standard deviation.
In contrast, previously seen tensions in the differential branchings fraction and angular observables of $b\to s\ell^+\ell^-$ decays persist.
In addition, a handful of new limits on lepton-flavor-violating decays have been set by LHCb and Belle, especially those involving $B^{0,\pm}$ and \Bs meson decays into a $\tau$ lepton.
Among the \CP violating observable in rare decays, the ``$K\pi$ \CP puzzle'' persists, and important new results have appeared in two- and three-body decays.
LHCb has produced many other results on a wide variety of decays, including those of $b$-baryons, for example, a first measurement of the photon polarization in $\Lambda_b\to\Lambda\gamma$ decays.

In the area of mixing and \CP\ violation in the charm sector, two highlights are the LHCb observation of dispersive mixing, 
i.e., a non-zero value for the mixing parameter 
$x\equiv \Delta M/\overline{\Gamma}$, and a high precision
measurement from LHCb of the observable $y^{}_{CP}$. 
A total of 63 measurements from 
LHCb, Belle, \babar, BES\,III, CLEO-c, CDF, FOCUS, and 
Fermilab E791 of 18 observables are input into
a global fit for 8-10 (depending on theoretical assumptions) 
mixing and \CP\ violation parameters. From this fit, the no-mixing hypothesis is excluded at a confidence level above $11\sigma$. 
The value $x\!=\!0$ is excluded with a statistical significance 
of $9.1\sigma$.
For the mixing parameter $y\equiv \Delta\Gamma/2\overline{\Gamma}$,
the value $y\!=\!0$ is excluded with a significance of $35\sigma$, 
and the precision on $y$ is improved by a factor of two 
from that of our previously published
result~\cite{HFLAV:2022esi}. The world average value 
for $y$ is positive, indicating that the \CP-even 
state has a shorter lifetime than the \CP-odd state;
this is similar to the $\Kz$--$\Kzb$ system. 
However, $x>0$ and thus the \CP-even 
state is heavier, in contrast to the $\Kz$--$\Kzb$ system.
The \CP\ violation parameters $|q/p|$ and $\phi$ are compatible with
\CP\ symmetry at the level of~$2.1\sigma$. Thus, there is no evidence
yet for {\it indirect\/} \CP\ violation arising from mixing
($|q/p|\neq 1$) or from a phase difference between the mixing 
amplitude and a direct decay amplitude ($\phi\neq 0$).
Measurements of the time-integrated \CP\ asymmetries in 
$D^0\ra K^+K^-/\pi^+\pi^-$ decays give $\Delta a^{\rm dir}_{\CP}=(-0.159\pm0.029)\%$. This value is similar to our previously
published result~\cite{HFLAV:2022esi}, which established 
direct \CP\ violation in singly Cabibbo-suppressed decays.
For $D^0\ra \pi^+\pi^-$ decays, although the precision 
of the measured \CP\ asymmetry has improved by more than 
a factor of two,
the impact on an $SU(3)$ sum rule for $D\ra \pi\pi$ decays 
is small. The sum rule remains compatible with zero 
at a precision of~0.3\%.

The most precise measurements of $|V^{}_{cd}|$ and $|V^{}_{cs}|$
are obtained from leptonic $D^+\ra \mu^+\nu$ and $D^+_s\ra\mu^+\nu/\tau^+\nu$
decays, respectively.
These measurements have theoretical uncertainties that arise from calculations of the decay constants.
However, thanks to improvements in lattice QCD calculations, the theory uncertainties are below ${\sim}20\%$ of the experimental uncertainties.
Averages of the branching fractions for the  decays $D^0 \to K^- \pi^+$, $D^0 \to K^+ K^-$, and $D^0 \to \pi^+ \pi^-$ must treat final
state radiation correctly and consistently across measurements for
the accuracy of the averages to match the measurement precision.
We update the measurements to account for improvements in
the simulation of final state radiation (e.g. interference
between photons radiated from the two charged particles).

Our average of the \mtau mass includes the recent measurements from the Belle~II and KEDR collaborations. In the \mtau branching
fraction fit, the recent Belle~II measurement of the ratio of the branching fractions for $\tau^-\to \mu^- \bar\nu_\mu \nu_\tau$ and $\tau^-\to e^- \bar\nu_e \nu_\tau$ has been added.
The newly added measurements slightly improve the precision of the lepton universality tests.
Furthermore, the modeling of the branching fraction
$\dxuse{Gamma805-td}$ has been improved, with a nuisance fit parameter for the branching fraction $\dxuse{Gamma980-td}$. 
Upper limits on lepton flavor violating \mtau decays are presented.
These upper limits are combined for channels in which searches from multiple experimental efforts have similar sensitivities. 

A small selection of highlights of the results described in
Sections~\ref{sec:life_mix}--\ref{sec:tau} are given in Tables~\ref{tab_summary_1}--\ref{tab_summary_3}.

\begin{table}
\caption{
  Selected world averages.
  Where two uncertainties are given the first is statistical and the second is systematic.
} %
\label{tab_summary_1}
\begin{tabular}{l c}
{\bf\boldmath \b-hadron lifetimes} &   \\
~~$\tau(\Bd)$         & \hflavTAUBD \\
~~$\tau(\Bu)$         & \hflavTAUBU \\
~~$\tau(\Bs) = 1/\Gs$  & \hflavTAUBSMEANCON \\
~~$\tau(B^0_{s\rm L})$ & \hflavTAUBSLCON \\
~~$\tau(B^0_{s\rm H})$  & \hflavTAUBSHCON \\
~~$\tau(\Bc)$         & \hflavTAUBC \\
~~$\tau(\Lb)$         & \hflavTAULB \\
~~$\tau(\Xibd)$       & \hflavTAUXBD  \\
~~$\tau(\Xibu)$       & \hflavTAUXBU  \\
~~$\tau(\Omegab)$     & \hflavTAUOB   \\
{\bf\boldmath \Bd\ and \Bs\ mixing \& \CP violation} &  \\
~~\dmd &  \hflavDMDWU \\
~~\DGGd  & \hflavSDGDGD \\
~~$|q_{\particle{d}}/p_{\particle{d}}|$ & \hflavQPD  \\
~~\dms  &  \hflavDMS \\
~~\DGs & \hflavDGSCON \\
~~$|q_{\particle{s}}/p_{\particle{s}}|$ & \hflavQPS   \\
~~\phiccbars  & \hflavPHISBCC rad\\
\end{tabular}
\end{table}

\begin{table}
\caption{
  Selected world averages.
  Where two uncertainties are given the first is statistical and the second is systematic.
} %
\label{tab_summary_2}
\begin{tabular}{lc}
{\bf\boldmath Semileptonic $b$-hadron decay parameters} & \\
 ~~${\cal B}(\Bzb\to D^{*+}\ell^-\nub_\ell)$ & $(4.90\pm 0.12)\%$\\
 ~~${\cal B}(\B^-\to D^{*0}\ell^-\nub_\ell)$ & $(5.53\pm 0.22)\%$\\
 ~~${\cal B}(\Bzb\to D^+\ell^-\nub_\ell)$ & $(2.12\pm 0.06)\%$\\
 ~~${\cal B}(\B^-\to D^0\ell^-\nub_\ell)$ & $(2.21\pm 0.06)\%$\\
 ~~${\cal B}(\Bzb \to\pi^+\ell^-\nub_\ell)$ & $(1.50\pm 0.05)\times 10^{-4}$\\
 ~~$\vcb$ from exclusive $B$, $B_s$ and $\Lambda_b$~decays & $(39.77\pm 0.46)\times 10^{-3}$\\
 ~~$\vub$ from exclusive $B$, $B_s$ and $\Lambda_b$~decays & $(3.43\pm 0.12)\times 10^{-3}$\\
\end{tabular}
\end{table}

\begin{table}
\caption{
  Selected world averages.
  Where two uncertainties are given the first is statistical and the second is systematic.
} %
\label{tab_summary_3}
\begin{tabular}{lc}
 {\bf\boldmath $D^0$ mixing and \CP violation } &   \\
 ~~$x$ &  $(0.41\,\pm 0.04)\%$  \\
 ~~$y$ &  $(0.65\,\pm 0.02)\%$  \\
~~$\delta^{}_{K\pi}$ &  $(11.4\,^{+3.5}_{-3.8})^\circ$  \\
 ~~$A^{}_D$ &  $(-0.77\,\pm 0.35)\%$  \\
 ~~$|q/p|$ & $0.994\,\pm 0.016$  \\
 ~~$\phi$ &  $(-2.6\,\pm 1.2)^\circ$  \\
 ~~$x^{}_{12}$ (no direct \CP violation in CF/DCS decays) 
                &  $(0.41\,\pm 0.04)\%$  \\
\end{tabular}
\end{table}

\clearpage

\mysection{Introduction}
\label{sec:intro}

Flavour dynamics plays an important role in elementary particle interactions. 
The accurate knowledge of properties of heavy flavour
hadrons, especially $b$ hadrons, plays an essential role in determination of the elements of the Cabibbo-Kobayashi-Maskawa (CKM)
quark-mixing matrix~\cite{Cabibbo:1963yz,Kobayashi:1973fv}. 
The operation of the \belle\ and \babar\ $e^+e^-$ $B$ factory 
experiments led to a large increase in the size of available 
$B$-meson, $D$-hadron and $\tau$-lepton samples, 
enabling dramatic improvement in the accuracies of related measurements.
The CDF and \dzero\ experiments at the Fermilab Tevatron 
have also provided important results in heavy flavour physics,
most notably in the $B^0_s$ sector.
In the $D$-meson sector, the dedicated $e^+e^-$ charm factory experiments
CLEO-c and BESIII have made significant contributions.
Run~I and Run~II of the CERN Large Hadron Collider delivered high luminosity, 
enabling the collection of even larger samples of $b$ and $c$ hadrons, and
thus a further leap in precision in many areas, at the ATLAS, CMS, and
(especially) LHCb experiments.  
With ongoing analyses of the LHC Run~II data, further improvements are  anticipated.

The Heavy Flavour Averaging Group (HFLAV)\footnote{
  The group was originally known by the acronym ``HFAG.''  
  Following feedback from the community, this was changed to HFLAV in 2017.
} 
was formed in 2002 to 
continue the activities of the LEP Heavy Flavour Steering 
Group~\cite{Abbaneo:2000ej_mod,*Abbaneo:2001bv_mod_cont}, which was responsible for calculating averages of measurements of $b$-flavour related quantities. 
HFLAV has evolved since its inception and currently consists of seven subgroups:
\begin{itemize}
\item the ``$B$ Lifetime and Oscillations'' subgroup provides 
averages for $b$-hadron lifetimes and various 
parameters governing $\Bz$--$\Bzb$ and $\Bs$--$\Bsb$ mixing and \CP violation;

\item the ``Unitarity Triangle Angles'' subgroup provides
averages for parameters associated with time-dependent $\CP$ 
asymmetries and $B \to DK$ decays, and resulting determinations 
of the angles of the CKM unitarity triangle;\footnote{Results from the Unitarity Triangle Angles group have not been updated for this iteration. See Ref.~\cite{HFLAV:2022esi} for the latest update.}

\item the ``Semileptonic $B$ Decays'' subgroup provides averages
for inclusive and exclusive measurements of $B$-decay branching fractions, and
subsequent determinations of the CKM matrix element magnitudes
$|V_{cb}|$ and $|V_{ub}|$;

\item the ``$B$ to Charm Decays'' subgroup provides averages of 
branching fractions for $b$-hadron decays to final states involving open 
charm or charmonium mesons, as well as branching fractions for $b$-hadron production in $\Upsilon(4S)$ and $\Upsilon(5S)$ decays;

\item the ``Rare $b$ Decays'' subgroup provides averages of branching 
fractions, $\CP$ asymmetries and other observables for charmless, radiative, 
leptonic, and baryonic $B$-meson and \b-baryon decays;

\item the ``Charm CP Violation and Oscillations'' subgroup provides averages of mixing, $\CP$-, and $T$-violation parameters in the $\Dz$--$\Dzb$ system;

\item the ``Charm Decays'' subgroup provides averages of charm-hadron branching fractions, properties of excited $D^{**}$ and $D^{}_{sJ}$ mesons, 
properties of charm baryons,
and the $D^+$ and $D^+_s$ decay constants $f^{}_{D}$ and~$f^{}_{D_s}$;

\item the ``Tau Physics'' subgroup provides averages for \mtau
  branching fractions using a global fit,  elaborates on the results
  to test lepton universality and to determine the CKM matrix element
  magnitude $|V_{us}|$, and lists and combines branching-fraction upper limits for \mtau lepton-flavour-violating decays.

\end{itemize}
Subgroups consist of representatives from experiments producing 
relevant results in that area, \ie, representatives from
\babar, \belle, \belletwo, BESIII,  LHCb, ATLAS, and CMS.

This article is an update of the last HFLAV publication, which used results available by March 2021~\cite{HFLAV:2022esi}. 
Here we report world averages using results available by September 2023.
In some cases, important new results made available later are included where possible.
In general, we use all publicly available results, including preliminary results that are supported by
written documentation, such as conference proceedings or publicly available reports from the collaborations.
However, we do not use preliminary results that remain unpublished 
for an extended period of time, or for which no publication is planned. 
Since HFLAV members are also members of the different collaborations, we exploit our close contact with analyzers to ensure that the
results are prepared in a form suitable for combinations.  

Section~\ref{sec:method} describes the methodology used for calculating
averages. In the averaging procedure, common input parameters used in 
the various analyses are adjusted (rescaled) to common values, and, 
where possible, known correlations are taken into account. 
Sections~\ref{sec:life_mix}--\ref{sec:tau} present world 
average values from each of the subgroups listed above. 
A complete listing of the averages and plots,
including updates since this document was prepared,
is available on the HFLAV web site~\cite{ref:hflavpage}.

\clearpage
\section{Averaging methodology} 
\label{sec:method} 

The main task of HFLAV is to combine independent but possibly
correlated measurements of a parameter to obtain the world's 
best estimate of that parameter's value and uncertainty. These
measurements are typically made by different experiments, or by the
same experiment using different data sets, or by the same
experiment using the same data but with different analysis methods.
In this section, the general approach adopted by HFLAV is outlined.
The software used to provide this is either the \textsc{COMBOS} package~\cite{Combos:1999}, the \textsc{HFLAVaveraging} package~\cite{HFLAVaveraging} or dedicated tools for some averages.

Our methodology focuses on the problem of combining measurements 
obtained with different assumptions about external (or ``nuisance'') 
parameters and with potentially correlated systematic uncertainties. Unless otherwise noted, we assume for our combinations that the quantities measured by experiments were performed in the asymptotic regime (large data samples), so that the measured estimates have a (one- or multi-dimensional) Gaussian likelihood function. We use $\boldsymbol{x}$ to represent a set of $n$ parameters and $\boldsymbol{x}_i$ to denote the $i$th set of measurements of those parameters. The covariance matrix for the measurement is $\boldsymbol{V}_{\!i}$.
In all fits, we ensure that $\boldsymbol{x}$ and $\boldsymbol{x}_i$ do not 
contain redundant information, \ie, they are vectors with $n$ 
elements that represents exactly $n$ parameters.
A $\chi^2$ statistic is constructed as
\begin{equation}
  \chi^2(\boldsymbol{x}) = \sum_i^N 
  \left( \boldsymbol{x}_i - \boldsymbol{x} \right)^{\rm T} 
  \boldsymbol{V}_{\!i}^{-1}  
  \left( \boldsymbol{x}_i - \boldsymbol{x} \right) \, ,
  \label{eq:averageEq}
\end{equation}
where the sum is over the $N$ independent determinations of the quantities
$\boldsymbol{x}$, typically coming from different experiments. This is the best linear unbiased estimator with minimum variance~\cite{aitken1936}
The results of the average are the central values $\boldsymbol{\hat{x}}$, 
which are the values of $\boldsymbol{x}$ at the minimum of
$\chi^2(\boldsymbol{x})$, and their covariance matrix
\begin{equation}
  \boldsymbol{\hat{V}}^{-1} = \sum_i^N \boldsymbol{V}_{\!i}^{-1},
\end{equation}
which is a generalisation of the one-dimensional estimate $\sigma^{-2} = \sum_i \sigma_i^{-2}$.

The value of $\chi^2(\boldsymbol{\hat{x}})$ provides a measure of the
consistency of the independent measurements of $\boldsymbol{x}$ after
accounting for the number of degrees of freedom ($\dof$), which is the 
difference $N - n$ between the number of measurements and the number of
fitted parameters.
The values of $\chi^2(\boldsymbol{\hat{x}})$ and $\dof$ are typically 
converted to a $p$-value and reported together with the 
averages. Unlike the Particle Data Group~\cite{PDG_2020}, when $\chi^2/\dof > 1$ 
we do not by default scale the resulting uncertainty.
Rather, we examine the systematic uncertainties of each measurement to better understand potential sources of the discrepancy.

In many cases, publications do not quote a direct measurement of a parameter of interest, but of a quantity that is a function of multiple parameters.
An example is the measurement of a ratio of branching fractions, from which a branching fraction of interest is determined using previous (and usually more precise) knowledge of the branching fraction of a "normalization mode". This leads to a correlation between the determinations of the two branching fractions that appear in the ratio.
In addition, if the same normalization mode is used for measurements of different branching fraction ratios, %
they too
become correlated. These correlations can be evaluated by performing a simultaneous fit to all averages involved. This is done by generalising Eq.~\ref{eq:averageEq} to the form
\begin{equation}
  \chi^2(\boldsymbol{p}) = \sum_i^N 
  \left( \boldsymbol{f}_i(\boldsymbol{p}) - \boldsymbol{x_i} \right)^{\rm T} 
  \boldsymbol{V}_{\!i}^{-1}  
  \left( \boldsymbol{f}_i(\boldsymbol{p}) - \boldsymbol{x_i} \right) \, ,
  \label{eq:averageEqDerived}
\end{equation}
where $\boldsymbol{p}$ are the fit parameters, including the quantities whose averages we want to determine,  $\boldsymbol{x}_i$ is the set of $i$th measurements (e.g., of branching fractions and branching-fraction ratios), and $\boldsymbol{f}_i$ is the dependence of the measured quantities $\boldsymbol{x}_i$ on the parameters $\boldsymbol{p}$.
This procedure is used for branching-fraction and related averages in Sections~\ref{sec:b2c} and ~\ref{sec:rare}.
An alternative approach, used in Section~\ref{sec:tau:br-fit}, is to construct the $\chi^2$ as in Eq.~(\ref{eq:averageEq}) and minimize it subject to a list of constraints implemented with Lagrange multipliers.
The two approaches are essentially identical, except that the covariance matrix is given in terms of $\boldsymbol{p}$ in the former and in terms of $\boldsymbol{x}_i$ in the latter.

If a special treatment is necessary in order to calculate an average, or 
if an approximation used in the calculation might not be sufficiently
accurate (\eg, assuming Gaussian uncertainties when the likelihood function 
exhibits non-Gaussian behavior), we point this out. 
Further modifications to the averaging procedures for non-Gaussian
situations are discussed in Sec.~\ref{sec:method:nonGaussian}.

\subsection{Treatment of correlated systematic uncertainties}
\label{sec:method:corrSysts} 

Consider two hypothetical measurements of a parameter $x$, which can
be summarized as
\begin{align*}
 & x_1 \pm \delta x_1 \pm \Delta x_{1,1} \pm \Delta x_{1,2} \ldots \\
 & x_2 \pm \delta x_2 \pm \Delta x_{2,1} \pm \Delta x_{2,2} \ldots \, ,
\end{align*}
where the $\delta x_k$ are statistical uncertainties and
the $\Delta x_{k,i}$ are contributions to the systematic
uncertainty. The simplest approach is to combine statistical 
and systematic uncertainties in quadrature
\begin{align*}
 & x_1 \pm \left(\delta x_1 \oplus \Delta x_{1,1} \oplus \Delta x_{1,2} \oplus \ldots\right) \\
 & x_2 \pm \left(\delta x_2 \oplus \Delta x_{2,1} \oplus \Delta x_{2,2} \oplus \ldots\right) \, ,
\end{align*}
and then perform a weighted average of $x_1$ and $x_2$ using their
combined uncertainties, treating the measurements as independent. This 
approach suffers from two potential problems that we try to address. 
First, the values $x_k$ may have been obtained using different
assumptions for nuisance parameters; \eg, different values of the \Bz
lifetime may have been used for different measurements of the
oscillation frequency $\deltamd$. The second potential problem 
is that some systematic uncertainties may be correlated
between measurements. For example, different measurements of 
$\deltamd$ may depend on the same branching fraction 
used to model a common background.

The above two problems are related. We can represent the systematic uncertainties as a set of nuisance parameters $y_i$
upon which $x_k$ depends. The uncertainty $\Delta y_i$, which is the uncertainty on $y_i$ coming from external measurements, contributes $\Delta x_{k,i}$ to the
systematic uncertainty on $x_i$. 
We thus use the values of $y_i$ and
$\Delta y_i$ assumed by each measurement in our averaging. 
To properly treat correlated systematic uncertainties among measurements,
requires decomposing the overall systematic uncertainties into correlated
and uncorrelated components. Correlated systematic uncertainties are those that depend on a shared nuisance  parameter, \eg a lifetime as mentioned above; uncorrelated systematic uncertainties do not share a nuisance parameter, \eg the statistical uncertainty resulting from independent limited size simulations of background components.
As different measurements often quote different types of systematic
uncertainties, achieving consistent definitions in order to properly 
treat correlations
requires close coordination between HFLAV and the experiments. 
In some cases, a group of
systematic uncertainties must be combined into a coarser
description in order to obtain an average that is consistent 
among measurements. Systematic uncertainties
that are uncorrelated with any other source of uncertainty are 
combined together with the statistical uncertainty, so that the only
systematic uncertainties treated explicitly are those that are
correlated with at least one other measurement via a consistently-defined
external parameter $y_i$.

\begin{figure}[!tb]
\centering
\includegraphics[width=3.0in, trim={0 0 4.0in 0},clip]{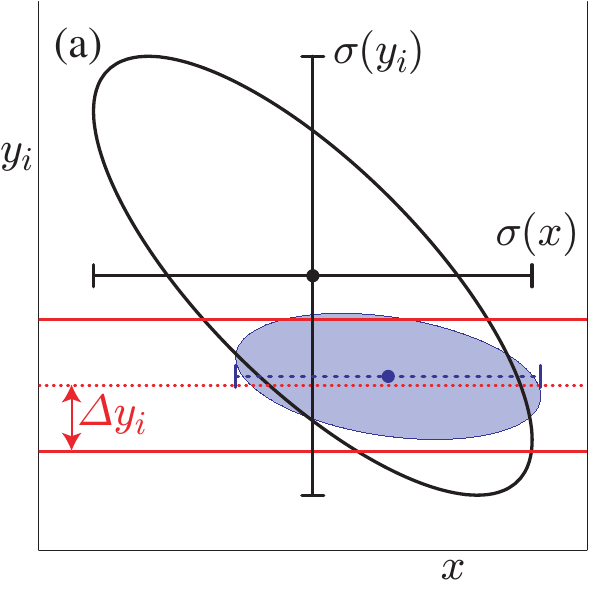}
\caption{
  Illustration of the possible dependence of a measured quantity $x$ on a
  nuisance parameter $y_i$.
  The plot compares the 68\% confidence level contours of a
  hypothetical measurement's unconstrained (large ellipse) and
  constrained (filled ellipse) likelihoods, using the Gaussian
  constraint on $y_i$ represented by the horizontal band. 
  The solid error bars represent the statistical uncertainties $\sigma(x)$ and
  $\sigma(y_i)$ of the unconstrained likelihood. 
  The dashed error bar shows the statistical uncertainty on $x$ from a
  constrained simultaneous fit to $x$ and $y_i$. 
}
\label{fig:singlefit}
\end{figure}

The fact that a measurement of $x$ is sensitive to $y_i$
indicates that, in principle, the data used to measure $x$ could
also be used for a simultaneous measurement of $x$ and $y_i$. This
is illustrated by the large contour in Fig.~\ref{fig:singlefit}(a).
However, there often exists an external measurement of $y_i$ with uncertainty $\Delta y_i$ (represented by the horizontal band in
Fig.~\ref{fig:singlefit}(a)) that is more precise than the constraint
$\sigma(y_i)$ from the $x$ data alone. In this case, the results presented in a publication can be from 
a simultaneous fit to $x$ and $y_i$, including the external 
measurement as a constraint, and obtain the filled $(x,y)$ contour and dashed 
one-dimensional estimate of $x$ shown in Fig.~\ref{fig:singlefit}(a). We call the fit without the external measurement \emph{unconstrained}, and the fit that include the external measurement is referred to as \emph{constrained}.

\begin{figure}[!tb]
\centering
\includegraphics[width=5.0in]{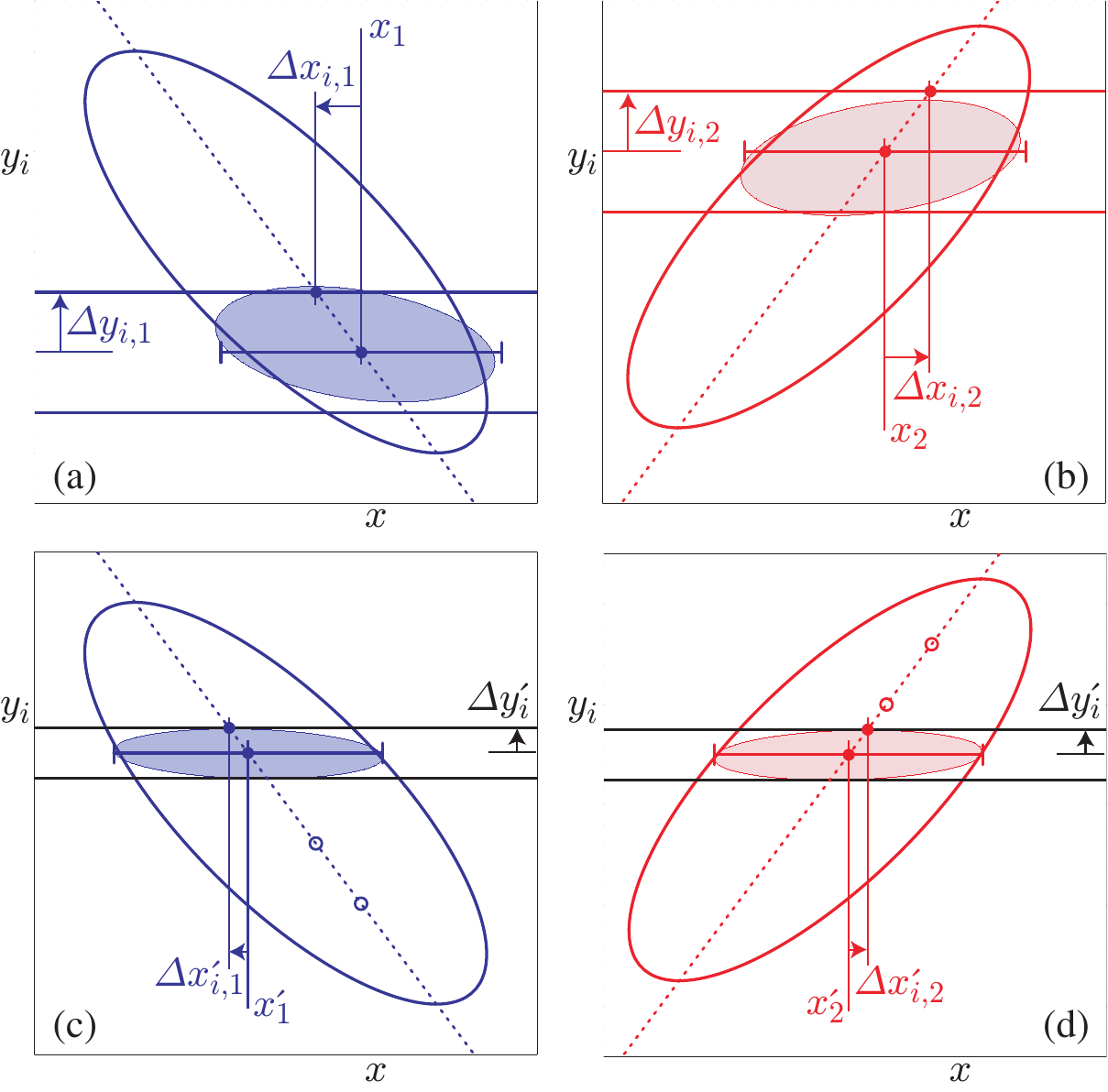}
\caption{
  Illustration of the HFLAV combination procedure for correlated systematic uncertainties.
  Upper plots (a) and (b) show examples of two individual measurements to be
  combined. 
  The large (filled) ellipses represent their unconstrained (constrained)
  iso-likelihood contours, while horizontal bands indicate the different assumptions about
  the value and uncertainty of $y_i$ used by each measurement. 
  The error bars show the results of the method described in the text for
  obtaining $x$ by performing fits with $y_i$ fixed to different values. 
  Lower plots (c) and (d) illustrate the adjustments to accommodate updated
  and consistent knowledge of $y_i$. 
  Open circles mark the central values of the unadjusted fits to $x$ with $y$
  fixed; these determine the dashed line used to obtain the adjusted values. 
}
\label{fig:multifit}
\end{figure}

To combine two or more measurements that share a systematic
uncertainty due to the same external parameter(s) $y_i$, the optimal solution is to take the unconstrained results from the publications and 
perform a constrained simultaneous fit of all measurements 
to obtain values of $x$ and $y_i$. 
Let us consider two statistically-independent measurements, 
$x_1 \pm (\delta x_1 \oplus \Delta x_{1,i})$ and 
$x_2\pm(\delta x_2\oplus \Delta x_{2,i})$, of 
the quantity $x$ as shown in Figs.~\ref{fig:multifit}(a,b).
For simplicity we consider only one correlated systematic 
uncertainty for each external parameter $y_i$.
Since the publications were made, our knowledge of $y_i$ will often have improved, causing 
the measurements of $x$ to shift to different central
values and have different uncertainties.

If the 
unconstrained likelihoods ${\cal L}_k(x,y_1,y_2,\ldots)$ for each of the measurements are
available, the exact method is to minimize the simultaneous likelihood
\begin{equation}
{\cal L}_{\text{comb}}(x,y_1,y_2,\ldots) \equiv \prod_k\,{\cal
  L}_k(x,y_1,y_2,\ldots)\,\prod_{i}\,{\cal 
  L}_i(y_i) \; ,
\end{equation}
with an independent Gaussian constraint
\begin{equation}
{\cal L}_i(y_i) = \exp\left[-\frac{1}{2}\,\left(\frac{y_i-y_i'}{\Delta
 y_i'}\right)^2\right]
\end{equation}
for each $y_i$.

However, most publications do not include the full likelihood, in which case we use an approximate method instead. The first step of our procedure is 
to adjust the values of each measurement to reflect the current 
best knowledge of the external parameters $y_i'$ and their
ranges $\Delta y_i'$, as illustrated in Figs.~\ref{fig:multifit}(c,d). 
We adjust the central values $x_k$ and correlated systematic uncertainties
$\Delta x_{k,i}$ linearly for each measurement (indexed by $k$) and each
external parameter (indexed by $i$):
\begin{align}
x_k' &= x_k + \sum_i\,\frac{\Delta x_{k,i}}{\Delta y_{k,i}}\left(y_i'-y_{k,i}\right)  \label{eq:shiftx} \\
\Delta x_{k,i}'&= \Delta x_{k,i} \frac{\Delta y_i'}{\Delta y_{k,i}} \label{eq:shiftDx} \, .
\end{align}
This procedure is exact in the limit that the unconstrained
likelihood of each measurement is Gaussian and the linear relationships in Eqs.~(\ref{eq:shiftx})and~(\ref{eq:shiftDx}) are valid.

\begin{figure}[!tb]
\centering
\includegraphics[width=3.0in]{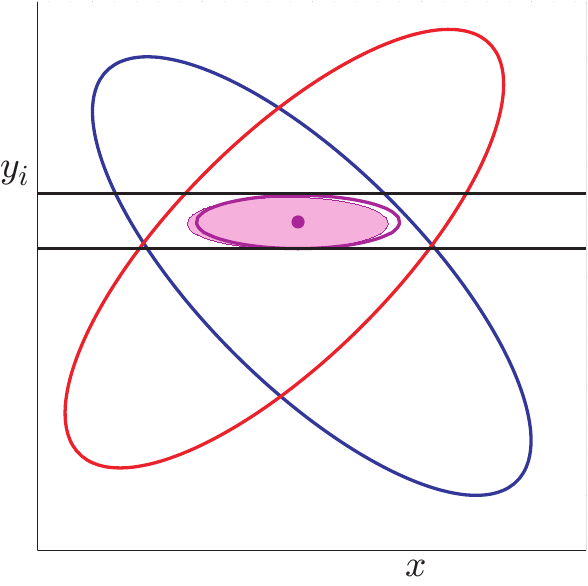}
\caption{
 Illustration of the combination of two hypothetical measurements of $x$
 using the method described in the text. 
 The ellipses represent the unconstrained likelihoods of each measurement,
 and the horizontal band represents the latest knowledge about $y_i$ that is
 used to adjust the individual measurements.
 The filled small ellipse shows the result of the exact method using 
 ${\cal L}_{\text{comb}}$, and the hollow small ellipse and dot show the
 result of the approximate method using $\chi^2_{\text{comb}}$. 
}
\label{fig:fit12}
\end{figure}

The second step is to combine the adjusted
measurements, $x_k'\pm (\delta x_k\oplus \Delta x_{k,1}'\oplus \Delta
x_{k,2}'\oplus\ldots)$ by constructing the goodness-of-fit statistic
\begin{equation}
\chi^2_{\text{comb}}(x,y_1,y_2,\ldots) \equiv \sum_k\,
\frac{1}{\delta x_k^2}\left[
x_k' - \left(x + \sum_i\,(y_i-y_i')\frac{\Delta x_{k,i}'}{\Delta y_i'}\right)
\right]^2 + \sum_i\,
\left(\frac{y_i - y_i'}{\Delta y_i'}\right)^2 \; .
\end{equation}
We minimize this $\chi^2$ to obtain the best values of $x$ and
$y_i$ and their uncertainties, as shown in Fig.~\ref{fig:fit12}. 
Although this method determines new values for
the $y_i$, we typically do not report them.
Since the $\Delta x_{i,k}$ reported
by each experiment are generally not intended for this purpose (for
example, they may represent a conservative upper limit rather than a
true reflection of a 68\% confidence level).

The results of the approximate method agree with the exact method 
when the ${\cal L}_k$ are 
Gaussian, $\Delta y_i' \ll \sigma(y_i)$ and the linear assumption for the approximate method is valid.

For averages where common sources of systematic uncertainty are important,
central values and uncertainties are rescaled to a common set of input 
parameters following the prescription above.
We use the most up-to-date values for common inputs, taking values  for experimental constraints from within HFLAV or from the Particle Data Group
when possible, and updated values of theoretical parameters from their publications.

\subsection{Treatment of unknown correlations}
\label{sec:method:unknownCorrelations} 

Another issue that needs careful treatment is that of unknown correlations
among measurements, \eg, due to use of the same decay model for 
intermediate states to calculate acceptances.
A common practice is to set the correlation
coefficient to unity to indicate full correlation. However, this is
not necessarily conservative and can result in 
an underestimated uncertainty on the average.  
The most conservative choice of correlation coefficient
between two measurements $i$ and $j$
is that which maximizes the uncertainty on $\hat{x}$
due to the pair of measurements,
\begin{equation}
\sigma_{\hat{x}(i,j)}^2 = \frac{\sigma_i^2\,\sigma_j^2\,(1-\rho_{ij}^2)}
   {\sigma_i^2 + \sigma_j^2 - 2\,\rho_{ij}\,\sigma_i\,\sigma_j} \, ,
\label{eq:correlij}
\end{equation}
with
\begin{equation}
\rho_{ij} =
\mathrm{min}\left(\frac{\sigma_i}{\sigma_j},\frac{\sigma_j}{\sigma_i}\right)
\, .
\label{eq:correlrho}
\end{equation}
This corresponds to setting 
$\sigma_{\hat{x}(i,j)}^2=\mathrm{min}(\sigma_i^2,\sigma_j^2)$.
Setting $\rho_{ij}=1$ when $\sigma_i\ne\sigma_j$ can lead to a significant
underestimate of the uncertainty on $\hat{x}$, as can be seen
from Eq.~(\ref{eq:correlij}). In the absence of better information on the correlation, we always use Eq.~(\ref{eq:correlij}).

\subsection{Treatment of asymmetric uncertainties}
\label{sec:method:nonGaussian} 

For measurements with no correlation between them and with Gaussian uncertainties, the usual estimator for the
average of a set of measurements is obtained by minimizing
\begin{equation}
  \chi^2(x) = \sum_k^N \frac{\left(x_k-x\right)^2}{\sigma^2_k} \, ,
\label{eq:chi2t}
\end{equation}
where $x_k$ is the $k$-th measured value of $x$ and $\sigma_k^2$ is the
variance of the distribution from which $x_k$ was drawn.  
The value $\hat{x}$ at minimum $\chi^2$ is the estimate for the parameter $x$.
The true $\sigma_k$ are unknown but typically the uncertainty as assigned by the
experiment $\sigma_k^{\rm raw}$ is used as an estimator for it.
However, caution is advised when $\sigma_k^{\rm raw}$
depends on the measured value $x_k$. 
Examples of this are multiplicative systematic uncertainties such as those
due to acceptance, or the $\sqrt{N}$
dependence of Poisson statistics for which $x_k \propto N$
and $\sigma_k \propto \sqrt{N}$.
Failing to account for this type of dependence when averaging leads to a
biased average. Such biases can be minimized
\begin{equation}
  \chi^2(x) = \sum_k^N \frac{\left(x_k-x\right)^2}{\sigma^2_k(\hat{x})} \,,
\label{eq:chi2that}
\end{equation}
where $\sigma_k(\hat{x})$ is the uncertainty on $x_k$ that includes 
the dependence of the uncertainty on the value measured.  As an example, 
consider the uncertainty due to detector acceptance, for which
$\sigma_k(\hat{x}) = (\hat{x} / x_k)\times\sigma_k^{\rm raw}$.
Inserting this into Eq.~(\ref{eq:chi2that}) leads to the solution
$$ 
\hat{x} = \frac{\sum_k^N x_k^3/(\sigma_k^{\rm raw})^2}
{\sum_k^N x_k^2/(\sigma_k^{\rm raw})^2} \, ,
$$
which is the correct behavior, \ie, every measurement is weighted by the inverse 
square of the fractional uncertainty $\sigma_k^{\rm raw}/x_k$.
When it is not possible to assess the dependence of $\sigma_k^{\rm raw}$ 
on $\hat{x}$ from the uncertainties quoted by the experiments, this dependence is ignored.  

Another example of a non-Gaussian likelihood function is when a measurement is given with asymmetric uncertainties. In general we symmetrize them by taking their linear average, however for branching fractions and asymmetries, we take asymmetric uncertainties into account through the use of Eq.~\ref{eq:averageEq} with a variable value for  
the $k$th diagonal element $V^{kk}$ of the covariance matrix for the measurement (dropping the measurement index $i$ for simplicity). We take $V^{kk} = (\sigma^k_{-})^2$ for $f^k(\boldsymbol{p}) - x^k < -\sigma_{k-}$ and  $V^{kk} = (\sigma^k_{+})^2$ for $f^k(\boldsymbol{p}) - x^k > \sigma^k_{+}$, where $\sigma^k_{-}$ ($\sigma^k_{+}$) are the left- (right-side) uncertainty quoted on the measurement of $x^k$, and $f^k$ is the $k$th element of $\boldsymbol{f}$. Between 
these regions,  $V^{kk}$ is interpolated linearly. While this will not fully recover the likelihood, it is the optimal solution when no further information is provided~\cite{Barlow:2004wg}.

\subsection{Splitting uncertainty for an average into components}
\label{sec:method:split} 

We carefully consider the various uncertainties
contributing to the overall uncertainty of an average. The 
covariance matrix describing the uncertainties of different 
measurements and their correlations is constructed, \ie,
$\boldsymbol{V} = 
\boldsymbol{V}_{\rm stat} + \boldsymbol{V}_{\rm sys} + 
\boldsymbol{V}_{\rm theory}$.
If the measurements are from independent data samples, then
$\boldsymbol{V}_{\rm stat}$ is diagonal, but
$\boldsymbol{V}_{\rm sys}$ and $\boldsymbol{V}_{\rm theory}$ 
may contain correlations.
The variance on the average $\hat{x}$ can be written as
\begin{eqnarray}
\sigma^2_{\hat{x}} 
 & = & 
\frac{1}{\sum_{i,j} \boldsymbol{V}^{-1}_{ij}} \ = \ 
\frac{ \sum_{i,j}
  \left(\boldsymbol{V}^{-1}\, 
    \boldsymbol{V} \, \boldsymbol{V}^{-1}\right)_{ij} }
{\left(\sum_{i,j} \boldsymbol{V}^{-1}_{ij}\right)^2} \\
& = & 
\frac{ \sum_{i,j}
  \left(\boldsymbol{V}^{-1}\, 
    \left[ \boldsymbol{V}_{\rm stat}+ \boldsymbol{V}_{\rm sys}+
      \boldsymbol{V}_{\rm theory} \right] \, \boldsymbol{V}^{-1}\right)_{ij} }
{\left(\sum_{i,j} \boldsymbol{V}^{-1}_{ij}\right)^2} 
\ = \ 
\sigma^2_{\text{stat}} + \sigma^2_{\text{sys}} + \sigma^2_{\text{th}} \, .
\end{eqnarray}
To calculate $\sigma^2_{\text{stat}}$ in the last step, the calculation is repeated without including $\boldsymbol{V}_{\rm stat}$ in $\boldsymbol{V}$ and this is then subtracted from the total. The same is done for the other two components.
This breakdown of uncertainties is provided in certain cases,
but usually only a single, total uncertainty is quoted for an average.

\clearpage
\newcommand{\fnB}{\ensuremath{f_{B\!\!\!\!/}}}
\newcommand{\Ufi}{\Upsilon(5S)}
\newcommand{\Ufo}{\Upsilon(4S)}
\newcommand{\bp}{B^+}
\newcommand{\bn}{B^0}
\newcommand{\jp}{J/\psi}
\newcommand{\kp}{K^+}
\newcommand{\kn}{K^0}

\mysection{\b-hadron production fractions}
\label{sec:fractions}

We consider here the relative fractions of the different \b-hadron
species produced in a specific process. These fractions are needed for
characterizing the signal composition in inclusive \b-hadron analyses,
predicting the background composition in exclusive analyses, and
converting observed event yields (or event yield ratios) into
branching fraction (or branching fraction ratio) measurements.
We distinguish here the following three $b$-hadron production
processes: $\Ufo$ decays, $\Ufi$ decays, and high-energy collisions
(including \Z decays).

\mysubsection{\b-hadron production fractions in $\Ufo$ decays}
\labs{fraction_Ups4S}

Pairs of $b\bar{b}$ quarks produced in $e^+e^-$ annihilations at the
energy of the $\Ufo$ resonance can hadronize into $B^+B^-$,
$B^0\bar{B}{}^0$ or final states without open-beauty mesons. The
latter are dominated by production of lower bottomonium levels
accompanied by light hadrons, e.g.\ $\Upsilon(1S)\pi^+\pi^-$ or
$\Upsilon(1S)\eta$. 
The corresponding production fractions satisfy
\begin{equation}
f^{+-}+f^{00}+\fnB=1.
\label{eq:sum_of_prod_frac_y4s}
\end{equation}
The measurements that are used for determining the production
fractions at the $\Ufo$ are listed in
Table~\ref{tab:y4s_inputs}.

The $f^{+-}/f^{00}$ measurements reported in Refs.~\cite{Alexander:2000tb,Aubert:2004rz,Belle:2022hka,Athar:2002mr} are based on a comparison of the yields of $\Bu$ and $\Bd$ decays to final states related by isospin transformation, e.g.\ $\Bu\to\jp\kp$ and $\Bd\to\jp\ks$, and an assumption that corresponding partial widths are equal. The theoretical
uncertainty of this assumption is considered only in the recent
measurement by Belle~\cite{Belle:2022hka}, where it is conservatively
estimated to be 4.4\%. We assume that the same theoretical uncertainty
is present in other measurements with quarkonium and $K^{(*)}$ in the
final state~\cite{Alexander:2000tb,Aubert:2004rz}, and that
these theoretical uncertainties are correlated among the different channels.

\begin{table}
  \caption{Measurements that are used for determining the production
    fractions at the $\Ufo$, corresponding methods, external
    parameters and experiments that performed them. }
\label{tab:y4s_inputs}
\begin{center}
\small
\begin{tabular}{lccr}
\hline
Measurement & Method & External parameters & Experiment \\
\hline
\noalign{\vskip 1mm}
$f^{+-}/f^{00}$ & & & \\
\noalign{\vskip 0.8mm}
$1.04\pm0.07\pm0.04$ & \particle{\jpsi K^{(*)}} & $\tau(\Bu)/\tau(\Bd)=1.066\pm0.024$ & CLEO, 2001 \cite{Alexander:2000tb} \\
\noalign{\vskip 0.8mm}
$1.06\pm0.02\pm0.03$ & \particle{(c\bar{c})K^{(*)}} & $\tau(\Bu)/\tau(\Bd)=1.086\pm0.017$ & \babar, 2005 \cite{Aubert:2004rz} \\
\noalign{\vskip 0.8mm}
$1.065\pm0.012\pm0.019\pm0.047$ & \particle{\jpsi K} & $\tau(\Bu)/\tau(\Bd)=1.076\pm0.004$ & \belle, 2023\cite{Belle:2022hka} \\
\noalign{\vskip 0.8mm}
$1.058\pm0.084\pm0.136$ & \particle{D^*\ell\nu} & $\tau(\Bu)/\tau(\Bd)=1.074\pm0.028$ & CLEO, 2002 \cite{Athar:2002mr} \\
\noalign{\vskip 0.8mm}
$1.01\pm0.03\pm0.09$ & Dilepton & $\tau(\Bu)/\tau(\Bd)=1.083\pm0.017$ & \belle, 2003 \cite{Belle:2002lms} \\
 &  & $\tau(\Bd)=(1.542\pm0.016)\,\mathrm{ps}$ & \\
$f^{+-}/f^{00} = 1.054\pm0.038$ & Average of\\
                                & the above \\
\hline
\noalign{\vskip 1mm}
$f^{00}$ &  &  &  \\
\noalign{\vskip 0.8mm}
$0.487\pm0.010\pm0.008$ &  &  & \babar, 2005 \cite{Aubert:2005bq} \\
\hline
\noalign{\vskip 1mm}
$\fnB$ &  &  &  \\
\noalign{\vskip 0.8mm}
$-0.0011\pm0.0143\pm0.0107$ & Semileptonic &  & CLEO, 1996~\cite{CLEO:1995umr} \\
 & decays &  \\
\noalign{\vskip 0.8mm}
$\geq0.00264\pm0.00021$ & Sum of known &  & \babar, \belle \\
 & channels &  & \cite{BaBar:2008xay,Belle:2017vat,Belle:2018boz,Belle:2015hnh} \\
\hline
\end{tabular}
\end{center}
\end{table}

In Ref.~\cite{Belle:2002lms}, the time dependence is
studied in events
where both $B$ mesons decay semileptonically, $B\to\ell\,X$. The $f^{+-}/f^{00}$ measurement is a byproduct of the
$B^0\bar{B}{}^0$ oscillation-frequency measurement. It is assumed that
$\Gamma(B^+\to\ell\,X)=\Gamma(B^0\to\ell\,X)$, which is expected to
hold with better than 1\% accuracy~\cite{Chay:1990da,Gronau:2006ei}, and the corresponding
theoretical uncertainty is neglected.
In Ref.~\cite{Aubert:2005bq}, the value of $f^{00}$ is determined by
comparing the numbers of events with one and with two reconstructed
$\Bd\to{}D^{*-}\ell^+\nu_\ell$ decays.
The measurement of a non-$B\bar{B}$ fraction in
Ref.~\cite{CLEO:1995umr} is based on the yields of single-lepton and
two-leptons events. Both of the above measurements do not rely on
model assumptions and do not depend upon external parameters.

At the $\Ufo$ energy, the \babar and Belle collaborations have observed production of five
final states containing bottomonium:
$\Upsilon(1S)\pi^+\pi^-$, $\Upsilon(1S)\eta$, $\Upsilon(1S)\eta'$,
$h_b(1P)\eta$ and
$\Upsilon(2S)\pi^+\pi^-$~\cite{BaBar:2008xay,Belle:2017vat,Belle:2018boz,Belle:2015hnh}.
Their total production fraction is
\begin{equation}
f_\mathrm{(b\bar{b})X} = 0.00264\pm0.00021,
\labe{eq:fnB}
\end{equation}
where the contribution of channels with $\pi^0\pi^0$ are included using isospin
relations. 
Since many transitions remain unexplored, we consider the value
in \Eq{eq:fnB} as a lower limit on $\fnB$.

We perform a fit to all measurements shown in
Table~\ref{tab:y4s_inputs} using Eq.~\eqref{eq:sum_of_prod_frac_y4s}
as a constraint. The values and uncertainties are corrected for a
change of the external parameters
$\tau(\Bu)/\tau(\Bd)$ and $\tau(\Bd)$, whose values at the
time of the publication of each measurement are listed in Table~\ref{tab:y4s_inputs}. The
current values are $\tau(\Bu)/\tau(\Bd)=1.076\pm0.004$ and
$\tau(\Bd)=(1.519\pm0.004)\,\mathrm{ps}$ (see
Sec.~\ref{sec:lifetime_ratio}). The corrections are applied with the method 
described in Sec.~\ref{sec:method:corrSysts}.

The fit is performed in two steps. First, we fit the $f^{+-}/f^{00}$
measurements taking into account correlations of the systematic
uncertainties due to $\tau(\Bu)/\tau(\Bd)$ and due to the
isospin-conservation assumption. The result, also shown in Table~\ref{tab:y4s_inputs}, is 
$f^{+-}/f^{00} = 1.054\pm0.038.$
%
Next, we perform the fit to this average and the measurements of $f^{00}$ and $\fnB$ shown in Table~\ref{tab:y4s_inputs}. 
It is assumed that all these measurements are uncorrelated. 
The positive error of $f_\mathrm{(b\bar{b})X}$ is set to infinity, to impose a constraint ${\fnB} \geq f_\mathrm{(b\bar{b})X}$. This constraint corresponds to the physics-motivated statement that the branching fraction of non-$B\bar B$ decay modes that have not yet been explored experimentally, defined as $\fnB^{\textrm{unexplored}} = {\fnB}\, - f_\mathrm{(b\bar{b})X}$, must be non-negative.
The results of the fit are:
\begin{equation}
f^{+-} = 0.5113^{+0.0073}_{-0.0108} \,,~~~
f^{00} = 0.4861^{+0.0074}_{-0.0080} \,,~~~ \fnB = 0.00264^{+0.0125}_{-0.0002} \,,~~~
\frac{f^{+-}}{f^{00}} = 1.052\pm0.031 \,.
\label{eq:prodfrac_y4s_results}
\end{equation}

\begin{figure}
\begin{center}
\includegraphics[width=0.53\textwidth]{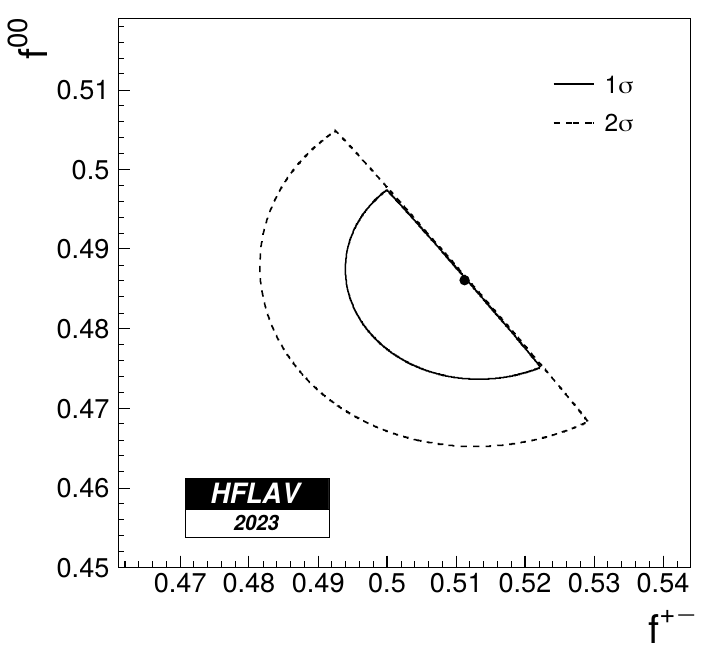}
\caption{The $f^{00}$ and $f^{+-}$ plane with the HFLAV fit
  results. The dot shows the central values, the solid and dashed curves show the $1\sigma$ (68\% CL) and $2\sigma$ (95\% CL)
  contours, respectively.}
\label{fig:f+-_vs_f00_220325_logo}
\end{center}
\end{figure}

Figure~\ref{fig:f+-_vs_f00_220325_logo} shows the plane of $f^{+-}$ vs.
$f^{00}$ with their central values and $1\sigma$ 
and $2\sigma$ contours. 
The strong correlation seen in the contours is a result of the constraint ${\fnB} \geq f_\mathrm{(b\bar{b})X}$.
Removing this constraint by adding $\fnB^{\textrm{unexplored}}$ as an unconstrained parameter in the fit results in larger and symmetric uncertainties on $f^{+-}$, $f^{00}$, and $f^{+-} / f^{00}$, equaling $\pm 0.0128$, $\pm 0.0090$, and $\pm 0.033$, respectively.

To test the ensemble behavior of the fit, we apply it to a set of parameterized experiments produced with the central values of the fit results shown in Eq.~\ref{eq:prodfrac_y4s_results} and the uncorrelated uncertainties of the measurements shown in Table~\ref{tab:y4s_inputs}.
Relative to the values shown in Eq.~(\ref{eq:prodfrac_y4s_results}), the fit values averaged over the parameterized experiments show biases of 
$\Delta f^{+-} = -0.0016$, $\Delta f^{00} = -0.0015$, and 
$\Delta (f^{+-} / f^{00}) = -0.0044$, due to the constraint ${\fnB} \geq f_\mathrm{(b\bar{b})X}$.
In terms of the fit uncertainty, the means of the pulls of these parameters are $-17\%$, $-19\%$, and $-14\%$, respectively.
Since the biases are small, we neglect them in our reported results.

The complex correlation between the values of $f^{+-}$ and $f^{00}$ cannot be represented by a correlation matrix. 
To obtain detailed information about the correlations, we suggest to reproduce our final fit. For this, one should express $\fnB=1-f^{+-}-f^{00}$ and fit the four measurements shown in Table 4 using two parameters, $f^{+-}$ and $f^{00}$.

\subsubsection{Discussion on measurements of $f^{+-}/f^{00}$}
Experimentally, the simplest method to determine
$f^{+-}/f^{00}$ is studying isospin-related decays of $\Bd$ and
$\Bu$. However, this method suffers from theoretical uncertainty due
to the assumption of isospin conservation, which is difficult to
control. Thus, it is important to perform measurements that give
model-independent results. Several proposals for such measurements are
presented in Refs.~\cite{Jung:2015yma,Bernlochner:2023bad}.

It is important to note that the ratio $f^{+-}/f^{00}$ varies with
center-of-mass energy due to the different phase space of $B^+B^-$ and $B^0\bar{B}{}^0$
pairs ($B^0$ is heavier than $B^+$, and phase space is proportional to
$p_B^3$), the electromagnetic interaction, and isospin violation in the
strong interaction~\cite{Bondar:2022kxv}. This may introduce an additional uncertainty in $f^{+-}/f^{00}$, which is not included in Eq.~\eqref{eq:prodfrac_y4s_results}, due to potential differences between the center-of-mass energies of the different experiments. Since the energy difference between the Belle and
Belle~II data samples collected at the $\Ufo$ is
$0.9\,\mathrm{MeV}$~\cite{Belle:2021lzm,Belle-II_BB_scan}, one can take the typical difference between the energies of the various
experiments to be $1\,\mathrm{MeV}$. In the \babar paper on
the $B^0$ and $B^+$ mass-difference measurement~\cite{BaBar:2008ikz},
it is assumed that the energy dependence of $f^{+-}/f^{00}$ is entirely
defined by the phase-space factor $(p_{B^+}/p_{B^0})^3$. In this case, the uncertainty in $f^{+-}/f^{00}$ due to different
energies among experiments is $\pm0.003$, which is negligibly small
relative to the total uncertainty shown in
Eq.~\eqref{eq:prodfrac_y4s_results}. 

However, $f^{+-}/f^{00}$ might have a much more
pronounced dependence on energy compared to the phase-space factor
$(p_{B^+}/p_{B^0})^3$, as
discussed in Ref.~\cite{Bondar:2022kxv, Belle-II:2024niz} (see also references therein). If so, the uncertainty due to differences in
the average collision energies between the experiments could be as
high as $\pm0.03$. Thus, it is of high interest to perform the
measurement of the energy dependence of $f^{+-}/f^{00}$. 

In addition, the different
energy spreads of the experiments can result in different measured
values of $f^{+-}/f^{00}$. Considering the typical energy spread in
the range $5-6\,\mathrm{MeV}$~\cite{Belle:2021lzm,Belle-II:2024niz},
we find that the corresponding uncertainty ranges from $\pm0.001$ to
$\pm0.01$, depending upon the hypotheses of the $f^{+-}/f^{00}$ energy
dependence. Thus, the uncertainty in $f^{+-}/f^{00}$ due to different
energy spreads is considerably lower than the uncertainty due to
different average energies in various experiments.

\mysubsection{\b-hadron production fractions at the $\Ufi$ energy}
\labs{fraction_Ups5S}

\newcommand{\DsX}{\ensuremath{D_s^{\pm} \, X}}
\newcommand{\Br}{\ensuremath{\mathcal{B}}}

\newcommand{\fsfive}{f^{\Ufi}_{s}}
\newcommand{\fudfive}{f^{\Ufi}_{u,d}}
\newcommand{\fnBfive}{f^{\Ufi}_{B\!\!\!\!/}}

Hadronic events produced in $e^+e^-$ collisions at the $\Ufi$ (also
known as $\Upsilon(10860)$) energy can be classified into three
categories: light-quark ($u$, $d$, $s$, $c$) continuum events,
$b\bar{b}$ continuum events (including $b\bar{b}\gamma$, etc., with
initial-state-radiation photons), and $\Ufi$ events. The latter two
cannot be distinguished on an event-by-event basis and are referred to as $b\bar{b}$ events in
the following. These $b\bar{b}$ events can hadronize into different
final states. We define $\fudfive$ to be the fraction of $b\bar{b}$
events at the $\Ufi$ energy with a pair of non-strange bottom mesons, namely, $B\bar{B}$,
$B\bar{B}{}^*$ or $B^*\bar{B}$, $B^*\bar{B}{}^*$, $B\bar{B}\pi$,
$B\bar{B}{}^*\pi$ or $B^*\bar{B}\pi$, $B^*\bar{B}{}^*\pi$, and
$B\bar{B}\pi\pi$, where $B$ denotes a $B^0$ or $B^+$ meson and
$\bar{B}$ denotes a $\bar{B}^0$ or $B^-$ meson. 
The contributions of these channels are listed in
Table~\ref{tab:y5s_B_channels}.
\begin{table}
\caption{ Visible cross sections of the
  $\ee\to{}B^{(*)}\bar{B}{}^{(*)}(\pi(\pi))$ processes measured by
  Belle at the $\Ufi$ resonance ($\sqrt{s}=10.866\,\gev$). The results
  of Ref.~\cite{Belle:2015upu} for
  $[B^{(*)}\bar{B}{}^{(*)}]^\pm\pi^\mp$ are multiplied by 1.5 to
  account for the contribution of $[B^{(*)}\bar{B}{}^{(*)}]^0\pi^0$. }
\renewcommand*{\arraystretch}{1.1}
\label{tab:y5s_B_channels}
\begin{center}
\small
\begin{tabular}{lccr}
\hline
Process & $\sigma^\mathrm{vis}$ (pb) & Ref. \\ 
\hline
$\ee\to$ & & \\
\hspace{7mm}$B\bar{B}$                    &  $33.3 \pm 1.2$ & \cite{Belle:2021lzm} \\
\hspace{7mm}$B\bar{B}{}^*+B^*\bar{B}$     &  $68.0 \pm 3.3$ & \cite{Belle:2021lzm} \\
\hspace{7mm}$B^*\bar{B}{}^*$              & $124.4 \pm 5.3$ & \cite{Belle:2021lzm} \\
\hspace{7mm}$B\bar{B}\pi$                 & $< 3.2$ & \cite{Belle:2015upu} \\
\hspace{7mm}$(B\bar{B}{}^*+B^*\bar{B})\pi$ & $16.8 \pm 2.3$ & \cite{Belle:2015upu} \\
\hspace{7mm}$B^*\bar{B}{}^*\pi$            & $8.4 \pm 1.5$ & \cite{Belle:2015upu} \\
\hspace{7mm}$B\bar{B}\pi\pi$              & not seen & \cite{Belle:2010hoy} \\
\hline
\end{tabular}
\end{center}
\end{table}
Similarly, we define
$\fsfive$ to be the fraction of $b\bar{b}$ events that hadronize into
a pair of strange bottom mesons, $B_s^0\bar{B}{}_s^0$,
$B_s^0\bar{B}{}_s^{*}$ or $B_s^{*}\bar{B}{}_s^0$, and
$B_s^{*}\bar{B}{}_s^{*}$. 
The production fractions for the latter two channels are $0.073\pm0.014$ and $0.870\pm0.017$, respectively~\cite{Belle:2012tsw}.
Note that the excited bottom-meson
states decay via $B^* \to B \gamma$ and $B_s^{*} \to B_s^0 \gamma$.
Lastly, $\fnBfive$ is defined to be the fraction of $b\bar{b}$ events
without open-bottom mesons in the final state. These events are
dominated by production of bottomonium accompanied by light hadrons.
By construction, these fractions satisfy
\begin{equation}
\fudfive + \fsfive + \fnBfive = 1 \,.
\label{eq:sum_frac_five}
\end{equation} 

The Belle collaboration measured $\fudfive$~\cite{Belle:2021lzm} based
on the formula
\begin{equation}
  \fudfive=\frac{N_B^{\Ufi}}{N_B^{\Ufo}}\;\frac{n_{b\bar{b}}^{\Ufo}}{n_{b\bar{b}}^{\Ufi}}\,,
  \label{eq:fudfi_formula}
\end{equation}
where $N_B^{\Ufi}$ and $N_B^{\Ufo}$ are the $B$ meson yields at $\Ufi$
and $\Ufo$, respectively, that were measured using five clean
$B$-decay channels; $n_{b\bar{b}}^{\Ufi}$ and $n_{b\bar{b}}^{\Ufo}$
are the $b\bar{b}$-event yields at $\Ufi$ and $\Ufo$, respectively,
determined by counting hadronic events and subtracting the continuum
contribution. In formula~\eqref{eq:fudfi_formula} it is assumed that the
non-$B\bar{B}$ fraction for $\Ufo$ decays is negligibly small. The obtained
value of $\fudfive$ is
\begin{equation}
  \fudfive=0.751\pm0.040\,,
  \label{eq:fudfive_value}
\end{equation}
where the uncertainty in $n_{b\bar{b}}^{\Ufi}$ dominates. This
measurement supersedes the previous Belle result reported in
Ref.~\cite{Belle:2010hoy}.

In 2023, Belle measured $\fsfive$ using inclusive production of $\Ds$
mesons~\cite{Belle:2023yfw}. The $\fsfive$ value was determined from
the formula
\begin{equation}
  \Br(\Ufi\to\DsX) = 2 \, \fsfive \, \Br(\Bs\to\DsX) + 2 \, \fudfive \, \Br(B\to\DsX),
  \label{eq:fsfive_formula}
\end{equation} 
where $\Br(\Ufi\to\DsX)$ is the $\Ds$ production fraction defined as
$N_{\DsX}^{\Ufi}/n_{b\bar{b}}^{\Ufi}$, where $N_{\DsX}^{\Ufi}$ is the number
of $\Ds$ mesons produced in $b\bar{b}$ events at the $\Ufi$
energy. The value of $N_{\DsX}^{\Ufi}$ was determined by counting $\Ds$
mesons and subtracting the continuum
contribution~\cite{Belle:2023yfw}. The inclusive branching fraction
$\Br(B\to\DsX)$ is defined as $\Br(\Ufo\to\DsX)/2$; it is measured
using the same approach as $\Br(\Ufi\to\DsX)$~\cite{Belle:2023yfw}.
The inclusive branching fraction $\Br(\Bs\to\DsX)$ was measured by
Belle to be $(60.2\pm5.8\pm2.3)\%$ using semileptonic tagging of the second $\Bs$ in the
event~\cite{Belle:2021qxu}. The
obtained value of $\fsfive$ is
\begin{equation}
  \fsfive = 0.230 \pm 0.002 \pm 0.028\,.
  \label{eq:fsfive_value}
\end{equation} 
This measurement supersedes previous Belle results reported in
Refs.~\cite{Belle:2012tsw,Belle:2021qxu}. It is important that all
parameters in Eq.~\eqref{eq:fsfive_formula} are measured by Belle, so
their systematic uncertainties partly cancel. The dominant
contribution to the uncertainty of $\fsfive$ originates from
$\Br(\Bs\to\DsX)$.

The CLEO collaboration performed measurements of $\Br(\Ufi\to\DsX)$
and $\fudfive$~\cite{Huang:2006em} that are in good agreement with
corresponding Belle results but have $3-4$ times lower
accuracy. We do not use the CLEO results to calculate the $\fsfive$
average, to ensure the partial cancellation of systematic uncertainties,
mentioned above.
The \babar collaboration measured $\fsfive$ using energy scan
data~\cite{Lees:2011ji}. This result relies on a model-dependent
estimation of $\Br(\Bs\to\phi\,X)$ and uses the constraint of
Eq.~\eqref{eq:sum_frac_five} neglecting $\fnBfive$. Therefore, we do
not use it in our average.

The Belle experiment observed the following bottomonium-production
processes at the $\Ufi$ energy: 
$e^+e^-\to\Upsilon(1S,2S,3S)\pi^+\pi^-$,
$\Upsilon(1S,2S,3S)\pi^0\pi^0$,
$\Upsilon(1S)K^+K^-$,
$h_b(1P,2P)\pi^+\pi^-$, 
$\chi_{b1,2}(1P)\pi^+\pi^-\pi^0$,
$\Upsilon_J(1D)\eta$ and 
$\Upsilon(2S)\eta$~\cite{Belle:2014vzn,
Belle:2013urd,
Belle:2011wqq,
Belle:2014sys,
Belle:2018hjt}.
The sum of the visible (i.e., uncorrected for initial-state radiation)
cross-sections into these final states, plus those of the unmeasured
final states $\Upsilon(1S)K^0\bar{K}{}^0$ and $h_b(1P,2P)\pi^0\pi^0$,
which are obtained by assuming isospin conservation, amounts to 
\begin{equation}
\sigma^{\rm vis}(e^+e^-\to (\b\bar{\b})X) = 16.1\pm1.5~{\rm pb} \,,
\end{equation}
where $(\b\bar{\b})=\Upsilon(1S,2S,3S)$, $\Upsilon_J(1D)$,
$h_b(1P,2P)$, $\chi_{b1,2}(1P)$; and $X=\pi\pi$, $\pi^+\pi^-\pi^0$, $KK$,
$\eta$. 
We divide this by the $\b\bar{\b}$ production cross section, 
$\sigma(e^+e^- \to \b\bar{\b}) = 340 \pm 16$~pb~\cite{Belle:2012tsw}, to obtain
\begin{equation}
f_{(b\bar{b})X}^{\Ufi} = 0.0473\pm0.0048 \,.
\label{eq:f_oni_y5s}
\end{equation}
This should be taken as a lower bound for $\fnBfive$.

To improve accuracy, Belle performed a fit to the values of
$\fudfive$, $\fsfive$ and $\fnBfive$ using
Eq.~\eqref{eq:sum_frac_five} as a constraint~\cite{Belle:2023yfw}. For
$\fnBfive$, Belle used the value of $f_{(b\bar{b})X}^{\Ufi}$, setting
its positive error to infinity. Correlations among the production
fractions were taken into account using the method described in
Sec.~\ref{sec:method:corrSysts} of this review. The obtained values are
\begin{eqnarray}
\fudfive &=& 0.732^{+0.020}_{-0.027} \,,\\
\fsfive  &=& 0.220^{+0.020}_{-0.021}  \,, \label{eq:5-Fractions-from-fit} \\
\fnBfive &=& 0.048^{+0.033}_{-0.006} \,,
\end{eqnarray}
where the strongly asymmetric uncertainty on $\fnBfive$ is due to the
one-sided constraint from the observed $(\b\bar{\b})X$ decays. These
results, together with their correlations, imply
\begin{eqnarray}
\fsfive/\fudfive  &=& 0.300^{+0.036}_{-0.035} \,.
\end{eqnarray}
Further methods to measure production fractions at $\Ufi$ are
discussed in Ref.~\cite{thesis_Louvot}.

\mysubsection{\b-hadron production fractions at high energy}
\labs{fractions_high_energy}
\labs{chibar}

In a previous publication~\cite{Amhis:2019ckw}, we presented two sets of averages, one including only measurements 
performed at LEP, and another including only measurements performed 
by CDF at the Tevatron.\footnote{The LHC production fractions results 
were still incomplete, lacking measurements of the production of 
weakly-decaying baryons heavier than \Lb.
} %
While the first set is well defined and is basically related to branching fractions of inclusive $Z$ decays, the other set is somewhat ill-defined, 
since it depends on the geometrical and kinematical acceptance of the experiments over which the measurements are integrated.
With the ever increasing precision in heavy flavour measurements, the \b-hadron fraction averages provided by HFLAV for high-energy hadron collisions are no longer of interest, since they are not directly transferable from one experiment to the other. We have, therefore, decided to no longer maintain these averages starting in Ref.{\cite{HFLAV:2022esi}. The interested reader should refer to Sec.~4.1.3 of our 2019 publication~\cite{Amhis:2019ckw}.

The relative fractions of \b-hadron types produced in $Z$ decays are universal and therefore still of interest. Since the averages we have reported in \Refx{Amhis:2019ckw} have remained stable over the last decade and new data are not expected until a future new electron-positron collider operates again at the $Z$ pole, they are not reported here. 

\clearpage

\mysection{Lifetimes and mixing parameters of \b hadrons}
\labs{life_mix}

Quantities such as \b-hadron production fractions, \b-hadron lifetimes, 
and neutral \B-meson oscillation frequencies were studied
in the 1990s at LEP and SLC, %
at DORIS~II and CESR, %
as well as at the Tevatron. %
This was followed by precise measurements of the \Bd and \Bu mesons
performed at the asymmetric \B factories, KEKB and PEPII, %
as well as measurements related 
to the other \b hadrons, in particular \Bs, \Bc and \Lb, 
performed at the upgraded Tevatron. %
Currently, the most precise measurements are coming from the 
ATLAS, CMS, LHCb, \belle and \belletwo experiments. %

In many cases, these basic quantities, in addition to being interesting by themselves,
are necessary ingredients for more refined measurements,
for example decay-time-dependent \CP-violating asymmetries.
Hence, some of the averages presented in this chapter are
used as input for the results given in subsequent chapters. 
In the past, many \b-hadron lifetime and mixing measurements had a significant dependence on the \b-hadron production fractions, which themselves depended on the lifetime and mixing measurements.
This circular coupling had to be dealt with carefully whenever inclusive or semiexclusive measurements of \b-hadron lifetime and mixing parameters were considered. In the past decade, this dependence has reduced to a negligible level, %
with increasingly precise exclusive measurements becoming available and dominating practically all averages.

In addition to %
lifetimes and oscillation frequencies, this chapter also deals with \CP violation
in the \Bd and \Bs mixing amplitudes, as well as the %
phase $\phiccbars$ %
that describes \CP violation in the interference between \Bs mixing and decay in $b\to c\bar{c}s$ transitions.
In the absence of new physics and subleading penguin contributions, this phase is equal to $-2\beta_s = -\arg\left[\left(V_{ts}V^*_{tb}\right)^2/\left(V_{cs}V^*_{cb}\right)^2\right]$.

Throughout this chapter, published results that have been superseded 
by subsequent publications are ignored (\ie, excluded from the averages). They are only referred to if necessary.

\mysubsection{\b-hadron lifetimes}
\labs{lifetimes}

Lifetime calculations are performed in the framework of
the Heavy Quark Expansion
(HQE) \cite{Shifman:1986mx,Chay:1990da,Bigi:1992su}.
In these calculations, 
the total decay rate of a hadron $H_b$ 
is expressed as a series of expectation values of operators of increasing dimension,
\begin{equation}
\Gamma_{H_b} = |{\rm CKM}|^2 \sum_{n,k} \frac{c_{nk}}{m_b^n}
\langle H_b|O_{nk}|H_b\rangle \,,
\labe{hqe}
\end{equation}
where $|{\rm CKM}|^2$ is the relevant combination of CKM matrix elements.
The coefficients $c_{nk}$
are calculated perturbatively~\cite{Wilson:1969zs}, 
\ie\ as a series in $\alpha_s(m_b)$.
The nonperturbative QCD effects are comprised in the matrix elements $\langle H_b| O_{nk} | H_b\rangle \propto \Lambda_{\rm QCD}^n$ of the operators $O_{nk}$. 
For a given dimension $n$, there are usually several operators, indicated by the index $k$. 
Hence the HQE predicts $\Gamma_{H_b}$ in the
form of an expansion in both $\Lambda_{\rm QCD}/m_b$ and
$\alpha_s(m_b)$, see, \textit{e.g.}, \Refs{Albrecht:2024oyn,Lenz:2015dra}.
The leading term in \Eq{hqe} corresponds to the weak
decay of a free \b quark.
At this order, all \b-flavored hadrons have the same lifetime. The concept of the HQE and first calculations of valence quark effects emerged in 1986~\cite{Shifman:1986mx}. In
the early 1990s experiments became sensitive enough to detect
lifetime differences among various $H_b$ species. 
The hypothetical existence of additional, non-powerlike, contributions to $\Gamma_{H_b}$, referred to as \emph{violation of quark-hadron duality}, is
not captured by the power series of the HQE~\cite{Shifman:2000jv,Bigi:2001ys}. The sizes of such terms can only be determined
experimentally, by confronting the HQE predictions with data. Possible violation of quark-hadron duality has been shown to be severely constrained by experimental results~\cite{Jubb:2016mvq,Albrecht:2024oyn}.
The matrix elements of~\Eq{hqe}
can be calculated using lattice QCD or QCD sum rules. In some cases they
can also be related to those appearing in other observables by utilizing
symmetries of QCD. One may reasonably expect that powers of
$\Lambda_{\rm QCD}/m_b\sim 0.1$ provide enough suppression that only the
first terms of the sum in \Eq{hqe} matter.  However, starting from
the third power the coefficients are enhanced by a factor of
$16\pi^2$, so that
these terms of order $16\pi^2 (\Lambda_{\rm QCD}/m_b)^3$~\cite{Voloshin:1999pz,*Guberina:1999bw,*Neubert:1996we,*Bigi:1997fj} do matter when comparing lifetimes of different $b$-hadrons. 
State-of-the-art calculations of these spectator effects 
to the $b$-hadrons  lifetimes predictions exist
in terms of both $\Lambda_{\rm QCD}/m_b$~\cite{Gabbiani:2004tp, Lenz:2011ti, Lenz:2006hd}
and $\alpha_s(m_b)$~\cite{Beneke:2002rj,Franco:2002fc,Lenz:2013aua}, with all subsequent theory
papers using these results. 
However, it was subsequently found  that the so-called Darwin contribution at order $(\Lambda_{\rm QCD}/m_b)^3$ might give very large numerical corrections, even without the $16\pi^2$ enhancement~\cite{Mannel:2020fts,Lenz:2020oce}.
The exact size of this effect depends on the value of the matrix elements of the Darwin operator, which is extracted from inclusive semileptonic $b\rightarrow c$ transitions, see Refs.~\cite{Finauri:2023kte,Bordone:2021oof,Bernlochner:2022ucr}. Moreover, there was recent progress in determining the non-perturbative matrix elements of the arising four-quark operators~\cite{Kirk:2017juj,King:2021jsq}.

Theoretical predictions are usually made for the ratios of the lifetimes
(with $\tau(\Bd)$ often chosen as the common denominator) rather than for the
individual lifetimes, since this leads to cancellation of several uncertainties.
The precision of the HQE calculations (see \textit{e.g.} \Refs{Gratrex:2023pfn,Lenz:2022rbq})
is in some instances already surpassed by the measurements,
\eg, in the case of $\tau(\Bu)/\tau(\Bd)$.  
Improvement in the precision of calculations requires progress
along two lines. First, better nonperturbative
matrix elements are needed. 
One expects precise calculations, especially from lattice QCD where significant advances have been made in the past decade. 
For the first promising steps in that direction, see~\Refs{Black:2023vju,Lin:2022fun}.
Second, the coefficients $c_{nk}$ must be calculated to higher
orders of $\alpha_s$. In particular, the $\alpha_s^2$ and $\alpha_s \Lambda_{\rm
QCD}/m_b $ contributions to the lifetime differences are needed to keep
up with the experimental precision.

The following important conclusions, which are in agreement with experimental observation, can be drawn from the HQE, even in its present state:
\begin{itemize}
\item The larger the mass of the heavy quark, the smaller the
  variation in the lifetimes among different hadrons containing this
  quark.
   This is %
   illustrated by the fact that lifetimes are rather
   similar in the \b sector, while they differ by large factors
   in the charm sector (see \textit{e.g.} Ref.~\cite{King:2021xqp}.
\item 
First corrections to the spectator model occur at order $\Lambda_{\rm QCD}^2/m_b^2$, leading to lifetime differences around one percent.
\item 
The dominant contribution to $\tau(B^+)/\tau(B^0)$ is of order
$16\pi^2 (\Lambda_{\rm QCD}/m_b)^3$ and typically amounts to several
percent, whereas $(\Lambda_{\rm QCD}/m_b)^3$ corrections due to the Darwin operator might have a crucial role for $\tau(B^0_s)/\tau(B^0)$.
\end{itemize}

\mysubsubsection{Overview of lifetime measurements}

This section gives  an overview of the types of \b-hadron lifetime  measurements, with details  given in subsequent sections. In most cases, the decay time of an $H_b$ state is estimated by measuring its flight
distance and dividing it by the relativistic factor $\beta\gamma c$.  Methods of accessing lifetime
information can roughly be divided into the following five categories:
\begin{enumerate}
\item {\bf\em Measurements in semileptonic decays of a specific
  {\boldmath $H_b$\unboldmath}}.  The virtual \particle{W} boson from \particle{\b\to Wc}
  produces a $\ell\nu_l$ pair (\particle{\ell=e,\mu}) in about 21\% of the
  cases.  The electron or muon from such decays provides a clean and efficient
  trigger signature.
  The \particle{c} quark and the $H_b$ spectator
  quark(s) combine into a charm hadron $H_c$,
  which is reconstructed in one or more exclusive decay channels.
  Identification of the $H_c$ species allows one to separate, at least
  statistically, different $H_b$ species.  The advantage of these
  measurements is in the sample size,
  which is usually larger than in the case of
  exclusively reconstructed hadronic $H_b$ decays (described  next). The main
  disadvantages are related to the difficulty of estimating the lepton+charm
  sample composition and to the reliance on Monte Carlo for
  the momentum (and hence $\beta\gamma$ factor) estimate.
\item {\bf\em Measurements in exclusively reconstructed hadronic decays}.
  These
  have the advantage of complete reconstruction of the decaying $H_b$ state, 
  which allows one to infer the decaying species, as well as to perform precise
  measurement of the $\beta\gamma$ factor.  Both lead to generally
  smaller systematic uncertainties than semileptonic decays.
  The downsides are smaller branching fractions and larger combinatorial
  backgrounds when the signal channel involves multihadron decays, such as $H_b\rightarrow H_c\pi(\pi\pi)$ with
  multibody $H_c$ decays. This problem is often more serious in a hadron collider environment, which has many hadrons and 
  a non-trivial underlying event.  Decays of the type $H_b\to \jpsi H_s$ are often used, as they are
  relatively clean and easy to trigger on due to the $\jpsi\to \ell^+\ell^-$
  signature.%
\item {\bf\em Measurements at asymmetric B factories}. 
  In the $\Ups\rightarrow B \bar{B}$ decay, the \B mesons (\Bu or \Bd) are
essentially at rest in the \Ups frame.  This makes direct lifetime
measurements impossible in experiments at symmetric-energy colliders, which produce the 
\Ups at rest. 
At asymmetric \B factories the \Ups meson is boosted,
resulting in the \B and \particle{\bar{B}} moving nearly parallel to each 
other with similar boosts. The decay time is inferred from the distance $\Delta z$        
separating the \B and \particle{\bar{B}} decay vertices along the boost axis 
and from the \Ups boost, which is known from the beam energies. In the \babar, 
\belle and \belletwo experiments, the \Ups boosts are 
$\beta \gamma \approx 0.55$, $0.43$ and $0.29$, respectively. They result
 in average \B decay lengths of approximately $250$, $190$, $130$~$\mu$m. 
While one \Bd or \Bu meson is fully reconstructed in a semileptonic or hadronic decay mode,
the other \B in the event is typically not fully reconstructed, in order to avoid loss of efficiency. Rather, only the position
of its decay vertex is determined from the remaining tracks in the event.
These measurements benefit from large sample sizes, but suffer from poor proper time 
resolution, comparable to the \B lifetime itself. The resolution is dominated by the 
uncertainty on the decay-vertex positions, which is typically 50~(100)~$\mu$m for a
fully (partially) reconstructed \B meson. 
With much larger samples in the future, 
the resolution and purity could be improved (and hence the systematics reduced)
by fully reconstructing both \B mesons in the event.
 
\item {\bf\em Measurement of lifetime ratios}.  This method, 
  initially applied 
  in the measurement of $\tau(\Bu)/\tau(\Bd)$, is now also used for other 
  \b-hadron species at the LHC. 
  The ratio of the lifetimes is extracted from the
  dependence of the ratio of the observed yields
  of two different \b-hadron species on the proper time, \ie, the time interval between the production 
and the decay in the rest frame of the \B meson. Subtle efficiency effects and systematic uncertainties are
  partially cancelled in the ratio by reconstructing both the decays in similar topologies. 
\item {\bf\em Inclusive (flavor-blind) measurements}.  Early, low-statistics
  measurements were aimed at extracting the lifetime from a mixture of
  \b-hadron decays, without distinguishing the decaying species.  Often,
 the exact $H_b$ composition was ill defined and analysis-dependent, and the $\beta\gamma$ factor was extracted from simulation. 
  Nowadays, it is of little interest as it is less fundamental than the precisely measured lifetimes of the individual species. As a result, we no longer review the inclusive \b-hadron lifetime measurements, the latest of which was published in 2004~\cite{Abdallah:2003sb}. 
The interested reader can refer to a previous publication~\cite{Amhis:2019ckw}.
\end{enumerate}

In some analyses, measurements of two [\eg, $\tau(\Bu)$ and
$\tau(\Bu)/\tau(\Bd)$] or three [\eg\ $\tau(\Bu)$,
$\tau(\Bu)/\tau(\Bd)$, and \dmd] quantities are combined.  This
introduces correlations among measurements.  Another source of
correlations among the measurements is represented by the systematic effects, which
could be common to a number of measurements in the same experiment or to an analysis technique across different
experiments.  When calculating the averages presented below, such known correlations are taken
into account.
\mysubsubsection{\Bd and \Bu lifetimes and their ratio}
\labs{taubd}
\labs{taubu}
\labs{lifetime_ratio}

After many years of dominating the \Bd and \Bu lifetime averages, the LEP experiments
yielded the scene to the asymmetric \B~factories and
the Tevatron experiments. The \B~factories have been very successful in
utilizing their potential --- in only a few years of running, \babar and,
to a greater extent, \belle and \belletwo, have struck a balance between the
statistical and the systematic uncertainties, with both being close to
(or even better than) an impressive 1\% level.  Meanwhile, CDF and
\dzero emerged as significant contributors to the field as the
Tevatron Run~II data flowed in. More recently, the LHC and \belletwo experiments came into play, matching the precision and, in case of LHCb and CMS, even improving it by around a factor two.
At present, we have three sets
of measurements (from LEP/SLC, the \B factories and Tevatron/LHC) performed in different environments, obtained using substantially
different techniques, and precise enough for cross-checking and comparison.

\begin{table}[!p]
\caption{Measurements of the \Bd lifetime with exclusive (excl.) or inclusive (incl.) decays, \B charge determination from the secondary vertex (sec. vtx), or partial reconstruction. See \Sec{lifetimes} for a detailed explanation of the method.}
\labt{lifebd}
\begin{center}
\begin{tabular}{lcccl} \hline
Experiment &Method                    &Data set &$\tau(\Bd)$ (ps)                  &Ref.\\
\hline
ALEPH  &\particle{D^{(*)} \ell}       &91--95 &$1.518\pm 0.053\pm 0.034$          &\cite{Barate:2000bs}\\
ALEPH  &Exclusive                     &91--94 &$1.25^{+0.15}_{-0.13}\pm 0.05$     &\cite{Buskulic:1996hy}\\
ALEPH  &Partial rec.\ $\pi^+\pi^-$    &91--94 &$1.49^{+0.17+0.08}_{-0.15-0.06}$   &\cite{Buskulic:1996hy}\\
DELPHI &\particle{D^{(*)} \ell}       &91--93 &$1.61^{+0.14}_{-0.13}\pm 0.08$     &\cite{Abreu:1995mc}\\
DELPHI &Charge sec.\ vtx              &91--93 &$1.63 \pm 0.14 \pm 0.13$           &\cite{Adam:1995mb}\\
DELPHI &Inclusive \particle{D^* \ell} &91--93 &$1.532\pm 0.041\pm 0.040$          &\cite{Abreu:1996gb}\\
DELPHI &Charge sec.\ vtx              &94--95 &$1.531 \pm 0.021\pm0.031$          &\cite{Abdallah:2003sb}\\
\end{tabular}
\end{center}
\end{table}

\begin{table}[p]
\centering
\caption{Measurements of the \Bu lifetime with exclusive (excl.) decays or with \B charge determination from the secondary vertex (sec. vtx). See \Sec{lifetimes} for a detailed explanation of the method.}
\labt{lifebu}
\begin{tabular}{lcccl} \hline
Experiment &Method                 &Dataset &$\tau(\Bu)$ (ps)                 &Ref.\\
\hline
ALEPH  &\particle{D^{(*)} \ell}    &91--95 &$1.648\pm 0.049\pm 0.035$          &\cite{Barate:2000bs}\\
ALEPH  &Exclusive                  &91--94 &$1.58^{+0.21+0.04}_{-0.18-0.03}$   &\cite{Buskulic:1996hy}\\
DELPHI &\particle{D^{(*)} \ell}    &91--93 &$1.61\pm 0.16\pm 0.12$             &\cite{Abreu:1995mc}$^a$\\
DELPHI &Charge sec.\ vtx           &91--93 &$1.72\pm 0.08\pm 0.06$             &\cite{Adam:1995mb}$^a$\\
DELPHI &Charge sec.\ vtx           &94--95 &$1.624\pm 0.014\pm 0.018$          &\cite{Abdallah:2003sb}\\
L3     &Charge sec.\ vtx           &94--95 &$1.66\pm  0.06\pm 0.03$            &\cite{Acciarri:1998uv}\\
OPAL   &\particle{D^{(*)} \ell}    &91--93 &$1.52 \pm 0.14\pm 0.09$            &\cite{Akers:1995pa}\\
\end{tabular}
\end{table}
\begin{table}[t]
\centering
\caption{Measurements of the ratio $\tau(\Bu)/\tau(\Bd)$ from exclusive (excl.) decays modes or with \B charge determination from the secondary vertex (sec. vtx). See \Sec{lifetimes} for a detailed explanation of the method.}
\labt{liferatioBuBd}
\begin{tabular}{lcccl} 
\hline
Experiment &Method                 &Dataset &Ratio $\tau(\Bu)/\tau(\Bd)$      &Ref.\\
\hline
ALEPH  &\particle{D^{(*)} \ell}    &91--95 &$1.085\pm 0.059\pm 0.018$          &\cite{Barate:2000bs}\\
ALEPH  &Exclusive                  &91--94 &$1.27^{+0.23+0.03}_{-0.19-0.02}$   &\cite{Buskulic:1996hy}\\
DELPHI &\particle{D^{(*)} \ell}    &91--93 &$1.00^{+0.17}_{-0.15}\pm 0.10$     &\cite{Abreu:1995mc}\\
DELPHI &Charge sec.\ vtx           &91--93 &$1.06^{+0.13}_{-0.11}\pm 0.10$     &\cite{Adam:1995mb}\\
DELPHI &Charge sec.\ vtx           &94--95 &$1.060\pm 0.021 \pm 0.024$         &\cite{Abdallah:2003sb}\\
L3     &Charge sec.\ vtx           &94--95 &$1.09\pm 0.07  \pm 0.03$           &\cite{Acciarri:1998uv}\\
OPAL   &\particle{D^{(*)} \ell}    &91--93 &$0.99\pm 0.14^{+0.05}_{-0.04}$     &\cite{Akers:1995pa}\\
\end{tabular}
\end{table}

The $\tau(\Bu)$, $\tau(\Bd)$ and $\tau(\Bu)/\tau(\Bd)$
measurements, and their averages, are summarized\unpublished{}{\footnote{%
We do not include the old unpublished measurements of \Refs{CDFnote7514:2005,CDFnote7386:2005,ATLAS-CONF-2011-092}.}}
in \Tablesss{lifebd}{lifebu}{liferatioBuBd}.
For the average of $\tau(\Bu)/\tau(\Bd)$ we use only direct measurements of this
ratio and not separate measurements of $\tau(\Bu)$ and
$\tau(\Bd)$.
In the context of the $\tau(\Bd)$ average, only systematic sources originating from \B-factories and LHC measurements are correlated in the averaging process within each experiment/machine. The specific sources that are correlated are outlined below:
\begin{itemize}
\item for the \B-factory measurements --- $\tau(\Bu)$, $\Delta m_d$, length scale, machine boost, detector alignment, detector resolution, beam spot, $B\bar{B}$ fraction and analysis bias (where applicable);
\item for the LHC measurements --- detector alignment, length scale and reconstruction effects.
\end{itemize}
A different approach is adopted for the average of $\tau(\Bu)$ and $\tau(\Bu)/\tau(\Bd)$, for which the following systematic sources are taken into account for correlation:
\begin{itemize}
\item for the SLC and LEP measurements --- \particle{D^{**}} branching fraction uncertainties~\cite{Abbaneo:2000ej_mod,*Abbaneo:2001bv_mod_cont},
estimation of the momentum  of \b mesons produced in \particle{Z^0} decays
(\b-quark fragmentation parameter $\langle X_E \rangle = 0.702 \pm 0.008$~\cite{Abbaneo:2000ej_mod,*Abbaneo:2001bv_mod_cont}),
\Bs and \b-baryon lifetimes (see \Secss{taubs}{taulb}),
and \b-hadron production fractions at high energy~\cite{Amhis:2019ckw}; 
\item for the \B-factory measurements --- detector alignment and length scale, machine boost, %
and sample composition (where applicable);
\item for the Tevatron and LHC measurements --- detector alignment, length scale and reconstruction effects.
\end{itemize}
The resultant averages are:
\begin{eqnarray}
\tau(\Bd) & = & \hflavTAUBD \,, \\
\tau(\Bu) & = & \hflavTAUBU \,, \\
\tau(\Bu)/\tau(\Bd) & = & \hflavRTAUBU \,.
\end{eqnarray}

\mysubsubsection{\Bs lifetimes}
\labs{taubs}

Like neutral kaons, neutral \B mesons contain
short- and long-lived components, since the
light (L) and heavy (H)
eigenstates %
differ not only
in their masses but also in their total decay widths. 
While in the \Bd system the decay width difference \DGd can be neglected, 
the \Bs system exhibits a significant value of the width difference
$\DGs = \Gamma_{s\rm L} - \Gamma_{s\rm H}$, where $\Gamma_{s\rm L}$ and $\Gamma_{s\rm H}$
are the total decay widths of the light eigenstate $\B^0_{s\rm L}$ and the heavy eigenstate $\B^0_{s\rm H}$, respectively.
The sign of \DGs is measured to be positive~\cite{Aaij:2012eq}, \ie,
$\B^0_{s\rm H}$ has a longer lifetime than $\B^0_{s\rm L}$. 
Specific measurements of \DGs and 
$\Gs = (\Gamma_{s\rm L} + \Gamma_{s\rm H})/2$, which are more involved than simple lifetime measurements,
are explained
and averaged in \Sec{DGs}, but the 
resulting averages for
$1/\Gamma_{s\rm L} = 1/(\Gs+\DGs/2)$, $1/\Gamma_{s\rm H}= 1/(\Gs-\DGs/2)$
and the mean \Bs lifetime, defined as $\tau(\Bs) = 1/\Gs$
are also quoted at the end of this section. 
Neglecting \CP violation in $\Bs-\Bsbar$ mixing, 
which is expected~\citehistory{Lenz:2019lvd,Jubb:2016mvq,Artuso:2015swg,Laplace:2002ik,Ciuchini:2003ww,Beneke:2003az}{Lenz:2019lvd,Jubb:2016mvq,Artuso:2015swg,*Lenz_hist,Laplace:2002ik,Ciuchini:2003ww,Beneke:2003az} and measured (see \Sec{qpd}) to be very
small, the mass eigenstates are also \CP eigenstates,
with the short-lived (light) %
state being \CP-even and the long-lived (heavy) %
state being \CP-odd~\cite{Aaij:2012eq}.

Many \Bs lifetime analyses, in particular the early 
ones performed before the nonzero value of \DGs was 
firmly established, ignore \DGs and fit the proper time 
distribution of a sample of \Bs candidates 
reconstructed in a certain final state $f$
with a model containing a single exponential function 
for the signal.
Such {\em effective lifetime} measurements, which we denote as $\tau_{\rm single}(\Bs\to f)$, are estimates of the expectation value  $\int_0^\infty t\,\Gamma(B_s(t)\to f) dt/\int_0^\infty \Gamma(B_s(t)\to f) dt$ of the %
decay time, where 
$\Gamma(B_s(t)\to f) $ is the total untagged time-dependent decay rate~\cite{Hartkorn:1999ga,Dunietz:2000cr,Fleischer:2011cw}. 
This expectation value may lie {\em a priori} anywhere
between $1/\Gamma_{s\rm L}$ %
and $1/\Gamma_{s\rm H}$, %
depending on the proportion of $B^0_{s\rm L}$ and $B^0_{s\rm H}$
in the final state $f$. 
More recent determinations of effective lifetimes may be interpreted as
measurements of the relative composition of 
$B^0_{s\rm L}$ and $B^0_{s\rm H}$
decaying to the final state $f$. 
\Table{lifebs} summarizes the effective 
lifetime measurements.

Averaging measurements of $\tau_{\rm single}(\Bs\to f)$
over several final states $f$ would yield a result 
corresponding to an ill-defined observable
when the proportions of $B^0_{s\rm L}$ and $B^0_{s\rm H}$
differ. 
Therefore, the effective \Bs lifetime measurements are broken down into
the following categories and averaged separately.

\begin{table}[t]
\centering
\caption{Measurements of the effective \Bs lifetimes obtained from single exponential fits (except for the $\jpsi \pi^+\pi^-$ result of Ref.~\cite{Aaij:2019mhf}, obtained from a time-dependent amplitude analysis).}
\labt{lifebs}
\vspace{-3mm}
\begin{tabular}{l@{}c@{}cc@{}rc@{}l} \hline
Experiment & \multicolumn{2}{c}{Final state $f$}           & \multicolumn{2}{c}{Data set} & $\tau_{\rm single}(\Bs\to f)$ (ps) & Ref. \\
\hline 
ALEPH  & \particle{D_s h}     & ill-defined & 91--95 & & $1.47\pm 0.14\pm 0.08$           & \cite{Barate:1997ua}          \\
DELPHI & \particle{D_s h}     & ill-defined & 91--95 & & $1.53^{+0.16}_{-0.15}\pm 0.07$   & \citehistory{Abreu:2000ev}{Abreu:2000ev,*Abreu:1996ep_hist} \\
OPAL   & \particle{D_s} incl. & ill-defined & 90--95 & & $1.72^{+0.20+0.18}_{-0.19-0.17}$ & \cite{Ackerstaff:1997ne}          \\ 
\vspace{-0.4cm} & & & & & \\
ALEPH  & \particle{D_s^- \ell^+}  & flavor-specific & 91--95 & & $1.54^{+0.14}_{-0.13}\pm 0.04$   & \cite{Buskulic:1996ei}          \\
CDF1   & \particle{D_s^- \ell^+}  & flavor-specific & 92--96 & & $1.36\pm 0.09 ^{+0.06}_{-0.05}$  & \cite{Abe:1998cj}           \\
DELPHI & \particle{D_s^- \ell^+}  & flavor-specific & 92--95 & & $1.42^{+0.14}_{-0.13}\pm 0.03$   & \cite{Abreu:2000sh}          \\
\end{tabular}
\end{table}

\afterpage{\clearpage}

\begin{itemize}

\item
{\bf\em \boldmath $\Bs\to D_s^{\mp} X$ decays}
include mostly flavor-specific decays but also decays 
with an unknown mixture of light and heavy components. 
Measurements performed with such inclusive states are
no longer used in our averages. 

\item 
{\bf\em Decays to flavor-specific final states}, 
\ie, decays to final states $f$ with decay amplitudes satisfying 
$A(\Bs\to f) \ne 0$, $A(\Bsbar\to \bar{f}) \ne 0$, 
$A(\Bs\to \bar{f}) = 0$ and $A(\Bsbar\to f)=0$. 
Since there are equal 
fractions of $B^0_{s\rm L}$ and $B^0_{s\rm H}$ at production time ($t=0$),
the corresponding effective lifetime,
called the {\em flavor-specific lifetime}, is equal to~\cite{Hartkorn:1999ga}
\begin{equation}
\tau_{\rm single}(\Bs\to \mbox{flavor specific})
 =  \frac{1/\Gamma_{s\rm L}^2+1/\Gamma_{s\rm H}^2}{1/\Gamma_{s\rm L}+1/\Gamma_{s\rm H}}
 = \frac{1}{\Gs} \,
\frac{{1+\left(\frac{\DGs}{2\Gs}\right)^2}}{{1-\left(\frac{\DGs}{2\Gs}\right)^2}
}\,.
\labe{fslife}
\end{equation}

Because of the fast $\Bs-\Bsbar$ oscillations, 
possible biases of the flavor-specific lifetime due to a
combination of $\Bs/\Bsbar$ production asymmetry,
\CP violation in the decay amplitudes ($|A(\Bs\to f)| \ne |A(\Bsbar\to \bar{f})|$), 
and \CP violation in $\Bs-\Bsbar$ mixing
($|q_{\particle{s}}/p_{\particle{s}}| \ne 1$, see~\Sec{mixing}) 
are strongly suppressed, by a factor $\sim x_s^2$ [where the definition of $x_s$ is given in \Eq{dm} and its value in \Eq{xs}].
The $\Bs/\Bsbar$ production asymmetry at LHCb and the \CP asymmetry due to mixing 
have been measured to be compatible with zero with a precision below 3\%~\cite{Aaij:2014bba} 
and 0.3\% [see \Eq{ASLS}], respectively. The corresponding effects on the flavor-specific lifetime, which therefore have a relative size of the order of $10^{-5}$ or smaller, can be neglected at the current level of experimental precision.
Under the assumption of no production asymmetry 
and no \CP violation in mixing, \Eq{fslife} is exact even for a flavor-specific decay with 
\CP violation in the decay amplitudes. Hence any flavor-specific decay 
mode can be used to measure the flavor-specific lifetime. 

The average of all flavor-specific 
\Bs lifetime measurements%
\unpublished{\citehistory{Buskulic:1996ei,Abe:1998cj,Abreu:2000sh,Ackerstaff:1997qi,Abazov:2014rua,Aaltonen:2011qsa,Aaij:2013bvd,Aaij:2014sua,Aaij:2014fia,Aaij:2017vqj}{Buskulic:1996ei,Abe:1998cj,Abreu:2000sh,Ackerstaff:1997qi,Abazov:2014rua,*Abazov:2006cb_hist,Aaltonen:2011qsa,Aaij:2013bvd,Aaij:2014sua,Aaij:2014fia,*Aaij:2012ns_hist,Aaij:2017vqj}}{\citehistory{Buskulic:1996ei,Abe:1998cj,Abreu:2000sh,Ackerstaff:1997qi,Abazov:2014rua,Aaltonen:2011qsa,Aaij:2013bvd,Aaij:2014sua,Aaij:2014fia,Aaij:2017vqj}{Buskulic:1996ei,Abe:1998cj,Abreu:2000sh,Ackerstaff:1997qi,Abazov:2014rua,*Abazov:2006cb_hist,Aaltonen:2011qsa,*Aaltonen:2011qsa_hist,Aaij:2013bvd,Aaij:2014sua,Aaij:2014fia,*Aaij:2012ns_hist,Aaij:2017vqj}}%
\unpublished{}{\footnote{%
An old unpublished measurement~\cite{CDFnote7757:2005} is not included.}}
is
\begin{equation}
\tau_{\rm single}(\Bs\to \mbox{flavor specific}) = \hflavTAUBSSL \,.
\labe{tau_fs}
\end{equation}

\item
{\bf\em 
{\boldmath The $\Bs \to \jpsi\phi$ \unboldmath} decay}
contains a well-measured mixture of \CP-even and \CP-odd states.
The published 
\particle{\Bs\to \jpsi\phi}
effective lifetime measurements~\cite{Abe:1997bd,Abazov:2004ce,Aaij:2014owa,Sirunyan:2017nbv} are combined  
into the average\unpublished{}{\footnote{%
The old unpublished measurements of Refs.~\citehistory{CDFnote8524:2007,ATLAS-CONF-2011-092}{CDFnote8524:2007,*CDFnote8524:2007_hist,ATLAS-CONF-2011-092} are not included.}}
$\tau_{\rm single}(\Bs\to \jpsi \phi) = \hflavTAUBSJF$. %
Analyses that separate the \CP-even and \CP-odd components in
this decay through a full angular study, outlined in \Sec{DGs},
provide directly precise measurements of $1/\Gs$ and $\DGs$ (see \Table{phisDGsGs}).

\item
{\bf\em 
{\boldmath The $\Bs \to \mu^+\mu^-$ \unboldmath} decay}
is predicted to be \CP-odd in the Standard Model.
Experimentally, a decay-time-dependent analysis can provide a measurement of the admixture of \CP-even and \CP-odd states. However, the statistical precision is currently too low to set any constraint on a possible \CP-even contribution.
Effective lifetime measurements have been published by 
LHCb~\citehistory{LHCb:2021vsc,*LHCb:2021awg}{LHCb:2021vsc,*LHCb:2021awg,*Aaij:2017vad_hist},
CMS~\cite{CMS:2022mgd} and ATLAS~\cite{ATLAS:2023trk}. As two of the measurements have asymmetric statistical errors, different averaging procedures have been tested and due to the current central values and errors, it has been found that all of them are giving results consistent with considering the bigger of the two errors in a classic average procedure. Hence we consider $2.07 \pm 0.29$~ps for the LHCb input, $1.83 \pm 0.23$~ps for the CMS input and $0.99 \pm 0.45$~ps for the ATLAS input.

\item
{\bf\em Decays to \boldmath\CP eigenstates} have also 
been measured.
These include the \CP-even modes 
$\Bs \to D_s^{(*)+}D_s^{(*)-}$ by ALEPH~\cite{Barate:2000kd},
$\Bs \to K^+ K^-$ by LHCb~\citehistory{Aaij:2012kn,Aaij:2014fia}{Aaij:2012kn,Aaij:2014fia,*Aaij:2012ns_hist}%
\unpublished{}{\footnote{An old unpublished measurement of the $\Bs \to K^+ K^-$
effective lifetime by CDF~\cite{Tonelli:2006np} is no longer considered.}},
$\Bs \to D_s^+D_s^-$ by LHCb~\cite{Aaij:2013bvd}
and $\Bs \to J/\psi \eta$ by LHCb~\cite{Aaij:2016dzn,LHCb-PAPER-2022-010}, as well as the \CP-odd modes 
$\Bs \to \jpsi f_0(980)$ by CDF~\cite{Aaltonen:2011nk}
and \dzero~\cite{Abazov:2016oqi},
$\Bs \to \jpsi \pi^+\pi^-$ by LHCb%
\footnote{The result of Ref.~\cite{Aaij:2019mhf} for the $\Bs \to \jpsi \pi^+\pi^-$ lifetime is not obtained through a single exponential fit of the lifetime distribution, but through a full time-dependent amplitude analysis, which concludes that the \CP-odd fraction is greater than 97\% at \CL{95} and yields $\Gamma_{s\rm H}-\Gamma_{d} = -0.050 \pm 0.004 \pm 0.004~{\rm ps}^{-1}$, where $1/\Gamma_{d}$ is the \Bd lifetime. Before being averaged with other determinations of the $\Bs \to \jpsi \pi^+\pi^-$ effective lifetime, this result is converted to a measurement of $1/\Gamma_{s\rm H}$ using the latest average of the \Bd lifetime.}%
~\citehistory{Aaij:2013oba,Aaij:2019mhf}{Aaij:2013oba,*LHCb:2011aa_hist,*LHCb:2012ad_hist,*LHCb:2011ab_hist,*Aaij:2012nta_hist,Aaij:2019mhf}
and CMS~\cite{Sirunyan:2017nbv},
and $\Bs \to \jpsi K^0_{\rm S}$ by LHCb~\cite{Aaij:2013eia} and CMS~\cite{CMS:2024cjq}.
If these 
decays are dominated by a single weak phase and if
\CP violation 
can be neglected, then $\tau_{\rm single}(\Bs \to \mbox{\CP-even}) = 1/\Gamma_{s\rm L}$ 
and  $\tau_{\rm single}(\Bs \to \mbox{\CP-odd}) = 1/\Gamma_{s\rm H}$ 
(see \Eqss{tau_KK_approx}{tau_Jpsif0_approx} for approximate relations in the presence of mixing-induced
\CP violation). 
However, not all these modes are pure \CP eigenstates:
a small \CP-odd component is most probably present 
in $\Bs \to D_s^{*+}D_s^{*-}$ decays~\cite{Aleksan:1993qp}. Furthermore, the decays
$\Bs \to K^+ K^-$ and $\Bs \to \jpsi K^0_{\rm S}$ %
may suffer from direct \CP violation due to interfering tree and loop amplitudes. 
The averages for the effective lifetimes obtained for decays to the
(nearly) pure \CP-even ($D_s^+D_s^-$, $\jpsi\eta$) and \CP-odd ($\jpsi f_0(980)$, $\jpsi \pi^+\pi^-$)
final states
are
\begin{eqnarray}
\tau_{\rm single}(\Bs \to \mbox{\CP-even}) & = & \hflavTAUBSSHORT \,,
\labe{tau_KK}
\\
\tau_{\rm single}(\Bs \to \mbox{\CP-odd}) & = & \hflavTAUBSLONG \,.
\labe{tau_Jpsif0}
\end{eqnarray}

\end{itemize}

As described in \Sec{DGs}, 
the effective lifetime averages of \Eqsss{tau_fs}{tau_KK}{tau_Jpsif0}
are used as constraints to improve the 
determination of $1/\Gs$ and \DGs obtained from the full angular analyses
of $\Bs\to \jpsi\phi$, $\Bs\to \psi(2S)\phi$ and $\Bs\to \jpsi K^+K^-$ decays. 
The resulting world averages for the \Bs lifetimes are
\begin{eqnarray}
\tau(B^0_{s\rm L}) = \frac{1}{\Gamma_{s\rm L}}
 = \frac{1}{\Gs+\DGs/2} & = & \hflavTAUBSLCON \,, \\
\tau(B^0_{s\rm H}) = \frac{1}{\Gamma_{s\rm H}}
 = \frac{1}{\Gs-\DGs/2} & = & \hflavTAUBSHCON \,, \\
\tau(\Bs) = \frac{1}{\Gs} = \frac{2}{\Gamma_{s\rm L}+\Gamma_{s\rm H}} & = & \hflavTAUBSMEANCON \,.
\labe{oneoverGs}
\end{eqnarray}

\mysubsubsection{\Bc lifetime}
\labs{taubc}

Early measurements of the \Bc meson lifetime,
from CDF~\unpublished{\cite{Abe:1998wi,Abulencia:2006zu}}{\cite{Abe:1998wi,CDFnote9294:2008,Abulencia:2006zu}} and \dzero~\cite{Abazov:2008rba},
use the semileptonic decay mode \particle{\Bc \to \jpsi \ell^+ \nu} 
and are based on a 
simultaneous fit to the mass and lifetime using the vertex formed
with the leptons from the decay of the \particle{\jpsi} and
the third lepton. Correction factors are used
to estimate the boost,  which cannot be measured directly due to the invisible neutrino.
Systematic uncertainties that are correlated among the measurements include the impact
of the uncertainty of the \Bc transverse-momentum spectrum on the correction
factors, the level of feed-down from $\psi(2S)$ decays, 
Monte Carlo modeling of the decay (estimated by varying the decay model from phase space
to the ISGW model), and uncertainties in the \Bc mass.
With more data, CDF2 was able to perform the first \Bc lifetime measurement
based on fully reconstructed
$\Bc \to J/\psi \pi^+$ decays~\cite{Aaltonen:2012yb},
which does not suffer from a missing neutrino.
More recent measurements at the LHC, both with  
\particle{\Bc \to \jpsi \mu^+ \nu} decays from LHCb~\cite{Aaij:2014bva} and 
\particle{\Bc \to \jpsi \pi^+} decays from LHCb~\cite{Aaij:2014gka} and CMS~\cite{Sirunyan:2017nbv},
achieve the highest level of precision. Two of them~\cite{Aaij:2014gka,Sirunyan:2017nbv} are made relative to the \Bu lifetime. Before averaging, they are scaled to our latest \Bu lifetime average, and the induced correlation is taken into account. 

All the measurements\unpublished{}{\footnote{We do not list (nor include in the average) an unpublished result from CDF2~\cite{CDFnote9294:2008}.}}
are summarized in 
\Table{lifebc}. The world average, dominated by the LHCb measurements, is
determined to be
\begin{equation}
\tau(\Bc) = \hflavTAUBC \,,
\end{equation}
which agrees well with the corresponding theory predictions in Refs.~\cite{Beneke:1996xe,Aebischer:2021ilm}.

\begin{table}[tb]
\centering
\caption{Measurements of the \Bc lifetime.}
\labt{lifebc}
\begin{tabular}{lccrcl} \hline
Experiment & Method                    & \multicolumn{2}{c}{Data set}  & $\tau(\Bc)$ (ps)
      & Ref.\\   \hline
CDF1       & \particle{\jpsi \ell} & 92--95 & 0.11 fb$^{-1}$ & $0.46^{+0.18}_{-0.16} \pm
 0.03$   & \cite{Abe:1998wi}  \\ 
CDF2       & \particle{\jpsi e} & 02--04 & 0.36 fb$^{-1}$ & $0.463^{+0.073}_{-0.065} \pm 0.036$   & \cite{Abulencia:2006zu} \\
 \dzero & \particle{\jpsi \mu} & 02--06 & 1.3 fb$^{-1}$  & $0.448^{+0.038}_{-0.036} \pm 0.032$
   & \cite{Abazov:2008rba}  \\
CDF2       & \particle{\jpsi \pi} & & 6.7 fb$^{-1}$ & $0.452 \pm 0.048 \pm 0.027$  & \cite{Aaltonen:2012yb} \\
LHCb & \particle{\jpsi \mu} & 2012 & 2 fb$^{-1}$  & $0.509 \pm 0.008 \pm 0.012$ & \cite{Aaij:2014bva}  \\
LHCb & \particle{\jpsi \pi} & 11--12 & 3 fb$^{-1}$  & $0.5134 \pm 0.0110 \pm 0.0057$ & \cite{Aaij:2014gka} \\
CMS  & \particle{\jpsi \pi} & 2012 & 19.7 fb$^{-1}$  & $0.541  \pm 0.026  \pm 0.014$ & \cite{Sirunyan:2017nbv} \\
\hline
  \multicolumn{2}{l}{Average} & &  &  \hflavTAUBCnounit
                 &    \\   \hline
\end{tabular}
\end{table}

\mysubsubsection{\Lb and other \b-baryon lifetimes}
\labs{taulb}

The first measurements of \b-baryon lifetimes, performed at LEP,
originate from two classes of partially reconstructed decays.
In the first class, decays with a fully
reconstructed \Lc baryon
and a lepton of opposite charge are used. These products are
likely to occur in the decay of \Lb baryons.
In the second class, more inclusive final states with a baryon
(\particle{p}, \particle{\bar{p}}, $\Lambda$, or $\bar{\Lambda}$) 
and a lepton have been used, and these final states can generally
arise from any \b baryon.  With the large \b-hadron samples available
at the Tevatron and the LHC, the most precise measurements of \b baryons now
come from fully reconstructed exclusive decays.

The following sources of correlated systematic uncertainties have 
been accounted for when averaging these measurements:
experimental time resolution within a given experiment, \b-quark
fragmentation distribution into weakly decaying \b baryons,
\Lb polarization, decay model,
and evaluation of the \b-baryon purity in the selected event samples.
In computing the averages,
the central values of the masses are scaled to 
$M(\Lb) = 5619.60 \pm 0.17\MeVcc$~\cite{PDG_2020}.

For measurements with partially reconstructed decays,
the meaning of the decay model
systematic uncertainties
and the correlation of these uncertainties between measurements
are not always clear.
Uncertainties related to the decay model are dominated by
assumptions on the fraction of $n$-body semileptonic decays.
To be conservative, it is assumed
that these are 100\% correlated whenever given as an uncertainty.
DELPHI varies the fraction of four-body decays from 0.0 to 0.3. 
In computing the average, the DELPHI
result is scaled to a value of $0.2 \pm 0.2$ for this fraction.
Furthermore,
the semileptonic decay results from LEP are scaled to a
$\Lb$ polarization of 
$-0.45^{+0.19}_{-0.17}$~\cite{Abbaneo:2000ej_mod,*Abbaneo:2001bv_mod_cont}
and a $b$ fragmentation parameter
$\langle x_E \rangle_b =0.702\pm 0.008$~\cite{ALEPH:2005ab}.

\renewcommand{\arraystretch}{0.95}
\begin{table}[!p]
\centering
\caption{Measurements of the \b-baryon lifetimes.
}
\labt{lifelb}
\begin{tabular}{lcccl} 
\hline
Experiment & Method    &Dataset& Lifetime (ps) & Ref. \\\hline
\multicolumn{5}{c}{\boldmath \b \textbf{baryon admixture}}\\
ALEPH  &$\Lambda\ell$         & 91--95 &$1.20 \pm 0.08 \pm 0.06$ & \cite{Barate:1997if}\\
DELPHI &$\Lambda\ell\pi$ vtx  & 91--94 &$1.16 \pm 0.20 \pm 0.08$        & \cite{Abreu:1999hu}$^b$\\
DELPHI &$\Lambda\mu$ i.p.     & 91--94 &$1.10^{+0.19}_{-0.17} \pm 0.09$ & \cite{Abreu:1996nt}$^b$ \\
DELPHI &\particle{p\ell}      & 91--94 &$1.19 \pm 0.14 \pm 0.07$        & \cite{Abreu:1999hu}$^b$\\
OPAL   &$\Lambda\ell$ i.p.    & 90--94 &$1.21^{+0.15}_{-0.13} \pm 0.10$ & \cite{Akers:1995ui}$^c$  \\
OPAL   &$\Lambda\ell$ vtx     & 90--94 &$1.15 \pm 0.12 \pm 0.06$        & \cite{Akers:1995ui}$^c$ \\
\end{tabular}
\end{table}
\renewcommand{\arraystretch}{1.0}

The list of all measurements is given in \Table{lifelb}.
We do not attempt to average measurements performed with $p\ell$ or 
$\Lambda\ell$ combinations, which select unknown mixtures of $b$ baryons. 
Measurements performed with $\Lc\ell$ or $\Lambda\ell^+\ell^-$
combinations can be assumed to correspond to semileptonic \Lb decays. 
Their average (\hflavTAULBS) is significantly different 
from the average using only measurements performed with
exclusively reconstructed hadronic \Lb decays (\hflavTAULBE). 
The latter is much more precise
and less prone to potential biases than the former. 
The discrepancy between the two averages is at the level of
$\hflavNSIGMATAULBEXCLSEMI\sigma$ 
and assumed to be due to a systematic effect in the 
semileptonic measurements, where the \Lb momentum is not determined directly, or to a rare statistical fluctuation.
Our final estimate of the \Lb lifetime is therefore based 
on the exclusive  measurements only. 
The CDF $\Lb \to \jpsi \Lambda$
lifetime result~\citehistory{Aaltonen:2014wfa}{Aaltonen:2014wfa,*Aaltonen:2014wfa_hist} 
is larger than the average of all other exclusive measurements
by $\hflavNSIGMATAULBCDFTWO\sigma$. 
It is nonetheless kept in the combination
without adjustment of input uncertainties.

For the strange \b baryons, we do not include the measurements based on
inclusive $\Xi^{\mp} \ell^{\mp}$
final states, which consist of a mixture of 
$\Xibd$ and $\Xibu$ baryons. Rather, we only use results obtained with 
fully reconstructed $\Xibd$, $\Xibu$ and $\Omegab$ baryons.

It should be noted that several $b$-baryon lifetime measurements from LHCb~%
\citehistory{Aaij:2014zyy,Aaij:2014lxa,Aaij:2014esa,Aaij:2016dls}{Aaij:2014zyy,*Aaij:2013oha_hist,Aaij:2014lxa,Aaij:2014esa,Aaij:2016dls,LHCb:2024lao}
were made with respect to the lifetime of another $b$ hadron
(\ie, the original measurement is that of a decay width difference). 
In particular, LHCb uses the $\Lb\to\Lc\pi^-$ decay as a normalisation channel for the $\tau(\Xibu)$ and $\tau(\Xibd)$ measurements, but does not use this decay for a measurement of $\tau(\Lb)$. Therefore, we average $\Lc$, $\Xibd$ and $\Xibu$ lifetime simultaneously, taking all relevant correlations into account and constraining $\tau(\Bd)$ to its average from \Sec{taubd}. We subsequently average the $\Omegab$ lifetime constraining the $\tau(\Xibd)$ to its value obtained in the previous step. We obtain
\begin{eqnarray}
\tau(\Lb) &=& \hflavTAULB \,,\\
\tau(\Xibd) &=& \hflavTAUXBD \,, \\
\tau(\Xibu) &=& \hflavTAUXBU \,, \\
\tau(\Omegab) &=& \hflavTAUOB \,. 
\end{eqnarray}

\mysubsubsection{Summary and comparison with theoretical predictions}
\labs{lifesummary}

Averages of lifetimes of specific \b-hadron species are collected
in \Table{sumlife}.
\begin{table}[t]
\centering
\caption{Summary of the lifetime averages for the different \b-hadron species.}
\labt{sumlife}
\begin{tabular}{lrc} \hline
\multicolumn{2}{l}{\b-hadron species} & Measured lifetime \\ \hline
\Bu &                       & \hflavTAUBU   \\
\Bd &                       & \hflavTAUBD   \\
\Bs & $1/\Gs~\, =$               & \hflavTAUBSMEANCON \\
~~ $B^0_{s\rm L}$ & $1/\Gamma_{s\rm L}=$  & \hflavTAUBSLCON \\
~~ $B^0_{s\rm H}$ & $1/\Gamma_{s\rm H}=$  & \hflavTAUBSHCON \\
\Bc     &                   & \hflavTAUBC   \\ 
\Lb     &                   & \hflavTAULB   \\
\Xibd   &                   & \hflavTAUXBD  \\
\Xibu   &                   & \hflavTAUXBU  \\
\Omegab &                   & \hflavTAUOB   \\
\hline
\end{tabular}
\end{table}
\begin{table}[t]
\centering
\caption{Experimental averages of \b-hadron lifetime ratios and
heavy-quark expansion (HQE) predictions.}
\labt{liferatio}
\begin{tabular}{lcc} \hline
Lifetime ratio & Experimental average & HQE prediction \\ \hline
$\tau(\Bu)/\tau(\Bd)$ & \hflavRTAUBU & $1.086 \pm 0.022$~\cite{Albrecht:2024oyn} \\
$\tau(\Bs)/\tau(\Bd)$ & \hflavRTAUBSMEANC & 
$1.028 \pm 0.011 $ or $1.003 \pm 0.006$~\cite{Albrecht:2024oyn} \\
$\tau(\Lb)/\tau(\Bd)$ & \hflavRTAULB & $0.955 \pm 0.014$~\cite{Albrecht:2024oyn} \\
$\tau(\Xibu)/\tau(\Xibd)$ & \hflavRTAUXBUXBD & $0.929 \pm 0.028$~\cite{Albrecht:2024oyn} \\
$\tau(\Omegab)/\tau(\Bd)$ &  \hflavRTAUOB & $1.081 \pm 0.042$~\cite{Albrecht:2024oyn}  \\
\hline
\end{tabular}
\end{table}
As described in the introduction to \Sec{lifetimes},
the HQE can be employed to explain the hierarchy 
$\tau(\Bc) \ll \tau(\Lb) < \tau(\Bs) \approx \tau(\Bd) < \tau(\Bu)$,
and to predict the ratios between lifetimes.
Recent predictions are compared to the measured 
lifetime ratios in \Table{liferatio}, where the experimental values 
have been computed as the ratio of our averages of the individual lifetimes. The ratios $\tau(\Lb)/\tau(\Bd), \tau(\Xibu)/\tau(\Xibd)$, and $\tau(\Bs)/\tau(\Bd)$ are computed taking into account the correlation between the numerators and denominators.
The ratio $\tau(\Bu)/\tau(\Bd)$ use only measurements of this ratio (see \Sec{lifetime_ratio}). The ratio $\tau(\Omegab)/\tau(\Bd)$ is computed neglecting the small correlation between both lifetimes.

The predictions of the ratio between the \Bu and \Bd lifetimes,
$1.086 \pm 0.022$~\cite{Albrecht:2024oyn}
is in good agreement with experiment.
The prediction of the ratio $\tau(\Bs)/\tau(\Bd)$ is less precise. 
Depending on the theoretical hypotheses, two solutions are proposed in~\Refx{Albrecht:2024oyn}:
\begin{equation}
    \frac{\tau(\Bs)}{\tau(\Bd)} = 1.028 \pm 0.011\,,
\end{equation}
\begin{equation}
    \frac{\tau(\Bs)}{\tau(\Bd)} = 1.003 \pm 0.006\,.
\end{equation}
These arise from the difference in the inputs in the calculation, taking the relevant parameters from \Refx{Bordone:2021oof} in the first and \Refx{Bernlochner:2022ucr} in the second solutions.
The second solution is much more compatible with the experimental result. 

The HQE prediction for the ratio $\tau(\Lb)/\tau(B^0) $ is now in excellent agreement with the data. Interestingly, in the 1990s a much
lower experimental value for  $\tau(\Lb)$ was measured, and the world average for the ratio at the time resulted to be almost $4\sigma$ below
the naive HQE prediction~\cite{Shifman:1986mx}. This large discrepancy triggered a considerable amount of interest in the theory community,
see \textit{e.g.}~\Refx{Bigi:2001ys,Albrecht:2024oyn}, 
and the validity of the HQE was put under scrutiny
\cite{altarelli:1996gt,cheng:1997xba,ito:1997qq}. Ultimately, the experimental value
of the $\Lb$ lifetime changed and the $\tau(\Lb)/\tau(B^0) $ ratio  were found to be in agreement with the HQE result.

There is good agreement for the 
$\tau(\Xibu)/\tau(\Xibd)$ ratio and for $\tau(\Omega_b^-)/\tau(B^0)$ ratio, 
for which the prediction is
based on the next-to-leading-order analysis of~\Refx{Gratrex:2023pfn}. It is interesting to note that for the latter ratio, the theory precision is even higher than the experimental one.
\mysubsection{Neutral \B-meson mixing}
\labs{mixing}

The $\Bd-\Bdbar$ and $\Bs-\Bsbar$ systems
both exhibit the phenomenon of particle-antiparticle mixing. For each of them, 
there are two mass eigenstates which are linear combinations of the two flavor states,
$B^0_q$ and $\bar{B}^0_q$, 
\begin{eqnarray}
| B^0_{q\rm L}\rangle &=& p_q |B^0_q \rangle +  q_q |\bar{B}^0_q \rangle \,, \\
| B^0_{q\rm H}\rangle &=& p_q |B^0_q \rangle -  q_q |\bar{B}^0_q \rangle  \,,
\end{eqnarray}
where the subscript $q=d$ is used for the  $B^0_d$ ($=\Bd$) meson and $q=s$ for the \Bs meson.
The heavier (lighter) of these mass states is denoted
$B^0_{q\rm H}$ ($B^0_{q\rm L}$),
with mass $m_{q\rm H}$ ($m_{q\rm L}$)
and total decay width $\Gamma_{q\rm H}$ ($\Gamma_{q\rm L}$). We define
\begin{eqnarray}
\Delta m_q = m_{q\rm H} - m_{q\rm L} \,, &~~~~&  x_q = \Delta m_q/\Gamma_q \,, \labe{dm} \\
\Delta \Gamma_q \, = \Gamma_{q\rm L} - \Gamma_{q\rm H} \,, ~ &~~~~&  y_q= \Delta\Gamma_q/(2\Gamma_q) \,, \labe{dg}
\end{eqnarray}
where 
$\Gamma_q = (\Gamma_{q\rm H} + \Gamma_{q\rm L})/2 =1/\bar{\tau}(B^0_q)$ 
is the average decay width. By definition,
$\Delta m_q$ is positive, and 
$\Delta \Gamma_q$ is expected to be positive within
the Standard Model.\footnote{
  \label{foot:life_mix:Eqdg}
  For reasons of symmetry in \Eqss{dm}{dg}, 
  $\Delta \Gamma$ is sometimes defined with the opposite sign. 
  The definition adopted in \Eq{dg} is the one used
  in most experimental papers and many phenomenology papers on \B physics.}

Four different time-dependent probabilities are needed to describe the 
evolution of a neutral \B meson that is produced as a flavor state and decays without
\CP violation to a flavor-specific final state. 
If \CPT is conserved, which  
will be assumed throughout, they can be written as 
\begin{equation}
\left\{
\begin{array}{rcl}
{\cal P}(B^0_q \to B^0_q) & = &  \frac{1}{2} e^{-\Gamma_q t} 
\left[ \cosh\!\left(\frac{1}{2}\Delta\Gamma_q t\right) + \cos\!\left(\Delta m_q t\right)\right]  \\
{\cal P}(B^0_q \to \bar{B}^0_q) & = &   \frac{1}{2} e^{-\Gamma_q t} 
\left[ \cosh\!\left(\frac{1}{2}\Delta\Gamma_q t\right) - \cos\!\left(\Delta m_q t\right)\right] 
\left|q_q/p_q\right|^2 \\
{\cal P}(\bar{B}^0_q \to B^0_q) & = &  \frac{1}{2} e^{-\Gamma_q t} 
\left[ \cosh\!\left(\frac{1}{2}\Delta\Gamma_q t\right) - \cos\!\left(\Delta m_q t\right)\right] 
\left|p_q/q_q\right|^2 \\
{\cal P}(\bar{B}^0_q \to\bar{B}^0_q) & = &  \frac{1}{2} e^{-\Gamma_q t}  
\left[ \cosh\!\left(\frac{1}{2}\Delta\Gamma_q t\right) + \cos\!\left(\Delta m_q t\right)\right] 
\end{array} \right. \,,
\labe{oscillations}
\end{equation}
where $t$ is the proper time of the system. 
At the \B factories BaBar and Belle (II), only the proper-time difference $\Delta t$ between the decays
of the two neutral \B mesons from the \Ups can be determined. However,
since the two \B mesons evolve coherently, meaning that they keep opposite flavors as long
as neither of them has decayed, the 
above formulas remain valid 
if $t$ is replaced with $\Delta t$ and the production flavor is replaced by the flavor 
at the time of the decay of the accompanying \B meson into a flavor-specific state.
As can be seen in the above expressions,
the mixing probabilities 
depend on three mixing observables:
$\Delta m_q$, $\Delta\Gamma_q$,
and $|q_q/p_q|^2$. In particular, \CP violation in mixing exists if $|q_q/p_q|^2 \ne 1$.
Another (nonindependent) observable often used to characterize \CP violation in the mixing 
is the so-called semileptonic asymmetry, as it is measured using \Bd semileptonic decays (see \Sec{qpd}),  defined as
\begin{equation} 
{\cal A}_{\rm SL}^q = 
\frac{|p_{\particle{q}}/q_{\particle{q}}|^2 - |q_{\particle{q}}/p_{\particle{q}}|^2}%
{|p_{\particle{q}}/q_{\particle{q}}|^2 + |q_{\particle{q}}/p_{\particle{q}}|^2} \,.
\labe{ASLq}
\end{equation} 
All  mixing observables depend on two complex numbers, $M^q_{12}$ and $\Gamma^q_{12}$, which are the off-diagonal elements of the $2\times 2$ mass and decay matrices describing the evolution of the $B^0_q-\bar{B}^0_q$ system. In the Standard Model, the quantity $|\Gamma^q_{12}/M^q_{12}|$ is small, of the order of $(m_b/m_t)^2$, where $m_b$ and $m_t$ are the bottom and top quark masses. The following relations hold to first order in $|\Gamma^q_{12}/M^q_{12}|$:
\begin{eqnarray}
\Delta m_q & = & 2 |M^q_{12}| \left[1 + {\cal O} \left(|\Gamma^q_{12}/M^q_{12}|^2 \right) \right] \,, \\
\Delta\Gamma_q & = & 2 |\Gamma^q_{12}| \cos\phi^q_{12} \left[1 + {\cal O} \left(|\Gamma^q_{12}/M^q_{12}|^2 \right) \right]   \,, \\
{\cal A}_{\rm SL}^q & = &  \Im \left(\Gamma^q_{12}/M^q_{12} \right) +
{\cal O} \left(|\Gamma^q_{12}/M^q_{12}|^2 \right) =
\frac{\Delta\Gamma_q}{\Delta m_q}\tan\phi^q_{12} +
{\cal O} \left(|\Gamma^q_{12}/M^q_{12}|^2 \right)  \,,
\labe{ALSq_tanphi2}
\end{eqnarray}
where 
\begin{equation}
\phi^q_{12} = \arg \left( -{M^q_{12}}/{\Gamma^q_{12}} \right) \,,
\labe{phi12}
\end{equation}
often called the mixing phase,
is the observable phase difference between $-M^q_{12}$ and $\Gamma^q_{12}$. 
It should be noted that the theoretical predictions for $\Gamma^q_{12}$ are based on a similar HQE as the lifetime predictions, see Refs.~\cite{Beneke:1996gn,Beneke:1998sy,Ciuchini:2003ww,Beneke:2003az,Gerlach:2022hoj},
for perturbative corrections, and
Refs.~\cite{Bazavov:2016nty,Grozin:2016uqy,Boyle:2018knm,King:2019lal,Dowdall:2019bea,Davies:2019gnp}
for non-perturbative matrix elements.

In the next sections we review in turn the experimental knowledge
on the \Bd decay-width and mass differences, 
the \Bs decay-width and mass differences,  
\CP violation in \Bd and \Bs mixing, and mixing-induced \CP violation in \Bs decays. 

\mysubsubsection{\Bd mixing parameters \DGd and \dmd}
\labs{DGd} \labs{dmd}

\begin{table}
\centering
\caption{Time-dependent measurements included in the \dmd average.
The results obtained from multidimensional fits involving also 
the \Bd (and \Bu) lifetime(s)
as free parameter(s)~\protect\citehistory{Aubert:2002sh,Aubert:2005kf,Abe:2004mz}{Aubert:2002sh,Aubert:2005kf,Abe:2004mz,*Abe:2002id_hist,*Tomura:2002qs_hist,*Hara:2002mq_hist} 
have been converted into one-dimensional measurements of \dmd.
All measurements have then been adjusted to a common set of physics
parameters before being combined.
The label \textit{rec.} refers to the decay channel used to reconstruct the signal \Bd, and \textit{tag} refers to the technique used to tag the initial flavour of that \Bd.}
\labt{dmd}
\vspace{-0.2cm}
\begin{tabular}{@{}rc@{}cc@{}c@{}cc@{}c@{}c@{}}
\hline
Experiment & \multicolumn{2}{c}{Method} & \multicolumn{3}{l}{\dmd in\invps}   
                                        & \multicolumn{3}{l}{\dmd in\invps}     \\[-0.8ex]
and Ref.   &  rec. & tag                & \multicolumn{3}{l}{before adjustment} 
                                        & \multicolumn{3}{l}{after adjustment} \\
\hline
ALEPH~\cite{Buskulic:1996qt}  & \particle{ \ell  } & \particle{ \Qjet  } & $  0.404 $ & $ \pm  0.045 $ & $ \pm  0.027 $ & & & \\ 
ALEPH~\cite{Buskulic:1996qt}  & \particle{ \ell  } & \particle{ \ell  } & $  0.452 $ & $ \pm  0.039 $ & $ \pm  0.044 $ & & & \\ 
ALEPH~\cite{Buskulic:1996qt}  & \multicolumn{2}{c}{above two combined} & $  0.422 $ & $ \pm  0.032 $ & $ \pm  0.026 $& $  0.439 $ & $ \pm  0.032 $ & $ \pm  0.019 $ \\ 
ALEPH~\cite{Buskulic:1996qt}  & \particle{ D^*  } & \particle{ \ell,\Qjet  } & $  0.482 $ & $ \pm  0.044 $ & $ \pm  0.024 $ & $  0.483 $ & $ \pm  0.044 $ & $ \pm  0.024 $ \\ 
DELPHI~\cite{Abreu:1997xq}  & \particle{ \ell  } & \particle{ \Qjet  } & $  0.493 $ & $ \pm  0.042 $ & $ \pm  0.027 $ & $  0.498 $ & $ \pm  0.041 $ & $ \pm  0.031 $ \\ 
DELPHI~\cite{Abreu:1997xq}  & \particle{ \pi^*\ell  } & \particle{ \Qjet  } & $  0.499 $ & $ \pm  0.053 $ & $ \pm  0.015 $ & $  0.500 $ & $ \pm  0.052 $ & $ \pm  0.033 $ \\
\hline
\end{tabular}
\end{table}

A large number of time-dependent \Bd--\Bdbar oscillation analyses
have been performed by the 
ALEPH, DELPHI, L3, OPAL, CDF, \dzero, \babar, \belle, \belletwo and  LHCb collaborations. 
The corresponding measurements of \dmd are summarized in 
\Table{dmd}\history{.}{,
where only the most recent results
are listed (\ie\ measurements superseded by more recent ones are omitted\unpublished{}{\footnote{
  \label{foot:life_mix:CDFnote8235:2006}
  Two old unpublished CDF2 measurements~\cite{CDFnote8235:2006,CDFnote7920:2005}
  are also omitted from our averages, \Table{dmd} and \Fig{dmd}.}}).}
It is notable that the systematic uncertainties are comparable to the statistical uncertainties;
they are often dominated by sample composition, mistag probability,
or \b-hadron lifetime contributions.
Before being combined, the measurements are adjusted to a 
common set of input values, including the averages of the 
 lifetimes given in this report 
(see \Secss{fractions}{lifetimes}) and the \b-hadron fractions
of Ref.~\cite{Amhis:2019ckw}.
Three of the \babar, four of the DELPHI, and two of the OPAL measurements have small statistical correlations among each other.
Taking into account the fraction of the dataset shared among these measurements, 
statistical correlations are fully accounted in the combination.
Systematic correlations arise both from common physics sources 
(fractions, lifetimes, branching fractions of \b hadrons), and from purely 
experimental or algorithmic effects (efficiency, resolution, flavor tagging, 
background description). Combining all published measurements
listed in \Table{dmd}
and accounting for all relevant correlations
yields $\dmd = \hflavDMDWfull$.
This measurement is in agreement with the latest Standard Model prediction of $0.535\pm0.021\;\invps$~\cite{Albrecht:2024oyn}.

In addition, ARGUS and CLEO have published 
measurements of the time-integrated mixing probability 
\chid~\cite{Albrecht:1992yd,*Albrecht:1993gr,Bartelt:1993cf,Behrens:2000qu}, 
which average to $\chid =\hflavCHIDU$.
Following \Refx{Behrens:2000qu}, 
the decay width difference \DGd could 
in principle be extracted from the
measured value of $\Gd=1/\tau(\Bd)$ and the above averages for 
\dmd and \chid 
(provided that \DGd has a negligible impact on 
the \dmd and $\tau(\Bd)$ analyses that have assumed $\DGd=0$), 
using the relation
\begin{equation}
\chid = \frac{\xd^2+\yd^2}{2(\xd^2+1)}
\,.
\labe{chid_definition}
\end{equation}
However, $\DGd/\Gd$ is too small and the knowledge of \chid too imprecise to provide useful sensitivity on $\DGd/\Gd$. 
Direct time-dependent studies provide much stronger constraints: 
$|\DGd|/\Gd < 18\%$ at \CL{95} from DELPHI~\cite{Abdallah:2002mr},
$-6.8\% < {\rm sign}({\rm Re} \lambda_{\CP}) \DGGd < 8.4\%$
at \CL{90} from \babar~\cite{Aubert:2003hd,*Aubert:2004xga},
and ${\rm sign}({\rm Re} \lambda_{\CP})\DGGd = (1.7 \pm 1.8 \pm 1.1)\%$~\cite{Higuchi:2012kx}
from Belle, 
where $\lambda_{\CP} = (q_{\particle{d}}/p_{\particle{d}}) A(\Bdbar\to f_{\CP})/A(\Bd\to f_{\CP})$
with $A$ %
denoting decays amplitudes to a \CP-even final state. 
The sensitivity to the overall sign of 
${\rm Re} \lambda_{\CP} \DGGd$ comes
from the use of \Bd decays to \CP eigenstates.
In addition,
LHCb has obtained $\DGGd=(-4.4 \pm 2.5 \pm 1.1)\%$~\cite{Aaij:2014owa}
by comparing measurements of the lifetime for $\Bd \to \jpsi K^{*0}$ and $\Bd \to \jpsi K^0_{\rm S}$
decays, following the method of Ref.~\cite{Gershon:2010wx}.
Using a similar method, ATLAS and CMS have measured %
$\DGGd=(-0.1 \pm 1.1 \pm 0.9)\%$~\cite{Aaboud:2016bro} and
$\DGGd=(+3.4 \pm 2.3 \pm 2.4)\%$~\cite{Sirunyan:2017nbv}, respectively.
Assuming ${\rm Re} \lambda_{\CP} > 0$, as expected from the global fits
of the unitarity triangle within the Standard Model~\cite{Charles:2015gya_mod,UTfit:2022hsi_mod},
a combination of these six results (after adjusting the DELPHI and \babar results to  
$1/\Gd=\tau(\Bd)=\hflavTAUBD$) yields
\begin{equation}
\DGGd  = \hflavSDGDGD \,.
\end{equation}
This average is consistent with zero and with the Standard Model prediction of $(3.97\pm0.90)\times 10^{-3}$~\cite{Artuso:2015swg}. 
An independent result, 
$\DGGd=(0.50 \pm 1.38)\%$\citehistory{Abazov:2013uma}{Abazov:2013uma,*Abazov:2011yk_hist,*Abazov:2010hv_hist,*Abazov:2010hj_hist,*Abazov:dimuon_hist},
was obtained by the \dzero collaboration 
from their measurements of the single muon and same-sign dimuon charge asymmetries,
under the interpretation that 
the observed asymmetries are due to \CP violation in neutral $B$-meson mixing and interference.
This indirect determination was called into question~\cite{Nierste_CKM2014}
and is therefore not included in the above average, 
as explained in \Sec{qpd}.%

Assuming $\DGd=0$ 
and using $1/\Gd=\tau(\Bd)=\hflavTAUBD$,
the \dmd and \chid results are combined through \Eq{chid_definition} 
to yield the 
world average
\begin{equation} 
\dmd = \hflavDMDWU \,,
\labe{dmd}
\end{equation} 
or, equivalently,
\begin{equation} 
\xd= \hflavXDWU ~~~ \mbox{and} ~~~ \chid=\hflavCHIDWU \,.  
\labe{chid}
\end{equation}
\Figure{dmd} compares the \dmd values obtained by the different experiments.

\begin{figure}
\begin{center}
\includegraphics[width=\textwidth]{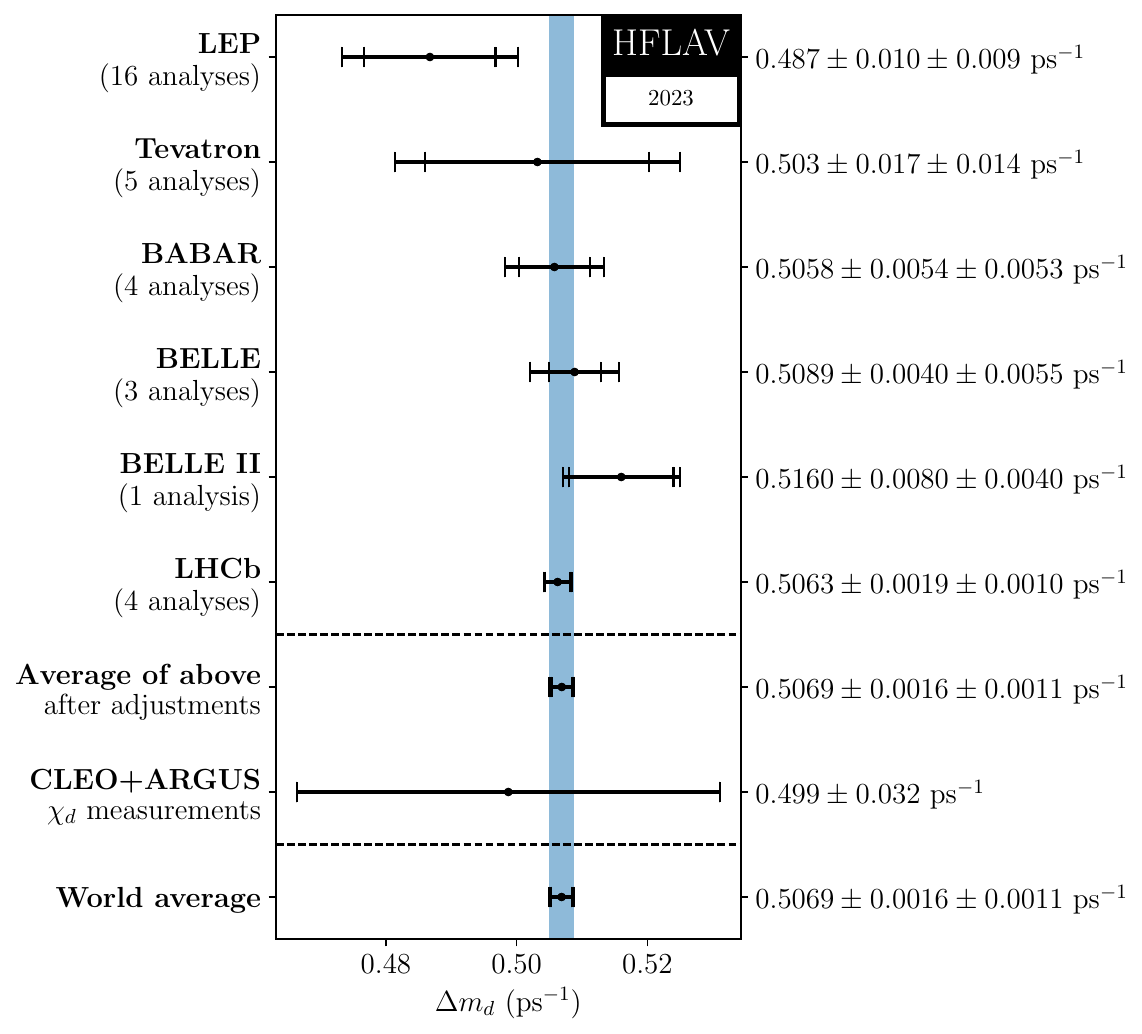}
\caption{The \Bd--\Bdbar oscillation frequency \dmd as measured by the different experiments. 
The LEP, Tevatron, \babar, \belle, \belletwo and LHCb averages are computed from the individual results 
listed in \Table{dmd} without performing any adjustments. The time-integrated measurements 
of \chid from the symmetric \B factory experiments ARGUS and CLEO are converted 
to a \dmd value using $\tau(\Bd)=\hflavTAUBD$. The two global averages are obtained 
after adjustments of all the individual \dmd results of \Table{dmd} (see text).}
\labf{dmd}
\end{center}
\end{figure}

The \chid and \xd averages are obtained by neglecting small statistical and systematic correlations between \dmd and $\tau(\Bd)$.

\mysubsubsection{\Bs mixing parameters \DGs and \dms}
\labs{DGs} \labs{dms}

The best sensitivity to \DGs is currently achieved 
by the time-dependent measurements
of the $\Bs\to\jpsi\phi$ (or more generally $\Bs\to (c\bar{c}) K^+K^-$) decay rates performed at
CDF~\citehistory{Aaltonen:2012ie}{Aaltonen:2012ie,*CDF:2011af_hist,*Aaltonen:2007he_hist,*Aaltonen:2007gf_hist},
\dzero~\citehistory{Abazov:2011ry}{Abazov:2011ry,*Abazov:2008af_hist,*Abazov:2007tx_hist},
ATLAS~\citehistory{Aad:2014cqa,Aad:2016tdj,Aad:2020jfw}{Aad:2014cqa,*Aad:2012kba_hist,Aad:2016tdj,Aad:2020jfw},
CMS~\cite{Khachatryan:2015nza,CMS-PAS-BPH-23-004}
and LHCb~\citehistory{Aaij:2014zsa,Aaij:2017zgz,Aaij:2016ohx, Aaij:2021mus,LHCb:2023sim}{Aaij:2014zsa,*Aaij:2013oba_supersede2,Aaij:2017zgz,*Aaij:2014zsa_partial_supersede,Aaij:2016ohx,Aaij:2021mus,LHCb:2023sim,*Aaij:2019vot_hist},
where the \CP-even and \CP-odd
amplitudes are statistically separated through a full angular analysis.
\unpublished{These}{With the exception of the first CMS analysis~\cite{CMS-PAS-BPH-11-006}%
\footnote{The CMS result of \Refx{CMS-PAS-BPH-11-006}
is statistically independent of that of
\Refx{Khachatryan:2015nza} but, since it has not
been published, it is not included in \Table{GsDGs} nor in our averages.},
these}
studies use both untagged and tagged \Bs\ candidates and 
are optimized for the measurement of the \CP-violating 
phase \phiccbars, defined later in \Sec{phasebs}.
The LHCb collaboration analyzed the $\Bs \to \jpsi K^+K^-$
decay, considering that the $K^+K^-$ system can be in a $P$-wave or $S$-wave state, 
and measured the dependence of the strong phase difference between the 
$P$-wave and $S$-wave amplitudes as a function of the $K^+K^-$ invariant
mass~\cite{Aaij:2012eq}. 
This allowed, for the first time, the unambiguous determination of the sign of 
$\DGs$, which was found to be positive at the $4.7\,\sigma$ level. 

The following averages present only the $\DGs > 0$ solutions. Two degenerate solutions, differing in the values of two of the measured strong phases, $\delta_{\perp}$ and $\delta_{\parallel}$, were found in the ATLAS Run~2 analysis~\cite{Aad:2020jfw}. These show minor differences in the \Bs lifetime and mixing parameters, so for simplicity, the following averages only use solution (a) of \Refx{Aad:2020jfw}.

The published results~\citehistory%
{Aaltonen:2012ie,Abazov:2011ry,Aad:2014cqa,Aad:2016tdj,Aad:2020jfw,Khachatryan:2015nza,CMS-PAS-BPH-23-004,Aaij:2014zsa,Aaij:2017zgz,Aaij:2016ohx,Aaij:2021mus,LHCb:2023sim}%
{Aaltonen:2012ie,*CDF:2011af_hist,*Aaltonen:2007he_hist,*Aaltonen:2007gf_hist,Abazov:2011ry,*Abazov:2008af_hist,*Abazov:2007tx_hist,Aad:2014cqa,*Aad:2012kba_hist,Aad:2016tdj,Aad:2020jfw,Khachatryan:2015nza,CMS-PAS-BPH-23-004,Aaij:2014zsa,*Aaij:2013oba_supersede2,Aaij:2017zgz,*Aaij:2014zsa_partial_supersede,Aaij:2016ohx,Aaij:2021mus,LHCb:2023sim,*Aaij:2019vot_hist}
are shown in \Table{GsDGs}. They are combined in a fit that includes all measured parameters and their correlations. These are $\phiccbars$, the direct {\it CP} violation parameter $|\lambda|=|(q_s/p_s)\bar{A}/A|$ ($\bar{A}$ and $A$ are the \Bsbar and \Bs decay amplitudes, respectively), 
$\Delta m_{s}$, polarization fractions and strong phases.
As detailed further in \Sec{phasebs}, the $\Bs\to\jpsi\Kp\Km$ measurements of ATLAS~\cite{Aad:2016tdj,Aad:2020jfw}, CMS~\citehistory{Khachatryan:2015nza,CMS-PAS-BPH-23-004}{Khachatryan:2015nza,*Sirunyan:2020vke} and LHCb~\citehistory{Aaij:2014zsa,LHCb:2023sim}{Aaij:2014zsa,LHCb:2023sim,*Aaij:2019vot_hist} are in tension at the level of several $\sigma$, driven by the time and angular parameters. To address this, the total uncertainty for each parameter in each $\Bs\to\jpsi\Kp\Km$ set of results by ATLAS, CDF, D0, CMS and LHCb is scaled up in a way such that the results are in an agreement at the $1\sigma$ level. For the parameters already in agreement, notably $\phi_{s}$, the uncertainties are not scaled. The covariance matrix is recomputed to preserve the correlations between the parameters. The resulting scale factors for \Gs and \DGs are \hflavGSCCKKSF and \hflavDGSCCKKSF.
The results, displayed as the red contours labeled ``$\Bs \to (c\bar{c}) KK$'' in \Fig{DGs}, are given in the first column of numbers of \Table{tabtauLH}.

\begin{table}
\caption{Measurements of \DGs and \Gs using
$\Bs\to\jpsi\Kp\Km$ and $\Bs\to\psi(2S)\Kp\Km$ decays.
Only the solution with $\DGs > 0$ is shown, since the two-fold ambiguity has been
resolved in \Refx{Aaij:2012eq}. The first error is due to 
statistics, the second one to systematics. The last line gives our average.}
\labt{GsDGs}
\begin{center}
\begin{tabular}{@{}l@{\,}l@{\,}rll@{}l@{}} 
\toprule
Exp.\ & Mode & Dataset
      & \multicolumn{1}{c}{\DGs (\!\!\invps)}
      & \multicolumn{1}{c}{\Gs  (\!\!\invps)}
      & Ref.\ \\
\midrule
CDF    & $\jpsi\Kp\Km$ & $9.6\invfb$
       & $+0.068\pm0.026\pm0.009$
       & $0.654\pm0.008\pm0.004$ %
       & \citehistory{Aaltonen:2012ie}{Aaltonen:2012ie,*CDF:2011af_hist,*Aaltonen:2007he_hist,*Aaltonen:2007gf_hist} \\
\dzero & $\jpsi\Kp\Km$ & $8.0\invfb$
       & $+0.163^{+0.065}_{-0.064}$ 
       & $0.693^{+0.018}_{-0.017}$
       & \citehistory{Abazov:2011ry}{Abazov:2011ry,*Abazov:2008af_hist,*Abazov:2007tx_hist} \\
\vspace{-0.3cm} & & & &  \\
ATLAS  & $\jpsi\Kp\Km$ & $4.9\invfb$
       & $+0.053 \pm0.021 \pm0.010$
       & $0.677 \pm0.007 \pm0.004$
       & \citehistory{Aad:2014cqa}{Aad:2014cqa,*Aad:2012kba_hist}  \\
ATLAS  & $\jpsi\Kp\Km$ & $14.3\invfb$
       & $+0.101 \pm0.013 \pm0.007$
       & $0.676 \pm0.004 \pm0.004$
       & \cite{Aad:2016tdj} \\
ATLAS  & $\jpsi\Kp\Km$ & $80.5\invfb$
       & $+0.0607 \pm0.0047 \pm0.0043$ 
       & $0.6687 \pm0.0015 \pm0.0022$ 
       & \cite{Aad:2020jfw} \\
ATLAS  & \multicolumn{2}{r}{above 3 combined}
       & $+0.0657 \pm0.0043 \pm0.0037$  
       & $0.6703 \pm0.0014 \pm0.0018$ 
       & \cite{Aad:2020jfw} \\
\end{tabular}
\end{center}
\end{table}

\begin{figure}
\begin{center}
\includegraphics[width=0.49\textwidth]{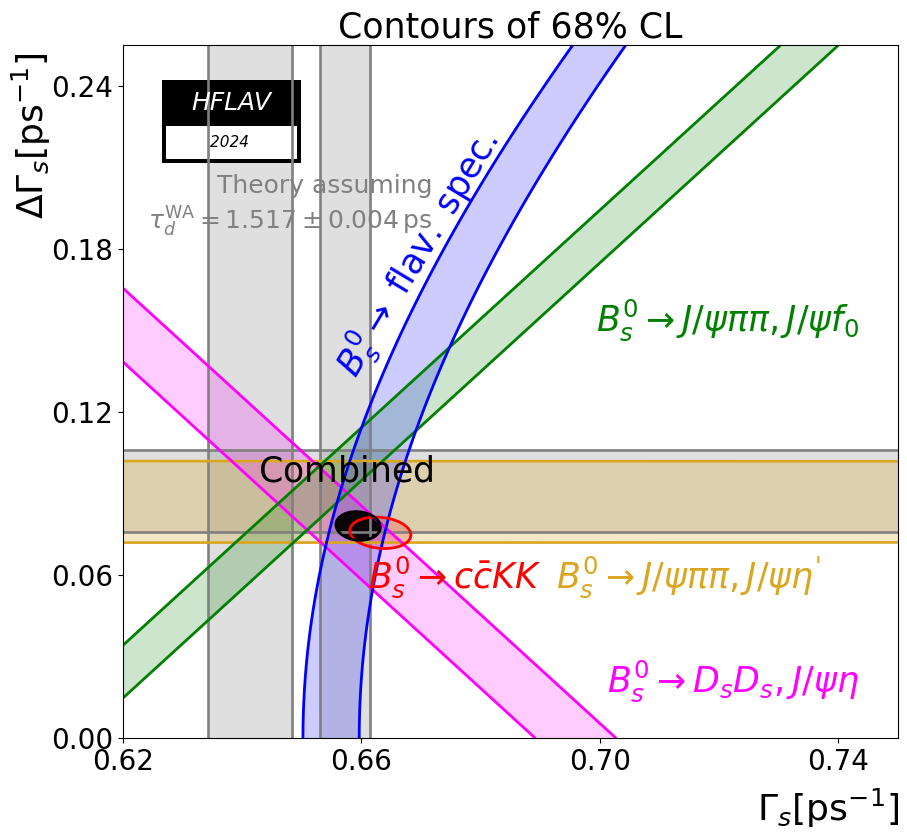}
\hfill
\includegraphics[width=0.49\textwidth]{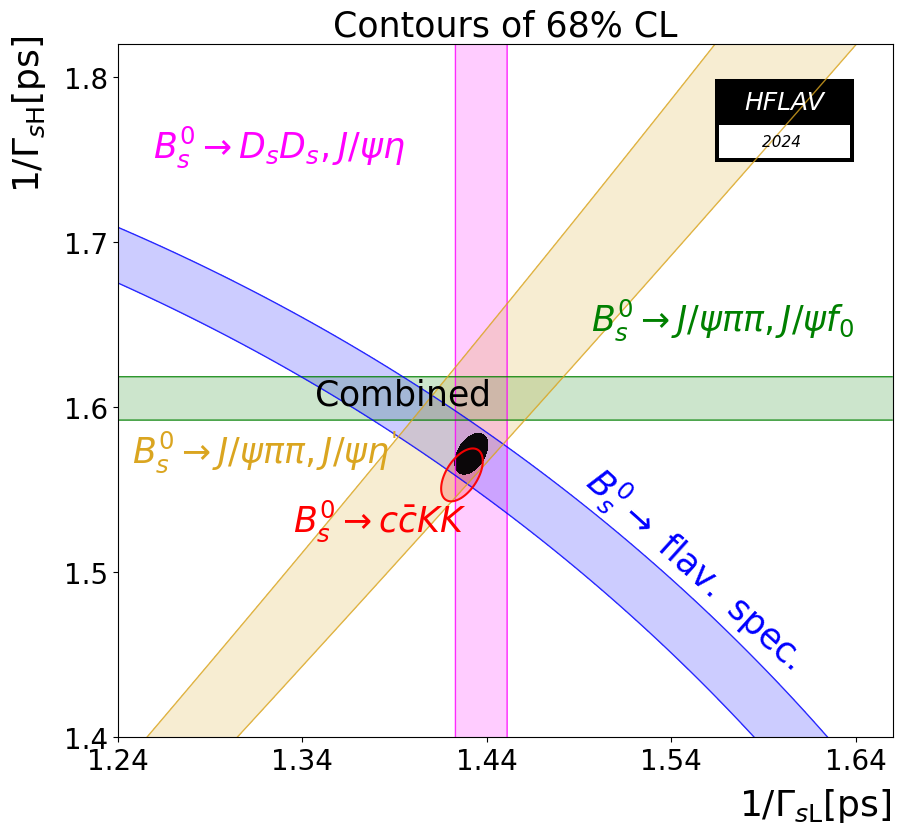}
\caption{Contours of 68\% CL shown in the $(\Gs,\,\DGs)$ plane on the lef tand in the $(1/\Gamma_{s\rm L},\,1/\Gamma_{s\rm H})$ plane on the right. The average of all $\Bs\to \jpsi K^+K^-$ and $\Bs \to \psi(2S)\Kp\Km$ results are shown as the red contour, where \Gs and \DGs are scaled by factors \hflavGSCCKKSF and \hflavDGSCCKKSF. The constraints given by the effective lifetime measurements of \Bs\ to flavor-specific, pure \CP-odd and pure \CP-even final states are shown as the blue, green and purple bands, respectively. The constraint on \DGs given by the measurement of the lifetime difference between $B_{s}^{0}\to J/\psi\pi^{+}\pi^{-}$ and $B_{s}^{0}\to J/\psi \eta^{'}$ is given by the golden bands.  The average taking all constraints into account is shown as the dark-filled contour. The light-gray bands are theory predictions. The horizontal band is $\DGs = +0.091 \pm 0.015~\hbox{ps}^{-1}$~\protect\citehistory{Albrecht:2024oyn}{Lenz:2011ti,Lenz:2006hd,Jubb:2016mvq,Artuso:2015swg,*Lenz_hist,Lenz:2019lvd,Albrecht:2024oyn} that assumes no new physics in \Bs\ mixing.  The vertical \Gs bands are calculated from the two solutions in \Refx{Albrecht:2024oyn} assuming the experimental world average for the \Bd lifetime, $\hflavTAUBD$. They are due to the two different sets of inputs to the theory calculation of $\tau(\Bs)/\tau(\Bd)$ according to \Refx{Bordone:2021oof} (toward lower \Gs values) and \Refx{Bernlochner:2022ucr} (toward higher \Gs values).}
\labf{DGs}
\end{center}
\end{figure}

\begin{table}
\caption{Averages of \DGs, $\Gs$ and related quantities, obtained from
$\Bs\to\jpsi\Kp\Km$ and $\Bs\to\psi(2S)\Kp\Km$ alone (first column),
adding the constraints from the effective lifetimes measured in pure \CP modes
$\Bs\to D_s^+D_s^-,J/\psi\eta$ and $\Bs \to \jpsi f_0(980), \jpsi \pi^+\pi^-$ (second column), adding the \DGs constraint from $B_{s}^{0}\to J/\psi\eta'$ and  $B_{s}^{0}\to J/\psi\pi^{+}\pi^{-}$,
and adding the constraint from the effective lifetime measured in flavor-specific modes $B_s^0\to D_s^-\ell^+\nu X$, 
$D_s^-\pi^+$, $D_s^-D^+$ (third column, recommended world averages).}
\labt{tabtauLH}
\begin{center}
\begin{tabular}{c|c|c|c}
\hline
& $\Bs\to (c\bar{c}) K^+K^-$ modes & $\Bs\to (c\bar{c}) K^+K^-$ modes & $\Bs\to (c\bar{c}) K^+K^-$ modes \\
& only (see \Table{GsDGs}) & + pure \CP modes & + pure \CP modes \\
&                          &                  & + flavor-specific modes \\
\hline
\Gs                & \hflavGS        &  \hflavGSCO        &  \hflavGSCON        \\
$1/\Gs$            & \hflavTAUBSMEAN &  \hflavTAUBSMEANCO &  \hflavTAUBSMEANCON \\
$1/\Gamma_{s\rm L}$ & \hflavTAUBSL    &  \hflavTAUBSLCO    &  \hflavTAUBSLCON    \\
$1/\Gamma_{s\rm H}$ & \hflavTAUBSH    &  \hflavTAUBSHCO    &  \hflavTAUBSHCON    \\
\DGs               & \hflavDGS       &  \hflavDGSCO       &  \hflavDGSCON       \\
\DGs/\Gs           & \hflavDGSGS     &  \hflavDGSGSCO     &  \hflavDGSGSCON     \\
$\rho(\Gs,\DGs)$   & \hflavRHOGSDGS  &  \hflavRHOGSDGSCO  &  \hflavRHOGSDGSCON  \\
\hline
\end{tabular}
\end{center}
\end{table}

An alternative approach, which is directly sensitive to first order in 
$\DGs/\Gs$, 
is to determine the effective lifetime of untagged \Bs\ candidates
decaying to %
pure \CP eigenstates; we use here measurements with
$\Bs \to D_s^+D_s^-$~\cite{Aaij:2013bvd}, 
$\Bs \to J/\psi \eta$~\cite{Aaij:2016dzn,LHCb-PAPER-2022-010},
$\Bs \to \jpsi f_0(980)$~\cite{Aaltonen:2011nk,Abazov:2016oqi}
and $\Bs\to \jpsi \pi^+\pi^-$~\citehistory{Aaij:2013oba,Aaij:2019mhf,Sirunyan:2017nbv}{Aaij:2013oba,*LHCb:2011aa_hist,*LHCb:2012ad_hist,*LHCb:2011ab_hist,*Aaij:2012nta_hist,Aaij:2019mhf} decays.
The precise extraction of $1/\Gs$ and $\DGs$
from such measurements, discussed in detail in \Refs{Hartkorn:1999ga,Dunietz:2000cr,Fleischer:2011cw}, 
requires additional information 
in the form of theoretical assumptions or
external inputs on weak phases and hadronic parameters. 
If $f$ denotes a final state into which both \Bs and \Bsbar can decay,
the ratio of the effective $\Bs \to f$
lifetime $\tau_{\rm single}$, found by
fitting the decay-time distribution to a single exponential, relative to the mean
\Bs lifetime is~\cite{Fleischer:2011cw}%
\footnote{%
\label{foot:life_mix:ADG-def}
The definition of $A_f^{\DG}$ given in \Eq{ADG} has the sign opposite to that given in \Refx{Fleischer:2011cw}.}
\begin{equation}
  \frac{\tau_{\rm single}(\Bs \to f)}{\tau(\Bs)} = \frac{1}{1-y_s^2} \left[ \frac{1 - 2A_f^{\DG} y_s + y_s^2}{1 - A_f^{\DG} y_s}\right ] \,,
\labe{tauf_fleisch}
\end{equation}
where
\begin{equation}
A_f^{\DG} = -\frac{2 \Re(\lambda_f)} {1+|\lambda_f|^2} \,.
\labe{ADG}
\end{equation}
To include the measurements of the effective
$\Bs \to D_s^+D_s^-$ (\CP-even), $\Bs \to \jpsi \eta$ (\CP-even), $\Bs \to \jpsi f_0(980)$ (\CP-odd) and
$\Bs \to \jpsi\pi^+\pi^-$ (\CP-odd) 
lifetimes as constraints in the \DGs fit,\footnote{%
\label{foot:life_mix:BKK}
The effective lifetimes measured in $\Bs\to K^+ K^-$ (mostly \CP-even) and  $\Bs \to \jpsi K_{\rm S}^0$ (mostly \CP-odd) are not used because we cannot quantify the penguin contributions in those modes.}
we neglect subleading penguin contributions and possible direct \CP violation. 
Explicitly, in \Eq{tauf_fleisch}, we set
$A_{\mbox{\scriptsize \CP-even}}^{\DG} = \cos \phiccbars$
and $A_{\mbox{\scriptsize \CP-odd}}^{\DG} = -\cos \phiccbars$.
Given the small value of $\phiccbars$, we have, to first order in $y_s$:
\begin{eqnarray}
\tau_{\rm single}(\Bs \to \mbox{\CP-even})
& \approx & \frac{1}{\Gamma_{s\rm L}} \left(1 + \frac{(\phiccbars)^2 y_s}{2} \right) \,,
\labe{tau_KK_approx}
\\
\tau_{\rm single}(\Bs \to \mbox{\CP-odd})
& \approx & \frac{1}{\Gamma_{s\rm H}} \left(1 - \frac{(\phiccbars)^2 y_s}{2} \right) \,.
\labe{tau_Jpsif0_approx}
\end{eqnarray}
The numerical inputs are taken from \Eqss{tau_KK}{tau_Jpsif0},
and the resulting averages, combined with the $\Bs\to\jpsi K^+K^-$ information,
are indicated in the second column of numbers of \Table{tabtauLH}. 

Information on \DGs is also obtained from the study of the
proper time distribution of untagged samples
of flavor-specific \Bs decays~\cite{Hartkorn:1999ga}, 
\eg semileptonic \Bs decays,
where
the flavor (\ie, \Bs or \Bsbar) at the time of decay is determined by
the decay products. Since there is
an equal mix of the heavy and light mass eigenstates at production time ($t=0$), the proper time distribution is a superposition 
of two exponential functions with decay constants
$\Gamma_{s\rm L}$ and $\Gamma_{s\rm H}$. %
This provides sensitivity to both $1/\Gs$ and 
$(\DGs/\Gs)^2$. Ignoring \DGs and fitting for 
a single exponential leads to an estimate of \Gs with a 
relative bias proportional to $(\DGs/\Gs)^2$, as shown in \Eq{fslife}. Furthermore, \DGs can be measured from the decay-width difference between \CP-even and \CP-odd $B_{s}^{0}$ decay final states, such as $B_{s}^{0}\to J/\psi \eta'$ and $B_{s}^{0}\to J/\psi \pi^{+}\pi^{-}$, respectively, as measured by LHCb~\cite{LHCb:2023xtc}. The result is added as a constraint, where it is assumed to be uncorrelated with previous $B_{s}^{0}\to J/\psi \pi^{+}\pi^{-}$ measurements, as the data overlap is expected to be small.
Including the constraint from the world-average flavor-specific \Bs 
lifetime, given in \Eq{tau_fs}, leads to the results shown in the last column 
of \Table{tabtauLH}.
These world averages are displayed as the dark-filled contours labeled ``Combined'' in the
plots of \Fig{DGs}. 
They correspond to the lifetime averages
$1/\Gs=\hflavTAUBSMEANCON$,
$1/\Gamma_{s\rm L}=\hflavTAUBSLCON$,
$1/\Gamma_{s\rm H}=\hflavTAUBSHCON$,
and to the decay-width difference
\begin{equation}
\DGs = \hflavDGSCON ~~~~\mbox{and} ~~~~~ \DGs/\Gs = \hflavDGSGSCON \,.
\labe{DGs_DGsGs}
\end{equation}
The good agreement with the Standard Model predictions 
$\DGs = +0.091 \pm 0.015~\hbox{ps}^{-1}$~\cite{Albrecht:2024oyn}
and $\DGs = +0.076 \pm 0.017~\hbox{ps}^{-1}$~\cite{Gerlach:2024qlz}
excludes significant quark-hadron duality violation in the HQE~\cite{Lenz:2012mb}. 
Estimates of $\DGs/\Gs$ obtained from measurements of the 
$\Bs \to D_s^{(*)+} D_s^{(*)-}$ branching fraction~\citehistory{Barate:2000kd,Esen:2010jq,Abazov:2008ig,Abulencia:2007zz,Aaltonen:2012mg}{Barate:2000kd,Esen:2010jq,Abazov:2008ig,*Abazov:2007rb_hist,Abulencia:2007zz,Aaltonen:2012mg}
are not used in the average,
since they are based on the questionable~\cite{Lenz:2011ti,Lenz:2006hd}
assumption that these decays account for all \CP-even final states.
The results of early lifetime analyses that attempted
to measure $\DGs/\Gs$~\citehistory{Acciarri:1998uv,Abreu:2000sh,Abreu:2000ev,Abe:1997bd}{Acciarri:1998uv,Abreu:2000sh,Abreu:2000ev,*Abreu:1996ep_hist,Abe:1997bd}
are not used either.

The probability of \Bs mixing has been known to be large for more than 20 years. 
Indeed the time-integrated measurements of the flavor blind measurement 
$\bar\chi = f^\prime_d \chi_d + f^\prime_s \chi_s$, 
where $f'_d$ and $f'_s$ are the fractions of \Bd and \Bs hadrons in a sample of
semileptonic \b-hadron decays\footnote{See Sec.~4.1.3 of our previous 
publication~\cite{Amhis:2019ckw}.}, %
when compared to our knowledge
of \chid and the \b-hadron fractions, indicated that 
\chis should be close to its maximal possible value of $1/2$.
Many searches of the time dependence of this mixing 
have been performed by ALEPH~\cite{Heister:2002gk}, %
DELPHI~\citehistory{Abreu:2000sh,Abreu:2000ev,Abdallah:2002mr,Abdallah:2003qga}{Abreu:2000sh,Abreu:2000ev,*Abreu:1996ep_hist,Abdallah:2002mr,Abdallah:2003qga}, %
OPAL~\cite{Abbiendi:1999gm,Abbiendi:2000bh},
SLD~\unpublished{\cite{Abe:2002ua,Abe:2002wfa}}{\cite{Abe:2002ua,Abe:2002wfa,Abe:2000gp}},
CDF (Run~I)~\cite{Abe:1998qj} and
\dzero~\cite{Abazov:2006dm}
but did not have enough statistical power
and proper time resolution to resolve 
the small period of the \Bs\ oscillations.

Oscillations of \Bs mesons were observed for the first time in 2006
by the CDF collaboration~\citehistory{Abulencia:2006ze}{Abulencia:2006ze,*Abulencia:2006mq_hist},
based on samples of flavor-tagged hadronic and semileptonic \Bs decays in flavor-specific final states, partially or fully reconstructed in 
$1\invfb$ of data collected during Tevatron's Run~II. 
\unpublished{}{
This was shortly followed by independent unpublished evidence obtained by the \dzero collaboration
with $2.4\invfb$ of
data~\cite{D0note5618:2008,*D0note5474:2007,*D0note5254:2006}.}
Since then, the LHCb collaboration obtained the most precise results using fully reconstructed 
$\Bs \to D_s^-\pi^+$ and $\Bs \to D_s^-\pi^+\pi^-\pi^+$ decays~\cite{Aaij:2011qx,Aaij:2013mpa,LHCb:2020qag,LHCb:2021moh}.
LHCb has also observed \Bs oscillations with semileptonic $\Bs \to D_s^-\mu^+ X$ decays~\cite{Aaij:2013gja}.
In addition, measurements with non flavor-specific final states have been performed with $\Bs\to\jpsi K^+K^-$ decays
by LHCb~\citehistory{Aaij:2014zsa,LHCb:2023sim}{Aaij:2014zsa,LHCb:2023sim,*Aaij:2019vot_hist} 
and CMS~\cite{CMS-PAS-BPH-23-004}.
The measurements of \dms are summarized in \Table{dms}. 

\begin{table}[t]
\caption{Measurements of \dms.}
\labt{dms}
\begin{center}
\begin{tabular}{ll@{}crl@{\,}l@{\,}ll} \hline
Experiment & Method           & \multicolumn{2}{c}{Data set} & \multicolumn{3}{c}{\dms (\!\!\invps)} & Ref. \\
\hline
CDF2   & \multicolumn{2}{l}{\particle{D_s^{(*)-} \ell^+ \nu}, \particle{D_s^{(*)-} \pi^+}, \particle{D_s^{-} \rho^+}}
       & 1 \invfb & $17.77$ & $\pm 0.10$ & $\pm 0.07~$
       & \citehistory{Abulencia:2006ze}{Abulencia:2006ze,*Abulencia:2006mq_hist} \\
\unpublished{}{%
\dzero & \particle{D_s^- \ell^+ X}, \particle{D_s^- \pi^+ X}
       &  & 2.4 \invfb & $18.53$ & $\pm 0.93$ & $\pm 0.30~$ 
       & \cite{D0note5618:2008,*D0note5474:2007,*D0note5254:2006}$^u$ \\
}%
LHCb   & \particle{D_s^- \pi^+}, \particle{D_s^- \pi^+\pi^-\pi^+}
       & 2010 & 0.036 \invfb & $17.63$ & $\pm 0.11$ & $\pm 0.02~$   
       & \cite{Aaij:2011qx} \\
LHCb   & \particle{D_s^- \mu^+ X}
       & 2011 & 1.0 \invfb & $17.93$ & $\pm 0.22$ & $\pm 0.15$ 
       & \cite{Aaij:2013gja}  \\
LHCb   & \particle{D_s^- \pi^+}
       & 2011 & 1.0 \invfb & $17.768$ & $\pm 0.023$ & $\pm 0.006$ 
       & \cite{Aaij:2013mpa}  \\
LHCb   & \particle{\jpsi K^+K^-}
       & 2011--2012 & 3.0 \invfb & $17.711$ & $^{+0.055}_{-0.057}$ & $\pm 0.011$ 
       & \citehistory{Aaij:2014zsa}{Aaij:2014zsa,*Aaij:2013oba_supersede2}  \\
LHCb   & \particle{\jpsi K^+K^-}
       & 2015--2018 & 6 \invfb & $17.743$ & 
       $\pm 0.033$ & $\pm 0.009$ 
       & \cite{LHCb:2023sim}  \\
LHCb   & above 2 combined
       & 2011--2018 & 9 \invfb & $17.730$ & 
        $\pm 0.029$ & %
       & \cite{LHCb:2023sim} \\
\end{tabular}
\end{center}
\end{table}

\begin{figure}
\begin{center}
\includegraphics[width=0.8\textwidth]{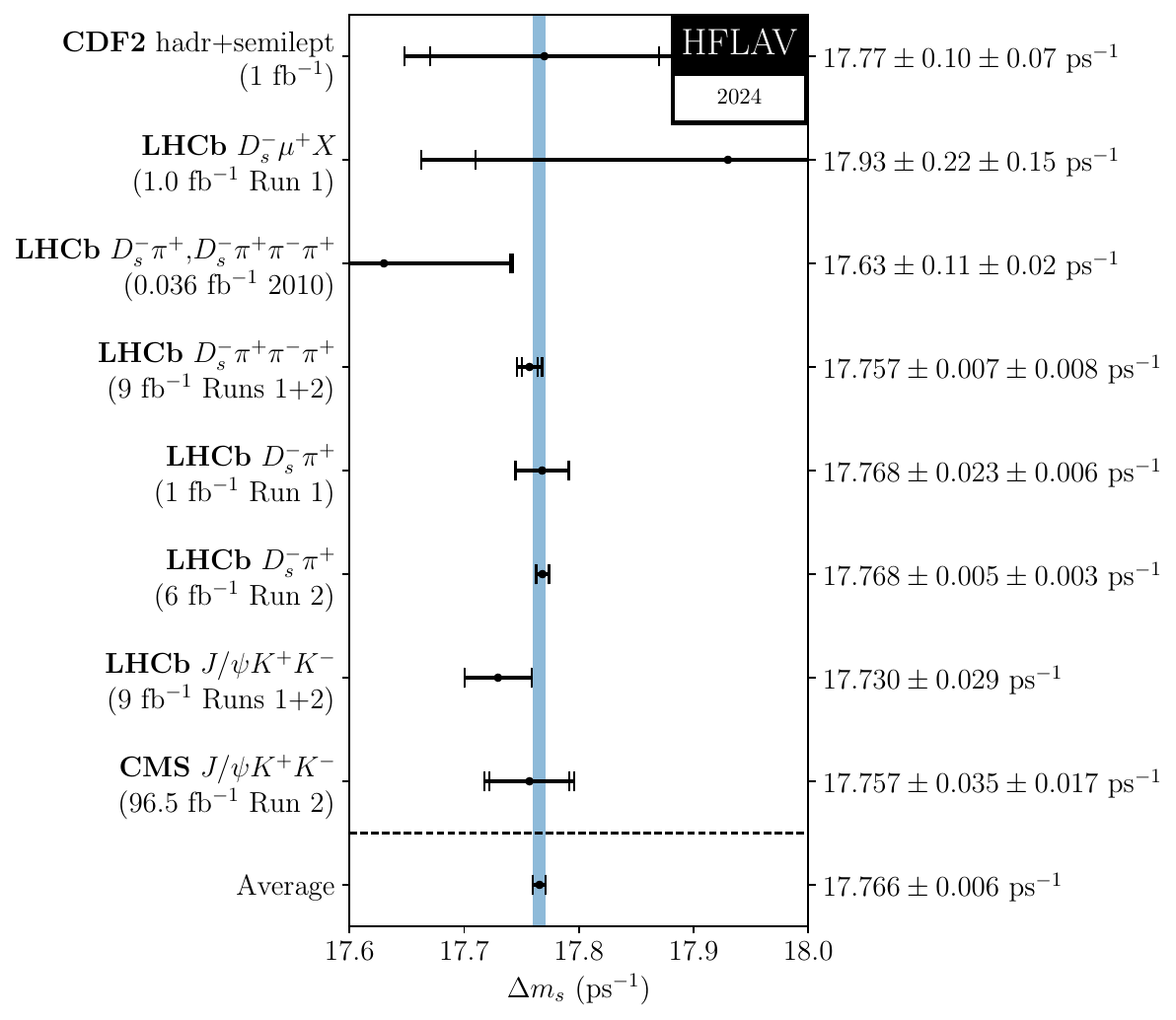}
\caption{%
Measurements of \dms, together with their average.} 
\label{dms}
\end{center}
\end{figure}

An average of all the published CDF2, LHCb and CMS results\unpublished{}{\footnote{
  \label{foot:life_mix:D0note5618:2008}
  We do not include the unpublished
  \dzero~\cite{D0note5618:2008,*D0note5474:2007,*D0note5254:2006} result in the average.}} yields 
\begin{equation}
\dms = \hflavDMSfull %
\labe{dms}
\end{equation}
and is illustrated in Fig.~\ref{dms}. The systematic uncertainties of the LHCb results due to the length scale (affecting all modes), the momentum scale ($\Bs \to D_s^-\pi^+$ Run~1 and Run~2, $\Bs \to D_s^-\pi^+\pi^-\pi^+$  Run~1 and $\Bs\to\jpsi K^+K^-$), the fit bias ($\Bs \to D_s^-\pi^+$ Run~1 and $\Bs \to D_s^-\pi^+\pi^-\pi^+$ Run~1) and the decay-time bias ($\Bs \to D_s^-\pi^+$ Run~1 and Run~2) are considered to be 100\% correlated. Furthermore, the CMS and LHCb measurements of \dms in $\Bs\to\jpsi K^+K^-$ decays are averaged using the measured central values and uncertainties of the full set of observables determined in these studies ($\phi_{s}$, \DGs, \Gs, $|\lambda|$, strong phases and polarization fractions) in order to account for their correlations with \dms.

The Standard Model prediction
$\dms = 18.23 \pm 0.63~\hbox{ps}^{-1}$~\cite{Albrecht:2024oyn} is consistent with the experimental value,
but has a much larger uncertainty, %
dominated by the uncertainty on the hadronic matrix elements 
determined in  \Refs{DiLuzio:2019jyq,Bazavov:2016nty,Grozin:2016uqy,Kirk:2017juj,Boyle:2018knm,King:2019lal,Dowdall:2019bea}.
The ratio $\DGs/\dms$ can be predicted more accurately to be $0.00499 \pm 0.00079$~\cite{Lenz:2019lvd,Davies:2019gnp,Lenz:2011ti,Lenz:2006hd,Artuso:2015swg,Albrecht:2024oyn,Gerlach:2022hoj}, %
in good agreement with the experimental determination of 
\begin{equation}
\DGs/\dms= \hflavRATIODGSDMS \,.
\end{equation}

Multiplying the \dms result of \Eq{dms} by the 
mean \Bs lifetime of \Eq{oneoverGs}, $1/\Gs=\hflavTAUBSMEANCON$,
yields
\begin{equation}
\xs %
= \hflavXS \,. \labe{xs}
\end{equation}
With $2\ys %
=\hflavDGSGSCON$ 
(see \Eq{DGs_DGsGs})
and under the assumption of no \CP violation in \Bs mixing,
this corresponds to
\begin{equation}
\chis = \frac{\xs^2+\ys^2}{2(\xs^2+1)} = \hflavCHIS \,. \labe{chis}
\end{equation}
The ratio 
\begin{equation}
\frac{\dmd}{\dms} = \hflavRATIODMDDMS \,, \labe{dmd_over_dms}
\end{equation}
of the \Bd and \Bs oscillation frequencies, 
obtained from \Eqss{dmd}{dms}, 
can be used to extract the following magnitude of the ratio of CKM matrix elements, 
\begin{equation}
\left|\frac{V_{td}}{V_{ts}}\right| =
\xi \sqrt{\frac{\dmd}{\dms}\frac{m(\Bs)}{m(\Bd)}} = 
\left\{ \begin{array}{l} 
\hflavVTDVTSLQCDfull\,\mbox{(lattice QCD)} \\ 
\hflavVTDVTSSRULfull\,\mbox{(sum rules)} \end{array} \right.
\,. \labe{Vtd_over_Vts}
\end{equation}
The first uncertainty is from experimental uncertainties 
(with the masses $m(\Bs)$ and $m(\Bd)$ taken from \Refx{PDG_2020}).
The second uncertainty arises from theoretical uncertainties 
in the estimation of the SU(3) flavor-symmetry breaking factor
$\xi %
= \hflavXILQCD$~\cite{Aoki:2019cca},
which is an average of three-flavor lattice QCD calculations dominated by the results of Ref.~\cite{Bazavov:2016nty}, 
or $\xi = \hflavXISRUL$~\cite{King:2019lal} obtained from sum rules.
Note that \Eq{Vtd_over_Vts} assumes that \dms and \dmd only receive 
Standard Model contributions.
\mysubsubsection{\CP violation in \Bd and \Bs mixing}
\labs{qpd} \labs{qps}

Evidence for \CP violation in \Bd mixing
has been searched for,
both with flavor-specific and inclusive \Bd decays, 
in samples where the initial 
flavor state is tagged. In the case of semileptonic 
(or other flavor-specific) decays, 
where the final state tag is 
also available, the asymmetry
\begin{equation} 
\ASLd = \frac{
N(\hbox{\Bdbar}(t) \to \ell^+      \nu_{\ell} X) -
N(\hbox{\Bd}(t)    \to \ell^- \bar{\nu}_{\ell} X) }{
N(\hbox{\Bdbar}(t) \to \ell^+      \nu_{\ell} X) +
N(\hbox{\Bd}(t)    \to \ell^- \bar{\nu}_{\ell} X) } 
\labe{ASLd}
\end{equation} 
has been measured, either in decay-time-integrated analyses at 
CLEO~\citehistory{Behrens:2000qu,Jaffe:2001hz}{Behrens:2000qu,Jaffe:2001hz,*Jaffe:2001hz_hist},
\babar~\citehistory{Lees:2014kep}{Lees:2014kep,*Lees:2014kep_hist},
CDF~\unpublished{\cite{Abe:1996zt}}{\cite{Abe:1996zt,CDFnote9015:2007}}
and \dzero~\citehistory{Abazov:2013uma}{Abazov:2013uma,*Abazov:2011yk_hist,*Abazov:2010hv_hist,*Abazov:2010hj_hist,*Abazov:dimuon_hist},
or in decay-time-dependent analyses at 
OPAL~\cite{Ackerstaff:1997vd}, ALEPH~\cite{Barate:2000uk}, 
\babar~\unpublished{%
\citehistory%
{Aubert:2003hd,*Aubert:2004xga,Lees:2013sua,Aubert:2006nf}%
{Aubert:2003hd,*Aubert:2004xga,Lees:2013sua,Aubert:2006nf,*Aubert:2002mn_hist}}{%
\citehistory%
{Aubert:2003hd,*Aubert:2004xga,Lees:2013sua,Aubert:2006nf}%
{Aubert:2003hd,*Aubert:2004xga,Lees:2013sua,*Margoni:2013qx_hist,*Aubert:2006sa_hist,Aubert:2006nf,*Aubert:2002mn_hist}}
and \belle~\cite{Nakano:2005jb}.
Note that the asymmetry of time-dependent decay rates in \Eq{ASLd} is 
related to  $|q_d/p_d|$ through \Eq{ASLq} and is therefore time-independent.
In the inclusive case, also investigated and published
by ALEPH~\cite{Barate:2000uk} and OPAL~\cite{Abbiendi:1998av},
no final state tag is used, and the asymmetry~\cite{Beneke:1996hv,*Dunietz:1998av}
\begin{equation} 
\frac{
N(\hbox{\Bd}(t) \to {\rm all}) -
N(\hbox{\Bdbar}(t) \to {\rm all}) }{
N(\hbox{\Bd}(t) \to {\rm all}) +
N(\hbox{\Bdbar}(t) \to {\rm all}) } 
\simeq
\ASLd \left[ \frac{\dmd}{2\Gd} \sin(\dmd \,t) - 
\sin^2\left(\frac{\dmd \,t}{2}\right)\right] 
\labe{ASLincl}
\end{equation} 
must be measured as a function of the proper time to extract information 
on \CP violation.
Furthermore, \dzero~\cite{Abazov:2012hha} and
LHCb~\cite{Aaij:2014nxa} have studied the time-dependence of the 
charge asymmetry of $B^0 \to D^{(*)-}\mu^+\nu_{\mu}X$ decays
without tagging the initial state,
which is equal to 
\begin{equation} 
\frac{N(D^{(*)-}\mu^+\nu_{\mu}X)-N(D^{(*)+}\mu^-\bar{\nu}_{\mu}X)}%
{N(D^{(*)-}\mu^+\nu_{\mu}X)+N(D^{(*)+}\mu^-\bar{\nu}_{\mu}X)} =
\ASLd \frac{1- \cos(\dmd \,t)}{2}
\labe{untagged_ASL}
\end{equation}
in absence of detection and production asymmetries. Note that \Eqss{ASLincl}{untagged_ASL} assume $\DGd =0$.

\begin{table}[t]
\caption{Measurements\footref{foot:life_mix:Abe:1996zt}
of \CP violation in \Bd mixing and their average
in terms of both \ASLd and $|q_{\particle{d}}/p_{\particle{d}}|$.
The individual results are listed as quoted in the original publications, 
or converted\footref{foot:life_mix:epsilon_B}
to an \ASLd value.
The ALEPH and OPAL %
results assume no \CP violation in \Bs mixing. The abbreviation "rec." stands for reconstructed. See section 5.1 for a detailed explanation of the method.}
\labt{qoverp}
\begin{center}
\resizebox{\textwidth}{!}{
\begin{tabular}{@{}rcl@{$\,\pm$}l@{$\pm$}ll@{$\,\pm$}l@{$\pm$}l@{}}
\toprule
Exp.\ \& Ref. & Method & \multicolumn{3}{c}{Measured \ASLd} 
                       & \multicolumn{3}{c}{Measured $|q_{\particle{d}}/p_{\particle{d}}|$} \\
\midrule
CLEO   \cite{Behrens:2000qu} & Partial hadronic rec. 
                             & $+0.017$ & 0.070 & 0.014 
                             & \multicolumn{3}{c}{} \\
CLEO   \citehistory{Jaffe:2001hz}{Jaffe:2001hz,*Jaffe:2001hz_hist}   & Dileptons 
                             & $+0.013$ & 0.050 & 0.005 
                             & \multicolumn{3}{c}{} \\
CLEO   \citehistory{Jaffe:2001hz}{Jaffe:2001hz,*Jaffe:2001hz_hist}   & Average of above two 
                             & $+0.014$ & 0.041 & 0.006 
                             & \multicolumn{3}{c}{} \\
\babar \cite{Aubert:2003hd,*Aubert:2004xga}   & Full hadronic rec. 
                             & \multicolumn{3}{c}{}  
                             & 1.029 & 0.013 & 0.011  \\
\babar \unpublished{\cite{Lees:2013sua}}{\citehistory{Lees:2013sua}{Lees:2013sua,*Margoni:2013qx_hist,*Aubert:2006sa_hist}} & Part.\ rec.\ $D^{*}X\ell\nu$ 
                             & $+0.0006$ & \multicolumn{2}{@{}l}{$0.0017 ^{+0.0038}_{-0.0032}$} 
                             & $0.99971$ & $0.00084$ & $0.00175$ \\ 
\babar \citehistory{Lees:2014kep}{Lees:2014kep,*Lees:2014kep_hist}  & Dileptons
                             & $-0.0039$ & 0.0035 & 0.0019 
                             & \multicolumn{3}{c}{} \\
\belle \cite{Nakano:2005jb}  & Dileptons 
                             & $-0.0011$ & 0.0079 & 0.0085 
                             & 1.0005 & 0.0040 & 0.0043 \\
\bottomrule
\end{tabular}
}
\end{center}
\end{table}

\Table{qoverp} summarizes the different measurements%
\footnote{
\label{foot:life_mix:Abe:1996zt}
A low-statistics result published by CDF using the Run~I data~\cite{Abe:1996zt}
\unpublished{is}{and an unpublished result by CDF using Run~II data~\cite{CDFnote9015:2007} are}
not included in our averages, nor in \Table{qoverp}.
}
of \ASLd and $|q_{\particle{d}}/p_{\particle{d}}|$. 
In all cases asymmetries compatible with zero have been found. 
A simple average of all measurements performed at the
\B factories~\unpublished%
{\citehistory%
{Behrens:2000qu,Jaffe:2001hz,Aubert:2003hd,*Aubert:2004xga,Lees:2013sua,Lees:2014kep,Nakano:2005jb}%
{Behrens:2000qu,Jaffe:2001hz,*Jaffe:2001hz_hist,Aubert:2003hd,*Aubert:2004xga,Lees:2013sua,Lees:2014kep,*Lees:2014kep_hist,Nakano:2005jb}}%
{\citehistory%
{Behrens:2000qu,Jaffe:2001hz,Aubert:2003hd,*Aubert:2004xga,Lees:2013sua,Lees:2014kep,Nakano:2005jb}%
{Behrens:2000qu,Jaffe:2001hz,*Jaffe:2001hz_hist,Aubert:2003hd,*Aubert:2004xga,Lees:2013sua,*Margoni:2013qx_hist,*Aubert:2006sa_hist,Lees:2014kep,*Lees:2014kep_hist,Nakano:2005jb}}
yields
$\ASLd = \hflavASLDB$.
Adding also the \dzero~\cite{Abazov:2012hha}
and LHCb~\cite{Aaij:2014nxa} measurements obtained with reconstructed 
semileptonic \Bd decays yields $\ASLd = \hflavASLDD$.
As discussed in more detail later in this section, 
the \dzero analysis with single muons and like-sign dimuons~\citehistory{Abazov:2013uma}{Abazov:2013uma,*Abazov:2011yk_hist,*Abazov:2010hv_hist,*Abazov:2010hj_hist,*Abazov:dimuon_hist}
separates the \Bd and \Bs contributions by exploiting the dependence on the muon impact parameter cut; including this 
\ASLd result from \dzero in the average yields
$\ASLd = \hflavASLDW$. %

All the other \Bd analyses performed at high energy, either at LEP or at the Tevatron,
did not separate the contributions from the \Bd and \Bs mesons.
Under the assumption of no \CP violation in \Bs mixing ($\ASLs =0$),
a number of these early analyses~\cite{Abazov:2006qw,Ackerstaff:1997vd,Barate:2000uk,Abbiendi:1998av}
report a measurement of $\ASLd$ or $|q_{\particle{d}}/p_{\particle{d}}|$ for the \Bd meson.
However, although we include them in \Table{qoverp}, these imprecise determinations no longer improve the world average of 
\ASLd.
Furthermore, the assumption makes sense within the Standard Model, 
since \ASLs is predicted to be about a factor 20 smaller than \ASLd~\cite{Lenz:2019lvd}, %
but may not be suitable in the presence of new physics. 

The Tevatron experiments
have measured linear combinations of 
\ASLd and \ASLs
using inclusive semileptonic decays of \b hadrons. CDF (Run~I) finds 
$\ASLb = +0.0015 \pm 0.0038 \mbox{(stat)} \pm 0.0020 \mbox{(syst)}$~\cite{Abe:1996zt},
and D0 obtains $\ASLb = -0.00496 \pm 0.00153 \mbox{(stat)} \pm 0.00072 \mbox{(syst)}$~\citehistory{Abazov:2013uma}{Abazov:2013uma,*Abazov:2011yk_hist,*Abazov:2010hv_hist,*Abazov:2010hj_hist,*Abazov:dimuon_hist}.
While the imprecise CDF result is compatible with no \CP violation%
\unpublished{}{\footnote{
  \label{foot:life_mix:CDFnote9015:2007}
  An unpublished measurement from CDF2, 
  $\ASLb = +0.0080 \pm 0.0090 \mbox{(stat)} \pm 0.0068 \mbox{(syst)}$~\cite{CDFnote9015:2007},
  more precise than the \dzero measurement,
  is also compatible with no \CP violation.
}},
the \dzero result, obtained by measuring
the single muon and like-sign dimuon charge asymmetries,
differs by 2.8 standard
deviations from the Standard Model expectation of
${\cal A}_{\rm SL}^{b,\rm SM} = (-2.3\pm 0.4) \times 10^{-4}$%
~\citehistory{Abazov:2013uma,Lenz:2011ti,Lenz:2006hd}{Abazov:2013uma,*Abazov:2011yk_hist,*Abazov:2010hv_hist,*Abazov:2010hj_hist,*Abazov:dimuon_hist,Lenz:2011ti,Lenz:2006hd}.
With a more sophisticated analysis in bins of the
muon impact parameters, \dzero conclude that the overall deviation of 
their measurements from the SM is at the level of $3.6\,\sigma$.
Interpreting the observed asymmetries
in bins of the muon impact parameters
in terms of \CP violation in $B$-meson mixing and in interference, 
and using 
the mixing parameters and the world-average \b-hadron production fractions 
of \Refx{Amhis:2012bh}, the \dzero collaboration
extracts~\citehistory{Abazov:2013uma}{Abazov:2013uma,*Abazov:2011yk_hist,*Abazov:2010hv_hist,*Abazov:2010hj_hist,*Abazov:dimuon_hist}
values for \ASLd and \ASLs and their correlation
coefficient\footnote{
\label{foot:life_mix:Abazov:2013uma}
In each impact parameter bin $i$ the measured same-sign dimuon asymmetry is interpreted as  
$A_i = K^s_i \ASLs + K^d_i \ASLd + \lambda K^{\rm int}_i \DGd/\Gd$, where the factors  $K^s_i$, $K^d_i$ and $K^{\rm int}_i$ are obtained by \dzero from Monte Carlo simulation. The \dzero publication~\citehistory{Abazov:2013uma}{Abazov:2013uma,*Abazov:2011yk_hist,*Abazov:2010hv_hist,*Abazov:2010hj_hist,*Abazov:dimuon_hist} assumes $\lambda=1$, but it has been demonstrated 
subsequently that $\lambda \le 0.49$~\cite{Nierste_CKM2014}. This particular point invalidates the $\DGd/\Gd$ result published by \dzero, but not the \ASLd and \ASLs results. As stated by \dzero, their \ASLd and \ASLs results assume the above expression for $A_i$, \ie\ that the observed asymmetries are due to \CP violation in $B$ mixing. As long as this assumption is not shown to be wrong (or withdrawn by \dzero), we include the \ASLd and \ASLs results in our world average.},
as shown in \Table{ASLs_ASLd}.
However, the various 
contributions to the total quoted uncertainties from this analysis and from the
external inputs are not given, so the adjustment of these results to different
or more recent values of the external inputs cannot (easily) be done. 

Finally, direct determinations of \ASLs,
also shown in \Table{ASLs_ASLd},
have been obtained by \dzero~\citehistory{Abazov:2012zz}{Abazov:2012zz,*Abazov:2009wg_hist,*Abazov:2007nw_hist}
and LHCb~\citehistory{Aaij:2016yze}{Aaij:2016yze,*Aaij:2013gta_hist}
from the time-integrated charge asymmetry of
untagged $\Bs \to D_s^- \mu^+\nu X$ decays.

Using a two-dimensional fit, all measurements of \ASLd and \ASLs obtained by 
\dzero and LHCb are combined with the 
\B-factory average of \Table{qoverp}. Correlations are taken into account as 
shown in \Table{ASLs_ASLd}.
The results, displayed graphically in \Fig{ASLs_ASLd}, are

\begin{table}[p]
\caption{Measurements of \CP violation in \Bd and \Bs mixing, together 
with their correlations $\rho(\ASLs,\ASLd)$
and their two-dimensional average. Only total errors are quoted.}
\labt{ASLs_ASLd}
\begin{center}
\begin{tabular}{rcccl}
\toprule
Exp.\ \& Ref.\ & Method & Measured \ASLd & Measured \ASLs & $\rho(\ASLd,\ASLs)$ \\
\midrule
\multicolumn{2}{l}{\B-factory average  of \Table{qoverp}}
       & \hflavASLDB & & \\ 
\dzero  \citehistory{Abazov:2012zz,Abazov:2012hha}{Abazov:2012zz,*Abazov:2009wg_hist,*Abazov:2007nw_hist,Abazov:2012hha} & $B_{(s)}^0 \to D_{(s)}^{(*)-} \mu^+\nu X$
       & $+0.0068 \pm 0.0047$ %
       & $-0.0112 \pm 0.0076$ %
       & ~~~$+0.$ \\
LHCb \citehistory{Aaij:2016yze,Aaij:2014nxa}{Aaij:2016yze,*Aaij:2013gta_hist,Aaij:2014nxa} & $B_{(s)}^0 \to D_{(s)}^{(*)-} \mu^+\nu X$
       & $-0.0002 \pm 0.0036$ %
       & $+0.0039 \pm 0.0033$ %
       & ~~~$+0.13$ \\
\multicolumn{2}{l}{Average of above}
       & \hflavASLDNOMU & \hflavASLSNOMU & ~~~\hflavRHOASLSASLDNOMU \\ 
\dzero  \citehistory{Abazov:2013uma}{Abazov:2013uma,*Abazov:2011yk_hist,*Abazov:2010hv_hist,*Abazov:2010hj_hist,*Abazov:dimuon_hist}  & muons \& dimuons
       & $-0.0062 \pm 0.0043$ %
       & $-0.0082 \pm 0.0099$ %
       & ~~~$-0.61$ \\          %
\midrule
\multicolumn{2}{l}{Average of all above}
       & \hflavASLD & \hflavASLS & ~~~\hflavRHOASLSASLD \\ 
\bottomrule
\end{tabular}
\end{center}
\end{table}

\begin{figure}[p]
\begin{center}
\includegraphics[width=0.8\textwidth]{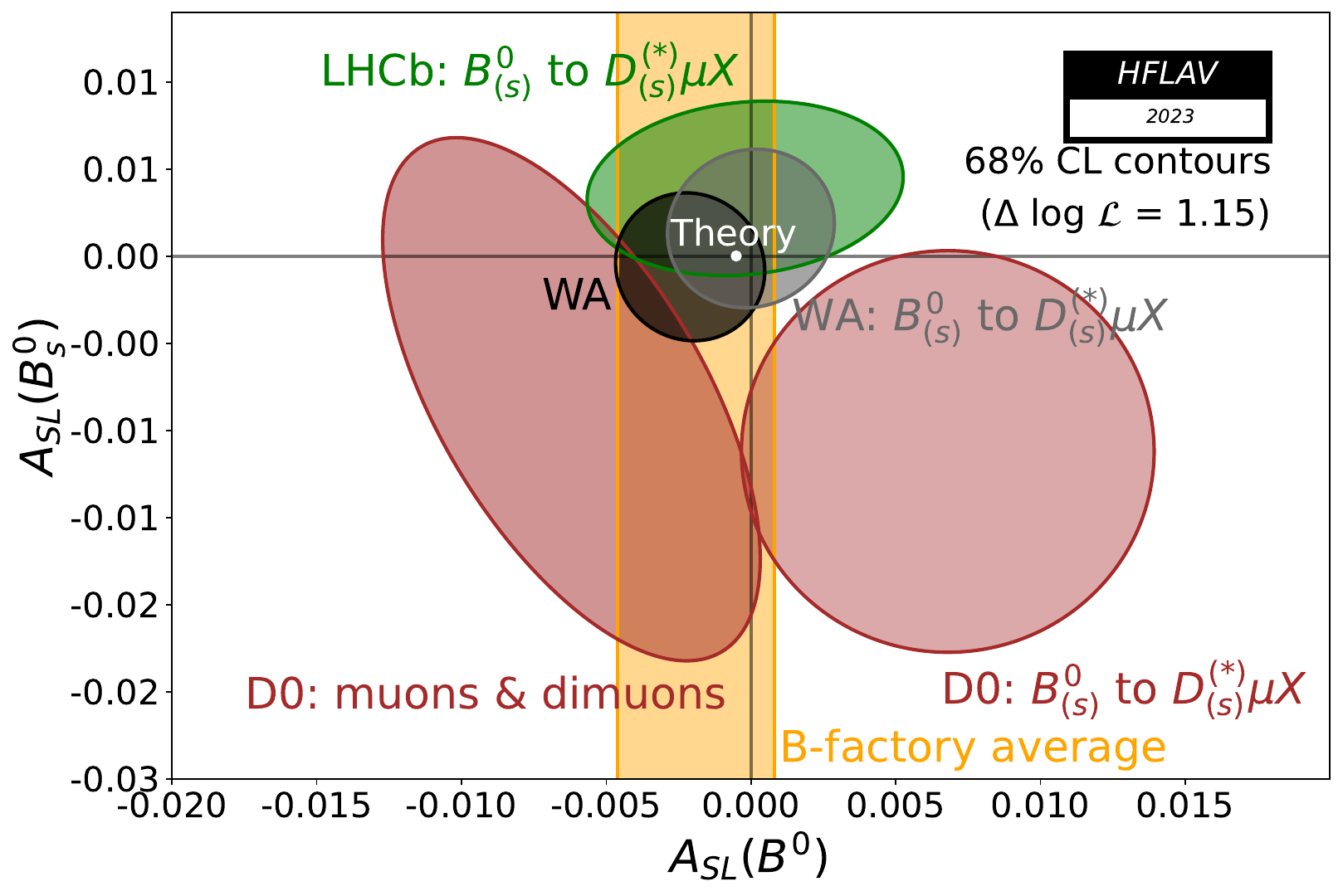}
\end{center}
\vspace{-5mm}
\caption{
Measurements of \ASLd and \ASLs listed in \Table{ASLs_ASLd}
(\B-factory average as the orange band, \dzero measurements as the brown ellipses, LHCb measurements as the green ellipse) 
together with their two-dimensional average (black ellipse). The grey ellipse is showing the average from \dzero and LHCb results using only untagged $B_{(s)}^0 \to D_{(s)}^{(*)-} \mu^+\nu X$ decays. 
The white point close to $(0,0)$ is the Standard Model prediction
of \Refx{Albrecht:2024oyn}. 
The prediction and the experimental world average are consistent with each other at the level of $\hflavASLDASLSNSIGMA\,\sigma$.}
\labf{ASLs_ASLd}
\end{figure}
\begin{eqnarray}
\ASLd & = & \hflavASLD ~~~ \Longleftrightarrow ~~~ |q_{\particle{d}}/p_{\particle{d}}| = \hflavQPD \,,
\labe{ASLD}
\\
\ASLs & = & \hflavASLS ~~~ \Longleftrightarrow ~~~ |q_{\particle{s}}/p_{\particle{s}}| = \hflavQPS \,,
\labe{ASLS}
\\
\rho(\ASLd , \ASLs) & = & \hflavRHOASLSASLD \,,
\labe{rhoASLDASLS}
\end{eqnarray}
where $\rho(\ASLd , \ASLs)$ is the correlation coefficient between the two measured parameters, and the relation between ${\cal A}_{\rm SL}^q$ and $|q_{\particle{q}}/p_{\particle{q}}|$ is given in \Eq{ASLq}.%
\footnote{
  \label{foot:life_mix:epsilon_B}
  Early analyses and %
  the PDG use the complex
  parameter $\epsilon_{\B} = (p_q-q_q)/(p_q+q_q)$ for the \Bd; if \CP violation in the mixing is small,
  $\ASLd \cong 4 {\rm Re}(\epsilon_{\B})/(1+|\epsilon_{\B}|^2)$ and the 
  average of \Eq{ASLD} corresponds to
  ${\rm Re}(\epsilon_{\B})/(1+|\epsilon_{\B}|^2)= \hflavREBD$.
}
However, the fit $\chi^2$ probability %
is only $\hflavCLPERCENTASLSASLD\%$.
This is mostly due to an overall discrepancy between the \dzero and 
LHCb averages at the level of $\hflavASLLHCBDZERONSIGMA\,\sigma$.
Since the assumptions underlying the inclusion of the \dzero muon results 
in the average\footref{foot:life_mix:Abazov:2013uma}
are somewhat controversial~\cite{Lenz_private_communication}, 
we also provide in  \Table{ASLs_ASLd} an average excluding these results.

The above averages show no evidence of \CP violation in \Bd or \Bs mixing.
They are consistent with the very small predictions of the Standard Model (SM), 
${\cal A}_{\rm SL}^{d,\rm SM} = -(5.1\pm 0.5)\times 10^{-4}$ and 
${\cal A}_{\rm SL}^{s,\rm SM} = +(2.2\pm 0.2)\times 10^{-5}$~\cite{Albrecht:2024oyn}. %
Given the current experimental uncertainties,
there is still significant room for a possible new physics contribution, in particular in the \Bs system. 
In this respect, the deviation of the \dzero dimuon
asymmetry~\citehistory{Abazov:2013uma}{Abazov:2013uma,*Abazov:2011yk_hist,*Abazov:2010hv_hist,*Abazov:2010hj_hist,*Abazov:dimuon_hist}
from expectation has generated significant interest.
However, the recent \ASLs and \ASLd results from LHCb 
are not precise enough yet to settle the issue.
It has been pointed out~\cite{DescotesGenon:2012kr}
that the \dzero dimuon result can be reconciled with the SM expectations
of \ASLs and \ASLd if there are non-SM sources of \CP violation
in the semileptonic decays of the $b$ and $c$ quarks. 
A Run~1 ATLAS study~\cite{Aaboud:2016bmk} of charge asymmetries
in muon+jets $t\bar{t}$ events, %
in which a \b-hadron decays semileptonically to a soft muon,
yields results with limited statistical precision, 
compatible both with the D0 dimuon asymmetry and with the SM predictions. 

At the more fundamental level, \CP violation in \Bs
mixing is caused by the weak phase difference $\phi^s_{12}$ defined in \Eq{phi12}.
The SM prediction for this phase is tiny~\citehistory{Jubb:2016mvq,Artuso:2015swg}{Jubb:2016mvq,Artuso:2015swg,*Lenz_hist},
\begin{equation}
\phi_{12}^{s,\rm SM} = \hflavPHISTWELVESM \,.
\labe{phis12SM}
\end{equation}
However, new physics in \Bs mixing could change the observed phase to
\begin{equation}
\phi^s_{12} = \phi_{12}^{s,\rm SM} + \phi_{12}^{s,\rm NP} \,.
\labe{phi12NP}
\end{equation}
Using \Eq{ALSq_tanphi2}, the current knowledge of \ASLs, \DGs and \dms, 
given in \Eqsss{ASLS}{DGs_DGsGs}{dms} respectively, yields an
experimental determination of $ \phi^s_{12}$,
\begin{equation}
\tan\phi^s_{12} = \ASLs \frac{\dms}{\DGs} = \hflavTANPHI \,,
\labe{tanphi12}
\comment{ %
from math import *
asls = -0.010460 ; easls = 0.006400
dms = 17.719032  ; edms = 0.042701
dgs = 0.0951919  ; edgs = 0.01362480381803716 
tanphi12 = asls*dms/dgs
etanphi12 = tanphi12*sqrt((easls/asls)**2+(edms/dms)**2+(edgs/dgs)**2)
print tanphi12, etanphi12
} %
\end{equation}
which represents only a very weak constraint on new physics constributions. 

\mysubsubsection{Mixing-induced \CP violation in \Bs decays}
\labs{phasebs}

\CP violation 
arising in the interference between $\Bs-\Bsbar$ mixing and decay is a very active field in which large experimental progress has been achieved in the last decade.
The main observable is the %
phase \phiccbars, 
which describes \CP violation in the interference between \Bs mixing and decay in $b \to c\bar{c}s$ transitions.

The golden mode for such studies is 
$\Bs \to \jpsi\Kp\Km$, where $\jpsi$ decays to a muon and an anti-muon and the \Kp and \Km originate predominantly from a $\phi$(1020) resonance, for which a full angular 
analysis of the decay products is performed to 
statistically separate the \CP-even and \CP-odd
contributions in the final state. As already mentioned in 
\Sec{DGs},
CDF~\citehistory{Aaltonen:2012ie}{Aaltonen:2012ie,*CDF:2011af_hist,*Aaltonen:2007he_hist,*Aaltonen:2007gf_hist},
\dzero~\citehistory{Abazov:2011ry}{Abazov:2011ry,*Abazov:2008af_hist,*Abazov:2007tx_hist},
ATLAS~\citehistory{Aad:2014cqa,Aad:2016tdj,Aad:2020jfw}{Aad:2014cqa,*Aad:2012kba_hist,Aad:2016tdj,Aad:2020jfw},
CMS~\citehistory{Khachatryan:2015nza,CMS-PAS-BPH-23-004}{Khachatryan:2015nza,*Sirunyan:2020vke,CMS-PAS-BPH-23-004}
and LHCb~\citehistory{Aaij:2014zsa,Aaij:2017zgz,Aaij:2016ohx,Aaij:2021mus,LHCb:2023sim}{Aaij:2014zsa,*Aaij:2013oba_supersede2,Aaij:2017zgz,*Aaij:2014zsa_partial_supersede,Aaij:2016ohx,Aaij:2021mus,LHCb:2023sim,*Aaij:2019vot_hist}
have used both untagged and tagged $\Bs \to \jpsi\Kp\Km$ (and more generally $\Bs \to (c\bar{c}) K^+K^-$) decays 
for the measurement of \phiccbars.
LHCb~\citehistory{Aaij:2014dka,Aaij:2019mhf}{Aaij:2014dka,*Aaij:2013oba_supersede,Aaij:2019mhf}
has used $\Bs \to \jpsi \pi^+\pi^-$ events, 
analyzed with a full amplitude model
including several $\pi^+\pi^-$ resonances (\eg, $f_0(980)$),
although the
$\jpsi \pi^+\pi^-$ final state had already been shown
to have a \CP-odd fraction
larger than 0.977 at \CL{95}~\cite{LHCb:2012ae}. 
In addition, LHCb has used the $\Bs \to \Dsp\Dsm$ channel~\cite{Aaij:2014ywt,LHCb:2024gkk} to measure \phiccbars.

All CDF, \dzero, ATLAS and CMS analyses provide 
two mirror solutions related by the transformation 
$(\DGs, \phiccbars) \to (-\DGs, \pi-\phiccbars)$. However, the
LHCb analysis of $\Bs \to \jpsi K^+K^-$ resolves this ambiguity and 
rules out the solution with negative \DGs~\cite{Aaij:2012eq},
a result in agreement with the Standard Model expectation.
Therefore, in what follows, we only consider the solution with $\DGs > 0$.

In the $\Bs\to\jpsi K^+K^-$ analyses, \phiccbars and \DGs 
come from a simultaneous fit that determines also the \Bs lifetime,
the longitudinal and perpendicular $\phi$ polarization amplitudes $|A_{0}|^{2}$ and $|A_{\perp}|^{2}$, the S-wave amplitude $|A_{S}|^2$,  and the strong phases.
While the correlation between \phiccbars and all other parameters is small,
the correlations between \DGs, \Gs and the polarization amplitudes are sizeable. Therefore the full set of parameters provided by the measurements are combined in a multidimensional fit that considers the correlations between them. The combination uses the single-experiment average provided by ATLAS~\cite{Aad:2020jfw}. Both CMS and LHCb measure $\Gamma_s-\Gamma_d$ instead of $\Gamma_s$. Therefore, in order to obtain a value for $\Gamma_s$ and combine it with the ATLAS result, we first perform an common average of CMS and LHCb results with the full data set from Run~1 and Run~2. This average uses an updated world average $\tau(B^{0})$ value (see \Table{sumlife}).

As second-order loop processes could have different contributions to \phiccbars, we perform two combinations. In the first one, 
we perform a combination of all the CDF~\citehistory{Aaltonen:2012ie}{Aaltonen:2012ie,*CDF:2011af_hist,*Aaltonen:2007he_hist,*Aaltonen:2007gf_hist},
\dzero~\citehistory{Abazov:2011ry}{Abazov:2011ry,*Abazov:2008af_hist,*Abazov:2007tx_hist},
ATLAS~\citehistory{Aad:2014cqa,Aad:2016tdj,Aad:2020jfw}{Aad:2014cqa,*Aad:2012kba_hist,Aad:2016tdj,Aad:2020jfw},
CMS~\citehistory{Khachatryan:2015nza,CMS-PAS-BPH-23-004}{Khachatryan:2015nza}
and LHCb~\citehistory%
{Aaij:2014zsa,Aaij:2014ywt,LHCb:2024gkk,Aaij:2017zgz,Aaij:2016ohx,Aaij:2014dka,Aaij:2019mhf,Aaij:2021mus,LHCb:2023sim}%
{Aaij:2014zsa,*Aaij:2013oba_supersede2,Aaij:2014ywt,LHCb:2024gkk,Aaij:2017zgz,*Aaij:2014zsa_partial_supersede,Aaij:2016ohx,Aaij:2014dka,*Aaij:2013oba_supersede,Aaij:2019mhf,Aaij:2021mus,LHCb:2023sim,*Aaij:2019vot_hist}
results listed in \Table{phisDGsGs}. The second one uses only the $\Bs\to \jpsi\Kp\Km$ measurements in the vicinity of the $\phi(1020)$ resonance~\citehistory%
{Aaltonen:2012ie,Abazov:2011ry,Aad:2014cqa,Aad:2016tdj,Khachatryan:2015nza,CMS-PAS-BPH-23-004,Aaij:2014zsa,Aad:2020jfw,Aaij:2021mus,LHCb:2023sim}%
{Aaltonen:2012ie,*CDF:2011af_hist,*Aaltonen:2007he_hist,*Aaltonen:2007gf_hist,Abazov:2011ry,*Abazov:2008af_hist,*Abazov:2007tx_hist,Aad:2014cqa,*Aad:2012kba_hist,Aad:2016tdj,Khachatryan:2015nza,CMS-PAS-BPH-23-004,Aaij:2014zsa,*Aaij:2013oba_supersede2,Aad:2020jfw,Aaij:2021mus,LHCb:2023sim,*Aaij:2019vot_hist}.

ATLAS~\cite{Aad:2020jfw} measures two solutions for the strong phases $\delta_{\perp}$ and $\delta_{\parallel}$. Using one or the other only leads to minor differences in the main parameters of interest, \phiccbars, \DGs and \Gs. For simplicity, in this average, only solution (a) is used. %
In previous averages, the values of $\Delta m_{s}$ and $|\lambda|$ were fixed to the ones used by the ATLAS measurement. These are found to have little impact on  \phiccbars, \DGs and \Gs, and are now floated in the fit. Different regions of the $m(KK)$ mass are used in each measurement and the corresponding $S$-wave fractions and strong phases are considered independent for each experiment.

Using the same approach as discussed in \Sec{DGs} to address the tension between the  $\Bs\to\jpsi\Kp\Km$ measurements of ATLAS~\cite{Aad:2016tdj,Aad:2020jfw}, CMS~\citehistory{Khachatryan:2015nza,CMS-PAS-BPH-23-004}{Khachatryan:2015nza,*Sirunyan:2020vke} and LHCb~\citehistory{Aaij:2014zsa,LHCb:2023sim}{Aaij:2014zsa,LHCb:2023sim,*Aaij:2019vot_hist}, the total uncertainty for each parameter in each $\Bs\to\jpsi\Kp\Km$ set of results by ATLAS, CDF, D0, CMS and LHCb is scaled up in a way that results in an agreement of $1\sigma$. The resulting scale factors are summarized in \Table{phisAllRes}.  

\begin{table}
\caption{Direct experimental measurements of \phiccbars and, wherever applicable, \DGs using $\jpsi K^+K^-$, $\psi(2S)\Kp\Km$, $\jpsi\pi^+\pi^-$ and $D_s^+D_s^-$ decays.
The first error is due to 
statistics, the second one to systematics. The last (last but one) line gives our averages, where the $\DGs$ uncertainties have been multiplied by \hflavDGSBCCSF (\hflavDGSJPSIKKSF) to account for inconsistencies between the $\Bs\to\jpsi\Kp\Km$ measurements. Only solution (a) of Ref.~\cite{Aad:2020jfw} is used.
}
\labt{phisDGsGs}
\begin{center}
\begin{tabular}{@{}ll@{\,}rll@{\,}r@{}} 
\hline
Exp.\ & Mode & Dataset & \multicolumn{1}{c}{\phiccbars (rad)}
                     & \multicolumn{1}{c}{\DGs (\!\!\invps)} & Ref.\ \\
\hline
CDF    & $\jpsi\Kp\Km$ & $9.6\invfb$
       & $[-0.60,\, +0.12]$, 68\% C.L. & $+0.068\pm0.026\pm0.009$
       & \protect\citehistory{Aaltonen:2012ie}{Aaltonen:2012ie,*CDF:2011af_hist,*Aaltonen:2007he_hist,*Aaltonen:2007gf_hist} \\
\dzero & $\jpsi\Kp\Km$ & $8.0\invfb$
       & $-0.55^{+0.38}_{-0.36}$ & $+0.163^{+0.065}_{-0.064}$ %
       & \citehistory{Abazov:2011ry}{Abazov:2011ry,*Abazov:2008af_hist,*Abazov:2007tx_hist} \\
\vspace{-0.3cm} & & & &  \\
ATLAS  & $\jpsi\Kp\Km$ & $4.9\invfb$
       & $+0.12 \pm 0.25 \pm 0.05$ & $+0.053 \pm0.021 \pm0.010$ %
       & \citehistory{Aad:2014cqa}{Aad:2014cqa,*Aad:2012kba_hist} \\
ATLAS  & $\jpsi\Kp\Km$ & $14.3\invfb$
       & $-0.110 \pm 0.082 \pm 0.042$ & $+0.101 \pm0.013 \pm0.007$ %
       & \cite{Aad:2016tdj} \\
ATLAS  & $\jpsi\Kp\Km$ & $80.5\invfb$
       & $-0.081 \pm 0.041 \pm 0.022$ & $+0.0607 \pm0.0047 \pm0.0043$ %
       & \cite{Aad:2020jfw} \\
ATLAS  & \multicolumn{2}{r}{above 3 combined} %
       & $-0.087 \pm 0.036 \pm 0.021$ & $+0.0657 \pm0.0043 \pm0.0037$ %
       & \cite{Aad:2020jfw} \\
\end{tabular}
\end{center}
\end{table}

\begin{figure}
\centering
\includegraphics[width=0.49\textwidth]{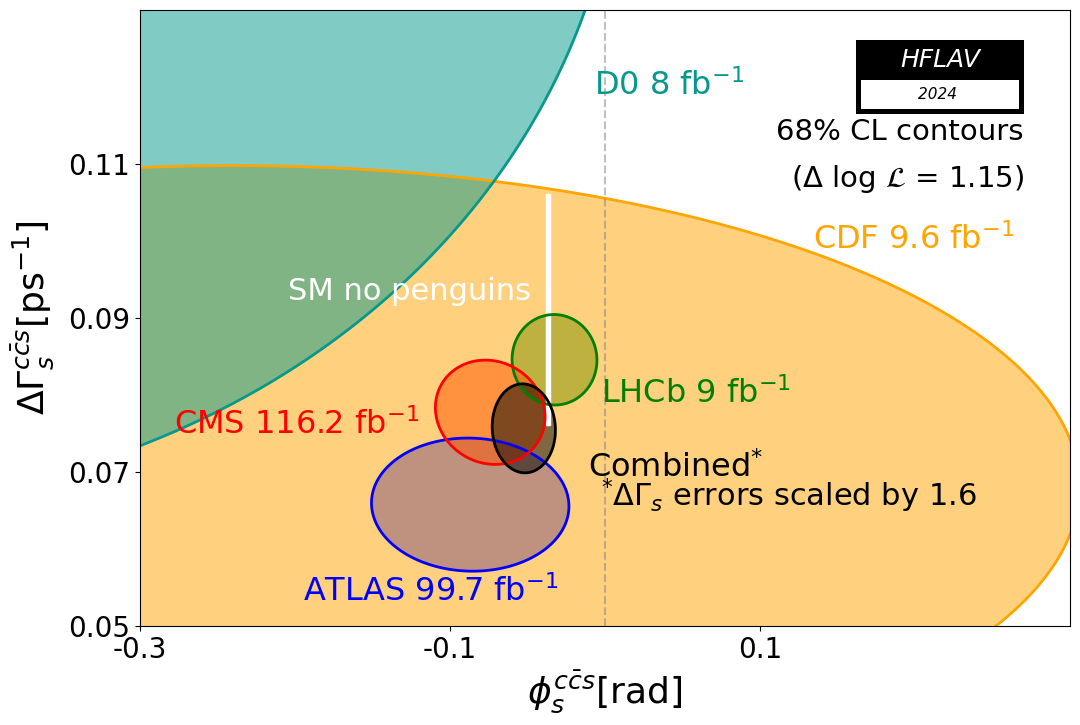}
\includegraphics[width=0.49\textwidth]{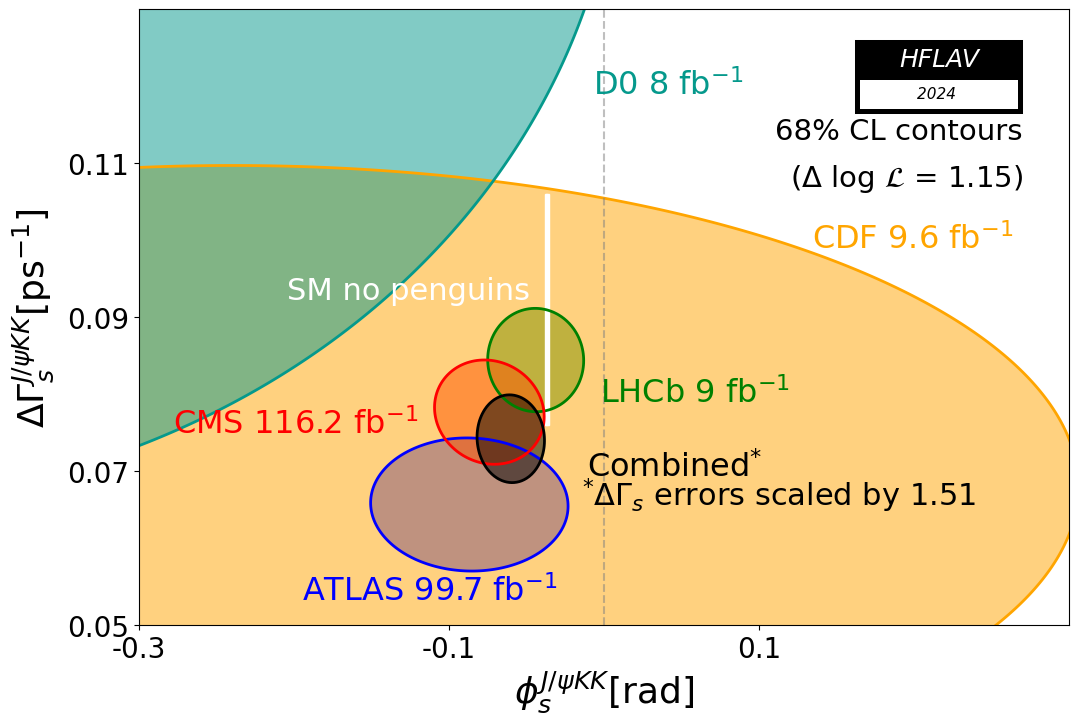}
\includegraphics[width=0.49\textwidth]{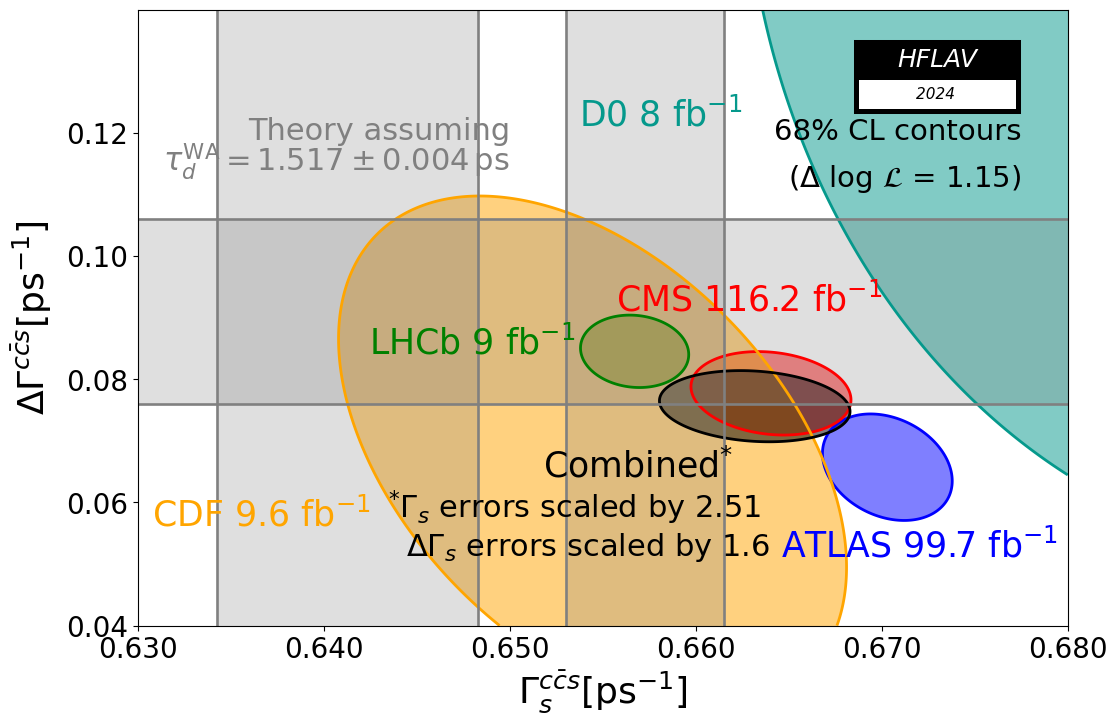}
\includegraphics[width=0.49\textwidth]{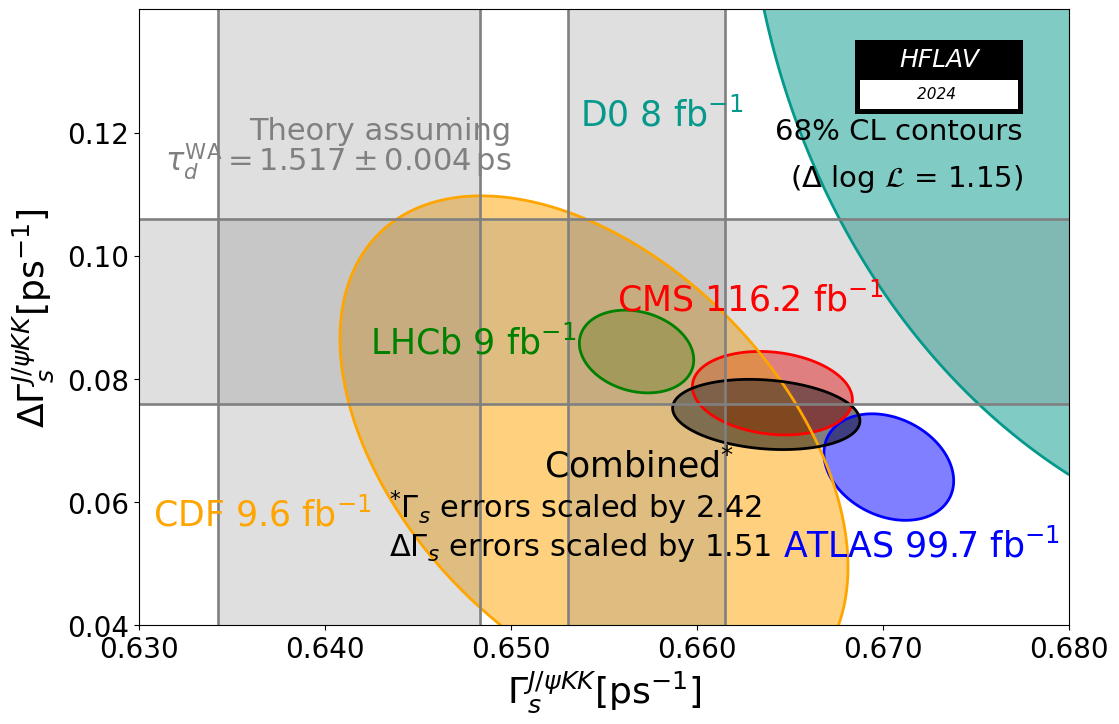}
\caption{
Regions at \CL{68} shown in the $(\phiccbars, \DGs)$ plane on the top and in the $(\Gs, \DGs)$ plane on the bottom, for individual experiments and their combination. The left plots 
are obtained from all 
CDF~\protect\citehistory{Aaltonen:2012ie}{Aaltonen:2012ie,*CDF:2011af_hist,*Aaltonen:2007he_hist,*Aaltonen:2007gf_hist},
\dzero~\protect\citehistory{Abazov:2011ry}{Abazov:2011ry,*Abazov:2008af_hist,*Abazov:2007tx_hist},
ATLAS~\protect\citehistory{Aad:2014cqa,Aad:2016tdj,Aad:2020jfw}{Aad:2014cqa,*Aad:2012kba_hist,Aad:2016tdj,Aad:2020jfw}, 
CMS~\protect\citehistory{Khachatryan:2015nza,CMS-PAS-BPH-23-004}{Khachatryan:2015nza,*Sirunyan:2020vke}
and LHCb~\protect\citehistory%
{Aaij:2014zsa,Aaij:2017zgz,Aaij:2016ohx,Aaij:2014dka,Aaij:2014ywt,LHCb:2024gkk,Aaij:2019mhf, Aaij:2021mus,LHCb:2023sim}%
{Aaij:2014zsa,*Aaij:2013oba_supersede2,Aaij:2017zgz,*Aaij:2014zsa_partial_supersede,Aaij:2016ohx,Aaij:2014dka,*Aaij:2013oba_supersede,Aaij:2014ywt,LHCb:2024gkk,Aaij:2019mhf,Aaij:2021mus,LHCb:2023sim,*Aaij:2019vot_hist}
measurements of  $\Bs\to \jpsi K^+K^-$, $\Bs\to \psi(2S) \Kp\Km$, $\Bs\to\jpsi\pi^+\pi^-$ and 
$\Bs\to D_s^+D_s^-$ decays, while the right plots are obtained from
$\Bs\to \jpsi\Kp\Km$ measurements only~\protect\citehistory%
{Aaltonen:2012ie,Abazov:2011ry,Aad:2014cqa,Aad:2016tdj,Khachatryan:2015nza,CMS-PAS-BPH-23-004,Aaij:2014zsa,Aad:2020jfw,Aaij:2021mus,LHCb:2023sim}%
{Aaltonen:2012ie,*CDF:2011af_hist,*Aaltonen:2007he_hist,*Aaltonen:2007gf_hist,Abazov:2011ry,*Abazov:2008af_hist,*Abazov:2007tx_hist,Aad:2014cqa,*Aad:2012kba_hist,Aad:2016tdj,Khachatryan:2015nza,*Sirunyan:2020vke,Aaij:2014zsa,*Aaij:2013oba_supersede2,Aad:2020jfw,Aaij:2021mus,LHCb:2023sim,*Aaij:2019vot_hist}.
The expectations within the Standard Model neglecting penguin contributions~\protect\citehistory{Charles:2015gya_mod,UTfit:2022hsi_mod,Albrecht:2024oyn}{Charles:2015gya_mod,Lenz:2011ti,*Lenz:2006hd,Jubb:2016mvq,Artuso:2015swg,*Lenz_hist,Lenz:2019lvd,UTfit:2022hsi_mod,Albrecht:2024oyn}
are shown as white rectangles in the top plots. The \Gs theory value in the bottom plots is calculated from \Refx{Albrecht:2024oyn} assuming the experimental world average for the \Bd lifetime, %
\hflavTAUBD. The two solutions in \Gs are due to the two different sets of inputs to the theory calculation of $\tau(\Bs)/\tau(\Bd)$ according to \Refx{Bordone:2021oof} (toward lower \Gs values) and \Refx{Bernlochner:2022ucr} (toward higher \Gs values).
}
\labf{DGs_phase}
\end{figure}
Given the increasing experimental precision of the LHC results, we have stopped using the two-dimensional $\DGs-\phiccbars$ histograms provided by the CDF and \dzero collaborations, and are now approximating them with two-dimensional Gaussian likelihoods. 

We obtain the individual and combined contours shown in \Fig{DGs_phase}. %
Maximizing the likelihood, we find, as summarized in \Table{phisDGsGs},  
\begin{eqnarray}
\DGs &=& \hflavDGSBCC \,, \\    
\phiccbars &=& \hflavPHISBCC \,.
\labe{phis}
\end{eqnarray}
This \DGs average is consistent but highly correlated with the average
of \Eq{DGs_DGsGs}. Our
final recommended average for \DGs is the one of \Eq{DGs_DGsGs}, which 
includes all available information on this quantity. Other averaged parameters are listed in \tabref{phisAllRes}, and their correlations are in \tabref{phisAllCorr}. For the sake of compactness, we exclude parameters that are used in single measurements.

\begin{table}
\caption{Results of the averaging procedure, including the fit results and the scale factors of the individual parameters, for all $b \to c\bar{c}s$ modes (second and third column) and for $\Bs\to\jpsi\Kp\Km$ modes only (fourth and fifth column). Only parameters measured by all three LHC experiments are scaled.}
\labt{phisAllRes}
\begin{center}
\begin{tabular}{lllll} 
\toprule
Parameter
& \multicolumn{2}{c}{ all $b\to c\bar{c}s$} & \multicolumn{2}{c}{$\Bs\to\jpsi \Kp\Km$}  \\
& fit result & scale factor & fit result & scale factor \\
\midrule
$\phi_{s}$ & \hflavPHISBCC & \hflavPHISBCCSF &\hflavPHISJPSIKK & \hflavPHISJPSIKKSF\\
$\left|\lambda\right|$  & \hflavLAMBBCC & -- & \hflavLAMBJPSIKK & --\\
$\Gamma_{s}$  & \hflavGSBCC & \hflavGSBCCSF & \hflavGSJPSIKK & \hflavGSJPSIKKSF\\
$\Delta\Gamma_{s}$    & \hflavDGSBCC  & \hflavDGSBCCSF & \hflavDGSJPSIKK & \hflavDGSJPSIKKSF \\
$\left|A_{\perp}^{J/\psi KK}\right|^2$  & \hflavAPERPSQBCC & \hflavAPERPBCCSF & \hflavAPERPSQJPSIKK & \hflavAPERPJPSIKKSF\\
$\left|A_{0}^{J/\psi KK}\right|^2$  & \hflavAZEROSQBCC  & \hflavAZEROBCCSF 
& \hflavAZEROSQJPSIKK & \hflavAZEROJPSIKKSF\\
$\delta^{J/\psi KK}_{\perp}$  & \hflavDPEBCC & \hflavDPEBCCSF & \hflavDPEJPSIKK & \hflavDPEJPSIKKSF\\
$\delta^{J/\psi KK}_{\parallel}$  & \hflavDPABCC & 
\hflavDPABCCSF& \hflavDPAJPSIKK & \hflavDPAJPSIKKSF \\
$\Delta m_s$ & \hflavDMSBCC & -- & \hflavDMSJPSIKK & --\\
\bottomrule

\end{tabular}
\end{center}
\end{table}

\begin{table}
\caption{Correlation tables of the averaged observables from the fit to all $b \to c\bar{c}s$ modes (top) and to the $\Bs\to\jpsi\Kp\Km$ modes only (bottom).}
\labt{phisAllCorr}
\begin{center}

\begin{tabular}{lrrrrrrrrr}
\toprule
Parameter & $\phi_{s}$ & $\left|\lambda^{J/\psi KK}\right|$ & $\Gamma_{s}$&  $\Delta\Gamma_{s}$  & $\left|A^{J/\psi KK}_{\perp}\right|^2$  & $\left|A^{J/\psi KK}_{0}\right|^2$ & $\delta^{J/\psi KK}_{\perp}$  & $\delta^{J/\psi KK}_{\parallel}$  & $\Delta m_{s}$ \\
\midrule$\phi_{s}$  & 1.00 & 0.03 & 0.00 & $-$0.04 & 0.01 & $-$0.01 & $-$0.01 & 0.00 & $-$0.01 \\ 
$\left|\lambda^{J/\psi KK}\right|$  &  & 1.00 & 0.00 & $-$0.01 & 0.03 & $-$0.03 & $-$0.11 & $-$0.06 & $-$0.11\\ 
 $\Gamma_{s}$   &  &  & 1.00 & $-$0.15 & 0.07 & $-$0.05 & $-$0.01 & $-$0.03 & 0.01 \\ 
 $\Delta\Gamma_{s}$  &  &  &  & 1.00 & $-$0.24 & 0.17 & $-$0.01 & 0.00 & 0.02 \\ 
$\left|A^{J/\psi KK}_{\perp}\right|^2$  &  &  &  &  & 1.00 & $-$0.73 & $-$0.04 & $-$0.05 & $-$0.02 \\ 
$\left|A^{J/\psi KK}_{0}\right|^2$  &  &  &  &  &  & 1.00 & 0.02 & 0.00 & 0.02\\ 
$\delta^{J/\psi KK}_{\perp}$  &  &  &  &  &  &  & 1.00 & 0.22 & 0.49\\ 
$\delta^{J/\psi KK}_{\parallel}$  &  &  &  &  &  &  &  & 1.00 & $-$0.09\\ 
$\Delta m_{s}$  &  &  &  &  &  &  &  &  & 1.00\\ 
\bottomrule
\end{tabular}
\begin{tabular}{lrrrrrrrrr} 
\toprule
Parameter & $\phi_{s}$ & $\left|\lambda^{J/\psi KK}\right|$ & $\Gamma_{s}$&  $\Delta\Gamma_{s}$  & $\left|A_{\perp}^{J/\psi KK}\right|^2$  & $\left|A_{0}^{J/\psi KK}\right|^2$ & $\delta_{\perp}$  & $\delta_{\parallel}$  & $\Delta m_{s}$ \\
\midrule
$\phi_{s}$  & 1.00 & 0.02 & 0.00 & $-$0.04 & 0.01 & $-$0.01 & 0.00 & 0.02 & 0.00\\ 
$\left|\lambda^{J/\psi KK}\right|$  &  & 1.00 &0.00 & $-$0.01 & 0.04 & $-$0.04 & $-$0.08 & $-$0.05 & $-$0.07\\
 $\Gamma_{s}$   &  &  & 1.00 & $-$0.18 & 0.08 & $-$0.06 & 0.00 & $-$0.02 & 0.01 \\ 
 $\Delta\Gamma_{s}$  &  &  &  & 1.00 & $-$0.25 & 0.18 & 0.00 & 0.01 & 0.03 \\ 
$\left|A_{\perp}^{J/\psi KK}\right|^2$  &  &  &  &  & 1.00 & $-$0.72 & $-$0.08 & $-$0.06 & $-$0.06 \\ 
$\left|A_{0}^{J/\psi KK}\right|^2$ &  &  &  &  &  & 1.00 & 0.05 & 0.00 & 0.05\\ 
$\delta^{J/\psi KK}_{\perp}$  &  &  &  &  &  &  & 1.00 & 0.26 & 0.40 \\ 
$\delta^{J/\psi KK}_{\parallel}$  &  &  &  &  &  &  &  & 1.00 & $-$0.12\\ 
$\Delta m_{s}$  &  &  &  &  &  &  &  &  & 1.00\\

\bottomrule

\end{tabular}

\end{center}
\end{table}

In the Standard Model and ignoring subleading penguin contributions, 
\phiccbars is expected to be equal to $-2\beta_s$, 
where
$\beta_s = \arg\left[-\left(V_{ts}V^*_{tb}\right)/\left(V_{cs}V^*_{cb}\right)\right]$ %
is a phase analogous to the angle $\beta$ of the usual CKM
unitarity triangle (aside from a sign change). %
An indirect determination via global fits to experimental data
gives%
\begin{equation}
(\phiccbars)^{\rm SM} = -2\beta_s =
\begin{cases}
\hflavPHISSM & \text{CKMfitter\cite{Charles:2015gya_mod}} \\
 \hflavPHISSMUTFIT & \text{UTfit\cite{UTfit:2022hsi_mod}}.
\end{cases}
\labe{phisSM}
\end{equation}
The average value of \phiccbars from \Eq{phis} is consistent with this
Standard Model expectation.
Penguin contributions to \phiccbars from $\Bs \to \jpsi \Kp\Km$ are calculated to be smaller than $0.021$ in magnitude~\citehistory{Frings:2015eva,Aaij:2014vda,LHCb:2015esn}{Frings:2015eva,Aaij:2014vda,Aaij:2015mea} but may become relevant if future measurements reduce the error in \Eq{phis}.
There are no reliable estimates of the  penguin contribution to $\Bs \to \jpsi f_0$.

From its measurements of time-dependent \CP violation in $\Bs \to K^+K^-$ decays, the LHCb collaboration has determined the 
\Bs mixing phase to be $-2\beta_s = -0.12^{+0.14}_{-0.12}$~\cite{Aaij:2014xba},
assuming a U-spin relation (with up to 50\% breaking effects) between the decay amplitudes of $\Bs \to K^+K^-$ 
and $\Bd \to \pi^+\pi^-$, and a value of the CKM angle $\gamma$  
of $(70.1 \pm7.1)^{\circ}$. This determination is compatible with, 
and less precise than, the world average of \phiccbars from \Eq{phis}.

New physics could contribute to \phiccbars. Assuming that new physics only 
enters in $M^s_{12}$ (rather than in $\Gamma^s_{12}$),
one can write~\cite{Lenz:2011ti,Lenz:2006hd}
\begin{equation}
\phiccbars = -  2\beta_s + \phi_{12}^{s,\rm NP} \,,
\end{equation}
where the new physics phase $\phi_{12}^{s,\rm NP}$ is the same as that appearing in \Eq{phi12NP}.
In this case
\begin{equation}
\phi^s_{12} = %
\phi_{12}^{s,\rm SM} +2\beta_s + \phiccbars = \hflavPHISTWELVE \,,
\end{equation}
where the numerical estimation was performed with the values of \Eqsss{phis12SM}{phisSM}{phis}. Note that, within the quoted precision, the same $\phi^s_{12}$ value is obtained with the $2\beta_s$ inputs from CKMfitter and UTfit.
Keeping in mind the approximation and assumption mentioned above,
this can serve as a reference value to which the measurement of \Eq{tanphi12} can be compared.

\clearpage

\section{Semileptonic $b$ hadron decays}
\label{sec:slbdecays}

In this chapter we present averages for semileptonic $b$~hadron decays,
\ie\ decays of the type $B\to X\ell\nu_\ell$, where $X$ refers to one
or more hadrons, $\ell$ to a charged lepton and $\nu_\ell$ to its associated
neutrino. Unless otherwise stated, $\ell$ stands for an electron
or a muon, lepton universality is assumed, and both charge
conjugate states are combined. Some averages assume isospin symmetry which is
explicitly mentioned at every instance.

Averages are presented separately for CKM favored $b\to c$ quark transitions
and CKM suppressed $b\to u$ modes. We further
distinguish \emph{exclusive} decays involving a specific meson ($X=D, D^*, \pi,
\rho,\dots$) from \emph{inclusive} decay modes, \ie\ the sum over all possible
hadronic states. Semileptonic decays proceed
via first order weak interactions and are well described in the framework of
the SM. Their decay rates are sensitive to the magnitude
squared of the CKM elements $V_{cb}$ and $V_{ub}$, the determination of which is
one of the primary goals for the study of these decays. Semileptonic decays
involving the $\tau$~lepton might be more sensitive to beyond-SM processes, because
the high $\tau$~mass can result in enhanced couplings 
to hypothetical new particles such as a charged Higgs boson or leptoquarks.

The technique for obtaining the averages follows the general HFLAV
procedure (see chapter~\ref{sec:method}) unless otherwise stated. More
information on the averages, in particular on the common input parameters,
is available on the HFLAV semileptonic webpage. %
In general, averages in this
section use experimental results available through autumn 2023.
\subsection{Exclusive CKM-favoured decays}
\label{slbdecays_b2cexcl}

\mysubsubsection{$\bar B\to D^*\ell^-\bar\nu_\ell$}
\label{slbdecays_dstarlnu}

The $\bar B\to D^*\ell^-\bar\nu_\ell$ decay rate is fully parameterized by the recoil variable  $w=v_B\cdot v_{D^*}$, the product of the four-velocities of the initial and final state
mesons and three decay angles $\theta_\ell$, $\theta_V$ and $\chi$, defined as follows (Fig.~\ref{fig:dslnu_angles}):
\begin{itemize}
\item The angle $\theta_\ell$ is the angle between the direction of the charged lepton and the direction opposite to the $B$~meson in the virtual $W$~boson rest frame,
\item The angle $\theta_V$ is the angle between the direction of the $D$~meson and the direction opposite to the $B$~meson in the $D^*$~meson rest frame, and
\item The angle $\chi$ is the azimuthal angle between the two decay planes spanned by the $W$~boson and $D^*$~meson decay products, and defined in the rest frame of the $B$~meson.
\end{itemize}
\begin{figure}
    \begin{center}
    \includegraphics[width=0.7\linewidth]{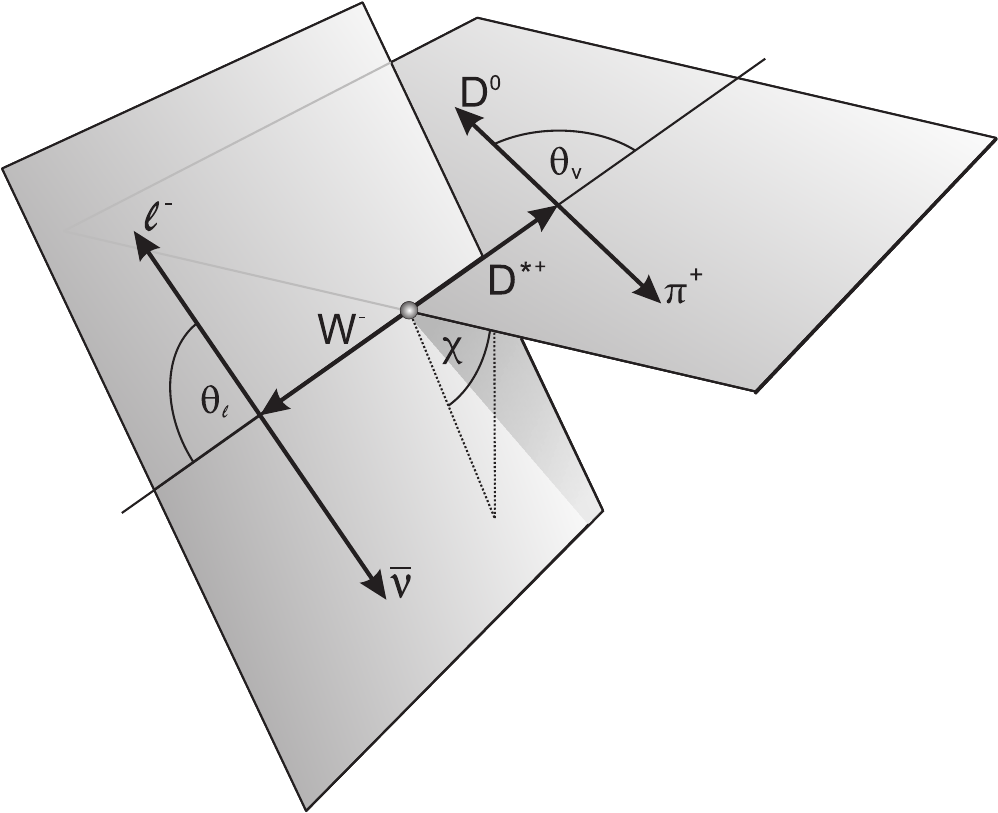}
    \end{center}
    \caption{Sketch of the helicity angles $\theta_\ell$, $\theta_V$ and $\chi$ in the decay $\bar B\to D^*\ell^-\bar\nu_\ell$.}\label{fig:dslnu_angles}
\end{figure}
Integrating over the angles, the differential decay rate for massless
fermions as a function of $w$ is given by (see, \eg,~\cite{Neubert:1993mb})
\begin{equation}
  \frac{d\Gamma(\bar B\to D^*\ell^-\bar\nu_\ell)}{dw} = \frac{G^2_\mathrm{F} m^3_{D^*}}{48\pi^3}(m_B-m_{D^*})^2\chi(w)\eta_\mathrm{EW}^2\mathcal{F}^2(w)\vcb^2~,
  \label{eq:gamma_dslnu}
\end{equation}
where $G_\mathrm{F}$ is Fermi's constant, $m_B$ and $m_{D^*}$ are the $B$ and
$D^*$ meson masses, $\chi(w)$ is a known phase-space factor, and
$\eta_\mathrm{EW}$ is a small electroweak correction~\cite{Sirlin:1981ie}.
Some authors also include a long-distance EM radiation effect (Coulomb
correction) in this factor.
The form factor $\mathcal{F}(w)$ for the $\bar B\to D^*\ell^-\bar\nu_\ell$
decay contains three independent functions, $h_{A_1}(w)$, $R_1(w)$ and $R_2(w)$,
\begin{eqnarray}
  &&\chi(w)\mathcal{F}^2(w)=\\
  && \phantom{\mathcal{F}^2(w)} h_{A_1}^2(w)\sqrt{w^2-1}(w+1)^2 \left\{2\left[\frac{1-2wr+r^2}{(1-r)^2}\right]\left[1+R^2_1(w)\frac{w-1}{w+1}\right]+\right. \nonumber\\
  && \phantom{\mathcal{F}^2(w)} \left.\left[1+(1-R_2(w))\frac{w-1}{1-r}\right]^2\right\}~, \nonumber
\end{eqnarray}
where $r=m_{D^*}/m_B$.

\mysubsubsubsection{Branching fraction}

First, we perform separate one-dimensional averages of the
$\BzbDstarlnu$ and $B^-\to D^{*0}\ell^-\bar\nu_\ell$
branching fractions. In the fit to the measurements listed in Tables~\ref{tab:dstarlnu} and \ref{tab:dstar0lnu}, external parameters (such as the branching fractions of charmed mesons) are constrained to their latest values 
and the following results are obtained
\begin{eqnarray}
  \cbf(\BzbDstarlnu) & = & (4.90\pm 0.01\pm 0.12)\%~, \label{eq:br_dstarlnu} \\
  \cbf(B^-\to D^{*0}\ell^-\bar\nu_\ell) & = & (5.53\pm 0.07\pm 0.21)\%~,
   \label{eq:br_dstar0lnu}
\end{eqnarray}
where the first uncertainty is statistical and the second one is systematic.
The results of these two fits are also shown in Fig.~\ref{fig:brdsl}.
\begin{table}[!htb]
  \caption{Average of the $\BzbDstarlnu$ branching fraction measurements.}
  \begin{center} 
\resizebox{0.99\textwidth}{!}{
\begin{tabular}{|l|c|c|}\hline
  Experiment & $\cbf(\BzbDstarlnu)$ [\%] (rescaled) &
  $\cbf(\BzbDstarlnu)$ [\%] (published)\\
  \hline\hline
  ALEPH~\cite{Buskulic:1996yq}
  & $5.29\pm 0.26_{\rm stat} \pm 0.33_{\rm syst}$
  & $5.53\pm 0.26_{\rm stat} \pm 0.52_{\rm syst}$\\
  OPAL incl~\cite{Abbiendi:2000hk}
  & $6.13\pm 0.28_{\rm stat} \pm 0.57_{\rm syst}$
  & $5.92\pm 0.27_{\rm stat} \pm 0.68_{\rm syst}$\\
  OPAL excl~\cite{Abbiendi:2000hk}
  & $5.11\pm 0.20_{\rm stat} \pm 0.36_{\rm syst}$
  & $5.11\pm 0.19_{\rm stat} \pm 0.49_{\rm syst}$\\
  DELPHI incl~\cite{Abreu:2001ic}
  & $4.94\pm 0.14_{\rm stat} \pm 0.35_{\rm syst}$
  & $4.70\pm 0.13_{\rm stat} \ {}^{+0.36}_{-0.31}\ {}_{\rm syst}$\\
  DELPHI excl~\cite{Abdallah:2004rz}
  & $5.05\pm 0.20_{\rm stat} \pm 0.42_{\rm syst}$
  & $5.90\pm 0.22_{\rm stat} \pm 0.50_{\rm syst}$\\
  CLEO~\cite{Adam:2002uw}
  & $5.85\pm 0.19_{\rm stat} \pm 0.35_{\rm syst}$
  & $6.09\pm 0.19_{\rm stat} \pm 0.40_{\rm syst}$\\
  Belle untagged~\cite{Belle:2018ezy}
  & $4.76\pm 0.02_{\rm stat} \pm 0.16_{\rm syst}$
  & $4.90\pm 0.02_{\rm stat} \pm 0.16_{\rm syst}$\\
  \babar\ untagged~\cite{Aubert:2006mb}
  & $4.35\pm 0.04_{\rm stat}\pm 0.33_{\rm syst}$
  & $4.69\pm 0.04_{\rm stat} \pm 0.34_{\rm syst}$\\
  \babar\ tagged~\cite{Aubert:vcbExcl}
  & $5.08\pm 0.16_{\rm stat}\pm 0.31_{\rm syst}$
  & $5.49\pm 0.16_{\rm stat} \pm 0.25_{\rm syst}$\\
  Belle II untagged~\cite{Belle-II:2023okj}
  & $4.83\pm 0.02_{\rm stat} \pm 0.20_{\rm syst}$
  & $4.922\pm 0.023_{\rm stat} \pm 0.220_{\rm syst}$\\
  Belle II tagged~\cite{Belle-II:2023jtw}
  & $5.19\pm 0.22_{\rm stat} \pm 0.40_{\rm syst}$
  & $5.27\pm 0.22_{\rm stat} \pm 0.38_{\rm syst}$\\
  
  \hline 
  {\bf Average} & \mathversion{bold}$4.90\pm 0.01_{\rm stat}\pm
  0.12_{\rm syst}$ & \mathversion{bold}$\chi^2/\dof = 15.3/10$ (CL=$12.2\%$)\\
  \hline 
\end{tabular}
}
\end{center}
\label{tab:dstarlnu}
\end{table}
\begin{table}[!htb]
\caption{Average of the $B^-\to D^{*0}\ell^-\bar\nu_\ell$ branching
  fraction measurements.}
\begin{center}
%
  \caption{Branching fractions of exclusive semileptonic $B$ decays:
    (a) $\BzbDstarlnu$ (Table~\ref{tab:dstarlnu}) and (b) $B^-\to
    D^{*0}\ell^-\bar\nu_\ell$ (Table~\ref{tab:dstar0lnu}).} \label{fig:brdsl}
  \end{center}
\end{figure}

We combine these two averages into a result for $\cbf(\BzbDstarlnu)$ assuming isospin symmetry. Therefore, we rescale the $B^+$ branching fractions by the lifetime ratio $\tau_{0+}=\tau(B^0)/\tau(B^+)$ and perform a combined 1-dimensional fit over the $\BzbDstarlnu$ and $B^-\to D^{*0}\ell^-\bar\nu_\ell$ branching fractions. To account for isospin breaking due to the Coulomb effect in $\BzbDstarlnu$ we inflate the uncertainty in $\tau_{0+}$ by $\tau_{0+}\alpha\pi\approx 0.02123$~\cite{Atwood:1989em}. The result reads
\begin{equation}
    \cbf(\BzbDstarlnu) = (4.90\pm 0.01\pm 0.11)\%~. \label{eq:br_dstarlnu_iso}
\end{equation}
The $\chi^2/$dof of the combination is 22.56/13 (CL=4.73\%).

\mysubsubsubsection{Average of differential distributions}
We perform an average of the normalized unfolded $\bar B\to D^*\ell^-\bar\nu_\ell$ differential measurements performed by the Belle~\cite{Belle:2023bwv} and Belle~II~\cite{Belle-II:2023okj} collaborations. We omit the Belle measurement in Ref.~\cite{Belle:2018ezy} because they do not provide unfolded spectra. %

We normalize each individual measurement and drop one data point in each marginal distribution to remove the singular values in the covariance matrix C. We then minimize a $\chi^2$ defined as
\begin{equation}
    \chi^2(\vec{\mu}) = \sum_\mathrm{meas.} \sum_x \sum_{i} \left( \left( \frac{\Delta \Gamma / \Gamma}{\Delta x} \right)_i - \mu_i \right) C^{-1}_{ij} \left( \left( \frac{\Delta \Gamma / \Gamma}{\Delta x} \right)_i - \mu_i \right) \,,
\end{equation}
where the first sum runs over all measurements, the second sum runs over the measured variables $x \in (w, \cos\theta_\ell, \cos\theta_V, \chi)$ and the third sum runs over the measured bins. The resulting p-value of the average is 15\% and indicates good agreement between the measurements. The redundant data point in each marginal distribution is recovered via $1 = \sum_\mathrm{bins} (\Delta \Gamma/\Gamma / \Delta x)_i$. The resulting spectrum is shown in Fig.~\ref{fig:BToDstarAverage}. The central values with uncertainties are given in Tab.~\ref{tab:BToDstAverage}. The corresponding correlation matrix of the averaged normalized $\bar B\to D^*\ell^-\bar\nu_\ell$ spectrum, together with the central values and uncertainties are available at~\cite{heavy_flavor_averaging_group_2024_12696548}.

\begin{figure}
    \centering
    \includegraphics[width=\linewidth]{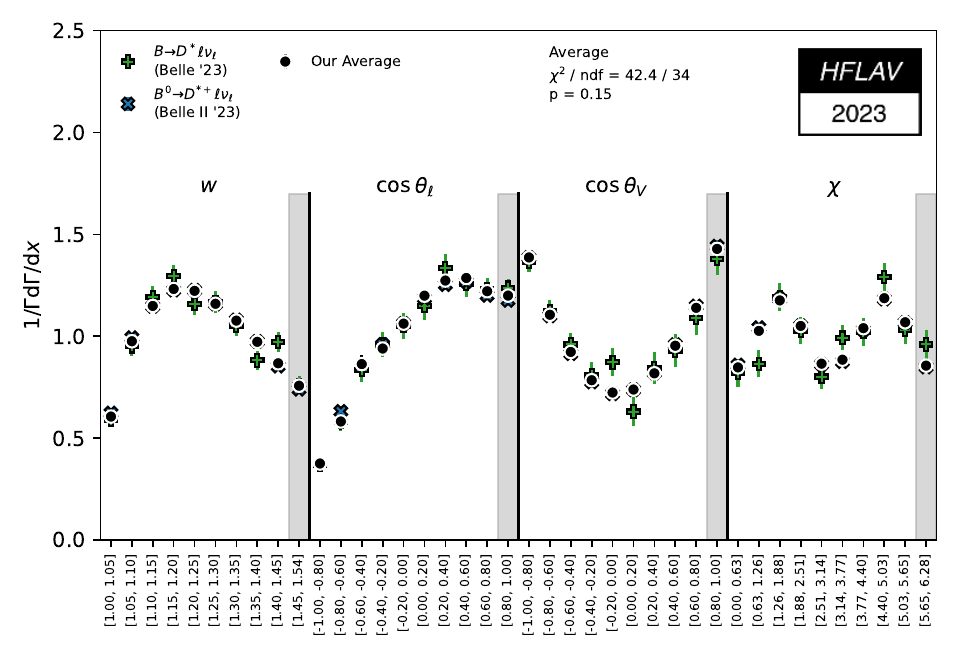}
    \caption{The averaged normalized $\bar B\to D^*\ell^-\bar\nu_\ell$ spectrum (black data points) together with the individual measurements by Belle~\cite{Belle:2023bwv} and Belle~II~\cite{Belle-II:2023okj} collaborations. The gray bands indicate the omitted data point in the average (see text).}
    \label{fig:BToDstarAverage}
\end{figure}

\begin{table}
    \centering
    \caption{The central values and uncertainties of the averaged normalized $\bar B\to D^*\ell^-\bar\nu_\ell$ spectrum.}
    \label{tab:BToDstAverage}
    \resizebox{\columnwidth}{!}{%
    \begin{tabular}{lrlrlrlr}
\toprule
$w$ bin & $1/\Gamma \Delta \Gamma / \Delta w$ & $\cos \theta_\ell$ bin & $1/\Gamma \Delta \Gamma / \Delta \cos \theta_\ell$ & $\cos \theta_V$ bin & $1/\Gamma \Delta \Gamma / \Delta \cos \theta_V$ & $\chi$ bin & $1/\Gamma \Delta \Gamma / \Delta \chi$ \\
\midrule
$[1.00, 1.05]$ & $0.0601\pm0.0017$ & $[-1.00, -0.80]$ & $0.0375\pm0.0034$ & $[-1.00, -0.80$] & $0.1387\pm0.0021$ & $[0.00, 0.63]$ & $0.0847\pm0.0020$ \\
$[1.05, 1.10]$ & $0.0968\pm0.0017$ & $[-0.80, -0.60]$ & $0.058\pm0.004$ & $[-0.80, -0.60$] & $0.1105\pm0.0015$ & $[0.63, 1.26]$ & $0.1026\pm0.0015$ \\
$[1.10, 1.15]$ & $0.1141\pm0.0017$ & $[-0.60, -0.40]$ & $0.086\pm0.005$ & $[-0.60, -0.40$] & $0.0923\pm0.0012$ & $[1.26, 1.88]$ & $0.1176\pm0.0017$ \\
$[1.15, 1.20]$ & $0.1223\pm0.0017$ & $[-0.40, -0.20]$ & $0.0940\pm0.0018$ & $[-0.40, -0.20$] & $0.0784\pm0.0010$ & $[1.88, 2.51]$ & $0.1051\pm0.0015$ \\
$[1.20, 1.25]$ & $0.1214\pm0.0016$ & $[-0.20, 0.00]$ & $0.1062\pm0.0017$ & $[-0.20, 0.00$] & $0.0723\pm0.0010$ & $[2.51, 3.14]$ & $0.0865\pm0.0014$ \\
$[1.25, 1.30]$ & $0.1151\pm0.0015$ & $[0.00, 0.20]$ & $0.1199\pm0.0017$ & $[0.00, 0.20$] & $0.0738\pm0.0010$ & $[3.14, 3.77]$ & $0.0885\pm0.0013$ \\
$[1.30, 1.35]$ & $0.1069\pm0.0015$ & $[0.20, 0.40]$ & $0.1273\pm0.0017$ & $[0.20, 0.40$] & $0.0818\pm0.0011$ & $[3.77, 4.40]$ & $0.1039\pm0.0015$ \\
$[1.35, 1.40]$ & $0.0966\pm0.0015$ & $[0.40, 0.60]$ & $0.1286\pm0.0017$ & $[0.40, 0.60$] & $0.0953\pm0.0013$ & $[4.40, 5.03]$ & $0.1187\pm0.0016$ \\
$[1.40, 1.45]$ & $0.0861\pm0.0014$ & $[0.60, 0.80]$ & $0.1221\pm0.0017$ & $[0.60, 0.80$] & $0.1139\pm0.0014$ & $[5.03, 5.65]$ & $0.1069\pm0.0014$ \\
$[1.45, 1.54]$ & $0.0805\pm0.0018$ & $[0.80, 1.00]$ & $0.1199\pm0.0018$ & $[0.80, 1.00$] & $0.1429\pm0.0024$ & $[5.65, 6.28]$ & $0.0856\pm0.0018$ \\
\bottomrule
\end{tabular}%
    }
\end{table}

\mysubsubsubsection{Extraction of $\vcb$ based on the BGL form factor}
To extract $\vcb$, we consider the model-independent parametrizations of the form factor functions $h_{A_1}(w)$, $R_1(w)$ and $R_2(w)$ contributing to the $\bar B\to D^*\ell^-\bar\nu_\ell$ differential decay rate by Boyd, Grinstein and Lebed~(BGL)~\cite{Boyd:1997kz,Grinstein:2017nlq,Bigi:2017njr}. We use the averaged normalized differential distributions in $w$, $\cos\theta_\ell$, $\cos\theta_V$, and $\chi$, the isospin average of the $B^0$ and $B^+$ absolute branching fractions, and the nonzero recoil lattice QCD calculation by Fermilab/MILC~\cite{FermilabLattice:2021cdg}, HPQCD~\cite{Harrison:2023dzh}, and JLQCD~\cite{Aoki:2023qpa} to extract the CKM matrix element $\vcb$. The Blasche factors in the BGL parameterization use
$[6.730, 6.736, 7.135, 7.142]\,\mathrm{GeV}$ for the axialvector poles and $[6.337, 6.899, 7.012, 7.280]\,\mathrm{GeV}$ for the vector poles~\cite{Eichten:1994gt}.\footnote{Including the fourth vector resonance in the spectrum is a choice by the authors, and is sometimes treated different in the literature.} The susceptibilities are $\chi_{1+}^T(0.33) = 5.28 \times 10^{-4}\,,\mathrm{GeV}^{-2}$ and $\tilde{\chi}_{1-}^T(0.33) = 3.07 \times 10^{-4}\,,\mathrm{GeV}^{-2}$.
The $\chi^2$ function is defined as
\begin{equation}
    \begin{aligned}
    \chi^2 &= \left(\left(\frac{\Delta\vec{\Gamma}}{\Gamma}\right)^{\kern-0.25em\rm m} \kern-0.5em{-} \left(\frac{\Delta\ \vec{\Gamma}({\vec{x}})}{\Gamma({\vec{x}})}\right)^{\kern-0.25em\rm p}\right) C^{-1}_\mathrm{exp} \left(\left(\frac{\Delta\vec{\Gamma}}{\Gamma}\right)^{\kern-0.25em\rm m} \kern-0.5em{-} \left(\frac{\Delta\ \vec{\Gamma}({\vec{x}})}{\Gamma({\vec{x}})}\right)^{\kern-0.25em\rm p}\right)^T \\
       & + (\Gamma^\mathrm{ext} - \Gamma^{\rm p}(\vec{x}))^2 / \sigma(\Gamma^\mathrm{ext})^2 \\
       & + (h_{X}^{\rm p}(\vec{x}) - h_{X}^{\rm LQCD}) C^{-1}_\mathrm{LQCD} (h_{X}^{\rm p}(\vec{x}) - h_{X}^{\rm LQCD}) \,,
    \end{aligned}
    \label{eq:BToDstChi2}
\end{equation}
with the averaged (predicted) normalized partial decay rates $(\Delta \vec \Gamma / \Gamma)^\mathrm{m (p)}$ in bins of $w$ and the helicity angles, where the predicted rate is a function of the form factor coefficients $\vec{x}$ and $|V_{cb}|$. The rate is calculated assuming the meson masses of $m_B = 5.28\,\mathrm{GeV}$ and $m_{D^*} = 2.01\,\mathrm{GeV}$, and the lepton as massless. The quantity $\Gamma^\mathrm{p(ext)}$ is the predicted (isospin averaged) absolute rate. The predicted (LQCD) form factors are $h_{X}^{\rm p(LQCD)}$. $C_\mathrm{exp}$ ($C_\mathrm{LQCD}$) is the covariance matrix of the experimental (lattice) data. The BGL coefficients $a_i$, $b_i$, $c_i$, and $d_i$ correspond to the form factors $g$, $f$, $\mathcal{F}_1$, $\mathcal{F}_2$ respectively. The notation $\mathrm{BGL}_{N_a N_b N_c N_d}$ corresponds to the number of free coefficients in the fit, and $c_0$ is always defined through the kinematical constraint. The number of BGL form factor coefficients is determined by a nested hypothesis test~\cite{Bernlochner:2019ldg}, and no unitarity bound is imposed. The impact on the uncertainty from the chosen truncation based on the NHT has not been studied in this report. We consider three scenarios defined by the amount of information used from LQCD: Include all nonzero recoil form factor information, include only the nonzero recoil form factor information on $h_{A_1}(w)$, or include only the zero-recoil point $\mathcal{F}(1) = h_{A_1}(1)$. In the latter case we perform a BGL$_{332}$ fit to the beyond-zero recoil lattice data and extrapolate to zero-coil to obtain $h_{A_1}(w)=0.895\pm 0.007$ which is subsequently used in the fit with data. These fits yield a $\chi^2 / \mathrm{ndf}$ of $72.0/60$, $52.5/44$, and $46.6/34$, respectively. The results of the fits are shown in Fig.~\ref{fig:BToDstarVcb} and the extracted values of $|V_{cb}|$ are given in Tab.~\ref{tab:BToDstarVcb}. Details on the three individual fits are given in Tab.~\ref{tab:BToDstVcb00}, \ref{tab:BToDstVcb01}, and \ref{tab:BToDstVcb02}.

\begin{figure}
    \centering
    \includegraphics[width=\linewidth]{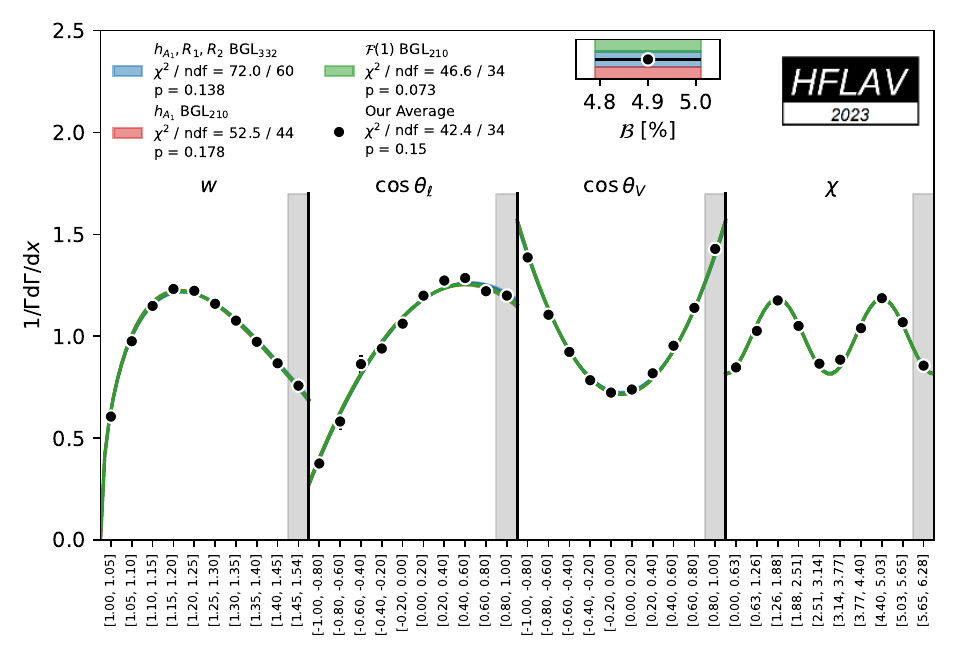}
    \caption{The fitted $\bar B\to D^*\ell^-\bar\nu_\ell$ differential decay rate from our averaged normalized differential distribution and our isospin average of the $\bar B\to D^*\ell^-\bar\nu_\ell$ branching fraction. We find consistent fitted shapes in all three considered scenarios.}
    \label{fig:BToDstarVcb}
\end{figure}

\begin{table}
    \centering
    \caption{Extracted $|V_{cb}|$ values from our averaged normalized differential distribution and our isospin average of the $\bar B\to D^*\ell^-\bar\nu_\ell$ branching fraction. For each lattice QCD input scenario we repeat the nested hypothesis test, which results in the chosen BGL parameterization (first column). We retrieve consistent $|V_{cb}|$ values in each scenario.}
    \label{tab:BToDstarVcb}
\begin{tabular}{ccccc}
        \toprule
        BGL & $|V_{cb}| \times 10^3$ & $|V_{cb}| \times 10^3 \times \mathcal{F}(1)$ & $\mathcal{F}(1)$  &  LQCD Input \\
        \midrule
        332 & $40.0 \pm 0.7$ & $36.0\pm 0.5$ & $0.900 \pm 0.008$& $h_{A_1}(w), R_1(w), R_2(w)$ \\
        210 & $40.2 \pm 0.6$ & $36.0\pm 0.4$ & $0.897 \pm 0.009$& $h_{A_1}(w)$ \\
        210 & $40.2 \pm 0.6$ & $36.0\pm 0.4$ & $0.895 \pm 0.007$& $h_{A_1}(1) = \mathcal{F}(1)$ \\
        \bottomrule
    \end{tabular}
\end{table}

\begin{table}
    \centering
    \caption{Fit result using all nonzero recoil form factors predicted by LQCD.}
    \label{tab:BToDstVcb00}
    \resizebox{\columnwidth}{!}{%
    \begin{tabular}{lrrrrrrrrrr}
\toprule
 & Value & Correlation &  &  &  &  &  &  &  &  \\
\midrule
$a_0 \times 10^2$ & $ 3.08\pm0.07 $ & 1.00 & -0.16 & -0.09 & 0.23 & 0.02 & -0.12 & 0.14 & -0.06 & -0.24 \\
$a_1 \times 10^2$ & $ -9.27\pm2.56 $ & -0.16 & 1.00 & -0.66 & -0.01 & 0.27 & -0.31 & 0.18 & -0.11 & -0.12 \\
$a_2 \times 10^2$ & $ 6.95\pm44.81 $ & -0.09 & -0.66 & 1.00 & 0.06 & -0.36 & 0.60 & -0.18 & 0.14 & 0.01 \\
$b_0 \times 10^2$ & $ 1.22\pm0.01 $ & 0.23 & -0.01 & 0.06 & 1.00 & -0.03 & 0.04 & 0.01 & -0.03 & -0.55 \\
$b_1 \times 10^2$ & $ 1.45\pm0.59 $ & 0.02 & 0.27 & -0.36 & -0.03 & 1.00 & -0.70 & 0.57 & -0.44 & -0.30 \\
$b_2 \times 10^2$ & $ -49.74\pm13.46 $ & -0.12 & -0.31 & 0.60 & 0.04 & -0.70 & 1.00 & -0.37 & 0.32 & 0.07 \\
$c_1 \times 10^2$ & $ -0.11\pm0.13 $ & 0.14 & 0.18 & -0.18 & 0.01 & 0.57 & -0.37 & 1.00 & -0.93 & -0.32 \\
$c_2 \times 10^2$ & $ 2.19\pm2.47 $ & -0.06 & -0.11 & 0.14 & -0.03 & -0.44 & 0.32 & -0.93 & 1.00 & 0.20 \\
$|V_{cb}| \times 10^3$ & $ 39.97\pm0.65 $ & -0.24 & -0.12 & 0.01 & -0.55 & -0.30 & 0.07 & -0.32 & 0.20 & 1.00 \\
\bottomrule
\end{tabular}
    }
\end{table}

\begin{table}
    \centering
    \caption{Fit result using the nonzero recoil form factor $h_{A_1}(w)$ predicted by LQCD.}
    \label{tab:BToDstVcb01}
    \begin{tabular}{lrrrrr}
\toprule
 & Value & Correlation & &   & \\
\midrule
$a_0 \times 10^2$      & $3.80\pm0.43$    & 1.00  & -0.98 & 0.08 & -0.05 \\
$a_1 \times 10^2$      & $-37.81\pm15.40$ & -0.98 & 1.00 & -0.02 & -0.01 \\
$b_0 \times 10^2$      & $1.21\pm0.01$    & 0.08  & -0.02 & 1.00 & -0.64 \\
$|V_{cb}| \times 10^3$ & $40.12\pm0.60$   & -0.05 & -0.01 & -0.64 & 1.00 \\
\bottomrule
\end{tabular}
\end{table}

\begin{table}
    \centering
    \caption{Fit result using the zero recoil form factor $h_{A_1}(1)=\mathcal{F}(1)$ predicted by LQCD.}
    \label{tab:BToDstVcb02}
    \begin{tabular}{lrrrrr}
\toprule
 & Value & Correlation &  &  &  \\
\midrule
$a_0 \times 10^2$ & $ 3.79\pm0.43$  & 1.00 & -0.98  & 0.07 & -0.04 \\
$a_1 \times 10^2$ & $ -37.74\pm15.37$  & -0.98 & 1.00  & -0.02 & -0.02 \\
$b_0 \times 10^2$ & $ 1.21\pm0.01$  & 0.07 & -0.02  & 1.00 & -0.57 \\
$|V_{cb}| \times 10^3$ & $ 40.19\pm0.56$  & -0.04 & -0.02  & -0.57 & 1.00 \\
\bottomrule
\end{tabular}
\end{table}

\mysubsubsection{$\bar B\to D\ell^-\bar\nu_\ell$}
\label{slbdecays_dlnu}

The differential decay rate for massless fermions as a function of $w$
(introduced in the previous section) is given by (see, \eg,~\cite{Neubert:1993mb})
\begin{equation}
  \frac{d\Gamma(\bar B\to D\ell^-\bar\nu_\ell)}{dw} = \frac{G^2_\mathrm{F} m^3_D}{48\pi^3}(m_B+m_D)^2(w^2-1)^{3/2}\eta_\mathrm{EW}^2\mathcal{G}^2(w)|V_{cb}|^2~,
\end{equation}
where $G_\mathrm{F}$ is Fermi's constant, and $m_B$ and $m_D$ are the $B$ and $D$
meson masses. Again, $\eta_\mathrm{EW}$ is the electroweak correction. In contrast to
$\bar B\to D^*\ell^-\bar\nu_\ell$, $\mathcal{G}(w)$ contains a single
form-factor function $f_+(w)$,
\begin{equation}
  \mathcal{G}^2(w) = \frac{4r}{(1+r)^2} f^2_+(w)~,
\end{equation}
where $r=m_D/m_B$.

\mysubsubsubsection{Branching fraction}

Separate one-dimensional averages of the
$\BzbDplnu$ and $B^-\to D^0\ell^-\bar\nu_\ell$ branching fractions are shown
in Tables~\ref{tab:dlnu} and \ref{tab:d0lnu}. We obtain
\begin{eqnarray}
  \cbf(\BzbDplnu) & = & (2.12\pm 0.02\pm 0.06)\%~, \label{eq:br_dlnu} \\
  \cbf(B^-\to D^0\ell^-\bar\nu_\ell) & = & (2.21\pm 0.02\pm 0.06)\%~,
   \label{eq:br_d0lnu}
\end{eqnarray}
where the first uncertainty is statistical and the second one is systematic.
These fits are also shown in Fig.~\ref{fig:brdl}.
\begin{table}[!htb]
\caption{Average of $\BzbDplnu$ branching fraction
  measurements.}
\begin{center}
\begin{tabular}{|l|c|c|}
  \hline
  Experiment
  & $\cbf(\BzbDplnu)$ [\%] (rescaled)
  & $\cbf(\BzbDplnu)$ [\%] (published)\\
  \hline \hline
  ALEPH~\cite{Buskulic:1996yq}
  & $2.16\pm 0.18_{\rm stat}\pm 0.35_{\rm syst}$
  & $2.35\pm 0.20_{\rm stat}\pm 0.44_{\rm syst}$\\
  CLEO~\cite{Bartelt:1998dq}
  & $2.10\pm 0.13_{\rm stat}\pm 0.15_{\rm syst}$
  & $2.20\pm 0.16_{\rm stat}\pm 0.19_{\rm syst}$\\
  \babar~\cite{Aubert:2009ac}
  & $2.15\pm 0.11_{\rm stat}\pm 0.14_{\rm syst}$
  & $2.23\pm 0.11_{\rm stat}\pm 0.11_{\rm syst}$\\
  \belle~\cite{Glattauer:2015teq}
  & $2.33\pm 0.04_{\rm stat}\pm 0.11_{\rm syst}$
  & $2.39\pm 0.04_{\rm stat}\pm 0.11_{\rm syst}$\\
  \belle~II~$De\nu$~\cite{Belle-II:2022ffa}
  & $1.99\pm 0.04_{\rm stat}\pm 0.08_{\rm syst}$
  & $1.99\pm 0.04_{\rm stat}\pm 0.08_{\rm syst}$\\
  \belle~II~$D\mu\nu$~\cite{Belle-II:2022ffa}
  & $2.03\pm 0.04_{\rm stat}\pm 0.08_{\rm syst}$
  & $2.03\pm 0.04_{\rm stat}\pm 0.09_{\rm syst}$\\
  \hline 
  {\bf Average}
  & \mathversion{bold}$2.12\pm 0.02_{\rm stat}\pm 0.06_{\rm syst}$
  & \mathversion{bold}$\chi^2/\dof = 6.09/5$ (CL=$29.8\%$)\\
  \hline 
\end{tabular}
\end{center}
\label{tab:dlnu}
\end{table}
\begin{table}[!htb]
\caption{Average of $B^-\to D^0\ell^-\bar\nu_\ell$ branching fraction
  measurements.}
\begin{center}
%
  \caption{Branching fractions of exclusive semileptonic $B$ decays:
    (a) $\BzbDplnu$ (Table~\ref{tab:dlnu}) and (b) $B^-\to
    D^0\ell^-\bar\nu_\ell$ (Table~\ref{tab:d0lnu}).} \label{fig:brdl}
  \end{center}
\end{figure}

Again, we also calculate an average for $\cbf(\BzbDplnu)$ assuming isospin symmetry: The $B^-$ branching fractions are rescaled by the lifetime ratio $\tau_{0-}=\tau(B^0)/\tau(B^-)$ and a combined 1-dimensional fit over the $\cbf(\BzbDplnu)$ and $\cbf(B^-\to D^0\ell^-\bar\nu_\ell)$ branching fractions is performed. To account for isospin breaking due to the Coulomb effect in $\BzbDplnu$ we inflate the uncertainty in $\tau_{0-}$ by $\tau_{0-}\alpha\pi\approx 0.02123$~\cite{Atwood:1989em}. The result reads
\begin{equation}
    \cbf(\BzbDplnu) = (2.11\pm 0.01\pm 0.05)\%~. \label{eq:br_dlnu_iso}
\end{equation}
The $\chi^2/$dof of the combination is 8.60/10 (CL=57.1\%).

\mysubsubsubsection{Extraction of $\vcb$ based on the BGL form factor}

A model-independent expression for the $\bar B\to D\ell^-\bar\nu_\ell$ form factor is again BGL. A BGL fit allows to include available lattice QCD data at non-zero recoil $w>1$~\cite{MILC:2015uhg,Na:2015kha} to improve the extrapolation to the zero recoil point~$w=1$. A $w$~spectrum of $\bar B\to D\ell^-\bar\nu_\ell$ has been published by BaBar~\cite{Aubert:2009ac} and Belle~\cite{Glattauer:2015teq}. As the BaBar result does not include the full error matrix of the $w$~spectrum, we perform the BGL fit to the Belle result alone.

We construct a $\chi^2$-function similar to Eq.~\ref{eq:BToDstChi2} to incorporate the normalized $w$~spectrum from Ref.~\cite{Glattauer:2015teq}, and include the isospin average of the $\bar B\to D\ell^-\bar\nu_\ell$ total branching fraction. We include the nonzero recoil lattice QCD inputs for the $f_0$ and $f_+$ form factors by Fermilab/MILC~\cite{MILC:2015uhg} and HPQCD~\cite{Na:2015kha, Monahan:2017uby} as synthetic data points and follow the recipe in Ref.~\cite{FlavourLatticeAveragingGroupFLAG:2021npn} for the correlation of the lattice data. Following Ref.~\cite{MILC:2015uhg}, the Blaschke factors in the BGL parameterization are set to unity, which might introduce some model-dependence. We impose the kinematic constraint $f_+(w_\mathrm{max}) = f_0(w_\mathrm{max})$ on the form factors and obtain
\begin{equation}
    |V_{cb}| = (38.9 \pm 0.7)\times 10^{-3}\, .
\end{equation}
The fitted distribution is shown in Fig.~\ref{fig:BToDVcb} and the form factor coefficients and correlations are listed in Tab.~\ref{tab:BToDVcb}.

\begin{figure}
    \centering
    \includegraphics[width=0.75\linewidth]{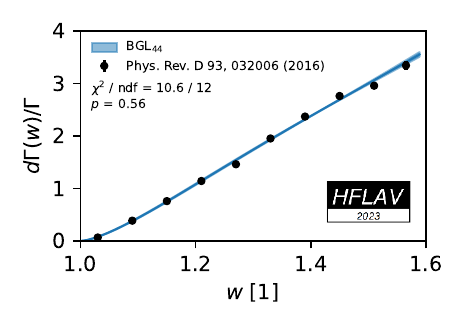}
    \caption{The fitted $\bar B\to D \ell^-\bar\nu_\ell$ differential decay rate and our isospin average of the $\bar B\to D \ell^-\bar\nu_\ell$ branching fraction.}
    \label{fig:BToDVcb}
\end{figure}

\begin{table}
    \centering
    \caption{Fit result for the $\bar B\to D\ell^-\bar\nu_\ell$ decay.}
    \label{tab:BToDVcb}
    \resizebox{\columnwidth}{!}{%
    \begin{tabular}{lrrrrrrrrrr}
\toprule
 & Value & Correlation &  &  &  &  &  &  &  &  \\
\midrule
$a_0^+ \times 10^2$ & $1.265\pm0.009$ & 1.00 & 0.02 & -0.12 & 0.00 & 0.78 & 0.17 & -0.07 & 0.00 & -0.50 \\
$a_1^+ \times 10^2$ & $-9.27\pm0.26$ & 0.02 & 1.00 & -0.69 & 0.04 & 0.08 & 0.35 & 0.31 & -0.02 & -0.41 \\
$a_2^+ \times 10^2$ & $29.1\pm3.4$ & -0.12 & -0.69 & 1.00 & -0.23 & -0.12 & -0.12 & -0.22 & 0.01 & 0.10 \\
$a_3^+ \times 10^2$ & $92\pm15$ & 0.00 & 0.04 & -0.23 & 1.00 & -0.00 & -0.02 & 0.01 & -0.00 & -0.03 \\
$a_0^0 \times 10^2$ & $1.145\pm0.008$ & 0.78 & 0.08 & -0.12 & -0.00 & 1.00 & -0.03 & -0.08 & 0.00 & -0.43 \\
$a_1^0 \times 10^2$ & $-5.96\pm0.27$ & 0.17 & 0.35 & -0.12 & -0.02 & -0.03 & 1.00 & -0.62 & 0.09 & -0.29 \\
$a_2^0 \times 10^2$ & $31\pm13$ & -0.07 & 0.31 & -0.22 & 0.01 & -0.08 & -0.62 & 1.00 & -0.25 & -0.08 \\
$a_3^0 \times 10^2$ & $90\pm90$ & 0.00 & -0.02 & 0.01 & -0.00 & 0.00 & 0.09 & -0.25 & 1.00 & 0.01 \\
$|V_{cb}| \times 10^3$ & $38.9\pm0.7$ & -0.50 & -0.41 & 0.10 & -0.03 & -0.43 & -0.29 & -0.08 & 0.01 & 1.00 \\
\bottomrule
\end{tabular}

    }
\end{table}

\mysubsubsection{$\bar{B}_s^0 \to D_s^{(*)+}\mu^-{\bar\nu}_\mu$}
\label{slbdecay_bs2dslnu}

LHCb has extracted $\vcb$ from semileptonic $\bar{B}_s^0$ decays for the first time~\cite{Aaij:2020hsi}. The measurement uses both $\bar{B}_s^0\rightarrow D_s^{+}\mu^-\bar\nu_{\mu}$ and $\bar{B}_s^0\rightarrow D_s^{*+}\mu^-\bar\nu_{\mu}$ decays using $3$~fb$^{-1}$ collected in 2011 and 2012. The value of $\vcb$ is determined from the observed yields of $\bar{B}_s^0$ decays normalized to those of $\bar{B}^0$ decays after correcting for the relative reconstruction and selection efficiencies, and considering the known relative $\bar{B}_s^0$ and $\bar{B}^0$ fragmentation fractions, $f_s/f_d$, in the LHCb acceptance. 

The normalization channels are $\bar{B}^0\to D^+\mu^-\bar\nu_{\mu}$ and $\bar{B}^0\to D^{*+}\mu^-\bar\nu_{\mu}$ decays. One of the key features of the analysis is that the $D^+$ is  reconstructed with the same decay mode of the $D_s^+$ ($D_{(s)}^+\to [K^+K^-]_{\phi}\pi^+$). With this choice the signal and the reference channels have the same particles in the final state and this minimizes the systematic uncertainties.  

The shape of the form factors are extracted as well, exploiting the kinematic variable $p_{\perp}(D_s)$, which is the component of the $D_s^+$ momentum perpendicular to the $\bar{B}_s^0$ flight direction. This variable is highly correlated with  $q^2$ and also slightly correlated with the helicity angles in the $\bar{B}_s^0\rightarrow D_s^{*+}\mu^-\bar\nu_{\mu}$ decay. The $D_s^{*+}$ is not explicitly reconstructed, but its contribution is disentangled kinematically from the $D_s$.

For the $\bar{B}_s^0\to D_s^+\mu^-\bar\nu_\mu$ decay, $\vcb$ is connected with the measured ratio of signal yields, $N_{\rm{sig}}$, and the normalization channel yields, $N_{\rm{ref}}$, through the relation 
\begin{equation}
\frac{N_{\rm{sig}}}{N_{\rm{ref}}}={\cal K}\tau_s \int{\frac{d{\Gamma}(\bar{B}_s^0\to D_s^+\mu^-\bar\nu_\mu)}{d w}}dw\nonumber \\
\end{equation}
\noindent where $\tau_s$ is the $\bar{B}_s^0$ lifetime, and the constant $\cal{K}$ depends on the external inputs as 
\begin{equation}
{\cal K}=\xi\frac{f_s}{f_d}\frac{{\cal B}(D_s^+\to K^+K^-\pi^+)}{{\cal B}(D^+\to K^+ K^-\pi^+)}  \frac{1}{{\cal B}(\bar{B}^0\to D^+\mu^-\bar\nu_{\mu})} \nonumber \\
\end{equation}
\noindent where $\xi$ is the efficiency ratio between the signal and the normalization. 
In the analogous expression for the $\bar{B}_s^0\to D_s^{*+}\mu^-\bar\nu_\mu$ decay, the integral of the decay width is done on the variables $(w, \cos\theta_\ell,\cos\theta_V,\chi)$, and there is an explicit dependence on the branching fraction of the $D^{*+}\to D^+\pi^0$ decay. The analysis takes advantage of the recent results from lattice on the $\bar{B}_s^0\to D_s^+$ and $\bar{B}_s^0\to D_s^{*+}$ form factor calculations. In particular for the $\bar{B}_s^0\to D_s^{*+}$, only the calculations at zero recoil, $h_{A1}^{B_s}(1)$ from Ref.~\cite{McLean:2019sds} is used.
Recently, HPQCD \cite{Harrison:2023dzh} provided calculations of form factors across for $\bar{B}_s^0\to D_s^{*+}\mu^-\bar\nu_{\mu}$  decays in the full $w$-range, but these results have not been utilized and would require a re-analysis of the LHCb data, to be properly exploited. For the $\bar{B}_s^0\to D_s^+\mu^-\bar\nu_{\mu}$ decay, the calculation of the $\bar{B}_s^0\to D_s^+$ form factors performed in the full $w$-range \cite{McLean:2019qcx} are used.

LHCb uses the BGL expansion both for $\bar{B}_s^0\rightarrow D_s^{+}\mu^-\bar\nu_{\mu}$ and $\bar{B}_s^0\rightarrow D_s^{*+}\mu^-\bar\nu_{\mu}$ decays. The form factors parameters are affected by large statistical uncertainty, but are consistent with the results from the $B$ decays.
The result for $\vcb$, updated with the most recent determination of $f_s/f_d$ from Ref.\cite{LHCb:2021qbv}, ${\cal B}(D_s^+\to K^+K^-\pi^+)$ and ${\cal B}(D^+\to K^+K^-\pi^+)$ from \cite{PDG_2022}, 
and ${\cal B}(\bar{B}^0\to D^{(*)+}\mu^-\bar\nu_{\mu})$ from Sections \ref{slbdecays_dlnu} and\ref{slbdecays_dstarlnu}, is
\begin{eqnarray}
|V_{cb}|_{\rm{BGL}} &=& (41.2\pm0.8\pm 0.9\pm 0.8)\times 10^{-3},\nonumber
\end{eqnarray}
\noindent where the first uncertainty is statistical, the second systematic and the third due to the limited knowledge of the external inputs, in particular the constant $f_s/f_d\times {\cal B}(D_s^+\to K^+K^-\pi^+)$. This result is in agreement with the exclusive determinations of $\vcb$ using the $\bar{B}^0$ and $B^-$ decays.

\mysubsubsection{$\bar{B} \to D^{(*)}\pi \ell^-\bar{\nu}_{\ell}$}
\label{slbdecays_dpilnu}

The average inclusive branching fractions for $\bar{B} \to D^{(*)}\pi\ell^-\bar{\nu}_{\ell}$ 
decays, where no constraint is applied to the mass of the $D^{(*)}\pi$ system, are determined by the
combination of the results provided in Table~\ref{tab:dpilnu} for 
$\bar{B}^0 \to D^0 \pi^+ \ell^-\bar{\nu}_{\ell}$, $\bar{B}^0 \to D^{*0} \pi^+\ell^-\bar{\nu}_{\ell}$,  $B^- \to D^+ \pi^- \ell^-\bar{\nu}_{\ell}$, and $B^- \to D^{*+} \pi^- \ell^-\bar{\nu}_{\ell}$ decays. For the $\bar{B}^0 \to D^0 \pi^+ \ell^-\bar{\nu}_{\ell}$ decays a veto to reject the $D^{*+}\to D^0 \pi^+$ decays is applied.
The measurements included in the average 
are scaled to a consistent set of input
parameters and their uncertainties. %
For both the \babar\ and Belle results, the $B$ semileptonic signal yields are
 extracted from a fit to the missing mass squared distribution for a sample of fully
 reconstructed \BB\ events.  
Figure~\ref{fig:brdpil} shows the measurements and the resulting average for the 
four decay modes.

\begin{table}[!htb]
\caption{Averages of the $B \to D^{(*)} \pi^- \ell^-\bar{\nu}_{\ell}$  branching fractions and individual results.}
\begin{center}

  \caption{Average branching fraction  of exclusive semileptonic $B$ decays
(a) $\bar{B}^0 \to D^0 \pi^+ \ell^-\bar{\nu}_{\ell}$, (b) $\bar{B}^0 \to D^{*0} \pi^+
\ell^-\bar{\nu}_{\ell}$, (c) $B^- \to D^+ \pi^-
\ell^-\bar{\nu}_{\ell}$, and (d) $B^- \to D^{*+} \pi^- \ell^-\bar{\nu}_{\ell}$.
The corresponding individual
  results are also shown.}
  \label{fig:brdpil}
 \end{center}
\end{figure}

\mysubsubsection{$\bar{B} \to D^{**} \ell^-\bar{\nu}_{\ell}$}
\label{slbdecays_dsslnu}

In this section we report results on $\bar{B} \to D^{**} \ell^-\bar{\nu}_{\ell}$ decays, where $D^{**}$ here denotes the lightest excited charm mesons above the $D$ and $D^{*}$ states. 
According to Heavy Quark Symmetry (HQS)~\cite{Isgur:1991wq}, the $D^{*}$ mesons with a charm and anti-quark with relative angular momentum $L=1$, form one doublet of states with angular momentum $j \equiv s_q + L= 3/2$  $\left[D_1(2420), D_2^*(2460)\right]$ and another doublet 
with $j=1/2$ $\left[D^*_0(2300), D_1'(2430)\right]$, where $s_q$ is the light quark spin \footnote{At present only these $L=1$ orbital excited states have been observed in the semileptonic $B$ meson decays, but in principle also radial $2S$ excitation and states with $L=2,3$,  observed in fully hadronic B decays, could contribute to semileptonic decays.}.
Parity and angular momentum conservation constrain the decays allowed for each state. The $D_1$ and $D_2^*$ 
states decay predominantly via D-wave to $D^*\pi$ and $D^{(*)}\pi$, respectively, and have small decay widths, 
while the $D_0^*$ and $D_1'$  states decay via S-wave to $D\pi$ and $D^*\pi$ and are very broad.
For the narrow states, the averages are determined by the
combination of the results provided in Table~\ref{tab:dss1lnu} and \ref{tab:dss2lnu} for 
$\cbf(B^- \to D_1^0\ell^-\bar{\nu}_{\ell})
\times \cbf(D_1^0 \to D^{*+}\pi^-)$ and $\cbf(B^- \to D_2^0\ell^-\bar{\nu}_{\ell})
\times \cbf(D_2^0 \to D^{*+}\pi^-)$. 
For the broad states, the averages are determined by the
combination of the results provided in Table~\ref{tab:dss1plnu} and \ref{tab:dss0lnu} for 
$\cbf(B^- \to D_1'^0\ell^-\bar{\nu}_{\ell})
\times \cbf(D_1'^0 \to D^{*+}\pi^-)$ and $\cbf(B^- \to D_0^{*0}\ell^-\bar{\nu}_{\ell})
\times \cbf(D_0^{*0} \to D^{+}\pi^-)$. 
The measurements are scaled to a consistent set of input
parameters and their uncertainties. %
The results are reported for $B^-$, and when measurements for both $B^0$ and $B^-$ are available, the combination assumes the isospin symmetry. 

For both the B-factory and the LEP and Tevatron results, the $B$ semileptonic 
signal yields are extracted from a fit to the invariant mass distribution of the $D^{(*)+}\pi^-$ system.
 The LEP and Tevatron measurements 
 are for the inclusive decays $\bar{B} \to D^{**}(D^*\pi^-)X \ell^- \bar{\nu}_{\ell}$. In the average with the results from the B-Factories, we use these measurements assuming that no particles are left in the $X$ system.
 The \babar tagged analysis of $\bar{B} \to D_2^* \ell^- \bar{\nu}_{\ell}$ was performed selecting $D_2^*\to D\pi$ decays.
 The \babar result reported in Table~\ref{tab:dss2lnu} is translated in a branching fraction for the $D_2^*\to D^*\pi$ decay mode assuming
 ${\cal B}(D_2^*\to D\pi)/{\cal B}(D_2^*\to D^*\pi)=1.52\pm 0.14$~\cite{PDG_2020}. 
The Belle tagged analysis in Ref.\cite{Belle:2022yzd}, reports the branching fraction of $\bar{B} \to D_2^* \ell^- \bar{\nu}_{\ell}$ separated for $D_2^*\to D\pi$  and $D_2^*\to D^*\pi$. The Belle results reported in Table~\ref{tab:dss2lnu} are obtained scaling all the decays to the $D_{2}^{*}\to D^{*}\pi$ decay mode assuming the ratio of branching fractions reported above. 
Figure~\ref{fig:brdssl} and ~\ref{fig:brdssl2} show the measurements and the resulting averages.

\begin{table}[!htb]
\caption{Published and rescaled individual measurements and their averages for the branching fraction $\cbf(B^- \to D_1^0\ell^-\bar{\nu}_{\ell})\times \cbf(D_1^0 \to D^{*+}\pi^-)$. 
}
\begin{center}
\resizebox{0.99\textwidth}{!}{
\begin{tabular}{l|c|c}\hline
Experiment                                 &$\cbf(B^- \to D_1^0(D^{*+}\pi^-)\ell^-\bar{\nu}_{\ell})
 [\%]$  &$\cbf(B^- \to D_1^0(D^{*+}\pi^-)\ell^-\bar{\nu}_{\ell})
 [\%]$  \\
                                                & (rescaled) & (published) \\

\hline\hline 
ALEPH ~\cite{Aleph:Dss}        &$0.436 \pm0.098_{\rm stat} \pm0.067_{\rm syst}$ 
 &$0.47 \pm0.10_{\rm stat} \pm0.07_{\rm syst}$ \\
OPAL  ~\cite{opal:Dss}         &$0.551 \pm0.210_{\rm stat} \pm0.100_{\rm syst}$  
&$0.70 \pm0.21_{\rm stat} \pm0.10_{\rm syst}$ \\
CLEO  ~\cite{cleo:Dss}         &$0.347 \pm0.085_{\rm stat} \pm0.056_{\rm syst}$ 
 &$0.373 \pm0.085_{\rm stat} \pm0.057_{\rm syst}$ \\
D0  ~\cite{D0:Dss}         &$0.213 \pm0.018_{\rm stat} \pm0.035_{\rm syst}$  
&$0.219 \pm0.018_{\rm stat} \pm0.035_{\rm syst}$ \\
\belle Tagged $B^-$ ~\cite{Belle:2022yzd}           &$0.249 \pm0.023_{\rm stat} \pm0.015_{\rm syst}$  
&$0.249 \pm0.023_{\rm stat} \pm0.015_{\rm syst}$ \\
\belle Tagged $B^0$ ~\cite{Belle:2022yzd}           &$0.329 \pm0.054_{\rm stat} \pm0.031_{\rm syst}$  
&$0.329 \pm0.054_{\rm stat} \pm0.031_{\rm syst}$ \\ 
\babar Tagged ~\cite{BaBar:2008dar}           &$0.273 \pm0.030_{\rm stat} \pm0.029_{\rm syst}$
&$0.29 \pm0.03_{\rm stat} \pm0.03_{\rm syst}$ \\
\babar Untagged $B^-$ ~\cite{BaBar:2008cfy}           &$0.290 \pm0.017_{\rm stat} \pm0.017_{\rm syst}$
&$0.30 \pm0.02_{\rm stat} \pm0.02_{\rm syst}$ \\
\babar Untagged $B^0$ ~\cite{BaBar:2008cfy}           &$0.279 \pm0.026_{\rm stat} \pm0.023_{\rm syst}$
&$0.30 \pm0.02_{\rm stat} \pm0.02_{\rm syst}$ \\
\hline
{\bf Average}                              &\mathversion{bold}$0.270 \pm0.015$ 
    &\mathversion{bold}$\chi^2/\dof = 9.0/8$ (CL=$34.0\%$)  \\
\hline 
\end{tabular}
}
\end{center}
\label{tab:dss1lnu}
\end{table}
\begin{table}[!htb]
\caption{Published and rescaled individual measurements and their averages for 
$\cbf(B^- \to D_2^0\ell^-\bar{\nu}_{\ell})\times \cbf(D_2^0 \to D^{*+}\pi^-)$. 
}
\begin{center}
\resizebox{0.99\textwidth}{!}{
\begin{tabular}{l|c|c}\hline
Experiment                                 &$\cbf(B^- \to D_2^0(D^{*+}\pi^-)\ell^-\bar{\nu}_{\ell})
 [\%]$  &$\cbf(B^- \to D_2^0(D^{*+}\pi^-)\ell^-\bar{\nu}_{\ell})
 [\%]$  \\
                                                & (rescaled) & (published) \\
\hline\hline 
CLEO  ~\cite{cleo:Dss}         &$0.054 \pm0.066_{\rm stat} \pm0.011_{\rm syst}$ 
 &$0.059 \pm0.066_{\rm stat} \pm0.011_{\rm syst}$ \\
D0  ~\cite{D0:Dss}         &$0.085 \pm0.018_{\rm stat} \pm0.020_{\rm syst}$  
&$0.088 \pm0.018_{\rm stat} \pm0.020_{\rm syst}$ \\

\belle Tagged $B^-$ ($D_2\to D^*\pi$)  ~\cite{Belle:2022yzd}  &$0.136 \pm0.026_{\rm stat} \pm0.009_{\rm syst}$  
&$0.36 \pm0.026_{\rm stat} \pm0.009_{\rm syst}$ \\

\belle Tagged $B^0$ ($D_2\to D^*\pi$)  ~\cite{Belle:2022yzd}           &$0.054 \pm0.043_{\rm stat} \pm0.011_{\rm syst}$  
&$0.054 \pm0.043_{\rm stat} \pm0.011_{\rm syst}$ \\

\belle Tagged $B^-$ ($D_2\to D\pi$)  ~\cite{Belle:2022yzd}           &$0.107 \pm0.007_{\rm stat} \pm0.011_{\rm syst}$  
&$0.107 \pm0.007_{\rm stat} \pm0.011_{\rm syst}$ \\

\belle Tagged $B^0$ ($D_2\to D^*\pi$)  ~\cite{Belle:2022yzd}           &$0.109 \pm0.011_{\rm stat} \pm0.011_{\rm syst}$  
&$0.109 \pm0.011_{\rm stat} \pm0.011_{\rm syst}$ \\

\babar Tagged ~\cite{BaBar:2008dar}           &$0.075 \pm0.013_{\rm stat} \pm0.009_{\rm syst}$
&$0.078 \pm0.013_{\rm stat} \pm0.010_{\rm syst}$ \\

\babar Untagged $B^-$ ~\cite{BaBar:2008cfy}           &$0.088 \pm0.009_{\rm stat} \pm0.007_{\rm syst}$
&$0.087 \pm0.013_{\rm stat} \pm0.007_{\rm syst}$ \\

\babar Untagged $B^0$ ~\cite{BaBar:2008cfy}           &$0.064 \pm0.010_{\rm stat} \pm0.004_{\rm syst}$
&$0.087 \pm0.013_{\rm stat} \pm0.007_{\rm syst}$ \\
\hline
{\bf Average}                              &\mathversion{bold}$0.085 \pm 0.006$ 
    &\mathversion{bold}$\chi^2/\dof = 12.9/8$ (CL=$11.5\%$)  \\
\hline 
\end{tabular}
}
\end{center}
\label{tab:dss2lnu}
\end{table}
\begin{table}[!htb]
\caption{
Published and rescaled individual measurements and their averages for 
$\cbf(B^- \to D_1^{'0}\ell^-\bar{\nu}_{\ell})\times \cbf(D_1^{'0} \to D^{*+}\pi^-)$. 
}
\begin{center}
\begin{tabular}{l|c|c}\hline
Experiment                                 &$\cbf(B^- \to D_1^{'0}(D^{*+}\pi^-)\ell^-\bar{\nu}_{\ell})
 [\%]$  &$\cbf(B^- \to D_1^{'0}(D^{*+}\pi^-)\ell^-\bar{\nu}_{\ell})
 [\%]$  \\
                                                & (rescaled) & (published) \\
\hline\hline 
DELPHI ~\cite{Abdallah:2005cx}        &$0.694 \pm0.170_{\rm stat} \pm0.176_{\rm syst}$ 
 &$0.83 \pm0.17_{\rm stat} \pm0.18_{\rm syst}$ \\
\belle  Tagged $B^+$~\cite{Belle:2022yzd}           &$0.138 \pm0.036_{\rm stat} \pm0.009_{\rm syst}$  
&$0.138 \pm0.036_{\rm stat} \pm0.009_{\rm syst}$ \\
\belle  Tagged $B^0$ ~\cite{Belle:2022yzd}           &$0.221 \pm0.073_{\rm stat} \pm0.027_{\rm syst}$  
& $0.221 \pm0.073_{\rm stat} \pm0.027_{\rm syst}$\\
\babar Tagged ~\cite{BaBar:2008dar}           &$0.254 \pm0.040_{\rm stat} \pm0.044_{\rm syst}$
&$0.27 \pm0.04_{\rm stat} \pm0.05_{\rm syst}$ \\
\hline
{\bf Average}                              &\mathversion{bold}$0.185 \pm 0.029$ 
    &\mathversion{bold}$\chi^2/\dof = 7.5/3$ (CL=$5\%$)  \\
\hline 
\end{tabular}
\end{center}
\label{tab:dss1plnu}
\end{table}
\begin{table}[!htb]
\caption{Published and rescaled individual measurements and their averages for  
$\cbf(B^- \to D_0^{*0}\ell^-\bar{\nu}_{\ell})\times \cbf(D_0^{*0} \to D^{+}\pi^-)$. }
\begin{center}
%
  \caption{Rescaled individual measurements and their averages for (a) 
  $\cbf(B^- \to D_1^0\ell^-\bar{\nu}_{\ell})
\times \cbf(D_1^0 \to D^{*+}\pi^-)$ and (b) $\cbf(B^- \to D_2^0\ell^-\bar{\nu}_{\ell})
\times \cbf(D_2^0 \to D^{*+}\pi^-)$.}
  \label{fig:brdssl}
 \end{center}
\end{figure}

\begin{figure}[!ht]
 \begin{center}
  \unitlength1.0cm %
  %
  \caption{Rescaled individual measurements and their averages for (a) 
  $\cbf(B^- \to D_1'^0\ell^-\bar{\nu}_{\ell})
\times \cbf(D_1'^0 \to D^{*+}\pi^-)$ and (b) $\cbf(B^- \to D_0^{*0}\ell^-\bar{\nu}_{\ell})
\times \cbf(D_0^{*0} \to D^{+}\pi^-)$.}
  \label{fig:brdssl2}
 \end{center}
\end{figure}

It is worth noticing that, while the results for the narrow resonances are consistent between the various experiments, the measurements for $B^- \to D_0^0\ell^-\bar{\nu}_{\ell}$ obtained by \babar \cite{BaBar:2008dar} and Belle \cite{Belle:2022yzd} with the hadronic tagging technique, are not compatible. Belle did not observe $\bar{B}^0  \to D_0^{*+}\ell^-\bar{\nu}_{\ell}$  and put an upper limit of $\cbf (\bar{B}^0  \to D_0^{*+}\ell^-\bar{\nu}_{\ell})<0.044\%$ at $90\%$ of C.L., while the corresponding mode for the $B^-$ is observed with $3.9\sigma$ of significance. The model used to describe the $D_0$ state in the existing measurements, has been questioned in recent publications, as seen in \cite{Albaladejo:2016lbb} and \cite{Du:2017zvv, Du:2020pui}.   
The results for $D_1'$ are instead consistent between \babar and Belle. 

\subsection{Inclusive CKM-favored decays}
\label{slbdecays_b2cincl}

\subsubsection{Global analysis of $\bar B\to X_c\ell^-\bar\nu_\ell$}

The semileptonic decay width $\Gamma(\bar B\to X_c\ell^-\bar\nu_\ell)$ is calculated in the framework of the operator production expansion (OPE)~\cite{Shifman:1986mx,Chay:1990da,Bigi:1992su}. The result is a double-expansion in $\Lambda_{\rm QCD}/m_b$ and $\alpha_s$, which depends on a number of non-perturbative
parameters. These parameters describe the dynamics of the $b$-quark inside the $B$~hadron and can be measured using spectral moments in inclusive $\bar B\to X_c\ell^-\bar\nu_\ell$ decays.

Recent theoretical work is done mostly in the kinetic scheme~\cite{Bigi:1996si}. Here, the non-perturbative parameters are the quark masses $m_b$ and $m_c$, $\mu^2_\pi$ and $\mu^2_G$ at $O(1/m^2_b)$, and $\rho^3_D$ and $\rho^3_{LS}$ at $O(1/m^3_b)$. To avoid proliferation of these parameters at $O(1/m^4_b)$, reparametrization invariance is proposed to link different operators in the Heavy-Quark expansion thus reducing the number of independent operators~\cite{Fael:2018vsp}. However, this limits the experimental input to certain spectral moments, such as the moments of leptonic invariant mass ($q^2$). An alternative expression is obtained in the $1S$~scheme~\cite{Bauer:2004ve}.

Measurements of $\langle E^n_\ell\rangle$, $\langle M^n_X\rangle$ and $\langle q^{2n}\rangle$ have been obtained by Belle~II~\cite{Belle-II:2022evt}, BaBar~\cite{Aubert:2009qda,Aubert:2004td}, Belle~\cite{Urquijo:2006wd,Schwanda:2006nf,Belle:2021idw}, CDF~\cite{Acosta:2005qh}, CLEO~\cite{Csorna:2004kp} and DELPHI~\cite{Abdallah:2005cx}. The moments of the $E^n_\ell$ and $M^n_X$~distributions are measured for different lower threshold values $E_\mathrm{cut}$ of $E_\ell$. For the moments of $q^{2n}$, the threshold is placed on $q^2$. As these measurements with successive $E_\mathrm{cut}$ values are highly correlated, global analyses typically discard a number of the experimental moments to avoid numerical issues in the fit. The region $E_\mathrm{cut}>1.5$~GeV is also avoided as the quark hadron duality assumption might not be applicable anymore.

\subsubsection*{Analysis using {\boldmath$E_\ell$ and $M_X$}~moments}

This analysis determines $|V_{cb}|$, the inclusive $\bar B\to X_c\ell^-\bar\nu_\ell$ branching fraction and the non-perturbative parameters of the kinetic scheme from a fit to the $\langle E^n_\ell\rangle$, $n=0,1,2,3$, and $\langle M^n_X\rangle$, $n=2,4,6$, moments measured by BaBar~\cite{Aubert:2009qda,Aubert:2004td}, Belle~\cite{Urquijo:2006wd,Schwanda:2006nf}, CDF~\cite{Acosta:2005qh}, CLEO~\cite{Csorna:2004kp} and DELPHI~\cite{Abdallah:2005cx} using the theoretical expressions at $O(1/m^3_b)$. Here, the zero order moment of the $E_\ell$~spectrum refers to the partial branching fractions. This global fit has been repeated several times~\cite{Gambino:2013rza,Alberti:2014yda,Bordone:2021oof} with Ref.~\cite{Bordone:2021oof} being the last iteration that will be reviewed here. As this analysis is known to constrain only a linear combination of $m_b$ and $m_c$, the FLAG 2019 averages, $\bar m_c(2~\mathrm{GeV})=1.093(8)$~GeV and $m_\mathrm{kin}(1~\mathrm{GeV})=4.565(19)$~GeV~\cite{flag:2019}, are used as constraints in the fit. Constraints are also placed on $\mu^2_G$ and $\rho^3_{LS}$. Theory uncertainties and correlations included in the fit have a significant numerical impact and are estimated using the procedure described in Ref.~\cite{Gambino:2013rza} with some updates: The theory uncertainty is modeled through variations of the non-perturbative parameters around fixed values ($\pm 7\%$ for $\mu^2_\pi$ and $\mu^2_G$, $\pm 20\%$ for $\rho^3_D$ and $\rho^3_{LS}$). An irreducible uncertainty of 4~MeV is applied to the quark masses $m_c$ and $m_b$, and $\alpha_s(m_b)$ is varied by $\pm 0.018$. The theoretical correlations for different central moments are considered to be zero while correlations of predictions of the same moment at different $E_\mathrm{cut}$ values are modeled by the proximity of the cuts.

Ref.~\cite{Bordone:2021oof} obtains
\begin{equation}
  |V_{cb}|=(42.16\pm 0.30(th)\pm 0.32(exp)\pm 0.25(\Gamma))\times 10^{-3}~,
\end{equation}
where the first uncertainty originates from the variations of the theory parameters, the second from the experimental input and the third from the expression of the semileptonic width. The inclusive semileptonic branching fraction is found to be
\begin{equation}
  \mathcal{B}(\bar B\to X_c\ell^-\bar\nu_\ell)=(10.66\pm 0.15)\%~.
\end{equation}
The $\chi^2$ of the fit divided by the number of degrees of freedom is 0.47.

\subsubsection*{Analysis using {\boldmath$q^2$}~moments}

Following the proposal in Ref.~\cite{Fael:2018vsp}, a first analysis using terms up to $O(1/m^4_b)$~\cite{Bernlochner:2022ucr} has been performed using the $\langle q^{2n}\rangle$~moments measured by Belle~\cite{Belle:2021idw} and Belle~II~\cite{Belle-II:2022evt}. Moments $\langle q^{2n}\rangle$ with $n=1,2,3,4$ and with different $q_\mathrm{cut}$~values for each moment are used. Similarly to Ref.~\cite{Bordone:2021oof}, the bottom and the charm quark masses are constrained as well as $\mu^2_G$ and $\rho^3_{LS}$. In addition, $\mu^2_\pi$ is constrained to $(0.43\pm 0.24)$~GeV$^2$ as this parameter enters the first order $q^2$~moments only quadratically. Theory uncertainties are again estimated by variation of the non-perturbative parameters and included in the fit.

In order to extract $|V_{cb}|$, information on the total semileptonic branching ratio is needed and the authors of Ref.~\cite{Bernlochner:2022ucr} perform a private average of available measurements,
\begin{equation}
   \mathcal{B}(\bar B\to X_c\ell^-\bar\nu_\ell)=(10.48\pm 0.13)\%~.
\end{equation}
With this normalization, the resulting $|V_{cb}|$ at N3LO precision of the decay rate is
\begin{equation}
  |V_{cb}|=(41.69\pm 0.63)\times 10^{-3}~.
\end{equation}
The $\chi^2/\mathrm{ndf.}=0.15$ indicates a good fit to the data. The numerical difference to the result of Ref.~\cite{Bordone:2021oof} stems mostly from the difference in the semileptonic branching fraction. If the $q^2$ analysis would be rescaled to $\mathcal{B}(\bar B\to X_c\ell^-\bar\nu_\ell)=(10.63\pm 0.19)\%$, the correspond value of $|V_{cb}|$ is $(41.99\pm 0.65)\times 10^{-3}$.

Despite this agreement, we note a tension for the non-perturbative parameter $\rho^3_D$ between Refs.~\cite{Bordone:2021oof} and \cite{Bernlochner:2022ucr}: The former analysis obtains $\rho^3_D=(0.185\pm 0.031)$~GeV$^3$ while the latter obtains $\rho^3_D=(0.03\pm 0.02)$~GeV$^3$ if only terms up to $O(1/m^3_b)$ are included. However, in the perturbative calculation both analyses do not contain the same orders in $\alpha_s$, which could affect the extraction of this parameter~\cite{Fael:2020iea,Fael:2020tow,Fael:2024gyw}.

\subsubsection*{Analysis using {\boldmath $E_\ell$, $M_X$ and $q^2$} moments}

Very recently, an updated calculation of the $\langle q^{2n}\rangle$~moments including $O(\alpha^2_s\beta_0)$ corrections has been published~\cite{Finauri:2023kte}. In the same publication this result is used to extend the analysis in Ref.~\cite{Bordone:2021oof} and include the $\langle q^{2n}\rangle$~moments measured by Belle~\cite{Belle:2021idw} and Belle~II~\cite{Belle-II:2022evt}. The results are
\begin{equation}
  |V_{cb}|=(41.97\pm 0.27(\mathrm{exp})\pm 0.31(\mathrm{th})\pm 0.25(\Gamma))\times 10^{-3}
\end{equation}
and
\begin{equation}
  \mathcal{B}(\bar B\to X_c\ell^-\bar\nu_\ell)=(10.63\pm 0.15)\%~,
\end{equation}
with $\chi^2/\mathrm{ndf.}=0.546$. The first uncertainty on $|V_{cb}|$ is experimental, the second originates from the variations of the theory parameters and the third from the expression of the semileptonic width.

We note that here the inclusion of the $\langle q^{2n}\rangle$~moments has no significant impact on the values of the non-perturbative parameters. In particular, $\rho^3_D$ changes from $(0.185\pm 0.031)$~GeV$^3$ to $0.167\pm 0.018$~GeV$^3$ when including the Belle and Belle II $\langle q^{2n}\rangle$~moments.

\subsubsection*{Average}

Due to the unknown correlations between the analyses in Refs.~\cite{Bordone:2021oof}, \cite{Bernlochner:2022ucr} and \cite{Finauri:2023kte}, we refrain from averaging the results for $|V_{cb}|$ and $\mathcal{B}(\bar B\to X_c\ell^-\bar\nu_\ell)$. Instead we use the result of Ref.~\cite{Finauri:2023kte} as our central value for $|V_{cb}|$ from inclusive decays $\bar B\to X_c\ell^-\bar\nu_\ell$,
\begin{equation}
  |V_{cb}|=(41.97\pm 0.48)\times 10^{-3}~.
\end{equation}
Note that this value is also very close to the arithmetic average of Refs.~\cite{Bordone:2021oof} and \cite{Bernlochner:2022ucr} assuming full correlation of the theoretical uncertainties. Consistently, we also take Ref.~\cite{Finauri:2023kte} for the central value of the charmed semileptonic branching ratio,
\begin{equation}
  \cbf(\bar B\to X_c\ell^-\bar\nu_\ell)=(10.63\pm 0.15)\%~.
\end{equation}
We calculate the charmless semileptonic branching ratio using Eq.~58 from Ref.~\cite{ref:BLNP} and the average BLNP value for $|V_{ub}|$ from Tab.~\ref{tab:bulnu} and obtain $\cbf(\bar B\to
X_u\ell^-\bar\nu_\ell)=(1.92\pm 0.21)\times 10^{-3}$. The total semileptonic branching fraction is thus
\begin{equation}
  \cbf(\bar B\to X\ell^-\bar\nu_\ell)=(10.82\pm 0.15)\%~.
\end{equation}

\subsection{Exclusive CKM-suppressed decays}
\label{slbdecays_b2uexcl}
In this section, we give results on exclusive charmless semileptonic branching fractions
and the determination of $\Vub$ based on \Btopilnu\ decays.
The measurements are based on two different event selections: tagged
events, in which the second $B$ meson in the event is fully (or partially)
reconstructed, and untagged events, for which the momentum
of the undetected neutrino is inferred from measurements of the total 
momentum sum of the detected particles and the knowledge of the initial state.

The LHCb experiment has reported a direct measurement of 
$|V_{ub}|/|V_{cb}|$ \cite{Aaij:2015bfa}, reconstructing the 
$\Lb\to p\mu\nu$ decays and normalizing the branching fraction to the $\Lb\to\Lc(\to pK\pi)\mu\nu$ decays. Recently LHCb reported also a measurement of $|V_{ub}|/|V_{cb}|$ \cite{aaij:2020nvo} using $B^0_s \to K\mu\nu$ decays normalized to $B^0_s\to D_s\mu\nu$ in two separate bins of $q^2$. 
We show a combination of $\Vub$ and $\Vcb$ using the LHCb constraints on $|V_{ub}|/|V_{cb}|$, 
the exclusive determination of $\Vub$ from \Btopilnu, and $\Vcb$ from $B\to D^*\ell\nu$, $B\to D\ell\nu$ and $B_s\to D_s^{(*)}\mu\nu$.

We also present branching fraction averages for 
$\Bz\to\rho\ell^+\nu$, $\Bp\to\omega\ell^+\nu$, $\Bp\to\eta\ell^+\nu$ and $\Bp\to\etapr\ell^+\nu$. Using the available measurements of the partial branching fractions of $\Bz\to\rho\ell^+\nu$, $\Bp\to\omega\ell^+\nu$ decays, we also present for the first time the combined $q^2$ spectrum for these two decays. 

\mysubsubsection{\Btopilnu\ branching fraction and $q^2$ spectrum}

We use the four most precise measurements of the differential \Btopilnu\ decay rate as a function of the four-momentum transfer squared, $q^2$, from \babar and Belle~\cite{Ha:2010rf,Sibidanov:2013rkk,delAmoSanchez:2010af,Lees:2012vv} to obtain an average $q^2$ spectrum and an average for the total branching fraction. The measurements are presented in Fig.~\ref{fig:B_to_pi_average}. From the two untagged \babar\ analyses~\cite{delAmoSanchez:2010af,Lees:2012vv}, the combined results for $B^0 \to \pi^- \ell^+ \nu$ and $B^+ \to \pi^0 \ell^+ \nu$ decays based on isospin symmetry are used. The hadronic-tag analysis by Belle~\cite{Sibidanov:2013rkk} provides results for $B^0 \to \pi^- \ell^+ \nu$ and $B^+ \to \pi^0 \ell^+ \nu$ separately, but not for the combination of both channels. In the untagged analysis by Belle~\cite{Ha:2010rf}, only $B^0 \to \pi^- \ell^+ \nu$ decays were measured. The experimental measurements use different binnings in $q^2$, but have matching bin edges, which allows them to be combined. 

To arrive at an average $q^2$ spectrum, a binned maximum-likelihood fit to determine the average partial branching fraction in each $q^2$ interval is performed, differentiating between common and individual uncertainties and correlations for the various measurements. Shared sources of systematic uncertainty of all measurements are included in the likelihood as nuisance parameters constrained assuming Gaussian distributions. The most important shared sources of uncertainty are due to continuum subtraction, the number of $B$-meson pairs (only correlated among measurement by the same experiment), tracking efficiency (only correlated among measurements by the same experiment), uncertainties from modelling the $b \to u \, \ell \, \bar\nu_\ell$ contamination, modelling of final state radiation, and contamination from $b \to c \, \ell \bar \nu_\ell$ decays. We avoid the d'Agostini bias, described in Ref.~\cite{DAgostini:1993arp}, by scaling the estimated averages with the nuisance parameters. The initial combination results in a poor $p$-value of 0.3\%, with the measurement of Ref.~\cite{delAmoSanchez:2010af} at high $q^2$ being in tension with the other determinations. This kinematic region is very sensitive to the modelling of other $b \to u \ell \bar \nu_\ell$ processes, and Ref.~\cite{delAmoSanchez:2010af} makes several assumptions, that no longer reflect the current knowledge.  To address this, we introduce an additional scaling parameter, as suggested by Ref.~\cite{Cowan:2018lhq}, which we allow to enlarge the uncertainties of the offending measurement. The scaling parameter is constrained using a log-normal function with a width parameter of 0.35. 

The resulting average is shown Figure~\ref{fig:B_to_pi_average} (left, black data points), and the resulting $p$-value is $40\%$.

\begin{figure} 
 \centering
 \includegraphics[width=\textwidth,page=1]{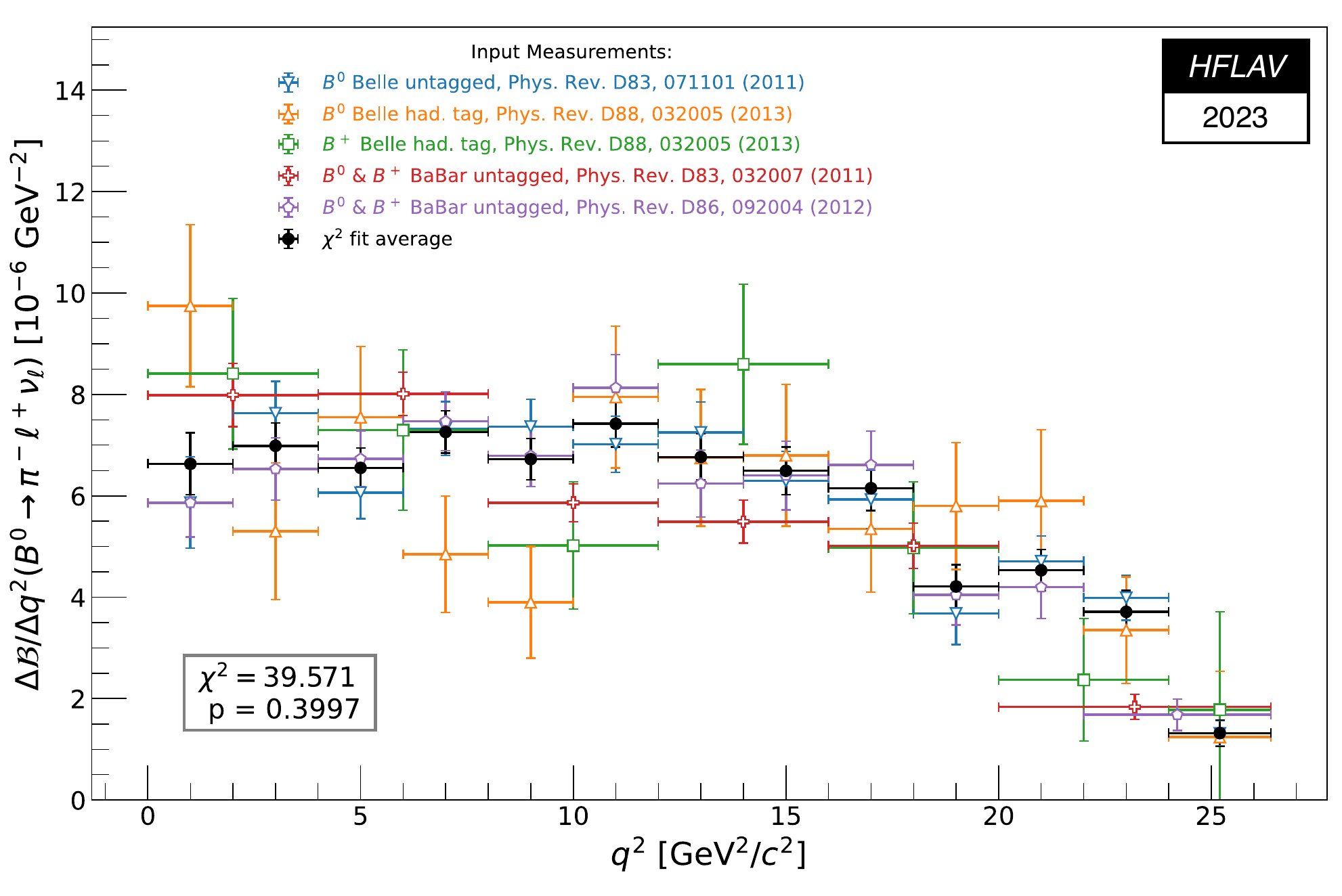}
\caption{The \Btopilnu\ $q^2$ spectrum measurements and the average spectrum obtained 
from the likelihood combination (shown in black). \label{fig:B_to_pi_average}}
\end{figure}

The partial branching fractions and the correlation matrix obtained from the likelihood fit are given in Tables \ref{tab:average} and ~\ref{tab:Cov}. The average for the total $B^0 \to \pi^- \ell^+ \nu_\ell$ branching fraction is obtained by summing up the partial branching fractions to give
\begin{align}
{\cal B}(B^0 \to \pi^- \ell^+ \nu_\ell) = (1.50 \pm 0.02_{\rm stat} \pm 0.05_{\rm syst}) \times 10^{-4}.
\end{align}

\begin{table}
\centering
\small
\caption{Partial $B^0 \to \pi^- \ell^+ \nu_\ell$ branching fractions per GeV$^2$ for the input measurements and the average
obtained from the likelihood fit. The uncertainties are the combined statistical and systematic uncertainties.\label{tab:average}}
\begin{tabular}{|c | c | c | c | c | c || c|}
\hline
\multirow{2}{2cm}{$\Delta q^2$ [GeV${}^2$] }&\multicolumn{6}{ c |}{ $\Delta \mathcal{B}(B^0 \to \pi^- \ell^+ \nu_\ell) / \Delta q^2 $\quad$ [10^{-7}]$}\\\cline{2-7}
& Belle & Belle & Belle & \babar & \babar & Average\\
& untagged & tagged & tagged & untagged & untagged & \\
& ($B^0$) & ($B^0$) & ($B^+$) & ($B^{0,+}$, 12 bins) & ($B^{0,+}$, 6 bins) & \\\hline
$ 0 - 2 $ & $ 58.7 \pm 12.9 $ & $ 97.5 \pm 16.7 $ &\multirow{2}{2cm}{ $ 84.1 \pm 15.5 $ } & $ 58.7 \pm 9.4 $ &\multirow{2}{2cm}{ $ 79.9 \pm 9.1 $} & $ 66.3 \pm 6.1 $\\
$ 2 - 4 $ & $ 76.3 \pm 8.0 $ & $ 53.0 \pm 13.8 $ & & $ 65.3 \pm 7.1 $ & & $ 69.9 \pm 4.5 $\\
$ 4 - 6 $ & $ 60.6 \pm 6.4 $ & $ 75.5 \pm 14.5 $ &\multirow{2}{2cm}{ $ 73.0 \pm 16.2 $} & $ 67.3 \pm 6.4 $ &\multirow{2}{2cm}{ $ 80.1 \pm 5.3 $ }& $ 65.5 \pm 3.9 $\\
$ 6 - 8 $ & $ 73.3 \pm 7.6 $ & $ 48.5 \pm 11.8 $ & & $ 74.7 \pm 7.1 $ & & $ 72.6 \pm 4.2 $\\
\hline
\end{tabular}
\end{table}

\begin{table}
\caption{Correlation matrix of the averaged partial branching fractions per GeV$^2$ in units of $10^{-14}$.\label{tab:Cov}}
\tiny
{\tiny\renewcommand{\arraystretch}{1.5}{%
\begin{tabular}{|c|c|c|c|c|c|c|c|c|c|c|c|c|c|}
\hline
$\Delta q^2$ [GeV$^2$] & 0-2 & 2-4 & 4-6 & 6-8 & 8-10 & 10-12 & 12-14 & 14-16 & 16-18 & 18-20 & 20-22 & 22-24 & 24-26,4 \\ \hline
0-2 & 1.0 & -0.009 & 0.233 & 0.176 & 0.136 & 0.17 & 0.145 & 0.184 & 0.159 & 0.133 & 0.167 & 0.15 & 0.106 \\
2-4 & & 1.0 & -0.091 & 0.279 & 0.126 & 0.154 & 0.139 & 0.144 & 0.126 & 0.091 & 0.127 & 0.102 & 0.072 \\
4-6 & & & 1.0 & -0.031 & 0.274 & 0.195 & 0.207 & 0.171 & 0.164 & 0.116 & 0.153 & 0.126 & 0.093 \\
6-8 & & & & 1.0 & -0.006 & 0.294 & 0.226 & 0.2 & 0.186 & 0.124 & 0.167 & 0.151 & 0.092 \\
8-10 & & & & & 1.0 & 0.033 & 0.297 & 0.171 & 0.182 & 0.12 & 0.138 & 0.138 & 0.082 \\
10-12 & & & & & & 1.0 & 0.097 & 0.254 & 0.219 & 0.153 & 0.172 & 0.173 & 0.103 \\
12-14 & & & & & & & 1.0 & -0.011 & 0.196 & 0.094 & 0.146 & 0.13 & 0.079 \\
14-16 & & & & & & & & 1.0 & 0.139 & 0.194 & 0.146 & 0.161 & 0.103 \\
16-18 & & & & & & & & & 1.0 & 0.076 & 0.149 & 0.13 & 0.087 \\
18-20 & & & & & & & & & & 1.0 & 0.043 & 0.111 & 0.057 \\
20-22 & & & & & & & & & & & 1.0 & 0.073 & 0.094 \\
22-24 & & & & & & & & & & & & 1.0 & -0.038 \\
24-26,4 & & & & & & & & & & & & & 1.0 \\
\hline
\end{tabular}}}
\end{table}

\mysubsubsection{\Vub from \Btopilnu}

The \Vub average can be determined from the averaged $q^2$ spectrum in combination with a prediction for
the normalization of the $\B \to \pi$ form factor.
The differential decay rate for light leptons ($e$, $\mu$) is given by
\begin{align}
 \Delta \Gamma =  \Delta \Gamma(q^2_{\rm low}, q^2_{\rm high})  = \int_{q^2_{\rm low}}^{q^2_{\rm high}} \text{d} q^2 \bigg[ \frac{8 \left| \vec p_\pi \right|  }{3} \frac{ G_F^2 \, \left| V_{ub} \right|^2 q^2 }{256 \, \pi^3 \, m_B^2}  H_0^2(q^2) \bigg]  \, ,
\end{align}
where $G_F$ is Fermi's constant, $\left| \vec p_\pi \right|$ is the magnitude of the three-momentum of the 
final state $\pi$ (a function of $q^2$), $m_B$ the $B^0$-meson mass, 
and $H_0(q^2)$ the only non-zero helicity amplitude. 
The helicity amplitude is a function of the form factor $f_+$, 
\begin{align}
 H_0 = \frac{2 m_B \, \left| \vec p_\pi \right| }{\sqrt{q^2}}\, f_+(q^2) .
\end{align} 
The form factor $f_{+}$ can be calculated with non-perturbative methods, but its general form can be constrained by the differential \Btopilnu\ spectrum. 
Here, we parametrize the form factor using the BCL parametrization truncated at the order $z^2$~\cite{Bourrely:2008za}.

The decay rate is proportional to $\Vub^2 |f_+(q^2)|^2$. Thus to extract \Vub one needs to determine $f_+(q^2)$
(at least at one value of $q^2$). In order to enhance the precision, a binned $\chi^2$ fit is performed
using a $\chi^2$ function of the form
\begin{align} \label{eq:chi2}
 \chi^2 & = \left( \vec{\cal B} - \Delta \vec{\Gamma} \, \tau \right)^T 
            C^{-1} 
            \left( \vec{\cal B} - \Delta \vec{\Gamma} \, \tau \right) + \chi^2_{\rm LQCD} 
\end{align}
where $C$ denotes the covariance matrix given in Table~\ref{tab:Cov}, $\vec{\cal B}$ is the vector of 
averaged partial branching fractions, and $\Delta \vec{\Gamma} \, \tau$ is the product of the vector of 
theoretical predictions of the partial decay rates and the $B^0$-meson lifetime. 
The form factor normalization is included in the fit by the extra term in Eq.~(\ref{eq:chi2}): 
$\chi_{\rm LQCD}$ uses the updated 2024 FLAG lattice average~\cite{FLAG_Webupdate} of the Fermilab/MILC, RBC/UKQCD and JLQCD lattice QCD calculations~\cite{Lattice:2015tia, Flynn:2015mha, Colquhoun:2022atw}. 
The resulting constraints are quoted directly in terms of the coefficients $b_j$ of the BCL parameterization 
and enter Eq.~(\ref{eq:chi2}) as
\begin{align}
 \chi^2_{\rm LQCD} & = \left( {\vec{b}} - {\vec{b}_{\rm LQCD}} \right)^T \, C_{\rm LQCD}^{-1} \,  \left( {\vec{b}} - {\vec{b}_{\rm LQCD}} \right) \, ,
\end{align}
with ${\vec{b}}$ the vector containing the free parameters of the $\chi^2$ fit 
constraining the $f_+$ and $f_0$ form factors, 
${\vec{b}_{\rm LQCD}}$ the averaged values from Ref.~\cite{FLAG_Webupdate}, 
and $C_{\rm LQCD}$ their covariance matrix. Note that the coefficients of the $f_0$ form factor, which is only relevant to describe final states with $\tau$ leptons, introduce additional information via their correlation to the expansion coefficients of $f_+$. Note that we impose the exact constraint of $f_{+}(q^2=0) = f_{0}(q^2=0)$. Further, FLAG is applying a scaling of the uncertainties due to disagreements between Ref.~\cite{Lattice:2015tia, Flynn:2015mha} with Ref.~\cite{Colquhoun:2022atw}. 

For the \Vub average we obtain
 \begin{align}
  \Vub & = \left( 3.75 \pm 0.06 \, _\text{exp} \pm 0.19 \, _\text{theo} \right) \times 10^{-3} \, .
 \end{align}
 The result of the fit is shown in Figure~\ref{fig:vub}. The $\chi^2$ probability of the fit is $37.7\%$. The best fit values for \Vub and the BCL parameters and their correlation matrix are given in Tables~\ref{tab:fitres2} and ~\ref{tab:fitcov2}.

\begin{figure} 
\centering
 \includegraphics[width=0.8\textwidth]{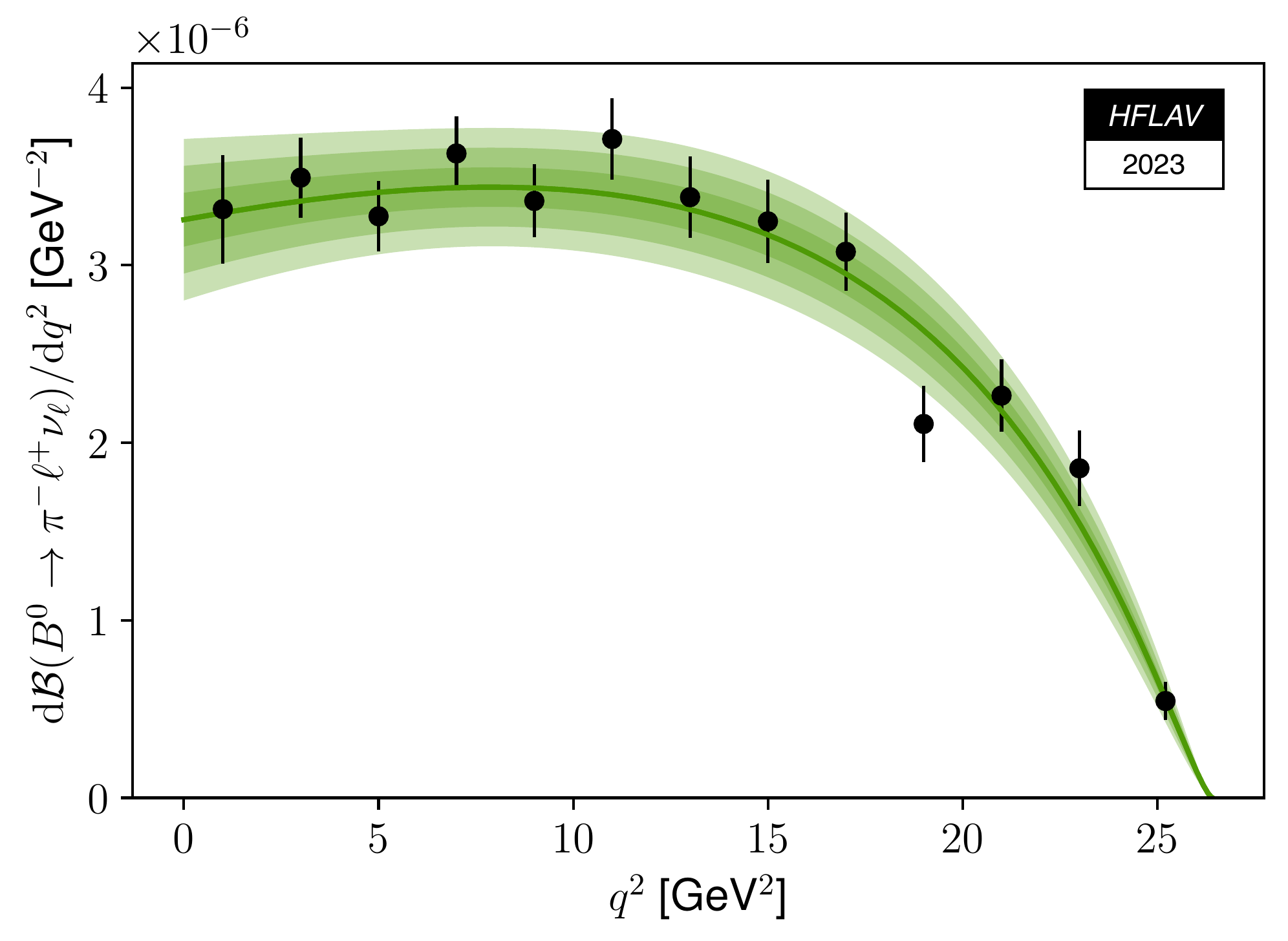}
\caption{Fit of the BCL parametrization to the averaged $q^2$ spectrum from \babar and Belle and the LQCD calculations. 
The error bands represent the $1~\sigma$ (dark green), $2~\sigma$ (green), and $3-\sigma$ (light green) uncertainties 
of the fitted spectrum. \label{fig:vub}}
\end{figure}

\begin{table}
\centering
\caption{Best fit values and uncertainties for the combined fit to data and LQCD results. \label{tab:fitres2}}
\begin{tabular}{c | c }
Parameter & Value\\\hline
$\Vub  $&$(3.75 \pm 0.20)\times 10^{-3}$\\
$b_0^+$   & $0.424 \pm 0.021$\\
$b_1^+$   & $-0.484 \pm 0.043$\\
$b_2^+$   &$-0.658 \pm 0.168$\\
$b_0^0$   &$0.561 \pm 0.024$\\
$b_1^0$   &$-1.393 \pm 0.070$\\
\end{tabular}
\end{table}

\begin{table}
\centering

\caption{Correlation matrix for the combined fit to data and LQCD. \label{tab:fitcov2}}

\begin{tabular}{c | c c c c c c}
Parameter & $\Vub$ & $b_0^+$ & $b_1^+$ & $b_2^+$ & $b_0^0$ & $b_1^0$ \\\hline
$\Vub$    & 1.000        & -0.913 & 0.164 & 0.308& -0.372 & -0.017 \\
$b_0^+$     & -0.913 &  1.000      & -0.345 & -0.344&  0.407&   -0.035  \\
$b_1^+$     & 0.164 &  -0.345 &  1.000      &  -0.597 &  -0.091 &   0.024 \\
$b_2^+$     &  0.308 &  -0.344& -0.597 &   1.000       &  -0.199 &  0.113  \\
$b_0^0$     &  -0.372&  0.407 & -0.091 & -0.199 & 1.000      &  -0.817 \\
$b_1^0$     & -0.017 & -0.035 &   0.024 &  0.113& -0.817 &   1.000         \\
\end{tabular}

\end{table}

\clearpage

\mysubsubsection{$B \to \rho \ell \nu_\ell$ and $B \to \omega \ell \nu_\ell$ branching fraction and $q^2$ spectrum} \label{sec:legacy-spectrum}

We report the branching fraction averages for $B \to V \ell \nu_\ell$, $V = \rho,\, \omega$. The measurements, adjusted for common inputs,  and their averages are listed in 
Tables~\ref{tab:rholnu},~\ref{tab:omegalnu}, and presented in Figures ~\ref{fig:xulnu1}.

In the $\Bp\to\rho^0\ell^+\nu$ average, both the $\Bz\to\rho^-\ell^+\nu$ and $\Bp\to\rho^0\ell^+\nu$ decays are used, 
where the $\Bz\to\rho^-\ell^+\nu$ are rescaled by $0.5\tau_{B^+}/\tau_{B^0}$ assuming the isospin symmetry.
The $\Bp\to\rho^0\ell^+\nu$ results show significant differences, in particular the \babar untagged analysis 
gives a branching fraction significantly lower (by about 3$\sigma$) than the Belle measurement based on the hadronic-tag. The difference is about 2$\sigma$ for the $\Bp\to\rho^0\ell^+\nu$ decay modes.

\begin{table}[!htb]
\begin{center}
\caption{Summary of exclusive determinations of $\Bp\to\rho^0\ell^+\nu$, rescaled to common inputs. The errors quoted
correspond to statistical and systematic uncertainties, respectively.}
\label{tab:rholnu}
\begin{small}
\begin{tabular}{lc}
\hline
& $\cbf [10^{-4}]$
\\
\hline\hline
CLEO (Untagged) $\rho^+$~\cite{Behrens:1999vv}
& $1.49\pm 0.22\pm 0.28\ $ 
\\ 
CLEO (Untagged) $\rho^+$~\cite{Adam:2007pv}
& $1.58\pm 0.20\pm 0.20\ $ 
\\ 
Belle (Hadronic Tag) $\rho^+$~\cite{Sibidanov:2013rkk}
& $1.73\pm 0.15\pm 0.13\ $
\\
Belle (Hadronic Tag) $\rho^0$~\cite{Sibidanov:2013rkk}
& $1.82\pm 0.10\pm 0.10\ $
\\
Belle (Semileptonic Tag) $\rho^+$~\cite{Hokuue:2006nr}
& $1.21\pm 0.29\pm 0.17\ $
\\
Belle (Semileptonic Tag) $\rho^0$~\cite{Hokuue:2006nr}
& $1.35\pm 0.23\pm 0.18\ $
\\
\babar (Untagged) $\rho^+$~\cite{delAmoSanchez:2010af}
& $1.05\pm 0.11\pm 0.21\ $
\\
\babar (Untagged) $\rho^0$~\cite{delAmoSanchez:2010af}
& $1.00\pm 0.10\pm 0.21\ $

\\  \hline
{\bf Average}
& \mathversion{bold}$1.58 \pm 0.11$
\\ 
\hline
\end{tabular}\\
\end{small}
\end{center}
\end{table}

\begin{table}[!htb]
\begin{center}
\caption{Summary of exclusive determinations of $\Bp\to\omega\ell^+\nu$, , rescaled to common inputs. The errors quoted
correspond to statistical and systematic uncertainties, respectively.}
\label{tab:omegalnu}
\begin{small}
%
 \caption{
 (a) Summary of exclusive determinations of $\cbf(\Bp\to\rho^0\ell^+\nu)$ and their average. Measurements
 of $\Bz\to \rho^{-}\ell^+\nu$ branching fractions have been scaled by $0.5\tau_{\Bp}/\tau_{\Bz}$ 
 in accordance with isospin symmetry.    
(b) Summary of exclusive determinations of $\Bp\to\omega\ell^+\nu$ and their average.
}
\label{fig:xulnu1}
\end{center}
\end{figure}

We use the most precise measurements of the differential $B \to V \ell \nu_\ell$, $V = \rho,\, \omega$ decay rates as a function of the four-momentum transfer $q^2$ published by \babar~\cite{delAmoSanchez:2010af,Lees:2012mq} and Belle~\cite{Sibidanov:2013rkk}. To obtain an averaged $q^2$ spectrum and averaged branching fractions, we perform a $\chi^2$~fit of the form 
\begin{equation}
\begin{aligned}
	\chi^2(\bm{\bar{x}}) &= \kern-4.5em\sum_{\kern4em m \in \{\text{Belle, \babar}\}} \kern-4.5em \Delta \vec{y}^T_m C^{-1}_m \Delta \vec{y}_m,\\
	\Delta \vec{y}_m &= \begin{pmatrix} 
	\vdots \\
	x_i^m - \kern-1.5em\sum\limits_{ \kern1em j > N_{i-1}}^{N_i} \kern-1em \bar{x}_j \\
	\vdots
	\end{pmatrix}\,,
\end{aligned}
\end{equation}
where $C_m$ is the covariance of the measurement and $x_i^m$ is the measured differential rate in bin $i$ multiplied by the corresponding bin width. 
Further, $\bm{\bar{x}}$ denotes the averaged spectrum and $(N_{i-1},N_i]$ the range of averaged bins used to map to the $i$th measured bin. 
The binning of the averaged spectrum is chosen to match the most granular experimental binning. 

For the average of the $B \to \omega \ell \bar\nu$ measurements from Belle and \babar , we again chose the binning of the most granular spectrum, in this case \babar's.
However the experimental spectra do not have a compatible binning in terms of matching bin boundaries.
In order to incorporate the Belle data and create an averaged spectrum, 
the LCSR fit results~\cite{Straub:2015ica} are used to create a model with which to split the second and fifth bin of the chosen binning, shown in black in Fig.~\ref{fig:legacy-spectra}.
To match the average bin onto a measurement without matching bin edges, the average bin $\bar{x}_i$, $i = 2$ or $5$, is split into two parts delimited by the lower bin edge, 
the $q^2$ value where the bin is split, and the upper bin edge. 
We label the two parts of the split bin as `left' and `right', respectively, in the following and define:
\begin{equation}
\begin{aligned}
	\bm{\bar{x}}_{i,\mathrm{left}} &= I_{i,\mathrm{left}}/I_i(1+\theta_i \varepsilon_{i,\mathrm{left}})\,, \\
	\bm{\bar{x}}_{i,\mathrm{right}} &= I_{i,\mathrm{right}}/I_i(1-\theta_i \varepsilon_{i,\mathrm{right}})\,,
\end{aligned}
\end{equation}
where $I_{i,\mathrm{left}}$ ($I_{i,\mathrm{right}}$) is the integral of the model function on the support of the left (right) part of the split bin, 
the sum $I_i = I_{i,\mathrm{left}} + I_{i,\mathrm{right}}$ is the integral over the entire bin, 
$\varepsilon_{i,\mathrm{left}}$ ($\varepsilon_{i,\mathrm{right}}$) the uncertainty of the integration given by the model uncertainty, 
and $\theta_i$ the nuisance parameter for the model dependence. 
We point out that the averaged spectrum does not depend on $\Vub$, as $\Vub$ cancels in the ratios $I_{i,\mathrm{left}} / I_i$ ($I_{i,\mathrm{right}} / I_i$).

The averaged spectra are shown in black in Fig.~\ref{fig:legacy-spectra} and tabulated in Table~\ref{tab:legacy-spectra}.

\begin{figure}
	\centering
	\includegraphics[width=0.49\linewidth]{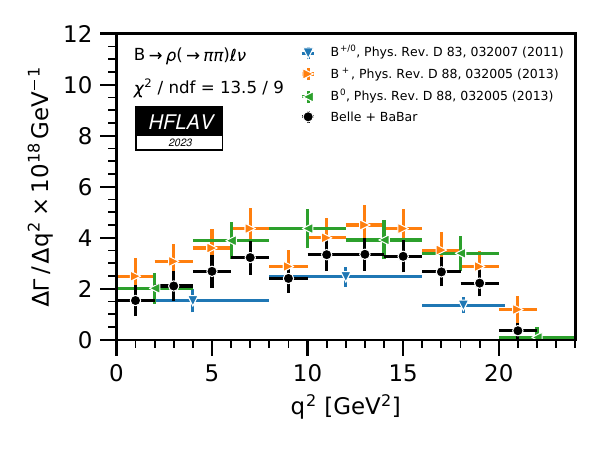}
	\includegraphics[width=0.49\linewidth]{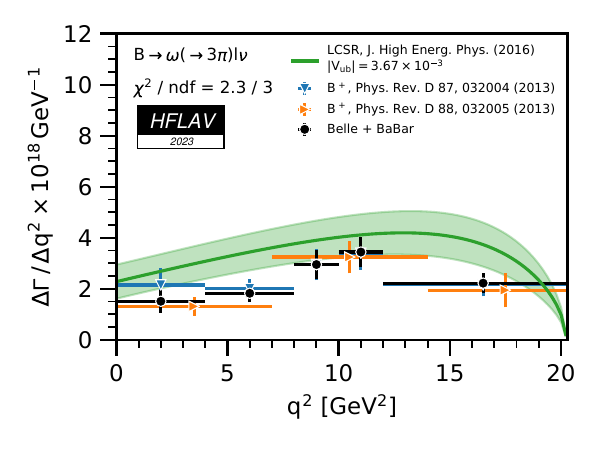}
	\caption{The averaged $q^2$ spectrum of the measurements listed in the text for the $\rho$ (left) and $\omega$ (right) final state on top of the latest Belle and \babar\ measurements. The isospin transformation is applied to the $B^0 \to \rho^-\ell^+\nu$ measurements. In the right figure we also show the model (green band) which was used to split the bins in the averaging procedure.}
	\label{fig:legacy-spectra}
\end{figure}

\begin{table}[tb]
\centering
	\caption{Averaged spectra. For the corresponding correlation matrices see Tables~\ref{tab:legacy-spectrum-correlation-rho} and~\ref{tab:legacy-spectrum-correlation-omega}.}
	\label{tab:legacy-spectra}
\begin{tabular}{cc}
		\hline
		\hline
		\multicolumn{2}{c}{$B \rightarrow \rho \ell \bar \nu$} \\
		$q^2$ bin & $\Delta \Gamma / \Delta q^2 \times 10^{18}$ [GeV$^{-1}$]\\
		\hline
$[0,\,2]$ & $1.54 \pm 0.62$ \\
$[2,\,4]$ & $2.11 \pm 0.60$ \\
$[4,\,6]$ & $2.68 \pm 0.65$ \\
$[6,\,8]$ & $3.22 \pm 0.67$ \\
$[8,\,10]$ & $2.40 \pm 0.56$ \\
$[10,\,12]$ & $3.34 \pm 0.65$ \\
$[12,\,14]$ & $3.35 \pm 0.65$ \\
$[14,\,16]$ & $3.27 \pm 0.63$ \\
$[16,\,18]$ & $2.66 \pm 0.57$ \\
$[18,\,20]$ & $2.22 \pm 0.52$ \\
$[20,\,22]$ & $0.35 \pm 0.32$ \\
		\hline
		\hline
	\end{tabular}
	\hspace{1em}
	\begin{tabular}{cc}
		\hline
		\hline
		\multicolumn{2}{c}{$B \rightarrow \omega \ell \bar \nu$} \\
		$q^2$ bin & $\Delta \Gamma / \Delta q^2 \times 10^{18}$ [GeV$^{-1}$]\\
		\hline
$[0,\,4]$ & $1.51 \pm 0.46$ \\
$[4,\,8]$ & $1.82 \pm 0.35$ \\
$[8,\,10]$ & $2.95 \pm 0.56$ \\
$[10,\,12]$ & $3.44 \pm 0.59$ \\
$[12,\,21]$ & $2.22 \pm 0.40$ \\
		\hline
		\multicolumn{2}{c}{Nuisance Parameters} \\
		\hline
$\theta_2$ & $-0.01 \pm 1.00$ \\
$\theta_5$ & $0.00 \pm 1.00$ \\
		\hline
		\hline
		\vspace{2.76em}
	\end{tabular}
\end{table}

\begin{table}[tbp]
\centering
	\caption{Correlation matrix of the averaged $B \to \rho \ell \nu_\ell$ spectrum.}
	\label{tab:legacy-spectrum-correlation-rho}
\scalebox{0.85}{
\begin{tabular}{crrrrrrrrrrr}
	\hline
	\hline	
	&   [0, 2] &   [2, 4] &   [4, 6] &   [6, 8] &   [8, 10] &   [10, 12] &   [12, 14] &   [14, 16] &   [16, 18] &   [18, 20] &   [20, 22] \\
	\hline
	{[0, 2]}   &     1.00 &    -0.30 &     0.03 &     0.01 &      0.09 &       0.09 &       0.09 &       0.09 &       0.08 &       0.08 &       0.02 \\
	{[2, 4]}   &    -0.30 &     1.00 &    -0.03 &     0.09 &      0.11 &       0.12 &       0.12 &       0.12 &       0.11 &       0.10 &       0.02 \\
	{[4, 6]}   &     0.03 &    -0.03 &     1.00 &    -0.18 &      0.13 &       0.13 &       0.15 &       0.14 &       0.13 &       0.12 &       0.03 \\
	{[6, 8]}   &     0.01 &     0.09 &    -0.18 &     1.00 &      0.06 &       0.18 &       0.18 &       0.18 &       0.16 &       0.14 &       0.04 \\
	{[8, 10]}  &     0.09 &     0.11 &     0.13 &     0.06 &      1.00 &      -0.21 &       0.05 &       0.04 &       0.12 &       0.10 &       0.03 \\
	{[10, 12]} &     0.09 &     0.12 &     0.13 &     0.18 &     -0.21 &       1.00 &      -0.00 &       0.07 &       0.15 &       0.13 &       0.04 \\
	{[12, 14]} &     0.09 &     0.12 &     0.15 &     0.18 &      0.05 &      -0.00 &       1.00 &      -0.16 &       0.14 &       0.12 &       0.04 \\
	{[14, 16]} &     0.09 &     0.12 &     0.14 &     0.18 &      0.04 &       0.07 &      -0.16 &       1.00 &       0.10 &       0.14 &       0.05 \\
	{[16, 18]} &     0.08 &     0.11 &     0.13 &     0.16 &      0.12 &       0.15 &       0.14 &       0.10 &       1.00 &      -0.27 &      -0.11 \\
	{[18, 20]} &     0.08 &     0.10 &     0.12 &     0.14 &      0.10 &       0.13 &       0.12 &       0.14 &      -0.27 &       1.00 &      -0.13 \\
	{[20, 22]} &     0.02 &     0.02 &     0.03 &     0.04 &      0.03 &       0.04 &       0.04 &       0.05 &      -0.11 &      -0.13 &       1.00 \\
	\hline
	\hline	
\end{tabular}
}
\end{table}

\begin{table}[tbp]
\centering
	\caption{Correlation matrix of the averaged $B \to \omega \ell \nu_\ell$ spectrum.}
	\label{tab:legacy-spectrum-correlation-omega}
\scalebox{0.85}{
\begin{tabular}{crrrrrrr}
	\hline
	\hline	
	&   [0, 4] &   [4, 8] &   [8, 10] &   [10, 12] &   [12, 21] &   $\theta_2$ &   $\theta_5$ \\
	\hline
	{[0, 4]}     &     1.00 &    -0.15 &      0.08 &       0.04 &       0.06 &        -0.01 &         0.00 \\
	{[4, 8]}     &    -0.15 &     1.00 &      0.09 &       0.09 &       0.15 &        -0.01 &        -0.00 \\
	{[8, 10]}    &     0.08 &     0.09 &      1.00 &      -0.01 &       0.12 &        -0.00 &        -0.00 \\
	{[10, 12]}   &     0.04 &     0.09 &     -0.01 &       1.00 &       0.15 &         0.00 &        -0.00 \\
	{[12, 21]}   &     0.06 &     0.15 &      0.12 &       0.15 &       1.00 &        -0.00 &        -0.00 \\
	$\theta_2$ &    -0.01 &    -0.01 &     -0.00 &       0.00 &      -0.00 &         1.00 &         0.00 \\
	$\theta_5$ &     0.00 &    -0.00 &     -0.00 &      -0.00 &      -0.00 &         0.00 &         1.00 \\
	\hline
	\hline	
\end{tabular}
}
\end{table}

\clearpage

\mysubsubsection{Other exclusive charmless semileptonic \B decays}

We report the branching fraction averages for $\Bp\to\eta\ell^+\nu$ and $\Bp\to\etapr\ell^+\nu$.  The measurements, , adjusted for common inputs, and their averages are listed in Tables ~\ref{tab:etalnu} and~\ref{tab:etaprimelnu}, 
and presented in Figure ~\ref{fig:xulnu2}. 
For $\Bp\to\eta\ell^+\nu$ decays, the agreement between the different measurements is good, while $\Bp\to\etapr\ell^+\nu$ shows a significant discrepancy between the old CLEO measurement and the \babar untagged 
analysis.

\begin{table}[!htb]
\begin{center}
\caption{Summary of exclusive determinations of $\Bp\to\eta\ell^+\nu$, rescaled to common inputs. 
The errors quoted correspond to statistical and systematic uncertainties, respectively.}
\label{tab:etalnu}
\begin{small}
\begin{tabular}{lc}
\hline
& $\cbf [10^{-4}]$
\\
\hline\hline
CLEO ~\cite{Gray:2007pw}
& $0.45\pm 0.23\pm 0.11\ $
\\
\babar\ (Untagged) ~\cite{Aubert:2008ct}
& $0.31\pm 0.06\pm 0.08\ $
\\ 
\babar\ (Semileptonic Tag) ~\cite{Aubert:2008bf}
& $0.64\pm 0.20\pm 0.04\ $
\\
\babar\ (Loose $\nu$-reco.) ~\cite{Lees:2012vv}
& $0.38\pm 0.05\pm 0.05\ $
\\  
\belle\ (Hadronic Tag) ~\cite{Beleno:2017cao}
& $0.42\pm 0.11\pm 0.09\ $
\\  
\belle\ (Untagged) ~\cite{Belle:2021hah}
& $0.283\pm 0.055\pm 0.034\ $
\\  
 \hline
{\bf Average}
& \mathversion{bold}$0.344 \pm 0.043 $
\\ 
\hline
\end{tabular}\\
\end{small}
\end{center}
\end{table}

\begin{table}[!htb]
\begin{center}
\caption{Summary of exclusive determinations of  $\Bp\to\eta'\ell^+\nu$, rescaled to common inputs. The errors quoted
correspond to statistical and systematic uncertainties, respectively.}
\label{tab:etaprimelnu}
\begin{small}
\begin{tabular}{lc}
\hline
& $\cbf [10^{-4}]$
\\
\hline\hline
CLEO ~\cite{Gray:2007pw} & $2.71\pm 0.80\pm 0.56\ $
\\
\babar\ (Semileptonic Tag) ~\cite{Aubert:2008bf}
& $0.04\pm 0.22\pm 0.04$, $(<0.47 ~~@~ 90\% C.L.)$
\\ 
\babar\ (Untagged) ~\cite{Lees:2012vv} & $0.24\pm 0.08\pm 0.03$
\\  
\belle\ (Hadronic Tag) ~\cite{Beleno:2017cao}
& $0.36\pm 0.27\pm 0.04\ $
\\  
\belle\ (Untagged) ~\cite{Belle:2021hah}
& $0.279\pm 0.129\pm 0.030\ $
\\  
 \hline
{\bf Average}
& \mathversion{bold}$0.249 \pm 0.067$
\\ 
\hline
\end{tabular}\\
\end{small}
\end{center}
\end{table}

\begin{figure}[!ht]
 \begin{center}
  \unitlength1.0cm %
  \begin{picture}(14.,10.0)  %
   \put( -1.5,  0.0){\includegraphics[width=9.0cm]{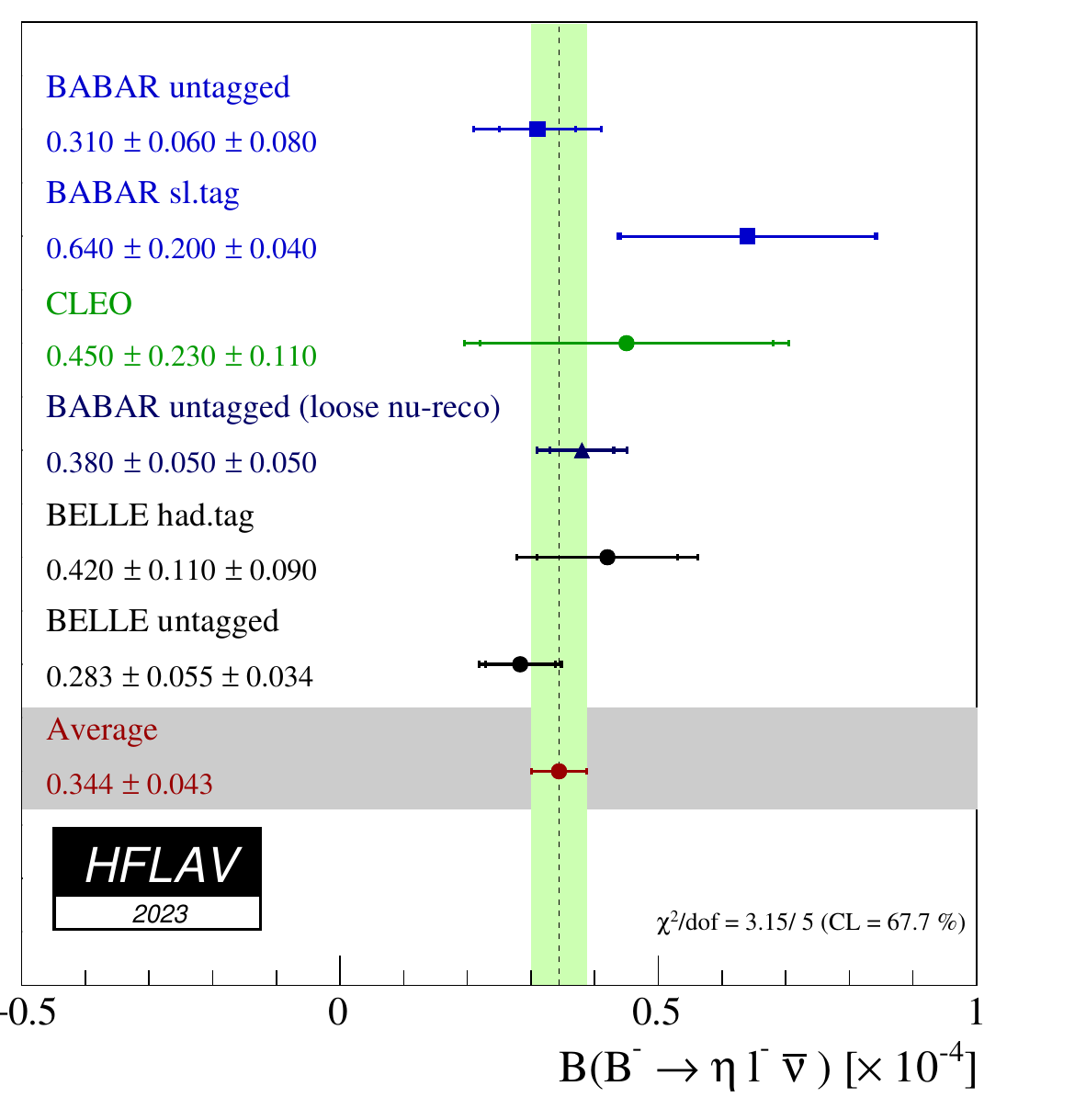}}
   \put( 7.5,  0.0){\includegraphics[width=9.0cm]{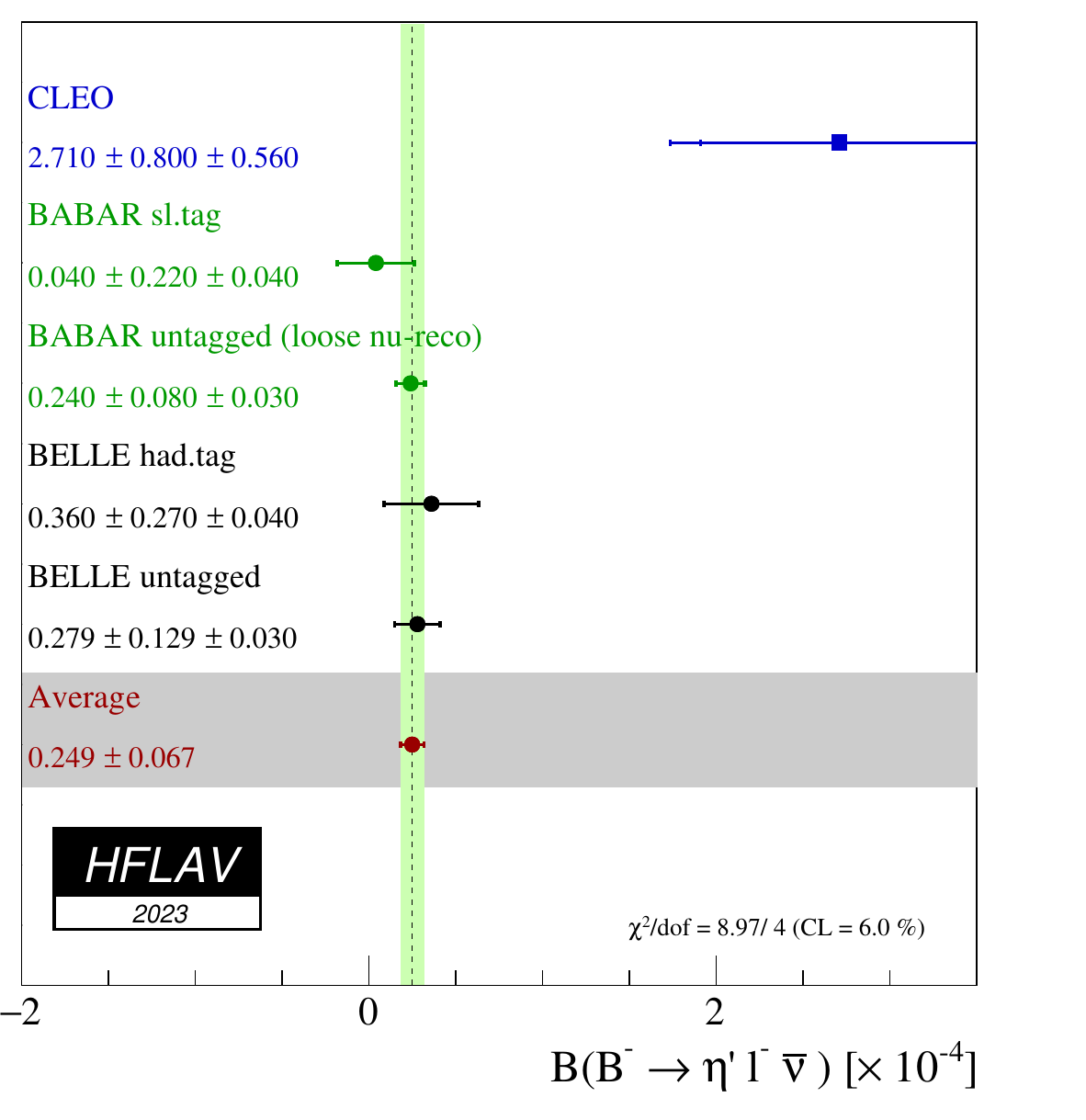}} 
   \put(  5.8,  8.5){{\large\bf a)}}     
   \put( 14.6,  8.5){{\large\bf b)}}
   
   \end{picture} \caption{
(a) Summary of exclusive determinations of $\cbf(\Bp\to\eta\ell^+\nu)$ and their average.
(b) Summary of exclusive determinations of $\cbf(\Bp\to\etapr\ell^+\nu)$ and their average.
}
\label{fig:xulnu2}
\end{center}
\end{figure}

\clearpage
\mysubsubsection{Direct measurements of $|V_{ub}|/|V_{cb}|$}
\label{slbdecay_vubvcb}

The \lhcb Collaboration published the direct measurement of the $|V_{ub}|/|V_{cb}|$ ratio using the $\Lambda_b\to p\mu^-\bar{\nu}_\mu$
and $\bar{B}_s^0\to K^+\mu^-\bar{\nu}_\mu$ decays, properly normalized.

\noindent $\boldsymbol{\Lambda_b\to p\mu^-\bar{\nu}_\mu}$ The first observation of the CKM-suppressed decay $\Lb\to p\mu^-\bar\nu_\mu$ 
has been obtained by \lhcb and published in Ref.\cite{Aaij:2015bfa}. The measurement of the ratio of partial branching fractions at high $q^2$
for $\Lb\to p\mu^-\bar\nu_\mu$ and $\Lb\to \Lc(\to pK\pi)\mu^-\bar\nu_\mu$ decays is
\begin{align}
R = \dfrac{{\cal B}(\Lb\to p\mu^-\bar\nu_\mu)_{q^2>15~GeV^2} }{{\cal B}(\Lb\to \Lc\mu^-\bar\nu_\mu)_{q^2>7~GeV^2} }=(1.00\pm 0.04\pm 0.08)\times 10^{-2}.
\end{align}

\noindent The ratio $R$ is proportional to $(|V_{ub}|/|V_{cb}|)^2$ and sensitive to the form factors 
of $\Lb\to p$ and $\Lb\to \Lc$ transitions that have to be computed with non-perturbative
methods, such as lattice QCD.
The uncertainty on ${\cal B}(\Lc\to p K \pi)$ is the largest source of systematic uncertainties
on $R$. Using the recent average of ${\cal B}(\Lc\to p K \pi)=(6.26\pm 0.29)\%$ \cite{PDG_2022}, the rescaled value for $R$ is  
\begin{align}
R = (0.92\pm 0.04\pm 0.06)\times 10^{-2}.
\end{align}
\noindent Using the precise lattice QCD prediction \cite{Detmold:2015aaa} of the form factors in the 
experimentally interesting $q^2$ region considered, we obtain
\begin{align}
\dfrac{|V_{ub}|}{|V_{cb}|} = 0.079\pm \, 0.003_\text{exp} \pm \, 0.004_\text{FF}
\end{align}

\noindent where the first uncertainty is the total experimental uncertainty, and the second one is due to the 
knowledge of the form factors.

\noindent $\boldsymbol{\bar{B}_s^0\to K^+\mu^-\bar{\nu}_\mu}$  
The \lhcb experiment also reported the first observation of the decay $\bar{B}_s^0\to K^+\mu^-\bar\nu_\mu$ and the measurements of its branching fraction normalised to the $\bar{B}_s^0\to D_s^+\mu^-\bar\nu_\mu$ decays \cite{aaij:2020nvo}. The measurement has been performed in two bins of $q^2$. The results of the partial branching fractions, adjusted for the recent average of ${\cal B}(D_s^+\to K^+ K^- \pi^-)=(5.37\pm 0.10)\%$ \cite{PDG_2022}, 
has been translated in measurements of $|V_{ub}|/|V_{cb}|$
using form factor calculation from LCSR for $q^2<7~\text{GeV}^2$ \cite{Khodjamirian:2017fxg}, 
and Lattice calculation for $q^2>7~\text{GeV}^2$ provided by FLAG \cite{FlavourLatticeAveragingGroupFLAG:2021npn}, which is based on the calculations by HPQCD \cite{Bouchard:2014ypa}, FNAL/MILC \cite{Bazavov:2019aom} and RBC/UKQCD \cite{Flynn:2023nhi}. The results are
\begin{eqnarray}
\dfrac{|V_{ub}|}{|V_{cb}|} &=& 0.0608\pm 0.0020_\text{exp} \pm \, 0.0030_\text{FF},~q^2<7~\text{GeV}^2, \\
\dfrac{|V_{ub}|}{|V_{cb}|} &=& 0.0860\pm 0.0038_\text{exp} \pm \, 0.0037_\text{FF},~q^2>7~\text{GeV}^2, 
\end{eqnarray} 

\noindent where the experimental uncertainties include also the uncertainties on the external inputs, and the last errors are due to the form factor calculations for both $B\to K$ and $B_s\to D_s$ decays. The discrepancy between the values of $|V_{ub}|/|V_{cb}|$ for the low and high $q^2$, requires further investigations. 
There are some recent efforts to combine LCSR and LQCD calculations, resulting in reduced discrepancies between low and high $q^2$, as shown in Ref.\cite{Bolognani:2023mcf}.

\subsection{Inclusive CKM-suppressed decays}
\label{slbdecays_b2uincl}
Measurements of $\bar{B} \to X_u \ell^- \bar\nu_\ell$  decays are very challenging because of background from the Cabibbo-favoured 
$\bar{B} \to X_c \ell^- \bar\nu_\ell$ decays, whose branching fraction is about 50 times larger than that of the signal.  Cuts designed to suppress this dominant background severely complicate the perturbative 
QCD calculations required to extract $\vub$.  Tight cuts necessitate parameterization of the so-called 
shape functions in order to describe the unmeasured regions of  phase space.  
We use several theoretical calculations to extract \vub~ and  
do not advocate the use of one method over another.
The authors of the different calculations have provided 
codes to compute the partial rates in limited regions of phase space covered by the measurements. 
Belle~\cite{ref:belle-multivariate} and \babar~\cite{Lees:2011fv,TheBABAR:2016lja} produced measurements that 
explore large portions of phase space and thus reduce %
theoretical 
uncertainties. 

In the averages reported in the following Sections (~\ref{subsec:blnp},~\ref{subsec:dge},~\ref{subsec:ggou}), the %
modelling of $\bar{B}\to X_c\ell^-\bar\nu_\ell$ and $\bar{B}\to X_u\ell^-\bar\nu_\ell$ decays and the theoretical
uncertainties are taken as fully correlated among all measurements.
Reconstruction-related uncertainties are taken as fully correlated within a given experiment.
Measurements of partial branching fractions for $\bar{B}\to X_u\ell^-\bar\nu_\ell$
transitions from $\Upsilon(4S)$ decays, together with the corresponding selected region, 
are given in Table~\ref{tab:BFbulnu}. 
We use all results published by \babar\ in Ref.~\cite{Lees:2011fv}, since the 
statistical correlations are given. 
The dependence of the quoted error on the measured value for each source of uncertainty 
is taken into account in the calculation of the averages.

It was first suggested by Neubert~\cite{Neubert:1993um} and later detailed by Leibovich, 
Low, and Rothstein (LLR)~\cite{Leibovich:1999xf} and Lange, Neubert and Paz (LNP)~\cite{Lange:2005qn}, 
that the uncertainty of
the leading shape functions can be eliminated by comparing inclusive rates for
$\bar{B}\to X_u\ell^-\bar\nu_\ell$ decays with the inclusive photon spectrum in $\bar{B}\to X_s\gamma$,
based on the assumption that the shape functions for transitions to light
quarks, $u$ or $s$, are the same to first order.
However, shape function uncertainties are only eliminated at the leading order
and they still enter via the signal models used for the determination of efficiency. 

No new measurements were performed since the previous release of this work. The $\vub$ values were recomputed from the measured partial branching fractions by recalculating the theoretical partial rates following an update of the heavy quark parameters. 
\begin{table}[!htb]
\caption{\label{tab:BFbulnu}
Summary of measurements  of partial branching
fractions for $\bar{B}\rightarrow X_u \ell^- \bar\nu_{\ell}$ decays.
The errors quoted on $\Delta\cbf$ correspond to
statistical and systematic uncertainties.
$E_e$ is the electron
energy in the $B$~rest frame, $p^*$ the lepton momentum in the
$B$~frame and $m_X$ is the invariant mass of the hadronic system. The
light-cone momentum $P_+$ is defined in the $B$ rest frame as
$P_+=E_X-|\vec p_X|$.
The $s_\mathrm{h}^{\mathrm{max}}$ variable is described in Refs.~\cite{ref:shmax,ref:babar-elq2}. }
\begin{center}
\begin{small}
\begin{tabular}{|llcl|}
\hline
Measurement & Accepted region &  $\Delta\cbf [10^{-4}]$ & Notes\\
\hline\hline
CLEO~\cite{ref:cleo-endpoint}
& $E_e>2.1\,\gev$ & $3.3\pm 0.2\pm 0.7$ &  \\ 
\babar~\cite{ref:babar-elq2}
& $E_e>2.0~\gev$, $s_\mathrm{h}^{\mathrm{max}}<3.5\,\mathrm{GeV}^2$ & $4.4\pm 0.4\pm 0.4$ & \\
\babar~\cite{TheBABAR:2016lja}
& $E_e>0.8\,\gev$  & $15.5\pm 0.8\pm 0.9$ &  Using the GGOU model\\

Belle~\cite{ref:belle-endpoint}
& $E_e>1.9\,\gev$  & $8.5\pm 0.4\pm 1.5$ & \\
\babar~\cite{Lees:2011fv}
& $M_X<1.7\,\gevcc, q^2>8\,\gevgevcccc$ & $6.9\pm 0.6\pm 0.4$ & 
\\
Belle~\cite{ref:belle-mxq2Anneal}
& $M_X<1.7\,\gevcc, q^2>8\,\gevgevcccc$ & $7.4\pm 0.9\pm 1.3$ & \\
\babar~\cite{Lees:2011fv}
& $P_+<0.66\,\gev$  & $9.9\pm 0.9\pm 0.8 $ & 
\\
\babar~\cite{Lees:2011fv}
& $M_X<1.7\,\gevcc$ & $11.6\pm 1.0\pm 0.8 $ &
\\ 
\babar~\cite{Lees:2011fv}
& $M_X<1.55\,\gevcc$ & $10.9\pm 0.8\pm 0.6 $ & 

\\ 
Belle~\cite{Belle:2021eni} 
& $p^*_{\ell} > 1~\gev/c$  & $15.9\pm 0.7\pm 1.6$ & \\

\babar~\cite{Lees:2011fv}
& ($M_X, q^2$) fit, $p^*_{\ell} > 1~\gev/c$  & $18.2\pm 1.3\pm 1.5$ & 
\\ 
\babar~\cite{Lees:2011fv}
& $p^*_{\ell} > 1.3~\gev/c$  & $15.5\pm 1.3\pm 1.4$ & 
\\ \hline
\end{tabular}\\
\end{small}
\end{center}
\end{table}

Three theoretical calculations are used, detailed in the following. Other calculation exist~\cite{Aglietti:2007ik,ref:BLL}, but they have not been utilised in this update. 

\subsubsection{BLNP}
\label{subsec:blnp}
Bosch, Lange, Neubert and Paz (BLNP)~\cite{ref:BLNP,
  ref:Neubert-new-1,ref:Neubert-new-2,ref:Neubert-new-3}
provide theoretical expressions for the triple
differential decay rate for $\bar{B}\to X_u \ell^- \bar\nu_\ell$ events, incorporating all known
contributions, while smoothly interpolating between the 
``shape-function region'' of large hadronic
energy and small invariant mass, and the ``OPE region'' in which all
hadronic kinematical variables scale with the $b$-quark mass. BLNP assign
uncertainties to the $b$-quark mass, which enters through the leading shape function, 
to sub-leading shape function forms, to possible weak annihilation
contribution, and to matching scales. 
The BLNP calculation uses the shape function renormalization scheme; the heavy quark parameters determined in $b \to c$ transitions    
from the global fit in the kinetic scheme, published in Ref.~\cite{Finauri:2023kte}, were therefore 
translated into the shape function scheme by using a prescription by Neubert 
\cite{Neubert:2004sp,Neubert:2005nt}. The resulting parameters are 
$m_b({\rm SF})=(4.600 \pm 0.012 \pm 0.018)~\gev$, 
$\mu_\pi^2({\rm SF})=(0.184 \pm 0.047 ^{+0.020}_{-0.040})~\gevcc$, 
where the second uncertainty is due to the scheme translation. 
The extracted values of \vub\, for each measurement along with their average are given in
Table~\ref{tab:bulnu} and illustrated in Fig.~\ref{fig:BLNP_DGE}(a). 
The total uncertainty is $\pm 5.3\%$ and is due to:
statistics ($\pm 1.5\%$),
detector effects ($\pm 1.7\%$),
$\bar{B}\to X_c \ell^- \bar\nu_\ell$ model ($^{+0.9}_{-1.0}\%$),
$\bar{B}\to X_u \ell^- \bar\nu_\ell$ model ($^{+1.6}_{-1.5}\%$),
heavy quark parameters ($\pm 2.4\%$),
SF functional form ($^{+0.1}_{-0.3}\%$),
sub-leading shape functions ($\pm 0.8\%$),
matching scales in BLNP $\mu,\mu_i,\mu_h$ ($\pm 3.6\%$), and
weak annihilation ($^{+0.0}_{-1.0}\%$).
The error assigned to the matching scales 
is the source of the largest uncertainty, while the
uncertainty due to HQE parameters ($b$-quark mass and $\mu_\pi^2({\rm SF}))$ is second. The uncertainty due to 
weak annihilation is assumed to be asymmetric, \ie\ it only tends to decrease \vub.

\begin{table}[!htb]
\caption{\label{tab:bulnu}
Summary of input parameters used by the different theory calculations,
corresponding inclusive determinations of $\vub$ and their average.
The errors quoted on \vub\ correspond to
experimental and theoretical uncertainties, respectively.}
\begin{center}
\resizebox{0.99\textwidth}{!}{

  \caption{Measurements of $\vub$ from inclusive semileptonic decays 
and their average based on the BLNP (a) and DGE (b) prescription. The
labels indicate the variables and selections used to define the
signal regions in the different analyses.
} \label{fig:BLNP_DGE}
 \end{center}
\end{figure}

\begin{figure}[!ht]
 \begin{center}
  \unitlength1.0cm %
   {\includegraphics[width=9.4cm]{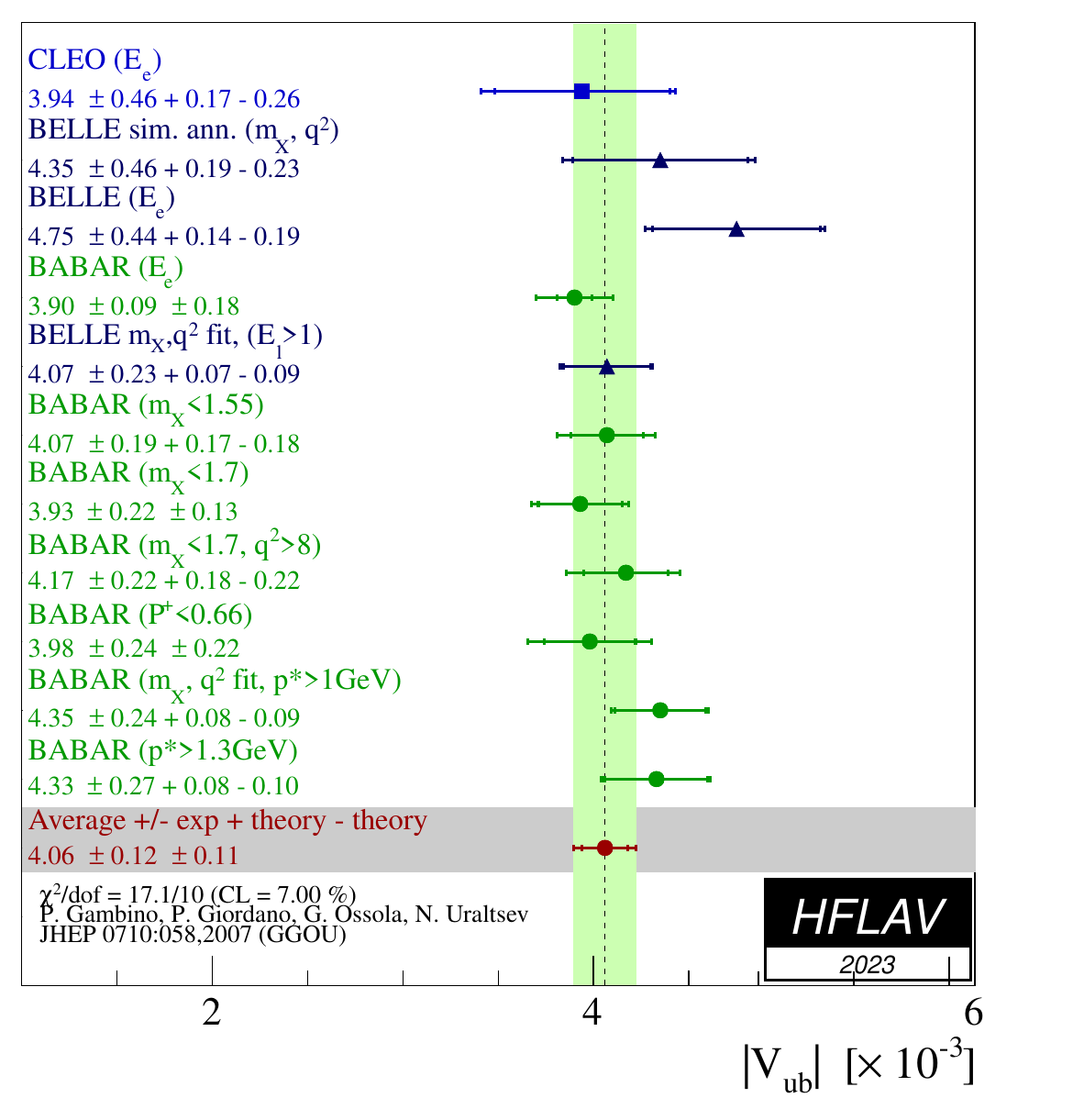}}
  \caption{Measurements of $\vub$ from inclusive semileptonic decays 
and their average based on the GGOU %
prescription. The
labels indicate the variables and selections used to define the
signal regions in the different analyses
.} \label{fig:GGOU_ADFR}
 \end{center}
\end{figure}

\subsubsection{DGE}
\label{subsec:dge}
Andersen and Gardi (Dressed Gluon Exponentiation, DGE)~\cite{ref:DGE} provide
a framework where the on-shell $b$-quark calculation, converted into hadronic variables, is
directly used as an approximation to the meson decay spectrum without
the use of a leading-power non-perturbative function (or, in other words,
a shape function). 
The DGE calculation uses the $\overline{MS}$ renormalization scheme. The heavy quark parameters determined in $b \to c$ transitions    
from the global fit in the kinetic scheme, published in Ref.~\cite{Finauri:2023kte}, were therefore 
translated into the $\overline{MS}$ scheme by using 
code provided by Einan Gardi (based on Refs.\cite{Blokland:2005uk,Gambino:2008fj}),
giving $m_b({\overline{MS}})=(4.206 \pm 0.040)~\gev$.
The extracted values
of \vub\, for each measurement along with their average are given in
Table~\ref{tab:bulnu} and illustrated in Fig.~\ref{fig:BLNP_DGE}(b).
The total error is $\pm3.3\%$, whose breakdown is:
statistics ($\pm1.5\%$),
detector effects ($\pm1.7\%$),
$\bar{B}\to X_c \ell^- \bar\nu_\ell$ model ($\pm0.7\%$),
$\bar{B}\to X_u \ell^- \bar\nu_\ell$ model ($\pm0.7\%$),
strong coupling $\alpha_s$ ($\pm0.3\%$),
$m_b$ ($\pm2.1\%$),
weak annihilation ($^{+0.0}_{-1.1}\%$),
matching scales in DGE ($^{+0.4}_{-0.5}\%$).
The largest contribution to the total error is due to the effect of the uncertainty 
on $m_b$. 
The uncertainty due to 
weak annihilation has been assumed to be asymmetric, \ie\ it only tends to decrease \vub.

\subsubsection{GGOU}
\label{subsec:ggou}
Gambino, Giordano, Ossola and Uraltsev (GGOU)~\cite{Gambino:2007rp} 
compute the triple differential decay rates of $\bar{B} \to X_u \ell^- \bar\nu_\ell$, 
including all perturbative and non--perturbative effects through $O(\alphas^2 \beta_0)$ 
and $O(1/m_b^3)$. 
The Fermi motion is parameterized in terms of a single lightcone function 
for each structure function and for any value of $q^2$, accounting for all subleading effects. 
The calculations are performed in the kinetic scheme, a framework characterized by a Wilsonian 
treatment with a hard cutoff $\mu \sim 1~\gev$.
GGOU have not included calculations for the ``($E_e,s^{\rm{max}}_h$)'' analysis~\cite{ref:babar-elq2}. 
The heavy quark parameters determined in $b \to c$ transitions    
from the global fit in the kinetic scheme of Ref.~\cite{Finauri:2023kte}, are used as inputs: 
$m_b^{\rm{kin}}=(4.573 \pm 0.012)~\gev$, 
$\mu_\pi^2=(0.454 \pm 0.043)~\gevcc$. 
The extracted values
of \vub\, for each measurement along with their average are given in
Table~\ref{tab:bulnu} and illustrated in Fig.~\ref{fig:GGOU_ADFR}(a).
The total error is $\pm3.9\%$ whose breakdown is:
statistics ($\pm1.4\%$),
detector effects ($\pm1.6\%$),
$\bar{B}\to X_c \ell^- \bar\nu_\ell$ model ($\pm0.9\%$),
$\bar{B}\to X_u \ell^- \bar\nu_\ell$ model ($\pm1.7\%$),
$\alpha_s$, $m_b$ and other non--perturbative parameters ($\pm1.6\%$), 
higher order perturbative and non--perturbative corrections ($\pm1.5\%$), 
modelling of the $q^2$ tail
($\pm1.3\%$), 
weak annihilations matrix element ($^{+0.0}_{-1.0}\%$), 
functional form of the distribution functions ($\pm0.2\%$).  
The leading uncertainties
on  \vub\ are both from theory, and are due to perturbative and non--perturbative
parameters and the modelling of the $q^2$ tail.
The uncertainty due to 
weak annihilation has been assumed to be asymmetric, \ie\ it only tends to decrease \vub.

\subsubsection{Summary}
The averages presented in several different
frameworks %
are presented in 
Table~\ref{tab:vubcomparison}.
In summary, we recognize that the experimental and theoretical uncertainties play out
differently between the schemes and the theoretical assumptions for the
theory calculations are different. Therefore, it is difficult to perform an average 
between the various determinations of \vub. 
Since the methodology is similar to that used to determine the 
inclusive \vcb\ average, we choose to quote as reference value the average determined 
by the GGOU calculation, which gives \vub $= (4.06 \pm 0.12 \pm 0.11) \times 10^{-3}$. 

\begin{table}[!htb]
\caption{\label{tab:vubcomparison}
Summary of inclusive determinations of $\vub$.
The errors quoted on \vub\ correspond to experimental and theoretical uncertainties.
}
\begin{center}
\begin{small}
\begin{tabular}{|lc|}
\hline
Framework
&  $\Vub [10^{-3}]$\\
\hline\hline
BLNP
& $4.25 \pm 0.13 {^{+0.19}_{-0.19}}$ \\ 
DGE
& $3.87 \pm 0.10 ^{+0.09}_{-0.10}$ \\
GGOU
& $4.06 \pm 0.12 ^{+0.11}_{-0.11}$ \\
\hline
\end{tabular}\\
\end{small}
\end{center}
\end{table}

\mysubsection{Combined extraction of $\Vub$ and $\Vcb$}

In this section we report the result of a 
combined fit for \Vub and \Vcb that includes the constraint from the averaged $\Vub/\Vcb$, and the determination of \Vub and \Vcb from exclusive $B$ meson decays.

The average of the $\Vub/\Vcb$ measurements from $\Lb\to p\mu^-\bar\nu_\ell$ and $\bar{B}_s^0\to K\mu^-\bar\nu_\ell$, using only results at high $q^2$ (based on LQCD), assuming the uncertainties due to trigger selection and tracking efficiency are fully correlated, is 
\begin{align}
    \dfrac{\Vub}{\Vcb}=0.0823\pm \, 0.0035
\end{align}

\noindent where the reported uncertainty includes both experimental and theoretical contributions. 
For the $\bar{B}_s^0\to K\mu^-\bar\nu_\ell$ measurements, we use only the result in the high $q^2$ region where various calculations of the form factors from LQCD are already available. Currently, there is only a single calculation based on LCSR at low $q^2$ and others are underway. Once the discrepancy between the two $q^2$ regions is resolved, utilizing both regions will give the best estimation of $\Vub/\Vcb$. 

The average of the \Vcb results from $\bar{B}\to D\ell^-\bar\nu_\ell$, $\bar{B}\to D^*\ell^-\bar\nu_\ell$ and $\bar{B}_s^0\to D_s^{(*)}\mu^-\bar\nu_\mu$, is 
\begin{align}
    \Vcb=(39.62 \pm\, 0.47) \times 10^{-3},
\end{align}
\noindent where the uncertainty also in this case includes both experimental and theoretical contributions. 

The combined fit for \Vub and \Vcb results in 
\begin{align}
 \Vub & = \left( 3.43 \pm 0.12 \right) \times 10^{-3}\,  \\
 \Vcb  & = \left( 39.77 \pm 0.46 \right) \times 10^{-3} \, \\
\rho(\Vub,\Vcb) & =0.239\,,
\end{align}
\noindent where the uncertainties in the inputs are considered uncorrelated. The $\chi^2$ of the fit is $3.9$ for $1$ d.o.f., corresponding to a $P(\chi^2)$ of 4.8\%. 
The fit result is shown in Fig.~\ref{fig:vubvc}. The difference of \Vub from the GGOU inclusive
result, $(4.06\pm 0.12^{+0.11}_{-0.11})\times 10^{-3}$ (Table~\ref{tab:bulnu}), taken as reference, is about $3\sigma$. Also \Vcb differs from the result for inclusive \Vcb from the global fit reported in Ref.~\cite{Finauri:2023kte}, $(41.97\pm 0.48)\times 10^{-3}$, by more than $3\sigma$.
\begin{figure} 
\centering
\includegraphics[width=1.0\textwidth]{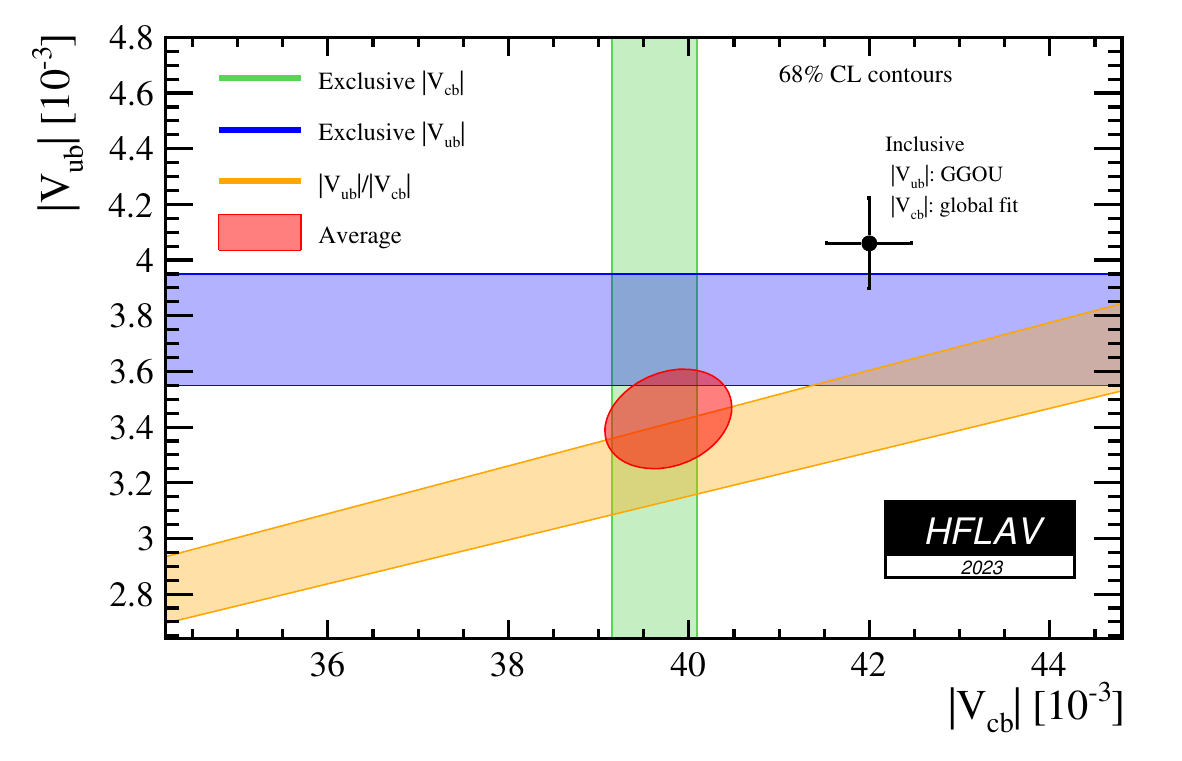}
 \caption{Combined average on \Vub and \Vcb including the LHCb measurements of $\Vub/\Vcb$ from $\Lb \to p\mu^-\bar\nu_\mu$ and $\bar{B}_s^0 \to K \mu^-\bar\nu_\mu$ decays, the exclusive $\Vub$ measurement from $\bar{B}\to \pi\ell^-\bar\nu_\ell$, and the \Vcb average from $\bar{B}\to D\ell^-\bar\nu_\ell$,  $\bar{B}\to D^*\ell^-\bar\nu_\ell$ and $\bar{B}_s^0\to D_s^{(*)}\mu^-\bar\nu_\mu$ measurements.  The point with the error bars corresponds to the inclusive \Vcb from from Ref.\cite{Finauri:2023kte}, and the  inclusive \Vub from GGOU calculation (Sec. \ref{subsec:ggou}). \label{fig:vubvc}}
\end{figure}

\mysubsection{$B\to D^{(*)}\tau \nu_\tau$ decays}
\label{slbdecays_b2dtaunu}

In the SM, the semileptonic decays are tree level processes which proceed via the coupling to the $W^{\pm}$ boson.
These couplings are assumed to be universal for all leptons and are well understood theoretically, (see Section 5.1 and 5.2).
This universality has been tested in purely leptonic and semileptonic $B$ meson decays involving a $\tau$ lepton, which might 
be sensitive to a hypothetical charged Higgs boson or other non-SM processes.

Compared to $B^-\to\tau^-\bar\nu_\tau$, the $B\to D^{(*)}\tau^- \bar\nu_\tau$ decay has advantages: the branching fraction is 
relatively high, because it is not Cabibbo-suppressed, and it is a three-body decay allowing access to many 
observables besides the branching fraction, such as $D^{(*)}$ momentum, $q^2$ distributions, and measurements of the 
$D^*$ and $\tau$ polarisations.

Experiments have measured two ratios of branching fractions defined as 
\begin{eqnarray}
{\cal R}(D)&=&\dfrac{ {\cal B}(\bar{B}\to D\tau^-\bar\nu_\tau) }{ {\cal B}(\bar{B}\to D\ell^-\bar\nu_\ell) },\\
{\cal R}(D^*)&=&\dfrac{ {\cal B}(\bar{B}\to D^*\tau^-\bar\nu_\tau) }{ {\cal B}(\bar{B}\to D^*\ell^-\bar\nu_\ell) } %
\end{eqnarray}
where $\ell$ refers either to electron or $\mu$. These ratios are independent of  $|V_{cb}|$ and to a large extent, also of 
the $B\to D^{(*)}$ form factors. As a consequence, the SM predictions for these ratios are quite precise:

\begin{itemize}
\item ${\cal R}(D)=0.296\pm 0.004$: where the central value and the uncertainty are obtained from an arithmetic average of the predictions from Refs.\cite{Bigi:2016mdz, Bordone:2019vic, Martinelli:2021onb, Bernlochner:2022ywh, Ray:2023xjn, FlavourLatticeAveragingGroupFLAG:2021npn}.  
All these predictions are based on calculations of form factors for $B\to D\ell\nu$ at non-zero recoil from FNAL/MILC~\cite{MILC:2015uhg} and HPQCD~\cite{Na:2015kha}. The FLAG Collaboration, has averaged these two calculations and provides a result of ${\cal R}(D)=0.2934\pm 0.0053$ with only lattice inputs and ${\cal R}(D)=0.2951\pm 0.0031$ using also experimental inputs from B-factories \cite{FlavourLatticeAveragingGroupFLAG:2021npn}. The latter is used in the prediction reported above.  
Some calculations use inputs from LCSR calculation at low $q^2$,  as in Refs.~\cite{Bordone:2019vic} and \cite{Ray:2023xjn}.  The prediction from Ref.\cite{Martinelli:2021onb} is based only on a re-analysis of the LQCD calculations. 

\item ${\cal R}(D^*)=0.254\pm 0.005$: where the central value and the uncertainty are obtained from an arithmetic average of the predictions from Refs.~\cite{BaBar:2019vpl,Bordone:2019vic, Gambino:2019sif, Martinelli:2021myh,Bernlochner:2022ywh,  Ray:2023xjn}.
These calculations are in good agreement with each other,  and consistent with older predictions.  They are based on  Lattice QCD calculations at non-zero recoil recently available, LCSR calculations at low $q^2$ and experimental inputs from Belle    \cite{Belle:2018ezy}.  Some predictions are the result of a combined fit of both ${\cal R}(D)$ and ${\cal R}(D^*)$,  like Ref.  \cite{Bordone:2019vic, Bernlochner:2022ywh, Ray:2023xjn}.  
The authors of Refs.\cite{Bordone:2019vic} and \cite{Ray:2023xjn}, obtain predictions with and without using experimental inputs.  In this average we use their results based on LQCD and LCSR only. 
The calculation in Ref.\cite{BaBar:2019vpl} is the result of the full angular analysis of $\bar{B}\to D^*\ell^-\bar\nu_\ell$ decay by \babar, and gives an independent prediction of ${\cal R}(D^*)=0.253\pm 0.005$, which is compatible with the predictions above.
\end{itemize}

\noindent 
The list of calculations included in the average SM prediction, are summarized in Table~\ref{tab:dtaunu_sm}.
\begin{table}[!htb]
\caption{Predictions for ${\cal R}(D^*)$ and ${\cal R}(D)$. Some calculations extract ${\cal R}(D^*)$ and ${\cal R}(D)$ from global fit, and provide
also a correlations, but these are ignored here.}
\begin{center}
\begin{tabular}{l|c|c}\hline
Input  &${\cal R}(D^*)$ & ${\cal R}(D)$ \\

\hline\hline 

D. Bigi,  P.  Gambino  ~\cite{Bigi:2016mdz}  &                                                             & $0.299 \pm0.003$  \\
\hspace{3mm} \textit{using:} FNAL/MILC~\cite{MILC:2015uhg}, HPQCD~\cite{Na:2015kha}, \belle~\cite{Glattauer:2015teq} & & \\
\hline
M.  Bordone,  M.  Jung, D.  van Dyk~\cite{Bordone:2019vic}      &$0.247 \pm0.006$  &$0.298 \pm0.003$    \\
\hspace{3mm} \textit{using:} FNAL/MILC, HPQCD~\cite{MILC:2015uhg,Na:2015kha,Aoki:2016frl}, & & \\
\hspace{3mm} QCDSR~\cite{Neubert:1992wq,Neubert:1992pn,Ligeti:1993hw}, LCSR~\cite{Gubernari:2018wyi} & & \\
\hline
G.  Martinelli, S.  Simula, L. Vittorio  ~\cite{Martinelli:2021onb}   &                               & $0.296 \pm0.008$   \\
\hspace{3mm} \textit{using:} FNAL/MILC~\cite{MILC:2015uhg} & & \\
\hline 
\end{tabular}
\end{center}
\label{tab:dtaunu_sm}
\end{table}

The first unquenched LQCD calculation of the $\bar{B}\to D^*\ell^-\bar\nu_\ell$ form factors at non-zero recoil was obtained by FNAL/MILC Collaboration and it predicts a value of ${\cal R}(D^*)=0.265\pm 0.013$ \cite{FermilabLattice:2021cdg}. More recently HPQCD and JLQCD collaborations released further calculations obtaining the predictions ${\cal R}(D^*)=0.273\pm 0.015$ \cite{Harrison:2023dzh} and ${\cal R}(D^*)=0.252\pm 0.022$ \cite{Aoki:2023qpa} respectively. These predictions, based only on LQCD, reduce the tensions with the experimental average, even if the larger uncertainty alleviates its significance. It is worth to mention that the results of joint fits between LQCD and experimental data are slightly in tension with these LQCD-only calculations. The result obtained by FNAL/MILC is ${\cal R}(D^*)=0.2492\pm 0.0012$ \cite{FermilabLattice:2021cdg}, while HPQCD obtained ${\cal R}(D^*)=0.2482\pm 0.0020$ \cite{Harrison:2023dzh}. 
Ongoing and future calculations of the form factors for $B\to D^{*}$ transitions, will provide higher precision.

On the experimental side, in the case of the leptonic $\tau$ decay, the ratios  ${\cal R}(D^{(*)})$ can be 
directly measured, and many systematic uncertainties cancel in the measurement. 
The $\bar{B}^0\to D^{*+}\tau^-\bar\nu_\tau$ decay was first observed by Belle~\cite{Matyja:2007kt} performing 
an "inclusive" reconstruction, which is based on the reconstruction of the $B_{\rm{tag}}$ from all the particles
of the events, other than the $D^{(*)}$ and the lepton candidate, without looking for any specific $B_{\rm{tag}}$ decay chain.  
Since then, both \babar and Belle have published improved measurements 
and have observed the $B^-\to D\tau^-\bar\nu_\tau$ decays~\cite{Aubert:2007dsa,Bozek:2010xy}.

The most powerful way to study these decays at the B-Factories exploits the hadronic or semileptonic $B_{\rm{tag}}$.
Using the full dataset and an improved hadronic $B_{\rm{tag}}$ selection, \babar measured~\cite{Lees:2012xj} ${\cal R}(D)=0.440\pm 0.058\pm 0.042$ and ${\cal R}(D^*)=0.332\pm 0.024\pm 0.018$, where decays to both $e^\pm$ and $\mu^\pm$ were summed, and results for $B^0$ and $B^-$ decays were combined in an isospin-constrained fit. 
Since then other measurements have been performed. Here we report the list of measurements included in the average, where measurements from the same experiment are identified by unique superscript characters:

\begin{description}

\item[BaBar] $R(D)$ and $R(D^\ast)$ with $\tau$ reconstructed in the $\tau^-\to \ell^- \bar{\nu}_{\ell} \nu_\tau$ mode ($\ell \in \{e, \mu\}$), and using the hadronic $B$-tagging approach, \cite{Lees:2012xj,Lees:2013uzd};

\item[Belle$^a$] $R(D)$ and $R(D^\ast)$ with $\tau$ reconstructed in the $\tau^-\to \ell^-  \bar{\nu}_{\ell} \nu_\tau$ mode ($\ell \in \{e, \mu\}$), and using the hadronic $B$-tagging approach, \cite{Huschle:2015rga};

\item[Belle$^b$] $R(D^\ast)$ and $\tau$ polarization with the $\tau$ reconstructed in hadronic $\tau^-\to\pi^-(\pi^0)\nu_\tau$ decay mode, and using the hadronic $B$-tagging, \cite{Hirose:2016wfn};

\item[Belle$^c$] $R(D)$ and $R(D^\ast)$ with $\tau^-\to\ell^-  \bar{\nu}_{\ell} \nu_\tau$, using the semileptonic B-tagging \cite{Belle:2019rba} (this measurement supersedes \cite{Sato:2016svk}, which measured only $R(D^\ast)$ using the same tagging approach);

\item[LHCb$^a$]  $R(D)$ and $R(D^\ast)$ with the $\tau^-$ reconstructed in $\tau^-\to \mu^-  \bar{\nu}_{\mu} \nu_\tau$ mode\cite{LHCb:2023zxo} (this measurement supersedes \cite{Aaij:2015yra}, which measured only $R(D^\ast)$ using the same data and similar analysis approach). 
 
\item[LHCb$^b$] $R(D^\ast)$ with $\tau^-$ reconstructed in the hadronic $\tau^-\to 3\pi^-(\pi^0)\nu_\tau$ mode, obtained combining the result from Ref.\cite{Aaij:2017uff} (based on Run 1 dataset) and the most recent from Ref.\cite{LHCb:2023uiv} (based on part of Run 2 dataset). These are direct measurements of the ratio $K(D^\ast)={\cal B}(\bar{B}^0 \to
D^\ast\tau^-\bar\nu_\tau )/{\cal B}(\bar{B}^0 \to D^\ast\pi^+\pi^-\pi^+)$, which is translated into a measurement of $R(D^\ast)$ using the
independently measured branching fractions of ${\cal B}(\bar{B}^0 \to D^\ast \pi^+\pi^-\pi^+)$ and ${\cal B}(\bar{B}^0 \to D^\ast \mu^-\bar\nu_\mu)$. 

\item[LHCb$^c$]  $R(D^+)$ and $R(D^{+\ast})$ with the $\tau^-$ reconstructed in $\tau^-\to \mu^-  \bar{\nu}_{\mu} \nu_\tau$ mode\cite{LHCb:2024jll}. 

 \item[Belle II] $R(D^\ast)$  with $\tau^-\to \ell^- \bar{\nu}_{\ell} \nu_\tau$, using the hadronic $B$-tagging  \cite{Belle-II:2024ami}.
\end{description}

\noindent The most important source of systematic uncertainties that are correlated among the different measurement  
is the $\bar{B}\to D^{**}$ background components, which are difficult to disentangle from the signal. In our average, the systematic uncertainties due to the $\bar{B}\to D^{**}$ composition and kinematics are considered fully correlated among the measurements. 

The results of the individual measurements, their averages and correlations are presented in Table \ref{tab:dtaunu} and Fig.\ref{fig:rds}. 
The combined results, projected separately on ${\cal R}(D)$ 
and ${\cal R}(D^*)$, are reported in Fig.\ref{fig:rd}(a) and  Fig.\ref{fig:rd}(b) respectively. 

The averaged ${\cal R}(D)$ and ${\cal R}(D^*)$ exceed the SM prediction given above, by 2.2$\sigma$ and 1.9$\sigma$, respectively. 
Considering the ${\cal R}(D)$ and ${\cal R}(D^*)$ total correlation of $-0.39$, the difference with respect to the 
SM is about 3.14$\sigma$, and the combined $\chi^2=12.78$ for 2 degrees of freedom corresponds to a $p$-value of $1.68\times 10^{-3}$, assuming Gaussian error distributions. 

An analogous measurement using $B^-_c\to J/\psi \mu^-\bar\nu_\mu$ decays has been performed by LHCb, leading to $R(J\psi)=0.71\pm 0.17_{\rm stat}\pm 0.18_{\rm syst}$ \cite{LHCb:2017vlu}, which lies $1.8\sigma$ above the most recent SM prediction obtained by HPQCD collaboration \cite{Harrison:2020nrv}. 
Recently LHCb reported the first observation of the $\Lambda_b^0\to \Lambda_c^+\tau^-\bar\nu_\tau$ decay \cite{LHCb:2022piu}, exploiting the three-prong hadronic $\tau^-$ decays. The resulting ratio of semileptonic branching fractions is ${\cal R}(\Lambda_c)=0.242\pm 0.026_{\rm stat}\pm 0.040_{\rm syst}\pm 0.059_{\rm ext}$, where the last term is due to the uncertainties on the external branching fractions measurement, in particular for the $\Lambda_b^0\to \Lambda_c^+\mu^-\bar\nu_\mu$ decay. This result is in agreement with the prediction of ${\cal R}(\Lambda_c)=0.324\pm 0.004$ from Ref.\cite{Bernlochner:2018bfn}. 
The first direct measurement of the ratio of inclusive semileptonic decays ${\cal R}(X_{\tau/\ell})={\cal B}(\bar{B}\to X\tau^-\bar\nu_\tau)/{\cal B}(\bar{B}\to X\ell^-\bar\nu_\ell)$ has been determined by Belle II \cite{Belle-II:2023aih}. The result, ${\cal R}(X_{\tau/\ell})=0.228\pm 0.016_{\rm stat}\pm 0.036_{\rm syst}$, is compatible with SM predictions,\cite{Ligeti:2021six,Rahimi:2022vlv}.
\begin{table}[!htb]
\caption{Measurements of ${\cal R}(D^*)$ and ${\cal R}(D)$, their correlations and the combined average. 
The superscript character is used to distinguish the measurements taken from the same experiment.}
\begin{center}
\resizebox{0.99\textwidth}{!}{
\begin{tabular}{l|c|c|c}\hline
Experiment  &${\cal R}(D^*)$ & ${\cal R}(D)$ & $\rho$ \\

\hline\hline 
\babar ~\cite{Lees:2012xj,Lees:2013uzd} &$0.332 \pm0.024_{\rm stat} \pm0.018_{\rm syst}$  &$0.440 \pm0.058_{\rm stat} \pm0.042_{\rm syst}$ & $-0.27$\\
Belle$^a$  ~\cite{Huschle:2015rga}         &$0.293 \pm0.038_{\rm stat} \pm0.015_{\rm syst}$  &$0.375 \pm0.064_{\rm stat} \pm0.026_{\rm syst}$ & $-0.49$ \\
Belle$^b$  ~\cite{Hirose:2016wfn}          &$0.270 \pm0.035_{\rm stat} \, {^{+0.028} _{-0.025}}_{\rm syst}$  & & \\
Belle$^c$   ~\cite{Belle:2019rba}            &$0.283 \pm0.018_{\rm stat} \pm0.014_{\rm syst}$  &   $0.307 \pm0.037_{\rm stat} \pm0.016_{\rm syst}$  & -0.51 \\
LHCb$^a$    ~\cite{LHCb:2023zxo}            &$0.281 \pm0.018_{\rm stat} \pm0.024_{\rm syst}$  & $0.441 \pm0.060_{\rm stat} \pm0.066_{\rm syst}$  & -0.43 \\
LHCb$^b$    ~\cite{Aaij:2017uff,Aaij:2017deq, LHCb:2023uiv} &$0.264 \pm0.011_{\rm stat} \pm0.018_{\rm syst}$  &   &\\
LHCb$^c$    ~\cite{LHCb:2024jll}            &$0.402 \pm0.081_{\rm stat} \pm0.085_{\rm syst}$  & $0.249 \pm0.043_{\rm stat} \pm0.047_{\rm syst}$  & -0.39 \\
Belle II ~\cite{Belle-II:2024ami}          &$0.262 \, {^{+0.041}_{-0.039}}_{\rm stat} \, {^{+0.028} _{-0.025}}_{\rm syst}$  & & \\
\hline
{\bf Average} &\mathversion{bold}$0.286 \pm0.012$ & \mathversion{bold}$0.342 \pm0.026$ & $-0.39$  \\
\hline 
\end{tabular}
}
\end{center}
\label{tab:dtaunu}
\end{table}

\begin{figure}
\centering
\includegraphics[width=1.0\textwidth]{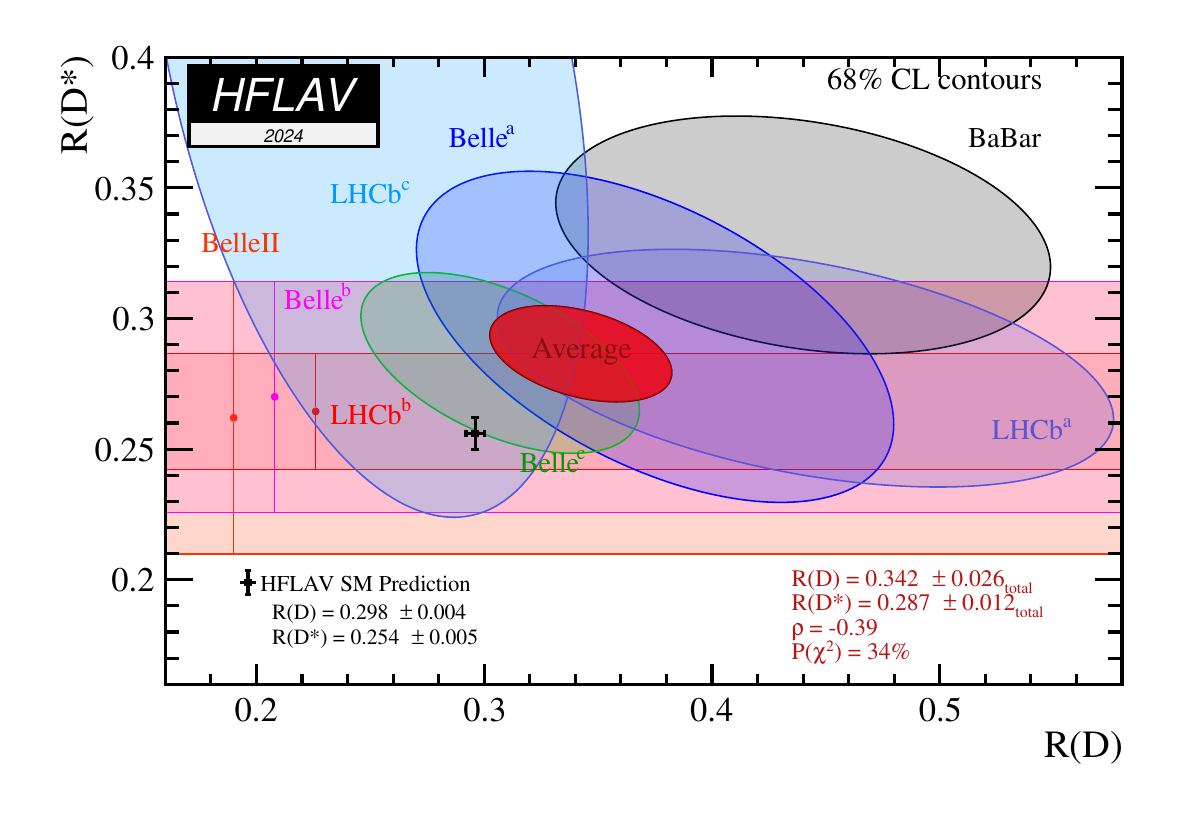}
\caption{Measurements of ${\cal R}(D)$ and ${\cal R}(D^*)$ listed in Table \ref{tab:dtaunu} and their two-dimensional average. The measurements from the same experiment are identified by the unique superscript characters as defined in the text and in Table.\ref{tab:dtaunu}. Contours correspond to 68\% CL ({\it i.e.} $\Delta \chi^2 = 2.3$). The black point with error bars, is the arithmetic average of the SM prediction for ${\cal R}(D^*)$ and ${\cal R}(D)$ reported in Table~\ref{tab:dtaunu_sm}. More information is given in the text. 
An average of these predictions and the experimental average deviate from each other by about 3.14$\sigma$. 
\label{fig:rds}}
\end{figure}

\begin{figure}[!ht]
  \begin{center}
  \unitlength 1.0cm %
  \begin{picture}(14.,11.0)
    \put(  -1.5, 0.0){\includegraphics[width=9.0cm]{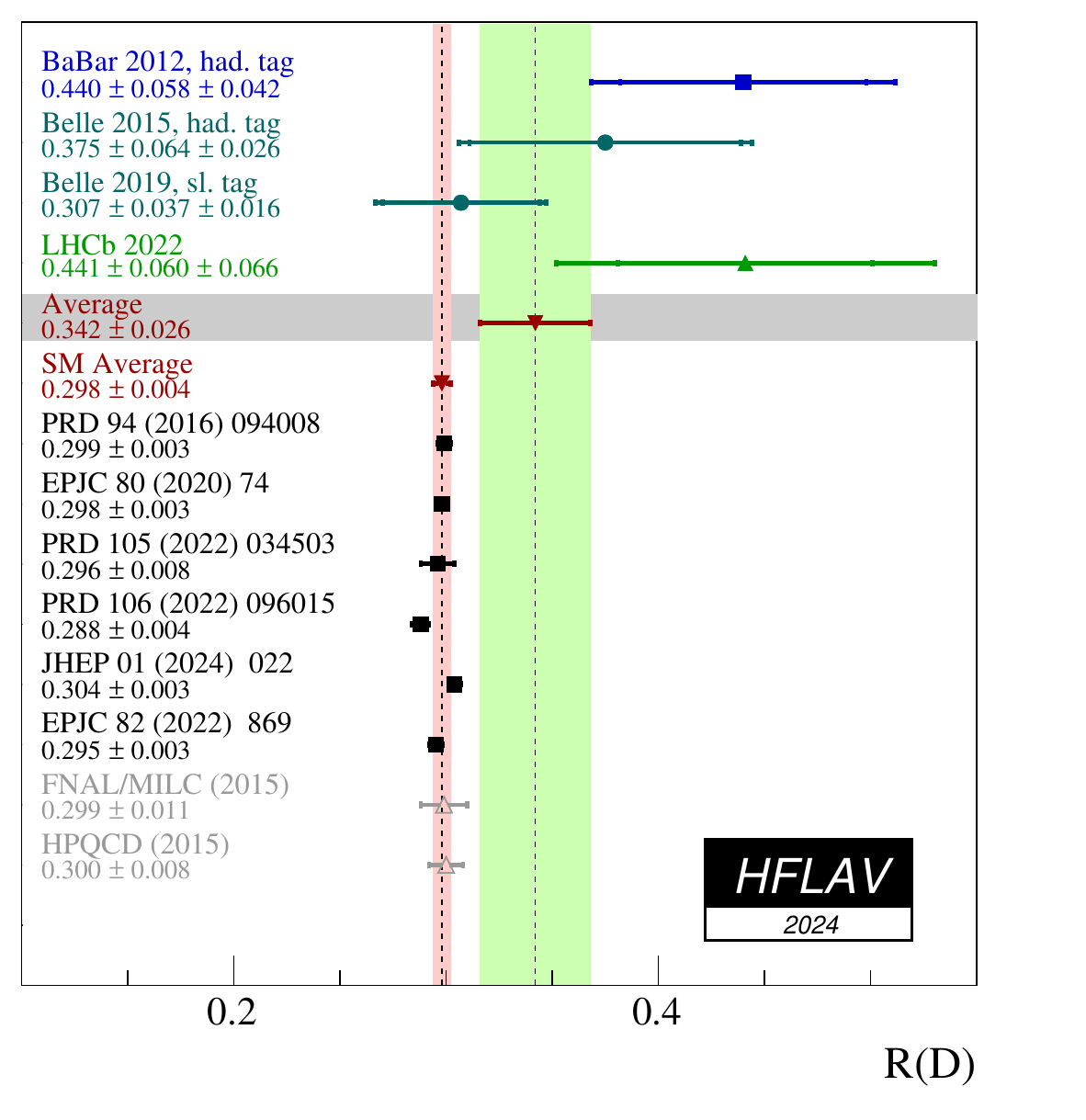}
    }
    \put(  7.5, 0.0){\includegraphics[width=9.0cm]{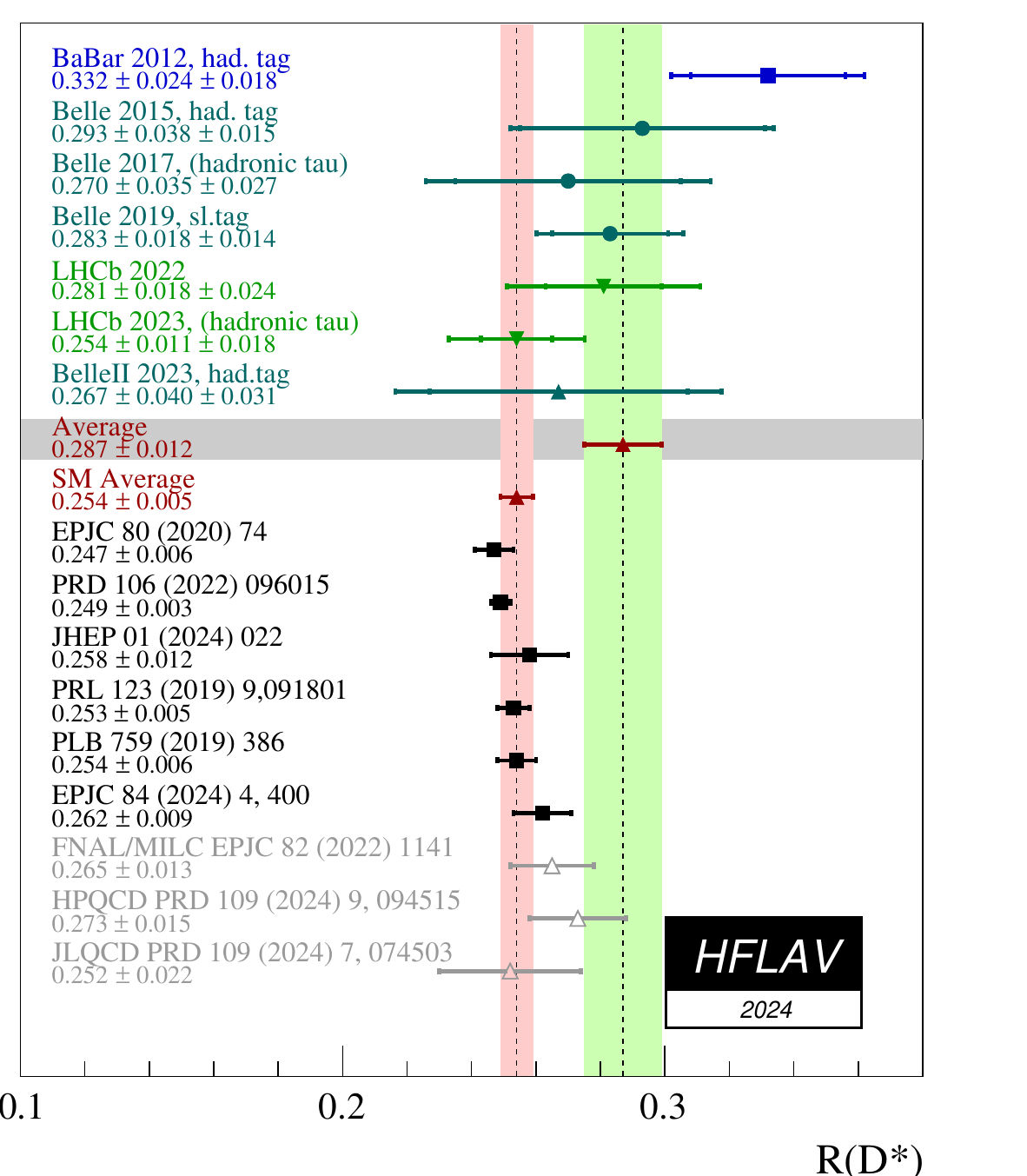}
    }
    \put(  5.8,  10.5){{\large\bf a)}}  
    \put( 14.7,  10.5){{\large\bf b)}}
  \end{picture}
  \caption{(a) Measurements of ${\cal R}(D)$ and (b) ${\cal R}(D^*)$. The green (right) vertical bands are the averages
  obtained from the combined fit. The red (left) vertical bands are the averages of the theoretical predictions obtained as explained in the text. The theory predictions represented in grey, are based on LQCD only and are shown for comparison but not included in the SM averages.} 
 \label{fig:rd}
 \end{center}
\end{figure}

\clearpage
\section{Decays of $b$-hadrons into open or hidden charm hadrons}
\label{sec:b2c}
Ground-state $B$ mesons and $b$ baryons dominantly decay to particles containing a charm quark via the $b \rightarrow c$ quark transition.
In this section, measurements of such decays to hadronic final states are summarized. %

Since hadronic $b\to c$ decays dominate the $b$-hadron widths, they are an important part of the experimental programme in heavy-flavour physics.
Hadronic $b\to c$ decays are often used, in particular at hadron colliders, as normalization modes for
measurements of rarer decays.
At B-factories, hadronic $b\to c$ decays are used for the tagging of $B$ mesons and a detailed knowledge is crucial for the optimization and calibration of the tagger performance.
In addition, they are the dominant background in many analyses.
To accurately model such backgrounds with simulated data, it is essential to have precise knowledge of the contributing decay modes.
In particular, with the expected increase in the data samples at LHCb and
Belle~II, the enhanced statistical sensitivity has to be matched by low
systematic uncertainties that arise from the limited understanding of the dominant $b$-hadron decay
modes. 
For multibody decays, knowledge of the distribution of decays across the
phase-space (\eg,\ the Dalitz plot density for three-body decays or the
polarization amplitudes for vector-vector final states) is required in
addition to the total branching fraction.

The large branching fractions of $b \to c$ decays 
make them ideal for studying the spectroscopy of both open and hidden charm hadrons.
In particular, they have been used to both discover and measure the properties
of exotic particles, such as the $X(3872)$~\cite{Choi:2003ue,Aaij:2015eva},
$Z(4430)^+$~\cite{Choi:2007wga,Aaij:2014jqa} and $P_c(4450)^+$~\cite{Aaij:2015tga} states.

In addition to the dominant $b \to c$ decays, there are several decays in this category that are expected
to be highly suppressed in the Standard Model.
These are of interest for probing particular decay amplitudes (\eg,\ the annihilation diagram,
which dominates the $B^- \to \Dsm \phi$ decay)
used to constrain effects in other hadronic decays, or in the search for new physics.
There are also open charm production modes that are dominated by $b \to u$ transitions, such as $\Bzb \to \Dsm \pip$, which are
mediated by the $W$ emission involving the $|V_{ub}|$ CKM matrix element.
Finally, $b \to c$ decays involving lepton flavour or number violation, such as $\Bp \to \Dm \ellp \ellp$, are
extremely suppressed in the Standard Model, and therefore provide highly
sensitive tests of new physics.

In this section, we give an exhaustive list of measured branching ratios of decay modes to
hadrons containing charm quarks.
The averaging procedure follows the methodology described in Chapter~\ref{sec:method}.
We perform fits of the likelihood function constructed from the quoted central values and uncertainties and do not adjust measurements as described in Section~\ref{sec:method:corrSysts}.
For the cases where more than one measurement is available, in total 76 fits are performed, with on average (maximally) 4.3 (188) parameters and 7.2 (303) measurements per fit.
Systematic uncertainties are taken as quoted without the scaling of multiplicative uncertainties  discussed in Section~\ref{sec:method:nonGaussian}.
Where available, correlations between measurements are taken into account.
We consider correlations not only between measurements of the same parameter, but also among parameters.
The correlations among parameters are given on the HFLAV web page on hadronic $B$ decays into open or hidden charm hadrons~\cite{ref:hflav-b2c}.
Limits are not included in the calculation of averages.
In case only limits are available the most stringent one is quoted.
If an insignificant measurement and a limit for the same parameter are provided in the same paper,
the former is quoted, so that it can be included in averages.
We also provide averages of the polarization amplitudes of $B$ meson decays to
vector-vector states. We do not currently provide detailed averages of
quantities obtained from Dalitz plot analyses, due to the complications
arising from the dependence on the model used.

The results are presented in subsections organized according to the type of decaying bottom
hadron: $\Bz$ (Sec.~\ref{sec:b2c:Bd}), $B^+$ (Sec.~\ref{sec:b2c:Bu}), $\Bz/B^+$ admixture (Sec.~\ref{sec:b2c:B}), $\Bs$ (Sec.~\ref{sec:b2c:Bs}), $B_c^+$ (Sec.~\ref{sec:b2c:Bc}), $b$ baryons (Sec.~\ref{sec:b2c:Bbaryon}).
For each subsection, the parameters $\boldsymbol{p}$ are arranged according to the final state
into the following groups: a single charmed meson, two charmed mesons, a
charmonium state, a charm baryon, or other states, e.g., $X(3872)$.
In our tables, the individual measurements and average of each parameter $p_j$ are shown in one row.
We quote numerical values of all direct measurements of a parameter $p_j$.
We also show numerical values derived from measurements of branching-fraction ratios $p_j/p_k$, performed with respect to the branching fraction $p_k$ of a normalization mode, as well as measurements of products $p_j p_k$ of the branching fraction of interest with those of daughter-particle decays.
In these cases, the quoted value and uncertainty of the measurement are determined with the fitted value of $p_k$, and the uncertainty of $p_k$ is included in the systematic uncertainty.
A footnote ``Using $p_k$'' is added in these cases.
Note that the fit uses $p_j/p_k$ or $p_j p_k$ directly and not the $p_j$ value that is quoted in the table.
The $p_j$ value is quoted to give a sense of the contribution of the measurement to the average.
When the measurement depends on $p_j$ in some other way, it is also included in our fit for $p_j$, but in the tables no derived value is shown.
Instead, the measured function $f$ of parameters is given in a footnote ``Measurement of $f$ used in the fit''.
In general we prefer a directly measured function of parameters over a derived parameter if both are quoted because we do not apply corrections for updated input parameters.

In most of the tables of this section the averages are compared to those from the Particle Data Group's 2022 Review of Particle Physics (PDG 2022)~\cite{PDG_2022} and 2023 update.
When this is done, the ‘‘Average'' column quotes the PDG averages in italics only if they differ from ours.
In general, such differences are due to different input parameters and measurements, differences in the averaging methods and different rounding conventions.
The fit $p$-value is quoted if it is below 1\%.
Input values that are not included in the PDG average are either new results published after the closing of PDG and before the closing of this report, September 2023, or results that do not quote a direct measurement of the parameter of interest and are therefore not considered in the PDG average.
Quoted upper limits are at 90\% confidence level (CL), unless mentioned otherwise.

The symbol $\mathcal{B}$ is used for branching ratios, and $f_X$ for the production fraction of hadron $X$ or $B$ meson with quark $X$ (see Section~\ref{sec:fractions_high_energy}).
The decay amplitudes for longitudinal, parallel, and perpendicular transverse polarization in pseudoscalar to vector-vector decays are denoted ${\cal{A}}_0$, ${\cal{A}}_\parallel$, and ${\cal{A}}_\perp$, respectively, and the definitions $\delta_\parallel = \arg({\cal{A}}_\parallel/{\cal{A}}_0)$ and $\delta_\perp = \arg({\cal{A}}_\perp/{\cal{A}}_0)$ are used for their relative phases.
For normalized P-wave amplitudes we use the notation $f_i = |{\cal{A}}_i|^2 / (|{\cal{A}}_0|^2 + |{\cal{A}}_\parallel|^2+ |{\cal{A}}_\perp|^2)$.
The inclusion of charge conjugate modes is always implied.

Following the approach used by the PDG~\cite{PDG_2020}, for decays that involve
neutral kaons we mainly quote results in terms of final states including
either a $\Kz$ or $\Kzb$ meson (instead of a \KS or \KL),
although the flavour of the neutral kaon is never determined experimentally.
The specification as \Kz or \Kzb simply follows the quark model
expectation for the dominant decay
and the inclusion of the conjugate final state neutral kaon is implied.
The exception is \Bs decays to \CP
eigenstates, where the width difference between the mass eigenstates (see
Sec.~\ref{sec:life_mix}) means that the measured branching fraction,
integrated over decay time, is specific to the final
state~\cite{DeBruyn:2012wj}. 
In such cases it is appropriate to quote the branching fraction for, \eg, $\Bsb \to
\jpsi \KS$ instead of $\Bsb \to \jpsi \Kzb$.

Most $B$-meson branching-fraction measurements assume $\Gamma(\Upsilon(4S) \to B^+B^-) = \Gamma(\Upsilon(4S) \to B^0\bar{B}^0)$.
While there is no evidence for isospin violation in $\Upsilon(4S)$ decays,
deviations from this assumption can be of the order of a few percent,
see Section~\ref{sec:fraction_Ups4S}.
As the effect is negligible for many averages, we take the quoted values without applying a correction 
or additional systematic uncertainty. However, we note that this can be relevant
for averages with  percent-level uncertainty.

\newlength{\hflavrowwidth}
\newcommand{\hflavrowdefshort}[4]{\def#1{#2 & #3 & #4}}
\newcommand{\hflavrowdeflong}[4]{\def#1{\multicolumn{3}{Sl}{#2}\\ & #3 & #4}}
\newcommand{\hflavrowdef}[4]{\settowidth{\hflavrowwidth}{\begin{tabular}{| l l l |} \\ #2 & #3 & #4 \\ \end{tabular}}\ifnum \hflavrowwidth>\textwidth \hflavrowdeflong{#1}{#2}{#3}{#4} \else \hflavrowdefshort{#1}{#2}{#3}{#4} \fi}

\subsection{Decays of $B^0$ mesons}
\label{sec:b2c:Bd}
Measurements of $B^0$ decays to charmed hadrons are summarized in Sections~\ref{sec:b2c:Bd_D} to~\ref{sec:b2c:Bd_other}.

\subsubsection{Decays to a single open charm meson}
\label{sec:b2c:Bd_D}
Averages of $B^0$ decays to a single open charm meson are shown in Tables~\ref{tab:b2charm_Bd_D_overview1}--\ref{tab:b2charm_Bd_D_overview9}.
\hflavrowdef{\hflavrowdefaaa}{${\cal B}(B^0 \to D^- \pi^+)$}{{\setlength{\tabcolsep}{0pt}
\begin{tabular}{m{6.5em}l} {BaBar \cite{Aubert:2006cd}\tnote{}\hphantom{\textsuperscript{}}} & { $2.55 \pm0.05 \pm0.16$ } \\ {BaBar \cite{Aubert:2006jc}\tnote{}\hphantom{\textsuperscript{}}} & { $3.03 \pm0.23 \pm0.23$ } \\ \multicolumn{2}{l}{Belle {\cite{Abe:2001waa}\tnote{1}\hphantom{\textsuperscript{1}}}} \\ \multicolumn{2}{l}{CDF {\cite{Abulencia:2005ia}\tnote{*,2}\hphantom{\textsuperscript{*,2}}}, {\cite{Abulencia:2006aa}\tnote{*,3}\hphantom{\textsuperscript{*,3}}}, {\cite{Abulencia:2006df}\tnote{*,4}\hphantom{\textsuperscript{*,4}}}} \\ \multicolumn{2}{l}{LHCb {\cite{Aaij:2011rj}\tnote{5}\hphantom{\textsuperscript{5}}}, {\cite{Aaij:2013qqa}\tnote{1}\hphantom{\textsuperscript{1}}}, {\cite{Aaij:2014jpa}\tnote{6}\hphantom{\textsuperscript{6}}}, {\cite{Aaij:2014jyk}\tnote{*,4}\hphantom{\textsuperscript{*,4}}}} \\ \end{tabular}}}{\begin{tabular}{l} $2.48 \pm0.11$ \\ $\mathit{\scriptstyle 2.51 \pm0.08}$\\ \end{tabular}}
\hflavrowdef{\hflavrowdefaab}{${\cal B}(B^0 \to D^- \pi^+ \pi^+ \pi^-)$}{{\setlength{\tabcolsep}{0pt}\begin{tabular}{m{6.5em}l} {LHCb \cite{Aaij:2011rj}\tnote{7}\hphantom{\textsuperscript{7}}} & { $5.91 \pm0.27 \pm0.59$ } \\ \multicolumn{2}{l}{CDF {\cite{Abulencia:2006aa}\tnote{*,8,9}\hphantom{\textsuperscript{*,8,9}}}} \\ \multicolumn{2}{l}{LHCb {\cite{Aaij:2011rj}\tnote{10}\hphantom{\textsuperscript{10}}}, {\cite{Aaij:2012bw}\tnote{11}\hphantom{\textsuperscript{11}}}} \\ \end{tabular}}}{\begin{tabular}{l} $5.78\,^{+0.62}_{-0.60}$ \\ $\mathit{\scriptstyle 5.98 \pm0.61}$\\ \end{tabular}}
\hflavrowdef{\hflavrowdefaac}{${\cal B}(B^0 \to D^*(2010)^- \pi^+)$}{{\setlength{\tabcolsep}{0pt}\begin{tabular}{m{6.5em}l} {Belle \cite{Belle:2022afp}\tnote{*}\hphantom{\textsuperscript{*}}} & { $2.62 \pm0.02 \pm0.09$ } \\ {Belle \cite{Kuzmin:2006mw}\tnote{*}\hphantom{\textsuperscript{*}}} & { $2.22 \pm0.04 \pm0.19$ } \\ {BaBar \cite{Aubert:2006cd}\tnote{}\hphantom{\textsuperscript{}}} & { $2.79 \pm0.08 \pm0.17$ } \\ {BaBar \cite{Aubert:2006jc}\tnote{}\hphantom{\textsuperscript{}}} & { $2.99 \pm0.23 \pm0.24$ } \\ \multicolumn{2}{l}{BaBar {\cite{BaBar:2005osj}\tnote{*,12}\hphantom{\textsuperscript{*,12}}}} \\ \multicolumn{2}{l}{Belle {\cite{Abe:2001waa}\tnote{*,12}\hphantom{\textsuperscript{*,12}}}} \\ \multicolumn{2}{l}{LHCb {\cite{Aaij:2013xca}\tnote{13}\hphantom{\textsuperscript{13}}}} \\ \end{tabular}}}{\begin{tabular}{l} $2.652 \pm0.064$ \\ $\mathit{\scriptstyle 2.738 \pm0.129}$\\ \end{tabular}}
\hflavrowdef{\hflavrowdefaad}{${\cal B}(B^0 \to D^*(2010)^- \pi^+ \pi^+ \pi^-)$}{{\setlength{\tabcolsep}{0pt}\begin{tabular}{m{6.5em}l} {LHCb \cite{Aaij:2013xca}\tnote{14}\hphantom{\textsuperscript{14}}} & { $7.00 \pm0.11 \pm0.38$ } \\ {BaBar \cite{TheBABAR:2016vzj}\tnote{}\hphantom{\textsuperscript{}}} & { $7.26 \pm0.11 \pm0.31$ } \\ {Belle \cite{Majumder:2004su}\tnote{}\hphantom{\textsuperscript{}}} & { $6.81 \pm0.23 \pm0.72$ } \\ \multicolumn{2}{l}{LHCb {\cite{Aaij:2013xca}\tnote{15,16}\hphantom{\textsuperscript{15,16}}}} \\ \end{tabular}}}{\begin{tabular}{l} $7.11 \pm0.24$ \\ $\mathit{\scriptstyle 7.21 \pm0.29}$\\ \end{tabular}}
\hflavrowdef{\hflavrowdefaae}{${\cal B}(B^0 \to \bar{D}^*(2007)^0 \pi^+ \pi^+ \pi^- \pi^-)$}{{\setlength{\tabcolsep}{0pt}\begin{tabular}{m{6.5em}l} {Belle \cite{Majumder:2004su}\tnote{}\hphantom{\textsuperscript{}}} & { $2.6 \pm0.5 \pm0.4$ } \\ \end{tabular}}}{\begin{tabular}{l} $2.60 \pm0.60$ \\ $\mathit{\scriptstyle 2.72 \pm0.50}$\\ \end{tabular}}
\hflavrowdef{\hflavrowdefaaf}{${\cal B}(B^0 \to D^*(2010)^- \pi^+ \pi^+ \pi^+ \pi^- \pi^-)$}{{\setlength{\tabcolsep}{0pt}\begin{tabular}{m{6.5em}l} {Belle \cite{Majumder:2004su}\tnote{}\hphantom{\textsuperscript{}}} & { $4.72 \pm0.59 \pm0.71$ } \\ \end{tabular}}}{\begin{tabular}{l} $4.72 \pm0.92$\\ \end{tabular}}
\hflavrowdef{\hflavrowdefaag}{${\cal B}(B^0 \to D^*(2010)^- \omega(782) \pi^+)$}{{\setlength{\tabcolsep}{0pt}\begin{tabular}{m{6.5em}l} {Belle \cite{Matvienko:2015gqa}\tnote{}\hphantom{\textsuperscript{}}} & { $2.31 \pm0.11 \pm0.14$ } \\ {BaBar \cite{Aubert:2006zb}\tnote{}\hphantom{\textsuperscript{}}} & { $2.88 \pm0.21 \pm0.31$ } \\ \end{tabular}}}{\begin{tabular}{l} $2.41 \pm0.16$ \\ $\mathit{\scriptstyle 2.46 \pm0.18}$\\ \end{tabular}}
\begin{table}[H]
\begin{center}
\begin{threeparttable}
\caption{Branching fractions for decays to a $D^{(*)}$ meson and one or more pions, I.}
\label{tab:b2charm_Bd_D_overview1}
%
 \\
\hline
\hflavrowdefaaa \\
 & & \\[-1ex]
\hflavrowdefaab \\
 & & \\[-1ex]
\hflavrowdefaac \\
 & & \\[-1ex]
\hflavrowdefaad \\
 & & \\[-1ex]
\hflavrowdefaae \\
 & & \\[-1ex]
\hflavrowdefaaf \\
 & & \\[-1ex]
\hflavrowdefaag \\
\hline
\hline
\end{tabular}
\begin{tablenotes}
\small
\item[*] Not included in PDG average.
\item[1] Measurement of ${\cal B}(B^0 \to D^- K^+) / {\cal B}(B^0 \to D^- \pi^+)$ used in our fit.
\item[2] Measurement of ${\cal B}(B^+ \to \bar{D}^0 \pi^+) / {\cal B}(B^0 \to D^- \pi^+)$ used in our fit.
\item[3] Measurement of ${\cal B}(B_s^0 \to D_s^- \pi^+) / {\cal B}(B^0 \to D^- \pi^+)$ used in our fit.
\item[4] Measurement of ${\cal B}(\Lambda_b^0 \to \Lambda_c^+ \pi^-) / {\cal B}(B^0 \to D^- \pi^+)$ used in our fit.
\item[5] Measurement of ${\cal B}(B^0 \to D^- \pi^+ \pi^+ \pi^-) / {\cal B}(B^0 \to D^- \pi^+)$ used in our fit.
\item[6] Measurement of ${\cal B}(B^0 \to D_s^- K^+) / {\cal B}(B^0 \to D^- \pi^+)$ used in our fit.
\item[7] Using ${\cal B}(B^0 \to D^- \pi^+)$.
\item[8] Using $f_s/f_d = 0.259 \pm 0.038$ from PDG 2006.
\item[9] Measurement of ${\cal B}(B_s^0 \to D_s^- \pi^+ \pi^+ \pi^-) / {\cal B}(B^0 \to D^- \pi^+ \pi^+ \pi^-)$ used in our fit.
\item[10] Measurement of ${\cal B}(B^0 \to D_1(2420)^- \pi^+) \times {\cal B}(D_1(2420)^+ \to D^+ \pi^+ \pi^-) / {\cal B}(B^0 \to D^- \pi^+ \pi^+ \pi^-)$ used in our fit.
\item[11] Measurement of ${\cal B}(B^0 \to D^- K^+ \pi^+ \pi^-) / {\cal B}(B^0 \to D^- \pi^+ \pi^+ \pi^-)$ used in our fit.
\item[12] Measurement of ${\cal B}(B^0 \to D^*(2010)^- K^+) / {\cal B}(B^0 \to D^*(2010)^- \pi^+)$ used in our fit.
\item[13] Measurement of ${\cal B}(B^0 \to D^*(2010)^- \pi^+ \pi^+ \pi^-) / {\cal B}(B^0 \to D^*(2010)^- \pi^+)$ used in our fit.
\item[14] Using ${\cal B}(B^0 \to D^*(2010)^- \pi^+)$.
\item[15] Measurement of ${\cal B}(B^0 \to D^*(2010)^- K^+ \pi^+ \pi^-) / {\cal B}(B^0 \to D^*(2010)^- \pi^+ \pi^+ \pi^-)$ used in our fit.
\item[16] Measurement of ${\cal B}(B^0 \to \bar{D}_1(2420)^0 \pi^+ \pi^-) \times {\cal B}(D_1(2420)^0 \to D^*(2010)^+ \pi^-) / {\cal B}(B^0 \to D^*(2010)^- \pi^+ \pi^+ \pi^-)$ used in our fit.
\end{tablenotes}
\end{threeparttable}
\end{center}
\end{table}

\hflavrowdef{\hflavrowdefaah}{${\cal B}(B^0 \to \bar{D}^0 \pi^0)$}{{\setlength{\tabcolsep}{0pt}
\begin{tabular}{m{6.5em}l} {BaBar \cite{Lees:2011gw}\tnote{}\hphantom{\textsuperscript{}}} & { $2.69 \pm0.09 \pm0.13$ } \\ {Belle \cite{Blyth:2006at}\tnote{}\hphantom{\textsuperscript{}}} & { $2.25 \pm0.14 \pm0.35$ } \\ \end{tabular}}}{\begin{tabular}{l} $2.62 \pm0.15$ \\ $\mathit{\scriptstyle 2.67 \pm0.09}$\\ \end{tabular}}
\hflavrowdef{\hflavrowdefaai}{${\cal B}(B^0 \to \bar{D}^*(2007)^0 \pi^0)$}{{\setlength{\tabcolsep}{0pt}\begin{tabular}{m{6.5em}l} {BaBar \cite{Lees:2011gw}\tnote{}\hphantom{\textsuperscript{}}} & { $3.05 \pm0.14 \pm0.28$ } \\ {Belle \cite{Blyth:2006at}\tnote{}\hphantom{\textsuperscript{}}} & { $1.39 \pm0.18 \pm0.26$ } \\ \end{tabular}}}{\begin{tabular}{l} $2.23 \pm0.22$ \\[-5pt] {\tiny p=0.02\%} \\ $\mathit{\scriptstyle 2.23 \pm0.57}$\\ \end{tabular}}
\hflavrowdef{\hflavrowdefaaj}{${\cal B}(B^0 \to \bar{D}^0 \pi^+ \pi^-)$}{{\setlength{\tabcolsep}{0pt}\begin{tabular}{m{6.5em}l} {LHCb \cite{Aaij:2015sqa}\tnote{1}\hphantom{\textsuperscript{1}}} & { $8.46 \pm0.14 \pm0.49$ } \\ {Belle \cite{Kuzmin:2006mw}\tnote{}\hphantom{\textsuperscript{}}} & { $8.4 \pm0.4 \pm0.8$ } \\ \multicolumn{2}{l}{LHCb {\cite{Aaij:2013pua}\tnote{2,3}\hphantom{\textsuperscript{2,3}}}, {\cite{LHCb:2018oeb}\tnote{4}\hphantom{\textsuperscript{4}}}} \\ \end{tabular}}}{\begin{tabular}{l} $8.50 \pm0.40$ \\ $\mathit{\scriptstyle 8.80 \pm0.46}$\\ \end{tabular}}
\hflavrowdef{\hflavrowdefaak}{${\cal B}(B^0 \to \bar{D}^*(2007)^0 \pi^+ \pi^-)$}{{\setlength{\tabcolsep}{0pt}\begin{tabular}{m{6.5em}l} {Belle \cite{Satpathy:2002js}\tnote{}\hphantom{\textsuperscript{}}} & { $6.2 \pm1.2 \pm1.8$ } \\ \multicolumn{2}{l}{LHCb {\cite{LHCb:2021jfy}\tnote{*,5}\hphantom{\textsuperscript{*,5}}}, {\cite{LHCb:2021jfy}\tnote{*,6}\hphantom{\textsuperscript{*,6}}}} \\ \end{tabular}}}{\begin{tabular}{l} $6.2 \pm2.1$ \\ $\mathit{\scriptstyle 6.2 \pm2.2}$\\ \end{tabular}}
\begin{table}[H]
\begin{center}
\begin{threeparttable}
\caption{Branching fractions for decays to a $D^{(*)}$ meson and one or more pions, II.}
\label{tab:b2charm_Bd_D_overview2}
%
 \\
\hline
\hflavrowdefaah \\
 & & \\[-1ex]
\hflavrowdefaai \\
 & & \\[-1ex]
\hflavrowdefaaj \\
 & & \\[-1ex]
\hflavrowdefaak \\
\hline
\hline
\end{tabular}
\begin{tablenotes}
\item[*] Not included in PDG average.
\item[1] In phase space region $M(\bar{D}^0\pi^+)>2.1$ GeV$/c^2$.
\item[2] Measurement of ${\cal B}(B_s^0 \to \bar{D}^0 K^- \pi^+) / {\cal B}(B^0 \to \bar{D}^0 \pi^+ \pi^-)$ used in our fit.
\item[3] Measurement of ${\cal B}(B^0 \to \bar{D}^0 K^+ \pi^-) / {\cal B}(B^0 \to \bar{D}^0 \pi^+ \pi^-)$ used in our fit.
\item[4] Measurement of ${\cal B}(B^0 \to \bar{D}^0 K^+ K^-) / {\cal B}(B^0 \to \bar{D}^0 \pi^+ \pi^-)$ used in our fit.
\item[5] Measurement of ${\cal B}(B^0 \to \bar{D}^*(2007)^0 K^+ \pi^-) / {\cal B}(B^0 \to \bar{D}^*(2007)^0 \pi^+ \pi^-)$ used in our fit.
\item[6] Measurement of ${\cal B}(B_s^0 \to \bar{D}^*(2007)^0 K^- \pi^+) / {\cal B}(B^0 \to \bar{D}^*(2007)^0 \pi^+ \pi^-)$ used in our fit.
\end{tablenotes}
\end{threeparttable}
\end{center}
\end{table}

\hflavrowdef{\hflavrowdefaal}{${\cal B}(B^0 \to \bar{D}^0 \rho^0(770))$}{{\setlength{\tabcolsep}{0pt}
\begin{tabular}{m{6.5em}l} {Belle \cite{Kuzmin:2006mw}\tnote{}\hphantom{\textsuperscript{}}} & { $3.19 \pm0.20 \pm0.45$ } \\ {Belle \cite{Satpathy:2002js}\tnote{1}\hphantom{\textsuperscript{1}}} & { $4.1 \pm2.0 \pm0.2$ } \\ \multicolumn{2}{l}{LHCb {\cite{Aaij:2011tz}\tnote{*,2}\hphantom{\textsuperscript{*,2}}}} \\ \end{tabular}}}{\begin{tabular}{l} $3.04\,^{+0.47}_{-0.46}$ \\ $\mathit{\scriptstyle 3.21 \pm0.21}$\\ \end{tabular}}
\hflavrowdef{\hflavrowdefaam}{${\cal B}(B^0 \to \bar{D}^*(2007)^0 \rho^0(770))$}{{\setlength{\tabcolsep}{0pt}\begin{tabular}{m{6.5em}l} {Belle \cite{Satpathy:2002js}\tnote{}\hphantom{\textsuperscript{}}} & { $< 5.1$ } \\ \end{tabular}}}{\begin{tabular}{l} $< 5.1$\\ \end{tabular}}
\hflavrowdef{\hflavrowdefaan}{${\cal B}(B^0 \to \bar{D}^0 \eta)$}{{\setlength{\tabcolsep}{0pt}\begin{tabular}{m{6.5em}l} {Belle \cite{Belle:2022ojy}\tnote{*}\hphantom{\textsuperscript{*}}} & { $2.66 \pm0.05 \pm0.02$ } \\ {BaBar \cite{Lees:2011gw}\tnote{}\hphantom{\textsuperscript{}}} & { $2.53 \pm0.09 \pm0.11$ } \\ \end{tabular}}}{\begin{tabular}{l} $2.644 \pm0.050$ \\ $\mathit{\scriptstyle 2.359 \pm0.317}$\\ \end{tabular}}
\hflavrowdef{\hflavrowdefaao}{${\cal B}(B^0 \to \bar{D}^*(2007)^0 \eta)$}{{\setlength{\tabcolsep}{0pt}\begin{tabular}{m{6.5em}l} {BaBar \cite{Lees:2011gw}\tnote{}\hphantom{\textsuperscript{}}} & { $2.69 \pm0.14 \pm0.23$ } \\ {Belle \cite{Blyth:2006at}\tnote{}\hphantom{\textsuperscript{}}} & { $1.4 \pm0.3 \pm0.3$ } \\ \end{tabular}}}{\begin{tabular}{l} $2.26 \pm0.22$ \\[-5pt] {\tiny p=0.58\%} \\ $\mathit{\scriptstyle 2.26 \pm0.61}$\\ \end{tabular}}
\hflavrowdef{\hflavrowdefaap}{${\cal B}(B^0 \to \bar{D}^0 \eta^\prime)$}{{\setlength{\tabcolsep}{0pt}\begin{tabular}{m{6.5em}l} {BaBar \cite{Lees:2011gw}\tnote{}\hphantom{\textsuperscript{}}} & { $1.48 \pm0.13 \pm0.07$ } \\ {Belle \cite{Schumann:2005ej}\tnote{}\hphantom{\textsuperscript{}}} & { $1.14 \pm0.20\,^{+0.10}_{-0.13}$ } \\ \end{tabular}}}{\begin{tabular}{l} $1.38 \pm0.12$ \\ $\mathit{\scriptstyle 1.38 \pm0.16}$\\ \end{tabular}}
\hflavrowdef{\hflavrowdefaaq}{${\cal B}(B^0 \to \bar{D}^*(2007)^0 \eta^\prime)$}{{\setlength{\tabcolsep}{0pt}\begin{tabular}{m{6.5em}l} {BaBar \cite{Lees:2011gw}\tnote{}\hphantom{\textsuperscript{}}} & { $1.48 \pm0.22 \pm0.13$ } \\ {Belle \cite{Schumann:2005ej}\tnote{}\hphantom{\textsuperscript{}}} & { $1.21 \pm0.34 \pm0.22$ } \\ \end{tabular}}}{\begin{tabular}{l} $1.40 \pm0.22$\\ \end{tabular}}
\hflavrowdef{\hflavrowdefaar}{${\cal B}(B^0 \to \bar{D}^0 \omega(782))$}{{\setlength{\tabcolsep}{0pt}\begin{tabular}{m{6.5em}l} {BaBar \cite{Lees:2011gw}\tnote{}\hphantom{\textsuperscript{}}} & { $2.57 \pm0.11 \pm0.14$ } \\ {Belle \cite{Blyth:2006at}\tnote{}\hphantom{\textsuperscript{}}} & { $2.37 \pm0.23 \pm0.28$ } \\ {LHCb \cite{Aaij:2015sqa}\tnote{}\hphantom{\textsuperscript{}}} & { $2.81 \pm0.72 \pm0.30$ } \\ \multicolumn{2}{l}{Belle {\cite{Satpathy:2002js}\tnote{*,3}\hphantom{\textsuperscript{*,3}}}} \\ \end{tabular}}}{\begin{tabular}{l} $2.53 \pm0.16$ \\ $\mathit{\scriptstyle 2.54 \pm0.16}$\\ \end{tabular}}
\hflavrowdef{\hflavrowdefaas}{${\cal B}(B^0 \to \bar{D}^*(2007)^0 \omega(782))$}{{\setlength{\tabcolsep}{0pt}\begin{tabular}{m{6.5em}l} {BaBar \cite{Lees:2011gw}\tnote{}\hphantom{\textsuperscript{}}} & { $4.55 \pm0.24 \pm0.39$ } \\ {Belle \cite{Blyth:2006at}\tnote{}\hphantom{\textsuperscript{}}} & { $2.29 \pm0.39 \pm0.40$ } \\ \end{tabular}}}{\begin{tabular}{l} $3.64 \pm0.35$ \\[-5pt] {\tiny p=0.18\%} \\ $\mathit{\scriptstyle 3.64 \pm1.11}$\\ \end{tabular}}
\hflavrowdef{\hflavrowdefaat}{${\cal B}(B^0 \to \bar{D}^0 f_2(1270))$}{{\setlength{\tabcolsep}{0pt}\begin{tabular}{m{6.5em}l} {LHCb \cite{Aaij:2015sqa}\tnote{}\hphantom{\textsuperscript{}}} & { $1.61 \pm0.11 \pm0.19$ } \\ {Belle \cite{Kuzmin:2006mw}\tnote{}\hphantom{\textsuperscript{}}} & { $1.2 \pm0.2 \pm0.4$ } \\ \end{tabular}}}{\begin{tabular}{l} $1.52 \pm0.19$ \\ $\mathit{\scriptstyle 1.56 \pm0.21}$\\ \end{tabular}}
\begin{table}[H]
\begin{center}
\begin{threeparttable}
\caption{Branching fractions for decays to a $D^{(*)0}$ meson and a light meson.}
\label{tab:b2charm_Bd_D_overview3}
%
 \\
\hline
\hflavrowdefaal \\
 & & \\[-1ex]
\hflavrowdefaam \\
 & & \\[-1ex]
\hflavrowdefaan \\
 & & \\[-1ex]
\hflavrowdefaao \\
 & & \\[-1ex]
\hflavrowdefaap \\
 & & \\[-1ex]
\hflavrowdefaaq \\
 & & \\[-1ex]
\hflavrowdefaar \\
 & & \\[-1ex]
\hflavrowdefaas \\
 & & \\[-1ex]
\hflavrowdefaat \\
\hline
\hline
\end{tabular}
\begin{tablenotes}
\item[*] Not included in PDG average.
\item[1] Using ${\cal B}(B^0 \to \bar{D}^0 \omega(782))$.
\item[2] Measurement of ${\cal B}(B_s^0 \to \bar{D}^0 \bar{K}^*(892)^0) / {\cal B}(B^0 \to \bar{D}^0 \rho^0(770))$ used in our fit.
\item[3] Measurement of ${\cal B}(B^0 \to \bar{D}^0 \rho^0(770)) / {\cal B}(B^0 \to \bar{D}^0 \omega(782))$ used in our fit.
\end{tablenotes}
\end{threeparttable}
\end{center}
\end{table}

\hflavrowdef{\hflavrowdefaau}{${\cal B}(B^0 \to D^- K^+)$}{{\setlength{\tabcolsep}{0pt}
\begin{tabular}{m{6.5em}l} {LHCb \cite{Aaij:2013qqa}\tnote{1}\hphantom{\textsuperscript{1}}} & { $2.040 \pm0.027 \pm0.112$ } \\ {Belle \cite{Abe:2001waa}\tnote{1}\hphantom{\textsuperscript{1}}} & { $1.69 \pm0.37 \pm0.19$ } \\ \end{tabular}}}{\begin{tabular}{l} $2.03 \pm0.12$ \\ $\mathit{\scriptstyle 2.05 \pm0.08}$\\ \end{tabular}}
\hflavrowdef{\hflavrowdefaav}{${\cal B}(B^0 \to D^*(2010)^- K^+)$}{{\setlength{\tabcolsep}{0pt}\begin{tabular}{m{6.5em}l} {Belle \cite{Belle:2022afp}\tnote{*}\hphantom{\textsuperscript{*}}} & { $2.22 \pm0.06 \pm0.08$ } \\ {BaBar \cite{BaBar:2005osj}\tnote{2}\hphantom{\textsuperscript{2}}} & { $2.058 \pm0.090 \pm0.091$ } \\ {Belle \cite{Abe:2001waa}\tnote{2}\hphantom{\textsuperscript{2}}} & { $1.96 \pm0.40 \pm0.17$ } \\ \end{tabular}}}{\begin{tabular}{l} $2.146 \pm0.079$ \\ $\mathit{\scriptstyle 2.117\,^{+0.155}_{-0.154}}$\\ \end{tabular}}
\hflavrowdef{\hflavrowdefaaw}{${\cal B}(B^0 \to D^- K^*(892)^+)$}{{\setlength{\tabcolsep}{0pt}\begin{tabular}{m{6.5em}l} {BaBar \cite{Aubert:2004at}\tnote{}\hphantom{\textsuperscript{}}} & { $4.6 \pm0.6 \pm0.5$ } \\ \end{tabular}}}{\begin{tabular}{l} $4.60 \pm0.78$ \\ $\mathit{\scriptstyle 4.46 \pm0.72}$\\ \end{tabular}}
\hflavrowdef{\hflavrowdefaax}{${\cal B}(B^0 \to D^*(2010)^- K^*(892)^+)$}{{\setlength{\tabcolsep}{0pt}\begin{tabular}{m{6.5em}l} {BaBar \cite{Aubert:2004at}\tnote{}\hphantom{\textsuperscript{}}} & { $3.2 \pm0.6 \pm0.3$ } \\ \end{tabular}}}{\begin{tabular}{l} $3.20 \pm0.67$ \\ $\mathit{\scriptstyle 3.30 \pm0.61}$\\ \end{tabular}}
\hflavrowdef{\hflavrowdefaay}{${\cal B}(B^0 \to D^- K^0 \pi^+)$}{{\setlength{\tabcolsep}{0pt}\begin{tabular}{m{6.5em}l} {BaBar \cite{Aubert:2004at}\tnote{}\hphantom{\textsuperscript{}}} & { $4.9 \pm0.7 \pm0.5$ } \\ \end{tabular}}}{\begin{tabular}{l} $4.90 \pm0.86$\\ \end{tabular}}
\hflavrowdef{\hflavrowdefaaz}{${\cal B}(B^0 \to D^*(2010)^- K^0 \pi^+)$}{{\setlength{\tabcolsep}{0pt}\begin{tabular}{m{6.5em}l} {BaBar \cite{Aubert:2004at}\tnote{}\hphantom{\textsuperscript{}}} & { $3.0 \pm0.7 \pm0.3$ } \\ \end{tabular}}}{\begin{tabular}{l} $3.00 \pm0.76$\\ \end{tabular}}
\hflavrowdef{\hflavrowdefaba}{${\cal B}(B^0 \to D^- K^+ \bar{K}^0)$}{{\setlength{\tabcolsep}{0pt}\begin{tabular}{m{6.5em}l} {Belle II \cite{Belle-II:2024xtf}\tnote{\dag}\hphantom{\textsuperscript{\dag}}} & { $0.82 \pm0.12 \pm0.05$ } \\ {Belle \cite{Drutskoy:2002ib}\tnote{}\hphantom{\textsuperscript{}}} & { $< 3.1$ } \\ \end{tabular}}}{\begin{tabular}{l} $0.82 \pm0.13$ \\ $\mathit{\scriptstyle < 3.10}$\\ \end{tabular}}
\hflavrowdef{\hflavrowdefabb}{${\cal B}(B^0 \to D^*(2010)^- K^+ \bar{K}^0)$}{{\setlength{\tabcolsep}{0pt}\begin{tabular}{m{6.5em}l} {Belle II \cite{Belle-II:2024xtf}\tnote{\dag}\hphantom{\textsuperscript{\dag}}} & { $0.91 \pm0.19 \pm0.05$ } \\ {Belle \cite{Drutskoy:2002ib}\tnote{}\hphantom{\textsuperscript{}}} & { $< 4.7$ } \\ \end{tabular}}}{\begin{tabular}{l} $0.91 \pm0.20$ \\ $\mathit{\scriptstyle < 4.70}$\\ \end{tabular}}
\hflavrowdef{\hflavrowdefabc}{${\cal B}(B^0 \to D^- \bar{K}^*(892)^0 K^+)$}{{\setlength{\tabcolsep}{0pt}\begin{tabular}{m{6.5em}l} {Belle \cite{Drutskoy:2002ib}\tnote{*}\hphantom{\textsuperscript{*}}} & { $8.8 \pm1.1 \pm1.5$ } \\ \end{tabular}}}{\begin{tabular}{l} $8.8 \pm1.9$ \\ \textit{\small none}\\ \end{tabular}}
\hflavrowdef{\hflavrowdefabd}{${\cal B}(B^0 \to D^*(2010)^- \bar{K}^*(892)^0 K^+)$}{{\setlength{\tabcolsep}{0pt}\begin{tabular}{m{6.5em}l} {Belle \cite{Drutskoy:2002ib}\tnote{*}\hphantom{\textsuperscript{*}}} & { $12.9 \pm2.2 \pm2.5$ } \\ \end{tabular}}}{\begin{tabular}{l} $12.9 \pm3.3$ \\ \textit{\small none}\\ \end{tabular}}
\hflavrowdef{\hflavrowdefabe}{${\cal B}(B^0 \to D^- K^+ \pi^+ \pi^-)$}{{\setlength{\tabcolsep}{0pt}\begin{tabular}{m{6.5em}l} {LHCb \cite{Aaij:2012bw}\tnote{3}\hphantom{\textsuperscript{3}}} & { $3.41 \pm0.64\,^{+0.47}_{-0.46}$ } \\ \end{tabular}}}{\begin{tabular}{l} $3.41\,^{+0.82}_{-0.76}$ \\ $\mathit{\scriptstyle 3.53 \pm0.81}$\\ \end{tabular}}
\hflavrowdef{\hflavrowdefabf}{${\cal B}(B^0 \to D^*(2010)^- K^+ \pi^+ \pi^-)$}{{\setlength{\tabcolsep}{0pt}\begin{tabular}{m{6.5em}l} {LHCb \cite{Aaij:2013xca}\tnote{4}\hphantom{\textsuperscript{4}}} & { $4.600 \pm0.263 \pm0.294$ } \\ \end{tabular}}}{\begin{tabular}{l} $4.60 \pm0.39$ \\ $\mathit{\scriptstyle 4.66 \pm0.41}$\\ \end{tabular}}
\begin{table}[H]
\begin{center}
\begin{threeparttable}
\caption{Branching fractions for decays to a $D^{(*)+}$ meson and one or more kaons.}
\label{tab:b2charm_Bd_D_overview4}
%
 \\
\hline
\hflavrowdefaau \\
 & & \\[-1ex]
\hflavrowdefaav \\
 & & \\[-1ex]
\hflavrowdefaaw \\
 & & \\[-1ex]
\hflavrowdefaax \\
 & & \\[-1ex]
\hflavrowdefaay \\
 & & \\[-1ex]
\hflavrowdefaaz \\
 & & \\[-1ex]
\hflavrowdefaba \\
 & & \\[-1ex]
\hflavrowdefabb \\
 & & \\[-1ex]
\hflavrowdefabc \\
 & & \\[-1ex]
\hflavrowdefabd \\
 & & \\[-1ex]
\hflavrowdefabe \\
 & & \\[-1ex]
\hflavrowdefabf \\
\hline
\hline
\end{tabular}
\begin{tablenotes}
\item[*] Not included in PDG average.
\item[\dag] Preliminary result.
\item[1] Using ${\cal B}(B^0 \to D^- \pi^+)$.
\item[2] Using ${\cal B}(B^0 \to D^*(2010)^- \pi^+)$.
\item[3] Using ${\cal B}(B^0 \to D^- \pi^+ \pi^+ \pi^-)$.
\item[4] Using ${\cal B}(B^0 \to D^*(2010)^- \pi^+ \pi^+ \pi^-)$.
\end{tablenotes}
\end{threeparttable}
\end{center}
\end{table}

\hflavrowdef{\hflavrowdefabg}{${\cal B}(B^0 \to \bar{D}^0 K^0)$}{{\setlength{\tabcolsep}{0pt}
\begin{tabular}{m{6.5em}l} {BaBar \cite{BaBar:2006ftr}\tnote{}\hphantom{\textsuperscript{}}} & { $5.3 \pm0.7 \pm0.3$ } \\ {Belle \cite{Belle:2022ojy}\tnote{*}\hphantom{\textsuperscript{*}}} & { $5.6 \pm1.2 \pm2.1$ } \\ \end{tabular}}}{\begin{tabular}{l} $5.33 \pm0.73$ \\ $\mathit{\scriptstyle 5.23\,^{+0.67}_{-0.66}}$\\ \end{tabular}}
\hflavrowdef{\hflavrowdefabh}{${\cal B}(B^0 \to \bar{D}^*(2007)^0 K^0)$}{{\setlength{\tabcolsep}{0pt}\begin{tabular}{m{6.5em}l} {BaBar \cite{BaBar:2006ftr}\tnote{}\hphantom{\textsuperscript{}}} & { $3.6 \pm1.2 \pm0.3$ } \\ {Belle \cite{Krokovny:2002ua}\tnote{}\hphantom{\textsuperscript{}}} & { $< 6.6$ } \\ \end{tabular}}}{\begin{tabular}{l} $3.6 \pm1.2$\\ \end{tabular}}
\hflavrowdef{\hflavrowdefabi}{${\cal B}(B^0 \to \bar{D}^0 K^+ \pi^-)$}{{\setlength{\tabcolsep}{0pt}\begin{tabular}{m{6.5em}l} {LHCb \cite{Aaij:2013pua}\tnote{1}\hphantom{\textsuperscript{1}}} & { $9.01 \pm0.59 \pm0.80$ } \\ {BaBar \cite{BaBar:2005osj}\tnote{2}\hphantom{\textsuperscript{2}}} & { $8.8 \pm1.5 \pm0.9$ } \\ \end{tabular}}}{\begin{tabular}{l} $8.96\,^{+0.88}_{-0.86}$ \\ $\mathit{\scriptstyle 8.80 \pm1.75}$\\ \end{tabular}%
}
\begin{table}[H]
\begin{center}
\begin{threeparttable}
\caption{Branching fractions for decays to a $D^{(*)0}$ meson and one or more kaons or a $\phi$.}
\label{tab:b2charm_Bd_D_overview5}
%
 \\
\hline
\hflavrowdefabg \\
 & & \\[-1ex]
\hflavrowdefabh \\
 & & \\[-1ex]
\hflavrowdefabi \\
 & & \\[-1ex]
\hflavrowdefabj \\
 & & \\[-1ex]
\hflavrowdefabk \\
 & & \\[-1ex]
\hflavrowdefabl \\
 & & \\[-1ex]
\hflavrowdefabm \\
 & & \\[-1ex]
\hflavrowdefabn \\
 & & \\[-1ex]
\hflavrowdefabo \\
 & & \\[-1ex]
\hflavrowdefabp \\
 & & \\[-1ex]
\hflavrowdefabq \\
 & & \\[-1ex]
\hflavrowdefabr \\
 & & \\[-1ex]
\hflavrowdefabs \\
 & & \\[-1ex]
\hflavrowdefabt \\
\hline
\hline
\end{tabular}
\begin{tablenotes}
\scriptsize
\item[*] Not included in PDG average.
\item[\dag] Preliminary result.
\item[1] Using ${\cal B}(B^0 \to \bar{D}^0 \pi^+ \pi^-)$.
\item[2] Excluding $D^*(2010)^+K^-$.
\item[3] Using ${\cal B}(K^*(892)^0 \to K^+ \pi^-)$.
\item[4] Measurement of ${\cal B}(B_s^0 \to \bar{D}^0 \bar{K}^*(892)^0) / {\cal B}(B^0 \to \bar{D}^0 K^*(892)^0)$ used in our fit.
\item[5] Measurement of ${\cal B}(B_s^0 \to \bar{D}^0 K^+ K^-) / {\cal B}(B^0 \to \bar{D}^0 K^+ K^-)$ used in our fit.
\item[6] Measurement of ${\cal B}(B^0 \to \bar{D}^0 \phi(1020)) {\cal B}(\phi(1020) \to K^+ K^-) / {\cal B}(B^0 \to \bar{D}^0 K^+ K^-)$ used in our fit.
\item[7] Measurement of ${\cal B}(B^0 \to \bar{D}^*(2007)^0 \phi(1020)) {\cal B}(\phi(1020) \to K^+ K^-) / {\cal B}(B^0 \to \bar{D}^0 K^+ K^-)$ used in our fit.
\item[8] Using ${\cal B}(B^0 \to \bar{D}^*(2007)^0 \pi^+ \pi^-)$.
\end{tablenotes}
\end{threeparttable}
\end{center}
\end{table}

\hflavrowdef{\hflavrowdefabu}{${\cal B}(B^0 \to D_s^+ \pi^-)$}{{\setlength{\tabcolsep}{0pt}
}
\begin{table}[H]
\begin{center}
\begin{threeparttable}
\caption{Branching fractions for decays to a $D_s^{(*)}$ meson.}
\label{tab:b2charm_Bd_D_overview6}
%
 \\
\hline
\hflavrowdefabu \\
 & & \\[-1ex]
\hflavrowdefabv \\
 & & \\[-1ex]
\hflavrowdefabw \\
 & & \\[-1ex]
\hflavrowdefabx \\
 & & \\[-1ex]
\hflavrowdefaby \\
 & & \\[-1ex]
\hflavrowdefabz \\
 & & \\[-1ex]
\hflavrowdefaca \\
 & & \\[-1ex]
\hflavrowdefacb \\
 & & \\[-1ex]
\hflavrowdefacc \\
 & & \\[-1ex]
\hflavrowdefacd \\
 & & \\[-1ex]
\hflavrowdeface \\
 & & \\[-1ex]
\hflavrowdefacf \\
 & & \\[-1ex]
\hflavrowdefacg \\
 & & \\[-1ex]
\hflavrowdefach \\
 & & \\[-1ex]
\hflavrowdefaci \\
\hline
\hline
\end{tabular}
\begin{tablenotes}
\footnotesize
\item[1] Using BaBar result for $\mathcal{B}(D_s^+ \to \phi \pi^+)$~\cite{Aubert:2005xu}
\item[2] Using ${\cal B}(B^0 \to D^- \pi^+)$.
\item[3] Using ${\cal B}(B_s^0 \to D_s^- K^+ \pi^+ \pi^-)$.
\end{tablenotes}
\end{threeparttable}
\end{center}
\end{table}

\hflavrowdef{\hflavrowdefacj}{${\cal{B}} ( B^{0} \to D^{**-} \pi^{+} )$\tnote{1}\hphantom{\textsuperscript{1}}}{{\setlength{\tabcolsep}{0pt}
}
\begin{table}[H]
\begin{center}
\begin{threeparttable}
\caption{Branching fractions for decays to excited $D$ mesons.}
\label{tab:b2charm_Bd_D_overview7}
%
\begin{tablenotes}
\item[1] $D^{**}$ refers to the sum of all the non-strange charm meson states with masses in the range 2.2 - 2.8 GeV$/c^2$
\end{tablenotes}
\end{threeparttable}
\end{center}
\end{table}

\hflavrowdef{\hflavrowdefack}{${\cal B}(B^0 \to \bar{D}_1(2420)^0 \pi^+ \pi^-) \times {\cal B}(D_1(2420)^0 \to D^*(2010)^+ \pi^-)$}{{\setlength{\tabcolsep}{0pt}%
}
\begin{table}[H]
\begin{center}
\begin{threeparttable}
\caption{Branching fractions for decays to excited $D_{s}$ mesons.}
\label{tab:b2charm_Bd_D_overview8}
%
 \\
\hline
\hflavrowdefack \\
 & & \\[-1ex]
\hflavrowdefacl \\
 & & \\[-1ex]
\hflavrowdefacm \\
 & & \\[-1ex]
\hflavrowdefacn \\
 & & \\[-1ex]
\hflavrowdefaco \\
 & & \\[-1ex]
\hflavrowdefacp \\
 & & \\[-1ex]
\hflavrowdefacq \\
 & & \\[-1ex]
\hflavrowdefacr \\
 & & \\[-1ex]
\hflavrowdefacs \\
 & & \\[-1ex]
\hflavrowdefact \\
 & & \\[-1ex]
\hflavrowdefacu \\
 & & \\[-1ex]
\hflavrowdefacv \\
\hline
\hline
\end{tabular}
\begin{tablenotes}
\item[*] Not included in PDG average.
\item[1] Using ${\cal B}(B^0 \to D^*(2010)^- \pi^+ \pi^+ \pi^-)$.
\item[2] Using ${\cal B}(B^0 \to D^- \pi^+ \pi^+ \pi^-)$.
\item[3] Using ${\cal B}(D_{s1}(2460)^+ \to D_s^+ \gamma)$.
\item[4] Using ${\cal B}(D_{s0}^*(2317)^+ \to D_s^+ \pi^0)$.
\end{tablenotes}
\end{threeparttable}
\end{center}
\end{table}

\hflavrowdef{\hflavrowdefacw}{${\cal B}(B^0 \to D^- p \bar{p} \pi^+)$}{{\setlength{\tabcolsep}{0pt}
}
\begin{table}[H]
\begin{center}
\begin{threeparttable}
\caption{Branching fractions for decays to a $D^{(*)}$ meson and baryons.}
\label{tab:b2charm_Bd_D_overview9}
%
 \\
\hline
\hflavrowdefacw \\
 & & \\[-1ex]
\hflavrowdefacx \\
 & & \\[-1ex]
\hflavrowdefacy \\
 & & \\[-1ex]
\hflavrowdefacz \\
 & & \\[-1ex]
\hflavrowdefada \\
 & & \\[-1ex]
\hflavrowdefadb \\
 & & \\[-1ex]
\hflavrowdefadc \\
 & & \\[-1ex]
\hflavrowdefadd \\
 & & \\[-1ex]
\hflavrowdefade \\
 & & \\[-1ex]
\hflavrowdefadf \\
 & & \\[-1ex]
\hflavrowdefadg \\
\hline
\hline
\end{tabular}
\end{threeparttable}
\end{center}
\end{table}

\subsubsection{Decays to two open charm mesons}
\label{sec:b2c:Bd_DD}
Averages of $B^0$ decays to two open charm mesons are shown in Tables~\ref{tab:b2charm_Bd_DD_overview1}--\ref{tab:b2charm_Bd_DD_overview5}.
\hflavrowdef{\hflavrowdefadh}{${\cal B}(B^0 \to D^+ D^-)$}{{\setlength{\tabcolsep}{0pt}
}
\begin{table}[H]
\begin{center}
\begin{threeparttable}
\caption{Branching fractions for decays to $D^{(*)}\bar{D}^{(*)}$.}
\label{tab:b2charm_Bd_DD_overview1}
%
 \\
\hline
\hflavrowdefadh \\
 & & \\[-1ex]
\hflavrowdefadi \\
 & & \\[-1ex]
\hflavrowdefadj \\
 & & \\[-1ex]
\hflavrowdefadk \\
 & & \\[-1ex]
\hflavrowdefadl \\
 & & \\[-1ex]
\hflavrowdefadm \\
\hline
\hline
\end{tabular}
\begin{tablenotes}
\item[*] Not included in PDG average.
\item[1] Measurement of ${\cal B}(B_s^0 \to D^+ D^-) / {\cal B}(B^0 \to D^+ D^-)$ used in our fit.
\item[2] Including the charge-conjugate final state.
\item[3] Measurement of ${\cal B}(B_s^0 \to D^*(2010)^+ D^*(2010)^-) / {\cal B}(B^0 \to D^*(2010)^+ D^*(2010)^-)$ used in our fit.
\item[4] Using ${\cal B}(B^+ \to D_s^+ \bar{D}^0)$.
\end{tablenotes}
\end{threeparttable}
\end{center}
\end{table}

\hflavrowdef{\hflavrowdefadn}{${\cal B}(B^0 \to D^*(2010)^- D^+ K^0)$}{{\setlength{\tabcolsep}{0pt}
}
\begin{table}[H]
\begin{center}
\begin{threeparttable}
\caption{Branching fractions for decays to two $D$ mesons and a kaon.}
\label{tab:b2charm_Bd_DD_overview2}
%
 \\
\hline
\hflavrowdefadn \\
 & & \\[-1ex]
\hflavrowdefado \\
 & & \\[-1ex]
\hflavrowdefadp \\
 & & \\[-1ex]
\hflavrowdefadq \\
 & & \\[-1ex]
\hflavrowdefadr \\
 & & \\[-1ex]
\hflavrowdefads \\
 & & \\[-1ex]
\hflavrowdefadt \\
 & & \\[-1ex]
\hflavrowdefadu \\
 & & \\[-1ex]
\hflavrowdefadv \\
 & & \\[-1ex]
\hflavrowdefadw \\
 & & \\[-1ex]
\hflavrowdefadx \\
\hline
\hline
\end{tabular}
\begin{tablenotes}
\item[*] Not included in PDG average.
\item[1] Measurement of ${\cal B}(B^0 \to D^*(2010)^+ D^*(2010)^- K^0) / 2$ used in our fit.
\item[2] Measurement of ${\cal B}(B^0 \to D_{s1}(2536)^+ D^*(2010)^-) / {\cal B}(B^0 \to D^*(2010)^+ D^*(2010)^- K^0) 2$ used in our fit.
\end{tablenotes}
\end{threeparttable}
\end{center}
\end{table}

\hflavrowdef{\hflavrowdefady}{${\cal B}(B^0 \to D_s^+ D^-)$}{{\setlength{\tabcolsep}{0pt}
}
\begin{table}[H]
\begin{center}
\begin{threeparttable}
\caption{Branching fractions for decays to $D_s^{(*)-}D^{(*)+}$.}
\label{tab:b2charm_Bd_DD_overview3}
%
 \\
\hline
\hflavrowdefady \\
 & & \\[-1ex]
\hflavrowdefadz \\
 & & \\[-1ex]
\hflavrowdefaea \\
 & & \\[-1ex]
\hflavrowdefaeb \\
\hline
\hline
\end{tabular}
\begin{tablenotes}
\item[\dag] Preliminary result.
\item[*] Not included in PDG average.
\item[1] Using ${\cal B}(D_s^+ \to \phi(1020) \pi^+)$.
\item[2] Measurement of ${\cal B}(B_s^0 \to D_s^- D^+) / {\cal B}(B^0 \to D_s^+ D^-)$ used in our fit.
\item[3] Measurement of ${\cal B}(B_s^0 \to D_s^+ D_s^-) / {\cal B}(B^0 \to D_s^+ D^-)$ used in our fit.
\item[4] Measurement of ${\cal B}(B^+ \to D_s^+ \bar{D}^0) / {\cal B}(B^0 \to D_s^+ D^-)$ used in our fit.
\item[5] At CL=95\,\%.
\item[6] Measurement of ${\cal B}(B^0 \to \Lambda_c^+ \bar{\Lambda}_c^-) / {\cal B}(B^0 \to D_s^+ D^-)$ used in our fit.
\end{tablenotes}
\end{threeparttable}
\end{center}
\end{table}

\hflavrowdef{\hflavrowdefaec}{${\cal B}(B^0 \to D_s^+ D_s^-)$}{{\setlength{\tabcolsep}{0pt}
}
\begin{table}[H]
\begin{center}
\begin{threeparttable}
\caption{Branching fractions for decays to $D_s^{(*)+}D_s^{(*)-}$.}
\label{tab:b2charm_Bd_DD_overview4}
%
}
\begin{table}[H]
\begin{center}
\begin{threeparttable}
\caption{Branching fractions for decays to excited $D_s$ mesons.}
\label{tab:b2charm_Bd_DD_overview5}
%
 \\
\hline
\hflavrowdefaef \\
 & & \\[-1ex]
\hflavrowdefaeg \\
 & & \\[-1ex]
\hflavrowdefaeh \\
 & & \\[-1ex]
\hflavrowdefaei \\
 & & \\[-1ex]
\hflavrowdefaej \\
 & & \\[-1ex]
\hflavrowdefaek \\
 & & \\[-1ex]
\hflavrowdefael \\
 & & \\[-1ex]
\hflavrowdefaem \\
\hline
\hline
\end{tabular}
\begin{tablenotes}
\item[*] Not included in PDG average.
\item[1] Using ${\cal B}(D_{s0}^*(2317)^+ \to D_s^+ \pi^0)$.
\item[2] Using ${\cal B}(D_{s1}(2460)^+ \to D_s^+ \gamma)$.
\item[3] Using ${\cal B}(D_{s1}(2460)^+ \to D_s^{*+} \pi^0)$.
\item[4] Using ${\cal B}(D_{s1}(2460)^+ \to D_s^+ \pi^+ \pi^-)$.
\item[5] Using ${\cal B}(D_{s1}(2536)^+ \to D^*(2007)^0 K^+)$.
\item[6] Using ${\cal B}(D_{s1}(2536)^+ \to D^*(2010)^+ K^0)$.
\item[7] Using ${\cal B}(B^0 \to D^*(2010)^+ D^*(2010)^- K^0)$.
\item[8] Measurement of ${\cal B}(B^0 \to D_{s1}(2536)^- D^+) ( {\cal B}(D_{s1}(2536)^+ \to D^*(2007)^0 K^+) + {\cal B}(D_{s1}(2536)^+ \to D^*(2010)^+ K^0) )$ used in our fit.
\item[9] Measurement of ${\cal B}(B^0 \to D_{s1}(2536)^- D^*(2010)^+) ( {\cal B}(D_{s1}(2536)^+ \to D^*(2007)^0 K^+) + {\cal B}(D_{s1}(2536)^+ \to D^*(2010)^+ K^0) )$ used in our fit.
\item[10] Measurement of ${\cal B}(B^0 \to D_{s1}(2536)^- D^*(2010)^+) BR_Ds1(2536)*_D*+_K0S$ used in our fit.
\end{tablenotes}
\end{threeparttable}
\end{center}
\end{table}

\subsubsection{Decays to charmonium states}
\label{sec:b2c:Bd_cc}
Averages of $B^0$ decays to charmonium states are shown in Tables~\ref{tab:b2charm_Bd_cc_overview1}--\ref{tab:b2charm_Bd_cc_overview6}.
\hflavrowdef{\hflavrowdefaen}{${\cal B}(B^0 \to J/\psi K^0)$}{{\setlength{\tabcolsep}{0pt}
}
\begin{table}[H]
\begin{center}
\resizebox{0.9\textwidth}{!}{
\begin{threeparttable}
\caption{Branching fractions for decays to $J/\psi$ and one kaon.}
\label{tab:b2charm_Bd_cc_overview1}
%
 \\
\hline
\hflavrowdefaen \\
 & & \\[-1ex]
\hflavrowdefaeo \\
 & & \\[-1ex]
\hflavrowdefaep \\
 & & \\[-1ex]
\hflavrowdefaeq \\
 & & \\[-1ex]
\hflavrowdefaer \\
 & & \\[-1ex]
\hflavrowdefaes \\
 & & \\[-1ex]
\hflavrowdefaet \\
 & & \\[-1ex]
\hflavrowdefaeu \\
 & & \\[-1ex]
\hflavrowdefaev \\
 & & \\[-1ex]
\hflavrowdefaew \\
 & & \\[-1ex]
\hflavrowdefaex \\
\hline
\hline
\end{tabular}
\begin{tablenotes}
\scriptsize
\item[*] Not included in PDG average.
\item[1] Measurement of ${\cal B}(B^0 \to \eta_c K^0) / {\cal B}(B^0 \to J/\psi K^0)$ used in our fit.
\item[2] Measurement of ${\cal B}(B^0 \to J/\psi K^*(892)^0) / {\cal B}(B^0 \to J/\psi K^0)$ used in our fit.
\item[3] Measurement of $0.5 {\cal B}(B^0 \to J/\psi K^0) {\cal B}(J/\psi \to p \bar{p} \pi^+ \pi^-)$ used in our fit.
\item[4] Measurement of $2 {\cal B}(B_s^0 \to J/\psi K^0_S) / {\cal B}(B^0 \to J/\psi K^0)$ used in our fit.
\item[5] Measurement of ${\cal B}(B^0 \to J/\psi K^0 \pi^+ \pi^-) / {\cal B}(B^0 \to J/\psi K^0)$ used in our fit.
\item[6] Measurement of ${\cal B}(B^0 \to \psi(2S) K^0) {\cal B}(\psi(2S) \to J/\psi \pi^+ \pi^-) / {\cal B}(B^0 \to J/\psi K^0)$ used in our fit.
\item[7] Measurement of $( {\cal B}(B^0 \to \eta_c K^+ \pi^-) {\cal B}(\eta_c \to p \bar{p}) ) / ( {\cal B}(B^0 \to J/\psi K^+ \pi^-) {\cal B}(J/\psi \to p \bar{p}) )$ used in our fit.
\item[8] Using ${\cal B}(B^0 \to J/\psi K^0)$.
\item[9] Measurement of ${\cal B}(B^0 \to \psi(2S) K^*(892)^0) / {\cal B}(B^0 \to J/\psi K^*(892)^0)$ used in our fit.
\item[10] Measurement of ${\cal B}(B_s^0 \to J/\psi K^0 K^- \pi^+ \mathrm{+c.c.}) / {\cal B}(B^0 \to J/\psi K^0 \pi^+ \pi^-)$ used in our fit.
\item[11] Using ${\cal B}(B^+ \to J/\psi \omega(782) K^+)$.
\item[12] Using ${\cal B}(B^+ \to J/\psi K^+)$.
\item[13] Measurement of ${\cal B}(B^0 \to J/\psi K^*(892)^0 K^+ K^-) / {\cal B}(B^0 \to J/\psi K^*(892)^0 \pi^+ \pi^-)$ used in our fit.
\end{tablenotes}
\end{threeparttable}
}
\end{center}
\end{table}

\hflavrowdef{\hflavrowdefaey}{${\cal B}(B^0 \to \psi(2S) K^0)$}{{\setlength{\tabcolsep}{0pt}
}
\begin{table}[H]
\begin{center}
\resizebox{0.85\textwidth}{!}{
\begin{threeparttable}
\caption{Branching fractions for decays to charmonium other than $J/\psi$ and one kaon.}
\label{tab:b2charm_Bd_cc_overview2}
%
 \\
\hline
\hflavrowdefaey \\
 & & \\[-1ex]
\hflavrowdefaez \\
 & & \\[-1ex]
\hflavrowdefafa \\
 & & \\[-1ex]
\hflavrowdefafb \\
 & & \\[-1ex]
\hflavrowdefafc \\
 & & \\[-1ex]
\hflavrowdefafd \\
 & & \\[-1ex]
\hflavrowdefafe \\
 & & \\[-1ex]
\hflavrowdefaff \\
\hline
\hline
\end{tabular}
\begin{tablenotes}
\scriptsize
\item[*] Not included in PDG average.
\item[1] Measurement of ${\cal B}(B^0 \to \psi(2S) K^*(892)^0) / {\cal B}(B^0 \to \psi(2S) K^0)$ used in our fit.
\item[2] Measurement of $0.5 {\cal B}(B^0 \to \psi(2S) K^0) {\cal B}(\psi(2S) \to p \bar{p} \pi^+ \pi^-)$ used in our fit.
\item[3] Measurement of ${\cal B}(B^0 \to \psi(2S) K^0) {\cal B}(\psi(2S) \to J/\psi \pi^+ \pi^-) / {\cal B}(B^0 \to J/\psi K^0)$ used in our fit.
\item[4] Using ${\cal B}(B^0 \to J/\psi K^*(892)^0)$.
\item[5] Using ${\cal B}(B^0 \to \psi(2S) K^0)$.
\item[6] Measurement of ${\cal B}(B_s^0 \to \psi(2S) \bar{K}^*(892)^0) / {\cal B}(B^0 \to \psi(2S) K^*(892)^0)$ used in our fit.
\item[7] Using ${\cal B}(B^+ \to \eta_c K^+)$.
\item[8] Using ${\cal B}(B^0 \to J/\psi K^0)$.
\item[9] Calculated using $\mathcal{B}(\eta_c \to p \bar{p})$
\item[10] Measurement of $0.5 {\cal B}(B^0 \to \eta_c K^0) {\cal B}(\eta_c \to p \bar{p} \pi^+ \pi^-)$ used in our fit.
\item[11] Measurement of ${\cal B}(B^0 \to \eta_c K^*(892)^0) / {\cal B}(B^0 \to \eta_c K^0)$ used in our fit.
\item[12] Using ${\cal B}(B^0 \to \eta_c K^0)$.
\item[13] Measurement of $( {\cal B}(B^0 \to \eta_c K^+ \pi^-) {\cal B}(\eta_c \to p \bar{p}) ) / ( {\cal B}(B^0 \to J/\psi K^+ \pi^-) {\cal B}(J/\psi \to p \bar{p}) )$ used in our fit.
\item[14] Using ${\cal B}(h_c \to \eta_c \gamma)$.
\item[15] Measurement of ${\cal B}(B^0 \to h_c K^*(892)^0) {\cal B}(h_c \to \eta_c \gamma) / {\cal B}(B^+ \to \eta_c K^+)$ used in our fit.
\item[16] Using ${\cal B}(\psi(3770) \to D^0 \bar{D}^0)$.
\item[17] Using ${\cal B}(\psi(3770) \to D^+ D^-)$.
\end{tablenotes}
\end{threeparttable}
}
\end{center}
\end{table}

\hflavrowdef{\hflavrowdefafg}{${\cal B}(B^0 \to \chi_{c0} K^0)$}{{\setlength{\tabcolsep}{0pt}
}
\begin{table}[H]
\begin{center}
\begin{threeparttable}
\caption{Branching fractions for decays to $\chi_c$ and one kaon.}
\label{tab:b2charm_Bd_cc_overview3}
%
 \\
\hline
\hflavrowdefafg \\
 & & \\[-1ex]
\hflavrowdefafh \\
 & & \\[-1ex]
\hflavrowdefafi \\
 & & \\[-1ex]
\hflavrowdefafj \\
 & & \\[-1ex]
\hflavrowdefafk \\
 & & \\[-1ex]
\hflavrowdefafl \\
 & & \\[-1ex]
\hflavrowdefafm \\
 & & \\[-1ex]
\hflavrowdefafn \\
 & & \\[-1ex]
\hflavrowdefafo \\
 & & \\[-1ex]
\hflavrowdefafp \\
\hline
\hline
\end{tabular}
\begin{tablenotes}
\item[1] Measurement of ${\cal B}(B^0 \to \chi_{c1} K^*(892)^0) / {\cal B}(B^0 \to \chi_{c1} K^0)$ used in our fit.
\item[2] Using ${\cal B}(B^0 \to \chi_{c1} K^0)$.
\end{tablenotes}
\end{threeparttable}
\end{center}
\end{table}

\hflavrowdef{\hflavrowdefafq}{${\cal B}(B^0 \to J/\psi \pi^0)$}{{\setlength{\tabcolsep}{0pt}
}
\begin{table}[H]
\begin{center}
\resizebox{0.95\textwidth}{!}{
\begin{threeparttable}
\caption{Branching fractions for decays to charmonium and light mesons.}
\label{tab:b2charm_Bd_cc_overview4}
%
 \\
\hline
\hflavrowdefafq \\
 & & \\[-1ex]
\hflavrowdefafr \\
 & & \\[-1ex]
\hflavrowdefafs \\
 & & \\[-1ex]
\hflavrowdefaft \\
 & & \\[-1ex]
\hflavrowdefafu \\
 & & \\[-1ex]
\hflavrowdefafv \\
 & & \\[-1ex]
\hflavrowdefafw \\
 & & \\[-1ex]
\hflavrowdefafx \\
 & & \\[-1ex]
\hflavrowdefafy \\
 & & \\[-1ex]
\hflavrowdefafz \\
 & & \\[-1ex]
\hflavrowdefaga \\
 & & \\[-1ex]
\hflavrowdefagb \\
 & & \\[-1ex]
\hflavrowdefagc \\
 & & \\[-1ex]
\hflavrowdefagd \\
 & & \\[-1ex]
\hflavrowdefage \\
 & & \\[-1ex]
\hflavrowdefagf \\
 & & \\[-1ex]
\hflavrowdefagg \\
 & & \\[-1ex]
\hflavrowdefagh \\
\hline
\hline
\end{tabular}
\begin{tablenotes}
\item[*] Not included in PDG average.
\item[1] Measurement of ${\cal B}(B^0 \to \psi(2S) \pi^+ \pi^-) / {\cal B}(B^0 \to J/\psi \pi^+ \pi^-)$ used in our fit.
\item[2] Non resonant only: $K^0_S$, $\rho$ and $f_2$ contributions have been subtracted out.
\item[3] Measurement of ${\cal B}(B^0 \to J/\psi \omega(782)) / {\cal B}(B^0 \to J/\psi \rho^0(770))$ used in our fit.
\item[4] Measurement of ${\cal B}(B_s^0 \to J/\psi \eta) / {\cal B}(B^0 \to J/\psi \rho^0(770))$ used in our fit.
\item[5] Measurement of ${\cal B}(B_s^0 \to J/\psi \eta^\prime) / {\cal B}(B^0 \to J/\psi \rho^0(770))$ used in our fit.
\item[6] Using ${\cal B}(B_s^0 \to J/\psi \eta)$.
\item[7] Using ${\cal B}(B_s^0 \to J/\psi \eta^\prime)$.
\item[8] Using ${\cal B}(B^0 \to J/\psi \rho^0(770))$.
\item[9] Measurement of ${\cal B}(B^0 \to J/\psi f_0(980)) BR_f0_pi+_pi-$ used in our fit.
\item[10] Using ${\cal B}(f_1(1285) \to \pi^+ \pi^+ \pi^- \pi^-)$.
\item[11] Using ${\cal B}(B^0 \to J/\psi K^*(892)^0 \pi^+ \pi^-)$.
\item[12] Using ${\cal B}(B^0 \to J/\psi \pi^+ \pi^-)$.
\end{tablenotes}
\end{threeparttable}
}
\end{center}
\end{table}

\hflavrowdef{\hflavrowdefagi}{${\cal B}(B^0 \to J/\psi \gamma)$}{{\setlength{\tabcolsep}{0pt}
}
\begin{table}[H]
\begin{center}
\begin{threeparttable}
\caption{Branching fractions for decays to  $J/\psi$ and photons, baryons, or heavy mesons.}
\label{tab:b2charm_Bd_cc_overview5}
%
}
\begin{table}[H]
\begin{center}
\begin{threeparttable}
\caption{Branching fraction ratios for decays to $J/\psi$.}
\label{tab:b2charm_Bd_cc_overview6}
%
\end{threeparttable}
\end{center}
\end{table}

\subsubsection{Decays to charm baryons}
\label{sec:b2c:Bd_baryon}
Averages of $B^0$ decays to charm baryons are shown in Tables~\ref{tab:b2charm_Bd_baryon_overview1}--\ref{tab:b2charm_Bd_baryon_overview2}.
\hflavrowdef{\hflavrowdefagm}{${\cal B}(B^0 \to \bar{\Lambda}_c^- p \pi^0)$}{{\setlength{\tabcolsep}{0pt}%
}
\begin{table}[H]
\begin{center}
\begin{threeparttable}
\caption{Branching fractions for decays to charm baryons, I.}
\label{tab:b2charm_Bd_baryon_overview1}
%
 \\
\hline
\hflavrowdefagm \\
 & & \\[-1ex]
\hflavrowdefagn \\
 & & \\[-1ex]
\hflavrowdefago \\
 & & \\[-1ex]
\hflavrowdefagp \\
 & & \\[-1ex]
\hflavrowdefagq \\
 & & \\[-1ex]
\hflavrowdefagr \\
 & & \\[-1ex]
\hflavrowdefags \\
 & & \\[-1ex]
\hflavrowdefagt \\
 & & \\[-1ex]
\hflavrowdefagu \\
\hline
\hline
\end{tabular}
\begin{tablenotes}
\item[*] Not included in PDG average.
\item[1] At CL=95\,\%.
\item[2] Using ${\cal B}(B^0 \to D_s^+ D^-)$.
\item[3] Using ${\cal B}(\Xi_c^+ \to \Xi^- \pi^+ \pi^+)$.
\end{tablenotes}
\end{threeparttable}
\end{center}
\end{table}

\hflavrowdef{\hflavrowdefagv}{${\cal B}(B^0 \to \bar{\Lambda}_c^- p K^+ K^-)$}{{\setlength{\tabcolsep}{0pt}
}
\begin{table}[H]
\begin{center}
\begin{threeparttable}
\caption{Branching fractions for decays to charm baryons, II.}
\label{tab:b2charm_Bd_baryon_overview2}
%
 \\
\hline
\hflavrowdefagv \\
 & & \\[-1ex]
\hflavrowdefagw \\
 & & \\[-1ex]
\hflavrowdefagx \\
 & & \\[-1ex]
\hflavrowdefagy \\
 & & \\[-1ex]
\hflavrowdefagz \\
 & & \\[-1ex]
\hflavrowdefaha \\
 & & \\[-1ex]
\hflavrowdefahb \\
 & & \\[-1ex]
\hflavrowdefahc \\
 & & \\[-1ex]
\hflavrowdefahd \\
 & & \\[-1ex]
\hflavrowdefahe \\
\hline
\hline
\end{tabular}
\begin{tablenotes}
\item[1] Measurement of ${\cal B}(B^+ \to \bar{\Lambda}_c^- p \pi^+) / {\cal B}(B^0 \to \bar{\Lambda}_c^- p)$ used in our fit.
\item[2] Using ${\cal B}(\Lambda_c^+ \to p K^- \pi^+)$.
\end{tablenotes}
\end{threeparttable}
\end{center}
\end{table}

\subsubsection{Decays to exotic states}
\label{sec:b2c:Bd_other}
Averages of $B^0$ decays to exotic states are shown in Tables~\ref{tab:b2charm_Bd_other_overview1}--\ref{tab:b2charm_Bd_other_overview4}.
\hflavrowdef{\hflavrowdefahf}{${\cal B}(B^0 \to X(3872) K^0)$}{{\setlength{\tabcolsep}{0pt}
}
\begin{table}[H]
\begin{center}
\begin{threeparttable}
\caption{Branching fractions for decays to $X(3872)$.}
\label{tab:b2charm_Bd_other_overview1}
%
 \\
\hline
\hflavrowdefahf \\
 & & \\[-1ex]
\hflavrowdefahg \\
 & & \\[-1ex]
\hflavrowdefahh \\
 & & \\[-1ex]
\hflavrowdefahi \\
\hline
\hline
\end{tabular}
\begin{tablenotes}
\item[*] Not included in PDG average.
\item[1] Using ${\cal B}(B^+ \to X(3872) K^+)$.
\item[2] Using ${\cal B}(X(3872) \to J/\psi \pi^+ \pi^-)$.
\item[3] Using ${\cal B}(X(3872) \to J/\psi \omega(782))$.
\item[4] Using ${\cal B}(X(3872) \to \bar{D}^*(2007)^0 D^0)$.
\end{tablenotes}
\end{threeparttable}
\end{center}
\end{table}

\hflavrowdeflong{\hflavrowdefahj}{${\cal B}(B^0 \to \psi_2(3823) K^0) \times {\cal B}(\psi_2(3823) \to \chi_{c1} \gamma)$}{{\setlength{\tabcolsep}{0pt}
}
\begin{table}[H]
\begin{center}
\begin{threeparttable}
\caption{Branching fractions for decays to neutral states other than $X(3872)$.}
\label{tab:b2charm_Bd_other_overview2}
%
\begin{tablenotes}
\item[*] Not included in PDG average.
\end{tablenotes}
\end{threeparttable}
\end{center}
\end{table}

\hflavrowdef{\hflavrowdefaho}{${\cal B}(B^0 \to X(3872)^- K^+)$}{{\setlength{\tabcolsep}{0pt}%
}
\begin{table}[H]
\begin{center}
\begin{threeparttable}
\caption{Branching fractions for decays to charged states.}
\label{tab:b2charm_Bd_other_overview3}
%
 \\
\hline
\hflavrowdefaho \\
 & & \\[-1ex]
\hflavrowdefahp \\
 & & \\[-1ex]
\hflavrowdefahq \\
 & & \\[-1ex]
\hflavrowdefahr \\
 & & \\[-1ex]
\hflavrowdefahs \\
 & & \\[-1ex]
\hflavrowdefaht \\
\hline
\hline
\end{tabular}
\begin{tablenotes}
\item[*] Not included in PDG average.
\item[1] The quoted amplitude fraction is multiplied by $\mathcal{B}(B^0 \to \psi(2S) K^+ \pi^-) = (5.80 \pm 0.39) \times 10^{-4}$
\end{tablenotes}
\end{threeparttable}
\end{center}
\end{table}

\hflavrowdef{\hflavrowdefahu}{$\dfrac{{\cal B}(B^0 \to Y(3940) K^0)}{{\cal B}(B^+ \to Y(3940) K^+)}$}{{\setlength{\tabcolsep}{0pt}\begin{tabular}{m{6.5em}l} {BaBar \cite{delAmoSanchez:2010jr}\tnote{}\hphantom{\textsuperscript{}}} & { $0.7\,^{+0.4}_{-0.3} \pm0.1$ } \\ \end{tabular}}}{\begin{tabular}{l} $0.70\,^{+0.41}_{-0.32}$\\ \end{tabular}}
\begin{table}[H]
\begin{center}
\begin{threeparttable}
\caption{Branching fraction ratios for decays to exotic states.}
\label{tab:b2charm_Bd_other_overview4}
\begin{tabular}{ Sl l l }
\hline
\hline
\textbf{Parameter} & \textbf{Measurements} & \begin{tabular}{l}\textbf{Average}\end{tabular} \\
\hline
\hflavrowdefahu \\
\hline
\hline
\end{tabular}
\end{threeparttable}
\end{center}
\end{table}

\subsection{Decays of $B^+$ mesons}
\label{sec:b2c:Bu}
Measurements of $B^+$ decays to charmed hadrons are summarized in Sections~\ref{sec:b2c:Bu_D} to~\ref{sec:b2c:Bu_other}.

\subsubsection{Decays to a single open charm meson}
\label{sec:b2c:Bu_D}
Averages of $B^+$ decays to a single open charm meson are shown in Tables~\ref{tab:b2charm_Bu_D_overview1}--\ref{tab:b2charm_Bu_D_overview7}.
\hflavrowdef{\hflavrowdefahv}{${\cal B}(B^+ \to \bar{D}^0 \pi^+)$}{{\setlength{\tabcolsep}{0pt}
}
\begin{table}[H]
\begin{center}
\begin{threeparttable}
\caption{Branching fractions for decays to a $\bar{D}^{(*)}$ meson and one or more light mesons.}
\label{tab:b2charm_Bu_D_overview1}
%
 \\
\hline
\hflavrowdefahv \\
 & & \\[-1ex]
\hflavrowdefahw \\
 & & \\[-1ex]
\hflavrowdefahx \\
 & & \\[-1ex]
\hflavrowdefahy \\
 & & \\[-1ex]
\hflavrowdefahz \\
 & & \\[-1ex]
\hflavrowdefaia \\
 & & \\[-1ex]
\hflavrowdefaib \\
 & & \\[-1ex]
\hflavrowdefaic \\
\hline
\hline
\end{tabular}
\begin{tablenotes}
\item[*] Not included in PDG average.
\item[1] Using ${\cal B}(B^0 \to D^- \pi^+)$.
\item[2] Using non-CP modes for the $D^0$.
\item[3] Measurement of ${\cal B}(B^+ \to \bar{D}^0 K^+) / {\cal B}(B^+ \to \bar{D}^0 \pi^+)$ used in our fit.
\item[4] Measurement of ${\cal B}(B^+ \to \bar{D}^0 \pi^+ \pi^+ \pi^-) / {\cal B}(B^+ \to \bar{D}^0 \pi^+)$ used in our fit.
\item[5] Measurement of ${\cal B}(B^+ \to \bar{D}^*(2007)^0 K^+) / {\cal B}(B^+ \to \bar{D}^*(2007)^0 \pi^+)$ used in our fit.
\item[6] Measurement of ${\cal B}(B^+ \to D^*(2010)^- K^+ \pi^+) / {\cal B}(B^+ \to D^*(2010)^- \pi^+ \pi^+)$ used in our fit.
\end{tablenotes}
\end{threeparttable}
\end{center}
\end{table}

\hflavrowdef{\hflavrowdefaid}{${\cal B}(B^+ \to \bar{D}^0 K^+)$}{{\setlength{\tabcolsep}{0pt}
}
\begin{table}[H]
\begin{center}
\resizebox{0.9\textwidth}{!}{
\begin{threeparttable}
\caption{Branching fractions for decays to a $\bar{D}^{(*)}$ meson and one or more kaons.}
\label{tab:b2charm_Bu_D_overview2}
%
 \\
\hline
\hflavrowdefaid \\
 & & \\[-1ex]
\hflavrowdefaie \\
 & & \\[-1ex]
\hflavrowdefaif \\
 & & \\[-1ex]
\hflavrowdefaig \\
 & & \\[-1ex]
\hflavrowdefaih \\
 & & \\[-1ex]
\hflavrowdefaii \\
 & & \\[-1ex]
\hflavrowdefaij \\
 & & \\[-1ex]
\hflavrowdefaik \\
 & & \\[-1ex]
\hflavrowdefail \\
 & & \\[-1ex]
\hflavrowdefaim \\
 & & \\[-1ex]
\hflavrowdefain \\
 & & \\[-1ex]
\hflavrowdefaio \\
 & & \\[-1ex]
\hflavrowdefaip \\
\hline
\hline
\end{tabular}
\begin{tablenotes}
\scriptsize
\item[\dag] Preliminary result.
\item[*] Not included in PDG average.
\item[1] Using ${\cal B}(B^+ \to \bar{D}^0 \pi^+)$.
\item[2] Using non-CP modes for the $D^0$.
\item[3] Measurement of ${\cal B}(B^+ \to D^0 K^+) / {\cal B}(B^+ \to \bar{D}^0 K^+)$ used in our fit.
\item[4] Statistical and systematic uncertainties are combined
\item[5] Measurement of ${\cal B}(B^+ \to \bar{D}^*(2007)^0 K^+) / {\cal B}(B^+ \to \bar{D}^0 K^+)$ used in our fit.
\item[6] Using ${\cal B}(B^+ \to \bar{D}^0 \pi^+ \pi^+ \pi^-)$.
\item[7] Measurement of ${\cal B}(B^+ \to \bar{D}^0 K^+ \bar{K}^0) / 2$ used in our fit.
\item[8] Using ${\cal B}(B^+ \to \bar{D}^*(2007)^0 \pi^+)$.
\item[9] Using ${\cal B}(B^+ \to \bar{D}^0 K^+)$.
\item[10] Measurement of ${\cal B}(B^+ \to \bar{D}^*(2007)^0 K^+ \bar{K}^0) / 2$ used in our fit.
\item[11] Using ${\cal B}(B^+ \to D^*(2010)^- \pi^+ \pi^+)$.
\end{tablenotes}
\end{threeparttable}
}
\end{center}
\end{table}

\hflavrowdef{\hflavrowdefaiq}{${\cal B}(B^+ \to D^0 K^+)$}{{\setlength{\tabcolsep}{0pt}
}
\begin{table}[H]
\begin{center}
\begin{threeparttable}
\caption{Branching fractions for decays to a $D^{(*)}$ meson.}
\label{tab:b2charm_Bu_D_overview3}
%
\begin{tablenotes}
\item[*] Not included in PDG average.
\item[1] Using ${\cal B}(B^+ \to \bar{D}^0 K^+)$.
\end{tablenotes}
\end{threeparttable}
\end{center}
\end{table}

\hflavrowdef{\hflavrowdefaiw}{${\cal{B}} ( B^{+} \to \bar{D}^{**0} \pi^{+} )$\tnote{1}\hphantom{\textsuperscript{1}}}{{\setlength{\tabcolsep}{0pt}%
}
\begin{table}[H]
\begin{center}
\begin{threeparttable}
\caption{Branching fractions for decays to excited $D$ mesons.}
\label{tab:b2charm_Bu_D_overview4}
%
 \\
\hline
\hflavrowdefaiw \\
 & & \\[-1ex]
\hflavrowdefaix \\
 & & \\[-1ex]
\hflavrowdefaiy \\
 & & \\[-1ex]
\hflavrowdefaiz \\
 & & \\[-1ex]
\hflavrowdefaja \\
 & & \\[-1ex]
\hflavrowdefajb \\
 & & \\[-1ex]
\hflavrowdefajc \\
 & & \\[-1ex]
\hflavrowdefajd \\
 & & \\[-1ex]
\hflavrowdefaje \\
 & & \\[-1ex]
\hflavrowdefajf \\
 & & \\[-1ex]
\hflavrowdefajg \\
\hline
\hline
\end{tabular}
\begin{tablenotes}
\footnotesize
\item[*] Not included in PDG average.
\item[1] $D^{**}$ refers to the sum of all the non-strange charm meson states with masses in the range 2.2 - 2.8 GeV$/c^2$
\item[2] Using ${\cal B}(B^+ \to \bar{D}^0 \pi^+ \pi^+ \pi^-)$.
\item[3] Non-$D^*$
\end{tablenotes}
\end{threeparttable}
\end{center}
\end{table}

\hflavrowdef{\hflavrowdefajh}{${\cal B}(B^+ \to D_s^- K^+ \pi^+)$}{{\setlength{\tabcolsep}{0pt}
}
\begin{table}[H]
\begin{center}
\begin{threeparttable}
\caption{Branching fractions for decays to $D_s^{(*)}$ mesons.}
\label{tab:b2charm_Bu_D_overview5}
%
 \\
\hline
\hflavrowdefajh \\
 & & \\[-1ex]
\hflavrowdefaji \\
 & & \\[-1ex]
\hflavrowdefajj \\
 & & \\[-1ex]
\hflavrowdefajk \\
 & & \\[-1ex]
\hflavrowdefajl \\
 & & \\[-1ex]
\hflavrowdefajm \\
 & & \\[-1ex]
\hflavrowdefajn \\
 & & \\[-1ex]
\hflavrowdefajo \\
\hline
\hline
\end{tabular}
\begin{tablenotes}
\item[1] Measurement of ${\cal B}(B^+ \to D_s^+ K^+ K^-) / ( {\cal B}(B^+ \to D_s^+ \bar{D}^0) {\cal B}(D^0 \to K^+ K^-) )$ used in our fit.
\end{tablenotes}
\end{threeparttable}
\end{center}
\end{table}

\hflavrowdef{\hflavrowdefajp}{${\cal B}(B^+ \to \bar{D}^0 p \bar{p} \pi^+)$}{{\setlength{\tabcolsep}{0pt}
}
\begin{table}[H]
\begin{center}
\begin{threeparttable}
\caption{Branching fractions for decays to a $D^{(*)}$ meson and baryons.}
\label{tab:b2charm_Bu_D_overview6}
%
}
\begin{table}[H]
\begin{center}
\begin{threeparttable}
\caption{Branching fractions of lepton number violating decays.}
\label{tab:b2charm_Bu_D_overview7}
%
\end{threeparttable}
\end{center}
\end{table}

\subsubsection{Decays to two open charm mesons}
\label{sec:b2c:Bu_DD}
Averages of $B^+$ decays to two open charm mesons are shown in Tables~\ref{tab:b2charm_Bu_DD_overview1}--\ref{tab:b2charm_Bu_DD_overview4}.
\hflavrowdef{\hflavrowdefaka}{${\cal B}(B^+ \to D^+ \bar{D}^0)$}{{\setlength{\tabcolsep}{0pt}%
}
\begin{table}[H]
\begin{center}
\begin{threeparttable}
\caption{Branching fractions for decays to $D^{(*)-}D^{(*)0}$.}
\label{tab:b2charm_Bu_DD_overview1}
%
 \\
\hline
\hflavrowdefaka \\
 & & \\[-1ex]
\hflavrowdefakb \\
 & & \\[-1ex]
\hflavrowdefakc \\
 & & \\[-1ex]
\hflavrowdefakd \\
\hline
\hline
\end{tabular}
\begin{tablenotes}
\item[*] Not included in PDG average.
\item[1] Measurement of $f_c \times {\cal B}(B_c^+ \to D^+ \bar{D}^0) / ( f_u {\cal B}(B^+ \to D^+ \bar{D}^0) )$ used in our fit.
\item[2] Measurement of $f_c \times {\cal B}(B_c^+ \to D^+ D^0) / ( f_u {\cal B}(B^+ \to D^+ \bar{D}^0) )$ used in our fit.
\item[3] Measurement of $f_c \times ( {\cal{B}} ( B_c^- \to D^{*-} D^0 ) \times {\cal{B}} (D^{*-}\to D^- (\pi^0,\gamma)) + {\cal{B}} (B_c^{-}\to D^- D^{*0}) ) / ( f_u {\cal B}(B^+ \to D^+ \bar{D}^0) )$ used in our fit.
\item[4] Measurement of $f_c \times ( {\cal{B}} ( B_c^- \to D^{*-} \bar{D}^0 ) \times {\cal{B}} (D^{*-}\to D^- (\pi^0,\gamma)) + {\cal{B}} (B_c^{-}\to D^- \bar{D}^{*0}) ) / ( f_u {\cal B}(B^+ \to D^+ \bar{D}^0) )$ used in our fit.
\item[5] Measurement of $f_c \times {\cal B}(B_c^+ \to D^*(2010)^+ \bar{D}^*(2007)^0) / ( f_u {\cal B}(B^+ \to D^+ \bar{D}^0) )$ used in our fit.
\item[6] Measurement of $f_c \times {\cal B}(B_c^+ \to D^*(2010)^+ D^*(2007)^0) / ( f_u {\cal B}(B^+ \to D^+ \bar{D}^0) )$ used in our fit.
\item[7] Measurement of ${\cal B}(B^+ \to D^+ \bar{D}^0) {\cal B}(D^+ \to K^- \pi^+ \pi^+) / ( {\cal B}(B^+ \to D_s^+ \bar{D}^0) {\cal B}(D_s^+ \to K^- K^+ \pi^+) )$ used in our fit.
\item[8] Measurement of ${\cal B}(B^+ \to D^*(2010)^+ \bar{D}^0) {\cal B}(D^*(2010)^+ \to D^0 \pi^+) {\cal B}(D^0 \to K^- \pi^+) / ( {\cal B}(B^+ \to D^+ \bar{D}^0) {\cal B}(D^+ \to K^- \pi^+ \pi^+) )$ used in our fit.
\end{tablenotes}
\end{threeparttable}
\end{center}
\end{table}

\hflavrowdef{\hflavrowdefake}{${\cal B}(B^+ \to \bar{D}^*(2007)^0 D^0 K^+)$}{{\setlength{\tabcolsep}{0pt}
}
\begin{table}[H]
\begin{center}
\begin{threeparttable}
\caption{Branching fractions for decays to two $D$ mesons and a kaon.}
\label{tab:b2charm_Bu_DD_overview2}
%
 \\
\hline
\hflavrowdefake \\
 & & \\[-1ex]
\hflavrowdefakf \\
 & & \\[-1ex]
\hflavrowdefakg \\
 & & \\[-1ex]
\hflavrowdefakh \\
 & & \\[-1ex]
\hflavrowdefaki \\
 & & \\[-1ex]
\hflavrowdefakj \\
 & & \\[-1ex]
\hflavrowdefakk \\
 & & \\[-1ex]
\hflavrowdefakl \\
 & & \\[-1ex]
\hflavrowdefakm \\
 & & \\[-1ex]
\hflavrowdefakn \\
 & & \\[-1ex]
\hflavrowdefako \\
 & & \\[-1ex]
\hflavrowdefakp \\
 & & \\[-1ex]
\hflavrowdefakq \\
 & & \\[-1ex]
\hflavrowdefakr \\
\hline
\hline
\end{tabular}
\begin{tablenotes}
\item[*] Not included in PDG average.
\item[1] Measurement of ${\cal B}(B^+ \to D_s^+ D_s^- K^+) / {\cal B}(B^+ \to D^+ D^- K^+)$ used in our fit.
\item[2] Using ${\cal B}(B^+ \to D^+ D^- K^+)$.
\end{tablenotes}
\end{threeparttable}
\end{center}
\end{table}

\hflavrowdef{\hflavrowdefaks}{${\cal B}(B^+ \to D_s^+ \bar{D}^0)$}{{\setlength{\tabcolsep}{0pt}
}
\begin{table}[H]
\begin{center}
\begin{threeparttable}
\caption{Branching fractions for decays to $D_s^{(*)-}D^{(*)+}$.}
\label{tab:b2charm_Bu_DD_overview3}
%
 \\
\hline
\hflavrowdefaks \\
 & & \\[-1ex]
\hflavrowdefakt \\
 & & \\[-1ex]
\hflavrowdefaku \\
 & & \\[-1ex]
\hflavrowdefakv \\
\hline
\hline
\end{tabular}
\begin{tablenotes}
\item[\dag] Preliminary result.
\item[*] Not included in PDG average.
\item[1] Using ${\cal B}(B^0 \to D_s^+ D^-)$.
\item[2] Using ${\cal B}(D_s^+ \to \phi(1020) \pi^+)$.
\item[3] Measurement of ${\cal B}(B_s^0 \to D^0 \bar{D}^0) / {\cal B}(B^+ \to D_s^+ \bar{D}^0)$ used in our fit.
\item[4] Measurement of ${\cal B}(B^0 \to D^0 \bar{D}^0) / {\cal B}(B^+ \to D_s^+ \bar{D}^0)$ used in our fit.
\item[5] Measurement of ${\cal B}(B^+ \to D_s^+ K^+ K^-) / ( {\cal B}(B^+ \to D_s^+ \bar{D}^0) {\cal B}(D^0 \to K^+ K^-) )$ used in our fit.
\item[6] Measurement of $f_c \times {\cal B}(B_c^+ \to D_s^+ \bar{D}^0) / ( f_u {\cal B}(B^+ \to D_s^+ \bar{D}^0) )$ used in our fit.
\item[7] Measurement of $f_c \times {\cal B}(B_c^+ \to D_s^+ D^0) / ( f_u {\cal B}(B^+ \to D_s^+ \bar{D}^0) )$ used in our fit.
\item[8] Measurement of $f_c \times {\cal B}(B_c^+ \to D_s^{*+} \bar{D}^0  +  D_s^+ \bar{D}^*(2007)^0) / ( f_u {\cal B}(B^+ \to D_s^+ \bar{D}^0) )$ used in our fit.
\item[9] Measurement of $f_c \times {\cal B}(B_c^+ \to D_s^{*+} D^0  +  D_s^+ D^*(2007)^0) / ( f_u {\cal B}(B^+ \to D_s^+ \bar{D}^0) )$ used in our fit.
\item[10] Measurement of $f_c \times {\cal B}(B_c^+ \to D_s^{*+} \bar{D}^*(2007)^0) / ( f_u {\cal B}(B^+ \to D_s^+ \bar{D}^0) )$ used in our fit.
\item[11] Measurement of $f_c \times {\cal B}(B_c^+ \to D_s^{*+} D^*(2007)^0) / ( f_u {\cal B}(B^+ \to D_s^+ \bar{D}^0) )$ used in our fit.
\item[12] Measurement of ${\cal B}(B^+ \to D^+ \bar{D}^0) {\cal B}(D^+ \to K^- \pi^+ \pi^+) / ( {\cal B}(B^+ \to D_s^+ \bar{D}^0) {\cal B}(D_s^+ \to K^- K^+ \pi^+) )$ used in our fit.
\end{tablenotes}
\end{threeparttable}
\end{center}
\end{table}

\hflavrowdef{\hflavrowdefakw}{${\cal B}(B^+ \to D_{s0}^*(2317)^+ \bar{D}^0)$}{{\setlength{\tabcolsep}{0pt}
}
\begin{table}[H]
\begin{center}
\begin{threeparttable}
\caption{Branching fractions for decays to excited $D_s$ mesons.}
\label{tab:b2charm_Bu_DD_overview4}
%
 \\
\hline
\hflavrowdefakw \\
 & & \\[-1ex]
\hflavrowdefakx \\
 & & \\[-1ex]
\hflavrowdefaky \\
 & & \\[-1ex]
\hflavrowdefakz \\
 & & \\[-1ex]
\hflavrowdefala \\
 & & \\[-1ex]
\hflavrowdefalb \\
\hline
\hline
\end{tabular}
\begin{tablenotes}
\item[*] Not included in PDG average.
\item[1] Using ${\cal B}(D_{s0}^*(2317)^+ \to D_s^+ \pi^0)$.
\item[2] Using ${\cal B}(D_{s1}(2460)^+ \to D_s^+ \gamma)$.
\item[3] Using ${\cal B}(D_{s1}(2460)^+ \to D_s^{*+} \pi^0)$.
\item[4] Using ${\cal B}(D_{s1}(2460)^+ \to D_s^+ \pi^+ \pi^-)$.
\item[5] Using ${\cal B}(D_{s1}(2536)^+ \to D^*(2007)^0 K^+)$.
\item[6] Using ${\cal B}(D_{s1}(2536)^+ \to D^*(2010)^+ K^0)$.
\item[7] Measurement of ${\cal B}(B^+ \to D_{s1}(2536)^+ \bar{D}^0) ( {\cal B}(D_{s1}(2536)^+ \to D^*(2007)^0 K^+) + {\cal B}(D_{s1}(2536)^+ \to D^*(2010)^+ K^0) )$ used in our fit.
\end{tablenotes}
\end{threeparttable}
\end{center}
\end{table}

\subsubsection{Decays to charmonium states}
\label{sec:b2c:Bu_cc}
Averages of $B^+$ decays to charmonium states are shown in Tables~\ref{tab:b2charm_Bu_cc_overview1}--\ref{tab:b2charm_Bu_cc_overview7}.
\hflavrowdef{\hflavrowdefalc}{${\cal B}(B^+ \to J/\psi K^+)$}{{\setlength{\tabcolsep}{0pt}
}
\begin{table}[H]
\begin{center}
\resizebox{0.85\textwidth}{!}{
\begin{threeparttable}
\caption{Branching fractions for decays to $J/\psi$ and one kaon.}
\label{tab:b2charm_Bu_cc_overview1}
%
 \\
\hline
\hflavrowdefalc \\
 & & \\[-1ex]
\hflavrowdefald \\
 & & \\[-1ex]
\hflavrowdefale \\
 & & \\[-1ex]
\hflavrowdefalf \\
 & & \\[-1ex]
\hflavrowdefalg \\
 & & \\[-1ex]
\hflavrowdefalh \\
 & & \\[-1ex]
\hflavrowdefali \\
 & & \\[-1ex]
\hflavrowdefalj \\
\hline
\hline
\end{tabular}
\begin{tablenotes}
\scriptsize
\item[*] Not included in PDG average.
\item[1] Using ${\cal B}(J/\psi \to p \bar{p} \pi^+ \pi^-)$.
\item[2] Using ${\cal B}(J/\psi \to p \bar{p})$.
\item[3] Using ${\cal B}(J/\psi \to \Lambda^0 \bar{\Lambda}^0)$.
\item[4] Measurement of ${\cal B}(B^+ \to \eta_c K^+) / {\cal B}(B^+ \to J/\psi K^+)$ used in our fit.
\item[5] Measurement of ${\cal B}(B^+ \to J/\psi \pi^+) / {\cal B}(B^+ \to J/\psi K^+)$ used in our fit.
\item[6] Measurement of ${\cal B}(B^+ \to J/\psi K^*(892)^+) / {\cal B}(B^+ \to J/\psi K^+)$ used in our fit.
\item[7] Measurement of ${\cal B}(B^+ \to \chi_{c0} K^+) / {\cal B}(B^+ \to J/\psi K^+)$ used in our fit.
\item[8] Measurement of ${\cal B}(B^+ \to J/\psi K_1(1270)^+) / {\cal B}(B^+ \to J/\psi K^+)$ used in our fit.
\item[9] Measurement of ${\cal B}(B^0 \to J/\psi K_1(1270)^0) / {\cal B}(B^+ \to J/\psi K^+)$ used in our fit.
\item[10] Measurement of $f_c \times {\cal B}(B_c^+ \to J/\psi \pi^+) / ( f_u {\cal B}(B^+ \to J/\psi K^+) )$ used in our fit.
\item[11] Measurement of ${\cal B}(B^+ \to \psi(2S) K^+) / {\cal B}(B^+ \to J/\psi K^+)$ used in our fit.
\item[12] $\sqrt{s} = 7$ TeV, $p_T > 4$ GeV and $2.5 < y < 4.5$
\item[13] Measurement of $( {\cal B}(B^+ \to \eta_c K^+) {\cal B}(\eta_c \to p \bar{p}) ) / ( {\cal B}(B^+ \to J/\psi K^+) {\cal B}(J/\psi \to p \bar{p}) )$ used in our fit.
\item[14] $\sqrt{s} = 8$ TeV, $0 < p_T < 20$ GeV and $2.0 < y < 4.5$
\item[15] Using ${\cal B}(B^+ \to J/\psi K^+)$.
\item[16] Measurement of ${\cal B}(B^+ \to J/\psi K_1(1400)^+) / {\cal B}(B^+ \to J/\psi K_1(1270)^+)$ used in our fit.
\item[17] Using ${\cal B}(B^+ \to J/\psi K_1(1270)^+)$.
\item[18] Measurement of ${\cal B}(B^0 \to J/\psi \omega(782) K^0) / {\cal B}(B^+ \to J/\psi \omega(782) K^+)$ used in our fit.
\item[19] Measurement of ${\cal B}(B^+ \to \chi_{c1}(4140) K^+) \times {\cal B}(\chi_{c1}(4140) \to J/\psi \phi(1020)) / {\cal B}(B^+ \to J/\psi \phi(1020) K^+)$ used in our fit.
\item[20] The quoted value is a fit fraction from a Dalitz plot fit.
\item[21] Measurement of ${\cal B}(B^+ \to \chi_{c1}(4274) K^+) \times {\cal B}(\chi_{c1}(4274) \to J/\psi \phi(1020)) / {\cal B}(B^+ \to J/\psi \phi(1020) K^+)$ used in our fit.
\end{tablenotes}
\end{threeparttable}
}
\end{center}
\end{table}

\hflavrowdef{\hflavrowdefalk}{${\cal B}(B^+ \to \psi(2S) K^+)$}{{\setlength{\tabcolsep}{0pt}
}
\begin{table}[H]
\begin{center}
\resizebox{0.75\textwidth}{!}{
\begin{threeparttable}
\caption{Branching fractions for decays to charmonium other than $J/\psi$ and one kaon.}
\label{tab:b2charm_Bu_cc_overview2}
%
 \\
\hline
\hflavrowdefalk \\
 & & \\[-1ex]
\hflavrowdefall \\
 & & \\[-1ex]
\hflavrowdefalm \\
 & & \\[-1ex]
\hflavrowdefaln \\
 & & \\[-1ex]
\hflavrowdefalo \\
 & & \\[-1ex]
\hflavrowdefalp \\
 & & \\[-1ex]
\hflavrowdefalq \\
 & & \\[-1ex]
\hflavrowdefalr \\
 & & \\[-1ex]
\hflavrowdefals \\
\hline
\hline
\end{tabular}
\begin{tablenotes}
\tiny
\item[*] Not included in PDG average.
\item[1] Using ${\cal B}(B^+ \to J/\psi K^+)$.
\item[2] Measurement of ${\cal B}(B^+ \to \psi(2S) K^*(892)^+) / {\cal B}(B^+ \to \psi(2S) K^+)$ used in our fit.
\item[3] Measurement of ${\cal B}(B^+ \to J/\psi \eta^\prime K^+) / {\cal B}(B^+ \to \psi(2S) K^+)$ used in our fit.
\item[4] Using ${\cal B}(B^+ \to \psi(2S) K^+)$.
\item[5] Using ${\cal B}(\psi(3770) \to D^0 \bar{D}^0)$.
\item[6] Using ${\cal B}(\psi(3770) \to D^+ D^-)$.
\item[7] Assumed $\mathcal{B}(\psi(3770) \to D^+ D^-) + \mathcal{B}(\psi(3770) \to D^0 \bar{D}^0) = 1$.
\item[8] Using ${\cal{B}} ( \eta_c \to K \bar{K} \pi)$.
\item[9] Using ${\cal B}(\eta_c \to p \bar{p})$.
\item[10] Using ${\cal B}(\eta_c \to p \bar{p} \pi^+ \pi^-)$.
\item[11] Using ${\cal B}(\eta_c \to \Lambda^0 \bar{\Lambda}^0)$.
\item[12] Measurement of ${\cal B}(B^0 \to \eta_c K^0) / {\cal B}(B^+ \to \eta_c K^+)$ used in our fit.
\item[13] Measurement of ${\cal B}(B^0 \to \eta_c K^*(892)^0) / {\cal B}(B^+ \to \eta_c K^+)$ used in our fit.
\item[14] Measurement of ${\cal B}(B^0 \to h_c K^*(892)^0) {\cal B}(h_c \to \eta_c \gamma) / {\cal B}(B^+ \to \eta_c K^+)$ used in our fit.
\item[15] Measurement of ${\cal B}(B^+ \to h_c K^+) {\cal B}(h_c \to \eta_c \gamma) / {\cal B}(B^+ \to \eta_c K^+)$ used in our fit.
\item[16] Measurement of $( {\cal B}(B^+ \to \eta_c K^+) {\cal B}(\eta_c \to p \bar{p}) ) / ( {\cal B}(B^+ \to J/\psi K^+) {\cal B}(J/\psi \to p \bar{p}) )$ used in our fit.
\item[17] Calculated using $\mathcal{B}(\eta_c \to p \bar{p})$
\item[18] Using ${\cal{B}} ( \eta_c(2S) \to K \bar{K} \pi$.
\end{tablenotes}
\end{threeparttable}
}
\end{center}
\end{table}

\hflavrowdef{\hflavrowdefalt}{${\cal B}(B^+ \to \chi_{c0} K^+)$}{{\setlength{\tabcolsep}{0pt}
}
\begin{table}[H]
\begin{center}
\begin{threeparttable}
\caption{Branching fractions for decays to $\chi_c$ and one kaon.}
\label{tab:b2charm_Bu_cc_overview3}
%
 \\
\hline
\hflavrowdefalt \\
 & & \\[-1ex]
\hflavrowdefalu \\
 & & \\[-1ex]
\hflavrowdefalv \\
 & & \\[-1ex]
\hflavrowdefalw \\
 & & \\[-1ex]
\hflavrowdefalx \\
 & & \\[-1ex]
\hflavrowdefaly \\
 & & \\[-1ex]
\hflavrowdefalz \\
 & & \\[-1ex]
\hflavrowdefama \\
 & & \\[-1ex]
\hflavrowdefamb \\
 & & \\[-1ex]
\hflavrowdefamc \\
 & & \\[-1ex]
\hflavrowdefamd \\
\hline
\hline
\end{tabular}
\begin{tablenotes}
\item[1] Using ${\cal B}(B^+ \to J/\psi K^+)$.
\item[2] Measurement of ${\cal B}(B^+ \to \chi_{c1} K^*(892)^+) / {\cal B}(B^+ \to \chi_{c1} K^+)$ used in our fit.
\item[3] Measurement of ${\cal B}(B^+ \to \chi_{c1} \pi^+) / {\cal B}(B^+ \to \chi_{c1} K^+)$ used in our fit.
\item[4] Using ${\cal B}(B^+ \to \chi_{c1} K^+)$.
\end{tablenotes}
\end{threeparttable}
\end{center}
\end{table}

\hflavrowdef{\hflavrowdefame}{${\cal B}(B^+ \to J/\psi \pi^+)$}{{\setlength{\tabcolsep}{0pt}
}
\begin{table}[H]
\begin{center}
\begin{threeparttable}
\caption{Branching fractions for decays to charmonium and light mesons.}
\label{tab:b2charm_Bu_cc_overview4}
%
 \\
\hline
\hflavrowdefame \\
 & & \\[-1ex]
\hflavrowdefamf \\
 & & \\[-1ex]
\hflavrowdefamg \\
 & & \\[-1ex]
\hflavrowdefamh \\
 & & \\[-1ex]
\hflavrowdefami \\
 & & \\[-1ex]
\hflavrowdefamj \\
\hline
\hline
\end{tabular}
\begin{tablenotes}
\item[1] Using ${\cal B}(B^+ \to J/\psi K^+)$.
\item[2] Non resonant only.
\item[3] Computed using PDG2004 value of $\mathcal{B}(\chi_{c0} \to \pi \pi)$.
\item[4] Using ${\cal B}(B^+ \to \chi_{c1} K^+)$.
\end{tablenotes}
\end{threeparttable}
\end{center}
\end{table}

\hflavrowdef{\hflavrowdefamk}{${\cal B}(B^+ \to J/\psi D^+)$}{{\setlength{\tabcolsep}{0pt}
}
\begin{table}[H]
\begin{center}
\begin{threeparttable}
\caption{Branching fractions for decays to $J/\psi$ and a heavy mesons.}
\label{tab:b2charm_Bu_cc_overview5}
%
}
\begin{table}[H]
\begin{center}
\begin{threeparttable}
\caption{Branching fractions for decays to $J/\psi$ and baryons.}
\label{tab:b2charm_Bu_cc_overview6}
%
}
\begin{table}[H]
\begin{center}
\begin{threeparttable}
\caption{Direct CP violation parameters for decays to $J/\psi$.}
\label{tab:b2charm_Bu_cc_overview7}
%
\end{threeparttable}
\end{center}
\end{table}

\subsubsection{Decays to charm baryons}
\label{sec:b2c:Bu_baryon}
Averages of $B^+$ decays to charm baryons are shown in Table~\ref{tab:b2charm_Bu_baryon_overview1}.
\hflavrowdef{\hflavrowdefams}{${\cal B}(B^+ \to \bar{\Lambda}_c^- p \pi^+)$}{{\setlength{\tabcolsep}{0pt}%
}
\begin{table}[H]
\begin{center}
\begin{threeparttable}
\caption{Branching fractions for decays to charm baryons.}
\label{tab:b2charm_Bu_baryon_overview1}
%
 \\
\hline
\hflavrowdefams \\
 & & \\[-1ex]
\hflavrowdefamt \\
 & & \\[-1ex]
\hflavrowdefamu \\
 & & \\[-1ex]
\hflavrowdefamv \\
 & & \\[-1ex]
\hflavrowdefamw \\
 & & \\[-1ex]
\hflavrowdefamx \\
 & & \\[-1ex]
\hflavrowdefamy \\
 & & \\[-1ex]
\hflavrowdefamz \\
\hline
\hline
\end{tabular}
\begin{tablenotes}
\item[*] Not included in PDG average.
\item[1] Using ${\cal B}(B^0 \to \bar{\Lambda}_c^- p)$.
\item[2] Measurement of ${\cal B}(B^+ \to \bar{\Sigma}_c(2455)^0 p) / {\cal B}(B^+ \to \bar{\Lambda}_c^- p \pi^+)$ used in our fit.
\item[3] Measurement of ${\cal B}(B^+ \to \bar{\Sigma}_c(2800)^0 p) / {\cal B}(B^+ \to \bar{\Lambda}_c^- p \pi^+)$ used in our fit.
\item[4] Using ${\cal B}(\Xi_c^0 \to \Xi^- \pi^+)$.
\item[5] Using ${\cal B}(\Xi_c(2930)^+ \to \Lambda_c^+ K^-)$.
\item[6] Using ${\cal B}(B^+ \to \bar{\Lambda}_c^- p \pi^+)$.
\end{tablenotes}
\end{threeparttable}
\end{center}
\end{table}

\subsubsection{Decays to exotic states}
\label{sec:b2c:Bu_other}
Averages of $B^+$ decays to exotic states are shown in Tables~\ref{tab:b2charm_Bu_other_overview1}--\ref{tab:b2charm_Bu_other_overview4}.
\hflavrowdef{\hflavrowdefana}{${\cal B}(B^+ \to X(3872) K^+)$}{{\setlength{\tabcolsep}{0pt}
}
\begin{table}[H]
\begin{center}
\begin{threeparttable}
\caption{Branching fractions for decays to exotic states.}
\label{tab:b2charm_Bu_other_overview1}
%
\begin{tablenotes}
\item[*] Not included in PDG average.
\item[1] Measurement of ${\cal B}(B^0 \to X(3872) K^0) / {\cal B}(B^+ \to X(3872) K^+)$ used in our fit.
\item[2] Using ${\cal B}(X(3872) \to J/\psi \pi^+ \pi^-)$.
\end{tablenotes}
\end{threeparttable}
\end{center}
\end{table}

\hflavrowdeflong{\hflavrowdefand}{${\cal B}(B^+ \to X(3872) K^+) \times {\cal B}(X(3872) \to D^*(2007)^0 \bar{D}^0)$}{{\setlength{\tabcolsep}{0pt}%
}
\begin{table}[H]
\begin{center}
\begin{threeparttable}
\caption{Product branching fractions for decays to $X(3872)$.}
\label{tab:b2charm_Bu_other_overview2}
%
 \\
\hline
\hflavrowdefand \\
 & & \\[-1ex]
\hflavrowdefane \\
 & & \\[-1ex]
\hflavrowdefanf \\
 & & \\[-1ex]
\hflavrowdefang \\
 & & \\[-1ex]
\hflavrowdefanh \\
 & & \\[-1ex]
\hflavrowdefani \\
 & & \\[-1ex]
\hflavrowdefanj \\
 & & \\[-1ex]
\hflavrowdefank \\
 & & \\[-1ex]
\hflavrowdefanl \\
\hline
\hline
\end{tabular}
\begin{tablenotes}
\item[*] Not included in PDG average.
\end{tablenotes}
\end{threeparttable}
\end{center}
\end{table}

\hflavrowdeflong{\hflavrowdefanm}{${\cal B}(B^+ \to \psi_2(3823) K^+) \times {\cal B}(\psi_2(3823) \to \chi_{c1} \gamma)$}{{\setlength{\tabcolsep}{0pt}
}
\begin{table}[H]
\begin{center}
\begin{threeparttable}
\caption{Product branching fractions to neutral states other than $X(3872)$.}
\label{tab:b2charm_Bu_other_overview3}
%
 \\
\hline
\hflavrowdefanm \\
 & & \\[-1ex]
\hflavrowdefann \\
 & & \\[-1ex]
\hflavrowdefano \\
 & & \\[-1ex]
\hflavrowdefanp \\
 & & \\[-1ex]
\hflavrowdefanq \\
 & & \\[-1ex]
\hflavrowdefanr \\
 & & \\[-1ex]
\hflavrowdefans \\
 & & \\[-1ex]
\hflavrowdefant \\
 & & \\[-1ex]
\hflavrowdefanu \\
\hline
\hline
\end{tabular}
\begin{tablenotes}
\item[*] Not included in PDG average.
\item[1] The quoted value is a fit fraction from a Dalitz plot fit.
\item[2] Using ${\cal B}(B^+ \to J/\psi \phi(1020) K^+)$.
\end{tablenotes}
\end{threeparttable}
\end{center}
\end{table}

\hflavrowdeflong{\hflavrowdefanv}{${\cal B}(B^+ \to X(3872)^+ K^0) \times {\cal B}(X(3872)^+ \to J/\psi \pi^+ \pi^0)$}{{\setlength{\tabcolsep}{0pt}\begin{tabular}{m{6.5em}l} {BaBar \cite{Aubert:2004zr}\tnote{}\hphantom{\textsuperscript{}}} & { $< 2.2$ } \\ \end{tabular}}}{\begin{tabular}{l} $< 2.2$ \\ $\mathit{\scriptstyle < 0.6}$\\ \end{tabular}}
\hflavrowdeflong{\hflavrowdefanw}{${\cal B}(B^+ \to Z_c(4430)^+ K^0) \times {\cal B}(Z_c(4430)^+ \to J/\psi \pi^+)$}{{\setlength{\tabcolsep}{0pt}\begin{tabular}{m{6.5em}l} {BaBar \cite{Aubert:2008aa}\tnote{}\hphantom{\textsuperscript{}}} & { $< 1.5$ } \\ \end{tabular}}}{\begin{tabular}{l} $< 1.5$\\ \end{tabular}}
\hflavrowdeflong{\hflavrowdefanx}{${\cal B}(B^+ \to Z_c(4430)^+ K^0) \times {\cal B}(Z_c(4430)^+ \to \psi(2S) \pi^+)$}{{\setlength{\tabcolsep}{0pt}\begin{tabular}{m{6.5em}l} {BaBar \cite{Aubert:2008aa}\tnote{}\hphantom{\textsuperscript{}}} & { $< 4.7$ } \\ \end{tabular}}}{\begin{tabular}{l} $< 4.7$\\ \end{tabular}}
\begin{table}[H]
\begin{center}
\begin{threeparttable}
\caption{Product branching fractions to charged exotic states.}
\label{tab:b2charm_Bu_other_overview4}
\begin{tabular}{ Sl l l }
\hline
\hline
\textbf{Parameter [$10^{-5}$]} & \textbf{Measurements} & \begin{tabular}{l}\textbf{Average $^\mathrm{HFLAV}_{\scriptscriptstyle PDG}$}\end{tabular} \\
\hline
\hflavrowdefanv \\
 & & \\[-1ex]
\hflavrowdefanw \\
 & & \\[-1ex]
\hflavrowdefanx \\
\hline
\hline
\end{tabular}
\end{threeparttable}
\end{center}
\end{table}

\subsection{Decays of admixtures of $B^0/B^+$ mesons}
\label{sec:b2c:B}
Measurements of $B^0/B^+$ decays to charmed hadrons are summarized in Sections~\ref{sec:b2c:B_DD} to~\ref{sec:b2c:B_other}.

\subsubsection{Decays to two open charm mesons}
\label{sec:b2c:B_DD}
Averages of $B^0/B^+$ decays to two open charm mesons are shown in Table~\ref{tab:b2charm_B_DD_overview1}.
\hflavrowdef{\hflavrowdefany}{${\cal B}(B \to D^0 \bar{D}^0 K \pi^0)$}{{\setlength{\tabcolsep}{0pt}
}
\begin{table}[H]
\begin{center}
\begin{threeparttable}
\caption{Branching fractions for decays to double charm.}
\label{tab:b2charm_B_DD_overview1}
%
\end{threeparttable}
\end{center}
\end{table}

\subsubsection{Decays to charmonium states}
\label{sec:b2c:B_cc}
Averages of $B^0/B^+$ decays to charmonium states are shown in Tables~\ref{tab:b2charm_B_cc_overview1}--\ref{tab:b2charm_B_cc_overview5}.
\hflavrowdef{\hflavrowdefanz}{$A_\parallel(B \to J/\psi K^*)$}{{\setlength{\tabcolsep}{0pt}%
}
\begin{table}[H]
\begin{center}
\begin{threeparttable}
\caption{Decay amplitudes for parallel transverse polarization of decays to charmonia.}
\label{tab:b2charm_B_cc_overview1}
%
}
\begin{table}[H]
\begin{center}
\begin{threeparttable}
\caption{Decay amplitudes for perpendicular transverse polarization of decays to charmonia.}
\label{tab:b2charm_B_cc_overview2}
%
}
\begin{table}[H]
\begin{center}
\begin{threeparttable}
\caption{Decay amplitudes for longitudinal polarization of decays to charmonia.}
\label{tab:b2charm_B_cc_overview3}
%
}
\begin{table}[H]
\begin{center}
\begin{threeparttable}
\caption{Relative phases of parallel transverse polarization decay amplitudes of decays to charmonia.}
\label{tab:b2charm_B_cc_overview4}
%
}
\begin{table}[H]
\begin{center}
\begin{threeparttable}
\caption{Relative phases of perpendicular transverse polarization decay amplitudes of decays to charmonia.}
\label{tab:b2charm_B_cc_overview5}
%
\end{threeparttable}
\end{center}
\end{table}

\subsubsection{Decays to exotic states}
\label{sec:b2c:B_other}
Averages of $B^0/B^+$ decays to exotic states are shown in Table~\ref{tab:b2charm_B_other_overview1}.
\hflavrowdef{\hflavrowdefaon}{${\cal B}(B \to Y(3940) K) \times {\cal B}(Y(3940) \to D^*(2007)^0 \bar{D}^*(2007)^0)$}{{\setlength{\tabcolsep}{0pt}%
}
\begin{table}[H]
\begin{center}
\begin{threeparttable}
\caption{Branching fractions for decays to exotic states.}
\label{tab:b2charm_B_other_overview1}
%
 \\
\hline
\hflavrowdefaon \\
 & & \\[-1ex]
\hflavrowdefaoo \\
\hline
\hline
\end{tabular}
\end{threeparttable}
\end{center}
\end{table}

\subsection{Decays of $B_s^0$ mesons}
\label{sec:b2c:Bs}
Measurements of $B_s^0$ decays to charmed hadrons are summarized in Sections~\ref{sec:b2c:Bs_D} to~\ref{sec:b2c:Bs_baryon}.

\subsubsection{Decays to a single open charm meson}
\label{sec:b2c:Bs_D}
Averages of $B_s^0$ decays to a single open charm meson are shown in Tables~\ref{tab:b2charm_Bs_D_overview1}--\ref{tab:b2charm_Bs_D_overview4}.
\hflavrowdef{\hflavrowdefaop}{${\cal B}(B_s^0 \to D_s^- \pi^+)$}{{\setlength{\tabcolsep}{0pt}
}
\begin{table}[H]
\begin{center}
\begin{threeparttable}
\caption{Branching fractions for decays to a $D_s^{(*)}$.}
\label{tab:b2charm_Bs_D_overview1}
%
 \\
\hline
\hflavrowdefaop \\
 & & \\[-1ex]
\hflavrowdefaoq \\
 & & \\[-1ex]
\hflavrowdefaor \\
 & & \\[-1ex]
\hflavrowdefaos \\
 & & \\[-1ex]
\hflavrowdefaot \\
 & & \\[-1ex]
\hflavrowdefaou \\
 & & \\[-1ex]
\hflavrowdefaov \\
 & & \\[-1ex]
\hflavrowdefaow \\
\hline
\hline
\end{tabular}
\begin{tablenotes}
\item[*] Not included in PDG average.
\item[1] Using ${\cal B}(B^0 \to D^- \pi^+)$.
\item[2] Measurement of ${\cal{B}} ( B_s^0 \to D_s^{\mp} K^{\pm} ) / {\cal B}(B_s^0 \to D_s^- \pi^+)$ used in our fit.
\item[3] Measurement of ${\cal B}(B_s^0 \to D_s^- \pi^+ \pi^+ \pi^-) / {\cal B}(B_s^0 \to D_s^- \pi^+)$ used in our fit.
\item[4] Measurement of ${\cal{B}} ( B_s^0 \to D_s^{*\mp} K^{\pm} ) / {\cal B}(B_s^0 \to D_s^{*-} \pi^+)$ used in our fit.
\item[5] Using $f_s/f_d = 0.259 \pm 0.038$ from PDG 2006.
\item[6] Using ${\cal B}(B^0 \to D^- \pi^+ \pi^+ \pi^-)$.
\item[7] Using ${\cal B}(B_s^0 \to D_s^- \pi^+)$.
\item[8] Measurement of ${\cal B}(B_s^0 \to D_s^- K^+ \pi^+ \pi^-) / {\cal B}(B_s^0 \to D_s^- \pi^+ \pi^+ \pi^-)$ used in our fit.
\item[9] Measurement of ${\cal B}(B_s^0 \to D_{s1}(2536)^- \pi^+) \times {\cal B}(D_{s1}(2536)^+ \to D_s^+ \pi^+ \pi^-) / {\cal B}(B_s^0 \to D_s^- \pi^+ \pi^+ \pi^-)$ used in our fit.
\item[10] Using ${\cal B}(B_s^0 \to D_s^{*-} \pi^+)$.
\item[11] Using ${\cal B}(B_s^0 \to D_s^- \pi^+ \pi^+ \pi^-)$.
\item[12] Measurement of ${\cal B}(B^0 \to D_s^- K^+ \pi^+ \pi^-) / {\cal B}(B_s^0 \to D_s^- K^+ \pi^+ \pi^-)$ used in our fit.
\end{tablenotes}
\end{threeparttable}
\end{center}
\end{table}

\hflavrowdef{\hflavrowdefaox}{${\cal B}(B_s^0 \to \bar{D}^0 \bar{K}^0)$}{{\setlength{\tabcolsep}{0pt}
}
\begin{table}[H]
\begin{center}
\begin{threeparttable}
\caption{Branching fractions for decays to a $D^{(*)}$.}
\label{tab:b2charm_Bs_D_overview2}
%
 \\
\hline
\hflavrowdefaox \\
 & & \\[-1ex]
\hflavrowdefaoy \\
 & & \\[-1ex]
\hflavrowdefaoz \\
 & & \\[-1ex]
\hflavrowdefapa \\
 & & \\[-1ex]
\hflavrowdefapb \\
 & & \\[-1ex]
\hflavrowdefapc \\
 & & \\[-1ex]
\hflavrowdefapd \\
 & & \\[-1ex]
\hflavrowdefape \\
 & & \\[-1ex]
\hflavrowdefapf \\
 & & \\[-1ex]
\hflavrowdefapg \\
\hline
\hline
\end{tabular}
\begin{tablenotes}
\item[*] Not included in PDG average.
\item[1] Using ${\cal B}(B^0 \to \bar{D}^0 K^*(892)^0)$.
\item[2] Using ${\cal B}(B^0 \to \bar{D}^0 \rho^0(770))$.
\item[3] Measurement of ${\cal B}(B_s^0 \to \bar{D}^0 \phi(1020)) / {\cal B}(B_s^0 \to \bar{D}^0 \bar{K}^*(892)^0)$ used in our fit.
\item[4] Using ${\cal B}(B_s^0 \to \bar{D}^0 \bar{K}^*(892)^0)$.
\item[5] Using ${\cal B}(B^0 \to \bar{D}^0 K^+ K^-)$.
\item[6] Using ${\cal B}(B^0 \to \bar{D}^0 \pi^+ \pi^-)$.
\item[7] Using ${\cal B}(B^0 \to \bar{D}^*(2007)^0 \pi^+ \pi^-)$.
\end{tablenotes}
\end{threeparttable}
\end{center}
\end{table}

\hflavrowdef{\hflavrowdefaph}{${\cal B}(B_s^0 \to D_{s1}(2536)^- \pi^+) \times {\cal B}(D_{s1}(2536)^+ \to D_s^+ \pi^+ \pi^-)$}{{\setlength{\tabcolsep}{0pt}
}
\begin{table}[H]
\begin{center}
\begin{threeparttable}
\caption{Branching fractions for decays to excited $D_s$ mesons.}
\label{tab:b2charm_Bs_D_overview3}
%
}
\begin{table}[H]
\begin{center}
\begin{threeparttable}
\caption{Longitudinal polarisation fraction for decays to a $D^{*}$.}
\label{tab:b2charm_Bs_D_overview4}
%
\end{threeparttable}
\end{center}
\end{table}

\subsubsection{Decays to two open charm mesons}
\label{sec:b2c:Bs_DD}
Averages of $B_s^0$ decays to two open charm mesons are shown in Table~\ref{tab:b2charm_Bs_DD_overview1}.
\hflavrowdef{\hflavrowdefapj}{${\cal B}(B_s^0 \to D_s^+ D_s^-)$}{{\setlength{\tabcolsep}{0pt}%
}
\begin{table}[H]
\begin{center}
\begin{threeparttable}
\caption{Branching fractions for decays to two $D_{(s)}^{(*)}$ mesons.}
\label{tab:b2charm_Bs_DD_overview1}
%
 \\
\hline
\hflavrowdefapj \\
 & & \\[-1ex]
\hflavrowdefapk \\
 & & \\[-1ex]
\hflavrowdefapl \\
 & & \\[-1ex]
\hflavrowdefapm \\
 & & \\[-1ex]
\hflavrowdefapn \\
 & & \\[-1ex]
\hflavrowdefapo \\
\hline
\hline
\end{tabular}
\begin{tablenotes}
\item[*] Not included in PDG average.
\item[1] Using ${\cal B}(B^0 \to D_s^+ D^-)$.
\item[2] Using $f_s/f_d = 0.269 \pm 0.033$ and $\mathcal{B}(B^0 \to D_s^+ D^-) = (7.2 \pm 0.8) \times 10^{-3}$.
\item[3] Measurement of ${\cal B}(B_s^0 \to D_s^+ D_s^-) + {\cal{B}} ( B_s^0 \to D_s^{*+} D_s^- + D_s^{*-} D_s^+ ) + {\cal B}(B_s^0 \to D_s^{*+} D_s^{*-})$ used in our fit.
\item[4] At CL=95\,\%.
\item[5] Measurement of ${\cal B}(B_s^0 \to \Lambda_c^+ \bar{\Lambda}_c^-) / {\cal B}(B_s^0 \to D_s^- D^+)$ used in our fit.
\item[6] Using ${\cal B}(B^0 \to D^+ D^-)$.
\item[7] Using ${\cal B}(B^+ \to D_s^+ \bar{D}^0)$.
\end{tablenotes}
\end{threeparttable}
\end{center}
\end{table}

\subsubsection{Decays to charmonium states}
\label{sec:b2c:Bs_cc}
Averages of $B_s^0$ decays to charmonium states are shown in Tables~\ref{tab:b2charm_Bs_cc_overview1}--\ref{tab:b2charm_Bs_cc_overview5}.
\hflavrowdef{\hflavrowdefapp}{${\cal B}(B_s^0 \to J/\psi \phi(1020))$}{{\setlength{\tabcolsep}{0pt}
}
\begin{table}[H]
\begin{center}
\begin{threeparttable}
\caption{Branching fractions for decays to $J/\psi$, I.}
\label{tab:b2charm_Bs_cc_overview1}
%
 \\
\hline
\hflavrowdefapp \\
 & & \\[-1ex]
\hflavrowdefapq \\
 & & \\[-1ex]
\hflavrowdefapr \\
 & & \\[-1ex]
\hflavrowdefaps \\
 & & \\[-1ex]
\hflavrowdefapt \\
 & & \\[-1ex]
\hflavrowdefapu \\
 & & \\[-1ex]
\hflavrowdefapv \\
 & & \\[-1ex]
\hflavrowdefapw \\
 & & \\[-1ex]
\hflavrowdefapx \\
\hline
\hline
\end{tabular}
\begin{tablenotes}
\footnotesize
\item[1] Measurement of ${\cal B}(B_s^0 \to J/\psi f_0(980)) \times {\cal B}(f_0(980) \to \pi^+ \pi^-) / {\cal B}(B_s^0 \to J/\psi \phi(1020)) / {\cal B}(\phi(1020) \to K^+ K^-)$ used in our fit.
\item[2] Measurement of ${\cal B}(B_s^0 \to \psi(2S) \phi(1020)) / {\cal B}(B_s^0 \to J/\psi \phi(1020))$ used in our fit.
\item[3] Measurement of ${\cal B}(B_s^0 \to J/\psi f_2^{\prime}(1525)) {\cal B}(f_2^{\prime}(1525) \to K^+ K^-) / {\cal B}(B_s^0 \to J/\psi \phi(1020)) / {\cal B}(\phi(1020) \to K^+ K^-)$ used in our fit.
\item[4] Measurement of ${\cal B}(B_s^0 \to J/\psi f_2^{\prime}(1525)) / {\cal B}(B_s^0 \to J/\psi \phi(1020))$ used in our fit.
\item[5] Measurement of ${\cal B}(B_s^0 \to J/\psi \phi(1020) \phi(1020)) / {\cal B}(B_s^0 \to J/\psi \phi(1020))$ used in our fit.
\item[6] Measurement of ${\cal B}(B_s^0 \to J/\psi \pi^+ \pi^-) / {\cal B}(B_s^0 \to J/\psi \phi(1020))$ used in our fit.
\item[7] Using ${\cal B}(B^0 \to J/\psi K^0 \pi^+ \pi^-)$.
\item[8] Using ${\cal B}(B^0 \to J/\psi \rho^0(770))$.
\item[9] Measurement of ${\cal B}(B^0 \to J/\psi \eta) / {\cal B}(B_s^0 \to J/\psi \eta)$ used in our fit.
\item[10] Measurement of ${\cal B}(B^0 \to J/\psi \eta^\prime) / {\cal B}(B_s^0 \to J/\psi \eta^\prime)$ used in our fit.
\item[11] Using ${\cal B}(B_s^0 \to J/\psi \phi(1020))$.
\item[12] Using $f_s/f_d = 0.269 \pm 0.033$.
\item[13] Measurement of $2 {\cal B}(B_s^0 \to J/\psi K^0_S)$ used in our fit.
\item[14] Measurement of $2 {\cal B}(B_s^0 \to J/\psi K^0_S) / {\cal B}(B^0 \to J/\psi K^0)$ used in our fit.
\end{tablenotes}
\end{threeparttable}
\end{center}
\end{table}

\hflavrowdef{\hflavrowdefapy}{${\cal B}(B_s^0 \to J/\psi \pi^+ \pi^-)$}{{\setlength{\tabcolsep}{0pt}
}
\begin{table}[H]
\begin{center}
\begin{threeparttable}
\caption{Branching fractions for decays to $J/\psi$, II.}
\label{tab:b2charm_Bs_cc_overview2}
%
 \\
\hline
\hflavrowdefapy \\
 & & \\[-1ex]
\hflavrowdefapz \\
 & & \\[-1ex]
\hflavrowdefaqa \\
 & & \\[-1ex]
\hflavrowdefaqb \\
 & & \\[-1ex]
\hflavrowdefaqc \\
 & & \\[-1ex]
\hflavrowdefaqd \\
 & & \\[-1ex]
\hflavrowdefaqe \\
 & & \\[-1ex]
\hflavrowdefaqf \\
\hline
\hline
\end{tabular}
\begin{tablenotes}
\item[1] Using ${\cal B}(B_s^0 \to J/\psi \phi(1020))$.
\item[2] Corrected for updates of ${\cal B}(\phi(1020) \to K^+ K^-)$.
\item[3] Measurement of ${\cal B}(B_s^0 \to \psi(2S) \pi^+ \pi^-) / {\cal B}(B_s^0 \to J/\psi \pi^+ \pi^-)$ used in our fit.
\item[4] Measurement of ${\cal B}(B_s^0 \to J/\psi f_0(980)) \times {\cal B}(f_0(980) \to \pi^+ \pi^-) / {\cal B}(B_s^0 \to J/\psi \phi(1020)) / {\cal B}(\phi(1020) \to K^+ K^-)$ used in our fit.
\item[5] Measurement of ${\cal B}(B_s^0 \to J/\psi f_0(500)) \times {\cal B}(f_0(500) \to \pi^+ \pi^-) / {\cal B}(B_s^0 \to J/\psi f_0(980)) \times {\cal B}(f_0(980) \to \pi^+ \pi^-)$ used in our fit.
\item[6] Using ${\cal B}(f_1(1285) \to \pi^+ \pi^+ \pi^- \pi^-)$.
\item[7] Measurement of ${\cal B}(B_s^0 \to J/\psi f_2^{\prime}(1525)) {\cal B}(f_2^{\prime}(1525) \to K^+ K^-) / {\cal B}(B_s^0 \to J/\psi \phi(1020)) / {\cal B}(\phi(1020) \to K^+ K^-)$ used in our fit.
\item[8] Using ${\cal B}(B_s^0 \to J/\psi f_0(980)) \times {\cal B}(f_0(980) \to \pi^+ \pi^-)$.
\end{tablenotes}
\end{threeparttable}
\end{center}
\end{table}

\hflavrowdef{\hflavrowdefaqg}{${\cal B}(B_s^0 \to \psi(2S) \phi(1020))$}{{\setlength{\tabcolsep}{0pt}
}
\begin{table}[H]
\begin{center}
\begin{threeparttable}
\caption{Branching fractions for decays to charmonium other than $J/\psi$.}
\label{tab:b2charm_Bs_cc_overview3}
%
 \\
\hline
\hflavrowdefaqg \\
 & & \\[-1ex]
\hflavrowdefaqh \\
 & & \\[-1ex]
\hflavrowdefaqi \\
 & & \\[-1ex]
\hflavrowdefaqj \\
\hline
\hline
\end{tabular}
\begin{tablenotes}
\item[*] Not included in PDG average.
\item[1] Using ${\cal B}(B_s^0 \to J/\psi \phi(1020))$.
\item[2] Measurement of ${\cal B}(B_s^0 \to X(3872) \phi(1020)) {\cal B}(X(3872) \to J/\psi \pi^+ \pi^-) / {\cal B}(B_s^0 \to \psi(2S) \phi(1020)) / {\cal B}(\psi(2S) \to J/\psi \pi^+ \pi^-)$ used in our fit.
\item[3] Measurement of ${\cal B}(B_s^0 \to J/\psi K^*(892)^0 \bar{K}^*(892)^0) {\cal B}(K^*(892)^0 \to K^+ \pi^-) **2 / {\cal B}(B_s^0 \to \psi(2S) \phi(1020)) / {\cal B}(\psi(2S) \to J/\psi \pi^+ \pi^-) / {\cal B}(\phi(1020) \to K^+ K^-)$ used in our fit.
\item[4] Using ${\cal B}(B^0 \to \psi(2S) K^+ \pi^-)$.
\item[5] Using ${\cal B}(B^0 \to \psi(2S) K^*(892)^0)$.
\item[6] Using ${\cal B}(B_s^0 \to J/\psi \pi^+ \pi^-)$.
\item[7] Measurement of ${\cal B}(B_s^0 \to X(3872) \pi^+ \pi^-) {\cal B}(X(3872) \to J/\psi \pi^+ \pi^-) / {\cal B}(B_s^0 \to \psi(2S) \pi^+ \pi^-) / {\cal B}(\psi(2S) \to J/\psi \pi^+ \pi^-)$ used in our fit.
\end{tablenotes}
\end{threeparttable}
\end{center}
\end{table}

\hflavrowdef{\hflavrowdefaqk}{$\dfrac{{\cal B}(B_s^0 \to \chi_{c2} K^+ K^-)}{{\cal B}(B_s^0 \to \chi_{c1} K^+ K^-)}$}{{\setlength{\tabcolsep}{0pt}
}
\begin{table}[H]
\begin{center}
\begin{threeparttable}
\caption{Branching fraction ratios for decays to charmonium.}
\label{tab:b2charm_Bs_cc_overview4}
%
}
\begin{table}[H]
\begin{center}
\begin{threeparttable}
\caption{Decay amplitudes and relative phases for decays to $J/\psi K^*$ (see Chapter 5 for decays to $J/\psi \phi$).}
\label{tab:b2charm_Bs_cc_overview5}
%
\end{threeparttable}
\end{center}
\end{table}

\subsubsection{Decays to charm baryons}
\label{sec:b2c:Bs_baryon}
Averages of $B_s^0$ decays to charm baryons are shown in Tables~\ref{tab:b2charm_Bs_baryon_overview1}--\ref{tab:b2charm_Bs_baryon_overview2}.
\hflavrowdef{\hflavrowdefaqp}{${\cal B}(B_s^0 \to \bar{\Lambda}_c^- \Lambda^0 \pi^+)$}{{\setlength{\tabcolsep}{0pt}%
}
\begin{table}[H]
\begin{center}
\begin{threeparttable}
\caption{Branching fractions for decays to one charm baryon.}
\label{tab:b2charm_Bs_baryon_overview1}
%
}
\begin{table}[H]
\begin{center}
\begin{threeparttable}
\caption{Branching fractions for decays to two charm baryons.}
\label{tab:b2charm_Bs_baryon_overview2}
%
 \\
\hline
\hflavrowdefaqq \\
\hline
\hline
\end{tabular}
\begin{tablenotes}
\item[1] At CL=95\,\%.
\item[2] Using ${\cal B}(B_s^0 \to D_s^- D^+)$.
\end{tablenotes}
\end{threeparttable}
\end{center}
\end{table}

\subsection{Decays of $B_c^+$ mesons}
\label{sec:b2c:Bc}
Measurements of $B_c^+$ decays to charmed hadrons are summarized in Sections~\ref{sec:b2c:Bc_D} to~\ref{sec:b2c:Bc_B}.

\subsubsection{Decays to a single open charm meson}
\label{sec:b2c:Bc_D}
Averages of $B_c^+$ decays to a single open charm meson are shown in Table~\ref{tab:b2charm_Bc_D_overview1}.
\hflavrowdef{\hflavrowdefaqr}{$f_c \times {\cal B}(B_c^+ \to D^0 K^+)$}{{\setlength{\tabcolsep}{0pt}
}
\begin{table}[H]
\begin{center}
\begin{threeparttable}
\caption{Branching fractions for decays to one charm meson.}
\label{tab:b2charm_Bc_D_overview1}
%
\begin{tablenotes}
\item[1] Using $f_u$.
\end{tablenotes}
\end{threeparttable}
\end{center}
\end{table}

\subsubsection{Decays to two open charm mesons}
\label{sec:b2c:Bc_DD}
Averages of $B_c^+$ decays to two open charm mesons are shown in Table~\ref{tab:b2charm_Bc_DD_overview1}.
\hflavrowdef{\hflavrowdefaqs}{$f_c \times {\cal B}(B_c^+ \to D_s^+ \bar{D}^0)$}{{\setlength{\tabcolsep}{0pt}%
}
\begin{table}[H]
\begin{center}
\begin{threeparttable}
\caption{Branching fractions for decays to two $D$ mesons.}
\label{tab:b2charm_Bc_DD_overview1}
%
 \\
\hline
\hflavrowdefaqs \\
 & & \\[-1ex]
\hflavrowdefaqt \\
 & & \\[-1ex]
\hflavrowdefaqu \\
 & & \\[-1ex]
\hflavrowdefaqv \\
 & & \\[-1ex]
\hflavrowdefaqw \\
 & & \\[-1ex]
\hflavrowdefaqx \\
 & & \\[-1ex]
\hflavrowdefaqy \\
 & & \\[-1ex]
\hflavrowdefaqz \\
 & & \\[-1ex]
\hflavrowdefara \\
 & & \\[-1ex]
\hflavrowdefarb \\
 & & \\[-1ex]
\hflavrowdefarc \\
 & & \\[-1ex]
\hflavrowdefard \\
\hline
\hline
\end{tabular}
\begin{tablenotes}
\footnotesize
\item[1] Measurement of $f_c \times {\cal B}(B_c^+ \to D_s^+ \bar{D}^0) / ( f_u {\cal B}(B^+ \to D_s^+ \bar{D}^0) )$ used in our fit.
\item[2] Measurement of $f_c \times {\cal B}(B_c^+ \to D_s^+ D^0) / ( f_u {\cal B}(B^+ \to D_s^+ \bar{D}^0) )$ used in our fit.
\item[3] Measurement of $f_c \times {\cal B}(B_c^+ \to D^+ \bar{D}^0) / ( f_u {\cal B}(B^+ \to D^+ \bar{D}^0) )$ used in our fit.
\item[4] Measurement of $f_c \times {\cal B}(B_c^+ \to D^+ D^0) / ( f_u {\cal B}(B^+ \to D^+ \bar{D}^0) )$ used in our fit.
\item[5] Measurement of $f_c \times {\cal B}(B_c^+ \to D_s^{*+} \bar{D}^0  +  D_s^+ \bar{D}^*(2007)^0) / ( f_u {\cal B}(B^+ \to D_s^+ \bar{D}^0) )$ used in our fit.
\item[6] Measurement of $f_c \times {\cal B}(B_c^+ \to D_s^{*+} D^0  +  D_s^+ D^*(2007)^0) / ( f_u {\cal B}(B^+ \to D_s^+ \bar{D}^0) )$ used in our fit.
\item[7] Measurement of $f_c \times {\cal B}(B_c^+ \to D_s^{*+} \bar{D}^*(2007)^0) / ( f_u {\cal B}(B^+ \to D_s^+ \bar{D}^0) )$ used in our fit.
\item[8] Measurement of $f_c \times {\cal B}(B_c^+ \to D_s^{*+} D^*(2007)^0) / ( f_u {\cal B}(B^+ \to D_s^+ \bar{D}^0) )$ used in our fit.
\item[9] Measurement of $f_c \times ( {\cal{B}} ( B_c^- \to D^{*-} D^0 ) \times {\cal{B}} (D^{*-}\to D^- (\pi^0,\gamma)) + {\cal{B}} (B_c^{-}\to D^- D^{*0}) ) / ( f_u {\cal B}(B^+ \to D^+ \bar{D}^0) )$ used in our fit.
\item[10] Measurement of $f_c \times ( {\cal{B}} ( B_c^- \to D^{*-} \bar{D}^0 ) \times {\cal{B}} (D^{*-}\to D^- (\pi^0,\gamma)) + {\cal{B}} (B_c^{-}\to D^- \bar{D}^{*0}) ) / ( f_u {\cal B}(B^+ \to D^+ \bar{D}^0) )$ used in our fit.
\item[11] Measurement of $f_c \times {\cal B}(B_c^+ \to D^*(2010)^+ D^*(2007)^0) / ( f_u {\cal B}(B^+ \to D^+ \bar{D}^0) )$ used in our fit.
\item[12] Measurement of $f_c \times {\cal B}(B_c^+ \to D^*(2010)^+ \bar{D}^*(2007)^0) / ( f_u {\cal B}(B^+ \to D^+ \bar{D}^0) )$ used in our fit.
\end{tablenotes}
\end{threeparttable}
\end{center}
\end{table}

\subsubsection{Decays to charmonium states}
\label{sec:b2c:Bc_cc}
Averages of $B_c^+$ decays to charmonium states are shown in Tables~\ref{tab:b2charm_Bc_cc_overview1}--\ref{tab:b2charm_Bc_cc_overview2}.
\hflavrowdef{\hflavrowdefare}{$f_c \times {\cal B}(B_c^+ \to J/\psi \pi^+)$}{{\setlength{\tabcolsep}{0pt}\begin{tabular}{m{6.5em}l} \multicolumn{2}{l}{CMS {\cite{Khachatryan:2014nfa}\tnote{1}\hphantom{\textsuperscript{1}}}} \\ \multicolumn{2}{l}{LHCb {\cite{Aaij:2012dd}\tnote{2,1}\hphantom{\textsuperscript{2,1}}}, {\cite{Aaij:2014ija}\tnote{3,1}\hphantom{\textsuperscript{3,1}}}} \\ \end{tabular}}}{\begin{tabular}{l} $2.79\,^{+0.11}_{-0.10}$\\ \end{tabular}}
\hflavrowdef{\hflavrowdefarf}{$f_c \times {\cal B}(B_c^+ \to \chi_{c0} \pi^+)$}{{\setlength{\tabcolsep}{0pt}\begin{tabular}{m{6.5em}l} {LHCb \cite{LHCb:2016utz}\tnote{4}\hphantom{\textsuperscript{4}}} & { $4.00\,^{+1.39}_{-1.22} \pm0.33$ } \\ \end{tabular}}}{\begin{tabular}{l} $4.0\,^{+1.4}_{-1.3}$\\ \end{tabular}}
\begin{table}[H]
\begin{center}
\begin{threeparttable}
\caption{Branching fractions for decays to charmonium.}
\label{tab:b2charm_Bc_cc_overview1}
\begin{tabular}{ Sl l l }
\hline
\hline
\textbf{Parameter [$10^{-6}$]} & \textbf{Measurements} & \begin{tabular}{l}\textbf{Average}\end{tabular} \\
\hline
\hflavrowdefare \\
 & & \\[-1ex]
\hflavrowdefarf \\
\hline
\hline
\end{tabular}
\begin{tablenotes}
\item[1] Measurement of $f_c \times {\cal B}(B_c^+ \to J/\psi \pi^+) / ( f_u {\cal B}(B^+ \to J/\psi K^+) )$ used in our fit.
\item[2] $\sqrt{s} = 7$ TeV, $p_T > 4$ GeV and $2.5 < y < 4.5$
\item[3] $\sqrt{s} = 8$ TeV, $0 < p_T < 20$ GeV and $2.0 < y < 4.5$
\item[4] Using $f_u$.
\end{tablenotes}
\end{threeparttable}
\end{center}
\end{table}

\hflavrowdef{\hflavrowdefarg}{$\dfrac{{\cal B}(B_c^+ \to J/\psi D_s^+)}{{\cal B}(B_c^+ \to J/\psi \pi^+)}$}{{\setlength{\tabcolsep}{0pt}
}
\begin{table}[H]
\begin{center}
\begin{threeparttable}
\caption{Branching fraction ratios for decays to charmonium.}
\label{tab:b2charm_Bc_cc_overview2}
%
 \\
\hline
\hflavrowdefarg \\
 & & \\[-1ex]
\hflavrowdefarh \\
 & & \\[-1ex]
\hflavrowdefari \\
 & & \\[-1ex]
\hflavrowdefarj \\
 & & \\[-1ex]
\hflavrowdefark \\
 & & \\[-1ex]
\hflavrowdefarl \\
 & & \\[-1ex]
\hflavrowdefarm \\
 & & \\[-1ex]
\hflavrowdefarn \\
 & & \\[-1ex]
\hflavrowdefaro \\
 & & \\[-1ex]
\hflavrowdefarp \\
 & & \\[-1ex]
\hflavrowdefarq \\
\hline
\hline
\end{tabular}
\end{threeparttable}
\end{center}
\end{table}

\subsubsection{Decays to a $B$ meson}
\label{sec:b2c:Bc_B}
Averages of $B_c^+$ decays to a $B$ meson are shown in Table~\ref{tab:b2charm_Bc_B_overview1}.
\hflavrowdef{\hflavrowdefarr}{$f_c \times {\cal B}(B_c^+ \to B_s^0 \pi^+)$}{{\setlength{\tabcolsep}{0pt}\begin{tabular}{m{6.5em}l} {LHCb \cite{LHCb:2013xlg}\tnote{1}\hphantom{\textsuperscript{1}}} & { $2.323 \pm0.304\,^{+0.291}_{-0.271}$ } \\ \end{tabular}}}{\begin{tabular}{l} $2.32\,^{+0.43}_{-0.39}$\\ \end{tabular}}
\begin{table}[H]
\begin{center}
\begin{threeparttable}
\caption{Branching fractions for decays to $B_s^{0}$ meson.}
\label{tab:b2charm_Bc_B_overview1}
\begin{tabular}{ Sl l l }
\hline
\hline
\textbf{Parameter [$10^{-4}$]} & \textbf{Measurements} & \begin{tabular}{l}\textbf{Average}\end{tabular} \\
\hline
\hflavrowdefarr \\
\hline
\hline
\end{tabular}
\begin{tablenotes}
\item[1] Using $f_s$.
\end{tablenotes}
\end{threeparttable}
\end{center}
\end{table}

\subsection{Decays of $b$ baryons}
\label{sec:b2c:Bbaryon}
Measurements of $b$ baryon decays to charmed hadrons are summarized in Sections~\ref{sec:b2c:Bbaryon_D} to~\ref{sec:b2c:Bbaryon_other}.

\subsubsection{Decays to a single open charm meson}
\label{sec:b2c:Bbaryon_D}
Averages of $b$ baryon decays to a single open charm meson are shown in Table~\ref{tab:b2charm_Bbaryon_D_overview1}.
\hflavrowdef{\hflavrowdefars}{${\cal B}(\Lambda_b^0 \to D^0 p K^-)$}{{\setlength{\tabcolsep}{0pt}\begin{tabular}{m{6.5em}l} {LHCb \cite{Aaij:2013pka}\tnote{1}\hphantom{\textsuperscript{1}}} & { $4.04 \pm0.44\,^{+0.44}_{-0.47}$ } \\ \multicolumn{2}{l}{LHCb {\cite{Aaij:2013pka}\tnote{2}\hphantom{\textsuperscript{2}}}} \\ \end{tabular}}}{\begin{tabular}{l} $4.04\,^{+0.65}_{-0.63}$ \\ $\mathit{\scriptstyle 4.55\,^{+0.75}_{-0.77}}$\\ \end{tabular}}
\hflavrowdef{\hflavrowdefart}{${\cal B}(\Lambda_b^0 \to D_s^- p)$}{{\setlength{\tabcolsep}{0pt}\begin{tabular}{m{6.5em}l} {LHCb \cite{LHCb:2022mzw}\tnote{*,3}\hphantom{\textsuperscript{*,3}}} & { $1.070 \pm0.042\,^{+0.099}_{-0.097}$ } \\ \end{tabular}}}{\begin{tabular}{l} $1.09 \pm0.10$ \\ \textit{\small none}\\ \end{tabular}}
\hflavrowdef{\hflavrowdefaru}{$\dfrac{f_{\Xi_b^0}}{f_{\Lambda_b^0}} \times {\cal B}(\Xi_b^0 \to D^0 p K^-)$}{{\setlength{\tabcolsep}{0pt}\begin{tabular}{m{6.5em}l} {LHCb \cite{Aaij:2013pka}\tnote{*,4}\hphantom{\textsuperscript{*,4}}} & { $1.78 \pm0.36 \pm0.37$ } \\ \end{tabular}}}{\begin{tabular}{l} $1.78\,^{+0.55}_{-0.49}$ \\ \textit{\small none}\\ \end{tabular}}
\begin{table}[H]
\begin{center}
\begin{threeparttable}
\caption{Branching fractions for decays to $D_{(s)}$ mesons.}
\label{tab:b2charm_Bbaryon_D_overview1}
\begin{tabular}{ Sl l l }
\hline
\hline
\textbf{Parameter [$10^{-5}$]} & \textbf{Measurements} & \begin{tabular}{l}\textbf{Average $^\mathrm{HFLAV}_{\scriptscriptstyle PDG}$}\end{tabular} \\
\hline
\hflavrowdefars \\
 & & \\[-1ex]
\hflavrowdefart \\
 & & \\[-1ex]
\hflavrowdefaru \\
\hline
\hline
\end{tabular}
\begin{tablenotes}
\item[*] Not included in PDG average.
\item[1] Using ${\cal B}(\Lambda_b^0 \to D^0 p \pi^-)$.
\item[2] Measurement of $\dfrac{f_{\Xi_b^0}}{f_{\Lambda_b^0}} \times {\cal B}(\Xi_b^0 \to D^0 p K^-) / {\cal B}(\Lambda_b^0 \to D^0 p K^-)$ used in our fit.
\item[3] Using ${\cal B}(\Lambda_b^0 \to \Lambda_c^+ \pi^-)$.
\item[4] Using ${\cal B}(\Lambda_b^0 \to D^0 p K^-)$.
\end{tablenotes}
\end{threeparttable}
\end{center}
\end{table}

\subsubsection{Decays to charmonium states}
\label{sec:b2c:Bbaryon_cc}
Averages of $b$ baryon decays to charmonium states are shown in Tables~\ref{tab:b2charm_Bbaryon_cc_overview1}--\ref{tab:b2charm_Bbaryon_cc_overview4}.
\hflavrowdef{\hflavrowdefarv}{${\cal B}(\Lambda_b^0 \to J/\psi p K^-)$}{{\setlength{\tabcolsep}{0pt}
}
\begin{table}[H]
\begin{center}
\begin{threeparttable}
\caption{$\Lambda_b^{0}$ branching fractions to charmonium.}
\label{tab:b2charm_Bbaryon_cc_overview1}
%
 \\
\hline
\hflavrowdefarv \\
 & & \\[-1ex]
\hflavrowdefarw \\
 & & \\[-1ex]
\hflavrowdefarx \\
 & & \\[-1ex]
\hflavrowdefary \\
 & & \\[-1ex]
\hflavrowdefarz \\
 & & \\[-1ex]
\hflavrowdefasa \\
 & & \\[-1ex]
\hflavrowdefasb \\
 & & \\[-1ex]
\hflavrowdefasc \\
 & & \\[-1ex]
\hflavrowdefasd \\
\hline
\hline
\end{tabular}
\begin{tablenotes}
\footnotesize
\item[*] Not included in PDG average.
\item[1] Measurement of ${\cal B}(\Lambda_b^0 \to J/\psi p \pi^-) / {\cal B}(\Lambda_b^0 \to J/\psi p K^-)$ used in our fit.
\item[2] Measurement of ${\cal B}(\Lambda_b^0 \to \psi(2S) p K^-) / {\cal B}(\Lambda_b^0 \to J/\psi p K^-)$ used in our fit.
\item[3] Measurement of ${\cal B}(\Lambda_b^0 \to J/\psi p K^- \pi^+ \pi^-) / {\cal B}(\Lambda_b^0 \to J/\psi p K^-)$ used in our fit.
\item[4] Measurement of ${\cal B}(\Lambda_b^0 \to P_c(4380)^+ \pi^-) / {\cal B}(\Lambda_b^0 \to J/\psi p K^-)$ used in our fit.
\item[5] Measurement of ${\cal B}(\Lambda_b^0 \to P_c(4457)^+ \pi^-) / {\cal B}(\Lambda_b^0 \to J/\psi p K^-)$ used in our fit.
\item[6] Measurement of ${\cal B}(\Lambda_b^0 \to Z_c(4200)^- p) / {\cal B}(\Lambda_b^0 \to J/\psi p K^-)$ used in our fit.
\item[7] Measurement of ${\cal B}(\Lambda_b^0 \to \chi_{c1} p K^-) / {\cal B}(\Lambda_b^0 \to J/\psi p K^-)$ used in our fit.
\item[8] Measurement of ${\cal B}(\Lambda_b^0 \to \chi_{c2} p K^-) / {\cal B}(\Lambda_b^0 \to J/\psi p K^-)$ used in our fit.
\item[9] Using ${\cal B}(\Lambda_b^0 \to J/\psi p K^-)$.
\item[10] Measurement of ${\cal B}(\Lambda_b^0 \to \psi(2S) \Lambda^0) / {\cal B}(\Lambda_b^0 \to J/\psi \Lambda^0)$ used in our fit.
\item[11] Measurement of $\dfrac{f_{\Xi_b^-}}{f_{\Lambda_b^0}} \times {\cal B}(\Xi_b^- \to J/\psi \Xi^-) / {\cal B}(\Lambda_b^0 \to J/\psi \Lambda^0)$ used in our fit.
\item[12] Measurement of $f_{\Omega_b^-} \times {\cal B}(\Omega_b^- \to J/\psi \Omega^-) / f_{\Lambda_b^0} \times {\cal B}(\Lambda_b^0 \to J/\psi \Lambda^0)$ used in our fit.
\item[13] Measurement of ${\cal B}(\Lambda_b^0 \to \psi(2S) p \pi^-) / {\cal B}(\Lambda_b^0 \to \psi(2S) p K^-)$ used in our fit.
\item[14] Measurement of ${\cal B}(\Lambda_b^0 \to X(3872) p K^-) {\cal B}(X(3872) \to J/\psi \pi^+ \pi^-) / {\cal B}(\Lambda_b^0 \to \psi(2S) p K^-) / {\cal B}(\psi(2S) \to J/\psi \pi^+ \pi^-)$ used in our fit.
\item[15] Using ${\cal B}(\Lambda_b^0 \to J/\psi \Lambda^0)$.
\item[16] Measurement of ${\cal B}(\Lambda_b^0 \to \chi_{c2} p K^-) / {\cal B}(\Lambda_b^0 \to \chi_{c1} p K^-)$ used in our fit.
\item[17] Using ${\cal B}(\Lambda_b^0 \to \chi_{c1} p K^-)$.
\end{tablenotes}
\end{threeparttable}
\end{center}
\end{table}

\hflavrowdef{\hflavrowdefase}{$\dfrac{f_{\Xi_b^-}}{f_{\Lambda_b^0}} \times {\cal B}(\Xi_b^- \to J/\psi \Xi^-)$}{{\setlength{\tabcolsep}{0pt}
}
\begin{table}[H]
\begin{center}
\begin{threeparttable}
\caption{$\Xi_b^{-}$ and $\Omega_b^{-}$ branching fractions to charmonium.}
\label{tab:b2charm_Bbaryon_cc_overview2}
%
\begin{tablenotes}
\item[*] Not included in PDG average.
\item[1] Using ${\cal B}(\Lambda_b^0 \to J/\psi \Lambda^0)$.
\item[2] Using $f_{\Lambda_b^0} \times {\cal B}(\Lambda_b^0 \to J/\psi \Lambda^0)$.
\end{tablenotes}
\end{threeparttable}
\end{center}
\end{table}

\hflavrowdef{\hflavrowdefasg}{$\alpha(\Lambda_b^0 \to J/\psi \Lambda^0)$}{{\setlength{\tabcolsep}{0pt}%
}
\begin{table}[H]
\begin{center}
\begin{threeparttable}
\caption{Parity-violating asymmetry in $\Lambda_b^{0}$ decays to charmonium.}
\label{tab:b2charm_Bbaryon_cc_overview3}
%
\begin{tablenotes}
\item[*] Not included in PDG average.
\end{tablenotes}
\end{threeparttable}
\end{center}
\end{table}

\hflavrowdef{\hflavrowdefash}{$P(\Lambda_b^0 \to J/\psi \Lambda^0)$}{{\setlength{\tabcolsep}{0pt}%
}
\begin{table}[H]
\begin{center}
\begin{threeparttable}
\caption{Transverse polarization of $\Lambda_b^{0}$ produced in $pp$ collisions.}
\label{tab:b2charm_Bbaryon_cc_overview4}
%
\begin{tablenotes}
\item[*] Not included in PDG average.
\end{tablenotes}
\end{threeparttable}
\end{center}
\end{table}

\subsubsection{Decays to charm baryons}
\label{sec:b2c:Bbaryon_baryon}
Averages of $b$ baryon decays to charm baryons are shown in Tables~\ref{tab:b2charm_Bbaryon_baryon_overview1}--\ref{tab:b2charm_Bbaryon_baryon_overview4}.
\hflavrowdef{\hflavrowdefasi}{${\cal B}(\Lambda_b^0 \to \Lambda_c^+ \pi^-)$}{{\setlength{\tabcolsep}{0pt}%
}
\begin{table}[H]
\begin{center}
\begin{threeparttable}
\caption{$\Lambda_b$ branching fractions.}
\label{tab:b2charm_Bbaryon_baryon_overview1}
%
 \\
\hline
\hflavrowdefasi \\
 & & \\[-1ex]
\hflavrowdefasj \\
 & & \\[-1ex]
\hflavrowdefask \\
 & & \\[-1ex]
\hflavrowdefasl \\
\hline
\hline
\end{tabular}
\begin{tablenotes}
\item[*] Not included in PDG average.
\item[1] Using ${\cal B}(B^0 \to D^- \pi^+)$.
\item[2] Measurement of ${\cal B}(\Lambda_b^0 \to \Lambda_c^+ \pi^+ \pi^- \pi^-) / {\cal B}(\Lambda_b^0 \to \Lambda_c^+ \pi^-)$ used in our fit.
\item[3] Measurement of ${\cal B}(\Lambda_b^0 \to \Lambda_c^+ K^-) / {\cal B}(\Lambda_b^0 \to \Lambda_c^+ \pi^-)$ used in our fit.
\item[4] Measurement of ${\cal B}(\Lambda_b^0 \to D^0 p \pi^-) {\cal B}(D^0 \to K^- \pi^+) / ( {\cal B}(\Lambda_b^0 \to \Lambda_c^+ \pi^-) {\cal B}(\Lambda_c^+ \to p K^- \pi^+) )$ used in our fit.
\item[5] Measurement of ${\cal B}(\Lambda_b^0 \to \Lambda_c^+ p \bar{p} \pi^-) / {\cal B}(\Lambda_b^0 \to \Lambda_c^+ \pi^-)$ used in our fit.
\item[6] Measurement of ${\cal B}(\Lambda_b^0 \to D_s^- p) / {\cal B}(\Lambda_b^0 \to \Lambda_c^+ \pi^-)$ used in our fit.
\item[7] Using ${\cal B}(\Lambda_b^0 \to \Lambda_c^+ \pi^-)$.
\item[8] Measurement of ${\cal B}(\Lambda_b^0 \to \Sigma_c(2455)^0 \pi^+ \pi^-) \times {\cal B}(\Sigma_c(2455)^0 \to \Lambda_c^+ \pi^-) / {\cal B}(\Lambda_b^0 \to \Lambda_c^+ \pi^+ \pi^- \pi^-)$ used in our fit.
\item[9] Measurement of ${\cal B}(\Lambda_b^0 \to \Sigma_c(2455)^{++} \pi^- \pi^-) \times {\cal B}(\Sigma_c(2455)^{++} \to \Lambda_c^+ \pi^+) / {\cal B}(\Lambda_b^0 \to \Lambda_c^+ \pi^+ \pi^- \pi^-)$ used in our fit.
\item[10] Measurement of ${\cal B}(\Lambda_b^0 \to \Lambda_c(2595)^+ \pi^-) \times {\cal B}(\Lambda_c(2595)^+ \to \Lambda_c^+ \pi^+ \pi^-) / {\cal B}(\Lambda_b^0 \to \Lambda_c^+ \pi^+ \pi^- \pi^-)$ used in our fit.
\item[11] Measurement of ${\cal B}(\Lambda_b^0 \to \Lambda_c(2625)^+ \pi^-) \times {\cal B}(\Lambda_c(2625)^+ \to \Lambda_c^+ \pi^+ \pi^-) / {\cal B}(\Lambda_b^0 \to \Lambda_c^+ \pi^+ \pi^- \pi^-)$ used in our fit.
\item[12] Measurement of ${\cal B}(\Lambda_b^0 \to \Sigma_c(2455)^0 p \bar{p}) \times {\cal B}(\Sigma_c(2455)^0 \to \Lambda_c^+ \pi^-) / {\cal B}(\Lambda_b^0 \to \Lambda_c^+ p \bar{p} \pi^-)$ used in our fit.
\item[13] Measurement of ${\cal B}(\Lambda_b^0 \to \Sigma_c(2520)^0 p \bar{p}) \times {\cal B}(\Sigma_c(2520)^0 \to \Lambda_c^+ \pi^-) / {\cal B}(\Lambda_b^0 \to \Lambda_c^+ p \bar{p} \pi^-)$ used in our fit.
\end{tablenotes}
\end{threeparttable}
\end{center}
\end{table}

\hflavrowdef{\hflavrowdefasm}{$\dfrac{f_{\Xi_b^-}}{f_{\Lambda_b^0}} \times {\cal B}(\Xi_b^- \to \Lambda_b^0 \pi^-)$}{{\setlength{\tabcolsep}{0pt}
}
\begin{table}[H]
\begin{center}
\begin{threeparttable}
\caption{$\Xi_b$ branching fractions.}
\label{tab:b2charm_Bbaryon_baryon_overview2}
%
\begin{tablenotes}
\item[*] Not included in PDG average.
\end{tablenotes}
\end{threeparttable}
\end{center}
\end{table}

\hflavrowdef{\hflavrowdefast}{$\dfrac{{\cal B}(\Lambda_b^0 \to \Lambda_c^+ D^-)}{{\cal B}(\Lambda_b^0 \to \Lambda_c^+ D_s^-)}$}{{\setlength{\tabcolsep}{0pt}%
}
\begin{table}[H]
\begin{center}
\begin{threeparttable}
\caption{Branching fraction ratios of $b$ baryons.}
\label{tab:b2charm_Bbaryon_baryon_overview3}
%
 \\
\hline
\hflavrowdefast \\
 & & \\[-1ex]
\hflavrowdefasu \\
 & & \\[-1ex]
\hflavrowdefasv \\
 & & \\[-1ex]
\hflavrowdefasw \\
\hline
\hline
\end{tabular}
\begin{tablenotes}
\item[*] Not included in PDG average.
\item[1] Measurement of $\dfrac{{\cal B}(\Xi_b^0 \to \Lambda_c^+ K^-)}{{\cal B}(\Xi_b^0 \to D^0 p K^-)} {\cal B}(\Lambda_c^+ \to p K^- \pi^+) / {\cal B}(D^0 \to K^- \pi^+)$ used in our fit.
\end{tablenotes}
\end{threeparttable}
\end{center}
\end{table}

\hflavrowdef{\hflavrowdefasn}{${\cal B}(\Lambda_b^0 \to \Sigma_c(2455)^0 p \bar{p}) \times {\cal B}(\Sigma_c(2455)^0 \to \Lambda_c^+ \pi^-)$}{{\setlength{\tabcolsep}{0pt}
}
\begin{table}[H]
\begin{center}
\begin{threeparttable}
\caption{Product branching fractions of $b$ baryons.}
\label{tab:b2charm_Bbaryon_baryon_overview4}
%
 \\
\hline
\hflavrowdefasn \\
 & & \\[-1ex]
\hflavrowdefaso \\
 & & \\[-1ex]
\hflavrowdefasp \\
 & & \\[-1ex]
\hflavrowdefasq \\
 & & \\[-1ex]
\hflavrowdefasr \\
 & & \\[-1ex]
\hflavrowdefass \\
\hline
\hline
\end{tabular}
\begin{tablenotes}
\item[1] Using ${\cal B}(\Lambda_b^0 \to \Lambda_c^+ p \bar{p} \pi^-)$.
\item[2] Using ${\cal B}(\Lambda_b^0 \to \Lambda_c^+ \pi^+ \pi^- \pi^-)$.
\end{tablenotes}
\end{threeparttable}
\end{center}
\end{table}

\subsubsection{Decays to exotic states}
\label{sec:b2c:Bbaryon_other}
Averages of $b$ baryon decays to $XYZP$ states are shown in Table~\ref{tab:b2charm_Bbaryon_other_overview1}.
\hflavrowdef{\hflavrowdefasx}{${\cal B}(\Lambda_b^0 \to P_c(4380)^+ \pi^-)$}{{\setlength{\tabcolsep}{0pt}
}
\begin{table}[H]
\begin{center}
\begin{threeparttable}
\caption{Branching fractions for decays to pentaquarks.}
\label{tab:b2charm_Bbaryon_other_overview1}
%
 \\
\hline
\hflavrowdefasx \\
 & & \\[-1ex]
\hflavrowdefasy \\
 & & \\[-1ex]
\hflavrowdefasz \\
 & & \\[-1ex]
\hflavrowdefata \\
 & & \\[-1ex]
\hflavrowdefatb \\
\hline
\hline
\end{tabular}
\begin{tablenotes}
\item[*] Not included in PDG average.
\item[1] Using ${\cal B}(\Lambda_b^0 \to J/\psi p K^-)$.
\item[2] Using ${\cal B}(\Lambda_b^0 \to P_c(4380)^+ K^-)$.
\item[3] Measurement of ${\cal B}(\Lambda_b^0 \to P_c(4380)^+ \pi^-) / {\cal B}(\Lambda_b^0 \to P_c(4380)^+ K^-)$ used in our fit.
\item[4] Using ${\cal B}(\Lambda_b^0 \to P_c(4457)^+ K^-)$.
\item[5] Measurement of ${\cal B}(\Lambda_b^0 \to P_c(4457)^+ \pi^-) / {\cal B}(\Lambda_b^0 \to P_c(4457)^+ K^-)$ used in our fit.
\end{tablenotes}
\end{threeparttable}
\end{center}
\end{table}

\clearpage

\mysection{$b$-hadron decays to charmless final states}

\label{sec:rare}

This section provides branching fractions (BF), polarization 
fractions, \CP\ asymmetries ($A_{\CP}$) and other observables of 
$b$-hadron decays to final states that do not contain charm hadrons or charmonium mesons\footnote{The treatment of intermediate charm or charmonium states differs between observables and sometimes among results for the same observable. In the latter case, when these results are averaged, we indicate the differences by footnotes.}, except for a few lepton-flavour- and lepton-number-violating decays reported in Sec.~\ref{sec:rare-radll}.
Four categories of $\Bz$ and $\Bp$ decays are reported: 
mesonic (\ie, final states containing only mesons), baryonic (hadronic final states with baryon-antibaryon pairs), radiative (including a photon or a lepton-antilepton pair) and semileptonic/leptonic (including/only leptons).
We also report measurements of $\Bs$, $\Bc$ and $b$-baryon decays, and measurements of final-state polarization in $b$-hadron decays.
As discussed in Sec.~\ref{sec:intro}, measurements included in our averages are those supported with public notes, including journal papers, 
conference-contributed papers, preprints or conference proceedings, except when a result has not led to a journal publication after an extended period of time.

The largest improvements since our last report~\cite{HFLAV:2022esi} have come from a
variety of new measurements from LHCb and Belle-II. This includes LHCb's update of the ratio $R_{K^*}$, which is now consistent with the standard model prediction at the level of $0.6$ standard deviations, as well as
the first evidence of the decay $B \to K \nu \overline{\nu}$
from Belle~II.

The averaging procedure follows the methodology described in Sec.~\ref{sec:method}.
We perform fits of the full likelihood function and do not use the approximation described in Sec.~\ref{sec:method:corrSysts}.
Thus, observables that are related to each other, e.g., by ratios of branching fractions, are fitted jointly.
Details about all the observables involved in each average, as well as the induced correlation coefficients and $p$-values, are available via clickable links from each average in the tables on our web page~\cite{HFLAV:rareURL}\footnote{Where available, other sources of correlations between measurements of the same observable and among different observables are also taken into account.}.
In total 404 fits are performed, with on average (maximally) 1.2 (37) parameters and 4.1 (20) measurements per fit.
In our tables, each group of rows contains the individual measurements and the average corresponding to a given parameter $p_j$.
The cases where the fits incorporate measurements that are functions of $p_j$, which are used as direct inputs to the fits, are indicated with a footnote.
In general, a value of $p_j$ is not quoted in the tables. There are two exceptions to this: a ratio of branching fractions, $p_j/p_k$, where $p_k$ is the branching fraction of a normalization mode, and a product, $p_j p_k$ of the branching fraction of interest with that of a daughter decay. In these two cases the numerical value of $p_j$, naively obtained using the known value of $p_k$, is quoted in the tables for reference, and the uncertainty on $p_k$ is included in the systematic uncertainty on $p_j$.

Systematic uncertainties are taken as quoted in the original publications, without the scaling of multiplicative uncertainties discussed in Sec.~\ref{sec:method:nonGaussian}. When several systematic uncertainties are given separately, we sum them in quadrature and quote a single systematic uncertainty. These cases are marked by a footnote.

If one or more experiments report a BF measurement with a significance of more than three standard deviations ($\sigma$), all available central values for that BF are used in our average. For BFs that do not satisfy this criterion, the most stringent limit is used.
Quoted upper limits are at 90\% confidence level (C.L.), unless mentioned otherwise.
For observables that are not BFs, such as $A_{\CP}$ or polarization fractions, we include in our averages all the available results, regardless of their significance.

Many of the branching fractions from \babar\ and Belle assume equal production 
of charged and neutral $B$-meson pairs. The best measurements to date show that this
is still a reasonable approximation (see Sec.~\ref{sec:fractions}), and thus, we do not correct for it and simply quote the results from the original publications.

At the end of some of the sections, we list results that were not
included in the tables. Typical cases are measurements of distributions, such as differential
branching fractions or longitudinal polarizations, which are measured in different
binning schemes by the different collaborations, and thus cannot be directly
used to obtain averages.

Observables obtained by Dalitz-plot analyses are marked by footnotes. In these analyses, different experimental collaborations often use different models, in particular for the nonresonant component. When applicable, we detail the model used for the nonresonant component in a footnote. In addition to this, Dalitz-plot analyses often yield multiple solutions. In this case, we take the results corresponding to the global minimum and follow the conclusions of the papers.

In most of the tables, the averages are compared to those from the PDG 2022 and 2023 updates~\cite{PDG_2022}. When this is done, the ‘‘Average'' column quotes the PDG averages (in italic characters) only if they differ from ours.
In general, such differences are due to different input parameters, differences in the averaging methods and different rounding conventions.
Unlike the PDG, no error scaling is applied in our averages when the fit $\chi^2$ is greater than 1. On the other hand, the fit $p$-value is quoted if it is below 1\%.
Input values that are not included in the PDG 2023 average are marked with footnotes. These are new results published since the closing of PDG 2023 and before the closing of this report in September 2023. Input values that were unpublished at the closing of this report (unpublished results are never included in the PDG averages) are also marked with footnotes.
Sections~\ref{sec:rare-charmless} and \ref{sec:rare-bary} provide compilations of branching fractions of $\Bz$ and $\Bp$ to mesonic and baryonic charmless final states, respectively. 
Sections~\ref{sec:rare-lb} and~\ref{sec:rare-bs} and~\ref{sec:rare-bc}  give branching fractions of $b$-baryon, \Bs-meson, and $\Bc$ meson decays to charmless final states, respectively~\footnote{Except for decays of $\Bc$ mesons to final states containing $\Bs$ mesons, which are quoted in Sec.~\ref{sec:b2c:Bc}.}.
In Sec.~\ref{sec:rare-radll} observables of interest are given for
radiative decays and FCNC) decays with leptons of \Bz and \Bp\ mesons, including
limits from searches for lepton-flavour/number-violating decays.
Finally, sections~\ref{sec:rare-acp} and \ref{sec:rare-polar} give \CP\ asymmetries and results of polarization measurements, respectively, in various $b$-hadron charmless decays.

\newpage
\mysubsection{Mesonic decays of \Bp\ and \Bz mesons}
\label{sec:rare-charmless}

This section provides branching fractions of charmless mesonic decays.
Tables~\ref{tab:rareDecays_charmlessBu_Bu_BR_S1} to~\ref{tab:rareDecays_charmlessBu_Bu_BR_NS3} are for \Bp\
and Tables~\ref{tab:rareDecays_charmlessBd_Bd_BR_S1} to \ref{tab:rareDecays_charmlessBd_Bd_BR_NS4}
are for \Bz\ mesons.
For both, decay modes with and without strange mesons in the final state appear in different tables.
Finally, Tables~\ref{tab:rareDecays_charmlessBu_Bu_RBR}
and~\ref{tab:rareDecays_charmlessBd_Bd_RBR}
detail several relative branching fractions of \Bp\ and \Bz\ decays, respectively.
Figure~\ref{fig:rare-mostprec} gives a graphic representation of a selection of high-precision branching fractions given in this section.

\hflavrowdef{\hflavrowdefaaa}{${\cal B}(B^+ \to K^0 \pi^+)$\tnote{1}\hphantom{\textsuperscript{1}}}{{\setlength{\tabcolsep}{0pt}
}
\begin{table}[H]
\begin{center}
\begin{threeparttable}
\caption{Branching fractions of charmless mesonic $B^+$ decays with strange mesons (part 1).}
\label{tab:rareDecays_charmlessBu_Bu_BR_S1}
%
 \\
\hline
\hflavrowdefaaa \\
 & & \\[-1ex]
\hflavrowdefaab \\
 & & \\[-1ex]
\hflavrowdefaac \\
 & & \\[-1ex]
\hflavrowdefaad \\
 & & \\[-1ex]
\hflavrowdefaae \\
 & & \\[-1ex]
\hflavrowdefaaf \\
 & & \\[-1ex]
\hflavrowdefaag \\
\hline
\hline
\end{tabular}
\begin{tablenotes}
\item[\dag] Preliminary result.
\item[*] Not included in PDG average.
\item[1] The PDG average is a result of a fit including input from other measurements.
\item[2] Measurement of ${\cal B}(B^+ \to K^+ \bar{K}^0) / {\cal B}(B^+ \to K^0 \pi^+)$ used in our fit.
\item[3] Measurement of ${\cal B}(B_s^0 \to \eta^\prime \eta^\prime) / {\cal B}(B^+ \to \eta^\prime K^+)$ used in our fit.
\item[4] Multiple systematic uncertainties are added in quadrature.
\end{tablenotes}
\end{threeparttable}
\end{center}
\end{table}

\hflavrowdef{\hflavrowdefaah}{${\cal B}(B^+ \to \eta K^+)$\tnote{1}\hphantom{\textsuperscript{1}}}{{\setlength{\tabcolsep}{0pt}
}
\begin{table}[H]
\begin{center}
\resizebox{1\textwidth}{!}{
\begin{threeparttable}
\caption{Branching fractions of charmless mesonic $B^+$ decays with strange mesons (part 2).}
\label{tab:rareDecays_charmlessBu_Bu_BR_S2}
%
 \\
\hline
\hflavrowdefaah \\
 & & \\[-1ex]
\hflavrowdefaai \\
 & & \\[-1ex]
\hflavrowdefaaj \\
 & & \\[-1ex]
\hflavrowdefaak \\
 & & \\[-1ex]
\hflavrowdefaal \\
 & & \\[-1ex]
\hflavrowdefaam \\
 & & \\[-1ex]
\hflavrowdefaan \\
 & & \\[-1ex]
\hflavrowdefaao \\
 & & \\[-1ex]
\hflavrowdefaap \\
 & & \\[-1ex]
\hflavrowdefaaq \\
 & & \\[-1ex]
\hflavrowdefaar \\
 & & \\[-1ex]
\hflavrowdefaas \\
 & & \\[-1ex]
\hflavrowdefaat \\
 & & \\[-1ex]
\hflavrowdefaau \\
\hline
\hline
\end{tabular}
\begin{tablenotes}
\item[1] The PDG uncertainty includes a scale factor.
\item[2] The PDG entry corresponds to ${\cal{B}}( B^+ \to \eta (K\pi)_0^{*+})$.
\item[3] Multiple systematic uncertainties are added in quadrature.
\item[4] Result extracted from Dalitz-plot analysis of $B^+ \to K^+ K^+ K^-$ decays.
\item[5] Result extracted from Dalitz-plot analysis of $B^+ \to K_S^0 K_S^0 K^+$ decays.
\end{tablenotes}
\end{threeparttable}
}
\end{center}
\end{table}

\hflavrowdef{\hflavrowdefaav}{${\cal B}(B^+ \to \omega(782) K^+)$\tnote{1}\hphantom{\textsuperscript{1}}}{{\setlength{\tabcolsep}{0pt}
}
\begin{table}[H]
\begin{center}
\resizebox{1\textwidth}{!}{
\begin{threeparttable}
\caption{Branching fractions of charmless mesonic $B^+$ decays with strange mesons (part 3).}
\label{tab:rareDecays_charmlessBu_Bu_BR_S3}
%
 \\
\hline
\hflavrowdefaav \\
 & & \\[-1ex]
\hflavrowdefaaw \\
 & & \\[-1ex]
\hflavrowdefaax \\
 & & \\[-1ex]
\hflavrowdefaay \\
 & & \\[-1ex]
\hflavrowdefaaz \\
 & & \\[-1ex]
\hflavrowdefaba \\
 & & \\[-1ex]
\hflavrowdefabb \\
 & & \\[-1ex]
\hflavrowdefabc \\
 & & \\[-1ex]
\hflavrowdefabd \\
 & & \\[-1ex]
\hflavrowdefabe \\
 & & \\[-1ex]
\hflavrowdefabf \\
\hline
\hline
\end{tabular}
\begin{tablenotes}
\item[\dag] Preliminary result.
\item[1] The measurement from the Dalitz-plot analysis of $B^+ \to K^+ \pi^+ \pi^-$ decays \cite{BaBar:2008lpx} was not included in this average. It is quoted as a separate entry.
\item[2] Result extracted from Dalitz-plot analysis of $B^+ \to K^+ \pi^+ \pi^-$ decays.
\item[3] Result extracted from Dalitz-plot analysis of $B^+ \to K_S^0 \pi^+ \pi^0$ decays.
\item[4] Multiple systematic uncertainties are added in quadrature.
\item[5] Using ${\cal B}(B^+ \to K^+ K^+ K^-)$.
\item[6] The total nonresonant contribution is obtained by combining a exponential nonresonant component with the effective-range part of the LASS lineshape.
\end{tablenotes}
\end{threeparttable}
}
\end{center}
\end{table}

\hflavrowdef{\hflavrowdefabg}{${\cal{B}}( B^+ \to \omega(782) K^+ \;(K^+ \pi^+ \pi^-) )$\tnote{1}\hphantom{\textsuperscript{1}}}{{\setlength{\tabcolsep}{0pt}
}
\begin{table}[H]
\begin{center}
\resizebox{1\textwidth}{!}{
\begin{threeparttable}
\caption{Branching fractions of charmless mesonic $B^+$ decays with strange mesons (part 4).}
\label{tab:rareDecays_charmlessBu_Bu_BR_S4}
%
 \\
\hline
\hflavrowdefabg \\
 & & \\[-1ex]
\hflavrowdefabh \\
 & & \\[-1ex]
\hflavrowdefabi \\
 & & \\[-1ex]
\hflavrowdefabj \\
 & & \\[-1ex]
\hflavrowdefabk \\
 & & \\[-1ex]
\hflavrowdefabl \\
 & & \\[-1ex]
\hflavrowdefabm \\
 & & \\[-1ex]
\hflavrowdefabn \\
 & & \\[-1ex]
\hflavrowdefabo \\
 & & \\[-1ex]
\hflavrowdefabp \\
 & & \\[-1ex]
\hflavrowdefabq \\
 & & \\[-1ex]
\hflavrowdefabr \\
 & & \\[-1ex]
\hflavrowdefabs \\
\hline
\hline
\end{tabular}
\begin{tablenotes}
\item[1] This result was not included in the main entry of ${\cal{B}}( B^+ \to \omega(782) K^+)$.
\item[2] Result extracted from Dalitz-plot analysis of $B^+ \to K^+ \pi^+ \pi^-$ decays.
\item[3] The PDG uncertainty includes a scale factor.
\item[4] Result extracted from Dalitz-plot analysis of $B^+ \to K_S^0 \pi^+ \pi^0$ decays.
\item[5] Multiple systematic uncertainties are added in quadrature.
\end{tablenotes}
\end{threeparttable}
}
\end{center}
\end{table}

\hflavrowdef{\hflavrowdefabt}{${\cal B}(B^+ \to K^- \pi^+ \pi^+)$}{{\setlength{\tabcolsep}{0pt}
}
\begin{table}[H]
\begin{center}
\begin{threeparttable}
\caption{Branching fractions of charmless mesonic $B^+$ decays with strange mesons (part 5).}
\label{tab:rareDecays_charmlessBu_Bu_BR_S5}
%
 \\
\hline
\hflavrowdefabt \\
 & & \\[-1ex]
\hflavrowdefabu \\
 & & \\[-1ex]
\hflavrowdefabv \\
 & & \\[-1ex]
\hflavrowdefabw \\
 & & \\[-1ex]
\hflavrowdefabx \\
 & & \\[-1ex]
\hflavrowdefaby \\
 & & \\[-1ex]
\hflavrowdefabz \\
 & & \\[-1ex]
\hflavrowdefaca \\
 & & \\[-1ex]
\hflavrowdefacb \\
 & & \\[-1ex]
\hflavrowdefacc \\
 & & \\[-1ex]
\hflavrowdefacd \\
 & & \\[-1ex]
\hflavrowdeface \\
 & & \\[-1ex]
\hflavrowdefacf \\
 & & \\[-1ex]
\hflavrowdefacg \\
 & & \\[-1ex]
\hflavrowdefach \\
\hline
\hline
\end{tabular}
\begin{tablenotes}
\item[1] Result extracted from Dalitz-plot analysis of $B^+ \to K_S^0 \pi^+ \pi^0$ decays.
\item[2] Multiple systematic uncertainties are added in quadrature.
\item[3] See also Ref.~\cite{Belle:2005fjn}.
\end{tablenotes}
\end{threeparttable}
\end{center}
\end{table}

\hflavrowdeflong{\hflavrowdefaci}{${\cal B}(B^+ \to b_1(1235)^0 K^+) \times {\cal B}(b_1(1235)^0 \to \omega(782) \pi^0)$}{{\setlength{\tabcolsep}{0pt}
}
\begin{table}[H]
\begin{center}
\begin{threeparttable}
\caption{Branching fractions of charmless mesonic $B^+$ decays with strange mesons (part 6).}
\label{tab:rareDecays_charmlessBu_Bu_BR_S6}
%
 \\
\hline
\hflavrowdefaci \\
 & & \\[-1ex]
\hflavrowdefacj \\
 & & \\[-1ex]
\hflavrowdefack \\
 & & \\[-1ex]
\hflavrowdefacl \\
 & & \\[-1ex]
\hflavrowdefacm \\
 & & \\[-1ex]
\hflavrowdefacn \\
 & & \\[-1ex]
\hflavrowdefaco \\
 & & \\[-1ex]
\hflavrowdefacp \\
 & & \\[-1ex]
\hflavrowdefacq \\
 & & \\[-1ex]
\hflavrowdefacr \\
\hline
\hline
\end{tabular}
\begin{tablenotes}
\item[1] The PDG average is a result of a fit including input from other measurements.
\item[2] Using ${\cal B}(B^+ \to K^0 \pi^+)$.
\item[3] PDG uses the BABAR result including the $\chi_{c0}$ intermediate state.
\item[4] Result extracted from Dalitz-plot analysis of $B^+ \to K_S^0 K_S^0 K^+$ decays.
\item[5] All charmonium resonances are vetoed. The analysis also reports ${\cal{B}}( B^{+} \to K^0_S K^0_S K^+ ) = (10.6 \pm 0.5 \pm 0.3) \times 10^{-6}$ including the $\chi_{c0}$ intermediate state.
\item[6] The nonresonant amplitude is modelled using a polynomial function of order 2.
\end{tablenotes}
\end{threeparttable}
\end{center}
\end{table}

\hflavrowdef{\hflavrowdefacs}{${\cal B}(B^+ \to K^+ K^- \pi^+)$}{{\setlength{\tabcolsep}{0pt}
}
\begin{table}[H]
\begin{center}
\resizebox{1\textwidth}{!}{
\begin{threeparttable}
\caption{Branching fractions of charmless mesonic $B^+$ decays with strange mesons (part 7).}
\label{tab:rareDecays_charmlessBu_Bu_BR_S7}
%
 \\
\hline
\hflavrowdefacs \\
 & & \\[-1ex]
\hflavrowdefact \\
 & & \\[-1ex]
\hflavrowdefacu \\
 & & \\[-1ex]
\hflavrowdefacv \\
 & & \\[-1ex]
\hflavrowdefacw \\
 & & \\[-1ex]
\hflavrowdefacx \\
 & & \\[-1ex]
\hflavrowdefacy \\
 & & \\[-1ex]
\hflavrowdefacz \\
 & & \\[-1ex]
\hflavrowdefada \\
 & & \\[-1ex]
\hflavrowdefadb \\
 & & \\[-1ex]
\hflavrowdefadc \\
\hline
\hline
\end{tabular}
\begin{tablenotes}
\item[*] Not included in PDG average.
\item[1] Using ${\cal B}(B^+ \to K^+ K^+ K^-)$.
\item[2] Also measured in bins of $m_{K^+K^-}$ and $m_{K^+\pi^-}$.
\item[3] LHCb uses a model of the nonresonant contribution obtained from a phenomenological description of the partonic interaction that produces the final state. This contribution is referred to as the single pole in the paper; see Ref.~\cite{LHCb:2019xmb} for details.
\item[4] Using ${\cal{B}}(B^+ \to K^+ K^- \pi^+)$.
\item[5] Result extracted from Dalitz-plot analysis of $B^+ \to K^+ K^- \pi^+$ decays.
\item[6] Measurement of $( {\cal B}(B^+ \to \bar{K}^*(892)^0 K^+) {\cal B}(K^*(892)^0 \to K \pi) 2/3 ) / {\cal{B}}(B^+ \to K^+ K^- \pi^+)$ used in our fit.
\item[7] Measurement of $( {\cal B}(B^+ \to \bar{K}_0^*(1430)^0 K^+) {\cal{B}}(K^*(1430) \to K \pi) 2/3 ) / {\cal{B}}(B^+ \to K^+ K^- \pi^+)$ used in our fit.
\item[8] LHCb uses a dedicated lineshape to take into account $\pi\pi \leftrightarrow KK$ rescattering, which is particularly significant in the region $1 < m_{KK} < 1.5~\text{GeV}/c^2$. See Ref.~\cite{LHCb:2019xmb} for details.
\item[9] The PDG uncertainty includes a scale factor.
\item[10] Result extracted from Dalitz-plot analysis of $B^+ \to K^+ K^+ K^-$ decays.
\item[11] Result extracted from Dalitz-plot analysis of $B^+ \to K_S^0 K_S^0 K^+$ decays.
\end{tablenotes}
\end{threeparttable}
}
\end{center}
\end{table}

\hflavrowdef{\hflavrowdefadd}{${\cal B}(B^+ \to K^+ K^+ K^-)$\tnote{1,2}\hphantom{\textsuperscript{1,2}}}{{\setlength{\tabcolsep}{0pt}
}
\begin{table}[H]
\begin{center}
\resizebox{0.95\textwidth}{!}{
\begin{threeparttable}
\caption{Branching fractions of charmless mesonic $B^+$ decays with strange mesons (part 8).}
\label{tab:rareDecays_charmlessBu_Bu_BR_S8}
%
 \\
\hline
\hflavrowdefadd \\
 & & \\[-1ex]
\hflavrowdefade \\
 & & \\[-1ex]
\hflavrowdefadf \\
 & & \\[-1ex]
\hflavrowdefadg \\
 & & \\[-1ex]
\hflavrowdefadh \\
 & & \\[-1ex]
\hflavrowdefadi \\
 & & \\[-1ex]
\hflavrowdefadj \\
\hline
\hline
\end{tabular}
\begin{tablenotes}
\item[\dag] Preliminary result.
\item[1] The PDG uncertainty includes a scale factor.
\item[2] Treatment of charmonium intermediate components differs between the results.
\item[3] Result extracted from Dalitz-plot analysis of $B^+ \to K^+ K^+ K^-$ decays.
\item[4] All charmonium resonances are vetoed, except for $\chi_{c0}$. The analysis also reports ${\cal{B}}( B^{+} \to K^+ K^+ K^- ) = (33.4 \pm 0.5 \pm 0.9) \times 10^{-6}$ excluding $\chi_{c0}$.
\item[5] Measurement of $\frac{f_{\Omega^{-}_{b}}}{f_{u}} \times {\cal{B}}(\Omega^{-}_{b} \to p  K^- K^-) / {\cal B}(B^+ \to K^+ K^+ K^-)$ used in our fit.
\item[6] Measurement of $\frac{f_{\Omega^{-}_{b}}}{f_{u}} \times {\cal{B}}(\Omega^{-}_{b} \to p  K^- \pi^-) / {\cal B}(B^+ \to K^+ K^+ K^-)$ used in our fit.
\item[7] Measurement of $\frac{f_{\Omega^{-}_{b}}}{f_{u}} \times {\cal{B}}(\Omega^{-}_{b} \to p  \pi^- \pi^-) / {\cal B}(B^+ \to K^+ K^+ K^-)$ used in our fit.
\item[8] Measurement of ${\cal B}(B^+ \to K^+ K^- \pi^+) / {\cal B}(B^+ \to K^+ K^+ K^-)$ used in our fit.
\item[9] Measurement of ${\cal B}(B^+ \to K^+ \pi^+ \pi^-) / {\cal B}(B^+ \to K^+ K^+ K^-)$ used in our fit.
\item[10] Measurement of ${\cal B}(B^+ \to \pi^+ \pi^+ \pi^-) / {\cal B}(B^+ \to K^+ K^+ K^-)$ used in our fit.
\item[11] The nonresonant amplitude is modelled using a polynomial function including S-wave and P-wave terms.
\end{tablenotes}
\end{threeparttable}
}
\end{center}
\end{table}

\hflavrowdef{\hflavrowdefadk}{${\cal B}(B^+ \to K^*(892)^+ K^+ K^-)$}{{\setlength{\tabcolsep}{0pt}
}
\begin{table}[H]
\begin{center}
\begin{threeparttable}
\caption{Branching fractions of charmless mesonic $B^+$ decays with strange mesons (part 9).}
\label{tab:rareDecays_charmlessBu_Bu_BR_S9}
%
 \\
\hline
\hflavrowdefadk \\
 & & \\[-1ex]
\hflavrowdefadl \\
 & & \\[-1ex]
\hflavrowdefadm \\
 & & \\[-1ex]
\hflavrowdefadn \\
 & & \\[-1ex]
\hflavrowdefado \\
 & & \\[-1ex]
\hflavrowdefadp \\
 & & \\[-1ex]
\hflavrowdefadq \\
 & & \\[-1ex]
\hflavrowdefadr \\
 & & \\[-1ex]
\hflavrowdefads \\
 & & \\[-1ex]
\hflavrowdefadt \\
 & & \\[-1ex]
\hflavrowdefadu \\
 & & \\[-1ex]
\hflavrowdefadv \\
 & & \\[-1ex]
\hflavrowdefadw \\
 & & \\[-1ex]
\hflavrowdefadx \\
 & & \\[-1ex]
\hflavrowdefady \\
 & & \\[-1ex]
\hflavrowdefadz \\
\hline
\hline
\end{tabular}
\begin{tablenotes}
\item[\dag] Preliminary result.
\item[1] The PDG uncertainty includes a scale factor.
\item[2] Combination of two final states of the $K^*(892)^{\pm}$, $K_S^0\pi^{\pm}$ and $K^{\pm}\pi^0$. In addition to the combined results, the paper reports separately the results for each individual final state.
\item[3] Measured in the $\phi \phi$ invariant mass range below the $\eta_c$ resonance ($m_{\phi \phi} < 2.85~\text{GeV}/c^2$).
\item[4] $h = \pi, K$.
\end{tablenotes}
\end{threeparttable}
\end{center}
\end{table}

\hflavrowdef{\hflavrowdefaea}{${\cal B}(B^+ \to \pi^+ \pi^0)$\tnote{1}\hphantom{\textsuperscript{1}}}{{\setlength{\tabcolsep}{0pt}
}
\begin{table}[H]
\begin{center}
\resizebox{1\textwidth}{!}{
\begin{threeparttable}
\caption{Branching fractions of charmless mesonic $B^+$ decays without strange mesons (part 1).}
\label{tab:rareDecays_charmlessBu_Bu_BR_NS1}
%
 \\
\hline
\hflavrowdefaea \\
 & & \\[-1ex]
\hflavrowdefaeb \\
 & & \\[-1ex]
\hflavrowdefaec \\
 & & \\[-1ex]
\hflavrowdefaed \\
 & & \\[-1ex]
\hflavrowdefaee \\
 & & \\[-1ex]
\hflavrowdefaef \\
 & & \\[-1ex]
\hflavrowdefaeg \\
 & & \\[-1ex]
\hflavrowdefaeh \\
 & & \\[-1ex]
\hflavrowdefaei \\
\hline
\hline
\end{tabular}
\begin{tablenotes}
\item[\dag] Preliminary result.
\item[*] Not included in PDG average.
\item[1] The PDG uncertainty includes a scale factor.
\item[2] Using ${\cal B}(B^+ \to K^+ K^+ K^-)$.
\item[3] Result extracted from Dalitz-plot analysis of $B^+ \to \pi^+ \pi^+ \pi^-$ decays.
\item[4] Charm and charmonium contributions are subtracted.
\item[5] Multiple systematic uncertainties are added in quadrature.
\item[6] This analysis uses three different approaches: isobar, $K$-matrix and quasi-model-independent, to describe the $S$-wave component. The results are taken from the isobar model with an additional error accounting for the different S-wave methods as reported in Appendix D of Ref.~\cite{LHCb:2019sus}.
\item[7] Using ${\cal{B}}(B^+ \to \pi^+ \pi^+ \pi^-)$.
\item[8] Result extracted from Dalitz-plot analysis of $B^+ \to K^+ K^- \pi^+$ decays.
\item[9] Using ${\cal{B}}(B^+ \to K^+ K^- \pi^+)$.
\end{tablenotes}
\end{threeparttable}
}
\end{center}
\end{table}

\hflavrowdef{\hflavrowdefaej}{${\cal{B}}(B^+ \to \pi^+ \pi^+ \pi^-)~S\rm{-wave}$}{{\setlength{\tabcolsep}{0pt}
}
\begin{table}[H]
\begin{center}
\resizebox{1\textwidth}{!}{
\begin{threeparttable}
\caption{Branching fractions of charmless mesonic $B^+$ decays without strange mesons (part 2).}
\label{tab:rareDecays_charmlessBu_Bu_BR_NS2}
%
 \\
\hline
\hflavrowdefaej \\
 & & \\[-1ex]
\hflavrowdefaek \\
 & & \\[-1ex]
\hflavrowdefael \\
 & & \\[-1ex]
\hflavrowdefaem \\
 & & \\[-1ex]
\hflavrowdefaen \\
 & & \\[-1ex]
\hflavrowdefaeo \\
 & & \\[-1ex]
\hflavrowdefaep \\
 & & \\[-1ex]
\hflavrowdefaeq \\
 & & \\[-1ex]
\hflavrowdefaer \\
 & & \\[-1ex]
\hflavrowdefaes \\
 & & \\[-1ex]
\hflavrowdefaet \\
 & & \\[-1ex]
\hflavrowdefaeu \\
 & & \\[-1ex]
\hflavrowdefaev \\
 & & \\[-1ex]
\hflavrowdefaew \\
\hline
\hline
\end{tabular}
\begin{tablenotes}
\item[*] Not included in PDG average.
\item[\dag] Preliminary result.
\item[1] LHCb accounts for the $S$-wave component using a model that comprises the coherent sum of a $\sigma$ pole and a rescattering amplitude. See Ref.~\cite{LHCb:2019jta} for details.
\item[2] This analysis uses three different approaches: isobar, $K$-matrix and quasi-model-independent, to describe the $S$-wave component. The results are taken from the isobar model with an additional error accounting for the different S-wave methods as reported in Appendix D of Ref.~\cite{LHCb:2019sus}.
\item[3] Multiple systematic uncertainties are added in quadrature.
\item[4] Using ${\cal{B}}(B^+ \to \pi^+ \pi^+ \pi^-)$.
\item[5] Result extracted from Dalitz-plot analysis of $B^+ \to \pi^+ \pi^+ \pi^-$ decays.
\item[6] The nonresonant amplitude is modelled using a sum of exponential functions.
\item[7] $X_{\pi^0 \pi^0}$ corresponds to a structure observed in Ref.~\cite{Belle:2022dgi}, likely arising due to multiple resonances.
\item[8] Measurement of $( {\cal B}(B^+ \to \omega(782) \pi^+) {\cal B}(\omega(782) \to \pi^+ \pi^-) ) / {\cal{B}}(B^+ \to \pi^+ \pi^+ \pi^-)$ used in our fit.
\end{tablenotes}
\end{threeparttable}
}
\end{center}
\end{table}

\hflavrowdef{\hflavrowdefaex}{${\cal B}(B^+ \to \eta \pi^+)$}{{\setlength{\tabcolsep}{0pt}
}
\begin{table}[H]
\begin{center}
\resizebox{1\textwidth}{!}{
\begin{threeparttable}
\caption{Branching fractions of charmless mesonic $B^+$ decays without strange mesons (part 3).}
\label{tab:rareDecays_charmlessBu_Bu_BR_NS3}
%
 \\
\hline
\hflavrowdefaex \\
 & & \\[-1ex]
\hflavrowdefaey \\
 & & \\[-1ex]
\hflavrowdefaez \\
 & & \\[-1ex]
\hflavrowdefafa \\
 & & \\[-1ex]
\hflavrowdefafb \\
 & & \\[-1ex]
\hflavrowdefafc \\
 & & \\[-1ex]
\hflavrowdefafd \\
 & & \\[-1ex]
\hflavrowdefafe \\
 & & \\[-1ex]
\hflavrowdefaff \\
 & & \\[-1ex]
\hflavrowdefafg \\
 & & \\[-1ex]
\hflavrowdefafh \\
 & & \\[-1ex]
\hflavrowdefafi \\
 & & \\[-1ex]
\hflavrowdefafj \\
 & & \\[-1ex]
\hflavrowdefafk \\
 & & \\[-1ex]
\hflavrowdefafl \\
 & & \\[-1ex]
\hflavrowdefafm \\
\hline
\hline
\end{tabular}
\begin{tablenotes}
\item[1] The PDG uncertainty includes a scale factor.
\item[2] Result extracted from Dalitz-plot analysis of $B^+ \to K^+ K^- \pi^+$ decays.
\item[3] Measurement of $( {\cal B}(B^+ \to \phi(1020) \pi^+) {\cal B}(\phi(1020) \to K^+ K^-) ) / {\cal{B}}(B^+ \to K^+ K^- \pi^+)$ used in our fit.
\item[4] CLEO assumes ${\cal{B}} (\Upsilon(4S) \to B^0 \overline{B}^0) = 0.43$. The result has been modified to account for a branching fraction of 0.50.
\end{tablenotes}
\end{threeparttable}
}
\end{center}
\end{table}

\hflavrowdef{\hflavrowdefaaa}{${\cal B}(B^0 \to K^+ \pi^-)$}{{\setlength{\tabcolsep}{0pt}
}
\begin{table}[H]
\begin{center}
\begin{threeparttable}
\caption{Branching fractions of charmless mesonic $B^0$ decays with strange mesons (part 1).}
\label{tab:rareDecays_charmlessBd_Bd_BR_S1}
%
 \\
\hline
\hflavrowdefaaa \\
 & & \\[-1ex]
\hflavrowdefaab \\
 & & \\[-1ex]
\hflavrowdefaac \\
 & & \\[-1ex]
\hflavrowdefaad \\
 & & \\[-1ex]
\hflavrowdefaae \\
 & & \\[-1ex]
\hflavrowdefaaf \\
 & & \\[-1ex]
\hflavrowdefaag \\
\hline
\hline
\end{tabular}
\begin{tablenotes}
\item[\dag] Preliminary result.
\item[*] Not included in PDG average.
\item[1] Measurement of $( {\cal B}(B_s^0 \to K^- \pi^+) / {\cal B}(B^0 \to K^+ \pi^-) ) \frac{f_s}{f_d}$ used in our fit.
\item[2] Measurement of $( {\cal B}(\Lambda_b^0 \to p \pi^-) / {\cal B}(B^0 \to K^+ \pi^-) ) ( f_{\Lambda^0_b} / f_{d} )$ used in our fit.
\item[3] Measurement of ${\cal B}(B^0 \to \pi^+ \pi^-) / {\cal B}(B^0 \to K^+ \pi^-)$ used in our fit.
\item[4] Measurement of $( {\cal B}(B_s^0 \to K^+ K^-) / {\cal B}(B^0 \to K^+ \pi^-) ) \frac{f_s}{f_d}$ used in our fit.
\item[5] Measurement of ${\cal B}(B^0 \to K^+ K^-) / {\cal B}(B^0 \to K^+ \pi^-)$ used in our fit.
\item[6] Measurement of $( {\cal B}(B_s^0 \to \pi^+ \pi^-) / {\cal B}(B^0 \to K^+ \pi^-) ) \frac{f_s}{f_d}$ used in our fit.
\item[7] Combination of time-integrated and time-dependent analyses using the best linear unbiased estimator Ref.~\cite{Valassi:2003mu}.
\item[8] The PDG uncertainty includes a scale factor.
\item[9] Measurement of ${\cal B}(\Lambda_b^0 \to \Lambda^0 \eta) / {\cal B}(B^0 \to \eta^\prime K^0)$ used in our fit.
\item[10] Measurement of ${\cal B}(\Lambda_b^0 \to \Lambda^0 \eta^\prime) / {\cal B}(B^0 \to \eta^\prime K^0)$ used in our fit.
\item[11] Multiple systematic uncertainties are added in quadrature.
\end{tablenotes}
\end{threeparttable}
\end{center}
\end{table}

\hflavrowdef{\hflavrowdefaah}{${\cal B}(B^0 \to \eta K^0)$}{{\setlength{\tabcolsep}{0pt}
}
\begin{table}[H]
\begin{center}
\begin{threeparttable}
\caption{Branching fractions of charmless mesonic $B^0$ decays with strange mesons (part 2).}
\label{tab:rareDecays_charmlessBd_Bd_BR_S2}
%
 \\
\hline
\hflavrowdefaah \\
 & & \\[-1ex]
\hflavrowdefaai \\
 & & \\[-1ex]
\hflavrowdefaaj \\
 & & \\[-1ex]
\hflavrowdefaak \\
 & & \\[-1ex]
\hflavrowdefaal \\
 & & \\[-1ex]
\hflavrowdefaam \\
 & & \\[-1ex]
\hflavrowdefaan \\
 & & \\[-1ex]
\hflavrowdefaao \\
 & & \\[-1ex]
\hflavrowdefaap \\
 & & \\[-1ex]
\hflavrowdefaaq \\
 & & \\[-1ex]
\hflavrowdefaar \\
 & & \\[-1ex]
\hflavrowdefaas \\
 & & \\[-1ex]
\hflavrowdefaat \\
 & & \\[-1ex]
\hflavrowdefaau \\
\hline
\hline
\end{tabular}
\begin{tablenotes}
\item[1] Multiple systematic uncertainties are added in quadrature.
\end{tablenotes}
\end{threeparttable}
\end{center}
\end{table}

\hflavrowdef{\hflavrowdefaav}{${\cal B}(B^0 \to \omega(782) K^*(892)^0)$}{{\setlength{\tabcolsep}{0pt}
}
\begin{table}[H]
\begin{center}
\resizebox{0.95\textwidth}{!}{
\begin{threeparttable}
\caption{Branching fractions of charmless mesonic $B^0$ decays with strange mesons (part 3).}
\label{tab:rareDecays_charmlessBd_Bd_BR_S3}
%
 \\
\hline
\hflavrowdefaav \\
 & & \\[-1ex]
\hflavrowdefaaw \\
 & & \\[-1ex]
\hflavrowdefaax \\
 & & \\[-1ex]
\hflavrowdefaay \\
 & & \\[-1ex]
\hflavrowdefaaz \\
 & & \\[-1ex]
\hflavrowdefaba \\
 & & \\[-1ex]
\hflavrowdefabb \\
 & & \\[-1ex]
\hflavrowdefabc \\
 & & \\[-1ex]
\hflavrowdefabd \\
 & & \\[-1ex]
\hflavrowdefabe \\
 & & \\[-1ex]
\hflavrowdefabf \\
 & & \\[-1ex]
\hflavrowdefabg \\
 & & \\[-1ex]
\hflavrowdefabh \\
 & & \\[-1ex]
\hflavrowdefabi \\
 & & \\[-1ex]
\hflavrowdefabj \\
\hline
\hline
\end{tabular}
\begin{tablenotes}
\item[\dag] Preliminary result.
\item[1] $0.755 < m_{K\pi} < 1.250 ~\text{GeV}/c^2$.
\item[2] Result extracted from Dalitz-plot analysis of $B^0 \to K^+ \pi^- \pi^0$ decays.
\item[3] Multiple systematic uncertainties are added in quadrature.
\item[4] The nonresonant amplitude is taken to be constant across the  Dalitz plane.
\item[5] $1.1 < m_{K\pi} < 1.6~\rm{GeV/c^2}$.
\item[6] $K^*_x$ stands for the possible candidates of $K^{*}(1410)$, $K^{*}_{0}(1430)$ and $K^{*}_{2}(1430)$
\end{tablenotes}
\end{threeparttable}
}
\end{center}
\end{table}

\hflavrowdef{\hflavrowdefabk}{${\cal B}(B^0 \to K^0 \pi^+ \pi^-)$\tnote{1,2}\hphantom{\textsuperscript{1,2}}}{{\setlength{\tabcolsep}{0pt}
}
\begin{table}[H]
\begin{center}
\resizebox{0.95\textwidth}{!}{
\begin{threeparttable}
\caption{Branching fractions of charmless mesonic $B^0$ decays with strange mesons (part 4).}
\label{tab:rareDecays_charmlessBd_Bd_BR_S4}
%
 \\
\hline
\hflavrowdefabk \\
 & & \\[-1ex]
\hflavrowdefabl \\
 & & \\[-1ex]
\hflavrowdefabm \\
 & & \\[-1ex]
\hflavrowdefabn \\
\hline
\hline
\end{tabular}
\begin{tablenotes}
\item[*] Not included in PDG average.
\item[1] The PDG average is a result of a fit including input from other measurements.
\item[2] Treatment of charmonium intermediate components differs between the results.
\item[3] Result extracted from Dalitz-plot analysis of $B^0 \to K_S^0 \pi^+ \pi^-$ decays.
\item[4] Multiple systematic uncertainties are added in quadrature.
\item[5] Measurement of ${\cal B}(\Lambda_b^0 \to p \bar{K}^0 \pi^-) / {\cal B}(B^0 \to K^0 \pi^+ \pi^-)$ used in our fit.
\item[6] Measurement of ${\cal B}(\Lambda_b^0 \to p K^0 K^-) / {\cal B}(B^0 \to K^0 \pi^+ \pi^-)$ used in our fit.
\item[7] Measurement of $\frac{f_{\Xi^{0}_{b}}}{f_d}{\cal{B}}(\Xi^{0}_{b} \to p \bar{K}^0 \pi^{-}) / {\cal B}(B^0 \to K^0 \pi^+ \pi^-)$ used in our fit.
\item[8] Measurement of $\frac{f_{\Xi^{0}_{b}}}{f_{d}}{\cal{B}}(\Xi^{0}_{b} \to p \bar{K}^0 K^{-}) / {\cal B}(B^0 \to K^0 \pi^+ \pi^-)$ used in our fit.
\item[9] Measurement of ${\cal B}(B^0 \to K^*(892)^0 \bar{K}^0 \mathrm{+c.c.}) / {\cal B}(B^0 \to K^0 \pi^+ \pi^-)$ used in our fit.
\item[10] Regions corresponding to $D$, $\Lambda_c^+$ and charmonium resonances are vetoed in this analysis.
\item[11] Measurement of ${\cal B}(B^0 \to K^0 K^+ \pi^- \mathrm{+c.c.}) / {\cal B}(B^0 \to K^0 \pi^+ \pi^-)$ used in our fit.
\item[12] Measurement of ${\cal B}(B^0 \to K^0 K^+ K^-) / {\cal B}(B^0 \to K^0 \pi^+ \pi^-)$ used in our fit.
\item[13] Measurement of ${\cal B}(B_s^0 \to K^0 \pi^+ \pi^-) / {\cal B}(B^0 \to K^0 \pi^+ \pi^-)$ used in our fit.
\item[14] Measurement of ${\cal B}(B_s^0 \to K^0 K^+ \pi^- \mathrm{+c.c.}) / {\cal B}(B^0 \to K^0 \pi^+ \pi^-)$ used in our fit.
\item[15] Measurement of ${\cal B}(B_s^0 \to K^0 K^+ K^-) / {\cal B}(B^0 \to K^0 \pi^+ \pi^-)$ used in our fit.
\item[16] The PDG uncertainty includes a scale factor.
\item[17] The nonresonant component is modelled as a flat contribution over the Dalitz plane.
\item[18] Using ${\cal{B}}(B^0 \to K^0 \pi^+ \pi^-)$.
\item[19] This value includes the flat NR component and the effective range of the LASS lineshape. The value corresponding to the flat component alone is also given in the article.
\item[20] The nonresonant component is modelled using a sum of two exponential functions.
\item[21] Result extracted from Dalitz-plot analysis of $B^0 \to K^+ \pi^- \pi^0$ decays.
\item[22] Measurement of ${\cal B}(B_s^0 \to K^*(892)^- \pi^+) / {\cal B}(B^0 \to K^*(892)^+ \pi^-)$ used in our fit.
\item[23] Measurement of ${\cal B}(B^0 \to K^*(892)^- K^+ \mathrm{+c.c.}) / {\cal B}(B^0 \to K^*(892)^+ \pi^-)$ used in our fit.
\item[24] Measurement of $( {\cal B}(B^0 \to K^*(892)^+ \pi^-) 2/3 ) / {\cal{B}}(B^0 \to K^0 \pi^+ \pi^-)$ used in our fit.
\end{tablenotes}
\end{threeparttable}
}
\end{center}
\end{table}

\hflavrowdef{\hflavrowdefabo}{${\cal B}(B^0 \to K_0^*(1430)^+ \pi^-)$\tnote{1}\hphantom{\textsuperscript{1}}}{{\setlength{\tabcolsep}{0pt}
}
\begin{table}[H]
\begin{center}
\resizebox{0.90\textwidth}{!}{
\begin{threeparttable}
\caption{Branching fractions of charmless mesonic $B^0$ decays with strange mesons (part 5).}
\label{tab:rareDecays_charmlessBd_Bd_BR_S5}
%
 \\
\hline
\hflavrowdefabo \\
 & & \\[-1ex]
\hflavrowdefabp \\
 & & \\[-1ex]
\hflavrowdefabq \\
 & & \\[-1ex]
\hflavrowdefabr \\
 & & \\[-1ex]
\hflavrowdefabs \\
 & & \\[-1ex]
\hflavrowdefabt \\
 & & \\[-1ex]
\hflavrowdefabu \\
 & & \\[-1ex]
\hflavrowdefabv \\
 & & \\[-1ex]
\hflavrowdefabw \\
\hline
\hline
\end{tabular}
\begin{tablenotes}
\item[*] Not included in PDG average.
\item[1] The PDG uncertainty includes a scale factor.
\item[2] Result extracted from Dalitz-plot analysis of $B^0 \to K_S^0 \pi^+ \pi^-$ decays.
\item[3] Multiple systematic uncertainties are added in quadrature.
\item[4] $1.1 < m_{K\pi} < 1.6~\rm{GeV/c^2}$.
\item[5] $K^*_x$ stands for the possible candidates of $K^{*}(1410)$, $K^{*}_{0}(1430)$ and $K^{*}_{2}(1430)$
\item[6] Using ${\cal{B}}(B^0 \to K^0 \pi^+ \pi^-)$.
\item[7] Using ${\cal B}(f_2(1270) \to \pi^+ \pi^-)$.
\end{tablenotes}
\end{threeparttable}
}
\end{center}
\end{table}

\hflavrowdef{\hflavrowdefabx}{${\cal B}(B^0 \to K^*(892)^0 \pi^0)$}{{\setlength{\tabcolsep}{0pt}
}
\begin{table}[H]
\begin{center}
\begin{threeparttable}
\caption{Branching fractions of charmless mesonic $B^0$ decays with strange mesons (part 6).}
\label{tab:rareDecays_charmlessBd_Bd_BR_S6}
%
 \\
\hline
\hflavrowdefabx \\
 & & \\[-1ex]
\hflavrowdefaby \\
 & & \\[-1ex]
\hflavrowdefabz \\
 & & \\[-1ex]
\hflavrowdefaca \\
 & & \\[-1ex]
\hflavrowdefacb \\
 & & \\[-1ex]
\hflavrowdefacc \\
 & & \\[-1ex]
\hflavrowdefacd \\
 & & \\[-1ex]
\hflavrowdeface \\
 & & \\[-1ex]
\hflavrowdefacf \\
 & & \\[-1ex]
\hflavrowdefacg \\
\hline
\hline
\end{tabular}
\begin{tablenotes}
\item[1] Result extracted from Dalitz-plot analysis of $B^0 \to K^+ \pi^- \pi^0$ decays.
\item[2] Result extracted from Dalitz-plot analysis of $B^0 \to K_S^0 \pi^+ \pi^-$ decays.
\item[3] Multiple systematic uncertainties are added in quadrature.
\item[4] Measurement of $( {\cal B}(B^0 \to K_2^*(1430)^+ \pi^-) {\cal B}(K_2^*(1430)^+ \to K \pi) \times 2/3 ) / {\cal{B}}(B^0 \to K^0 \pi^+ \pi^-)$ used in our fit.
\item[5] Measurement of $( {\cal B}(B^0 \to K^*(1680)^+ \pi^-) {\cal B}(K^*(1680)^+ \to K \pi) \times 2/3) / {\cal{B}}(B^0 \to K^0 \pi^+ \pi^-)$ used in our fit.
\item[6] $0.75 < m_{K\pi} < 1.20~\rm{GeV/c^2}$.
\item[7] $0.55 < m_{\pi\pi} < 1.20~\rm{GeV/c^2}$.
\item[8] The PDG uncertainty includes a scale factor.
\end{tablenotes}
\end{threeparttable}
\end{center}
\end{table}

\hflavrowdef{\hflavrowdefach}{${\cal B}(B^0 \to K_1(1270)^+ \pi^-)$}{{\setlength{\tabcolsep}{0pt}
}
\begin{table}[H]
\begin{center}
\resizebox{1\textwidth}{!}{
\begin{threeparttable}
\caption{Branching fractions of charmless mesonic $B^0$ decays with strange mesons (part 7).}
\label{tab:rareDecays_charmlessBd_Bd_BR_S7}
%
 \\
\hline
\hflavrowdefach \\
 & & \\[-1ex]
\hflavrowdefaci \\
 & & \\[-1ex]
\hflavrowdefacj \\
 & & \\[-1ex]
\hflavrowdefack \\
 & & \\[-1ex]
\hflavrowdefacl \\
 & & \\[-1ex]
\hflavrowdefacm \\
 & & \\[-1ex]
\hflavrowdefacn \\
 & & \\[-1ex]
\hflavrowdefaco \\
 & & \\[-1ex]
\hflavrowdefacp \\
 & & \\[-1ex]
\hflavrowdefacq \\
 & & \\[-1ex]
\hflavrowdefacr \\
 & & \\[-1ex]
\hflavrowdefacs \\
 & & \\[-1ex]
\hflavrowdefact \\
 & & \\[-1ex]
\hflavrowdefacu \\
 & & \\[-1ex]
\hflavrowdefacv \\
 & & \\[-1ex]
\hflavrowdefacw \\
 & & \\[-1ex]
\hflavrowdefacx \\
\hline
\hline
\end{tabular}
\begin{tablenotes}
\item[1] Multiple systematic uncertainties are added in quadrature.
\item[2] Using ${\cal B}(B^0 \to K^+ \pi^-)$.
\item[3] Regions corresponding to $D$, $\Lambda_c^+$ and charmonium resonances are vetoed in this analysis.
\item[4] Using ${\cal B}(B^0 \to K^0 \pi^+ \pi^-)$.
\item[5] Using ${\cal B}(B^0 \to K^*(892)^+ \pi^-)$.
\item[6] $0.75 < m_{K\pi} < 1.20~\rm{GeV/c^2.}$
\end{tablenotes}
\end{threeparttable}
}
\end{center}
\end{table}

\hflavrowdef{\hflavrowdefacy}{${\cal B}(B^0 \to K^+ K^- \pi^0)$}{{\setlength{\tabcolsep}{0pt}
}
\begin{table}[H]
\begin{center}
\resizebox{0.95\textwidth}{!}{
\begin{threeparttable}
\caption{Branching fractions of charmless mesonic $B^0$ decays with strange mesons (part 8).}
\label{tab:rareDecays_charmlessBd_Bd_BR_S8}
%
 \\
\hline
\hflavrowdefacy \\
 & & \\[-1ex]
\hflavrowdefacz \\
 & & \\[-1ex]
\hflavrowdefada \\
 & & \\[-1ex]
\hflavrowdefadb \\
 & & \\[-1ex]
\hflavrowdefadc \\
 & & \\[-1ex]
\hflavrowdefadd \\
 & & \\[-1ex]
\hflavrowdefade \\
 & & \\[-1ex]
\hflavrowdefadf \\
 & & \\[-1ex]
\hflavrowdefadg \\
 & & \\[-1ex]
\hflavrowdefadh \\
 & & \\[-1ex]
\hflavrowdefadi \\
\hline
\hline
\end{tabular}
\begin{tablenotes}
\item[\dag] Preliminary result.
\item[*] Not included in PDG average.
\item[1] Regions corresponding to $D$, $\Lambda_c^+$ and charmonium resonances are vetoed in this analysis.
\item[2] Using ${\cal B}(B^0 \to K^0 \pi^+ \pi^-)$.
\item[3] Result extracted from Dalitz-plot analysis of $B^0 \to K_S^0 K^+ K^-$ decays.
\item[4] All charmonium resonances are vetoed, except for $\chi_{c0}$. The analysis also reports ${\cal{B}}( B^{0} \to K^0 K^+ K^- ) = (25.4 \pm 0.9 \pm 0.8) \times 10^{-6}$ excluding $\chi_{c0}$.
\item[5] Measurement of $( {\cal B}(\Lambda_b^0 \to \Lambda^0 \phi(1020))  / {\cal B}(B^0 \to \phi(1020) K^0) ) ( f_{\Lambda^0_b} / f_{d} ) \times 2$ used in our fit.
\item[6] Multiple systematic uncertainties are added in quadrature.
\item[7] Measurement of ${\cal B}(B_s^0 \to K^0 \bar{K}^0) / {\cal B}(B^0 \to \phi(1020) K^0)$ used in our fit.
\item[8] The nonresonant amplitude is modelled using a polynomial function including S-wave and P-wave terms.
\end{tablenotes}
\end{threeparttable}
}
\end{center}
\end{table}

\hflavrowdef{\hflavrowdefadj}{${\cal B}(B^0 \to K^0_S K^0_S K^0_S)$\tnote{1}\hphantom{\textsuperscript{1}}}{{\setlength{\tabcolsep}{0pt}
}
\begin{table}[H]
\begin{center}
\resizebox{1\textwidth}{!}{
\begin{threeparttable}
\caption{Branching fractions of charmless mesonic $B^0$ decays with strange mesons (part 9).}
\label{tab:rareDecays_charmlessBd_Bd_BR_S9}
%
 \\
\hline
\hflavrowdefadj \\
 & & \\[-1ex]
\hflavrowdefadk \\
 & & \\[-1ex]
\hflavrowdefadl \\
 & & \\[-1ex]
\hflavrowdefadm \\
 & & \\[-1ex]
\hflavrowdefadn \\
 & & \\[-1ex]
\hflavrowdefado \\
\hline
\hline
\end{tabular}
\begin{tablenotes}
\item[1] The PDG uncertainty includes a scale factor.
\item[2] Result extracted from Dalitz-plot analysis of $B^0 \to K_S^0 K_S^0 K_S^0$ decays.
\item[3] Multiple systematic uncertainties are added in quadrature.
\item[4] The nonresonant amplitude is modelled using an exponential function.
\item[5] $0.75<m_{K\pi}<1.20~\rm{GeV/c^2.}$
\end{tablenotes}
\end{threeparttable}
}
\end{center}
\end{table}

\hflavrowdef{\hflavrowdefadp}{${\cal B}(B^0 \to K^*(892)^0 K^+ K^-)$}{{\setlength{\tabcolsep}{0pt}
}
\begin{table}[H]
\begin{center}
\resizebox{1\textwidth}{!}{
\begin{threeparttable}
\caption{Branching fractions of charmless mesonic $B^0$ decays with strange mesons (part 10).}
\label{tab:rareDecays_charmlessBd_Bd_BR_S10}
%
 \\
\hline
\hflavrowdefadp \\
 & & \\[-1ex]
\hflavrowdefadq \\
 & & \\[-1ex]
\hflavrowdefadr \\
 & & \\[-1ex]
\hflavrowdefads \\
 & & \\[-1ex]
\hflavrowdefadt \\
 & & \\[-1ex]
\hflavrowdefadu \\
 & & \\[-1ex]
\hflavrowdefadv \\
 & & \\[-1ex]
\hflavrowdefadw \\
 & & \\[-1ex]
\hflavrowdefadx \\
 & & \\[-1ex]
\hflavrowdefady \\
 & & \\[-1ex]
\hflavrowdefadz \\
 & & \\[-1ex]
\hflavrowdefaea \\
\hline
\hline
\end{tabular}
\begin{tablenotes}
\item[\dag] Preliminary result.
\item[*] Not included in PDG average.
\item[1] Multiple systematic uncertainties are added in quadrature.
\item[2] Measurement of ${\cal B}(B_s^0 \to \phi(1020) \bar{K}^*(892)^0) / {\cal B}(B^0 \to \phi(1020) K^*(892)^0)$ used in our fit.
\item[3] Measurement of ${\cal B}(B_s^0 \to K^*(892)^0 \bar{K}^*(892)^0) / {\cal B}(B^0 \to \phi(1020) K^*(892)^0)$ used in our fit.
\item[4] Measurement of ${\cal B}(B_s^0 \to \phi(1020) \phi(1020)) / {\cal B}(B^0 \to \phi(1020) K^*(892)^0)$ used in our fit.
\item[5] Measurement of ${\cal B}(B^0 \to \rho^0(770) \rho^0(770)) / {\cal B}(B^0 \to \phi(1020) K^*(892)^0)$ used in our fit.
\item[6] $0.70<m_{K\pi}<1.70~\rm{GeV/c^2.}$
\item[7] The PDG uncertainty includes a scale factor.
\item[8] Using ${\cal B}(B_s^0 \to K^*(892)^0 \bar{K}^*(892)^0)$.
\end{tablenotes}
\end{threeparttable}
}
\end{center}
\end{table}

\hflavrowdef{\hflavrowdefaeb}{${\cal B}(B^0 \to K_0^*(1430)^0 \pi^+ K^-)$}{{\setlength{\tabcolsep}{0pt}
}
\begin{table}[H]
\begin{center}
\resizebox{1\textwidth}{!}{
\begin{threeparttable}
\caption{Branching fractions of charmless mesonic $B^0$ decays with strange mesons (part 11).}
\label{tab:rareDecays_charmlessBd_Bd_BR_S11}
%
 \\
\hline
\hflavrowdefaeb \\
 & & \\[-1ex]
\hflavrowdefaec \\
 & & \\[-1ex]
\hflavrowdefaed \\
 & & \\[-1ex]
\hflavrowdefaee \\
 & & \\[-1ex]
\hflavrowdefaef \\
 & & \\[-1ex]
\hflavrowdefaeg \\
 & & \\[-1ex]
\hflavrowdefaeh \\
 & & \\[-1ex]
\hflavrowdefaei \\
 & & \\[-1ex]
\hflavrowdefaej \\
 & & \\[-1ex]
\hflavrowdefaek \\
 & & \\[-1ex]
\hflavrowdefael \\
 & & \\[-1ex]
\hflavrowdefaem \\
 & & \\[-1ex]
\hflavrowdefaen \\
\hline
\hline
\end{tabular}
\begin{tablenotes}
\item[1] $0.70<m_{K\pi}<1.70~\rm{GeV/c^2.}$
\item[2] The PDG uncertainty includes a scale factor.
\item[3] Measured in the $\phi \phi$ invariant mass range below the $\eta_c$ resonance ($m_{\phi \phi} < 2.85~\text{GeV}/c^2$).
\end{tablenotes}
\end{threeparttable}
}
\end{center}
\end{table}

\hflavrowdef{\hflavrowdefaeo}{${\cal B}(B^0 \to \pi^+ \pi^-)$}{{\setlength{\tabcolsep}{0pt}
}
\begin{table}[H]
\begin{center}
\begin{threeparttable}
\caption{Branching fractions of charmless mesonic $B^0$ decays without strange mesons (part 1).}
\label{tab:rareDecays_charmlessBd_Bd_BR_NS1}
%
 \\
\hline
\hflavrowdefaeo \\
 & & \\[-1ex]
\hflavrowdefaep \\
 & & \\[-1ex]
\hflavrowdefaeq \\
 & & \\[-1ex]
\hflavrowdefaer \\
 & & \\[-1ex]
\hflavrowdefaes \\
 & & \\[-1ex]
\hflavrowdefaet \\
 & & \\[-1ex]
\hflavrowdefaeu \\
 & & \\[-1ex]
\hflavrowdefaev \\
 & & \\[-1ex]
\hflavrowdefaew \\
 & & \\[-1ex]
\hflavrowdefaex \\
 & & \\[-1ex]
\hflavrowdefaey \\
 & & \\[-1ex]
\hflavrowdefaez \\
\hline
\hline
\end{tabular}
\begin{tablenotes}
\item[\dag] Preliminary result.
\item[*] Not included in PDG average.
\item[1] Using ${\cal B}(B^0 \to K^+ \pi^-)$.
\item[2] The PDG uncertainty includes a scale factor.
\end{tablenotes}
\end{threeparttable}
\end{center}
\end{table}

\hflavrowdef{\hflavrowdefafa}{${\cal B}(B^0 \to \omega(782) \eta^\prime)$}{{\setlength{\tabcolsep}{0pt}
}
\begin{table}[H]
\begin{center}
\begin{threeparttable}
\caption{Branching fractions of charmless mesonic $B^0$ decays without strange mesons (part 2).}
\label{tab:rareDecays_charmlessBd_Bd_BR_NS2}
%
 \\
\hline
\hflavrowdefafa \\
 & & \\[-1ex]
\hflavrowdefafb \\
 & & \\[-1ex]
\hflavrowdefafc \\
 & & \\[-1ex]
\hflavrowdefafd \\
 & & \\[-1ex]
\hflavrowdefafe \\
 & & \\[-1ex]
\hflavrowdefaff \\
 & & \\[-1ex]
\hflavrowdefafg \\
 & & \\[-1ex]
\hflavrowdefafh \\
 & & \\[-1ex]
\hflavrowdefafi \\
 & & \\[-1ex]
\hflavrowdefafj \\
 & & \\[-1ex]
\hflavrowdefafk \\
 & & \\[-1ex]
\hflavrowdefafl \\
 & & \\[-1ex]
\hflavrowdefafm \\
 & & \\[-1ex]
\hflavrowdefafn \\
\hline
\hline
\end{tabular}
\begin{tablenotes}
\item[1] $400<m_{\pi^+\pi^-}<1600~\rm{MeV/c^2.}$
\item[2] Multiple systematic uncertainties are added in quadrature.
\end{tablenotes}
\end{threeparttable}
\end{center}
\end{table}

\hflavrowdef{\hflavrowdefafo}{${\cal B}(B^0 \to \pi^+ \pi^- \pi^0)$}{{\setlength{\tabcolsep}{0pt}
}
\begin{table}[H]
\begin{center}
\begin{threeparttable}
\caption{Branching fractions of charmless mesonic $B^0$ decays without strange mesons (part 3).}
\label{tab:rareDecays_charmlessBd_Bd_BR_NS3}
%
 \\
\hline
\hflavrowdefafo \\
 & & \\[-1ex]
\hflavrowdefafp \\
 & & \\[-1ex]
\hflavrowdefafq \\
 & & \\[-1ex]
\hflavrowdefafr \\
 & & \\[-1ex]
\hflavrowdefafs \\
 & & \\[-1ex]
\hflavrowdefaft \\
 & & \\[-1ex]
\hflavrowdefafu \\
 & & \\[-1ex]
\hflavrowdefafv \\
 & & \\[-1ex]
\hflavrowdefafw \\
 & & \\[-1ex]
\hflavrowdefafx \\
\hline
\hline
\end{tabular}
\begin{tablenotes}
\item[1] Result extracted from Dalitz-plot analysis of $B^0 \to \pi^+ \pi^- \pi^0$ decays.
\item[2] $0.52 < m_{\pi^+\pi^-} < 1.15~\text{GeV}/c^2$.
\item[3] $0.55 < m_{\pi^+\pi^-} < 1.050~\text{GeV}/c^2$.
\item[4] Using ${\cal B}(B^0 \to \phi(1020) K^*(892)^0)$.
\end{tablenotes}
\end{threeparttable}
\end{center}
\end{table}

\hflavrowdef{\hflavrowdefafy}{${\cal B}(B^0 \to a_1(1260)^+ \pi^- \mathrm{+c.c.})$\tnote{1}\hphantom{\textsuperscript{1}}}{{\setlength{\tabcolsep}{0pt}
}
\begin{table}[H]
\begin{center}
\begin{threeparttable}
\caption{Branching fractions of charmless mesonic $B^0$ decays without strange mesons (part 4).}
\label{tab:rareDecays_charmlessBd_Bd_BR_NS4}
%
 \\
\hline
\hflavrowdefafy \\
 & & \\[-1ex]
\hflavrowdefafz \\
 & & \\[-1ex]
\hflavrowdefaga \\
 & & \\[-1ex]
\hflavrowdefagb \\
 & & \\[-1ex]
\hflavrowdefagc \\
 & & \\[-1ex]
\hflavrowdefagd \\
 & & \\[-1ex]
\hflavrowdefage \\
 & & \\[-1ex]
\hflavrowdefagf \\
 & & \\[-1ex]
\hflavrowdefagg \\
 & & \\[-1ex]
\hflavrowdefagh \\
 & & \\[-1ex]
\hflavrowdefagi \\
 & & \\[-1ex]
\hflavrowdefagj \\
 & & \\[-1ex]
\hflavrowdefagk \\
 & & \\[-1ex]
\hflavrowdefagl \\
 & & \\[-1ex]
\hflavrowdefagm \\
 & & \\[-1ex]
\hflavrowdefagn \\
\hline
\hline
\end{tabular}
\begin{tablenotes}
\item[\dag] Preliminary result.
\item[1] The PDG uncertainty includes a scale factor.
\end{tablenotes}
\end{threeparttable}
\end{center}
\end{table}

\hflavrowdef{\hflavrowdefafn}{$\dfrac{{\cal B}(B^+ \to K^+ K^- \pi^+)}{{\cal B}(B^+ \to K^+ K^+ K^-)}$}{{\setlength{\tabcolsep}{0pt}
}
\begin{table}[H]
\begin{center}
\begin{threeparttable}
\caption{Relative branching fractions of mesonic $B^+$ decays.}
\label{tab:rareDecays_charmlessBu_Bu_RBR}
%
}
\begin{table}[H]
\begin{center}
\begin{threeparttable}
\caption{Relative branching fractions of mesonic $B^0$ decays.}
\label{tab:rareDecays_charmlessBd_Bd_RBR}
%
 \\
\hline
\hflavrowdefago \\
 & & \\[-1ex]
\hflavrowdefagp \\
 & & \\[-1ex]
\hflavrowdefagq \\
 & & \\[-1ex]
\hflavrowdefagr \\
 & & \\[-1ex]
\hflavrowdefags \\
 & & \\[-1ex]
\hflavrowdefagt \\
 & & \\[-1ex]
\hflavrowdefagu \\
 & & \\[-1ex]
\hflavrowdefagv \\
 & & \\[-1ex]
\hflavrowdefagw \\
 & & \\[-1ex]
\hflavrowdefagx \\
 & & \\[-1ex]
\hflavrowdefagy \\
\hline
\hline
\end{tabular}
\begin{tablenotes}
\item[1] Regions corresponding to $D$, $\Lambda_c^+$ and charmonium resonances are vetoed in this analysis.
\item[2] Multiple systematic uncertainties are added in quadrature.
\end{tablenotes}
\end{threeparttable}
\end{center}
\end{table}

\clearpage
\begin{figure}[htbp!]
\centering
\includegraphics[width=0.8\textwidth]{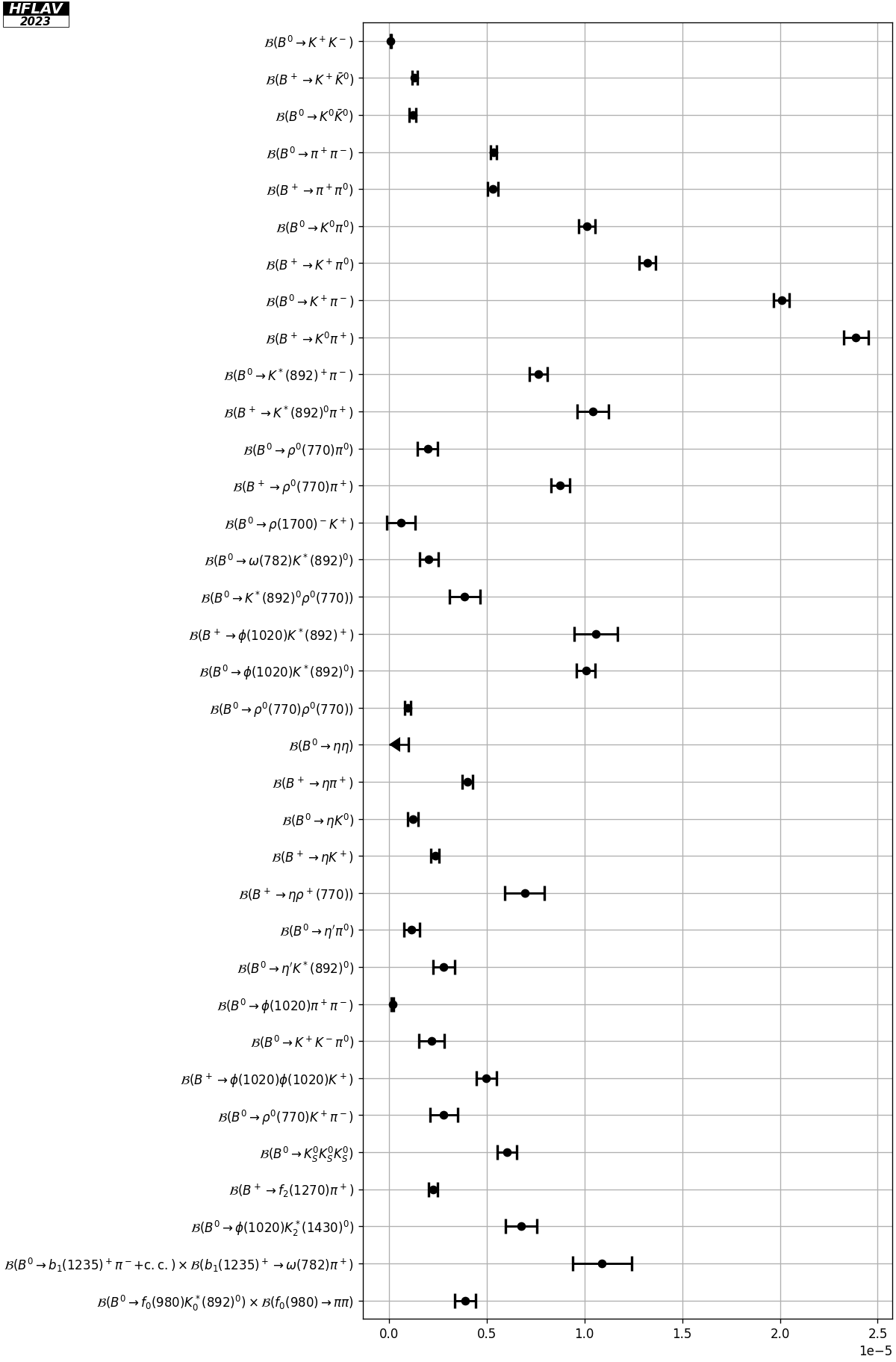}
\caption{A selection of high-precision measurements of Branching Fractions of $B$-meson decays into  charmless mesonic final states.}
\label{fig:rare-mostprec}
\end{figure}

\mysubsection{Baryonic decays of \Bp\ and \Bz mesons}
\label{sec:rare-bary}

This section provides branching fractions of charmless baryonic decays of \Bp\ and \Bz mesons in Tables~\ref{tab:rareDecays_baryonic_Bu_BR_1}-\ref{tab:rareDecays_baryonic_Bu_BR_3} and~\ref{tab:rareDecays_baryonic_Bd_BR_1}-\ref{tab:rareDecays_baryonic_Bd_BR_2}, respectively. Relative branching fractions are given in Table~\ref{tab:rareDecays_baryonic_RBR_1}.
Figures~\ref{fig:rare-baryns} and~\ref{fig:rare-bary} show graphic representations of a selection of results given in this section.

\hflavrowdef{\hflavrowdefaaa}{${\cal B}(B^+ \to p \bar{n} \pi^0)$}{{\setlength{\tabcolsep}{0pt}
}
\begin{table}[H]
\begin{center}
\resizebox{1\textwidth}{!}{
\begin{threeparttable}
\caption{Branching fractions of charmless baryonic $B^+$ decays (part 1).}
\label{tab:rareDecays_baryonic_Bu_BR_1}
%
 \\
\hline
\hflavrowdefaaa \\
 & & \\[-1ex]
\hflavrowdefaab \\
 & & \\[-1ex]
\hflavrowdefaac \\
 & & \\[-1ex]
\hflavrowdefaad \\
 & & \\[-1ex]
\hflavrowdefaae \\
 & & \\[-1ex]
\hflavrowdefaaf \\
\hline
\hline
\end{tabular}
\begin{tablenotes}
\item[1] The charmonium mass regions are vetoed.
\item[2] Charmonium decays to $p \overline{p}$ have been statistically subtracted.
\item[3] Measurement of ${\cal{B}}( B^+ \to p \overline{p} \pi^+ ),~m_{p\overline{p}} < 2.85~{\rm GeV/c^2} / ( {\cal B}(B^+ \to J/\psi \pi^+) {\cal B}(J/\psi \to p \bar{p}) )$ used in our fit.
\item[4] $m_{\pi^+ \pi^0}<1.3~{\rm GeV/c^2}.$
\end{tablenotes}
\end{threeparttable}
}
\end{center}
\end{table}

\hflavrowdef{\hflavrowdefaag}{${\cal B}(B^+ \to p \bar{p} K^+)$\tnote{1}\hphantom{\textsuperscript{1}}}{{\setlength{\tabcolsep}{0pt}
}
\begin{table}[H]
\begin{center}
\begin{threeparttable}
\caption{Branching fractions of charmless baryonic $B^+$ decays (part 2).}
\label{tab:rareDecays_baryonic_Bu_BR_2}
%
 \\
\hline
\hflavrowdefaag \\
 & & \\[-1ex]
\hflavrowdefaah \\
 & & \\[-1ex]
\hflavrowdefaai \\
 & & \\[-1ex]
\hflavrowdefaaj \\
 & & \\[-1ex]
\hflavrowdefaak \\
 & & \\[-1ex]
\hflavrowdefaal \\
 & & \\[-1ex]
\hflavrowdefaam \\
 & & \\[-1ex]
\hflavrowdefaan \\
 & & \\[-1ex]
\hflavrowdefaao \\
 & & \\[-1ex]
\hflavrowdefaap \\
 & & \\[-1ex]
\hflavrowdefaaq \\
 & & \\[-1ex]
\hflavrowdefaar \\
\hline
\hline
\end{tabular}
\begin{tablenotes}
\item[1] The PDG uncertainty includes a scale factor.
\item[2] The charmonium mass regions are vetoed.
\item[3] Charmonium decays to $p \overline{p}$ have been statistically subtracted.
\item[4] Measurement of ${\cal{B}}( B^+ \to p \overline{p} K^+ ),~m_{p\overline{p}} < 2.85~{\rm GeV/c^2} / ( {\cal B}(B^+ \to J/\psi K^+) {\cal B}(J/\psi \to p \bar{p}) )$ used in our fit.
\item[5] Pentaquark candidate.
\item[6] Measurement of $( {\cal{B}}( B^+ \to p \overline{\Lambda}(1520)) {\cal B}(\bar{\Lambda(1520)} \to K^+ p) ) / ( {\cal B}(B^+ \to J/\psi K^+) {\cal B}(J/\psi \to p \bar{p}) )$ used in our fit.
\item[7] The charmonium mass region has been vetoed.
\end{tablenotes}
\end{threeparttable}
\end{center}
\end{table}

\hflavrowdef{\hflavrowdefaas}{${\cal B}(B^+ \to p \bar{\Lambda}^0 \pi^+ \pi^-)$}{{\setlength{\tabcolsep}{0pt}
}
\begin{table}[H]
\begin{center}
\begin{threeparttable}
\caption{Branching fractions of charmless baryonic $B^+$ decays (part 3).}
\label{tab:rareDecays_baryonic_Bu_BR_3}
%
 \\
\hline
\hflavrowdefaas \\
 & & \\[-1ex]
\hflavrowdefaat \\
 & & \\[-1ex]
\hflavrowdefaau \\
 & & \\[-1ex]
\hflavrowdefaav \\
 & & \\[-1ex]
\hflavrowdefaaw \\
 & & \\[-1ex]
\hflavrowdefaax \\
 & & \\[-1ex]
\hflavrowdefaay \\
 & & \\[-1ex]
\hflavrowdefaaz \\
 & & \\[-1ex]
\hflavrowdefaba \\
 & & \\[-1ex]
\hflavrowdefabb \\
 & & \\[-1ex]
\hflavrowdefabc \\
 & & \\[-1ex]
\hflavrowdefabd \\
 & & \\[-1ex]
\hflavrowdefabe \\
 & & \\[-1ex]
\hflavrowdefabf \\
\hline
\hline
\end{tabular}
\begin{tablenotes}
\item[1] The charmonium mass regions are vetoed.
\item[2] $m_{\Lambda^0\overline{\Lambda^0}} < 2.85 ~\text{GeV}/c^2$.
\end{tablenotes}
\end{threeparttable}
\end{center}
\end{table}

\hflavrowdef{\hflavrowdefabg}{${\cal B}(B^0 \to p \bar{p})$}{{\setlength{\tabcolsep}{0pt}
}
\begin{table}[H]
\begin{center}
\resizebox{0.95\textwidth}{!}{
\begin{threeparttable}
\caption{Branching fractions of charmless baryonic $B^0$ decays (part 1).}
\label{tab:rareDecays_baryonic_Bd_BR_1}
%
 \\
\hline
\hflavrowdefabg \\
 & & \\[-1ex]
\hflavrowdefabh \\
 & & \\[-1ex]
\hflavrowdefabi \\
 & & \\[-1ex]
\hflavrowdefabj \\
 & & \\[-1ex]
\hflavrowdefabk \\
 & & \\[-1ex]
\hflavrowdefabl \\
 & & \\[-1ex]
\hflavrowdefabm \\
 & & \\[-1ex]
\hflavrowdefabn \\
 & & \\[-1ex]
\hflavrowdefabo \\
\hline
\hline
\end{tabular}
\begin{tablenotes}
\item[*] Not included in PDG average.
\item[1] Run I and run II combination.
\item[2] Multiple systematic uncertainties are added in quadrature.
\item[3] $m_{p\overline{p}} < 2.85~\rm{GeV/}c^2$.
\item[4] $0.46 < m_{\pi^+ \pi^-} < 0.53~\rm{GeV/c^2}$ invariant mass region has been excluded.
\item[5] The charmonium mass region has been vetoed.
\item[6] Charmonium decays to $p \overline{p}$ have been statistically subtracted.
\item[7] Pentaquark candidate.
\end{tablenotes}
\end{threeparttable}
}
\end{center}
\end{table}

\hflavrowdef{\hflavrowdefabp}{${\cal B}(B^0 \to p \bar{p} K^+ K^-)$}{{\setlength{\tabcolsep}{0pt}
}
\begin{table}[H]
\begin{center}
\begin{threeparttable}
\caption{Branching fractions of charmless baryonic $B^0$ decays (part 2).}
\label{tab:rareDecays_baryonic_Bd_BR_2}
%
 \\
\hline
\hflavrowdefabp \\
 & & \\[-1ex]
\hflavrowdefabq \\
 & & \\[-1ex]
\hflavrowdefabr \\
 & & \\[-1ex]
\hflavrowdefabs \\
 & & \\[-1ex]
\hflavrowdefabt \\
 & & \\[-1ex]
\hflavrowdefabu \\
 & & \\[-1ex]
\hflavrowdefabv \\
 & & \\[-1ex]
\hflavrowdefabw \\
 & & \\[-1ex]
\hflavrowdefabx \\
 & & \\[-1ex]
\hflavrowdefaby \\
 & & \\[-1ex]
\hflavrowdefabz \\
 & & \\[-1ex]
\hflavrowdefaca \\
 & & \\[-1ex]
\hflavrowdefacb \\
 & & \\[-1ex]
\hflavrowdefacc \\
 & & \\[-1ex]
\hflavrowdefacd \\
\hline
\hline
\end{tabular}
\begin{tablenotes}
\item[*] Not included in PDG average.
\item[1] $m_{p\overline{p}} < 2.85~\rm{GeV/}c^2$.
\item[2] Multiple systematic uncertainties are added in quadrature.
\item[3] The charmonium mass regions are vetoed.
\item[4] CLEO assumes ${\cal{B}} (\Upsilon(4S) \to B^0 \overline{B}^0) = 0.43$. The result has been modified to account for a branching fraction of 0.50.
\item[5] Measurement of ${\cal B}(B^0 \to p \bar{p} p \bar{p}) / ( {\cal B}(B^0 \to J/\psi K^*(892)^0) {\cal B}(J/\psi \to p \bar{p}) {\cal B}(K^*(892)^0 \to K \pi) \times 2/3  )$ used in our fit.
\end{tablenotes}
\end{threeparttable}
\end{center}
\end{table}

\hflavrowdef{\hflavrowdeface}{$\frac{ {\cal{B}}(B^+ \to p \overline{p} \pi^+, m_{p \overline{p}} < 2.85~\mathrm{GeV/c^2}) }{ {\cal{B}}(B^+ \to J/\psi \pi^+) \times {\cal{B}} (J/\psi \to p \bar{p} ) }$}{{\setlength{\tabcolsep}{0pt}
}
\begin{table}[H]
\begin{center}
\begin{threeparttable}
\caption{Baryonic relative branching fractions.}
\label{tab:rareDecays_baryonic_RBR_1}
%
 \\
\hline
\hflavrowdeface \\
 & & \\[-1ex]
\hflavrowdefacf \\
 & & \\[-1ex]
\hflavrowdefacg \\
 & & \\[-1ex]
\hflavrowdefach \\
 & & \\[-1ex]
\hflavrowdefaci \\
 & & \\[-1ex]
\hflavrowdefacj \\
 & & \\[-1ex]
\hflavrowdefack \\
 & & \\[-1ex]
\hflavrowdefacl \\
\hline
\hline
\end{tabular}
\begin{tablenotes}
\item[1] Includes contribution where $p \overline{p}$ is produced in charmonium decays.
\item[2] $m_{p\overline{p}} < 2.85~\rm{GeV/}c^2$.
\end{tablenotes}
\end{threeparttable}
\end{center}
\end{table}

\begin{figure}[htbp!]
\centering
\includegraphics[width=0.8\textwidth]{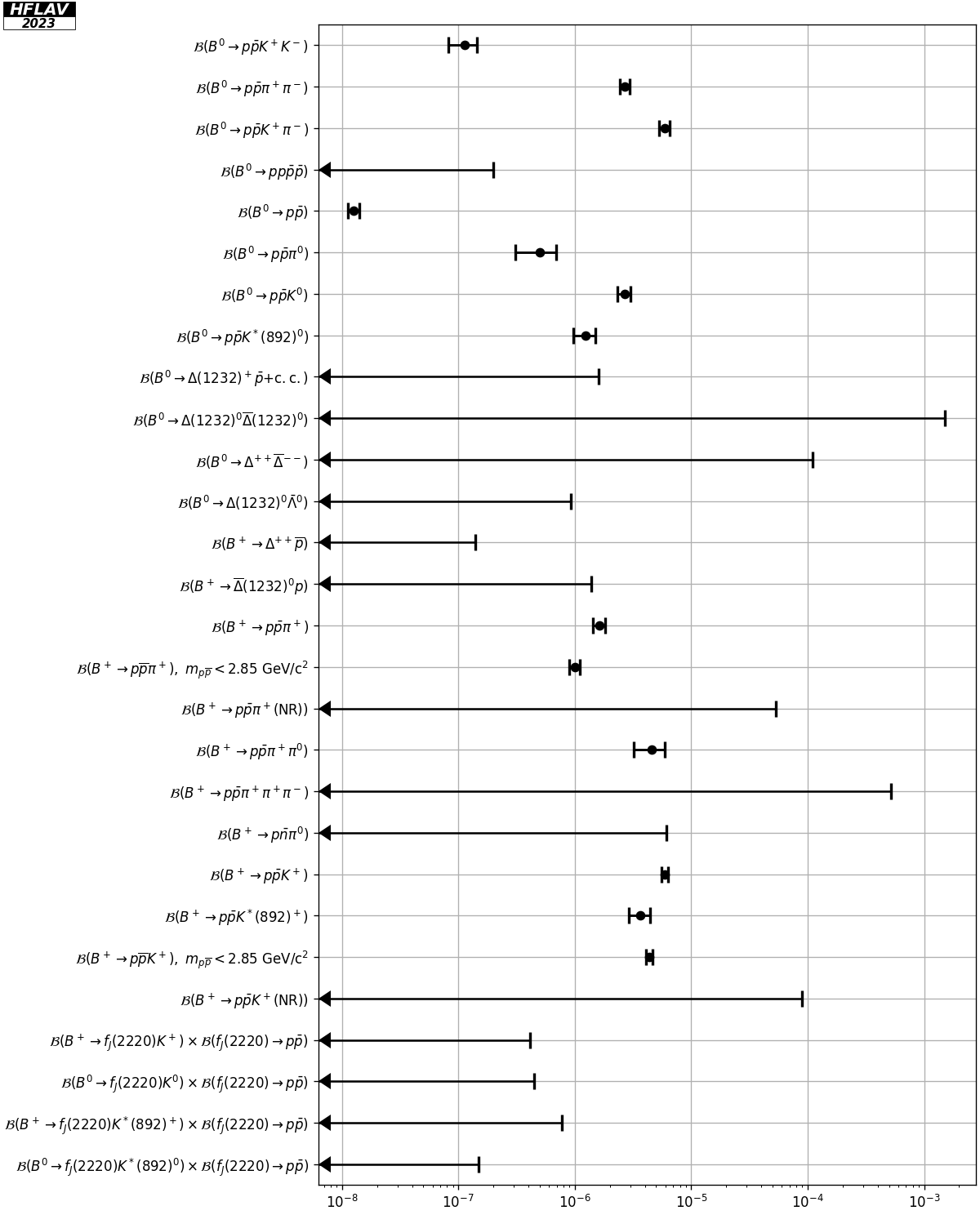}
\caption{Branching fractions of charmless $B^+$ and $B^0$ decays into nonstrange baryons.}
\label{fig:rare-baryns}
\end{figure}

\begin{figure}[htbp!]
\centering
\includegraphics[width=0.8\textwidth]{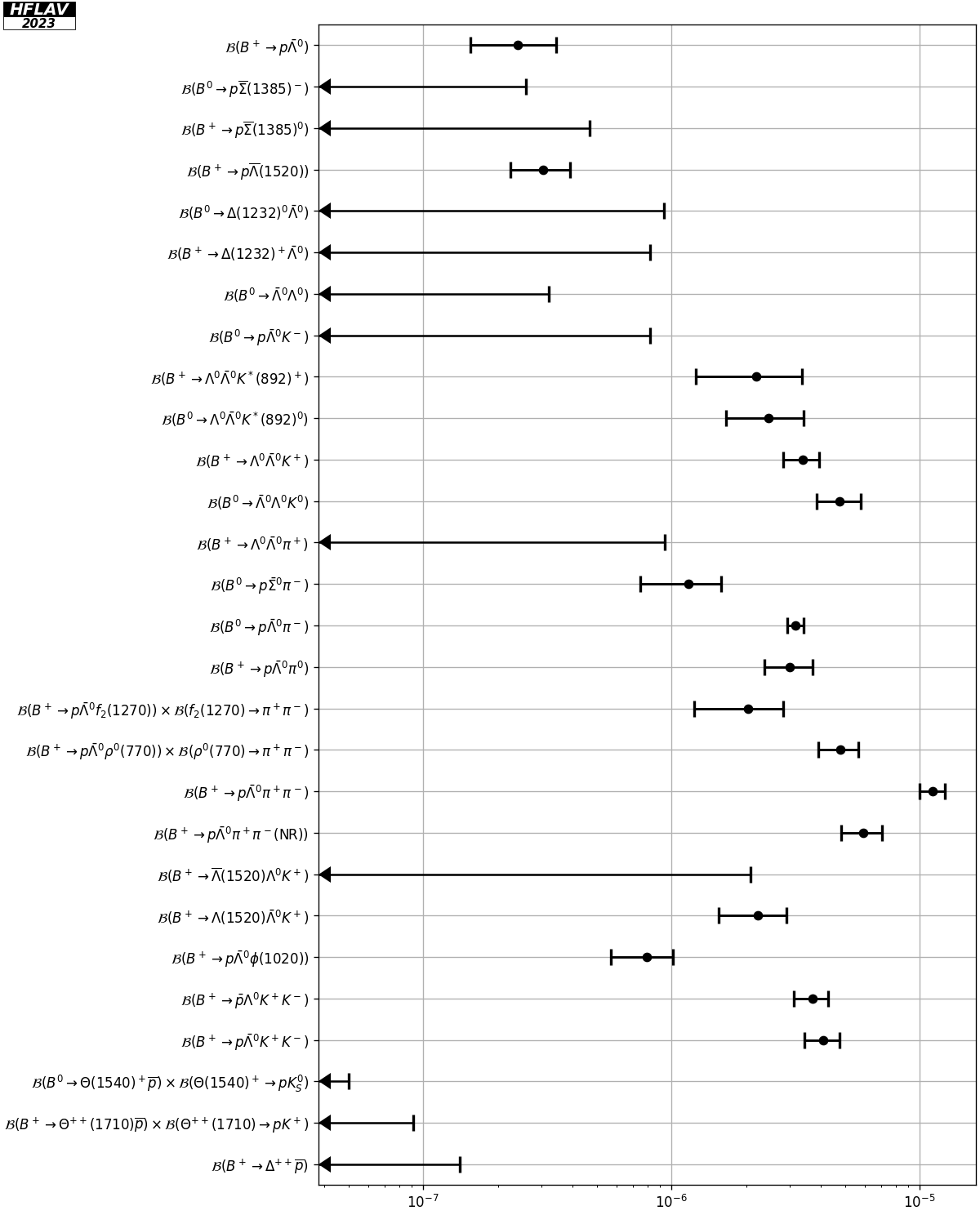}
\caption{Branching fractions of charmless $B^+$ and $B^0$ decays into strange baryons.}
\label{fig:rare-bary}
\end{figure}

\noindent Measurements that are not included in the tables:
\begin{itemize}
\item In Ref.~\cite{Belle:2021gmc}, Belle searches for $B^0$ mesons decaying into a final state containing a $\Lambda$ baryon and missing energy. Upper limits on the branching fractions are set in the range $2.1-3.8 \times 10^{-5}$.
\end{itemize}

\clearpage

\mysubsection{Decays of \b baryons}
\label{sec:rare-lb}

A compilation of branching fractions of $\Lb$ baryon decays is given in Tables~\ref{tab:rareDecays_bBaryons_Lb_BR_1} to~\ref{tab:rareDecays_bBaryons_Lb_BR_3}. Table~\ref{tab:rareDecays_bBaryons_Lb_PBR} provides the partial branching fractions of $\Lb\to\Lambda\mup\mun$ decays in intervals of $q^2=m^2(\mu^+\mu^-)$.  Compilations of branching fractions of $\Xi^{0}_{b}$, $\Xi^{-}_{b}$ and $\Omega^{-}_{b}$ baryon decays are given in Tables~\ref{tab:rareDecays_bBaryons_Xib_Xib0}, \ref{tab:rareDecays_bBaryons_Xib_Xib-}, and~\ref{tab:rareDecays_bBaryons_Omegab_BR}, respectively.
Finally, ratios of branching fractions of $\Lb$, $\Xi^{0}_{b}$, $\Xi^{-}_{b}$ and $\Omega^{-}_{b}$ baryon decays are detailed in Tables~\ref{tab:rareDecays_bBaryons_Lb_RBR_1}, %
to~\ref{tab:rareDecays_bBaryons_Omegab_RBR}.
Figure~\ref{fig:rare-lb} shows a graphic representation of branching fractions of $\Lb$ decays.

\hflavrowdef{\hflavrowdefaaa}{${\cal B}(\Lambda_b^0 \to p \bar{K}^0 \pi^-)$}{{\setlength{\tabcolsep}{0pt}
}
\begin{table}[H]
\begin{center}
\begin{threeparttable}
\caption{Branching fractions of charmless $\Lambda_b^0$ decays (part 1).}
\label{tab:rareDecays_bBaryons_Lb_BR_1}
%
 \\
\hline
\hflavrowdefaaa \\
 & & \\[-1ex]
\hflavrowdefaab \\
 & & \\[-1ex]
\hflavrowdefaac \\
 & & \\[-1ex]
\hflavrowdefaad \\
 & & \\[-1ex]
\hflavrowdefaae \\
 & & \\[-1ex]
\hflavrowdefaaf \\
 & & \\[-1ex]
\hflavrowdefaag \\
 & & \\[-1ex]
\hflavrowdefaah \\
\hline
\hline
\end{tabular}
\begin{tablenotes}
\item[1] Multiple systematic uncertainties are added in quadrature.
\item[2] Using ${\cal B}(B^0 \to K^0 \pi^+ \pi^-)$.
\item[3] The PDG average is a result of a fit including input from other measurements.
\item[4] Using ${\cal B}(\Lambda_b^0 \to p K^-)$.
\item[5] Measurement of $( {\cal B}(\Lambda_b^0 \to p \pi^-) / {\cal B}(B^0 \to K^+ \pi^-) ) ( f_{\Lambda^0_b} / f_{d} )$ used in our fit.
\item[6] Measurement of ${\cal B}(\Lambda_b^0 \to p \pi^-) / {\cal B}(\Lambda_b^0 \to p K^-)$ used in our fit.
\item[7] Using ${\cal B}(\Lambda_b^0 \to J/\psi \Lambda^0)$.
\item[8] Measurement of ${\cal B}(\Lambda_b^0 \to p \pi^- \mu^+ \mu^-) / ( {\cal B}(\Lambda_b^0 \to J/\psi p \pi^-) {\cal B}(J/\psi \to \mu^+ \mu^-) )$ used in our fit.
\item[9] Measured in the $m^2_{\ell^+\ell^-}$ bin $[0.1, 6.0]~\rm{GeV^2/c^4}$ and for $m_{pK}<2.6~\rm{GeV/c^2}$.
\item[10] Using ${\cal B}(\Lambda_b^0 \to J/\psi p K^-)$.
\end{tablenotes}
\end{threeparttable}
\end{center}
\end{table}

\hflavrowdef{\hflavrowdefaai}{${\cal B}(\Lambda_b^0 \to \Lambda^0 \gamma)$}{{\setlength{\tabcolsep}{0pt}
}
\begin{table}[H]
\begin{center}
\begin{threeparttable}
\caption{Branching fractions of charmless $\Lambda_b^0$ decays (part 2).}
\label{tab:rareDecays_bBaryons_Lb_BR_2}
%
 \\
\hline
\hflavrowdefaai \\
 & & \\[-1ex]
\hflavrowdefaaj \\
 & & \\[-1ex]
\hflavrowdefaak \\
 & & \\[-1ex]
\hflavrowdefaal \\
 & & \\[-1ex]
\hflavrowdefaam \\
 & & \\[-1ex]
\hflavrowdefaan \\
 & & \\[-1ex]
\hflavrowdefaao \\
\hline
\hline
\end{tabular}
\begin{tablenotes}
\item[1] Measurement of $( {\cal B}(\Lambda_b^0 \to \Lambda^0 \gamma) / {\cal B}(B^0 \to K^*(892)^0 \gamma) ) \frac{f_{\Lambda^0_b}}{f_d}$ used in our fit.
\item[2] Using ${\cal B}(B^0 \to \eta^\prime K^0)$.
\item[3] Measurement of ${\cal B}(\Lambda_b^0 \to \Lambda^0 \pi^+ \pi^-) / ( {\cal B}(\Lambda_b^0 \to \Lambda_c^+ \pi^-) {\cal B}(\Lambda_c^+ \to \Lambda^0 \pi^+) )$ used in our fit.
\item[4] Measurement of ${\cal B}(\Lambda_b^0 \to \Lambda^0 K^+ \pi^-) / ( {\cal B}(\Lambda_b^0 \to \Lambda_c^+ \pi^-) {\cal B}(\Lambda_c^+ \to \Lambda^0 \pi^+) )$ used in our fit.
\item[5] Measurement of ${\cal B}(\Lambda_b^0 \to \Lambda^0 K^+ K^-) / ( {\cal B}(\Lambda_b^0 \to \Lambda_c^+ \pi^-) {\cal B}(\Lambda_c^+ \to \Lambda^0 \pi^+) )$ used in our fit.
\item[6] Measurement of $( {\cal B}(\Lambda_b^0 \to \Lambda^0 \phi(1020))  / {\cal B}(B^0 \to \phi(1020) K^0) \times 2) ( f_{\Lambda^0_b} / f_{d} )$ used in our fit.
\end{tablenotes}
\end{threeparttable}
\end{center}
\end{table}

\hflavrowdef{\hflavrowdefaap}{${\cal B}(\Lambda_b^0 \to p \pi^+ \pi^- \pi^-)$}{{\setlength{\tabcolsep}{0pt}\begin{tabular}{m{6.5em}l} \multicolumn{2}{l}{LHCb {\cite{LHCb:2017gjj}\tnote{1,2,3}\hphantom{\textsuperscript{1,2,3}}}} \\ \end{tabular}}}{\begin{tabular}{l} $21.1\,^{+2.4}_{-2.3}$ \\ $\mathit{\scriptstyle 20.9 \pm2.1}$\\ \end{tabular}}
\hflavrowdef{\hflavrowdefaaq}{${\cal B}(\Lambda_b^0 \to p K^- K^+ \pi^-)$}{{\setlength{\tabcolsep}{0pt}\begin{tabular}{m{6.5em}l} \multicolumn{2}{l}{LHCb {\cite{LHCb:2017gjj}\tnote{2,4}\hphantom{\textsuperscript{2,4}}}} \\ \end{tabular}}}{\begin{tabular}{l} $4.06\,^{+0.66}_{-0.61}$ \\ $\mathit{\scriptstyle 4.03 \pm0.60}$\\ \end{tabular}}
\hflavrowdef{\hflavrowdefaar}{${\cal B}(\Lambda_b^0 \to p K^- \pi^+ \pi^-)$}{{\setlength{\tabcolsep}{0pt}\begin{tabular}{m{6.5em}l} \multicolumn{2}{l}{LHCb {\cite{LHCb:2017gjj}\tnote{2,5}\hphantom{\textsuperscript{2,5}}}} \\ \end{tabular}}}{\begin{tabular}{l} $50.5\,^{+5.6}_{-5.3}$ \\ $\mathit{\scriptstyle 50.0 \pm4.9}$\\ \end{tabular}}
\hflavrowdef{\hflavrowdefaas}{${\cal B}(\Lambda_b^0 \to p K^- K^+ K^-)$}{{\setlength{\tabcolsep}{0pt}\begin{tabular}{m{6.5em}l} \multicolumn{2}{l}{LHCb {\cite{LHCb:2017gjj}\tnote{2,6}\hphantom{\textsuperscript{2,6}}}} \\ \end{tabular}}}{\begin{tabular}{l} $12.6\,^{+1.5}_{-1.4}$ \\ $\mathit{\scriptstyle 12.5 \pm1.3}$\\ \end{tabular}}
\begin{table}[H]
\begin{center}
\begin{threeparttable}
\caption{Branching fractions of charmless $\Lambda_b^0$ decays (part 3).}
\label{tab:rareDecays_bBaryons_Lb_BR_3}
\begin{tabular}{ Sl l l }
\hline
\hline
\textbf{Parameter [$10^{-6}$]} & \textbf{Measurements} & \begin{tabular}{l}\textbf{Average $^\mathrm{HFLAV}_{\scriptscriptstyle PDG}$}\end{tabular} \\
\hline
\hflavrowdefaap \\
 & & \\[-1ex]
\hflavrowdefaaq \\
 & & \\[-1ex]
\hflavrowdefaar \\
 & & \\[-1ex]
\hflavrowdefaas \\
\hline
\hline
\end{tabular}
\begin{tablenotes}
\item[1] Vetoes on charm and charmonium resonances are applied.
\item[2] Multiple systematic uncertainties are added in quadrature.
\item[3] Measurement of ${\cal B}(\Lambda_b^0 \to p \pi^+ \pi^- \pi^-) / ( {\cal B}(\Lambda_b^0 \to \Lambda_c^+ \pi^-) {\cal B}(\Lambda_c^+ \to p K^- \pi^+) )$ used in our fit.
\item[4] Measurement of ${\cal B}(\Lambda_b^0 \to p K^- K^+ \pi^-) / ( {\cal B}(\Lambda_b^0 \to \Lambda_c^+ \pi^-) {\cal B}(\Lambda_c^+ \to p K^- \pi^+) )$ used in our fit.
\item[5] Measurement of ${\cal B}(\Lambda_b^0 \to p K^- \pi^+ \pi^-) / ( {\cal B}(\Lambda_b^0 \to \Lambda_c^+ \pi^-) {\cal B}(\Lambda_c^+ \to p K^- \pi^+) )$ used in our fit.
\item[6] Measurement of ${\cal B}(\Lambda_b^0 \to p K^- K^+ K^-) / ( {\cal B}(\Lambda_b^0 \to \Lambda_c^+ \pi^-) {\cal B}(\Lambda_c^+ \to p K^- \pi^+) )$ used in our fit.
\end{tablenotes}
\end{threeparttable}
\end{center}
\end{table}

\hflavrowdeflong{\hflavrowdefaat}{$m_{\mu^+\mu^-}^2 < 2.0~\rm{GeV^2/c^4}$}{{\setlength{\tabcolsep}{0pt}
}
\begin{table}[H]
\begin{center}
\begin{threeparttable}
\caption{Partial branching fractions of $\Lambda_b^0 \to \Lambda \mu^+ \mu^-$ decays in intervals of $m^2_{\mu^+\mu^-}$.}
\label{tab:rareDecays_bBaryons_Lb_PBR}
%
}
\begin{table}[H]
\begin{center}
\begin{threeparttable}
\caption{Branching fractions of charmless $\Xi^{0}_{b}$ decays.}
\label{tab:rareDecays_bBaryons_Xib_Xib0}
%
 \\
\hline
\hflavrowdefabs \\
 & & \\[-1ex]
\hflavrowdefabt \\
 & & \\[-1ex]
\hflavrowdefabu \\
 & & \\[-1ex]
\hflavrowdefabv \\
 & & \\[-1ex]
\hflavrowdefabw \\
 & & \\[-1ex]
\hflavrowdefabx \\
 & & \\[-1ex]
\hflavrowdefaby \\
 & & \\[-1ex]
\hflavrowdefabz \\
\hline
\hline
\end{tabular}
\begin{tablenotes}
\item[1] Using ${\cal B}(B^0 \to K^0 \pi^+ \pi^-)$.
\item[2] Multiple systematic uncertainties are added in quadrature.
\item[3] Measurement of $\frac{f_{\Xi^{0}_{b}}}{f_{\Lambda^{0}_{b}}}{\cal{B}}(\Xi^{0}_{b} \to p  K^- \pi^+ \pi^-) / ( {\cal B}(\Lambda_b^0 \to \Lambda_c^+ \pi^-) {\cal B}(\Lambda_c^+ \to p K^- \pi^+) )$ used in our fit.
\item[4] Measurement of $\frac{f_{\Xi^{0}_{b}}}{f_{\Lambda^{0}_{b}}}{\cal{B}}(\Xi^{0}_{b} \to p  K^- K^- \pi^+) / ( {\cal B}(\Lambda_b^0 \to \Lambda_c^+ \pi^-) {\cal B}(\Lambda_c^+ \to p K^- \pi^+) )$ used in our fit.
\item[5] Measurement of $\frac{f_{\Xi^{0}_{b}}}{f_{\Lambda^{0}_{b}}}{\cal{B}}(\Xi^{0}_{b} \to p  K^+ K^- K^-) / ( {\cal B}(\Lambda_b^0 \to \Lambda_c^+ \pi^-) {\cal B}(\Lambda_c^+ \to p K^- \pi^+) )$ used in our fit.
\end{tablenotes}
\end{threeparttable}
\end{center}
\end{table}

\hflavrowdef{\hflavrowdefacf}{${\cal B}(\Xi_b^- \to \Sigma(1385)^0 K^-)$}{{\setlength{\tabcolsep}{0pt}
}
\begin{table}[H]
\begin{center}
\begin{threeparttable}
\caption{Branching fractions of charmless $\Xi^{-}_{b}$ decays.}
\label{tab:rareDecays_bBaryons_Xib_Xib-}
%
 \\
\hline
\hflavrowdefacf \\
 & & \\[-1ex]
\hflavrowdefacg \\
 & & \\[-1ex]
\hflavrowdefach \\
 & & \\[-1ex]
\hflavrowdefaci \\
 & & \\[-1ex]
\hflavrowdefacj \\
 & & \\[-1ex]
\hflavrowdefack \\
\hline
\hline
\end{tabular}
\begin{tablenotes}
\item[1] Result extracted from Dalitz plot analysis of $\Xi_b^- \to p K^- K^-$.
\item[2] Multiple systematic uncertainties are added in quadrature.
\end{tablenotes}
\end{threeparttable}
\end{center}
\end{table}

\hflavrowdef{\hflavrowdefacr}{$\frac{f_{\Omega^{-}_{b}}}{f_{u}} \times {\cal{B}}(\Omega^{-}_{b} \to p  K^- K^-)$}{{\setlength{\tabcolsep}{0pt}
}
\begin{table}[H]
\begin{center}
\begin{threeparttable}
\caption{Branching fractions of charmless $\Omega^{-}_{b}$ decays.}
\label{tab:rareDecays_bBaryons_Omegab_BR}
%
}
\begin{table}[H]
\begin{center}
\resizebox{1\textwidth}{!}{
\begin{threeparttable}
\caption{Relative branching fractions of $\Lambda_b^0$ decays (part 1).}
\label{tab:rareDecays_bBaryons_Lb_RBR_1}
%
 \\
\hline
\hflavrowdefaaz \\
 & & \\[-1ex]
\hflavrowdefaba \\
 & & \\[-1ex]
\hflavrowdefabb \\
 & & \\[-1ex]
\hflavrowdefabc \\
 & & \\[-1ex]
\hflavrowdefabd \\
 & & \\[-1ex]
\hflavrowdefabe \\
 & & \\[-1ex]
\hflavrowdefabf \\
 & & \\[-1ex]
\hflavrowdefabg \\
 & & \\[-1ex]
\hflavrowdefabh \\
\hline
\hline
\end{tabular}
\end{threeparttable}
}
\end{center}
\end{table}

\hflavrowdefshort{\hflavrowdefabi}{$\frac{ {\cal{B}}( \Lambda_b^0 \to p \pi^- \pi^+ \pi^- ) }{ {\cal{B}}( \Lambda_b^0 \to \Lambda_c^+ \pi^-)\times {\cal{B}}( \Lambda_c^+ \to p K^- \pi^+) }$}{{\setlength{\tabcolsep}{0pt}
}
\begin{table}[H]
\begin{center}
\resizebox{1\textwidth}{!}{
\begin{threeparttable}
\caption{Relative branching fractions of $\Lambda_b^0$ decays (part 2).}
\label{tab:rareDecays_bBaryons_Lb_RBR_2}
%
 \\
\hline
\hflavrowdefabi \\
 & & \\[-1ex]
\hflavrowdefabj \\
 & & \\[-1ex]
\hflavrowdefabk \\
 & & \\[-1ex]
\hflavrowdefabl \\
 & & \\[-1ex]
\hflavrowdefabm \\
 & & \\[-1ex]
\hflavrowdefabn \\
 & & \\[-1ex]
\hflavrowdefabo \\
 & & \\[-1ex]
\hflavrowdefabp \\
 & & \\[-1ex]
\hflavrowdefabq \\
 & & \\[-1ex]
\hflavrowdefabr \\
\hline
\hline
\end{tabular}
\begin{tablenotes}
\item[1] Multiple systematic uncertainties are added in quadrature.
\item[2] Measured in the $m^2_{\ell^+\ell^-}$ bin $[0.1, 6.0]~\rm{GeV^2/c^4}$ and for $m_{pK}<2.6~\rm{GeV/c^2}$.
\end{tablenotes}
\end{threeparttable}
}
\end{center}
\end{table}

\hflavrowdef{\hflavrowdefaca}{$\frac{f_{\Xi^{0}_{b}}}{f_{d}} \times \frac{ {\cal{B}}( \Xi^{0}_{b} \to p \overline{K^0} \pi^- ) }{ {\cal{B}}( B^0 \to K^0 \pi^+ \pi^-) }$}{{\setlength{\tabcolsep}{0pt}
}
\begin{table}[H]
\begin{center}
\begin{threeparttable}
\caption{Relative branching fractions of $\Xi_{b}^0$ decays.}
\label{tab:rareDecays_bBaryons_Xib_Xib0_RBR}
%
\begin{tablenotes}
\item[1] Multiple systematic uncertainties are added in quadrature.
\end{tablenotes}
\end{threeparttable}
\end{center}
\end{table}

\hflavrowdefshort{\hflavrowdefacl}{$\frac{f_{\Xi^{-}_{b}}}{f_u} \frac{ {\cal{B}}(\Xi^{-}_{b} \to p  K^- K^-) }{ {\cal{B}}(B ^- \to K^+  K^- K^-) }$}{{\setlength{\tabcolsep}{0pt}%
}
\begin{table}[H]
\begin{center}
\begin{threeparttable}
\caption{Relative branching fractions of $\Xi_{b}^-$ decays.}
\label{tab:rareDecays_bBaryons_Xib_Xib-_RBR}
%
}
\begin{table}[H]
\begin{center}
\begin{threeparttable}
\caption{Relative branching fractions of $\Omega^{-}_{b}$ decays.}
\label{tab:rareDecays_bBaryons_Omegab_RBR}
%
 \\
\hline
\hflavrowdefacu \\
 & & \\[-1ex]
\hflavrowdefacv \\
 & & \\[-1ex]
\hflavrowdefacw \\
\hline
\hline
\end{tabular}
\end{threeparttable}
\end{center}
\end{table}

\begin{figure}[h!]
\centering
\includegraphics[width=0.8\textwidth]{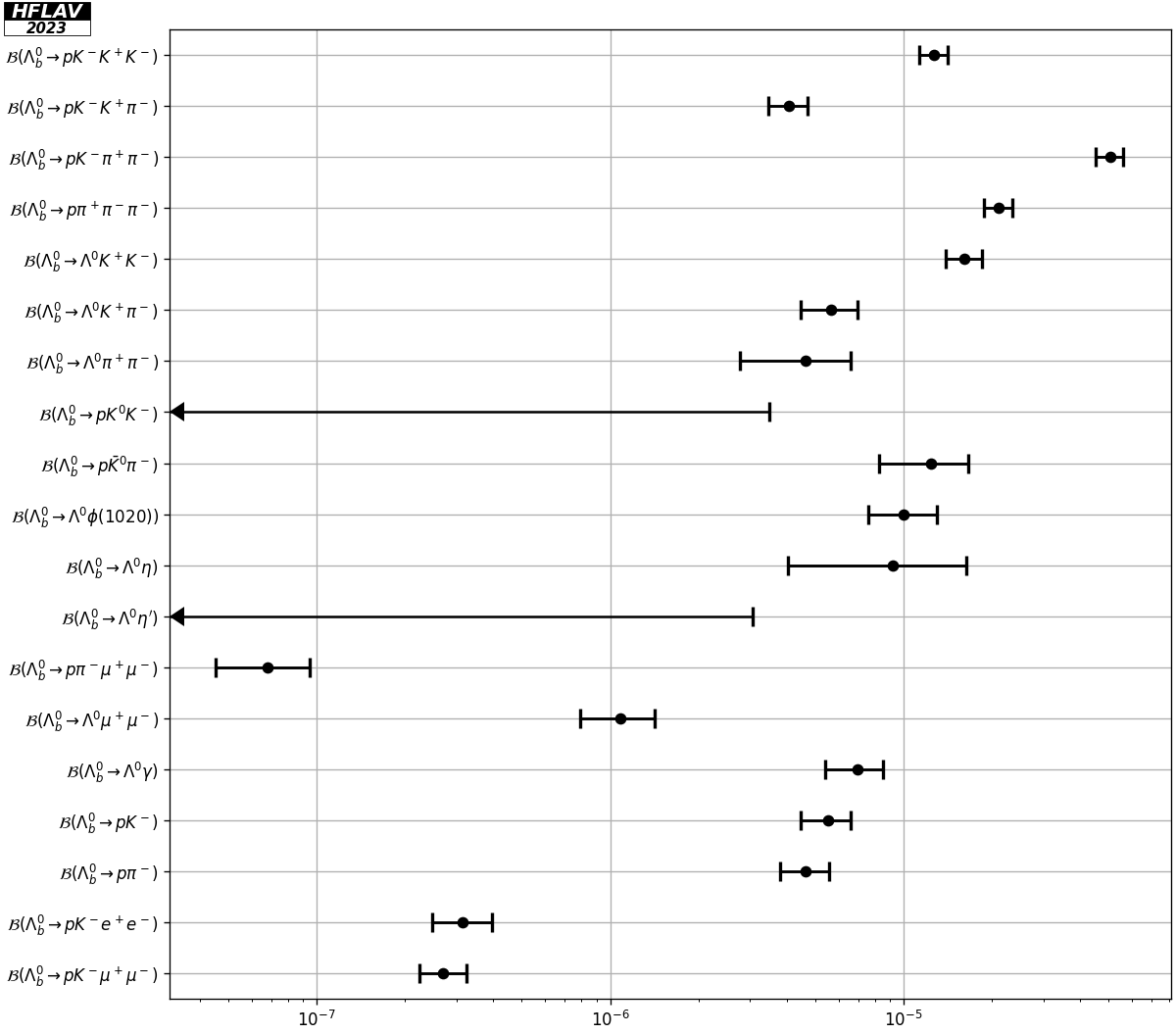}
\caption{Branching fractions of charmless $\Lb$ decays.}
\label{fig:rare-lb}
\end{figure}

\newpage
\noindent Measurements that are not included in the tables:
\begin{itemize}
\item In Ref.~\cite{LHCb:2018jna}, LHCb measures angular observables of the decay $\Lb \to \Lambda \mu^+\mu^-$, including the lepton-side, hadron-side and combined forward-backward asymmetries of the decay in the low recoil region $15<m^2(\ell\ell)<20\ \gevgevcccc$.
\item In Ref.~\cite{LHCb:2017vth}, LHCb performs a search for baryon-number-violating $\Xi^0_b$ oscillations and set an upper limit of $\omega<0.08\ {\rm ps^{-1}}$ on the oscillation rate.
\item In Ref.~\cite{LHCb:2021byf}, LHCb measures the photon polarization in $\Lambda_b \to \Lambda \gamma$ decays to be $\alpha_\gamma = 0.82\,^{+0.17}_{-0.26}\,^{+0.04}_{-0.13}$.
\end{itemize}

\mysubsection{Decays of \Bs mesons}
\label{sec:rare-bs}

Tables~\ref{tab:rareDecays_Bs_BR_1} to~\ref{tab:rareDecays_Bs_BR_6} and~\ref{tab:rareDecays_Bs_RBR_1} to~\ref{tab:rareDecays_Bs_RBR_2}
detail branching fractions and relative branching fractions of \Bs meson decays, respectively. 
Figures~\ref{fig:rare-bsleptonic} and~\ref{fig:rare-bs} show graphic representations of a selection of results given in this section.
\newpage

\hflavrowdef{\hflavrowdefaaa}{${\cal B}(B_s^0 \to \pi^+ \pi^-)$}{{\setlength{\tabcolsep}{0pt}
}
\begin{table}[H]
\begin{center}
\begin{threeparttable}
\caption{Branching fractions of charmless $B^0_s$ decays (part 1).}
\label{tab:rareDecays_Bs_BR_1}
%
 \\
\hline
\hflavrowdefaaa \\
 & & \\[-1ex]
\hflavrowdefaab \\
 & & \\[-1ex]
\hflavrowdefaac \\
 & & \\[-1ex]
\hflavrowdefaad \\
 & & \\[-1ex]
\hflavrowdefaae \\
 & & \\[-1ex]
\hflavrowdefaaf \\
 & & \\[-1ex]
\hflavrowdefaag \\
 & & \\[-1ex]
\hflavrowdefaah \\
 & & \\[-1ex]
\hflavrowdefaai \\
 & & \\[-1ex]
\hflavrowdefaaj \\
 & & \\[-1ex]
\hflavrowdefaak \\
 & & \\[-1ex]
\hflavrowdefaal \\
 & & \\[-1ex]
\hflavrowdefaam \\
\hline
\hline
\end{tabular}
\begin{tablenotes}
\item[*] Not included in PDG average.
\item[1] Measurement of $( {\cal B}(B_s^0 \to \pi^+ \pi^-) / {\cal B}(B^0 \to K^+ \pi^-) ) \frac{f_s}{f_d}$ used in our fit.
\item[2] Using $f_{s}$.
\item[3] Using ${\cal B}(B^+ \to \eta^\prime K^+)$.
\item[4] Multiple systematic uncertainties are added in quadrature.
\item[5] $400<m_{\pi^+\pi^-}<1600~\rm{MeV/c^2.}$
\item[6] Using ${\cal B}(B^0 \to \phi(1020) K^*(892)^0)$.
\item[7] Using ${\cal B}(B_s^0 \to J/\psi \phi(1020))$.
\end{tablenotes}
\end{threeparttable}
\end{center}
\end{table}

\hflavrowdef{\hflavrowdefaan}{${\cal B}(B_s^0 \to K^- \pi^+)$}{{\setlength{\tabcolsep}{0pt}
}
\begin{table}[H]
\begin{center}
\begin{threeparttable}
\caption{Branching fractions of charmless $B^0_s$ decays (part 2).}
\label{tab:rareDecays_Bs_BR_2}
%
 \\
\hline
\hflavrowdefaan \\
 & & \\[-1ex]
\hflavrowdefaao \\
 & & \\[-1ex]
\hflavrowdefaap \\
 & & \\[-1ex]
\hflavrowdefaaq \\
 & & \\[-1ex]
\hflavrowdefaar \\
 & & \\[-1ex]
\hflavrowdefaas \\
\hline
\hline
\end{tabular}
\begin{tablenotes}
\item[1] Measurement of $( {\cal B}(B_s^0 \to K^- \pi^+) / {\cal B}(B^0 \to K^+ \pi^-) ) \frac{f_s}{f_d}$ used in our fit.
\item[2] Multiple systematic uncertainties are added in quadrature.
\item[3] Measurement of $( {\cal B}(B_s^0 \to K^+ K^-) / {\cal B}(B^0 \to K^+ \pi^-) ) \frac{f_s}{f_d}$ used in our fit.
\item[4] Using ${\cal B}(B^0 \to \phi(1020) K^0)$.
\item[5] Regions corresponding to $D$, $\Lambda_c^+$ and charmonium resonances are vetoed in this analysis.
\item[6] Using ${\cal B}(B^0 \to K^0 \pi^+ \pi^-)$.
\item[7] Using ${\cal B}(B^0 \to K^*(892)^+ \pi^-)$.
\end{tablenotes}
\end{threeparttable}
\end{center}
\end{table}

\hflavrowdef{\hflavrowdefaat}{${\cal B}(B_s^0 \to K^*(892)^+ K^- \mathrm{+c.c.})$}{{\setlength{\tabcolsep}{0pt}
}
\begin{table}[H]
\begin{center}
\resizebox{1\textwidth}{!}{
\begin{threeparttable}
\caption{Branching fractions of charmless $B^0_s$ decays (part 3).}
\label{tab:rareDecays_Bs_BR_3}
%
 \\
\hline
\hflavrowdefaat \\
 & & \\[-1ex]
\hflavrowdefaau \\
 & & \\[-1ex]
\hflavrowdefaav \\
 & & \\[-1ex]
\hflavrowdefaaw \\
 & & \\[-1ex]
\hflavrowdefaax \\
 & & \\[-1ex]
\hflavrowdefaay \\
 & & \\[-1ex]
\hflavrowdefaaz \\
 & & \\[-1ex]
\hflavrowdefaba \\
 & & \\[-1ex]
\hflavrowdefabb \\
 & & \\[-1ex]
\hflavrowdefabc \\
 & & \\[-1ex]
\hflavrowdefabd \\
 & & \\[-1ex]
\hflavrowdefabe \\
 & & \\[-1ex]
\hflavrowdefabf \\
\hline
\hline
\end{tabular}
\begin{tablenotes}
\item[*] Not included in PDG average.
\item[1] Result extracted from Dalitz-plot analysis of $B^0_s \to K_S^0 K^+ \pi^-$ decays.
\item[2] Multiple systematic uncertainties are added in quadrature.
\item[3] Using ${\cal{B}}(B^0 \to K^0 \pi^+ \pi^-)$.
\item[4] Regions corresponding to $D$, $\Lambda_c^+$ and charmonium resonances are vetoed in this analysis.
\item[5] Using ${\cal B}(B^0 \to K^0 \pi^+ \pi^-)$.
\item[6] Using ${\cal B}(B^0 \to \phi(1020) K^*(892)^0)$.
\item[7] Measurement of ${\cal B}(B^0 \to K^*(892)^0 \bar{K}^*(892)^0) / {\cal B}(B_s^0 \to K^*(892)^0 \bar{K}^*(892)^0)$ used in our fit.
\end{tablenotes}
\end{threeparttable}
}
\end{center}
\end{table}

\hflavrowdef{\hflavrowdefabg}{${\cal B}(B_s^0 \to p \bar{p})$}{{\setlength{\tabcolsep}{0pt}
}
\begin{table}[H]
\begin{center}
\resizebox{1\textwidth}{!}{
\begin{threeparttable}
\caption{Branching fractions of charmless $B^0_s$ decays (part 4).}
\label{tab:rareDecays_Bs_BR_4}
%
 \\
\hline
\hflavrowdefabg \\
 & & \\[-1ex]
\hflavrowdefabh \\
 & & \\[-1ex]
\hflavrowdefabi \\
 & & \\[-1ex]
\hflavrowdefabj \\
 & & \\[-1ex]
\hflavrowdefabk \\
 & & \\[-1ex]
\hflavrowdefabl \\
 & & \\[-1ex]
\hflavrowdefabm \\
\hline
\hline
\end{tabular}
\begin{tablenotes}
\item[*] Not included in PDG average.
\item[1] $m_{p\overline{p}} < 2.85~\rm{GeV/}c^2$.
\item[2] Multiple systematic uncertainties are added in quadrature.
\item[3] Using ${\cal B}(B^0 \to K^*(892)^0 \gamma)$.
\end{tablenotes}
\end{threeparttable}
}
\end{center}
\end{table}

\hflavrowdef{\hflavrowdefabn}{${\cal B}(B_s^0 \to e^+ e^-)$}{{\setlength{\tabcolsep}{0pt}
}
\begin{table}[H]
\begin{center}
\resizebox{0.95\textwidth}{!}{
\begin{threeparttable}
\caption{Branching fractions of charmless $B^0_s$ decays (part 5).}
\label{tab:rareDecays_Bs_BR_5}
%
 \\
\hline
\hflavrowdefabn \\
 & & \\[-1ex]
\hflavrowdefabo \\
 & & \\[-1ex]
\hflavrowdefabp \\
 & & \\[-1ex]
\hflavrowdefabq \\
 & & \\[-1ex]
\hflavrowdefabr \\
 & & \\[-1ex]
\hflavrowdefabs \\
 & & \\[-1ex]
\hflavrowdefabt \\
 & & \\[-1ex]
\hflavrowdefabu \\
 & & \\[-1ex]
\hflavrowdefabv \\
\hline
\hline
\end{tabular}
\begin{tablenotes}
\item[*] Not included in PDG average.
\item[1] PDG shows the result obtained at 95\% CL.
\item[2] $m_{\mu^+\mu^-}>4.9~\text{GeV}/c^2$.
\item[3] At CL=95\,\%.
\item[4] The mass windows corresponding to $\phi$ and charmonium resonances decaying to $\mu \mu$ are vetoed.
\item[5] $a$ is a promptly decaying scalar particle with a mass of 1 $\text{GeV}/c^2$ 
\item[6] The PDG uncertainty includes a scale factor.
\item[7] Treatment of charmonium intermediate components differs between the results.
\item[8] Multiple systematic uncertainties are added in quadrature.
\item[9] Using ${\cal B}(B_s^0 \to J/\psi \phi(1020))$.
\item[10] $0.5 < m_{\pi^+\pi^-} <  1.3~\rm{GeV/}c^2$.
\item[11] Measurement of ${\cal B}(B_s^0 \to \pi^+ \pi^- \mu^+ \mu^-) / ( {\cal B}(B^0 \to J/\psi K^*(892)^0) {\cal B}(J/\psi \to \mu^+ \mu^-) {\cal B}(K^*(892)^0 \to K \pi) \times 2/3 )$ used in our fit.
\end{tablenotes}
\end{threeparttable}
}
\end{center}
\end{table}

\hflavrowdef{\hflavrowdefabw}{${\cal B}(B_s^0 \to e^+ \mu^- \mathrm{+c.c.})$}{{\setlength{\tabcolsep}{0pt}
}
\begin{table}[H]
\begin{center}
\resizebox{1\textwidth}{!}{
\begin{threeparttable}
\caption{Branching fractions of charmless $B^0_s$ decays (part 6).}
\label{tab:rareDecays_Bs_BR_6}
%
 \\
\hline
\hflavrowdefabw \\
 & & \\[-1ex]
\hflavrowdefabx \\
 & & \\[-1ex]
\hflavrowdefaby \\
 & & \\[-1ex]
\hflavrowdefabz \\
 & & \\[-1ex]
\hflavrowdefaca \\
 & & \\[-1ex]
\hflavrowdefacb \\
 & & \\[-1ex]
\hflavrowdefacc \\
 & & \\[-1ex]
\hflavrowdefacd \\
 & & \\[-1ex]
\hflavrowdeface \\
 & & \\[-1ex]
\hflavrowdefacf \\
\hline
\hline
\end{tabular}
\begin{tablenotes}
\item[*] Not included in PDG average.
\item[\dag] Preliminary result.
\item[1] PDG shows the result obtained at 95\% CL.
\item[2] Multiple systematic uncertainties are added in quadrature.
\item[3] Using ${\cal B}(B_s^0 \to J/\psi \phi(1020))$.
\item[4] $m_{X_{ss}} < 2.4~\text{GeV}/c^2$
\item[5] Using $f_{s}$.
\item[6] black
\item[7] Measurement of ${\cal B}(B_s^0 \to p \bar{p} p \bar{p}) / ( {\cal B}(B_s^0 \to J/\psi \phi(1020)) {\cal B}(J/\psi \to p \bar{p}) {\cal B}(\phi(1020) \to K^+ K^-)  )$ used in our fit.
\end{tablenotes}
\end{threeparttable}
}
\end{center}
\end{table}

\hflavrowdef{\hflavrowdefacg}{$\frac{f_s}{f_d}  \frac{\mathcal{B}(B^0_s\rightarrow\pi^+\pi^-)}{\mathcal{B}(B^0\rightarrow K^+\pi^-)}$}{{\setlength{\tabcolsep}{0pt}
}
\begin{table}[H]
\begin{center}
\begin{threeparttable}
\caption{Relative branching fractions of $B^0_s$ decays (part 1).}
\label{tab:rareDecays_Bs_RBR_1}
%
 \\
\hline
\hflavrowdefacg \\
 & & \\[-1ex]
\hflavrowdefach \\
 & & \\[-1ex]
\hflavrowdefaci \\
 & & \\[-1ex]
\hflavrowdefacj \\
 & & \\[-1ex]
\hflavrowdefack \\
 & & \\[-1ex]
\hflavrowdefacl \\
 & & \\[-1ex]
\hflavrowdefacm \\
 & & \\[-1ex]
\hflavrowdefacn \\
 & & \\[-1ex]
\hflavrowdefaco \\
 & & \\[-1ex]
\hflavrowdefacp \\
 & & \\[-1ex]
\hflavrowdefacq \\
 & & \\[-1ex]
\hflavrowdefacr \\
 & & \\[-1ex]
\hflavrowdefacs \\
\hline
\hline
\end{tabular}
\begin{tablenotes}
\item[1] The PDG average is a result of a fit including input from other measurements.
\item[2] Multiple systematic uncertainties are added in quadrature.
\item[3] Regions corresponding to $D$, $\Lambda_c^+$ and charmonium resonances are vetoed in this analysis.
\end{tablenotes}
\end{threeparttable}
\end{center}
\end{table}

\hflavrowdef{\hflavrowdefact}{$\dfrac{{\cal B}(B_s^0 \to p \bar{p} K^+ \pi^-)}{{\cal B}(B^0 \to p \bar{p} K^+ \pi^-)}$}{{\setlength{\tabcolsep}{0pt}
}
\begin{table}[H]
\begin{center}
\begin{threeparttable}
\caption{Relative branching fractions of $B^0_s$ decays (part 2).}
\label{tab:rareDecays_Bs_RBR_2}
%
 \\
\hline
\hflavrowdefact \\
 & & \\[-1ex]
\hflavrowdefacu \\
 & & \\[-1ex]
\hflavrowdefacv \\
 & & \\[-1ex]
\hflavrowdefacw \\
 & & \\[-1ex]
\hflavrowdefacx \\
 & & \\[-1ex]
\hflavrowdefacy \\
 & & \\[-1ex]
\hflavrowdefacz \\
 & & \\[-1ex]
\hflavrowdefada \\
 & & \\[-1ex]
\hflavrowdefadb \\
 & & \\[-1ex]
\hflavrowdefadc \\
\hline
\hline
\end{tabular}
\begin{tablenotes}
\item[1] $m_{p\overline{p}} < 2.85~\rm{GeV/}c^2$.
\item[2] Multiple systematic uncertainties are added in quadrature.
\item[3] $0.5 < m_{\pi^+\pi^-} <  1.3~\rm{GeV/}c^2$.
\end{tablenotes}
\end{threeparttable}
\end{center}
\end{table}

\newpage
\noindent Measurements that are not included in the tables (the definitions of observables can be found in the corresponding experimental papers):
\begin{itemize}
\item In Ref.~\cite{LHCb:2021zwz}, LHCb reports the differential $\Bs \to \phi\mu^+\mu^-$ branching fraction in bins of $m^2(\mu^+\mu^-)$.
\item In Ref.~\cite{LHCb:2021xxq}, LHCb performs an angular analysis of $\Bs \to \phi\mu^+\mu^-$ decays and reports  $F_L$, $S_3$, $S_4$, $S_7$, $A_5$, $A_{FB}^{CP}$, $A_8$ and $A_9$ in bins of $m^2(\mu^+\mu^-)$.
\item  In Ref. \cite{LHCb:2016oeh}, LHCb reports the photon polarization in $\Bs \to \phi \gamma$ decays.
\item We do not perform the average of the branching fraction of $B\to\mu^+\mu^-$ decays, which is taken care of by the LHC Heavy Flavour Working Group~\cite{LHC_HF_WG}, taking into account the correlations between the $\Bz \to \mu^+\mu^-$ and $\Bs \to \mu^+\mu^-$ branching fractions. The latest results from ATLAS, CMS and LHCb are in Refs.~\cite{ATLAS:2018cur,CMS:2022mgd,LHCb:2021awg2}, respectively.
\end{itemize}

\begin{figure}[htbp!]
\centering
\includegraphics[width=0.8\textwidth]{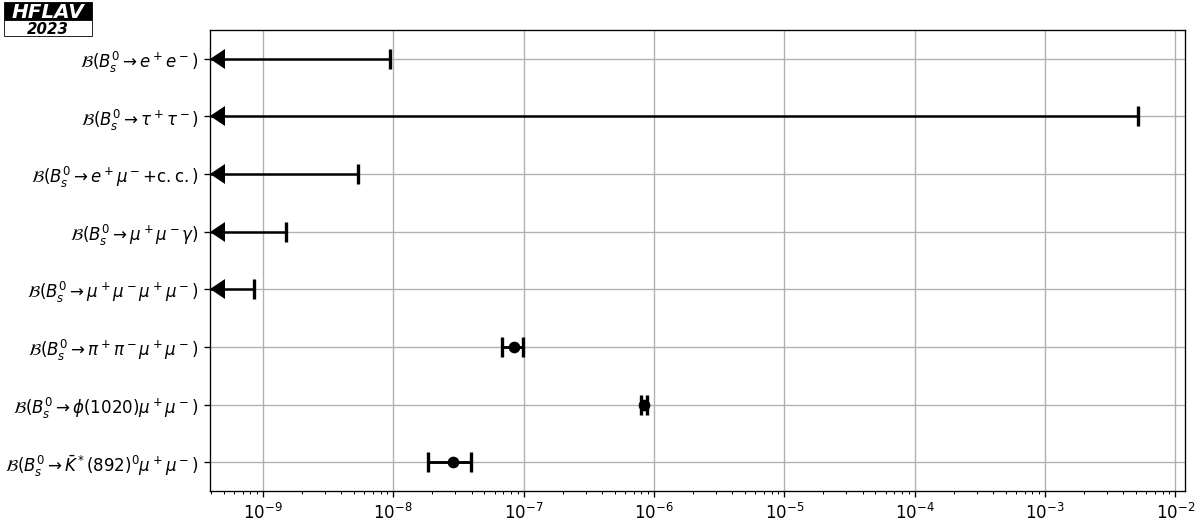}
\caption{Branching fractions of charmless leptonic $\Bs$ decays.}
\label{fig:rare-bsleptonic}
\end{figure}

\begin{figure}[ht!]
\centering
\includegraphics[width=0.8\textwidth]{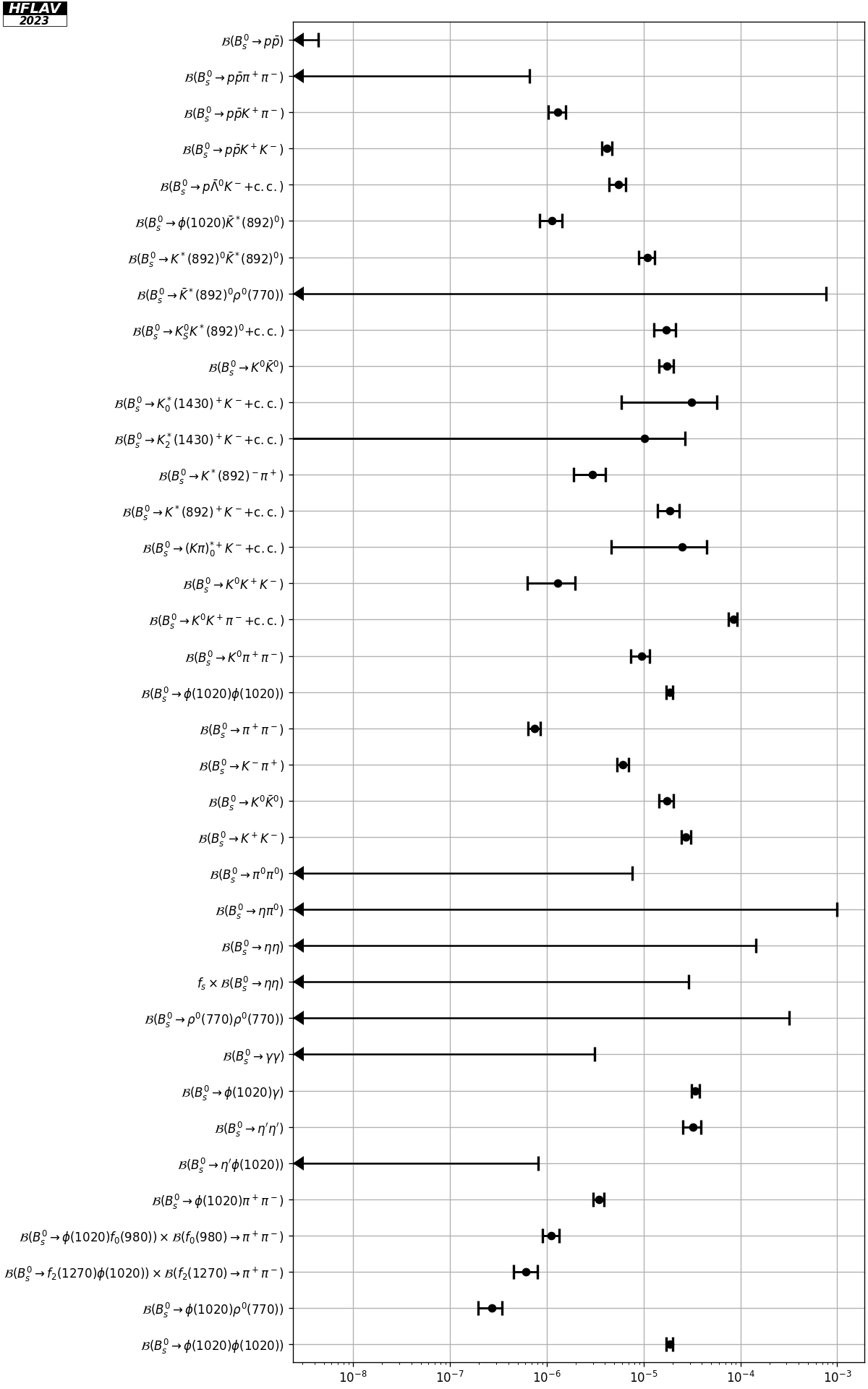}
\caption{Branching fractions of charmless nonleptonic $\Bs$ decays.}
\label{fig:rare-bs}
\end{figure}

\clearpage

\mysubsection{Decays of $\Bc$ mesons}
\label{sec:rare-bc}

Table~\ref{tab:rareDecays_Bc_BR}
details branching fractions and ratios of branching fractions of $\Bc$ meson decays to
charmless hadronic final states, except for decays to final states containing $\Bs$ mesons that are quoted in Sec.~\ref{sec:b2c:Bc}.

\hflavrowdef{\hflavrowdefaaa}{${\cal{B}}(B^{+}_{c} \to p \bar{p}\pi^+)\times \frac{\it{f}_c}{\it{f}_u}$ [$10^{-8}$]}{{\setlength{\tabcolsep}{0pt}\begin{tabular}{m{6.5em}l} {LHCb \cite{LHCb:2016fkv}\tnote{1}\hphantom{\textsuperscript{1}}} & { $< 2.8$ } \\ \end{tabular}}}{\begin{tabular}{l} $< 2.8$\\ \end{tabular}}
\hflavrowdef{\hflavrowdefaab}{$\frac{\mathcal{B}(B^{+}_{c} \to K^+ K_S^0)}{\mathcal{B}(B^+\to K_S^0 \pi^+)} \times \frac{\it{f}_c}{\it{f}_u}$ [$10^{-2}$]}{{\setlength{\tabcolsep}{0pt}\begin{tabular}{m{6.5em}l} {LHCb \cite{LHCb:2013vip}\tnote{}\hphantom{\textsuperscript{}}} & { $< 5.8$ } \\ \end{tabular}}}{\begin{tabular}{l} $< 5.8$\\ \end{tabular}}
\hflavrowdef{\hflavrowdefaac}{${\cal B}(B_c^+ \to K^+ \bar{K}^0)$\tnote{2}\hphantom{\textsuperscript{2}} [$10^{-4}$]}{{\setlength{\tabcolsep}{0pt}\begin{tabular}{m{6.5em}l} {LHCb \cite{LHCb:2013vip}\tnote{}\hphantom{\textsuperscript{}}} & { $< 4.6$ } \\ \end{tabular}}}{\begin{tabular}{l} $< 4.6$\\ \end{tabular}}
\hflavrowdef{\hflavrowdefaad}{${\cal{B}}(B^{+}_{c} \to K^+ K^- \pi^+) \times \frac{\it{f}_c}{\it{f}_u}$ [$10^{-7}$]}{{\setlength{\tabcolsep}{0pt}\begin{tabular}{m{6.5em}l} {LHCb \cite{LHCb:2016utz}\tnote{3}\hphantom{\textsuperscript{3}}} & { $< 1.50$ } \\ \end{tabular}}}{\begin{tabular}{l} $< 1.5$\\ \end{tabular}}
\begin{table}[H]
\begin{center}
\begin{threeparttable}
\caption{Branching fractions and relative branching fractions of $B^{+}_{c}$ decays.}
\label{tab:rareDecays_Bc_BR}
\begin{tabular}{ Sl l l }
\hline
\hline
\textbf{Parameter} & \textbf{Measurements} & \begin{tabular}{l}\textbf{Average}\end{tabular} \\
\hline
\hflavrowdefaaa \\
 & & \\[-1ex]
\hflavrowdefaab \\
 & & \\[-1ex]
\hflavrowdefaac \\
 & & \\[-1ex]
\hflavrowdefaad \\
\hline
\hline
\end{tabular}
\begin{tablenotes}
\item[1] Measured in the region $_{p\overline{p}} < 2.85~\rm{GeV/c^2}$, $p_T(B)<20~\rm{GeV/c}$ and $2.0 < y(B) < 4.5$.
\item[2] Derived from the ratio in the previous entry using ${\cal B} (B^+ \to K^0 \pi^+) = (23.97 \pm 0.53 \pm 0.71)\times 10^{-6}$, $f_u=0.33$ and $f_c=0.001$.
\item[3] Measured in the annihilation region $m_{K^+\pi^+} < 1.834~\rm{GeV/c^2}$, and in the fiducial region $p_T(B) < 20~\rm{GeV/c}$ and $2.0 < y(B) < 4.5$
\end{tablenotes}
\end{threeparttable}
\end{center}
\end{table}

\mysubsection{Rare decays of \Bz and \Bp\ mesons with photons and/or leptons}
\label{sec:rare-radll}

This section reports different observables
for radiative decays, lepton-flavour/number-violating (LFV/LNV) decays and flavour-changing-neutral-current (FCNC) decays with leptons of \Bz and \Bp\ mesons.
In all decays listed in this section, charmonium intermediate states are vetoed.
Tables~\ref{tab:rareDecays_radll_Bu_BR_1} to~\ref{tab:rareDecays_radll_Bu_BR_5}, \ref{tab:rareDecays_radll_Bd_BR_1} to~\ref{tab:rareDecays_radll_Bd_BR_5} and~\ref{tab:rareDecays_radll_Badmix_BR_1} to~\ref{tab:rareDecays_radll_Badmix_BR_3} provide compilations of branching fractions of radiative and FCNC decays with leptons of \Bp\ mesons, \Bz\ mesons and their admixture, respectively.
Tables~\ref{tab:rareDecays_radll_Bd_BR_5} and~\ref{tab:rareDecays_radll_Badmix_BR_3} also include LFV/LNV decays.
Tables~\ref{tab:rareDecays_radll_Leptonic_BR_1} and~\ref{tab:rareDecays_radll_Leptonic_BR_2} contain branching fractions of leptonic and radiative-leptonic $\Bp$ and $\Bz$ decays.
These are followed by Tables~\ref{tab:rareDecays_radll_RBR_1} to~\ref{tab:rareDecays_radll_RBR_3}, which give relative branching fractions of $\Bp$ and $\Bz$ decays, then Table~\ref{tab:rareDecays_radll_Bincl_0}, which gives a compilation of inclusive decays. In the modes listed in Table~\ref{tab:rareDecays_radll_Bincl_0}, the radiated particle is a gluon, which is an exception in this section.
Table~\ref{tab:rareDecays_radll_IA_0} contains isospin asymmetry measurements.
Finally, Tables~\ref{tab:rareDecays_radll_LFN_LFV_BR_1} to~\ref{tab:rareDecays_radll_LFN_LFV_BR_3} and~\ref{tab:rareDecays_radll_LFN_LFV_BR_4}  provide compilations of branching fractions of \Bp\ and \Bz\ mesons to lepton-flavour/number-violating final states, respectively. 
The average of 
Figures \ref{fig:rare-btosll} to~\ref{fig:rare-neutrino} show graphic representations of a selection of results given in this section.

\hflavrowdef{\hflavrowdefaaa}{${\cal B}(B^+ \to K^*(892)^+ \gamma)$\tnote{1}\hphantom{\textsuperscript{1}}}{{\setlength{\tabcolsep}{0pt}
}
\begin{table}[H]
\begin{center}
\begin{threeparttable}
\caption{Branching fractions of $B^+$ radiative and FCNC decays with leptons (part 1).}
\label{tab:rareDecays_radll_Bu_BR_1}
%
 \\
\hline
\hflavrowdefaaa \\
 & & \\[-1ex]
\hflavrowdefaab \\
 & & \\[-1ex]
\hflavrowdefaac \\
 & & \\[-1ex]
\hflavrowdefaad \\
 & & \\[-1ex]
\hflavrowdefaae \\
 & & \\[-1ex]
\hflavrowdefaaf \\
 & & \\[-1ex]
\hflavrowdefaag \\
\hline
\hline
\end{tabular}
\begin{tablenotes}
\item[\dag] Preliminary result.
\item[1] The PDG uncertainty includes a scale factor.
\item[2] $m_{K\pi} < 2.0~\text{GeV}/c^2$.
\item[3] $0.79 < m_{K\pi} < 1.0~\text{GeV}/c^2$.
\item[4] Multiple systematic uncertainties are added in quadrature.
\item[5] $1 < m_{K\pi\pi} < 2~\text{GeV}/c^2$.
\item[6] $m_{K\eta^{(\prime)}} < 3.25 ~\text{GeV}/c^2$.
\item[7] $m_{K\eta} < 2.4 ~\text{GeV}/c^2$.
\item[8] $m_{K\eta^\prime} < 3.4 ~\text{GeV}/c^2$
\item[9] $m_{\phi K} < 3.0 ~\text{GeV}/c^2$.
\end{tablenotes}
\end{threeparttable}
\end{center}
\end{table}

\hflavrowdef{\hflavrowdefaah}{${\cal B}(B^+ \to K^+ \pi^- \pi^+ \gamma)$\tnote{1}\hphantom{\textsuperscript{1}}}{{\setlength{\tabcolsep}{0pt}
}
\begin{table}[H]
\begin{center}
\resizebox{0.85\textwidth}{!}{
\begin{threeparttable}
\caption{Branching fractions of $B^+$ radiative and FCNC decays with leptons (part 2).}
\label{tab:rareDecays_radll_Bu_BR_2}
%
 \\
\hline
\hflavrowdefaah \\
 & & \\[-1ex]
\hflavrowdefaai \\
 & & \\[-1ex]
\hflavrowdefaaj \\
 & & \\[-1ex]
\hflavrowdefaak \\
 & & \\[-1ex]
\hflavrowdefaal \\
 & & \\[-1ex]
\hflavrowdefaam \\
 & & \\[-1ex]
\hflavrowdefaan \\
 & & \\[-1ex]
\hflavrowdefaao \\
 & & \\[-1ex]
\hflavrowdefaap \\
 & & \\[-1ex]
\hflavrowdefaaq \\
 & & \\[-1ex]
\hflavrowdefaar \\
 & & \\[-1ex]
\hflavrowdefaas \\
 & & \\[-1ex]
\hflavrowdefaat \\
\hline
\hline
\end{tabular}
\begin{tablenotes}
\item[1] The PDG uncertainty includes a scale factor.
\item[2] $m_{K\pi\pi} < 1.8~\text{GeV}/c^2$.
\item[3] $1 < m_{K\pi\pi} < 2~\text{GeV}/c^2$.
\item[4] $m_{K\pi\pi} < 2.4~\text{GeV}/c^2$.
\item[5] This corresponds to the $(K \pi)$ $S$-wave obtained with LASS parameterisation~\cite{Aston:1987ir}.
\item[6] $m_{K\pi} < 1.6 ~\text{GeV}/c^2$.
\item[7]  $1.25 < m_{K\pi} < 1.6~\text{GeV}/c^2$ and $m_{K\pi\pi} < 2.4~\text{GeV}/c^2$.
\item[8] Multiple systematic uncertainties are added in quadrature.
\end{tablenotes}
\end{threeparttable}
}
\end{center}
\end{table}

\hflavrowdef{\hflavrowdefaau}{${\cal B}(B^+ \to \rho^+(770) \gamma)$}{{\setlength{\tabcolsep}{0pt}
}
\begin{table}[H]
\begin{center}
\begin{threeparttable}
\caption{Branching fractions of $B^+$ radiative and FCNC decays with leptons (part 3).}
\label{tab:rareDecays_radll_Bu_BR_3}
%
}
\begin{table}[H]
\begin{center}
\begin{threeparttable}
\caption{Branching fractions of $B^+$ radiative and FCNC decays with leptons (part 4).}
\label{tab:rareDecays_radll_Bu_BR_4}
%
 \\
\hline
\hflavrowdefaax \\
 & & \\[-1ex]
\hflavrowdefaay \\
 & & \\[-1ex]
\hflavrowdefaaz \\
 & & \\[-1ex]
\hflavrowdefaba \\
\hline
\hline
\end{tabular}
\begin{tablenotes}
\item[1] Treatment of charmonium intermediate components differs between the results.
\item[2] LHCb  also reports the branching fraction  in bins of $m^2_{\ell^+\ell^-}$.
\item[3] Measurement of ${\cal B}(B^+ \to \pi^+ \mu^+ \mu^-) / ( {\cal B}(B^+ \to J/\psi K^+) {\cal B}(J/\psi \to \mu^+ \mu^-) )$ used in our fit.
\end{tablenotes}
\end{threeparttable}
\end{center}
\end{table}

\hflavrowdef{\hflavrowdefabb}{${\cal B}(B^+ \to K^+ \ell^+ \ell^-)$\tnote{1}\hphantom{\textsuperscript{1}}}{{\setlength{\tabcolsep}{0pt}
}
\begin{table}[H]
\begin{center}
\resizebox{0.95\textwidth}{!}{
\begin{threeparttable}
\caption{Branching fractions of $B^+$ radiative and FCNC decays with leptons (part 5).}
\label{tab:rareDecays_radll_Bu_BR_5}
%
 \\
\hline
\hflavrowdefabb \\
 & & \\[-1ex]
\hflavrowdefabc \\
 & & \\[-1ex]
\hflavrowdefabd \\
 & & \\[-1ex]
\hflavrowdefabe \\
 & & \\[-1ex]
\hflavrowdefabf \\
 & & \\[-1ex]
\hflavrowdefabg \\
 & & \\[-1ex]
\hflavrowdefabh \\
 & & \\[-1ex]
\hflavrowdefabi \\
 & & \\[-1ex]
\hflavrowdefabj \\
 & & \\[-1ex]
\hflavrowdefabk \\
 & & \\[-1ex]
\hflavrowdefabl \\
 & & \\[-1ex]
\hflavrowdefabm \\
 & & \\[-1ex]
\hflavrowdefabn \\
\hline
\hline
\end{tabular}
\begin{tablenotes}
\item[*] Not included in PDG average.
\item[1] Treatment of charmonium intermediate components differs between the results.
\item[2] The PDG uncertainty includes a scale factor.
\item[3] Only muons are used.
\item[4] Using ${\cal B}(B^+ \to \psi(2S) K^+)$.
\item[5] Using ${\cal B}(B^+ \to J/\psi \phi(1020) K^+)$.
\end{tablenotes}
\end{threeparttable}
}
\end{center}
\end{table}

\hflavrowdef{\hflavrowdefabo}{${\cal B}(B^0 \to K^*(892)^0 \gamma)$\tnote{1}\hphantom{\textsuperscript{1}}}{{\setlength{\tabcolsep}{0pt}
}
\begin{table}[H]
\begin{center}
\begin{threeparttable}
\caption{Branching fractions of $B^0$ radiative and FCNC decays with leptons (part 1).}
\label{tab:rareDecays_radll_Bd_BR_1}
%
 \\
\hline
\hflavrowdefabo \\
 & & \\[-1ex]
\hflavrowdefabp \\
 & & \\[-1ex]
\hflavrowdefabq \\
 & & \\[-1ex]
\hflavrowdefabr \\
 & & \\[-1ex]
\hflavrowdefabs \\
 & & \\[-1ex]
\hflavrowdefabt \\
 & & \\[-1ex]
\hflavrowdefabu \\
 & & \\[-1ex]
\hflavrowdefabv \\
 & & \\[-1ex]
\hflavrowdefabw \\
\hline
\hline
\end{tabular}
\begin{tablenotes}
\item[\dag] Preliminary result.
\item[*] Not included in PDG average.
\item[1] The PDG uncertainty includes a scale factor.
\item[2] Measurement of ${\cal B}(B_s^0 \to \phi(1020) \gamma) / {\cal B}(B^0 \to K^*(892)^0 \gamma)$ used in our fit.
\item[3] Measurement of $( {\cal B}(\Lambda_b^0 \to \Lambda^0 \gamma) / {\cal B}(B^0 \to K^*(892)^0 \gamma) ) \frac{f_{\Lambda^0_b}}{f_d}$ used in our fit.
\item[4] $m_{K\pi} < 2.0~\text{GeV}/c^2$.
\item[5] $0.78 < m_{K\pi} < 1.1~\text{GeV}/c^2$.
\item[6] $0.82 < m_{K\pi} < 1.0~\text{GeV}/c^2$.
\item[7]  $1.25 < m_{K\pi} < 1.6~\text{GeV}/c^2$.
\item[8] $m_{K\eta^{(\prime)}} < 3.25 ~\text{GeV}/c^2$.
\item[9] $m_{K\eta} < 2.4 ~\text{GeV}/c^2$.
\item[10] $m_{K\eta^\prime} < 3.4 ~\text{GeV}/c^2$
\item[11] $m_{\phi K} < 3.0 ~\text{GeV}/c^2$.
\item[12] $X(214)$ is searched for in the mass range $[212, 300]~\text{MeV}/c^2$.
\end{tablenotes}
\end{threeparttable}
\end{center}
\end{table}

\hflavrowdef{\hflavrowdefabx}{${\cal B}(B^0 \to K^0 \pi^+ \pi^- \gamma)$}{{\setlength{\tabcolsep}{0pt}
}
\begin{table}[H]
\begin{center}
\begin{threeparttable}
\caption{Branching fractions of $B^0$ radiative and FCNC decays with leptons (part 2).}
\label{tab:rareDecays_radll_Bd_BR_2}
%
 \\
\hline
\hflavrowdefabx \\
 & & \\[-1ex]
\hflavrowdefaby \\
 & & \\[-1ex]
\hflavrowdefabz \\
 & & \\[-1ex]
\hflavrowdefaca \\
 & & \\[-1ex]
\hflavrowdefacb \\
 & & \\[-1ex]
\hflavrowdefacc \\
 & & \\[-1ex]
\hflavrowdefacd \\
 & & \\[-1ex]
\hflavrowdeface \\
 & & \\[-1ex]
\hflavrowdefacf \\
\hline
\hline
\end{tabular}
\begin{tablenotes}
\item[1] $m_{K\pi\pi} < 1.8~\text{GeV}/c^2$.
\item[2] $1 < m_{K\pi\pi} < 2~\text{GeV}/c^2$.
\item[3] Measured in bins of $m_{K^0_S K^0_S}$. We report the result for the full range, $1.0~\text{GeV}/c^2 < m_{K^0_S K^0_S} < 3.0~\text{GeV}/c^2$.
\item[4] Measured in bins of $m_{K^0_S K^0_S}$. We report the result for the full range, $1.00~\text{GeV}/c^2 < m_{K^0_S K^0_S} < 1.44~\text{GeV}/c^2$.
\item[5] Measured in bins of $m_{K^0_S K^0_S}$. We report the result for the full range, $1.44~\text{GeV}/c^2 < m_{K^0_S K^0_S} < 1.63~\text{GeV}/c^2$.
\end{tablenotes}
\end{threeparttable}
\end{center}
\end{table}

\hflavrowdef{\hflavrowdefacg}{${\cal B}(B^0 \to \rho^0(770) \gamma)$}{{\setlength{\tabcolsep}{0pt}
}
\begin{table}[H]
\begin{center}
\begin{threeparttable}
\caption{Branching fractions of $B^0$ radiative and FCNC decays with leptons (part 3).}
\label{tab:rareDecays_radll_Bd_BR_3}
%
 \\
\hline
\hflavrowdefacg \\
 & & \\[-1ex]
\hflavrowdefach \\
 & & \\[-1ex]
\hflavrowdefaci \\
 & & \\[-1ex]
\hflavrowdefacj \\
 & & \\[-1ex]
\hflavrowdefack \\
 & & \\[-1ex]
\hflavrowdefacl \\
 & & \\[-1ex]
\hflavrowdefacm \\
 & & \\[-1ex]
\hflavrowdefacn \\
\hline
\hline
\end{tabular}
\begin{tablenotes}
\item[1] $X(214)$ is searched for in the mass range $[212, 300]~\text{MeV}/c^2$.
\item[2] Treatment of charmonium intermediate components differs between the results.
\end{tablenotes}
\end{threeparttable}
\end{center}
\end{table}

\hflavrowdef{\hflavrowdefaco}{${\cal B}(B^0 \to \eta \ell^+ \ell^-)$}{{\setlength{\tabcolsep}{0pt}
}
\begin{table}[H]
\begin{center}
\begin{threeparttable}
\caption{Branching fractions of $B^0$ radiative and FCNC decays with leptons (part 4).}
\label{tab:rareDecays_radll_Bd_BR_4}
%
 \\
\hline
\hflavrowdefaco \\
 & & \\[-1ex]
\hflavrowdefacp \\
 & & \\[-1ex]
\hflavrowdefacq \\
 & & \\[-1ex]
\hflavrowdefacr \\
 & & \\[-1ex]
\hflavrowdefacs \\
 & & \\[-1ex]
\hflavrowdefact \\
 & & \\[-1ex]
\hflavrowdefacu \\
 & & \\[-1ex]
\hflavrowdefacv \\
 & & \\[-1ex]
\hflavrowdefacw \\
 & & \\[-1ex]
\hflavrowdefacx \\
 & & \\[-1ex]
\hflavrowdefacy \\
 & & \\[-1ex]
\hflavrowdefacz \\
 & & \\[-1ex]
\hflavrowdefada \\
\hline
\hline
\end{tabular}
\begin{tablenotes}
\item[*] Not included in PDG average.
\item[1] Treatment of charmonium intermediate components differs between the results.
\item[2] Only muons are used.
\item[3] Multiple systematic uncertainties are added in quadrature.
\end{tablenotes}
\end{threeparttable}
\end{center}
\end{table}

\hflavrowdef{\hflavrowdefadb}{${\cal B}(B^0 \to \pi^+ \pi^- \mu^+ \mu^-)$}{{\setlength{\tabcolsep}{0pt}
}
\begin{table}[H]
\begin{center}
\begin{threeparttable}
\caption{Branching fractions of $B^0$ radiative and FCNC decays with leptons (part 5).}
\label{tab:rareDecays_radll_Bd_BR_5}
%
 \\
\hline
\hflavrowdefadb \\
 & & \\[-1ex]
\hflavrowdefadc \\
 & & \\[-1ex]
\hflavrowdefadd \\
 & & \\[-1ex]
\hflavrowdefade \\
 & & \\[-1ex]
\hflavrowdefadf \\
 & & \\[-1ex]
\hflavrowdefadg \\
 & & \\[-1ex]
\hflavrowdefadh \\
 & & \\[-1ex]
\hflavrowdefadi \\
 & & \\[-1ex]
\hflavrowdefadj \\
 & & \\[-1ex]
\hflavrowdefadk \\
 & & \\[-1ex]
\hflavrowdefadl \\
\hline
\hline
\end{tabular}
\begin{tablenotes}
\item[\dag] Preliminary result.
\item[1] The mass windows corresponding to $\phi$ and charmonium resonances decaying to $\mu \mu$ are vetoed.
\item[2] $0.5 < m_{\pi^+\pi^-} <  1.3~\rm{GeV/}c^2$.
\item[3] Measurement of ${\cal B}(B^0 \to \pi^+ \pi^- \mu^+ \mu^-) / ( {\cal B}(B^0 \to J/\psi K^*(892)^0) {\cal B}(J/\psi \to \mu^+ \mu^-) {\cal B}(K^*(892)^0 \to K \pi) \times 2/3 )$ used in our fit.
\item[4] LHCb also reports an upper limit at $2.3\times 10^{-9}$ excluding the $\phi$ and charmonium regions.
\end{tablenotes}
\end{threeparttable}
\end{center}
\end{table}

\hflavrowdef{\hflavrowdefadm}{${\cal B}(B \to K \eta \gamma)$}{{\setlength{\tabcolsep}{0pt}
}
\begin{table}[H]
\begin{center}
\resizebox{0.75\textwidth}{!}{
\begin{threeparttable}
\caption{Branching fractions of radiative, FCNC decays with leptons and LFV/LNV decays of $B^{\pm}/B^0$ admixture (part 1).}
\label{tab:rareDecays_radll_Badmix_BR_1}
%
 \\
\hline
\hflavrowdefadm \\
 & & \\[-1ex]
\hflavrowdefadn \\
 & & \\[-1ex]
\hflavrowdefado \\
 & & \\[-1ex]
\hflavrowdefadp \\
 & & \\[-1ex]
\hflavrowdefadq \\
 & & \\[-1ex]
\hflavrowdefadr \\
 & & \\[-1ex]
\hflavrowdefads \\
 & & \\[-1ex]
\hflavrowdefadt \\
 & & \\[-1ex]
\hflavrowdefadu \\
 & & \\[-1ex]
\hflavrowdefadv \\
 & & \\[-1ex]
\hflavrowdefadw \\
\hline
\hline
\end{tabular}
\begin{tablenotes}
\item[\dag] Preliminary result.
\item[1] $m_{K\eta} < 2.4 ~\text{GeV}/c^2$.
\item[2] Measurement extrapolated to $E_{\gamma} > 1.6 ~\text{GeV}$ using the method from Ref.~\cite{Buchmuller:2005zv}.
\item[3] The systematic error includes a shape-function systematic of 0.01.
\item[4] The systematic error includes a shape-function systematic of 0.02.
\item[5] The systematic error includes a shape-function systematic of 0.04.
\item[6] The systematic error includes a shape-function systematic of 0.06.
\item[7] The PDG uncertainty includes a scale factor.
\item[8] Belle uses $m_{\ell^+\ell^-} > 0.2~\text{GeV}/c^2$, BABAR uses $m_{\ell^+\ell^-} > 0.1 ~\text{GeV}/c^2$.
\item[9] Treatment of charmonium intermediate components differs between the results.
\item[10] Multiple systematic uncertainties are added in quadrature.
\end{tablenotes}
\end{threeparttable}
}
\end{center}
\end{table}

\hflavrowdef{\hflavrowdefadx}{${\cal B}(B \to \pi \ell^+ \ell^-)$\tnote{1}\hphantom{\textsuperscript{1}}}{{\setlength{\tabcolsep}{0pt}
}
\begin{table}[H]
\begin{center}
\begin{threeparttable}
\caption{Branching fractions of radiative, FCNC decays with leptons and LFV/LNV decays of $B^{\pm}/B^0$ admixture (part 2).}
\label{tab:rareDecays_radll_Badmix_BR_2}
%
 \\
\hline
\hflavrowdefadx \\
 & & \\[-1ex]
\hflavrowdefady \\
 & & \\[-1ex]
\hflavrowdefadz \\
 & & \\[-1ex]
\hflavrowdefaea \\
 & & \\[-1ex]
\hflavrowdefaeb \\
 & & \\[-1ex]
\hflavrowdefaec \\
 & & \\[-1ex]
\hflavrowdefaed \\
 & & \\[-1ex]
\hflavrowdefaee \\
 & & \\[-1ex]
\hflavrowdefaef \\
\hline
\hline
\end{tabular}
\begin{tablenotes}
\item[\dag] Preliminary result.
\item[1] Treatment of charmonium intermediate components differs between the results.
\item[2] The PDG uncertainty includes a scale factor.
\item[3] $J/\psi$ and $\psi(2S)$ regions are vetoed.
\item[4] $m_{e^+e^-}>0.14~\text{GeV}/c^2$.
\end{tablenotes}
\end{threeparttable}
\end{center}
\end{table}

\hflavrowdef{\hflavrowdefaeg}{${\cal B}(B \to K \nu \bar{\nu})$}{{\setlength{\tabcolsep}{0pt}
}
\begin{table}[H]
\begin{center}
\begin{threeparttable}
\caption{Branching fractions of radiative, FCNC decays with leptons and LFV/LNV decays of $B^{\pm}/B^0$ admixture (part 3).}
\label{tab:rareDecays_radll_Badmix_BR_3}
%
}
\begin{table}[H]
\begin{center}
\begin{threeparttable}
\caption{Branching fractions of leptonic and radiative-leptonic $B^+$ and $B^0$ decays (part 1).}
\label{tab:rareDecays_radll_Leptonic_BR_1}
%
 \\
\hline
\hflavrowdefaeo \\
 & & \\[-1ex]
\hflavrowdefaep \\
 & & \\[-1ex]
\hflavrowdefaeq \\
 & & \\[-1ex]
\hflavrowdefaer \\
 & & \\[-1ex]
\hflavrowdefaes \\
 & & \\[-1ex]
\hflavrowdefaet \\
 & & \\[-1ex]
\hflavrowdefaeu \\
 & & \\[-1ex]
\hflavrowdefaev \\
 & & \\[-1ex]
\hflavrowdefaew \\
\hline
\hline
\end{tabular}
\begin{tablenotes}
\item[1] The PDG uncertainty includes a scale factor.
\item[2]  $E_\gamma > 1~\text{GeV}$.
\end{tablenotes}
\end{threeparttable}
\end{center}
\end{table}

\hflavrowdef{\hflavrowdefaex}{${\cal B}(B^0 \to \mu^+ \mu^- \gamma)$}{{\setlength{\tabcolsep}{0pt}
}
\begin{table}[H]
\begin{center}
\begin{threeparttable}
\caption{Branching fractions of leptonic and radiative-leptonic $B^+$ and $B^0$ decays (part 2).}
\label{tab:rareDecays_radll_Leptonic_BR_2}
%
 \\
\hline
\hflavrowdefaex \\
 & & \\[-1ex]
\hflavrowdefaey \\
 & & \\[-1ex]
\hflavrowdefaez \\
 & & \\[-1ex]
\hflavrowdefafa \\
 & & \\[-1ex]
\hflavrowdefafb \\
 & & \\[-1ex]
\hflavrowdefafc \\
 & & \\[-1ex]
\hflavrowdefafd \\
 & & \\[-1ex]
\hflavrowdefafe \\
\hline
\hline
\end{tabular}
\begin{tablenotes}
\item[*] Not included in PDG average.
\item[1] The mass windows corresponding to $\phi$ and charmonium resonances decaying to $\mu \mu$ are vetoed.
\item[2] At CL=95\,\%.
\item[3] $a$ is a promptly decaying scalar particle with a mass of 1 $\text{GeV}/c^2$ 
\item[4] PDG shows the result obtained at 95\% CL.
\item[5]  $E_\gamma > 0.5 ~\text{GeV}$.
\item[6]  $E_\gamma > 1.2 ~\text{GeV}$.
\end{tablenotes}
\end{threeparttable}
\end{center}
\end{table}

\hflavrowdef{\hflavrowdefaff}{$\frac{ {\cal{B}}(B^+ \to \pi^+ \mu^+ \mu^-) }{ {\cal{B}}(B^+ \to K^+ \mu^+ \mu^-) },~1.0 < m^2_{\ell^+\ell^-} < 6.0~\rm{GeV^2/c^4}$}{{\setlength{\tabcolsep}{0pt}
}
\begin{table}[H]
\begin{center}
\begin{threeparttable}
\caption{Relative branching fractions of $B$ radiative and FCNC decays with leptons (part 1).}
\label{tab:rareDecays_radll_RBR_1}
%
 \\
\hline
\hflavrowdefaff \\
 & & \\[-1ex]
\hflavrowdefafg \\
 & & \\[-1ex]
\hflavrowdefafh \\
 & & \\[-1ex]
\hflavrowdefafi \\
 & & \\[-1ex]
\hflavrowdefafj \\
 & & \\[-1ex]
\hflavrowdefafk \\
 & & \\[-1ex]
\hflavrowdefafl \\
 & & \\[-1ex]
\hflavrowdefafm \\
 & & \\[-1ex]
\hflavrowdefafn \\
 & & \\[-1ex]
\hflavrowdefafo \\
 & & \\[-1ex]
\hflavrowdefafp \\
\hline
\hline
\end{tabular}
\begin{tablenotes}
\item[\dag] Preliminary result.
\item[1] For the other bins see the article.
\end{tablenotes}
\end{threeparttable}
\end{center}
\end{table}

\hflavrowdefshort{\hflavrowdefafq}{$\frac{ {\cal{B}}(B \to K^* \mu^+ \mu^-) }{ {\cal{B}}(B \to K^* e^+ e^-) },~\rm{Full}~m^2_{\ell^+\ell^-}~\rm{range}$}{{\setlength{\tabcolsep}{0pt}
}
\begin{table}[H]
\begin{center}
\begin{threeparttable}
\caption{Relative branching fractions of $B$ radiative and FCNC decays with leptons (part 2).}
\label{tab:rareDecays_radll_RBR_2}
%
 \\
\hline
\hflavrowdefafq \\
 & & \\[-1ex]
\hflavrowdefafr \\
 & & \\[-1ex]
\hflavrowdefafs \\
 & & \\[-1ex]
\hflavrowdefaft \\
 & & \\[-1ex]
\hflavrowdefafu \\
 & & \\[-1ex]
\hflavrowdefafv \\
 & & \\[-1ex]
\hflavrowdefafw \\
 & & \\[-1ex]
\hflavrowdefafx \\
 & & \\[-1ex]
\hflavrowdefafy \\
 & & \\[-1ex]
\hflavrowdefafz \\
 & & \\[-1ex]
\hflavrowdefaga \\
 & & \\[-1ex]
\hflavrowdefagb \\
 & & \\[-1ex]
\hflavrowdefagc \\
\hline
\hline
\end{tabular}
\end{threeparttable}
\end{center}
\end{table}

\hflavrowdef{\hflavrowdefagd}{$\dfrac{{\cal B}(B^0 \to K^*(892)^0 \gamma)}{{\cal B}(B_s^0 \to \phi(1020) \gamma)}$}{{\setlength{\tabcolsep}{0pt}\begin{tabular}{m{6.5em}l} {LHCb \cite{LHCb:2012quo}\tnote{1}\hphantom{\textsuperscript{1}}} & { $1.23 \pm0.06 \pm0.11$ } \\ {Belle \cite{Belle:2017hum}\tnote{1}\hphantom{\textsuperscript{1}}} & { $1.10 \pm0.16 \pm0.20$ } \\ \end{tabular}}}{\begin{tabular}{l} $1.21 \pm0.11$\\ \end{tabular}}
\hflavrowdef{\hflavrowdefage}{$\dfrac{{\cal B}(B^0 \to \mu^+ \mu^-)}{{\cal B}(B_s^0 \to \mu^+ \mu^-)}$ [$10^{-2}$]}{{\setlength{\tabcolsep}{0pt}\begin{tabular}{m{6.5em}l} {LHCb \cite{LHCb:2021awg2}\tnote{}\hphantom{\textsuperscript{}}} & { $< 8.1$ } \\ \end{tabular}}}{\begin{tabular}{l} $< 8.1$\\ \end{tabular}}
\hflavrowdef{\hflavrowdefagf}{$\dfrac{{\cal B}(B^0 \to \phi(1020) \mu^+ \mu^-)}{{\cal B}(B_s^0 \to \phi(1020) \mu^+ \mu^-)}$ [$10^{-3}$]}{{\setlength{\tabcolsep}{0pt}\begin{tabular}{m{6.5em}l} {LHCb \cite{LHCb:2022sjo}\tnote{2}\hphantom{\textsuperscript{2}}} & { $< 4.4$ } \\ \end{tabular}}}{\begin{tabular}{l} $< 4.4$\\ \end{tabular}}
\hflavrowdef{\hflavrowdefagg}{$\frac{{\cal{B}}( B^{0} \to \pi^+ \pi^- \mu^+ \mu^-)}{{\cal{B}}( B^{0} \to J/\psi K^{*0} ) \times {\cal{B}}( J/\psi \to \mu^+ \mu^- ) \times {\cal{B}}( K^{*0} \to K^+ \pi^- )}$ [$10^{-4}$]}{{\setlength{\tabcolsep}{0pt}\begin{tabular}{m{6.5em}l} {LHCb \cite{LHCb:2014yov}\tnote{3,4}\hphantom{\textsuperscript{3,4}}} & { $4.1 \pm1.0 \pm0.3$ } \\ \end{tabular}}}{\begin{tabular}{l} $4.1 \pm1.0$\\ \end{tabular}}
\begin{table}[H]
\begin{center}
\begin{threeparttable}
\caption{Relative branching fractions of $B$ radiative and FCNC decays with leptons (part 3).}
\label{tab:rareDecays_radll_RBR_3}
\begin{tabular}{ Sl l l }
\hline
\hline
\textbf{Parameter} & \textbf{Measurements} & \begin{tabular}{l}\textbf{Average}\end{tabular} \\
\hline
\hflavrowdefagd \\
 & & \\[-1ex]
\hflavrowdefage \\
 & & \\[-1ex]
\hflavrowdefagf \\
 & & \\[-1ex]
\hflavrowdefagg \\
\hline
\hline
\end{tabular}
\begin{tablenotes}
\item[1] Multiple systematic uncertainties are added in quadrature.
\item[2] $\phi$ and charmonium regions excluded from the dimuon spectrum.
\item[3] The mass windows corresponding to $\phi$ and charmonium resonances decaying to $\mu \mu$ are vetoed.
\item[4] $0.5 < m_{\pi^+\pi^-} <  1.3~\rm{GeV/}c^2$.
\end{tablenotes}
\end{threeparttable}
\end{center}
\end{table}

\hflavrowdef{\hflavrowdefagh}{${\cal{B}}( B \to \eta X )$}{{\setlength{\tabcolsep}{0pt}\begin{tabular}{m{6.5em}l} {Belle \cite{Belle:2009udg}\tnote{1}\hphantom{\textsuperscript{1}}} & { $2.610 \pm0.300\,^{+0.440}_{-0.740}$ } \\ {CLEO \cite{CLEO:1998noh}\tnote{2}\hphantom{\textsuperscript{2}}} & { $< 4.400$ } \\ \end{tabular}}}{\begin{tabular}{l} $2.61\,^{+0.53}_{-0.80}$\\ \end{tabular}}
\hflavrowdef{\hflavrowdefagi}{${\cal{B}}( B \to \eta^{\prime} X )$}{{\setlength{\tabcolsep}{0pt}\begin{tabular}{m{6.5em}l} {BaBar \cite{BaBar:2004ntz}\tnote{3}\hphantom{\textsuperscript{3}}} & { $3.90 \pm0.80 \pm0.90$ } \\ {CLEO \cite{CLEO:2003iqk}\tnote{3}\hphantom{\textsuperscript{3}}} & { $4.60 \pm1.10 \pm0.60$ } \\ \end{tabular}}}{\begin{tabular}{l} $4.24 \pm0.87$\\ \end{tabular}}
\hflavrowdef{\hflavrowdefagj}{${\cal{B}}( B \to K^+ X )$}{{\setlength{\tabcolsep}{0pt}\begin{tabular}{m{6.5em}l} {BaBar \cite{BaBar:2010yju}\tnote{4}\hphantom{\textsuperscript{4}}} & { $< 1.87$ } \\ \end{tabular}}}{\begin{tabular}{l} $< 1.9$\\ \end{tabular}}
\hflavrowdef{\hflavrowdefagk}{${\cal{B}}( B \to K^0 X )$}{{\setlength{\tabcolsep}{0pt}\begin{tabular}{m{6.5em}l} {BaBar \cite{BaBar:2010yju}\tnote{4}\hphantom{\textsuperscript{4}}} & { $1.95\,^{+0.51}_{-0.45} \pm0.50$ } \\ \end{tabular}}}{\begin{tabular}{l} $1.95 \pm0.69$ \\ $\mathit{\scriptstyle 1.95\,^{+0.71}_{-0.67}}$\\ \end{tabular}}
\hflavrowdef{\hflavrowdefagl}{${\cal{B}}( B \to \pi^+ X )$}{{\setlength{\tabcolsep}{0pt}\begin{tabular}{m{6.5em}l} {BaBar \cite{BaBar:2010yju}\tnote{5}\hphantom{\textsuperscript{5}}} & { $3.72\,^{+0.50}_{-0.47} \pm0.59$ } \\ \end{tabular}}}{\begin{tabular}{l} $3.72 \pm0.76$ \\ $\mathit{\scriptstyle 3.72\,^{+0.77}_{-0.75}}$\\ \end{tabular}}
\begin{table}[H]
\begin{center}
\begin{threeparttable}
\caption{Branching fractions of $B^+$/$B^0 \to q\overline{q}^{\prime}$ gluon decays.}
\label{tab:rareDecays_radll_Bincl_0}
\begin{tabular}{ Sl l l }
\hline
\hline
\textbf{Parameter [$10^{-4}$]} & \textbf{Measurements} & \begin{tabular}{l}\textbf{Average $^\mathrm{HFLAV}_{\scriptscriptstyle PDG}$}\end{tabular} \\
\hline
\hflavrowdefagh \\
 & & \\[-1ex]
\hflavrowdefagi \\
 & & \\[-1ex]
\hflavrowdefagj \\
 & & \\[-1ex]
\hflavrowdefagk \\
 & & \\[-1ex]
\hflavrowdefagl \\
\hline
\hline
\end{tabular}
\begin{tablenotes}
\item[1] $0.4<m_X<2.6~\text{GeV}/c^2$.
\item[2]  $ 2.1 < p_\eta < 2.7 ~\text{GeV}/c$. 
\item[3] $2.0 < p^*(\eta^\prime) < 2.7~\text{GeV}/c$.
\item[4] $p^*(K) < 2.34~\text{GeV}/c$.
\item[5] $p^*(\pi^+)<2.36~\text{GeV}/c$.
\end{tablenotes}
\end{threeparttable}
\end{center}
\end{table}

\hflavrowdef{\hflavrowdefagm}{$\Delta_{0^-}(B \to X_s\gamma)$}{{\setlength{\tabcolsep}{0pt}
}
\begin{table}[H]
\begin{center}
\begin{threeparttable}
\caption{Isospin asymmetry in $B$ mesons radiative and FCNC decays with leptons. In some of the $B$-factory results it is assumed that $\mathcal{B}(\Upsilon(4S) \to B^+ B^-) = \mathcal{B}(\Upsilon(4S) \to B^0 \overline{B}{}^0)$, and in others a measured value of the ratio of branching fractions is used. See original papers for details. The averages quoted here are computed naively and should be treated with caution.}
\label{tab:rareDecays_radll_IA_0}
%
 \\
\hline
\hflavrowdefagm \\
 & & \\[-1ex]
\hflavrowdefagn \\
 & & \\[-1ex]
\hflavrowdefago \\
 & & \\[-1ex]
\hflavrowdefagp \\
 & & \\[-1ex]
\hflavrowdefagq \\
 & & \\[-1ex]
\hflavrowdefagr \\
 & & \\[-1ex]
\hflavrowdefags \\
\hline
\hline
\end{tabular}
\begin{tablenotes}
\item[1] $m_{X_s} < 2.8 ~\text{GeV}/c^2$.
\item[2] Multiple systematic uncertainties are added in quadrature.
\item[3] $E_\gamma>2.2~\text{GeV}$.
\item[4] The PDG uncertainty includes a scale factor.
\item[5] Only muons are used, $1.1<m^2_{\ell^+\ell^-}<6.0~\rm{GeV^2/c^4}$.
\item[6] $1.0<m^2_{\ell^+\ell^-}<6.0~\rm{GeV^2/c^4}.$
\item[7] $m^2_{\ell^+\ell^-}<8.68~\rm{GeV^2/c^4}.$
\item[8] $0.1<m^2_{\ell^+\ell^-}<7.02~\rm{GeV^2/c^4}.$
\end{tablenotes}
\end{threeparttable}
\end{center}
\end{table}

\hflavrowdef{\hflavrowdefagt}{${\cal B}(B^+ \to \pi^+ e^+ \mu^- \mathrm{+c.c.})$}{{\setlength{\tabcolsep}{0pt}
}
\begin{table}[H]
\begin{center}
\begin{threeparttable}
\caption{Branching fractions of semileptonic $B^+$ decays to LFV and LNV final states (part 1).}
\label{tab:rareDecays_radll_LFN_LFV_BR_1}
%
 \\
\hline
\hflavrowdefagt \\
 & & \\[-1ex]
\hflavrowdefagu \\
 & & \\[-1ex]
\hflavrowdefagv \\
 & & \\[-1ex]
\hflavrowdefagw \\
 & & \\[-1ex]
\hflavrowdefagx \\
 & & \\[-1ex]
\hflavrowdefagy \\
 & & \\[-1ex]
\hflavrowdefagz \\
 & & \\[-1ex]
\hflavrowdefaha \\
 & & \\[-1ex]
\hflavrowdefahb \\
 & & \\[-1ex]
\hflavrowdefahc \\
 & & \\[-1ex]
\hflavrowdefahd \\
 & & \\[-1ex]
\hflavrowdefahe \\
 & & \\[-1ex]
\hflavrowdefahf \\
 & & \\[-1ex]
\hflavrowdefahg \\
 & & \\[-1ex]
\hflavrowdefahh \\
 & & \\[-1ex]
\hflavrowdefahi \\
\hline
\hline
\end{tabular}
\begin{tablenotes}
\item[*] Not included in PDG average.
\end{tablenotes}
\end{threeparttable}
\end{center}
\end{table}

\hflavrowdef{\hflavrowdefahj}{${\cal B}(B^+ \to K^*(892)^+ e^+ \mu^-)$}{{\setlength{\tabcolsep}{0pt}
}
\begin{table}[H]
\begin{center}
\begin{threeparttable}
\caption{Branching fractions of semileptonic $B^+$ decays to LFV and LNV final states (part 2).}
\label{tab:rareDecays_radll_LFN_LFV_BR_2}
%
}
\begin{table}[H]
\begin{center}
\begin{threeparttable}
\caption{Branching fractions of semileptonic $B^+$ decays to LFV and LNV final states (part 3).}
\label{tab:rareDecays_radll_LFN_LFV_BR_3}
%
 \\
\hline
\hflavrowdefahs \\
 & & \\[-1ex]
\hflavrowdefaht \\
 & & \\[-1ex]
\hflavrowdefahu \\
 & & \\[-1ex]
\hflavrowdefahv \\
 & & \\[-1ex]
\hflavrowdefahw \\
 & & \\[-1ex]
\hflavrowdefahx \\
 & & \\[-1ex]
\hflavrowdefahy \\
 & & \\[-1ex]
\hflavrowdefahz \\
 & & \\[-1ex]
\hflavrowdefaia \\
 & & \\[-1ex]
\hflavrowdefaib \\
 & & \\[-1ex]
\hflavrowdefaic \\
 & & \\[-1ex]
\hflavrowdefaid \\
 & & \\[-1ex]
\hflavrowdefaie \\
 & & \\[-1ex]
\hflavrowdefaif \\
 & & \\[-1ex]
\hflavrowdefaig \\
 & & \\[-1ex]
\hflavrowdefaih \\
\hline
\hline
\end{tabular}
\begin{tablenotes}
\item[1] At CL=95\,\%.
\end{tablenotes}
\end{threeparttable}
\end{center}
\end{table}

\hflavrowdef{\hflavrowdefaii}{${\cal B}(B^0 \to K^*(892)^0 e^- \mu^+)$}{{\setlength{\tabcolsep}{0pt}
}
\begin{table}[H]
\begin{center}
\begin{threeparttable}
\caption{Branching fractions of semileptonic $B^0$ decays to LFV and LNV final states.}
\label{tab:rareDecays_radll_LFN_LFV_BR_4}
%
 \\
\hline
\hflavrowdefaii \\
 & & \\[-1ex]
\hflavrowdefaij \\
 & & \\[-1ex]
\hflavrowdefaik \\
 & & \\[-1ex]
\hflavrowdefail \\
 & & \\[-1ex]
\hflavrowdefaim \\
 & & \\[-1ex]
\hflavrowdefain \\
 & & \\[-1ex]
\hflavrowdefaio \\
 & & \\[-1ex]
\hflavrowdefaip \\
 & & \\[-1ex]
\hflavrowdefaiq \\
 & & \\[-1ex]
\hflavrowdefair \\
 & & \\[-1ex]
\hflavrowdefais \\
\hline
\hline
\end{tabular}
\begin{tablenotes}
\item[\dag] Preliminary result.
\item[*] Not included in PDG average.
\item[1] PDG shows the result obtained at 95\% CL.
\end{tablenotes}
\end{threeparttable}
\end{center}
\end{table}

\clearpage
\noindent Measurements that are not included in the tables (the definitions of observables can be found in the corresponding experimental papers):
\begin{itemize}
\item In Ref. \cite{LHCb:2014vnw}, LHCb reports the up-down asymmetries in bins of the $K\pi\pi\gamma$ mass of the $B^+ \to K^+ \pi^- \pi^+ \gamma$ decay.
\item For the $B \to K \ell^- \ell^+$ channel, LHCb measures $F_H$ and $A_{\rm FB}$ in 17 (5) bins of $m^2(\ell^+ \ell^-)$  for the $K^+$ (\ks) final state \cite{LHCb:2014auh}.
Belle  measures $F_L$ and $A_{\rm FB}$ in 6 $m^2(\ell^+ \ell^-)$ \cite{Belle:2009zue}.%
\item For the  $B \to K^{*} \ell^- \ell^+$ analyses, partial branching fractions and angular observables in bins of $m^2(\ell^+ \ell^-)$ are also available:
\begin{itemize}
\item   $B^0 \to K^{*0} e^- e^+$ : LHCb reports $F_L$, $A_T^{(2)}$, $A_T^{\rm Im}$, $A_T^{\rm Re}$ in the $[0.0008, 0.257]~\gevgevcccc$ bin of $m^2(\ell^+ \ell^-)$ putting constraints on the $B\to K^{*0} \gamma$ photon polarization \cite{LHCb:2020dof}. In Ref. \cite{LHCb:2013pra}, LHCb  determines the branching fraction in the dilepton mass region $[0.0009,1.0]~\gevgevcccc$.
\item   $B \to K^{*} \ell^- \ell^+$ : Belle  measures $F_L$, $A_{\rm FB}$, isospin asymmetry in 6 $m^2(\ell^+ \ell^-)$ bins \cite{Belle:2009zue} and $P_4'$, $P_5'$, $P_6'$, $P_8'$ in 4 $m^2(\ell^+ \ell^-)$ bins \cite{Belle:2016xuo}. In a more recent paper~\cite{Belle:2016fev}, they report measurements of $P_4'$ and $P_5'$, separately for $\ell=\mu$ or $e$, in 4 $m^2(\ell^+ \ell^-)$ bins and in the region $[1, 6]~\gevgevcccc$. The measurements use both $\Bz$ and $\Bp$ decays. They also measure the LFU observables $Q_i = P_i^{\mu}-P_i^e$, for $i=4,5$.
\babar\ reports $F_L$, $A_{\rm FB}$, $P_2$ in 5 $m^2(\ell^+ \ell^-)$ bins \cite{BaBar:2015wkg}.
\item   $B^0 \to K^{*0} \mu^- \mu^+$ : LHCb  measures $F_L$, $A_{\rm FB}$, $S_3-S_9$, $A_3-A_9$, $P_1-P_3$, $P_4'-P_8'$ in 8 $m^2(\ell^+ \ell^-)$ bins \cite{LHCb:2015svh}. An updated measurement of the \CP-averaged observables  is presented in Ref. \cite{LHCb:2020lmf}.
CMS measures $F_L$ and $A_{\rm FB}$ in 7 $m^2(\ell^+ \ell^-)$ bins \cite{CMS:2015bcy}, as well as $P_1,P_5'$  \cite{CMS:2017rzx}. ATLAS measures $F_L$, $S_{3,4,5,7,8}$ and $P_{1,4,5,6,8}'$ in  6 $m^2(\ell^+ \ell^-)$ bins \cite{ATLAS:2018gqc}.
\item $B^+ \to K^{*+} \mu^- \mu^+$: LHCb reports the full set of \CP-averaged angular observables in 8 $m^2(\ell^+ \ell^-)$ bins \cite{LHCb:2020gog}. CMS measures $F_L$ and $A_{\rm FB}$ in 3 $m^2(\ell^+ \ell^-)$ bins \cite{CMS:2020oqb}.
\end{itemize}
\item $B \to X_s \ell^- \ell^+$ (where $X_s$ is a hadronic system with an $s$ quark): Belle  measures $A_{\rm FB}$ in bins of $m^2(\ell^+ \ell^-)$ with a sum of 10 exclusive final states \cite{Belle:2014owz}.
\item $B^0 \to K^+ \pi^- \mu^+ \mu^-$, with $1330 < m(K^+ \pi^-) < 1530~\gevcc$: LHCb measures the partial branching fraction in bins of  $m^2(\mu^+ \mu^-)$ in the range $[0.1,8.0]~\gevgevcccc$, and reports angular moments \cite{LHCb:2016eyu}.
\item In Ref.~\cite{LHCb:2016due}, LHCb  measures the phase difference between the short- and long-distance contributions to the $\Bp \to K^+\mu^+\mu^-$ decay. The measurement is based on the analysis of the dimuon mass distribution in the regions of the $J/\psi$ and $\psi(2S)$ resonances and far from their poles, to probe long and short distance effects, respectively.
\item In Ref.~\cite{CMS:2018qih}, CMS performs the study of the angular distribution of the $\Bp \to K^+\mu^+\mu^-$ channel and measures, in 7 $m^2(\mu^+ \mu^-)$ bins, $A_{\rm FB}$ and the contribution $F_{\rm H}$ from the pseudoscalar, scalar and tensor amplitudes to the decay.
\item In Ref.~\cite{LHCb:2015nkv}, LHCb performs a search for a hidden-sector boson  $\chi$ decaying into two muons in $B^0 \to K^{*0} \mu^+\mu^-$ decays. Results are given as function of mass and lifetime in the range  $214<m(\chi)<4350\ \mevcc$and $0<\tau(\chi)<1000$~ps.
\item In Ref.~\cite{LHCb:2016awg}, LHCb  performs a search for a hypothetical new scalar particle $\chi$, assumed to have a narrow width, through the decay $\Bp \to K^+ \chi(\mu^+\mu^-)$ in the ranges of mass $250<m(\chi)<4700\ \mevcc$ and lifetime $0.1<\tau(\chi)<1000$~ps. Upper limits are given as a function of $m(\chi)$ and $\tau(\chi)$. %
\item In Ref.~\cite{LHCb:2023ptw} LHCb reports the differential branching fraction of $\Lambda^0_b \to \Lambda(1520)\mu^+\mu^-$ in 5 $m(\mu^+\mu^-)$ intervals in the range $[0.1,17]~\gevgevcccc$. It also reports the branching fraction in the interval $1.1<m^2(\mu^+ \mu^-)<6~\gevgevcccc$.
\item In Ref.~\cite{Belle-II:2023ueh}, Belle-II performs a search for long-lived spin-0 particles ($S$) in $B$-meson decays mediated by a $b \to s$ quark transition. They set model-independent upper limits, at the level of $10^{-7}$, on the products of branching fractions ${\cal B}(B^0 \to K^*(892)^0S) \times {\cal B}(S \to x^+x^-$) and ${\cal B}(B^+ \to K^+S) \times {\cal B}(S \to x^+x^-)$, where $x^+x^-$ indicates $e^+e^-$, $\mu^+\mu^-$, $\pi^+\pi^-$, or $K^+K^-$, as functions of the $S$ mass and lifetime.
\item We do not perform the average of the branching fraction of $B\to\mu^+\mu^-$ decays, which is taken care of by the LHC Heavy Flavour Working Group~\cite{LHC_HF_WG}, taking into account the correlations between the $\Bz \to \mu^+\mu^-$ and $\Bs \to \mu^+\mu^-$ branching fractions. The latest results from ATLAS, CMS and LHCb are in Refs.~\cite{ATLAS:2018cur,CMS:2022mgd,LHCb:2021awg2}, respectively.
\end{itemize}

\begin{figure}[htbp!]
\centering
\includegraphics[width=0.8\textwidth]{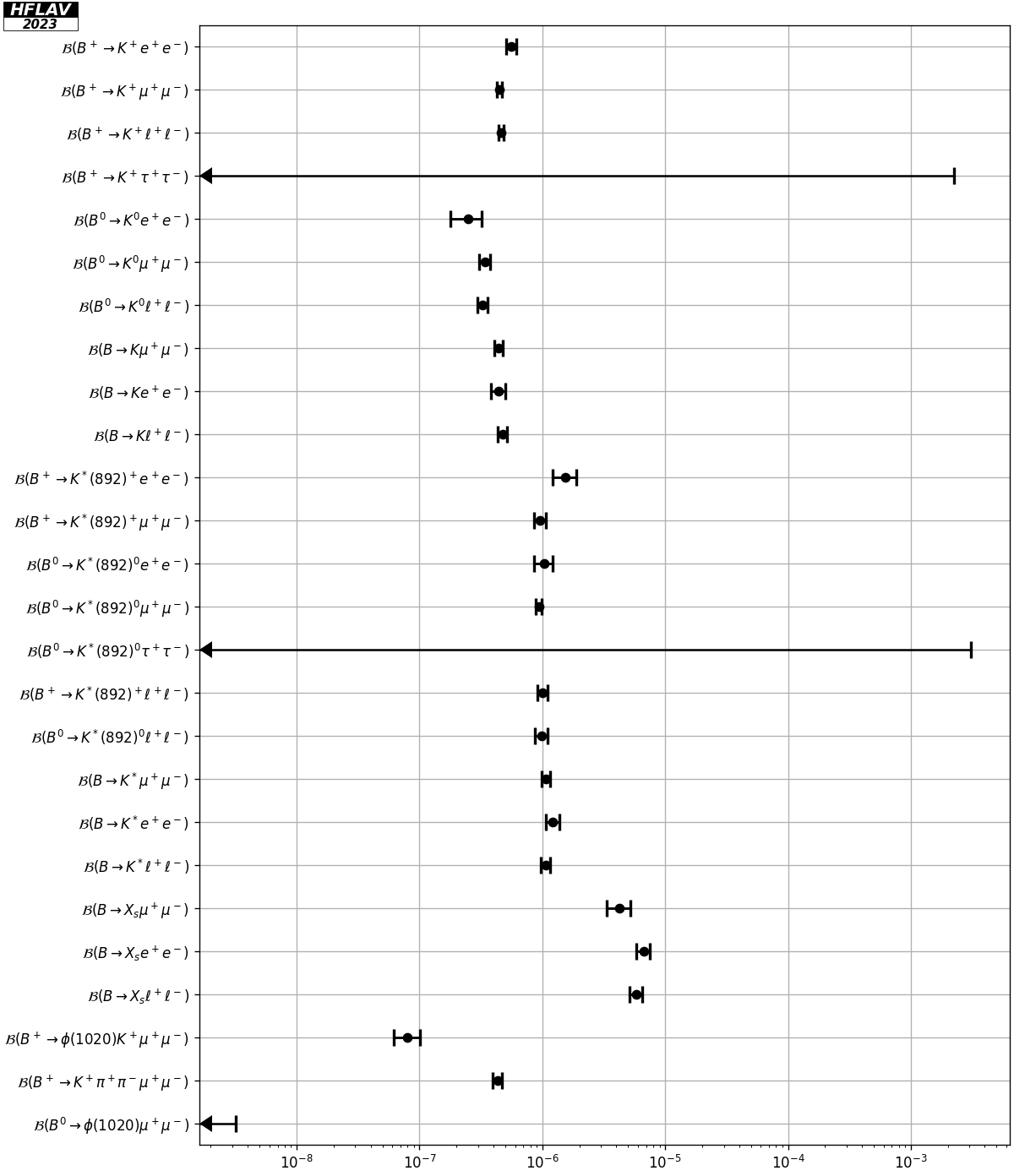}
\caption{Branching fractions of $B^+$ and $B^0$ decays of the type $b\to s\ell^{+}\ell^{-}$.}
\label{fig:rare-btosll}
\end{figure}

\begin{figure}[htbp!]
\centering
\includegraphics[width=0.8\textwidth]{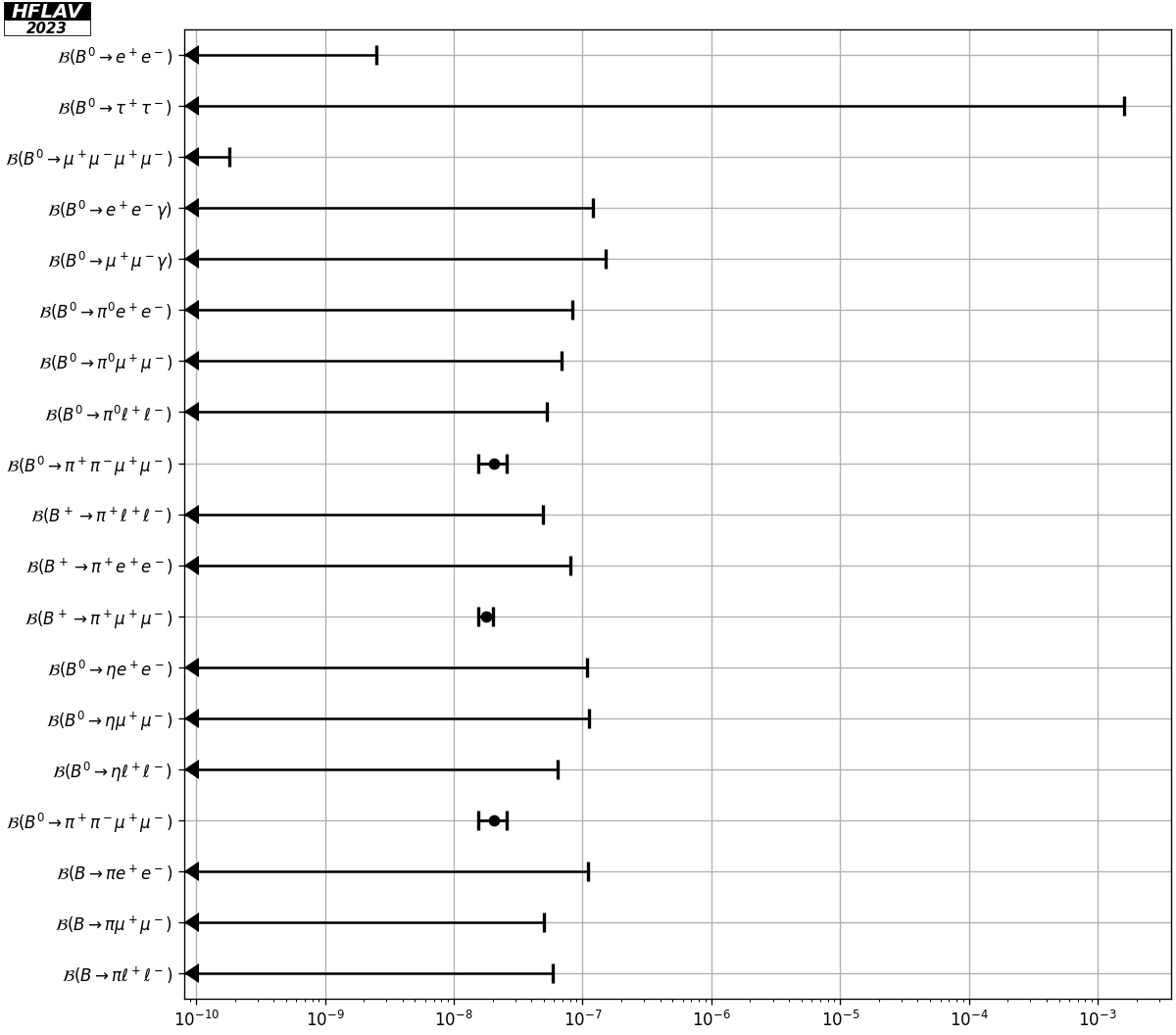}
\caption{Branching fractions of $B^+$ and $B^0$ decays of the type $b\to u \ell^{+}\ell^{-}$, purely leptonic and leptonic radiative.}
\label{fig:rare-btodllg}
\end{figure}

\begin{figure}[htbp!]
\centering
\includegraphics[width=1.0\textwidth]{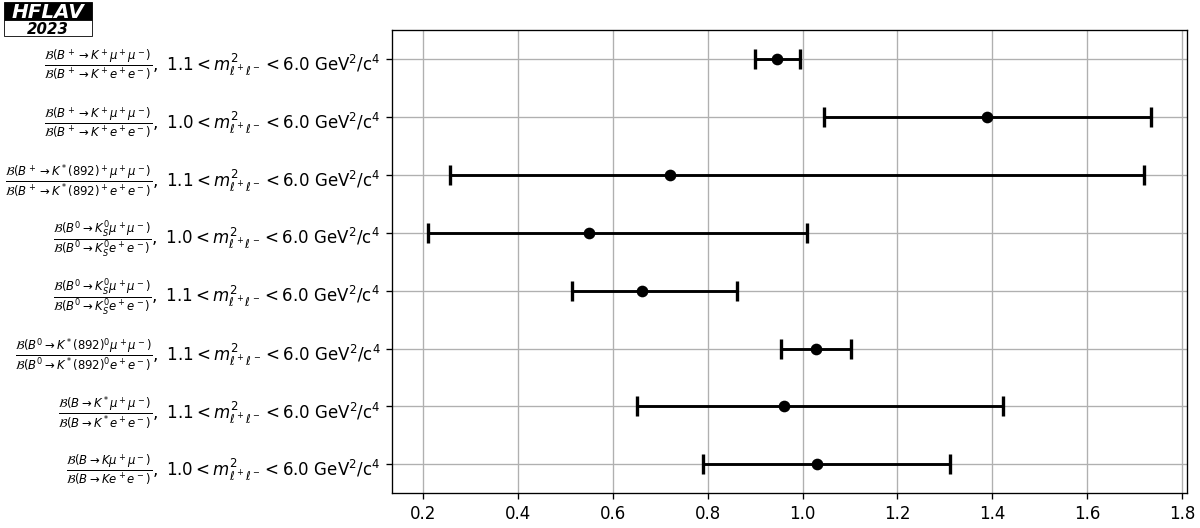}
\caption{Compilation of $R_K^{(\ast)}$ ratios in the low dilepton invariant-mass region. These are ratios between branching fractions of $B$-meson decays to $K^{(\ast)}\mu^+\mu^-$ and $K^{(\ast)}e^+e^-$, which  provide information on lepton universality.}     
\label{fig:rare-RK_RKst}
\end{figure}

\begin{figure}[htbp!]
\centering
\includegraphics[width=0.8\textwidth]{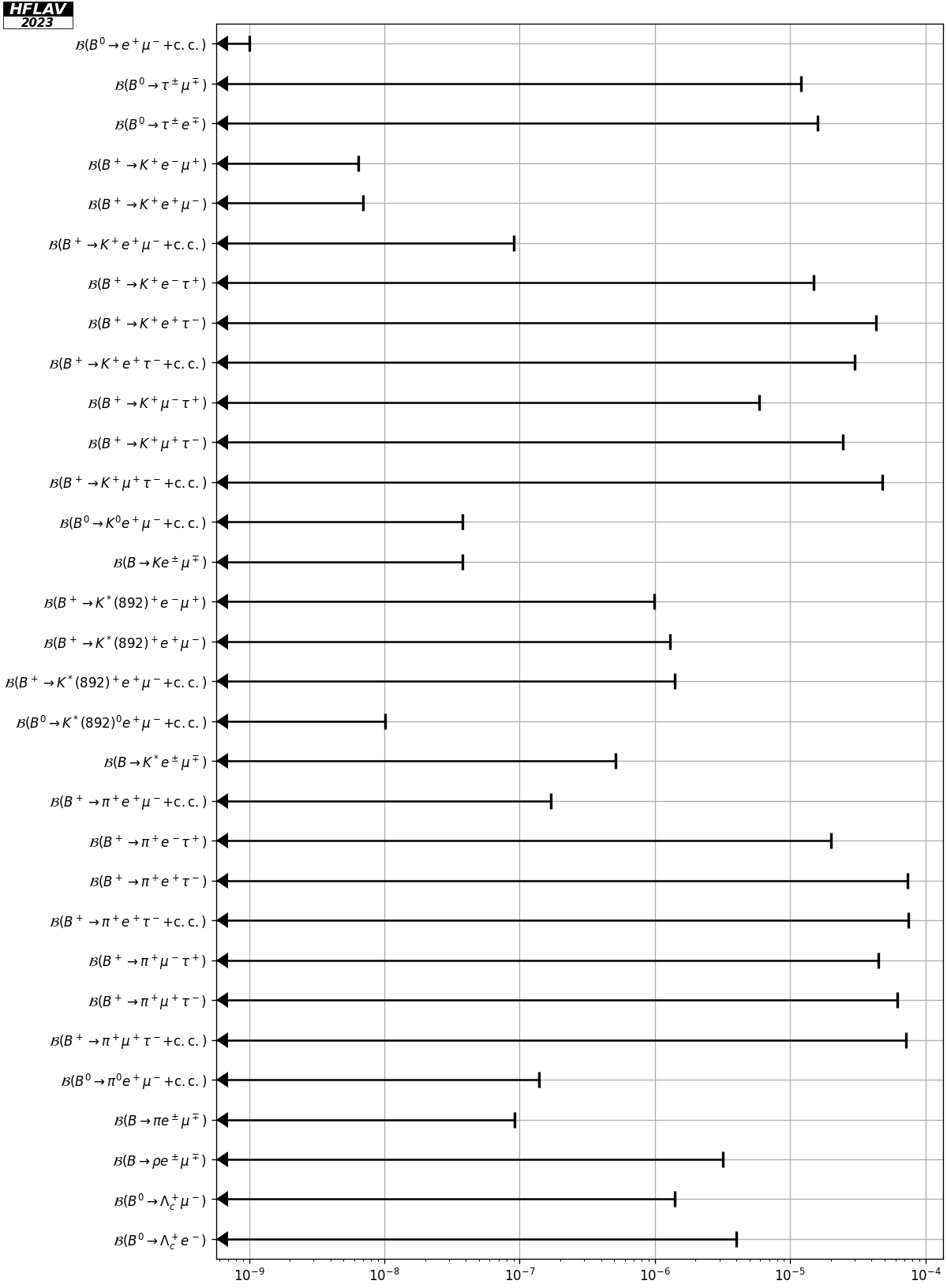}
\caption{Limits on branching fractions of lepton-flavour-violating $B^+$ and $B^0$ decays.}
\label{fig:rare-leptonflavourviol}
\end{figure}

\begin{figure}[htbp!]
\centering
\includegraphics[width=0.8\textwidth]{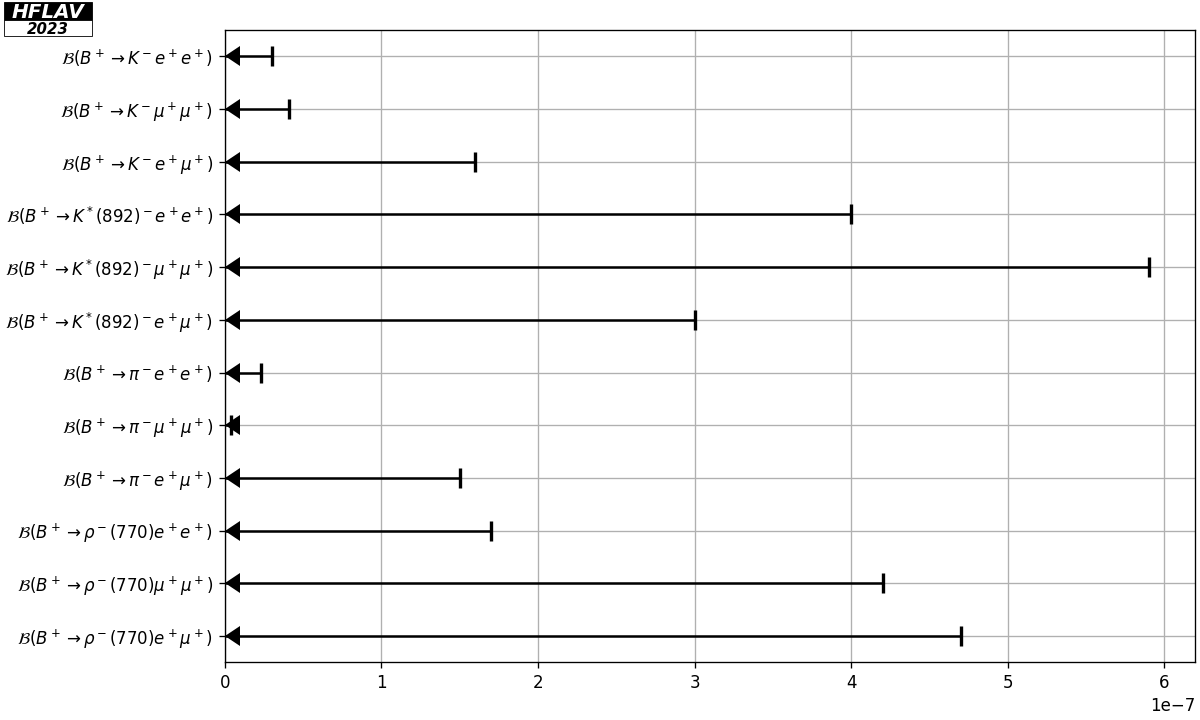}
\caption{Limits on branching fractions of lepton-number-violating $B^+$ and $B^0$ decays.}
\label{fig:rare-leptonnumberviol}
\end{figure}

\clearpage

\begin{figure}[htbp!]
\centering
\includegraphics[width=0.8\textwidth]{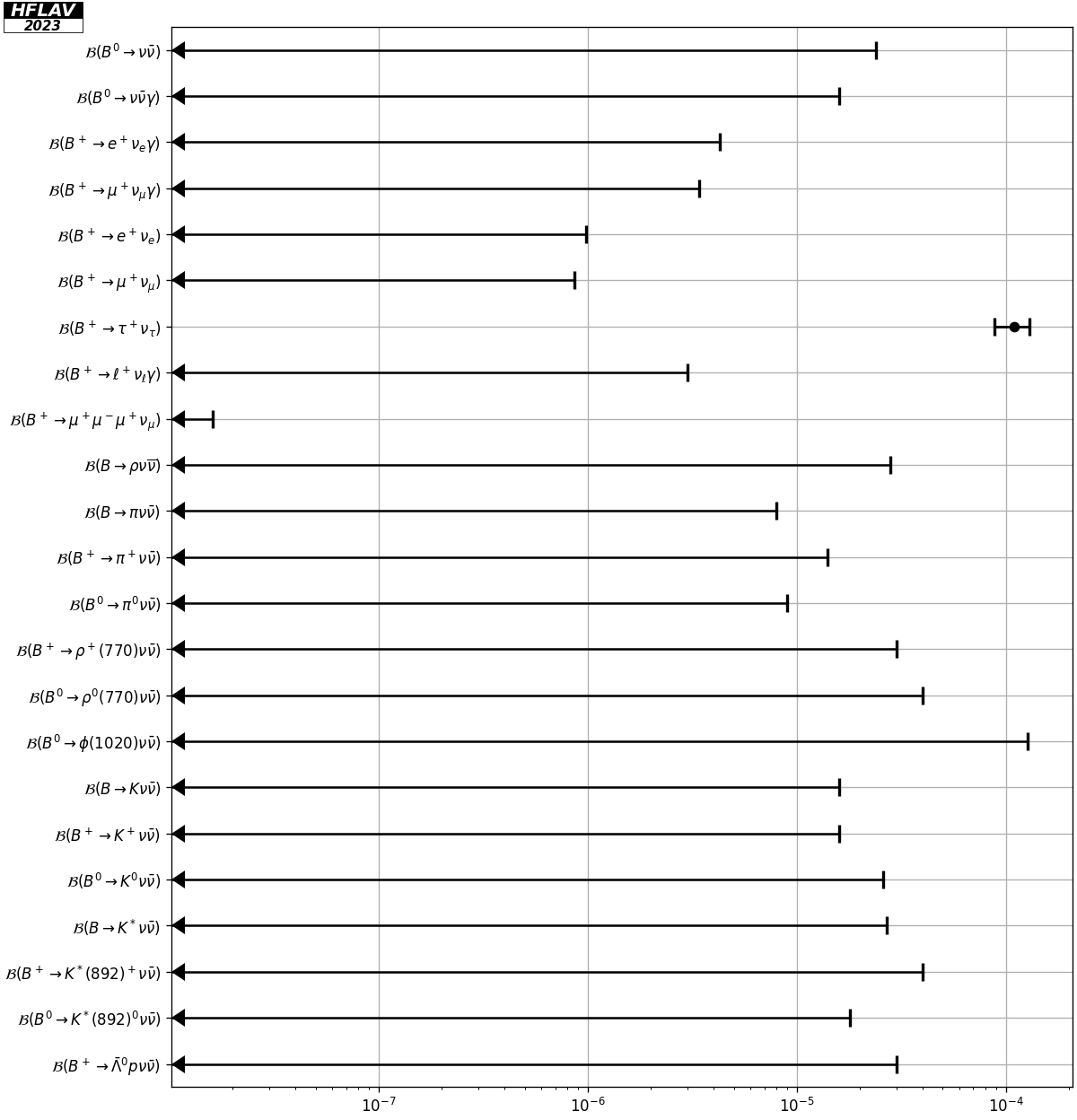}
\caption{Branching fractions of charmless $B$ decays with neutrinos.}
\label{fig:rare-neutrino}
\end{figure}

\mysubsection{\CP asymmetries in \b-hadron decays}
\label{sec:rare-acp}

This section contains, in Tables~\ref{tab:rareDecays_ACP_Bu_1} to~\ref{tab:rareDecays_ACP_Other_Xib},
compilations of \CP\ asymmetries in decays of various \b-hadrons: \Bp, \Bz
mesons, $\Bpm/\Bz$ admixtures, \Bs mesons and finally \Lb baryons.
The \CP asymmetry is defined as
\begin{equation}
	A_{\CP} = \frac{N_b - N_{\bbar}}{N_b + N_{\bbar}},
\end{equation}
where $N_b$ ($N_{\bbar}$) is the number of hadrons containing a $b$ ($\bbar$) quark
decaying into a specific final state (the \CP-conjugate state).
Figure~\ref{fig:rare-acpselect} shows a graphic representation of a selection of results given in this section.

\newpage

\hflavrowdef{\hflavrowdefaaa}{$A_\mathrm{CP}(B^+ \to K^0_S \pi^+)$}{{\setlength{\tabcolsep}{0pt}
}
\begin{table}[H]
\begin{center}
\begin{threeparttable}
\caption{$C\!P$ asymmetries of charmless hadronic $B^+$ decays (part 1).}
\label{tab:rareDecays_ACP_Bu_1}
%
 \\
\hline
\hflavrowdefaaa \\
 & & \\[-1ex]
\hflavrowdefaab \\
 & & \\[-1ex]
\hflavrowdefaac \\
 & & \\[-1ex]
\hflavrowdefaad \\
 & & \\[-1ex]
\hflavrowdefaae \\
 & & \\[-1ex]
\hflavrowdefaaf \\
 & & \\[-1ex]
\hflavrowdefaag \\
 & & \\[-1ex]
\hflavrowdefaah \\
 & & \\[-1ex]
\hflavrowdefaai \\
 & & \\[-1ex]
\hflavrowdefaaj \\
\hline
\hline
\end{tabular}
\begin{tablenotes}
\item[\dag] Preliminary result.
\item[1] Multiple systematic uncertainties are added in quadrature.
\end{tablenotes}
\end{threeparttable}
\end{center}
\end{table}

\hflavrowdef{\hflavrowdefaak}{$A_\mathrm{CP}(B^+ \to \omega(782) K^+)$}{{\setlength{\tabcolsep}{0pt}
}
\begin{table}[H]
\begin{center}
\begin{threeparttable}
\caption{$C\!P$ asymmetries of charmless hadronic $B^+$ decays (part 2).}
\label{tab:rareDecays_ACP_Bu_2}
%
 \\
\hline
\hflavrowdefaak \\
 & & \\[-1ex]
\hflavrowdefaal \\
 & & \\[-1ex]
\hflavrowdefaam \\
 & & \\[-1ex]
\hflavrowdefaan \\
 & & \\[-1ex]
\hflavrowdefaao \\
 & & \\[-1ex]
\hflavrowdefaap \\
 & & \\[-1ex]
\hflavrowdefaaq \\
 & & \\[-1ex]
\hflavrowdefaar \\
\hline
\hline
\end{tabular}
\begin{tablenotes}
\item[\dag] Preliminary result.
\item[1] Result extracted from Dalitz-plot analysis of $B^+ \to K^+ \pi^+ \pi^-$ decays.
\item[2] Multiple systematic uncertainties are added in quadrature.
\item[3] Result extracted from Dalitz-plot analysis of $B^+ \to K_S^0 \pi^+ \pi^0$ decays.
\item[4] Treatment of charmonium intermediate components differs between the results.
\item[5] Using run II dataset, corresponding to an integrated luminosity of $5.9 ~\mathrm{fb}^{-1}$ collected at a center-of-mass energy of 13 TeV (2015 to 2018).
\item[6] Also measured in several invariant mass regions.
\item[7] Using run I dataset, corresponding to an integrated luminosity of $3.0 ~\mathrm{fb}^{-1}$ collected at a center-of-mass energy of 7 TeV (2011) and 8 TeV (2012).
\item[8] The nonresonant amplitude is modelled using a polynomial function including S-wave and P-wave terms.
\end{tablenotes}
\end{threeparttable}
\end{center}
\end{table}

\hflavrowdef{\hflavrowdefaas}{$A_\mathrm{CP}(B^+ \to f_0(980) K^+)$}{{\setlength{\tabcolsep}{0pt}
}
\begin{table}[H]
\begin{center}
\begin{threeparttable}
\caption{$C\!P$ asymmetries of charmless hadronic $B^+$ decays (part 3).}
\label{tab:rareDecays_ACP_Bu_3}
%
 \\
\hline
\hflavrowdefaas \\
 & & \\[-1ex]
\hflavrowdefaat \\
 & & \\[-1ex]
\hflavrowdefaau \\
 & & \\[-1ex]
\hflavrowdefaav \\
 & & \\[-1ex]
\hflavrowdefaaw \\
 & & \\[-1ex]
\hflavrowdefaax \\
 & & \\[-1ex]
\hflavrowdefaay \\
 & & \\[-1ex]
\hflavrowdefaaz \\
 & & \\[-1ex]
\hflavrowdefaba \\
 & & \\[-1ex]
\hflavrowdefabb \\
 & & \\[-1ex]
\hflavrowdefabc \\
 & & \\[-1ex]
\hflavrowdefabd \\
 & & \\[-1ex]
\hflavrowdefabe \\
 & & \\[-1ex]
\hflavrowdefabf \\
 & & \\[-1ex]
\hflavrowdefabg \\
\hline
\hline
\end{tabular}
\begin{tablenotes}
\item[1] Result extracted from Dalitz-plot analysis of $B^+ \to K^+ \pi^+ \pi^-$ decays.
\item[2] Multiple systematic uncertainties are added in quadrature.
\item[3] Result extracted from Dalitz-plot analysis of $B^+ \to K^+ K^+ K^-$ decays.
\item[4] Result extracted from Dalitz-plot analysis of $B^+ \to K_S^0 \pi^+ \pi^0$ decays.
\item[5] $X_{\pi^0 \pi^0}$ corresponds to a structure observed in Ref.~\cite{Belle:2022dgi}, likely arising due to multiple resonances.
\end{tablenotes}
\end{threeparttable}
\end{center}
\end{table}

\hflavrowdef{\hflavrowdefabh}{$A_\mathrm{CP}(B^+ \to a_1(1260)^+ K^0)$}{{\setlength{\tabcolsep}{0pt}
}
\begin{table}[H]
\begin{center}
\begin{threeparttable}
\caption{$C\!P$ asymmetries of charmless hadronic $B^+$ decays (part 4).}
\label{tab:rareDecays_ACP_Bu_4}
%
 \\
\hline
\hflavrowdefabh \\
 & & \\[-1ex]
\hflavrowdefabi \\
 & & \\[-1ex]
\hflavrowdefabj \\
 & & \\[-1ex]
\hflavrowdefabk \\
 & & \\[-1ex]
\hflavrowdefabl \\
 & & \\[-1ex]
\hflavrowdefabm \\
 & & \\[-1ex]
\hflavrowdefabn \\
 & & \\[-1ex]
\hflavrowdefabo \\
 & & \\[-1ex]
\hflavrowdefabp \\
 & & \\[-1ex]
\hflavrowdefabq \\
 & & \\[-1ex]
\hflavrowdefabr \\
 & & \\[-1ex]
\hflavrowdefabs \\
\hline
\hline
\end{tabular}
\begin{tablenotes}
\item[1] Treatment of charmonium intermediate components differs between the results.
\item[2] $A_{C\!P}$ is also measured in bins of $m_{K^0_S K^0_S}$
\item[3] Result extracted from Dalitz-plot analysis of $B^0 \to K_S^0 K^+ K^-$ decays.
\item[4] Using run II dataset, corresponding to an integrated luminosity of $5.9 ~\mathrm{fb}^{-1}$ collected at a center-of-mass energy of 13 TeV (2015 to 2018).
\item[5] Also measured in several invariant mass regions.
\item[6] Multiple systematic uncertainties are added in quadrature.
\item[7] Using run I dataset, corresponding to an integrated luminosity of $3.0 ~\mathrm{fb}^{-1}$ collected at a center-of-mass energy of 7 TeV (2011) and 8 TeV (2012).
\item[8] Also measured in bins of $m_{K^+K^-}$.
\item[9] LHCb uses a model of the nonresonant contribution obtained from a phenomenological description of the partonic interaction that produces the final state. This contribution is referred to as the single pole in the paper; see Ref.~\cite{LHCb:2019xmb} for details.
\item[10] Result extracted from Dalitz-plot analysis of $B^+ \to K^+ K^- \pi^+$ decays.
\end{tablenotes}
\end{threeparttable}
\end{center}
\end{table}

\hflavrowdef{\hflavrowdefabt}{$A_\mathrm{CP}(B^+ \to K^+ K^+ K^-)$}{{\setlength{\tabcolsep}{0pt}
}
\begin{table}[H]
\begin{center}
\begin{threeparttable}
\caption{$C\!P$ asymmetries of charmless hadronic $B^+$ decays (part 5).}
\label{tab:rareDecays_ACP_Bu_5}
%
 \\
\hline
\hflavrowdefabt \\
 & & \\[-1ex]
\hflavrowdefabu \\
 & & \\[-1ex]
\hflavrowdefabv \\
 & & \\[-1ex]
\hflavrowdefabw \\
 & & \\[-1ex]
\hflavrowdefabx \\
 & & \\[-1ex]
\hflavrowdefaby \\
 & & \\[-1ex]
\hflavrowdefabz \\
 & & \\[-1ex]
\hflavrowdefaca \\
\hline
\hline
\end{tabular}
\begin{tablenotes}
\item[\dag] Preliminary result.
\item[1] Using run II dataset, corresponding to an integrated luminosity of $5.9 ~\mathrm{fb}^{-1}$ collected at a center-of-mass energy of 13 TeV (2015 to 2018).
\item[2] Also measured in several invariant mass regions.
\item[3] Multiple systematic uncertainties are added in quadrature.
\item[4] Using run I dataset, corresponding to an integrated luminosity of $3.0 ~\mathrm{fb}^{-1}$ collected at a center-of-mass energy of 7 TeV (2011) and 8 TeV (2012).
\item[5] Result extracted from Dalitz-plot analysis of $B^+ \to K^+ K^+ K^-$ decays.
\item[6] Combination of two final states of the $K^*(892)^{\pm}$, $K_S^0\pi^{\pm}$ and $K^{\pm}\pi^0$. In addition to the combined results, the paper reports separately the results for each individual final state.
\item[7] Measured in the $\phi \phi$ invariant mass range below the $\eta_c$ resonance ($m_{\phi \phi} < 2.85~\text{GeV}/c^2$).
\end{tablenotes}
\end{threeparttable}
\end{center}
\end{table}

\hflavrowdef{\hflavrowdefacb}{$A_\mathrm{CP}(B^+ \to K^*(892)^+ \gamma)$}{{\setlength{\tabcolsep}{0pt}
}
\begin{table}[H]
\begin{center}
\begin{threeparttable}
\caption{$C\!P$ asymmetries of charmless hadronic $B^+$ decays (part 6).}
\label{tab:rareDecays_ACP_Bu_6}
%
 \\
\hline
\hflavrowdefacb \\
 & & \\[-1ex]
\hflavrowdefacc \\
 & & \\[-1ex]
\hflavrowdefacd \\
 & & \\[-1ex]
\hflavrowdeface \\
 & & \\[-1ex]
\hflavrowdefacf \\
 & & \\[-1ex]
\hflavrowdefacg \\
 & & \\[-1ex]
\hflavrowdefach \\
\hline
\hline
\end{tabular}
\begin{tablenotes}
\item[1] $m_{K\pi} < 2.0~\text{GeV}/c^2$.
\item[2] $0.79 < m_{K\pi} < 1.0~\text{GeV}/c^2$.
\item[3] $m_{X_s} < 2.8 ~\text{GeV}/c^2$.
\item[4] $m_{K\eta} < 2.4 ~\text{GeV}/c^2$.
\item[5] $m_{K\eta^{(\prime)}} < 3.25 ~\text{GeV}/c^2$.
\item[6] $1.4 \leq E_{\gamma}^{*} \leq 3.4~\text{GeV}/c^2$, where $E_{\gamma}^{*}$ is the photon energy in the center-of-mass frame.
\item[7] $m_{\phi K} < 3.0 ~\text{GeV}/c^2$.
\end{tablenotes}
\end{threeparttable}
\end{center}
\end{table}

\hflavrowdef{\hflavrowdefaci}{$A_\mathrm{CP}(B^+ \to \pi^+ \pi^0)$}{{\setlength{\tabcolsep}{0pt}
}
\begin{table}[H]
\begin{center}
\resizebox{0.95\textwidth}{!}{
\begin{threeparttable}
\caption{$C\!P$ asymmetries of charmless hadronic $B^+$ decays (part 7).}
\label{tab:rareDecays_ACP_Bu_7}
%
 \\
\hline
\hflavrowdefaci \\
 & & \\[-1ex]
\hflavrowdefacj \\
 & & \\[-1ex]
\hflavrowdefack \\
 & & \\[-1ex]
\hflavrowdefacl \\
 & & \\[-1ex]
\hflavrowdefacm \\
 & & \\[-1ex]
\hflavrowdefacn \\
 & & \\[-1ex]
\hflavrowdefaco \\
 & & \\[-1ex]
\hflavrowdefacp \\
 & & \\[-1ex]
\hflavrowdefacq \\
 & & \\[-1ex]
\hflavrowdefacr \\
\hline
\hline
\end{tabular}
\begin{tablenotes}
\item[\dag] Preliminary result.
\item[1] Treatment of charmonium intermediate components differs between the results.
\item[2] Using run II dataset, corresponding to an integrated luminosity of $5.9 ~\mathrm{fb}^{-1}$ collected at a center-of-mass energy of 13 TeV (2015 to 2018).
\item[3] Also measured in several invariant mass regions.
\item[4] Multiple systematic uncertainties are added in quadrature.
\item[5] Using run I dataset, corresponding to an integrated luminosity of $3.0 ~\mathrm{fb}^{-1}$ collected at a center-of-mass energy of 7 TeV (2011) and 8 TeV (2012).
\item[6] Result extracted from Dalitz-plot analysis of $B^+ \to \pi^+ \pi^+ \pi^-$ decays.
\item[7] This analysis uses three different approaches: isobar, $K$-matrix and quasi-model-independent, to describe the $S$-wave component. The $A_{C\!P}$ results are taken from the isobar model with an additional error accounting for the different S-wave methods as reported in Appendix D of Ref.~\cite{LHCb:2019sus}.
\item[8] Result extracted from Dalitz-plot analysis of $B^+ \to K^+ K^- \pi^+$ decays.
\item[9] The nonresonant amplitude is modelled using a sum of exponential functions.
\end{tablenotes}
\end{threeparttable}
}
\end{center}
\end{table}

\hflavrowdef{\hflavrowdefacs}{$A_\mathrm{CP}(B^+ \to \rho^+(770) \rho^0(770))$}{{\setlength{\tabcolsep}{0pt}
}
\begin{table}[H]
\begin{center}
\resizebox{1\textwidth}{!}{
\begin{threeparttable}
\caption{$C\!P$ asymmetries of charmless hadronic $B^+$ decays (part 8).}
\label{tab:rareDecays_ACP_Bu_8}
%
 \\
\hline
\hflavrowdefacs \\
 & & \\[-1ex]
\hflavrowdefact \\
 & & \\[-1ex]
\hflavrowdefacu \\
 & & \\[-1ex]
\hflavrowdefacv \\
 & & \\[-1ex]
\hflavrowdefacw \\
 & & \\[-1ex]
\hflavrowdefacx \\
 & & \\[-1ex]
\hflavrowdefacy \\
 & & \\[-1ex]
\hflavrowdefacz \\
 & & \\[-1ex]
\hflavrowdefada \\
\hline
\hline
\end{tabular}
\begin{tablenotes}
\item[\dag] Preliminary result.
\item[1] Result extracted from Dalitz-plot analysis of $B^+ \to \pi^+ \pi^+ \pi^-$ decays.
\item[2] This analysis uses three different approaches: isobar, $K$-matrix and quasi-model-independent, to describe the $S$-wave component. The $A_{C\!P}$ results are taken from the isobar model with an additional error accounting for the different S-wave methods as reported in Appendix D of Ref.~\cite{LHCb:2019sus}.
\item[3] Multiple systematic uncertainties are added in quadrature.
\end{tablenotes}
\end{threeparttable}
}
\end{center}
\end{table}

\hflavrowdef{\hflavrowdefadb}{$A_\mathrm{CP}(B^+ \to p \bar{p} \pi^+)$}{{\setlength{\tabcolsep}{0pt}
}
\begin{table}[H]
\begin{center}
\resizebox{0.95\textwidth}{!}{
\begin{threeparttable}
\caption{$C\!P$ asymmetries of charmless hadronic $B^+$ decays (part 9).}
\label{tab:rareDecays_ACP_Bu_9}
%
\begin{tablenotes}
\item[1] Treatment of charmonium intermediate components differs between the results.
\end{tablenotes}
\end{threeparttable}
}
\end{center}
\end{table}

\hflavrowdef{\hflavrowdefadh}{$A_\mathrm{CP}(B^+ \to K^+ \ell^+ \ell^-)$}{{\setlength{\tabcolsep}{0pt}%
}
\begin{table}[H]
\begin{center}
\resizebox{0.95\textwidth}{!}{
\begin{threeparttable}
\caption{$C\!P$ asymmetries of charmless hadronic $B^+$ decays (part 10).}
\label{tab:rareDecays_ACP_Bu_10}
%
 \\
\hline
\hflavrowdefadh \\
 & & \\[-1ex]
\hflavrowdefadi \\
 & & \\[-1ex]
\hflavrowdefadj \\
 & & \\[-1ex]
\hflavrowdefadk \\
 & & \\[-1ex]
\hflavrowdefadl \\
 & & \\[-1ex]
\hflavrowdefadm \\
 & & \\[-1ex]
\hflavrowdefadn \\
\hline
\hline
\end{tabular}
\begin{tablenotes}
\item[1] $A_{C\!P}$ is also measured in bins of $m_{\mu^+ \mu^-}$
\item[2] Mass regions corresponding to $\phi$, $J/\psi$ and $\psi(2S)$ are vetoed.
\item[3] Mass regions corresponding to $J/\psi$ and $\psi(2S)$ are vetoed.
\end{tablenotes}
\end{threeparttable}
}
\end{center}
\end{table}

\hflavrowdef{\hflavrowdefado}{$A_\mathrm{CP}(B^0 \to \pi^0 \pi^0)$}{{\setlength{\tabcolsep}{0pt}
}
\begin{table}[H]
\begin{center}
\begin{threeparttable}
\caption{$C\!P$ asymmetries of charmless hadronic $B^0$ decays (part 1).}
\label{tab:rareDecays_ACP_Bd_1}
%
 \\
\hline
\hflavrowdefado \\
 & & \\[-1ex]
\hflavrowdefadp \\
 & & \\[-1ex]
\hflavrowdefadq \\
 & & \\[-1ex]
\hflavrowdefadr \\
 & & \\[-1ex]
\hflavrowdefads \\
 & & \\[-1ex]
\hflavrowdefadt \\
 & & \\[-1ex]
\hflavrowdefadu \\
 & & \\[-1ex]
\hflavrowdefadv \\
 & & \\[-1ex]
\hflavrowdefadw \\
 & & \\[-1ex]
\hflavrowdefadx \\
 & & \\[-1ex]
\hflavrowdefady \\
 & & \\[-1ex]
\hflavrowdefadz \\
 & & \\[-1ex]
\hflavrowdefaea \\
\hline
\hline
\end{tabular}
\begin{tablenotes}
\item[\dag] Preliminary result.
\item[1] LHCb combines results of the 1.9~fb$^{-1}$ run 2 data analysis with those based on Run 1 dataset \cite{LHCb:2018pff}. The full statistical and systematic covariance matrices are used in the combination.
\item[2] Combination of time-integrated and time-dependent analyses using the best linear unbiased estimator Ref.~\cite{Valassi:2003mu}.
\end{tablenotes}
\end{threeparttable}
\end{center}
\end{table}

\hflavrowdef{\hflavrowdefaeb}{$A_\mathrm{CP}(B^0 \to K^+ \pi^- \pi^0)$}{{\setlength{\tabcolsep}{0pt}
}
\begin{table}[H]
\begin{center}
\begin{threeparttable}
\caption{$C\!P$ asymmetries of charmless hadronic $B^0$ decays (part 2).}
\label{tab:rareDecays_ACP_Bd_2}
%
 \\
\hline
\hflavrowdefaeb \\
 & & \\[-1ex]
\hflavrowdefaec \\
 & & \\[-1ex]
\hflavrowdefaed \\
 & & \\[-1ex]
\hflavrowdefaee \\
 & & \\[-1ex]
\hflavrowdefaef \\
 & & \\[-1ex]
\hflavrowdefaeg \\
 & & \\[-1ex]
\hflavrowdefaeh \\
 & & \\[-1ex]
\hflavrowdefaei \\
 & & \\[-1ex]
\hflavrowdefaej \\
 & & \\[-1ex]
\hflavrowdefaek \\
 & & \\[-1ex]
\hflavrowdefael \\
 & & \\[-1ex]
\hflavrowdefaem \\
 & & \\[-1ex]
\hflavrowdefaen \\
\hline
\hline
\end{tabular}
\begin{tablenotes}
\item[\dag] Preliminary result.
\item[1] Result extracted from Dalitz-plot analysis of $B^0 \to K^+ \pi^- \pi^0$ decays.
\item[2] The nonresonant amplitude is taken to be constant across the  Dalitz plane.
\item[3] Result extracted from Dalitz-plot analysis of $B^0 \to K_S^0 \pi^+ \pi^-$ decays.
\item[4] Multiple systematic uncertainties are added in quadrature.
\item[5] The official HFLAV average includes results from time-dependent analyses and is given in the section Measurements related to Unitarity Triangle angles.
\end{tablenotes}
\end{threeparttable}
\end{center}
\end{table}

\hflavrowdef{\hflavrowdefaeo}{$A_\mathrm{CP}(B^0 \to K^*(892)^0 \pi^+ \pi^-)$}{{\setlength{\tabcolsep}{0pt}
}
\begin{table}[H]
\begin{center}
\begin{threeparttable}
\caption{$C\!P$ asymmetries of charmless hadronic $B^0$ decays (part 3).}
\label{tab:rareDecays_ACP_Bd_3}
%
}
\begin{table}[H]
\begin{center}
\begin{threeparttable}
\caption{$C\!P$ asymmetries of charmless hadronic $B^0$ decays (part 4).}
\label{tab:rareDecays_ACP_Bd_4}
%
}
\begin{table}[H]
\begin{center}
\begin{threeparttable}
\caption{$C\!P$ asymmetries of charmless hadronic $B^0$ decays (part 5).}
\label{tab:rareDecays_ACP_Bd_5}
%
 \\
\hline
\hflavrowdefafc \\
 & & \\[-1ex]
\hflavrowdefafd \\
 & & \\[-1ex]
\hflavrowdefafe \\
 & & \\[-1ex]
\hflavrowdefaff \\
 & & \\[-1ex]
\hflavrowdefafg \\
 & & \\[-1ex]
\hflavrowdefafh \\
\hline
\hline
\end{tabular}
\begin{tablenotes}
\item[1] Treatment of charmonium intermediate components differs between the results.
\item[2] $A_{C\!P}$ is also measured in bins of $m_{\mu^+ \mu^-}$
\item[3] Mass regions corresponding to $\phi$, $J/\psi$ and $\psi(2S)$ are vetoed.
\item[4] Mass regions corresponding to $J/\psi$ and $\psi(2S)$ are vetoed.
\end{tablenotes}
\end{threeparttable}
\end{center}
\end{table}

\hflavrowdef{\hflavrowdefafi}{$A_\mathrm{CP}(B \to K^* \gamma)$}{{\setlength{\tabcolsep}{0pt}
}
\begin{table}[H]
\begin{center}
\begin{threeparttable}
\caption{$C\!P$ asymmetries of charmless hadronic decays of $B^{\pm}/B^0$ admixture.}
\label{tab:rareDecays_ACP_Badmix_1}
%
 \\
\hline
\hflavrowdefafi \\
 & & \\[-1ex]
\hflavrowdefafj \\
 & & \\[-1ex]
\hflavrowdefafk \\
 & & \\[-1ex]
\hflavrowdefafl \\
 & & \\[-1ex]
\hflavrowdefafm \\
 & & \\[-1ex]
\hflavrowdefafn \\
 & & \\[-1ex]
\hflavrowdefafo \\
 & & \\[-1ex]
\hflavrowdefafp \\
 & & \\[-1ex]
\hflavrowdefafq \\
\hline
\hline
\end{tabular}
\begin{tablenotes}
\item[1] $m_{X_s} < 2.8 ~\text{GeV}/c^2$.
\item[2] $0.6 < m_{X_s} < 2.0 ~\text{GeV}/c^2$.
\item[3] $E^{*}_{\gamma} \geq 2.1~\text{GeV}$ where $E^{*}_{\gamma}$ is the photon energy in the center-of-mass frame.
\item[4] $2.1 < E^{*}_{\gamma} < 2.8~\text{GeV}$ where $E^{*}_{\gamma}$ is the photon energy in the center-of-mass frame.
\item[5] $0.4<m_X<2.6~\text{GeV}/c^2$.
\end{tablenotes}
\end{threeparttable}
\end{center}
\end{table}

\hflavrowdef{\hflavrowdefafr}{$A_{C\!P}( B^{0}_{s} \to \pi^+ K^- )$}{{\setlength{\tabcolsep}{0pt}
}
\begin{table}[H]
\begin{center}
\begin{threeparttable}
\caption{$C\!P$ asymmetries of charmless hadronic $B^0_s$ decays.}
\label{tab:rareDecays_ACP_Other_Bs}
%
\begin{tablenotes}
\item[1] LHCb combines results of the 1.9~fb$^{-1}$ run 2 data analysis with those based on Run 1 dataset \cite{LHCb:2018pff}. The full statistical and systematic covariance matrices are used in the combination.
\end{tablenotes}
\end{threeparttable}
\end{center}
\end{table}

\hflavrowdef{\hflavrowdefafs}{$A_\mathrm{CP}(\Lambda_b^0 \to p \pi^-)$}{{\setlength{\tabcolsep}{0pt}%
}
\begin{table}[H]
\begin{center}
\begin{threeparttable}
\caption{$C\!P$ asymmetries of charmless hadronic $\Lambda^0_b$ decays.}
\label{tab:rareDecays_ACP_Other_Lb}
%
}
\begin{table}[H]
\begin{center}
\begin{threeparttable}
\caption{$C\!P$ asymmetries of charmless hadronic $\Xi_b^-$ decays.}
\label{tab:rareDecays_ACP_Other_Xib}
%
 \\
\hline
\hflavrowdefafx \\
 & & \\[-1ex]
\hflavrowdefafy \\
 & & \\[-1ex]
\hflavrowdefafz \\
 & & \\[-1ex]
\hflavrowdefaga \\
 & & \\[-1ex]
\hflavrowdefagb \\
 & & \\[-1ex]
\hflavrowdefagc \\
\hline
\hline
\end{tabular}
\end{threeparttable}
\end{center}
\end{table}

\noindent Measurements that are not included in the tables (the definitions of observables can be found in the corresponding experimental papers):
\begin{itemize}

\item In Ref.~\cite{LHCb:2016yco},
LHCb reports the triple-product asymmetries ($a_{C\!P}^{\hat T-odd}$,  $a_{P}^{\hat T-odd}$) for the decays $\Lb \to p\pi^-\pi^+\pi^-$ and $\Lb \to p\pi^- K^+ K^-$.
\item In Ref.~\cite{LHCb:2017slr}, LHCb reports $a_{C\!P}^{\hat T-odd}$,  $a_{P}^{\hat T-odd}$ and $\Delta(A_{C\!P}) = A_{C\!P}(\Lambda_b^0 \to p K^- \mu^+ \mu^-) - A_{C\!P}(\Lambda_b^0 \to p K^- J/\psi)$.
\item In Ref.~\cite{LHCb:2018fpt},
LHCb reports $a_{C\!P}^{\hat T-odd}$ and  $a_{P}^{\hat T-odd}$ for the decays $\Lb \to p K^-\pi^+\pi^-$,  $\Lb \to p K^- K^+ K^-$ and $\Xi^{0}_{b} \to p K^- K^- \pi^+$.
\item In Ref.~\cite{LHCb:2019jyj} LHCb measures differences of \CP asymmetries between $\Lb$ and $\Xi_b^0$ charmless decays into a proton and three charged mesons and the decays to the same final states with an intermediate charmed baryon.

\end{itemize}

\begin{figure}[htbp!]
\centering
\includegraphics[width=0.8\textwidth]{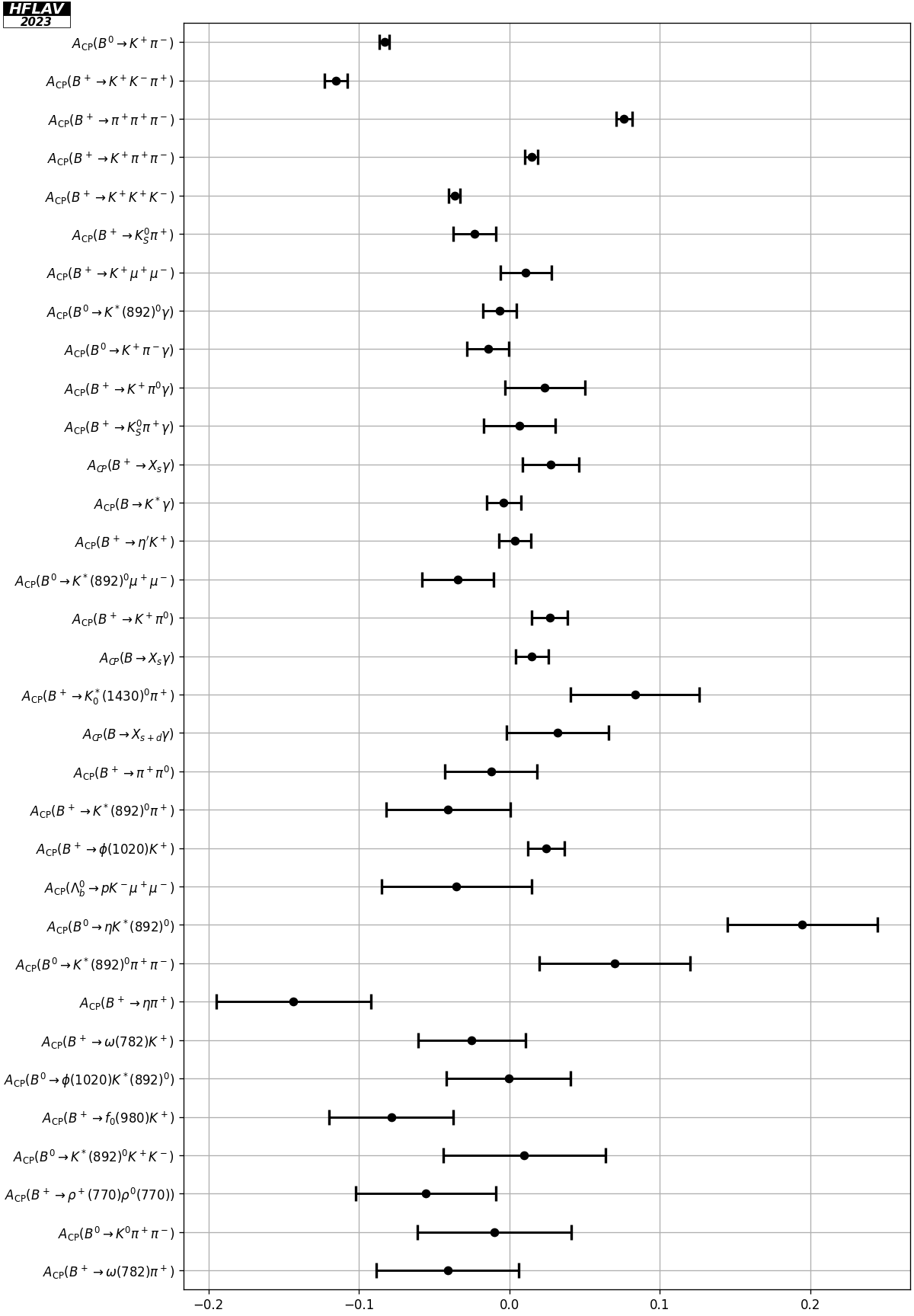}
\caption{A selection among the most precise direct \CP\ asymmetries ($A_{\CP}$) measured in charmless $B^+$ and $B^0$ decay modes.}
\label{fig:rare-acpselect}
\end{figure}

\clearpage

\mysubsection{Polarization measurements in \b-hadron decays}
\label{sec:rare-polar}

In this section, compilations of polarization measurements in \b-hadron decays
are given. Tables~\ref{tab:rareDecays_VVpol_Bu_FL}, \ref{tab:rareDecays_VVpol_Bd_FL}, and ~\ref{tab:rareDecays_VVpol_Bs_FL} detail measurements
of the longitudinal polarization fraction $f_L$ in \Bp\, \Bz, and \Bs decays, respectively.
They are followed by Tables~\ref{tab:rareDecays_VVpol_AA_Bu}, \ref{tab:rareDecays_VVpol_AA_Bd} and ~\ref{tab:rareDecays_VVpol_AA_Bs}, which list polarization fractions and \CP\ parameters measured in full
angular analyses of $\Bp$, \Bz and \Bs decays. 
Figures~\ref{fig:rare-polar} and~\ref{fig:rare-polarbs} show graphic representations of a selection of results shown in this section.

Most of the final states considered in the tables are pairs of vector mesons and thus, we detail below the corresponding definitions. For specific definitions, for example regarding vector-tensor final states or vector recoiling against dispin-half states, please refer to the articles.
In the decay of a pseudoscalar meson into two vector mesons, momentum conservation allows for three helicity configurations: $H_0, H_{\pm 1}$. They can be expressed in terms of longitudinal polarisation amplitudes, $A_0 = H_0$, and transverse polarization amplitudes, $A_{\perp} = (H_{+1}-H_{-1})/\sqrt{2}$ and $A_{\parallel} = (H_{+1}+H_{-1})/\sqrt{2}$. The corresponding amplitudes for the charge conjugate decays are denoted $\overline{A_0}$, $\overline{A_{\parallel}}$, and $\overline{A_{\perp}}$. Using the definition 
\begin{equation}
	F_{k=0, \parallel, \perp} = \frac{ |A_k|^2 }{ |A_{0}|^2 + |A_{\perp}|^2 + |A_{\parallel}|^2 },~~
	\overline{F}_{k=0, \parallel, \perp} = \frac{|\overline{A_k}|^2}{|\overline{A_{0}}|^2 + |\overline{A_{\perp}}|^2 + |\overline{A_{\parallel}}|^2 },
\end{equation}
the following $C\!P$ conserving and $C\!P$ violating observables, which are used in our tables, are defined:
\begin{equation}
	f_{k=0, \parallel, \perp} = \frac{1}{2}(F_k + \overline{F_k}),~~
	A_{C\!P}^{k=0, \perp} = \frac{ F_k - \overline{F_k} }{ F_k + \overline{F_k} }.
\end{equation}
Note that, in the literature, $f_0$ and $f_L$ are used interchangeably to denote the longitudinal polarization fraction.

\hflavrowdef{\hflavrowdefaaa}{$f_L(B^+ \to \omega(782) K^*(892)^+)$}{{\setlength{\tabcolsep}{0pt}
}
\begin{table}[H]
\begin{center}
\begin{threeparttable}
\caption{Longitudinal polarization fraction, $f_L$, in $B^+$ decays.}
\label{tab:rareDecays_VVpol_Bu_FL}
%
 \\
\hline
\hflavrowdefaaa \\
 & & \\[-1ex]
\hflavrowdefaab \\
 & & \\[-1ex]
\hflavrowdefaac \\
 & & \\[-1ex]
\hflavrowdefaad \\
 & & \\[-1ex]
\hflavrowdefaae \\
 & & \\[-1ex]
\hflavrowdefaaf \\
 & & \\[-1ex]
\hflavrowdefaag \\
 & & \\[-1ex]
\hflavrowdefaah \\
 & & \\[-1ex]
\hflavrowdefaai \\
 & & \\[-1ex]
\hflavrowdefaaj \\
 & & \\[-1ex]
\hflavrowdefaak \\
\hline
\hline
\end{tabular}
\begin{tablenotes}
\item[\dag] Preliminary result.
\item[1] Combination of two final states of the $K^*(892)^{\pm}$, $K_S^0\pi^{\pm}$ and $K^{\pm}\pi^0$. In addition to the combined results, the paper reports separately the results for each individual final state.
\item[2] See also Ref.~\cite{Belle:2005fjn}.
\end{tablenotes}
\end{threeparttable}
\end{center}
\end{table}

\hflavrowdef{\hflavrowdefaal}{$f_L(B^0 \to \omega(782) K^*(892)^0)$}{{\setlength{\tabcolsep}{0pt}
}
\begin{table}[H]
\begin{center}
\begin{threeparttable}
\caption{Longitudinal polarization fraction, $f_L$, in $B^0$ decays.}
\label{tab:rareDecays_VVpol_Bd_FL}
%
 \\
\hline
\hflavrowdefaal \\
 & & \\[-1ex]
\hflavrowdefaam \\
 & & \\[-1ex]
\hflavrowdefaan \\
 & & \\[-1ex]
\hflavrowdefaao \\
 & & \\[-1ex]
\hflavrowdefaap \\
 & & \\[-1ex]
\hflavrowdefaaq \\
 & & \\[-1ex]
\hflavrowdefaar \\
 & & \\[-1ex]
\hflavrowdefaas \\
 & & \\[-1ex]
\hflavrowdefaat \\
 & & \\[-1ex]
\hflavrowdefaau \\
 & & \\[-1ex]
\hflavrowdefaav \\
 & & \\[-1ex]
\hflavrowdefaaw \\
 & & \\[-1ex]
\hflavrowdefaax \\
 & & \\[-1ex]
\hflavrowdefaay \\
\hline
\hline
\end{tabular}
\begin{tablenotes}
\item[\dag] Preliminary result.
\item[1] The PDG uncertainty includes a scale factor.
\item[2] The charmonium mass regions are vetoed.
\item[3] $m_{\Lambda^0\overline{\Lambda^0}} < 2.85 ~\text{GeV}/c^2$.
\end{tablenotes}
\end{threeparttable}
\end{center}
\end{table}

\hflavrowdef{\hflavrowdefaaz}{$f_L(B_s^0 \to \phi(1020) \phi(1020))$}{{\setlength{\tabcolsep}{0pt}
}
\begin{table}[H]
\begin{center}
\begin{threeparttable}
\caption{Longitudinal polarization fraction, $f_L$, in $B^0_s$ decays.}
\label{tab:rareDecays_VVpol_Bs_FL}
%
}
\begin{table}[H]
\begin{center}
\begin{threeparttable}
\caption{Results of full angular analyses of $B^+$ decays.}
\label{tab:rareDecays_VVpol_AA_Bu}
%
\begin{tablenotes}
\item[1] Combination of two final states of the $K^*(892)^{\pm}$, $K_S^0\pi^{\pm}$ and $K^{\pm}\pi^0$. In addition to the combined results, the paper reports separately the results for each individual final state.
\end{tablenotes}
\end{threeparttable}
\end{center}
\end{table}

\hflavrowdef{\hflavrowdefabg}{$f_\perp(B^0 \to \phi(1020) K^*(892)^0)$}{{\setlength{\tabcolsep}{0pt}%
}
\begin{table}[H]
\begin{center}
\begin{threeparttable}
\caption{Results of full angular analyses of $B^0$ decays.}
\label{tab:rareDecays_VVpol_AA_Bd}
%
\begin{tablenotes}
\item[1] The PDG uncertainty includes a scale factor.
\end{tablenotes}
\end{threeparttable}
\end{center}
\end{table}

\hflavrowdef{\hflavrowdefabi}{$f_\perp(B_s^0 \to \phi(1020) \phi(1020))$}{{\setlength{\tabcolsep}{0pt}%
}
\begin{table}[H]
\begin{center}
\begin{threeparttable}
\caption{Results of full angular analyses of $B^0_s$ decays.}
\label{tab:rareDecays_VVpol_AA_Bs}
%
 \\
\hline
\hflavrowdefabi \\
 & & \\[-1ex]
\hflavrowdefabj \\
 & & \\[-1ex]
\hflavrowdefabk \\
 & & \\[-1ex]
\hflavrowdefabl \\
\hline
\hline
\end{tabular}
\end{threeparttable}
\end{center}
\end{table}

\clearpage
\noindent Measurements that are not included in the tables (the definitions of observables can be found in the corresponding experimental papers):
\begin{itemize}
\item In the angular analysis of $B^0 \to \phi K^*(892)^0$ decays~\cite{LHCb:2014xzf}, in addition to the results quoted in Table~\ref{tab:rareDecays_VVpol_AA_Bd}, LHCb reports observables related to the $S$-wave component contributing the final state $K^+K^-K^+\pi^-$: $f_S(K\pi)$, $f_S(KK)$, $\delta_s(K\pi)$, $\delta_s(KK)$, ${\cal{A}}_S(K\pi)^{C\!P}$, ${\cal{A}}_S(KK)^{C\!P}$, $\delta_S(K\pi)^{C\!P}$, $\delta_S(KK)^{C\!P}$.
\item In the amplitude analysis of $B_s^0 \to \phi \phi$ decays, in addition to the results quoted in Table~\ref{tab:rareDecays_VVpol_AA_Bs}, LHCb, in Ref.~\cite{LHCb:2019jgw}, extracts the $C\!P$-violating phase $\phi_s^{s\bar{s}s}$ and the $C\!P$-violating parameter $|\lambda|$ from a decay-time-dependent and polarisation independent fit. The $C\!P$-violating phases $\phi_{s, \parallel}$ and $\phi_{s, \perp}$ are obtained in a polarisation-dependent fit. A time-integrated fit is performed to extract the triple-product asymmetries $A_U$ and $A_V$. CDF, in Ref.~\cite{CDF:2011kwp} also reports the triple-product asymmetries $A_U$ and $A_V$.
\item  In Ref.~\cite{LHCb:2017rwt}, LHCb presents a flavor-tagged, decay-time-dependent amplitude analysis of $\B^0_s \to (K^+\pi^-)(K^-\pi^+)$ decays in the $K^\pm\pi^\mp$ mass range from 750 to 1600~MeV/$c^2$. The paper includes measurements of 19 \CP-averaged amplitude parameters corresponding to scalar, vector and tensor final states as well as the first measurement of the $C\!P$-violating phase $\phi^{d\bar{d}}_s$.
\item Reference~\cite{LHCb:2018hsm} presents an amplitude analysis of $B^0 \to \rho K^*(892)^0$ realised by LHCb. Scalar ($S$) and vector ($V$) contributions to the final state $(\pi^+\pi^+)(K^+\pi^-)$ are considered through partial waves sharing the same angular dependence ($VV$, $SS$, $SV$, $VS$) and the corresponding amplitudes are extracted for each case. Triple product asymmetries are also reported.
\end{itemize}

\begin{figure}[hp!]
\centering
\includegraphics[width=0.8\textwidth]{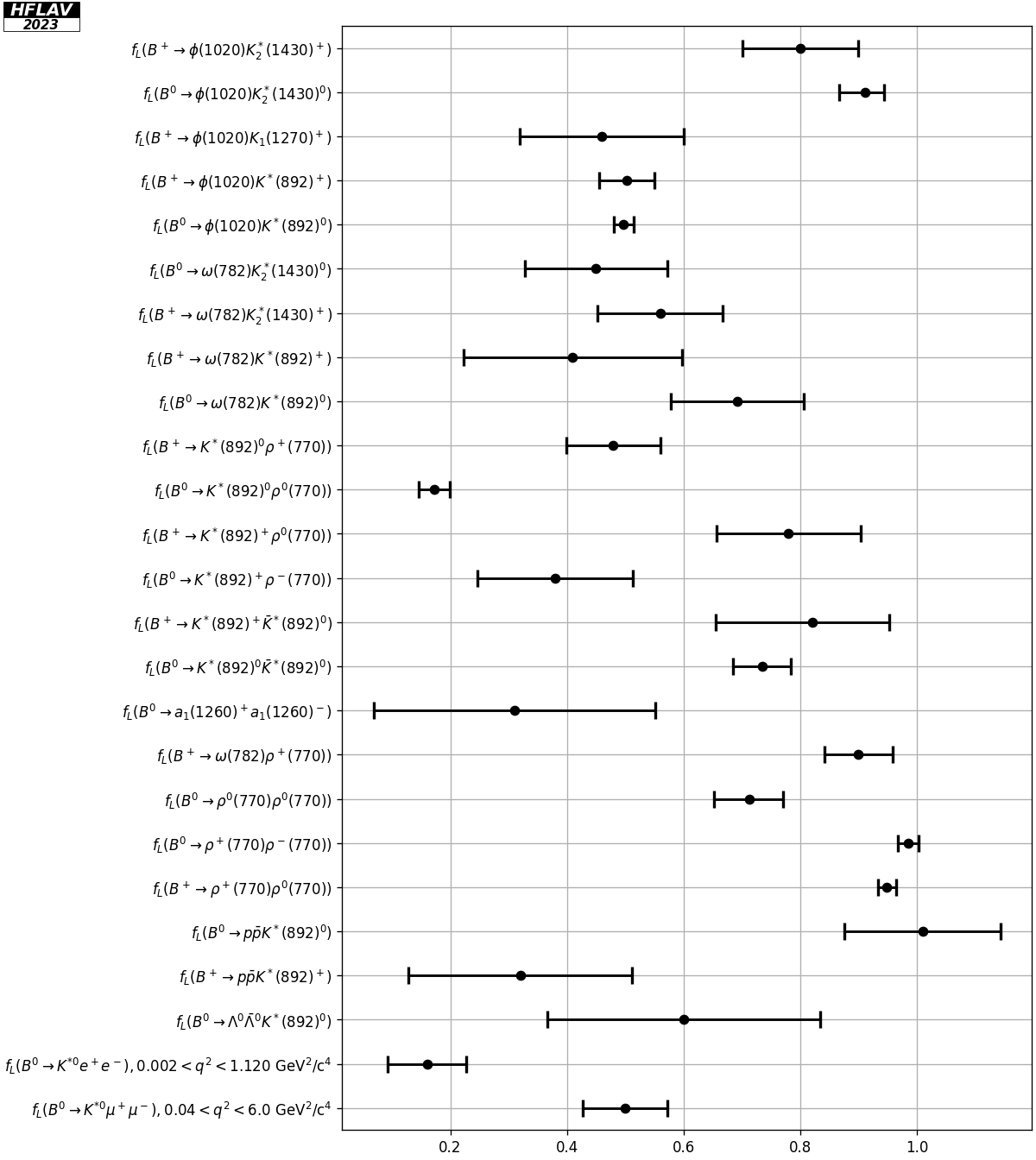}
\caption{Longitudinal polarization fraction in charmless $B$ decays.}
\label{fig:rare-polar}
\end{figure}

\begin{figure}[hbp!]
\centering
\includegraphics[width=0.8\textwidth]{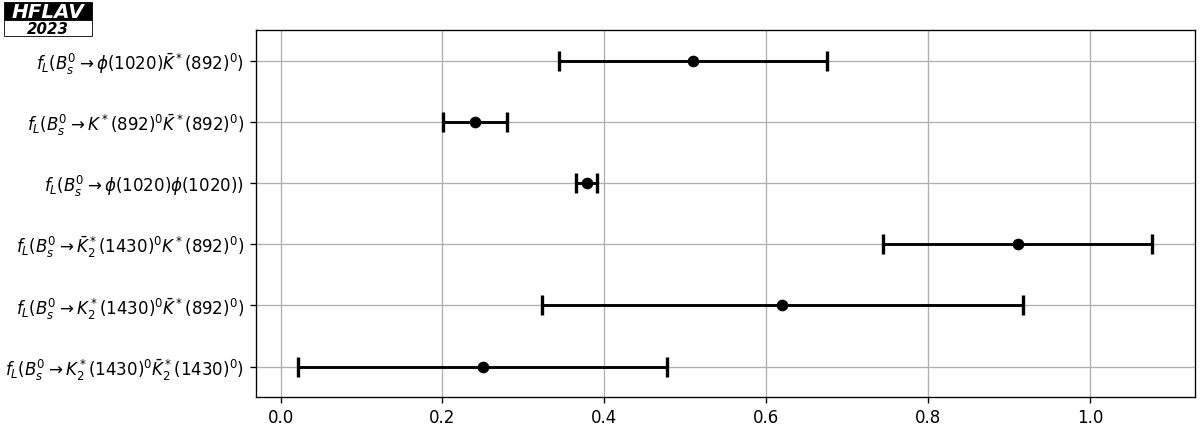}
\caption{Longitudinal polarization fraction in charmless $\Bs$ decays.}
\label{fig:rare-polarbs}
\end{figure}

\clearpage
\section{Charm mixing and \CP violation}
\label{sec:charm_cpv_oscillations}

\def\kbar{\overline{K}{}^{\,0}}
\def\dbar{\overline{D}{}^{\,0}}
\def\bbar{\overline{B}{}^{\,0}}
\def\cp{$\CP$}
\def\cpv{$CPV$}
\def\ra{\!\rightarrow\!}
\def\ddbar{$D^0$-$\dbar$}
\def\ycp{$y^{}_{\CP}$}

\def\dklnu{$D^0\ra K^{(*)+}\ell^-\bar{\nu}$}
\def\dkpi{$D^0\ra K^+\pi^-$}
\def\dkk{$D^0\ra K^+K^-$}
\def\dpipi{$D^0\ra\pi^+\pi^-$}
\def\dkkpp{$D^0\ra K^+K^-/\pi^+\pi^-$}
\def\dkspp{$D^0\ra K^0_S\,\pi^+\pi^-$}
\def\dppp{$D^0\ra \pi^0\,\pi^+\pi^-$}
\def\dkskk{$D^0\ra K^0_S\,K^+ K^-$}
\def\dkppp{$D^0\ra K^+\pi^-\pi^+\pi^-$}

\def\dsphipi{$D^+_s\ra\phi\,\pi^+$}

\def\lcp{\Lambda_c^+}

\def\simge{\mathrel{%
   \rlap{\raise 0.511ex \hbox{$>$}}{\lower 0.511ex \hbox{$\sim$}}}}
\def\simle{\mathrel{
   \rlap{\raise 0.511ex \hbox{$<$}}{\lower 0.511ex \hbox{$\sim$}}}}

\newcommand{\Dnan}{\ensuremath{D_0^\ast(2300)^0}}
\newcommand{\Dtan}{\ensuremath{D_2^\ast(2460)^0}}
\newcommand{\Don}{\ensuremath{D_1(2420)^{0}}}
\newcommand{\Dopn}{\ensuremath{D_1(2430)^{0}}}
\newcommand{\Dnap}{\ensuremath{D_0^\ast(2300)^\pm}}
\newcommand{\Dtap}{\ensuremath{D_2^\ast(2460)^\pm}}
\newcommand{\Dop}{\ensuremath{D_1(2420)^{\pm}}}
\newcommand{\Dopp}{\ensuremath{D_1(2430)^{\pm}}}

\newcommand{\Dsa}{\ensuremath{D_s^{\ast\pm}}}
\newcommand{\Dsna}{\ensuremath{D_{s0}^\ast(2317)^{\pm}}}
\newcommand{\Dsop}{\ensuremath{D_{s1}(2460)^{\pm}}}
\newcommand{\Dso}{\ensuremath{D_{s1}(2536)^{\pm}}}
\newcommand{\Dst}{\ensuremath{D_{s2}(2573)^{\pm}}}
\newcommand{\Dsts}{\ensuremath{D_{sJ}(2700)^{\pm}}}
\newcommand{\Dste}{\ensuremath{D_{sJ}(2860)^{\pm}}}
\newcommand{\Dstsi}{\ensuremath{D_{sJ}(2632)^{\pm}}}

\newcommand{\citep}{\cite}

\newcommand{\kst}{K^*(892)^0}
\newcommand{\akst}{\overline{K}^*(892)^0}
\newcommand{\kstp}{K^*(1410)^0}
\newcommand{\akstp}{\overline{K}^*(1410)^0}
\newcommand{\kstd}{K^*_2(1430)^0}
\newcommand{\akstd}{\overline{K}^*_2(1430)^0}

\newcommand{\ksts}{K^*_0(1430)^0}
\newcommand{\aksts}{\overline{K}^*_0(1430)^0}

\newcommand{\ds}{D_{s}}
\newcommand{\dsp}{D_{s}^+}
\newcommand{\dsm}{D_{s}^-}
\newcommand{\dspm}{D_{s}^{\pm}}
\newcommand{\dsmunu}{\ds^+\to\mu^+\nu_{\mu}}
\newcommand{\dsellnu}{\ds^+\to\ell^+\nu_{\ell}}
\newcommand{\br}{{\cal B}}
\newcommand{\ellnu}{\ell^+\nu_{\ell}}
\newcommand{\enu}{e^+\nu_{e}}
\newcommand{\munu}{\mu^+\nu_{\mu}}
\newcommand{\taunu}{\tau^+\nu_{\tau}}
\newcommand{\taumunu}{\tau^+(\mu^+)\nu_{\tau}}
\newcommand{\tauenu}{\tau^+(e^+)\nu_{\tau}}
\newcommand{\taupinuCharm}{\tau^+(\pi^+)\nu_{\tau}}
\newcommand{\taurhonu}{\tau^+(\rho^+)\nu_{\tau}}

\subsection{\emph{$D^0$-$\dbar$} mixing and \emph{\cp}\ violation}
\label{sec:charm:mixcpv}

\subsubsection{Introduction}

The first evidence for $D^0$-$\dbar$ oscillations, 
indicating mixing, was obtained in 2007 
by Belle~\cite{Staric:2007dt} and \babar~\cite{Aubert:2007wf}.
These results were confirmed later that year by CDF~\cite{Aaltonen:2007uc}
and, in 2013 with improved precision, by LHCb~\cite{LHCb:2012zll}.
There are now numerous measurements of $D^0$-$\dbar$ mixing 
with various levels of sensitivity. 
In 2021, LHCb measured the mixing parameters $x$ and $y$ (see below) with 
much higher precision~\cite{LHCb:2021ykz} than that of previous 
measurements. All these measurements, plus others, are input into a ``global fit'' reported here to determine world-average values for 
mixing parameters, \cp\ violation (\cpv) parameters, and strong 
phase differences.

Our notation is as follows. We use the phase convention 
$\CP|D^0\rangle=-|\dbar\rangle$ and 
$\CP|\dbar\rangle=-|D^0\rangle$~\cite{Bergmann:2000id}
and denote the mass eigenstates as
\begin{eqnarray} 
D^{}_1 & = & p|D^0\rangle-q|\dbar\rangle \\ 
D^{}_2 & = & p|D^0\rangle+q|\dbar\rangle\,.
\end{eqnarray}
Our phase convention corresponds to the following:
for $p = q$ (i.e., no indirect \CP\ violation), 
$D^{}_1$ is \cp-even and $D^{}_2$ is \cp-odd.
The mixing parameters are defined as 
$x\equiv(m^{}_1-m^{}_2)/\Gamma$ and 
$y\equiv (\Gamma^{}_1-\Gamma^{}_2)/(2\Gamma)$, where 
$m^{}_1,\,m^{}_2$ and $\Gamma^{}_1,\,\Gamma^{}_2$ are
the masses and decay widths, respectively, of the 
mass eigenstates, and $\Gamma\equiv (\Gamma^{}_1+\Gamma^{}_2)/2$. 

We perform a global fit to determine central values and uncertainties for ten underlying parameters. These 
parameters, in addition to $x$ and $y$, are as follows:
\begin{itemize}
\item \cpv\ parameters $|q/p|$ and ${\rm Arg}(q/p)\equiv \phi$, which give rise to indirect \cpv\ (see Sec.~\ref{sec:cp_asym} 
for a discussion of indirect and direct \cpv). Here we assume that indirect \cpv\ (when present) is ``universal,'' i.e., independent 
of the final state $f$ in the decay $D^0\ra f$.
\item direct \cpv\ asymmetries 
\begin{eqnarray*}
A^{}_D & \equiv & \frac{\Gamma(D^0\ra K^+\pi^-)-\Gamma(\dbar\ra K^-\pi^+)}
{\Gamma(D^0\ra K^+\pi^-)+\Gamma(\dbar\ra K^-\pi^+)} \\ \\
A^{}_K & \equiv & \frac{\Gamma(D^0\ra K^+ K^-)-\Gamma(\dbar\ra K^- K^+)}
{\Gamma(D^0\ra K^+ K^-)+\Gamma(\dbar\ra K^- K^+)} \\ \\
A^{}_\pi & \equiv & \frac{\Gamma(D^0\ra\pi^+\pi^-)-\Gamma(\dbar\ra\pi^-\pi^+)}
{\Gamma(D^0\ra\pi^+\pi^-)+\Gamma(\dbar\ra\pi^-\pi^+)}\,,
\end{eqnarray*}
where, as indicated, the decay rates are for the pure $D^0$ and $\dbar$ flavour eigenstates.
\item the ratio of doubly Cabibbo-suppressed (DCS) to Cabibbo-favored (CF) decay rates
\begin{eqnarray*}
R^{}_D & \equiv & \frac{\Gamma(D^0\ra K^+\pi^-)+\Gamma(\dbar\ra K^-\pi^+)}
{\Gamma(D^0\ra K^-\pi^+)+\Gamma(\dbar\ra K^+\pi^-)}\,.
\end{eqnarray*}
\item the strong phase difference $\delta$ between the amplitudes
${\cal A}(\dbar\ra K^-\pi^+)$ and ${\cal A}(D^0\ra K^-\pi^+)$; and 
\item the strong phase difference $\delta^{}_{K\pi\pi}$ between the amplitudes 
${\cal A}(\dbar\ra K^-\rho^+)$ and ${\cal A}(D^0\ra K^-\rho^+)$.
\end{itemize}

There are 63 measurements of 18 observables used in the fit. The
measurements are based on the following Cabibbo-suppressed 
decays:\footnote{Charge-conjugate modes are implicitly 
included.} \dklnu, \dkk, \dpipi, \dkpi, 
$D^0\ra K^+\pi^-\pi^0$, %
\dkspp, \dppp, \dkskk, and \dkppp.
The fit also uses measurements of mixing parameters and strong phases
determined using quantum correlations and double-tagged branching fractions 
measured at the $\psi(3770)$ resonance. The relationships between measured 
observables and fitted parameters are given in Table~\ref{tab:relationships}.
Correlations among measured observables are accounted for by using covariance 
matrices provided by the experimental collaborations. Uncertainties are 
assumed to be Gaussian, and systematic uncertainties among different 
experiments are assumed to be uncorrelated unless specific 
correlations have been identified.
We have compared this method with a second method that adds
together three-dimensional log-likelihood functions for $x$, 
$y$, and $\delta$ obtained from several independent measurements;
this combination accounts for non-Gaussian uncertainties.
When both methods are applied to the same set of 
measurements, equivalent results are obtained. 
We have furthermore compared the results to 
those obtained from an independent fit based 
on the {\sc GammaCombo} framework~\cite{LHCb:2016mag} 
and found excellent agreement.

\begin{table}
\renewcommand{\arraycolsep}{0.02in}
\renewcommand{\arraystretch}{1.3}
\begin{center}
\caption{\label{tab:relationships}
Left column: decay modes used to determine the fitted parameters 
$x$, $y$, $\delta$, $\delta^{}_{K\pi\pi}$, $R^{}_D$, 
$A^{}_D$, $A^{}_K$, $A^{}_\pi$, $|q/p|$, and $\phi$.
Middle column: measured observables for each decay mode. 
Right column: relationships between the measured observables and the fitted
parameters. The symbol $\langle t\rangle$ denotes the mean reconstructed 
decay time for $D^0\ra K^+K^-$ or $D^0\ra\pi^+\pi^-$ decays.}
\vspace*{6pt}
\footnotesize
\resizebox{0.99\textwidth}{!}{
 \\
\hline
\end{tabular}
}
\end{center}
\end{table}

Mixing in the heavy flavour $B^0$ and $B^0_s$ systems
is governed by a short-distance box diagram. In the $D^0$ 
system, however, this box diagram is both doubly-Cabibbo-suppressed 
and GIM-suppressed~\cite{Glashow:1970gm}, and thus the short-distance 
mixing rate is tiny. As a consequence, $D^0$-$\dbar$ mixing is dominated 
by long-distance processes. These are difficult to calculate, and 
theoretical estimates for $x$ and $y$ range over three orders 
of magnitude, up to the percent 
level~\cite{Bigi:2000wn,Petrov:2003un,Petrov:2004rf,Falk:2004wg}.

Most experimental analyses besides that of the 
$\psi(3770)\ra \overline{D}D$ measurements identify the flavour 
of the $D^0$ or $\dbar$ when produced by reconstructing the decay
$D^{*+}\ra D^0\pi^+$ or $D^{*-}\ra\dbar\pi^-$. The charge
of the pion, which has low momentum in the lab frame relative 
to that of the $D^0$ daughters and is often referred to as the 
``soft'' pion, identifies the $D^0$ flavour. For $D^{*+}\ra D^0\pi^+$, 
$M^{}_{D^*}-M^{}_{D^0}-M^{}_{\pi^+}\equiv Q\approx 6~\mevcc$; 
this value is close to the kinematic threshold. Thus, analyses 
typically require that the reconstructed $Q$ be less than 
some value (e.g., 20~\mevcc) to suppress backgrounds. 
In some analyses, LHCb identifies the flavour of
the $D^0$ by partially reconstructing $\overline{B}\ra D^{(*)}\mu^- X$
and $B\ra\overline{D}{}^{(*)}\mu^+ X$ decays; in this case the charge 
of the $\mu^\mp$ identifies the flavour of the $D^0$ or $\dbar$.

For time-dependent measurements, the $D^0$ decay time 
is calculated as $t = M^{}_{D^0} (\vec{d}\cdot\hat{p})/p$, 
where $\vec{d}$ is the displacement vector from the
$D^0$ production vertex to the $D^0$ decay vertex,
$\hat{p}$ is the direction of the $D^0$ momentum, 
and $p$ is its magnitude.
At $e^+e^-$ experiments, the $D^{*+}$ vertex position is typically 
taken as the intersection of the $D^0$ momentum vector 
with the beam spot profile. At $pp$ and $\bar{p}p$ experiments, 
the $D^{*+}$ decay vertex is taken to be at a primary interaction vertex.

\subsubsection{Input measurements}

The global fit determines central values and uncertainties 
for the above ten parameters by minimizing a $\chi^2$ statistic.
All input measurements are listed in 
Tables~\ref{tab:rm_semi}-\ref{tab:observables3}.
In the $D^0\ra K^+\pi^-\pi^0$ 
Dalitz plot analysis~\cite{Aubert:2008zh}, the phases of
intermediate resonances are fitted relative to the phase of
${\cal A}(D^0\ra K^+\rho^-)$, and the phases of resonances in
$\dbar\ra K^+\pi^-\pi^0$ are fitted relative to the phase of
${\cal A}(\dbar\ra K^+\rho^-)$. The phase difference 
$\delta^{}_{K\pi\pi} \equiv 
{\rm Arg}[{\cal A}(\dbar\ra K^+\rho^-)/{\cal A}(D^0\ra K^+\rho^-)]$
cannot be determined from these 
fits but can be constrained in the global 
fit and is included as a fitted parameter.

There are three observables input to the fit that are 
themselves world averages of relevent measurements:
\begin{eqnarray}
R^{}_M & = & \frac{x^2+y^2}{2} \\ 
 & & \nonumber \\
y^{}_{\CP} & = & 
\frac{1}{2}\left(\left|\frac{q}{p}\right| + \left|\frac{p}{q}\right|\right)
y\cos\phi - 
\frac{1}{2}\left(\left|\frac{q}{p}\right| - \left|\frac{p}{q}\right|\right)
x\sin\phi \\
 & & \nonumber \\
A^{}_\Gamma & = & 
\frac{1}{2}\left(\left|\frac{q}{p}\right| - \left|\frac{p}{q}\right|\right)
y\cos\phi - 
\frac{1}{2}\left(\left|\frac{q}{p}\right| + \left|\frac{p}{q}\right|\right)
x\sin\phi\,. 
\end{eqnarray} 
The world average for $R^{}_M$ is calculated from measurements of 
\dklnu\ decays relative to $D^0\rightarrow K^{(*)-}\ell^+\nu$ decays, 
as listed in Tab.~\ref{tab:rm_semi}. A measurement of $R^{}_M$ 
using $D^0\ra K^+\pi^-\pi^+\pi^-$ decays~\cite{LHCb:2016zmn} is input
separately to the global fit. The world averages for $y^{}_{CP}$ and 
$A^{}_\Gamma$ are calculated from measurements of 
\dkk, \dpipi, $D^0\rightarrow K_S^0 h~(h=\pi^0\!,\eta,\omega)$, 
$D^0\rightarrow K_S^0 \pi^0\pi^0$, and $D^0\rightarrow K_S^0 K^+K^-$ 
decays, as listed in Tabs.~\ref{tab:yCP} and~\ref{tab:Agamma}.
The most recent measurement of $y^{}_{CP}$ from LHCb is sufficiently precise such
that the small value of $y^{K\pi}_{CP}$ for Cabibbo-favored $D^0\rightarrow K^-\pi^+$ decays,
which is also used for the measurement, must be taken into 
account (i.e, subtracted)~\cite{Pajero:2021jev}.
The $y^{K\pi}_{CP}$ correction we use for the global fit is from Ref.~\cite{Schwartz:2022egt}.

\begin{table}
\caption{\label{tab:rm_semi} Measurements of $R^{}_M = (x^2+y^2)/2$ 
from ``wrong-sign'' semileptonic \dklnu\ decays measured relative to
``right-sign'' $D^0\rightarrow K^{(*)-}\ell^+\nu$ decays.
The weighted average is used in the global fit.}
\vspace*{-8pt}
\footnotesize
\begin{center}
\renewcommand{\arraystretch}{1.3}
\renewcommand{\arraycolsep}{0.02in}

\end{center}
\end{table}

\begin{table}
\caption{\label{tab:yCP} Measurements of $y^{}_{CP}$ using 
singly Cabibbo-suppressed decays \dkk, \dpipi, 
$D^0\rightarrow K_S^0 h~(h=\pi^0\!,\eta,\omega)$, 
$D^0\rightarrow K_S^0 \pi^0\pi^0$, and $D^0\rightarrow K_S^0 K^+K^-$.
The weighted average is used in the global fit.}
\vspace*{-8pt}
\footnotesize
\begin{center}
\renewcommand{\arraystretch}{1.2}
\renewcommand{\tabcolsep}{3pt}
%
\end{center}
\end{table}

\begin{table}
\caption{\label{tab:Agamma} Measurements of $A^{}_\Gamma$ 
using singly Cabibbo-suppressed decays \dkk\ and \dpipi.
The weighted average is used in the global fit.}
\vspace*{-8pt}
\footnotesize
\begin{center}
\renewcommand{\arraystretch}{1.2}
\renewcommand{\arraycolsep}{0.02in}
%
 \\
\hline
{\bf Weighted average} & & $0.089\pm 0.113$ &  \\
\hline
\end{tabular}
\end{center}
\end{table}

The \dkpi\ measurements used are from 
Belle~\cite{Zhang:2006dp,Ko:2014qvu}, 
\babar~\cite{Aubert:2007wf}, CDF~\cite{Aaltonen:2013pja}, 
and LHCb~\cite{LHCb:2016qsa,Aaij:2017urz}; earlier measurements are 
either superseded or have significantly less precision and are not used.
The observables from \dkspp\ decays are measured in two ways:
assuming \cp\ conservation ($D^0$ and $\dbar$ decays combined),
and allowing for \cp\ violation ($D^0$ and $\dbar$ decays
fitted separately). The no-\cpv\ measurements are from 
Belle~\cite{Peng:2014oda}, \babar~\cite{delAmoSanchez:2010xz},
and LHCb~\cite{Aaij:2015xoa}; for the \cpv-allowed case, only
Belle~\cite{Peng:2014oda} and LHCb~\cite{Aaij:2019jot,LHCb:2021ykz}
measurements are available. 
The $D^0\ra K^+\pi^-\pi^0$, $D^0\ra K^0_S K^+ K^-$, 
and $D^0\ra \pi^0\,\pi^+\pi^-$ results are from 
\babar~\cite{Aubert:2008zh,Lees:2016gom}; 
the $D^0\ra K^+\pi^-\pi^+\pi^-$ results are from LHCb~\cite{LHCb:2016zmn}; and
the $\psi(3770)\ra\overline{D}D$ results are from CLEOc~\cite{Asner:2012xb}.
Measurements of the strong phase $\delta$ by BES\,III~\cite{BESIII:2014rtm,BESIII:2022qkh} 
based on $\psi(3770)\ra\overline{D}D$ events use HFLAV's world averages for 
$R^{}_D$ and $y$ as external inputs; thus, we do not include these 
results in the global fit. However, a BES\,III measurement~\cite{BESIII:2022qkh} 
of the asymmetry between $CP$-even-tagged and $CP$-odd-tagged $D^0\ra K^-\pi^+$ 
decays, denoted $A_{K\pi}^{CP}$, is included. To lowest order in the mixing parameters, 
$A_{K\pi}^{CP} = (2\sqrt{R^{}_D}\cos\delta + y)/(1 + R)$, where
$R = R^{}_D + \sqrt{R^{}_D}(y\cos\delta - x\sin\delta) + (x^2+y^2)/2$.

\begin{table}
\renewcommand{\arraystretch}{1.3}
\renewcommand{\arraycolsep}{0.02in}
\renewcommand{\tabcolsep}{0.05in}
\caption{\label{tab:observables1}
Measurements used in the global fit, except those from
time-dependent \dkpi\ measurements 
(listed in Table~\ref{tab:observables2})
and those from direct \cpv\ measurements
(listed in Table~\ref{tab:observables3}).
}
\vspace*{6pt}
\footnotesize
\resizebox{0.99\textwidth}{!}{

}
\end{table}

\begin{table}
\renewcommand{\arraystretch}{1.3}
\renewcommand{\arraycolsep}{0.02in}
\caption{\label{tab:observables2}
Time-dependent \dkpi\ observables used for the global fit.
The observables $R^+_D$ and $R^-_D$ are related to parameters 
$R^{}_D$ and $A^{}_D$ via $R^\pm_D = R^{}_D (1\pm A^{}_D)$.}
\vspace*{6pt}
\footnotesize
\begin{center}
%
 \right\}$ \\
\hline
\end{tabular}
\end{center}
\end{table}

\begin{table}
\renewcommand{\arraystretch}{1.3}
\renewcommand{\arraycolsep}{0.02in}
\caption{\label{tab:observables3}
Measurements of time-integrated \cp\ asymmetries. The observable 
$A^{}_{\CP}(f)= [\Gamma(D^0\ra f)-\Gamma(\dbar\ra f)]/
[\Gamma(D^0\ra f)+\Gamma(\dbar\ra f)]$. The symbol
$\Delta\langle t\rangle$ denotes the difference
between the mean reconstructed decay times for 
$D^0\ra K^+K^-$ and $D^0\ra\pi^+\pi^-$ decays due to 
different trigger and reconstruction efficiencies.}
\vspace*{6pt}
\footnotesize
\begin{center}
\resizebox{\textwidth}{!}{
\begin{tabular}{l|ccc}
\hline
{\bf Mode} & \textbf{Observable} & {\bf Values} & 
                  {\boldmath $\Delta\langle t\rangle/\tau^{}_D$} \\
\hline
\begin{tabular}{c}
$D^0\ra h^+ h^-$~\cite{Aubert:2007if} \\
(\babar\ 386 fb$^{-1}$)
\end{tabular} & 
\begin{tabular}{c}
$A^{}_{\CP}(K^+K^-)$ \\
$A^{}_{\CP}(\pi^+\pi^-)$ 
\end{tabular} & 
\begin{tabular}{c}
$(+0.00 \pm 0.34 \pm 0.13)\%$ \\
$(-0.24 \pm 0.52 \pm 0.22)\%$ 
\end{tabular} &
0 \\
\hline
\begin{tabular}{c}
$D^0\ra h^+ h^-$~\cite{Belle:2008ddg} \\
(Belle 540~fb$^{-1}$)
\end{tabular} & 
\begin{tabular}{c}
$A^{}_{\CP}(K^+K^-)$ \\
$A^{}_{\CP}(\pi^+\pi^-)$ 
\end{tabular} & 
\begin{tabular}{c}
$(-0.43 \pm 0.30 \pm 0.11)\%$ \\
$(+0.43 \pm 0.52 \pm 0.12)\%$ 
\end{tabular} &
0 \\
\hline
\begin{tabular}{c}
$D^0\ra h^+ h^-$~\cite{cdf_public_note_10784,Collaboration:2012qw} \\
(CDF 9.7~fb$^{-1}$)
\end{tabular} & 
\begin{tabular}{c}
$A^{}_{\CP}(K^+K^-)-A^{}_{\CP}(\pi^+\pi^-)$ \\
$A^{}_{\CP}(K^+K^-)$ \\
$A^{}_{\CP}(\pi^+\pi^-)$ 
\end{tabular} & 
\begin{tabular}{c}
$(-0.62 \pm 0.21 \pm 0.10)\%$ \\
$(-0.32 \pm 0.21)\%$ \\
$(+0.31 \pm 0.22)\%$ 
\end{tabular} &
$0.27 \pm 0.01$ \\
\hline
\begin{tabular}{c}
$D^0\ra h^+ h^-$~\cite{LHCb:2019hro} \\
(LHCb 8.9~fb$^{-1}$, \\
$D^{*+}\ra D^0\pi^+$ + \\
\ $\overline{B}\ra D^0\mu^- X$ \\
\ \ tags combined)
\end{tabular} & 
$A^{}_{\CP}(K^+K^-)-A^{}_{\CP}(\pi^+\pi^-)$ &
$(-0.154 \pm 0.029)\%$ &
$0.115 \pm 0.002$ \\
\hline
\begin{tabular}{c}
$D^0\ra K^+ K^-$~\cite{LHCb:2022lry} \\
\ \ (LHCb 5.7~fb$^{-1}$, \\
\ \ \ \ $D^{*+}\ra D^0\pi^+$ tag)
\end{tabular} & $A^{}_{\CP}(K^+K^-)$ & $(0.068\pm 0.054\pm 0.016)\%$ &
\begin{tabular}{c}
$\frac{\displaystyle \langle t\rangle^{}_K}{\displaystyle \tau^{}_D} = 
\frac{\displaystyle (701.5\pm 1.1)}{\displaystyle (410.3\pm 1.0)}$ \\
 \hskip0.40in $= 1.7097\pm 0.0050$ \end{tabular} \\
\hline
\end{tabular}
}
\end{center}
\end{table}

For each set of correlated observables, we construct a difference 
vector $\vec{V}$ between the measured values and those calculated 
from the fitted parameters using the relations of 
Table~\ref{tab:relationships}.
For example, for $D^0\ra K^0_S\,\pi^+\pi^-$ decays,
$\vec{V}=(\Delta x,\Delta y,\Delta |q/p|,\Delta \phi)$,
where $\Delta x \equiv x^{}_{\rm measured} - x^{}_{\rm fitted}$
and similarly for $\Delta y, \Delta |q/p|$, and $\Delta \phi$.
The contribution of a set of observables to the fit $\chi^2$ 
is calculated as $\vec{V}\cdot (M^{-1})\cdot\vec{V}^T$, 
where $M^{-1}$ is the inverse of the covariance matrix 
for the measured observables. Covariance matrices are 
constructed from the correlation coefficients among the 
observables. These correlation coefficients are furnished
by the experiments and listed in 
Tables~\ref{tab:observables1}-\ref{tab:observables2}.

\subsubsection{Fit results}

The global fitter uses {\sc MINUIT} with the MIGRAD minimizer, 
and all uncertainties are obtained from MINOS~\cite{MINUIT:webpage}. 
Three categories of fits are performed:
\begin{enumerate}
\item [ 1)] 
{\bf One-parameter description of indirect CP violation} \\
This parametrization is based on two simplifications: 
{\it (1)}\ subleading amplitudes in CF and DCS decays are neglected; and 
{\it (2)}\ subleading amplitudes in 
singly Cabibbo-suppressed (SCS) decays are neglected in indirect 
$CP$ violation observables, as their contribution is suppressed by 
the mixing parameters $x$ and $y$. These simplifications imply there 
is no direct $CP$ violation in CF or DCS decays ($A_D = 0$), and that  
all indirect $CP$ violation can be parameterized in terms 
of the phase difference between $M^{}_{12}$ and $\Gamma^{}_{12}$, 
the off-diagonal elements of the mass and decay matrices,
respectively, for the $D^0$-$\dbar$ system. This difference,
${\rm Arg}(M^{}_{12}/\Gamma^{}_{12})\equiv \phi^{}_{12}$, 
is independent of phase convention. For the phase convention 
in which the dominant $\Delta (U\mbox{-spin}) =2$ 
amplitudes are aligned with the real axis, 
${\rm Arg}(\Gamma^{}_{12}) = 0$, and thus $\phi^{}_{12} = {\rm Arg}(M^{}_{12})$.
Neglecting subleading amplitudes, it can be shown that
$\tan\phi = (1-|q/p|^2)/(1+|q/p|^2)(x/y)$, i.e., only three of the four
parameters $(x,y,|q/p|,\phi)$ are independent.
This relation was first derived by Ciuchini et al.~\cite{Ciuchini:2007cw}
and later, independently, by Kagan and Sokoloff~\cite{Kagan:2009gb}. Defining 
alternative mixing parameters $x^{}_{12}\equiv 2|M^{}_{12}|/\Gamma$ and 
$y^{}_{12}\equiv |\Gamma^{}_{12}|/\Gamma$, the traditional parameters 
$(x, y, |q/p|, \phi)$ can be expressed in terms of 
$(x^{}_{12}, y^{}_{12}, \phi^{}_{12})$ as follows~\cite{Kagan:2009gb,Kagan:2020vri}:
\begin{eqnarray}
x & = &
\left[\frac{x_{12}^2 - y_{12}^2 + \sqrt{(x_{12}^2 + y_{12}^2)^2 - 
4x_{12}^2 y_{12}^2\sin^2\phi^{}_{12}}}{2}\right]^{1/2} \label{eqn:xrelation} \\ \nonumber \\
y & = &
\left[\frac{y_{12}^2 - x_{12}^2 + \sqrt{(x_{12}^2 + y_{12}^2)^2 - 
4x_{12}^2 y_{12}^2\sin^2\phi^{}_{12}}}{2}\right]^{1/2} \label{eqn:yrelation} \\ \nonumber \\
\left|\frac{q}{p}\right| & = & 
\left(\frac{\displaystyle x_{12}^2 + y_{12}^2 + 2x^{}_{12} y^{}_{12}\sin\phi^{}_{12}}
{\displaystyle x_{12}^2 + y_{12}^2 - 2x^{}_{12} y^{}_{12}\sin\phi^{}_{12}}\right)^{1/4} 
\label{eqn:qoprelation} \\ \nonumber \\
\tan 2\phi & = & 
-\frac{\displaystyle \sin 2\phi^{}_{12}}{\displaystyle \cos 2\phi^{}_{12} + (y^{}_{12}/x^{}_{12})^2}\,.
\label{eqn:phirelation} 
\end{eqnarray}
Based on this formalism, three separate fits are performed:
\begin{itemize}
\item [ 1a)] 
float $x$, $y$, and $\phi$ and use the Ciuchini/Kagan formula to derive $|q/p|$; 
this yields (MINOS) errors for $(x,y,\phi)$.
\item [ 1b)] 
float $x$, $y$, and $|q/p|$ and use the Ciuchini/Kagan formula to derive $\phi$; 
this yields (MINOS) errors for $(x,y,|q/p|)$.
\item [ 1c)] 
fit for the parameters $x^{}_{12}$, $y^{}_{12}$, and $\phi^{}_{12}$.
\end{itemize}
\item [ 2)] 
{\bf Two-parameter description of indirect CP violation} \\
In this parameterization, subleading amplitudes in CF and DCS decays are 
neglected as in Fit \#1, and thus $A^{}_D= 0$. However, subleading amplitudes 
in SCS decays are taken into account in indirect $CP$ violation observables. There 
is one simplification: final-state-dependent effects for $D^0\rightarrow K^+K^-$ and 
$D^0\rightarrow\pi^+\pi^-$ are neglected in $A^{}_\Gamma$, as they cancel at 
leading-order in $U$-spin breaking. The subleading SCS amplitudes contribute to both 
$M^{}_{12}$ and $\Gamma^{}_{12}$, and thus ${\rm Arg}(\Gamma^{}_{12})$ can be nonzero. 
To account for this, we fit for four parameters: $(x^{}_{12}, y^{}_{12}, \phi^M_2, \phi^\Gamma_2)$, 
where $\phi^M_2$ and $\phi^\Gamma_2$ are the phases of 
$M^{}_{12}$ and $\Gamma^{}_{12}$, respectively, 
relative to that of the dominant $\Delta (U\mbox{-spin}) =2$
dispersive and absorptive amplitudes. This parameterization 
is discussed by Kagan and Silvestrini~\cite{Kagan:2020vri}. 
The relationships between $(x, y, |q/p|)$ and 
$(x^{}_{12}, y^{}_{12}, \phi^M_2, \phi^\Gamma_2)$
remain as given in Eqs.~(\ref{eqn:xrelation})-(\ref{eqn:qoprelation}),
now with $\phi^{}_{12}=\phi_2^M - \phi_2^\Gamma$, while Eq.~(\ref{eqn:phirelation}) is replaced by
\begin{eqnarray*}
\tan 2\phi & = & 
-\frac{\displaystyle x^2_{12}\sin 2\phi^M_2 + y^2_{12}\sin 2\phi^\Gamma_2}
{\displaystyle x^2_{12}\cos 2\phi^M_2 + y^2_{12}\cos 2\phi^\Gamma_2}\,.
\end{eqnarray*}
\item [ 3)] {\bf Allowing all $CP$ violation} \\
We allow for direct $CP$ violation in DCS decays by relaxing the constraint $A^{}_D=0$. 
Thus, we now fit for all ten parameters: $x$, $y$, $\delta$, $R^{}_D$, $A^{}_D$, 
$\delta^{}_{K\pi\pi}$, $|q/p|$, $\phi$, $A^{}_K$, and~$A^{}_\pi$.
\end{enumerate}

All fit results are listed in Table~\ref{tab:results}. 
The $\chi^2$ for the most general Fit~\#3 
(all \cpv\ allowed) is 65.4 for $63-10=53$ degrees 
of freedom. Individual contributions to the $\chi^2$ 
are listed in Table~\ref{tab:results_chi2}. Confidence 
contours in the two dimensions $(x,y)$ or $(|q/p|,\phi)$ are 
obtained by finding the minimum $\chi^2$ for each fixed point in the 
two-dimensional plane. The resulting $1\sigma$--$5\sigma$ contours 
are shown in Fig.~\ref{fig:contours_ndcpv} for Fits~\#1 and \#2, 
and in Fig.~\ref{fig:contours_cpv} for Fit~\#3.
These contours are determined from the increase of the
$\chi^2$ above the overall minimum value, 
$\Delta\chi^2=\chi^2-\chi^2_{\rm min}$.
For the all-\cpv-allowed Fit~\#3, 
$\Delta\chi^2=2631$ at the no-mixing point $(x,y)\!=\!(0,0)$;
this corresponds to a statistical significance for two degrees 
of freedom of greater than $11.5\sigma$, and 
the no-mixing hypothesis is excluded at this high level. 
In the $(|q/p|,\phi)$ plot (Fig.~\ref{fig:contours_cpv}, bottom),
$\Delta\chi^2=6.49$ at the no-\cpv\ point $(|q/p|,\phi)\!=\!(1,0)$;
this corresponds to a statistical significance of~$2.1\sigma$.
For Fit~\#2, $\Delta\chi^2=3.40$ at the no-\cpv\ point 
$(\phi_2^M, \phi_2^\Gamma)\!=\!(0,0)$; this corresponds 
to a statistical significance of only~$1.3\sigma$. 
This significance is less than that 
for Fit~\#3 because the constraint $A^{}_D\!=\!0$ increases the 
minimum $\chi^2$ value.

\begin{table}
\renewcommand{\arraystretch}{1.3}
\begin{center}
\caption{\label{tab:results}
Results of the global fit for different assumptions regarding \cpv.
The $\chi^2$/d.o.f.\ are considered satisfactory, although care should be taken 
when interpreting them in terms of probability due to unknown systematic uncertainties.}
\vspace*{6pt}
\footnotesize
\resizebox{0.99\textwidth}{!}{

}
\end{center}
\end{table}

\begin{table}
\renewcommand{\arraystretch}{1.3}
\begin{center}
\caption{\label{tab:results_chi2}
Individual contributions to the $\chi^2$ for the 
all-\cpv-allowed Fit~\#3.}
\vspace*{6pt}
\footnotesize
%
\end{center}
\end{table}

One-dimensional likelihood curves for individual parameters 
are obtained by finding the minimum $\chi^2$ for fixed values of 
the parameter of interest. The resulting functions 
$\Delta\chi^2=\chi^2-\chi^2_{\rm min}$ ($\chi^2_{\rm min}$
is the overall minimum value) are shown in Fig.~\ref{fig:1dlikelihood}.
The points where $\Delta\chi^2=3.84$ determine 95\% C.L. intervals 
for the parameters. These intervals are listed in Table~\ref{tab:results}.
The value of $\Delta\chi^2$ at $x\!=\!0$ is 82.8 (Fig.~\ref{fig:1dlikelihood}, 
upper left), and the value of $\Delta\chi^2$ at $y\!=\!0$ is 1227 
(Fig.~\ref{fig:1dlikelihood}, upper right). These correspond to statistical 
significances of $9.1\sigma$ and $35\sigma$, respectively. These 
large values demonstrate that neutral $D$ mesons undergo both dispersive 
($\Delta M\neq 0$) and absorptive ($\Delta\Gamma\neq 0$) mixing.

\begin{figure}
\begin{center}
\vbox{
\includegraphics[width=84mm]{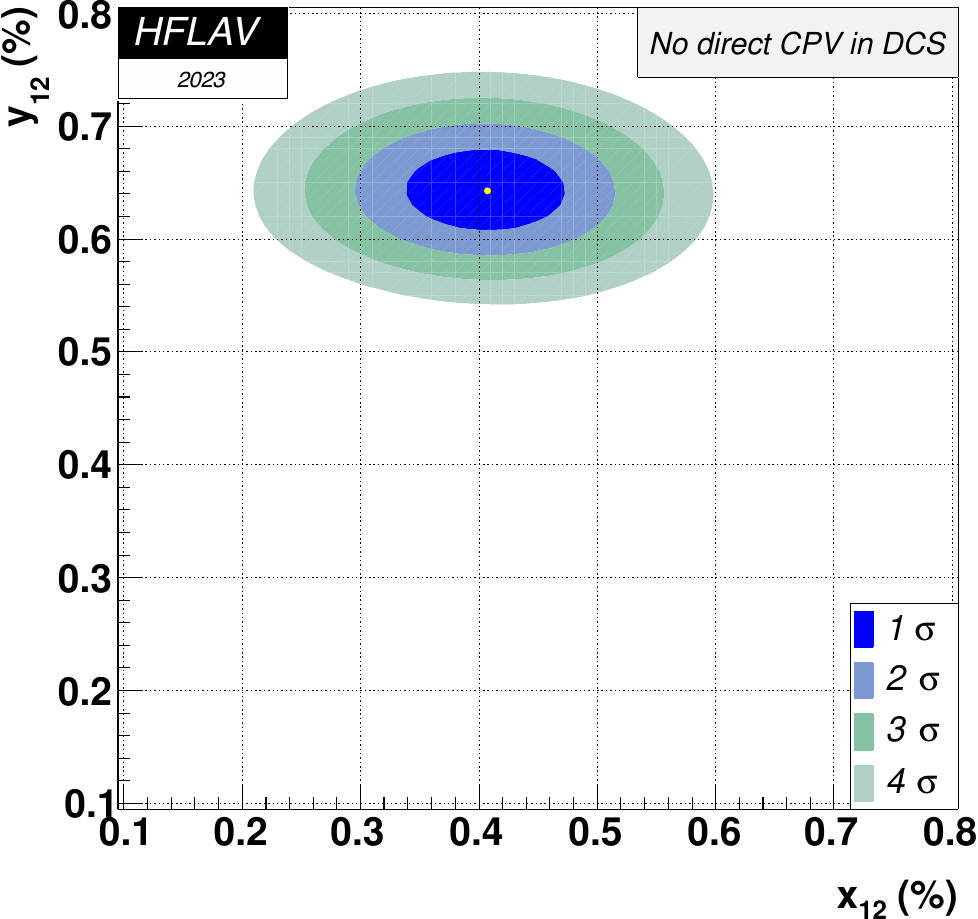}
\includegraphics[width=84mm]{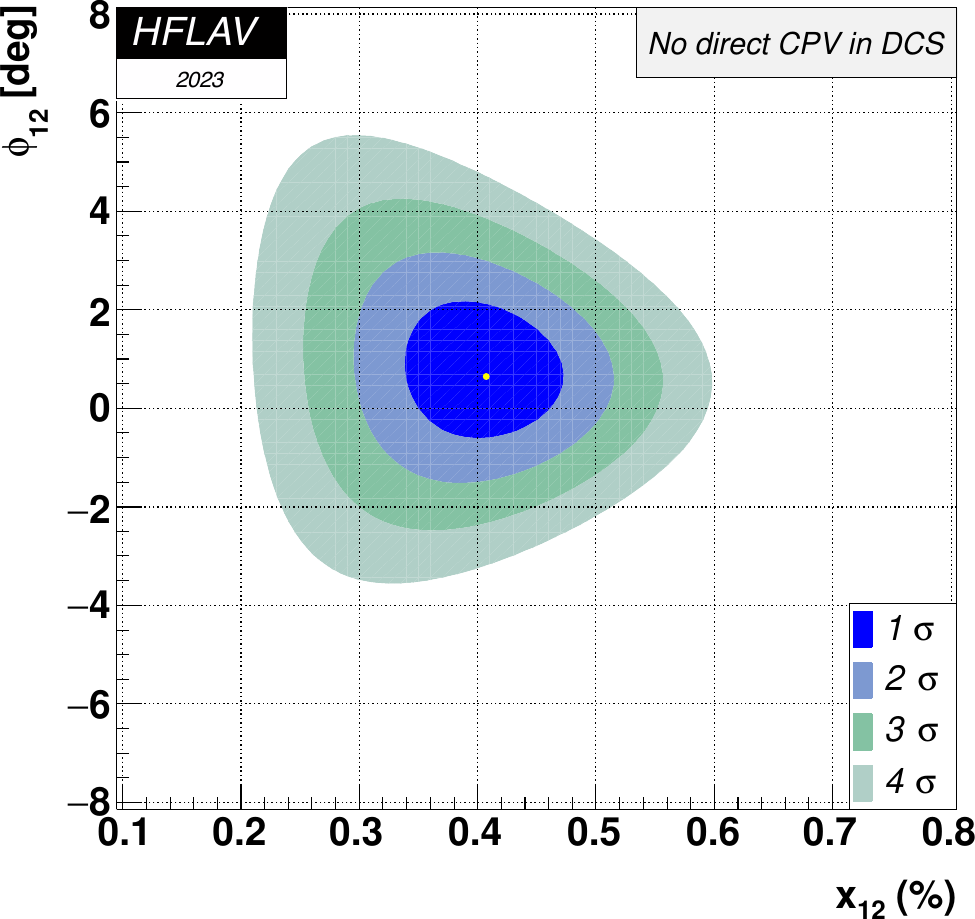}
\vskip0.20in
\includegraphics[width=84mm]{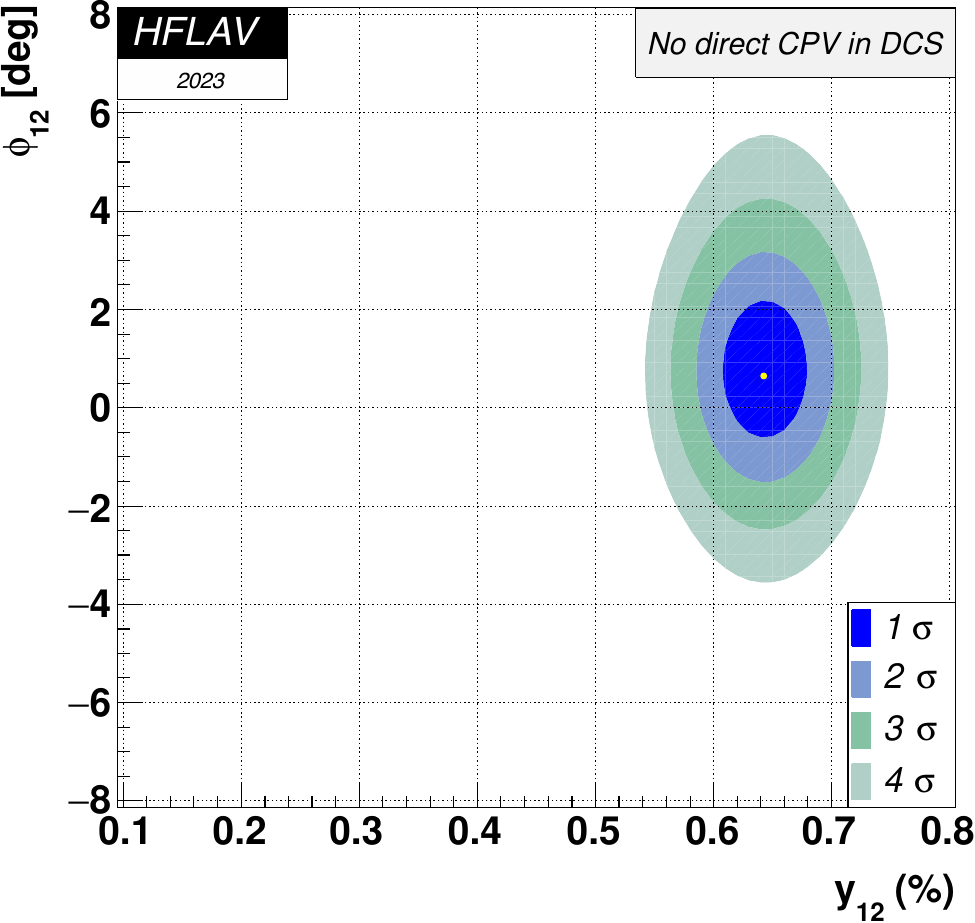}
\includegraphics[width=84mm]{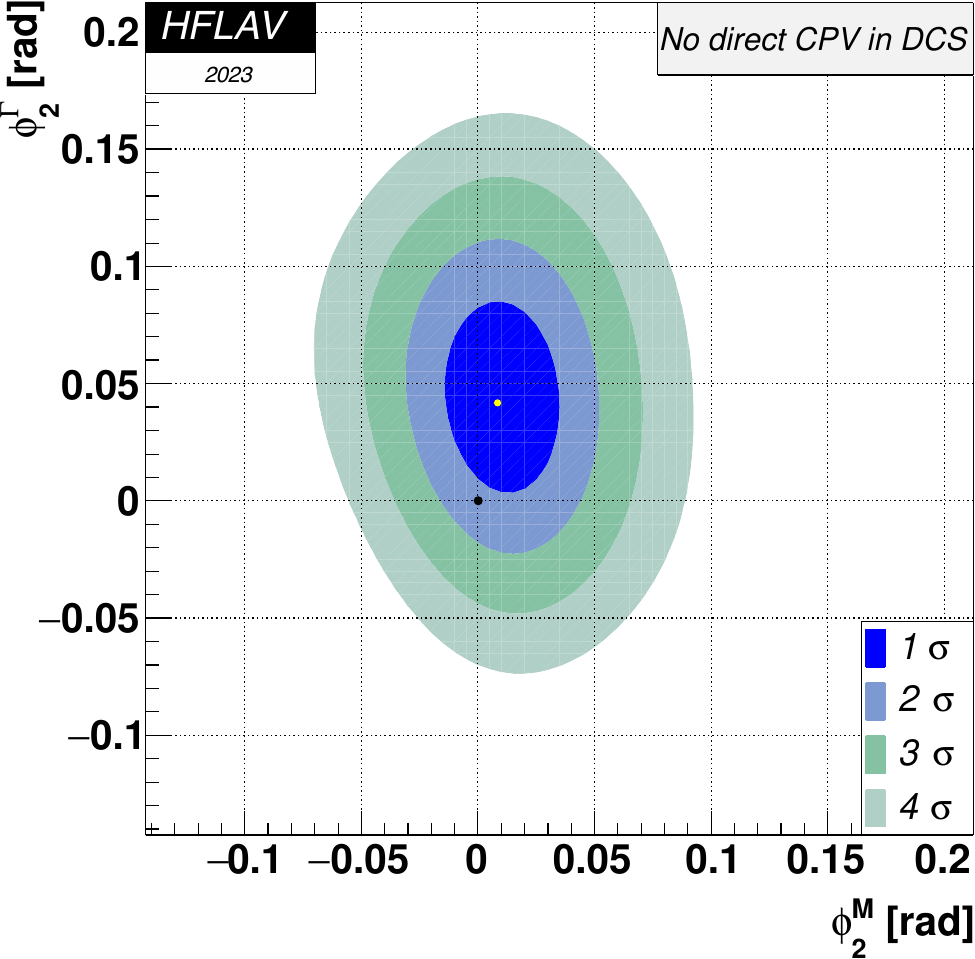}
}
\end{center}
\caption{\label{fig:contours_ndcpv}
Two-dimensional contours for parameters 
$(x^{}_{12},y^{}_{12})$ (top left), 
$(x^{}_{12},\phi^{}_{12})$ (top right), and 
$(y^{}_{12},\phi^{}_{12})$ (bottom left), resulting from Fit~\#1.
Contours for $(\phi^{M}_2,\phi^{\Gamma}_2)$ resulting 
from Fit~\#2 are shown in the bottom right.
}
\end{figure}

\begin{figure}
\begin{center}
\vbox{
\includegraphics[width=94mm]{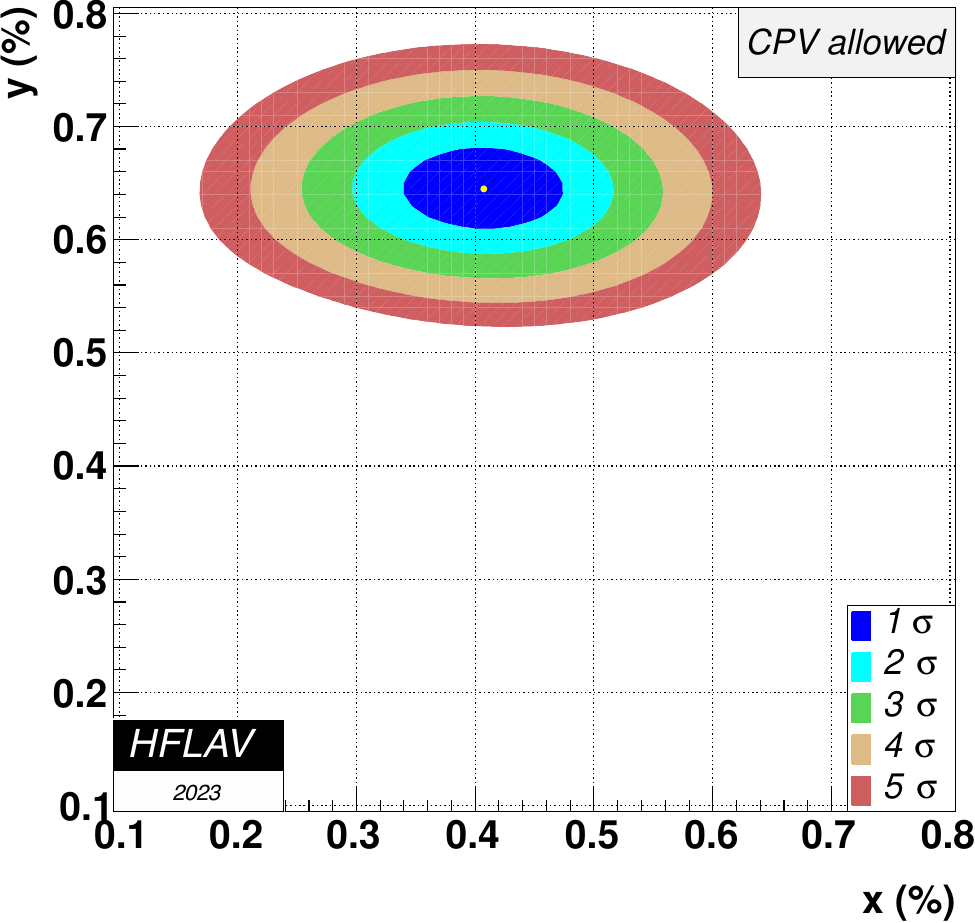}
\vskip0.20in
\includegraphics[width=94mm]{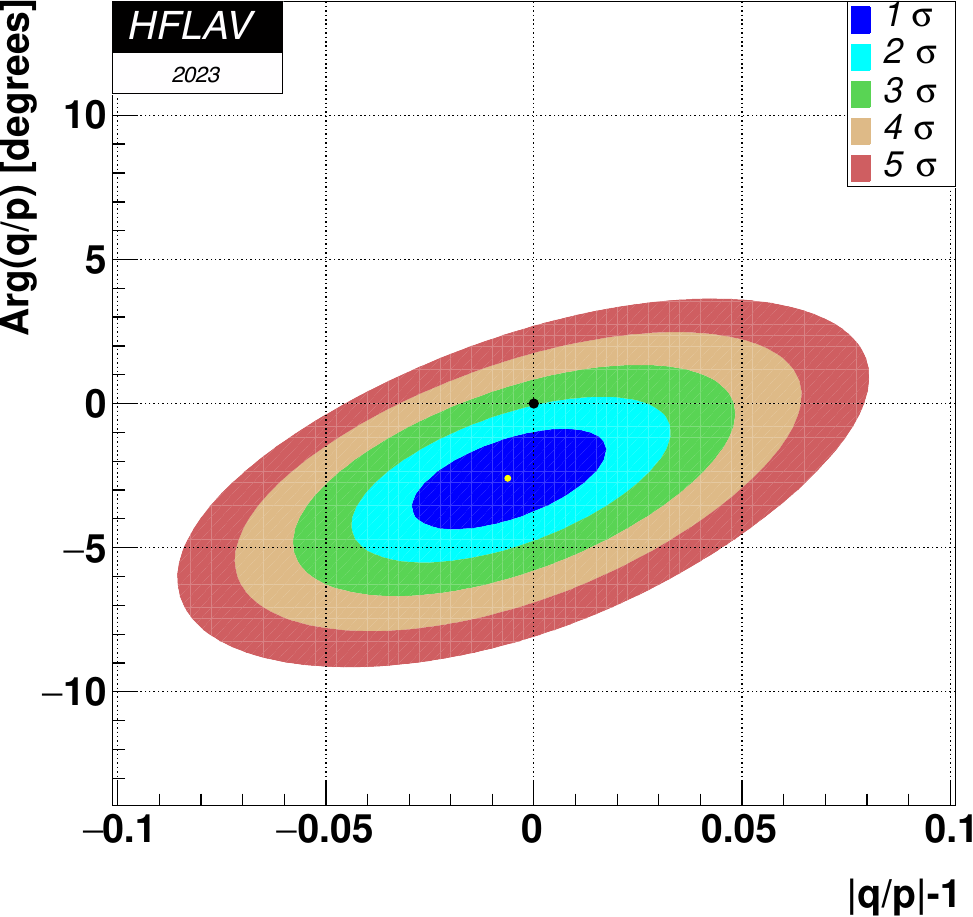}
}
\end{center}
\caption{\label{fig:contours_cpv}
Two-dimensional contours for parameters $(x,y)$ (upper) 
and $(|q/p|-1,\phi)$ (lower), allowing for \cpv\ (Fit~\#3).}
\end{figure}

\begin{figure}
\begin{center}
\vbox{
\includegraphics[width=68mm]{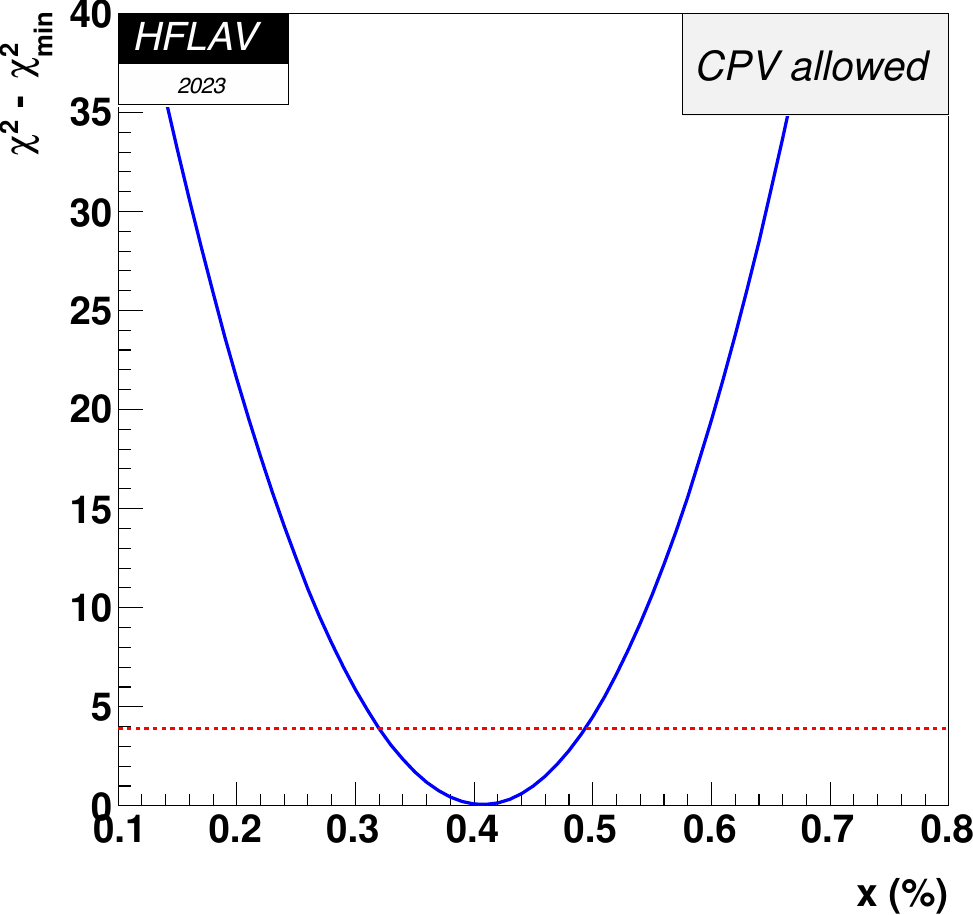}
\includegraphics[width=68mm]{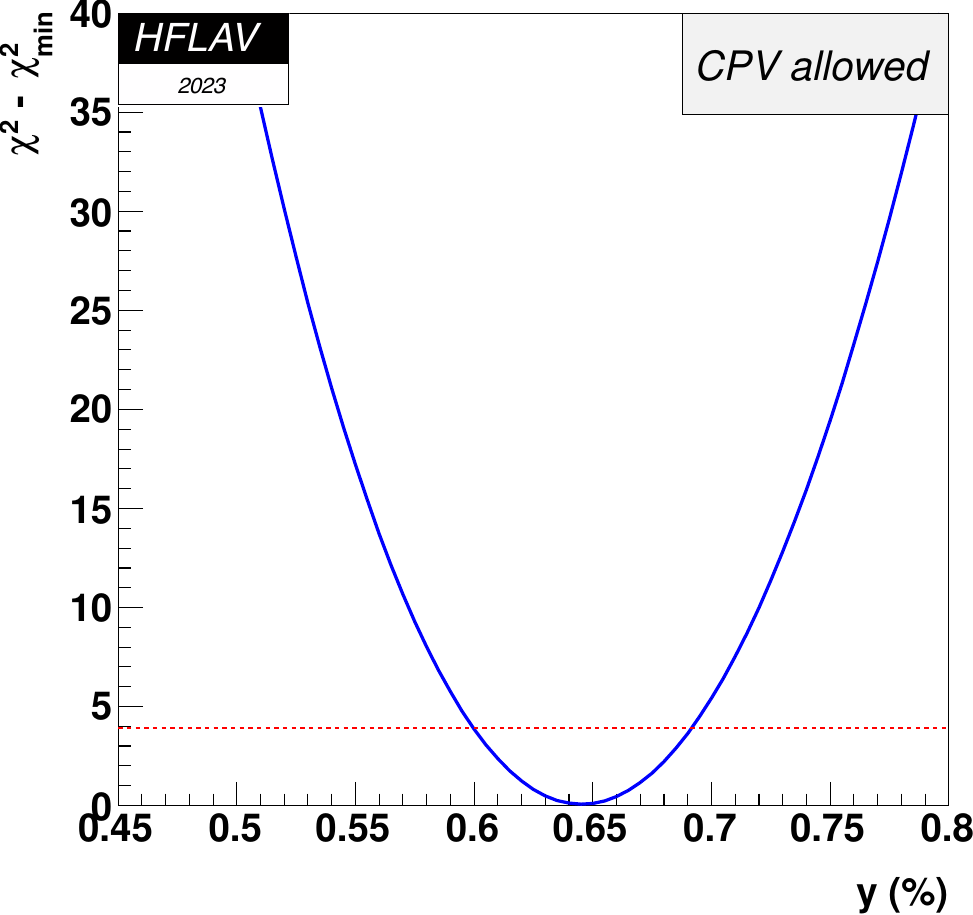}
\vskip0.20in
\includegraphics[width=68mm]{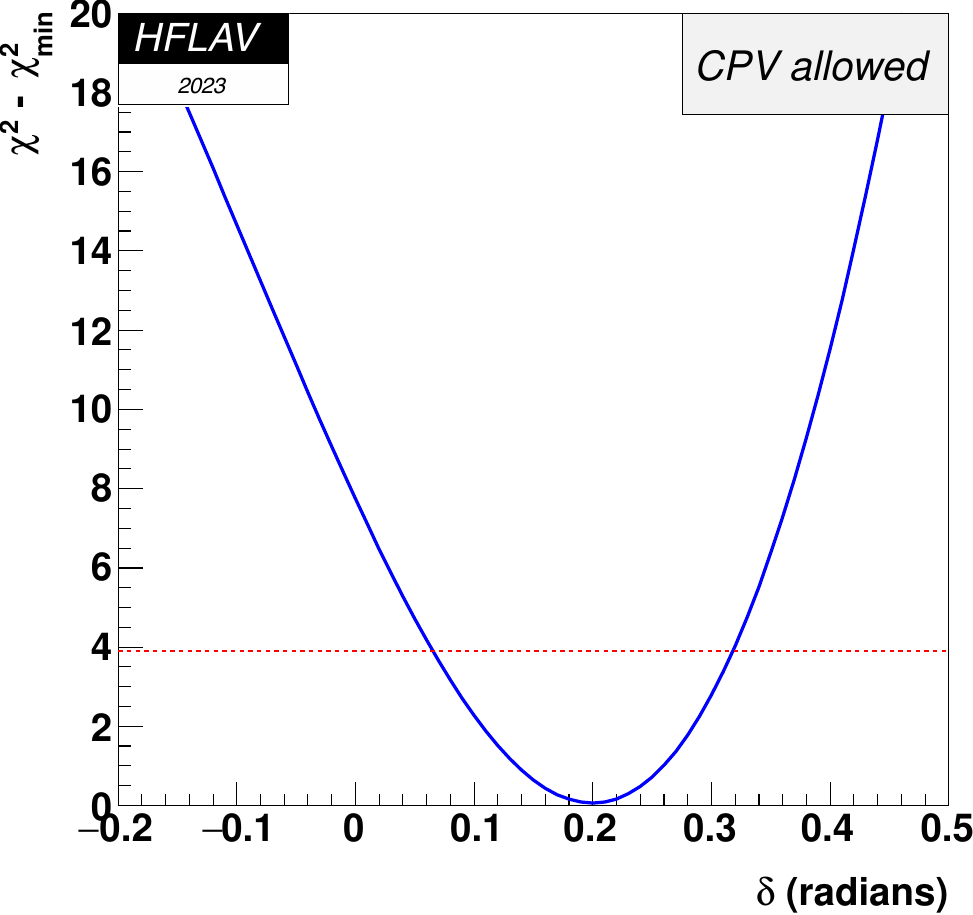}
\includegraphics[width=68mm]{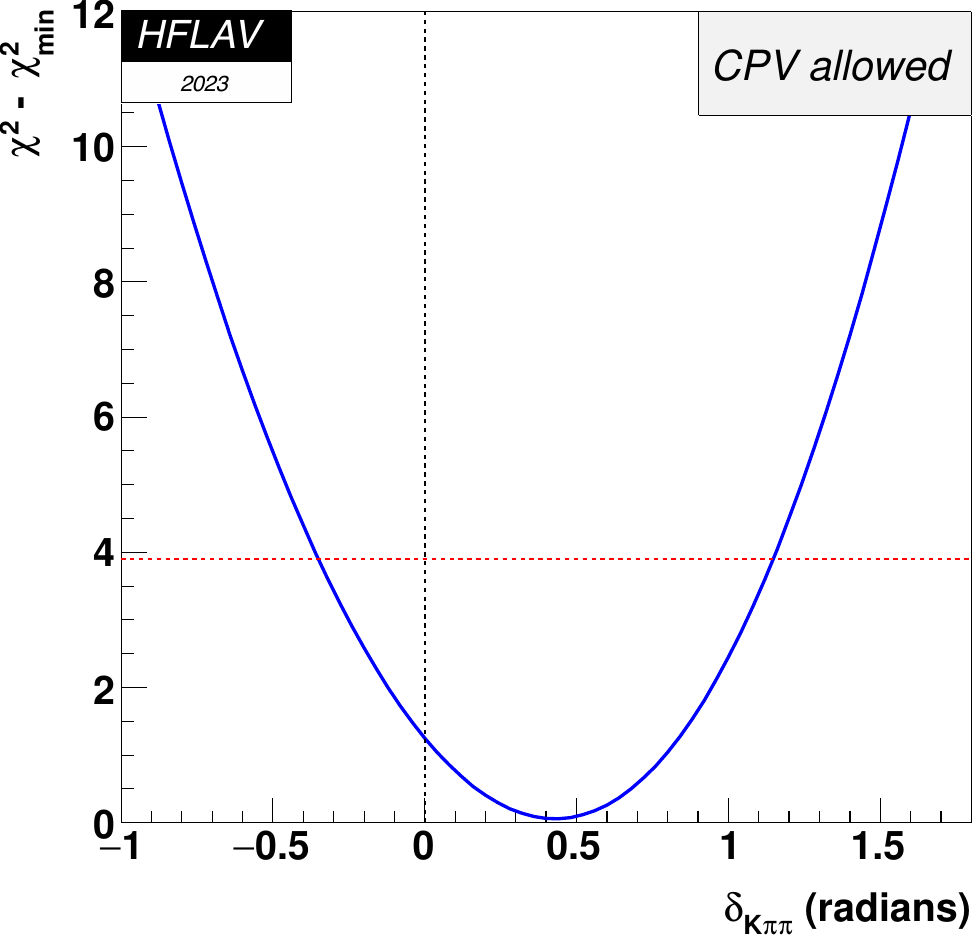}
\vskip0.20in
\includegraphics[width=68mm]{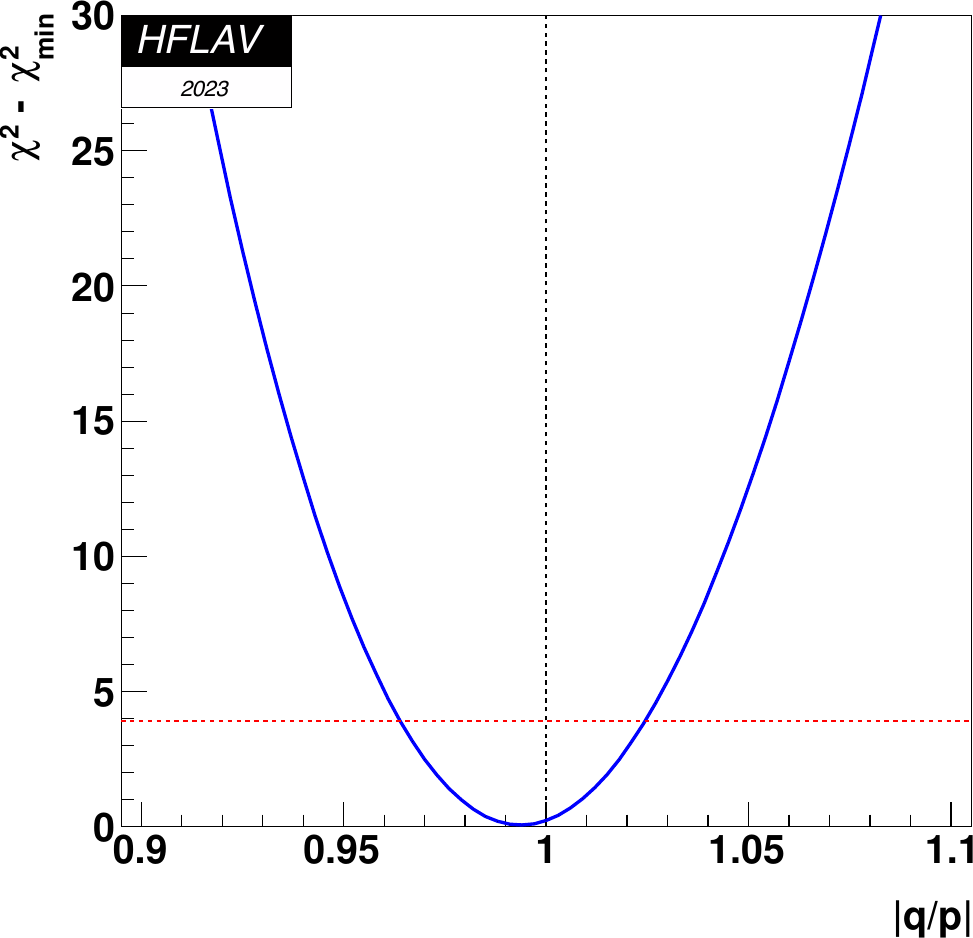}
\includegraphics[width=68mm]{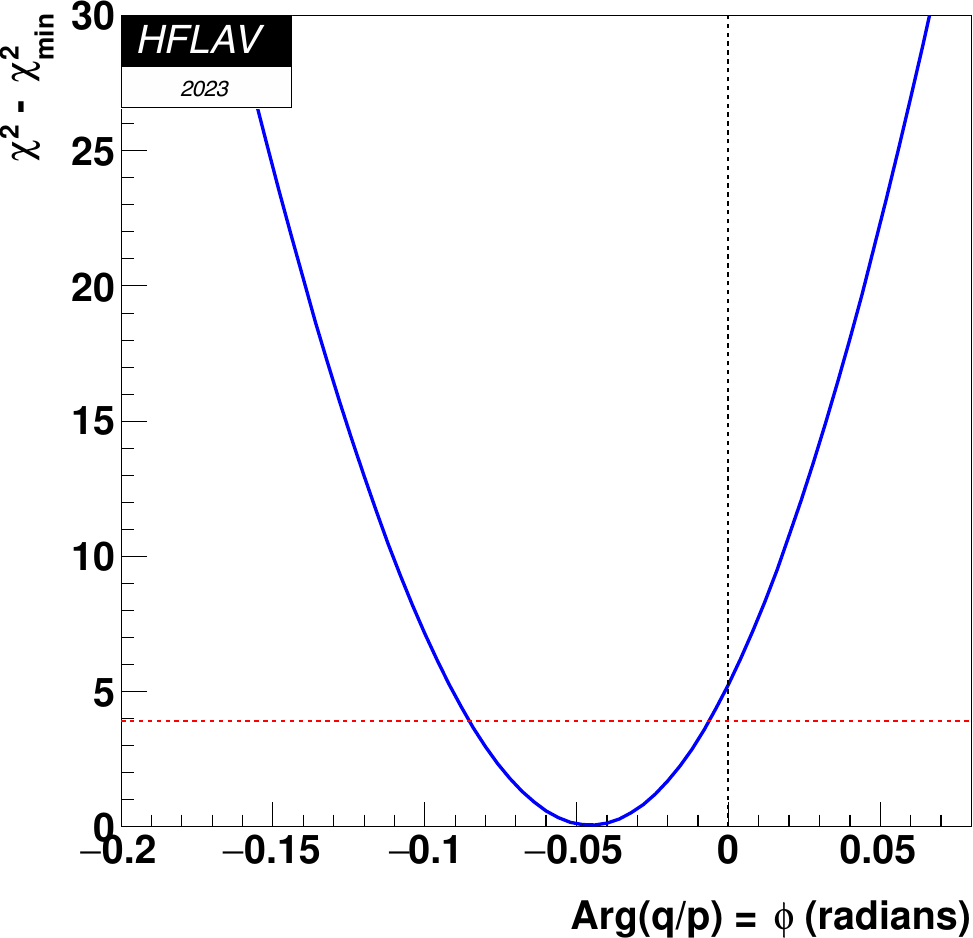}
}
\end{center}
\caption{\label{fig:1dlikelihood}
The function $\Delta\chi^2=\chi^2-\chi^2_{\rm min}$ for parameters
$x,\,y,\,\delta,\,\delta^{}_{K\pi\pi},\,|q/p|$, and $\phi$,
from Fit~\#3. The points where $\Delta\chi^2=3.84$ (denoted 
by dashed horizontal lines) determine 95\% C.L. intervals. }
\end{figure}

\subsubsection{Conclusions}

From the results listed in Table~\ref{tab:results}
and shown in Figs.~\ref{fig:contours_cpv} and \ref{fig:1dlikelihood},
we conclude the following:
\begin{itemize}
\item 
The experimental data consistently demonstrate $D^0$-$\dbar$ mixing. 
The no-mixing point $x\!=\!y\!=\!0$ is excluded at more 
than~$11.5\sigma$. 
The parameter $x$ differs from zero with a significance of $9.1\sigma$, 
and $y$ differs from zero with a significance of $35\sigma$. 
Within the SM, mixing at the observed level is too large to be accounted 
for by short-distance amplitudes and must be dominated by long-distance 
processes, which are difficult to calculate.
\item 
Since \ycp\ is positive, the mostly \cp-even state is shorter-lived,
as in the $K^0$-$\kbar$ system. However, since $x$ is positive, 
the mostly \cp-even state is heavier, unlike in the $K^0$-$\kbar$ system.
\item 
There is no significant evidence for indirect \cpv\ arising from 
$D^0$-$\dbar$ mixing ($|q/p|\neq 1$) or from a phase difference between
the mixing amplitude and a direct decay amplitude ($\phi\neq 0$). 
The fitted values for these parameters differ from the
no-\cpv\ case with a significance of $2.1\sigma$, 
and more data is needed to establish whether indirect \cpv\ is present. 
We note that small {\it direct\/} \cpv\ (at the level of 0.15\%) 
has been observed in time-integrated $D^0\ra K^+K^-\!\!,\,\pi^+\pi^-$ 
decays by LHCb~\cite{LHCb:2019hro}, as listed in Table~\ref{tab:observables3}. Several theory calculations 
indicate this value is consistent with Standard Model expectations, 
although new physics contributions cannot be excluded. Direct \cp\ 
asymmetries are discussed in Sec.~\ref{sec:cp_asym}.
\end{itemize}

\clearpage
\mysubsection{\CP\ asymmetries}\label{sec:cp_asym}

\mysubsubsection{Introduction}\label{sec:cp_asym:intro}

One manifestation of \CP\ violation is a difference in decay
rates between that of a particle and that of its \CP-conjugate 
anti-particle~\cite{Bigi:2000yz}. 
Such phenomena can be classified into two broad categories: 
{\it direct\/} \CP\ violation and 
{\it indirect\/} \CP\ violation~\cite{Nir:1999mg}. 

Direct \CP\ violation refers to charm-changing $\Delta C\!=\!1$
processes and can occur in both charged and neutral charm hadron decays. 
It results from interference between at least two different decay amplitudes, \eg, a 
penguin amplitude and a tree amplitude, that have different weak and strong phases. 
The weak phase difference between the interfering amplitudes ($\Delta\phi$) has opposite signs for $D\ra f$ 
and $\bar{D}\ra\bar{f}$ decays, while the strong phase difference ($\Delta\delta$)
has the same sign. As a result, squaring the total amplitudes 
to obtain the decay rates gives an interference term proportional to
$\cos(\Delta\phi + \Delta\delta)$ for $D\ra f$ decays, and proportional to
$\cos(-\Delta\phi + \Delta\delta)$ for $\bar{D}\ra\bar{f}$ decays. Thus, the 
decay rates differ. This difference is time-independent and can be measured in time-integrated measurements.

In the Standard Model (SM), the strong-phase difference can arise
due to differences in the final-state interactions (FSI)\cite{Buccella:1994nf}, isospin 
amplitudes, intermediate-resonance contributions, or partial waves of the interfering decay amplitudes.
A difference in weak phases arises from different CKM vertex factors, as 
is often the case for tree and penguin diagrams. Within the SM, direct 
\CP\ violation is expected only in singly Cabibbo-suppressed (SCS) charm 
decays, as only these decays receive a non-negligible contribution from the penguin amplitude. At a short distance, the $c\to u$ penguin amplitude is severely suppressed by CKM factors and the GIM mechanism~\cite{Glashow:1970gm}. Long-distance effects can lift the GIM cancellation between $d$- and $s$-quark loops, and increase the penguin contribution. Direct \CP\ violation depends on the decay mode, 
and the \CP\ asymmetries can reach the permille level. 

Indirect \CP\ violation 
refers to $\Delta C\!=\!2$ processes and arises in $D^0$ decays due to 
$D^0$-$\dbar$ mixing. It can occur as an asymmetry in the mixing itself, or result from interference between a decay amplitude following mixing and a 
non-mixed amplitude. Thus, the final state must be reachable from both 
$D^0$ and $\overline{D}{}^0$ decays. Within the SM,  
charm indirect \CP\ violation is expected 
to be universal, \ie, independent of final state. Current experimental limits 
on indirect \CP violation are discussed in Sec.~\ref{sec:charm:mixcpv}. 

The time-integrated \CP\ asymmetry $A_{\CP}$ is defined as the difference 
between $D$ and $\bar{D}$ partial widths to a final state $f$ divided by 
their sum:
\begin{eqnarray}  
A_{\CP} & = & \frac{\Gamma(D\to f)-\Gamma(\bar{D}\to f)}
{\Gamma(D\to f)+\Gamma(\bar{D}\to f)}\,.
\end{eqnarray}
In the case of $D^+$ and $D^+_s$ decays, $A^{}_{\CP}$ measures 
direct \CP\ violation; in the case of $D^0$ decays, $A^{}_{\CP}$ 
measures direct and indirect \CP\ violation combined 
(see also Sec.~\ref{sec:charm:cpvdir}). Given experimental 
constraints on $A_\Gamma$ (see Table~\ref{tab:Agamma}),
a contribution from indirect \CP\ violation 
would be negligible compared to current $A_{\CP}$ sensitivities.

In multi-body decays, which usually proceed via intermediate states, strong phase differences typically vary over the phase space. Thus, local \CP\ asymmetries, 
\ie, those corresponding to a local region of phase space or those involving specific intermediate states, can offer better sensitivity to \CP\ violation than a global asymmetry. Probing the multi-body phase space is often done 
in a model-dependent way by employing a Dalitz-plot analysis or, more generally, an amplitude analysis, separately for $D$ and $\bar{D}$ decays. A \CP\ asymmetry 
is then determined for each contributing amplitude. The \cp-violating 
observables are asymmetries in the magnitudes and phases of \cp-conjugate 
amplitudes, as well as asymmetries in the amplitude fit fractions. 

For multi-body decays, some experiments use model-independent 
techniques to search for local \CP\ asymmetries. One approach, the so-called Miranda technique  
(see Refs.~\cite{Aubert:2008yd, Bediaga:2009tr}) uses a binned 
$\chi^2$ approach to compare the relative density in a bin 
of phase space for $D\to f$ with that of the \CP-conjugate decay.
Another technique (the {\it Energy Test} technique~\cite{doi:10.1080/00949650410001661440})
uses a test statistic variable ($T$) to determine the average distance between
events in phase space. If the distribution of events in two \CP-conjugate samples
are identical (the \CP-symmetric case), $T$ will fluctuate around a value close to zero.
Both techniques yield a $p$-value for the no-\CP\ violation hypothesis and
identifies \CP-asymmetric phase-space regions.

\mysubsubsection{Nuisance asymmetries}\label{sec:cp_asym:nuisance}

In each experiment, care must be taken to correct for production 
and detection asymmetries, as they can reach the percent level. 
To take into account differences in production rates between 
$D$ and $\bar{D}$, which would affect the number of respective 
decays observed, some experiments (such as E791 and FOCUS) normalize 
$A_{\CP}$ to that measured in a Cabibbo-favored (CF) mode. 
This method relies on the absence of \CP\ violation in this so-called normalization mode. Explicitly, the \CP\ asymmetry is calculated as
\begin{eqnarray}
A_{\CP} &=&\frac{\eta(D)-\eta(\bar{D})}{\eta(D)+\eta(\bar{D})}\,,
\end{eqnarray}
where (considering, for example, $D^0 \to K^+K^-$)
\begin{eqnarray}
 \eta(D) &=& \frac{N(D^0 \rightarrow K^+K^-)}{N(D^0 \rightarrow K^-\pi^+)}\,, \\
 \eta(\bar{D}) & = & \frac{N(\dbar\rightarrow K^-K^+)}
{N(\dbar\rightarrow K^+\pi^-)}\,,
\end{eqnarray}
and $N(D\!\rightarrow\!f)$ is the number of $D\!\rightarrow\!f$ decays reconstructed.
This method has the additional advantage that most corrections due to 
reconstruction inefficiencies cancel out, reducing systematic uncertainties. Possible detection asymmetries between the $K^-\pi^+$ and $K^+\pi^+$ systems, which are introduced in this approach without further corrections, are negligible compared to the limited statistical sensitivities of these past experiments. 

Other experiments (such as Belle and LHCb) determine $A_{\CP}$ via the relation
\begin{eqnarray}
A_{\rm meas} & = & A_{\CP} + A_{\rm prod} + A_{\rm det}\,,
\end{eqnarray}
where $A_{\rm meas}$ is the measured (raw) asymmetry, $A_{\rm prod}$ is 
the asymmetry in the charm hadron production, and $A_{\rm det}$ 
is due to a difference in detection efficiencies between positively 
and negatively charged hadrons.
The production asymmetry at the LHC arises from a charge asymmetry of 
the colliding particles: in $pp$ collisions more charm baryons are
produced than anti-baryons, and, as a result, charm mesons are 
less abundantly produced than anti-charm mesons. 
Though not yet experimentally confirmed~\cite{LHCb:2012fb}, \cite{LHCb:2012fb}, 
such a production asymmetry is expected to be dependent on the kinematics 
of the produced charm hadrons. 
The production asymmetry in $e^+e^-$ 
collisions appears as a forward-backward (FB) asymmetry caused by an 
interference of the off-shell photon and $Z^0$ contributions. The detection 
asymmetries typically arise from differences in hadron interactions 
with detector material. In particular, the interaction cross sections 
for $K^+$ and $K^-$ significantly differ, with the differences being 
dependent on the kaon momentum. 

The $B$-factory strategy to separate the production and \CP\ 
asymmetries relies on the former being odd, while the latter being even, 
with respect to the center-of-mass production polar angle ($\theta^*$). 
The $A_{\rm meas}$ is measured in $|\cos \theta^*|$ bins and subsequently 
averaged; this removes the $A_{\rm prod}$ contribution. The soft $\pi$ detection asymmetry is estimated by Belle by comparing $A_{\rm meas}$ for $D^*$-tagged $D^0 \to K^- \pi^+$ decays with that for untagged $D^0 \to K^- \pi^+$ decays, both measured in $|\cos \theta^*|$ bins. 

At LHCb, the production asymmetry is removed by measuring $A_{\rm meas}$ for
$D^*$-tagged $D^0 \to K^- \pi^+$ decays; this also corrects for the 
soft $\pi$ detection asymmetry.  
Subsequently, $D^+ \to K^- \pi^+ \pi^+$ decays are used to correct for 
the detection asymmetry introduced by the $K^- \pi^+$ system itself, and  
$D^+ \to K_S \pi^+$ decays are then used to remove the asymmetries in $D^+$ 
production and $\pi^+$ detection. 
Finally, the asymmetry related to the neutral kaon~\cite{Grossman:2012aa}, 
\ie, from regeneration and different interactions of $K^0$ and $\kbar$ 
with the detector, as well as from \CP\ violation occurring in 
the $K^0 \text{-} \kbar$ mixing, is calculated. Put together, this gives
\begin{equation}
A_{\CP} (K^+K^-) = A_{\rm meas}(K^+K^-) - A_{\rm meas} (K^- \pi^+) 
+ A_{\rm meas}(K^- \pi^+ \pi^+) - A_{\rm meas}(K_S \pi^+) 
+ A(\kbar \text{-} K^0).
\label{eq:nuisance:ACP_KK}
\end{equation}
For some decays, typically the ones with lower statistics, one corrects 
for nuisance asymmetries by measuring $A_{\CP}$ relative to 
some well-measured reference channel, for instance
\begin{equation}
A_{\CP} (D^+\to \eta^{'}\pi^+) = A_{\rm meas}(D^+\to \eta^{'}\pi^+) 
- A_{\rm meas} (D^+\to \phi \pi^+)+ A_{\CP}(D^+\to \phi \pi^+).
\end{equation}
The uncertainty of the reference $A_{\CP}$, here $A_{\CP}(D^+\to \phi \pi^+)$, is treated as an external-input uncertainty.
$A_{\CP}(D^+\to \phi \pi^+)$  itself is precisely measured by LHCb %
exploiting $D^+ \to K_S \pi^+$ reference channel, in which \cp \ violation is expected only due to the neutral kaon. Therefore one gets  
\begin{equation}
A_{\CP}(D^+\to \phi \pi^+) = A_{\rm meas}(D^+\to  \phi \pi^+) 
- A_{\rm meas} (D^+\to K_S \pi^+)+ A(\kbar \text{-} K^0).
\end{equation}

BESIII performed measurements of $A_{\CP}$ for $D_{(s)}^+$ decays using $e^+e^-$ collision data collected at the $D_{(s)}\bar{D}_{(s)}$ threshold. Employing the double-tag technique, where both charm mesons produced are reconstructed, results in quite a limited statistical sensitivity. Therefore any impact of the production asymmetry, expected to be smaller than at the $B$-factories, would have a negligible impact and is not corrected for.

\mysubsubsection{Measurements for charmed mesons}\label{sec:cp_asym:mesons}

Measurements of $A_{\CP}$ {\it differences}, 
denoted $\Delta A_{\CP}$, are often easier 
to interpret theoretically than individual 
$A_{\CP}$ measurements. The most important 
difference is that for $D^0\to K^+K^-$ and $D^0\to \pi^+\pi^-$ decays, which is discussed 
in Sec.~\ref{sec:charm:cpvdir}. Notably, its measurement by LHCb, 
$\Delta A_{\CP} = (-15.4 \pm 2.9) \times 10^{-4}$, constitutes
the first observation of \CP violation in the charm sector~\cite{LHCb:2019hro}.  We note that, in the
limit of $U$-spin symmetry, direct \CP\ violation in $D^0\to K^+K^-$ and 
$D^0\to \pi^+\pi^-$ decays is expected to have equal magnitude but opposite
sign~\cite{Brod:2012ud}; 
thus the measurement of $\Delta A_{\CP}$ would {\it double} the effect. The 
measured $\Delta A_{CP}$ value is at the upper end of the SM predictions. This triggers many interpretations, some involve New Physics origins~\cite{Chala:2019fdb}, \cite{Dery:2019ysp}, \cite{Bause:2020obd}, while others explain the experimental observation by enhanced long-distance effects~\cite{Grossman:2019xcj}, 
\cite{Buccella:2019kpn}, \cite{Cheng:2019ggx}, \cite{Schacht:2021jaz}, \cite{Bediaga:2022sxw}, \cite{Pich:2023kim}, \cite{Gavrilova:2023fzy}.

Results of measurements of $A^{}_{\CP}$ for $D^+$, $D^0$ and $D_s^+$ 
decays are 
listed in Tables~\ref{tab:cp_charged}, \ref{tab:cp_charged2}, \ref{tab:cp_neutral}, \ref{tab:cp_neutral2}, \ref{tab:cp_neutral_rare}, \ref{tab:cp_neutral_binned}, \ref{tab:cp_ds}, and \ref{tab:cp_ds2}. Overall, \CP\ asymmetries have been measured for more than 60 charm 
decay modes, and in several modes the uncertainty on $A^{}_{\CP}$ is well below 
$5 \times 10^{-3}$. \CP\ violation has been observed only in the difference of $A_CP$ asymmetries of the modes 
$D^0\to K^+K^-$ and $D^0\to\pi^+\pi^-$. %
The most recent LHCb measurement~\cite{LHCb:2022lry} of the individual \CP\ asymmetries for these decays leads to an evidence of \CP\ violation in $D^0\to\pi^+\pi^-$, $A^{}_{\CP}(D^0\to\pi^+\pi^-) = (+2.32 \pm 0.61) \times 10^{-3}$. Interestingly, the sum of the direct \CP \ asymmetries for $D^0\to K^+K^-$ and $D^0\to\pi^+\pi^-$ decays is non-zero at $2.7 \,\sigma$, giving an evidence for $U$-spin symmetry breaking. 

Modes with a single $K^{}_S$ meson 
in the final state can exhibit a \CP\ asymmetry due to \CP\ violation
in $K^0$-$\kbar$ mixing~\cite{Grossman:2012aa}; \ie, the rate for 
$\kbar\ra K^{}_S$ differs slightly from that for $K^0\ra K^{}_S$. 
The \CP\ asymmetry observed for the mode $D^+ \to K^{}_S\,\pi^+$ (see Table~\ref{tab:cp_charged})
is consistent with that expected from $K^0\text{-}\kbar$ 
mixing~\cite{Grossman:2012aa}. Thus, it is not attributed to 
direct \CP\ violation in charm but to the $K^0\text{-}\kbar$ mixing.
For modes indicated with a $K^0$ or $\kbar$ 
in the final state,  the table entries are already corrected for this effect. This correction, denoted in Eq.~\ref{eq:nuisance:ACP_KK} as $A(\kbar \text{-} K^0)$, is at the level of $10^{-4} \div 10^{-3}$, and has the negative sign. For a given experiment, the value of this correction depends on the experimental conditions, such as measured distributions of the momentum and decay-time of $K^{}_S$ mesons. 
Some of the BESIII measurements are for final states involving $K_L$~\cite{BESIII:2018pku}, \cite{BESIII:2019kfh}, which makes them unique. 
The asymmetry for the DCS decay $D^0 \to K^+\pi^-$ is not included 
in these tables, as it is a by-product of charm-mixing measurements 
and thus is discussed in Sec.~\ref{sec:charm:mixcpv} (where it is 
referred to as $A_D$).

The asymmetries for three- 
and four-body decays (Tables~ \ref{tab:cp_charged2}, \ref{tab:cp_neutral2}, and \ref{tab:cp_ds2}) are given for their observed final state, \ie, 
resonant substructure is implicitly included but not considered 
separately. Most asymmetries measured for three- and four-body 
channels are still only global asymmetries. The reported model-independent tests, which attempt to probe 
the decay phase space, yield $p$-values typically at the level of 
a few percent or higher and thus are consistent with no \CP\ violation. 
The lowest $p$-value of 0.6\%, corresponding to a significance for 
\CP\ violation of $2.7\sigma$, is obtained for the $P$-odd (parity-odd)
test of $D^0 \to \pi^+\pi^-\pi^+\pi^-$ decay~\cite{Aaij:2013aa}. This
implies that the effect, if not a statistical fluctuation, originates
in a $P$-odd amplitude such as $D^0 \to [\rho^0 \rho^0]_{L=1}$.
For $D^0 \to K^+K^-\pi^+\pi^-$ decay~\cite{Aaij:2018nis},
a model-dependent amplitude analysis was performed, and
\CP\ asymmetries were measured for 25 intermediate amplitudes.
The uncertainties on these asymmetries ranged from 1\% to 15\%
and were dominated by statistical errors.
No significant \CP\ violation was observed, and the most
significant asymmetry of $2.8\sigma$ was observed for the
phase of the $P$-odd amplitude $D^0 \to [\phi(1020) \rho(1450)^0]_{L=1}$. 
\cp\ violation arising through $P$ violation is 
discussed further in Sec.~\ref{sec:todd_asym}. LHCb is still analyzing many multi-body decays, some with yields of the order of $10^8$ events. 

\CP\ asymmetries have also been measured for decays classified as rare:
the radiative modes $D^0 \to V \gamma$, with $V=\ \bar{K}^{*0}, \ \phi(1020), 
\ \rho^{0}$, as well as the di-muon decays $D^0 \to \pi^+ \pi^- \mu^+ \mu^-$
and $D^0 \to K^+ K^- \mu^+ \mu^-$ (see Table~\ref{tab:cp_neutral_rare}). 
For the di-muon modes, in addition to their global asymmetries listed 
in Table~\ref{tab:cp_neutral2},
\CP\ asymmetries in bins of the di-muon invariant mass
have been measured by LHCb for the ranges with significant signal yields. They are given in Table~\ref{tab:cp_neutral_binned}. Asymmetries for mass regions away from 
$\mu^+\mu^-$ production via 
$\eta$, $\rho\text{-}\omega$ or $\phi$ decays still have
very limited sensitivities, with uncertainties ranging from 12\% to 26\%. These 
non-resonant regions are particularly important for New Physics 
searches (see Sec.~\ref{sec:charm:rare}).

\begin{table}[!htb]
\renewcommand{\arraystretch}{1.4}
\caption{\CP\ asymmetries 
$A^{}_{\CP}= [\Gamma(D^+)-\Gamma(D^-)]/[\Gamma(D^+)+\Gamma(D^-)]$
for two-body $D^\pm$ decays. For each entry, the first uncertainty is statistical, 
and the second is systematic. %
\label{tab:cp_charged}}
\footnotesize
\begin{center}

\end{center} 
\end{table}

\begin{table}[!htb]
\renewcommand{\arraystretch}{1.4}
\caption{\CP\ asymmetries 
$A^{}_{\CP}= [\Gamma(D^+)-\Gamma(D^-)]/[\Gamma(D^+)+\Gamma(D^-)]$
for three- and four-body $D^\pm$ decays. For each entry, the first uncertainty is statistical,
and the second (if quoted) is systematic.
\label{tab:cp_charged2}}
\footnotesize
\begin{center}
%
\end{center} 
\end{table}

\begin{table}[!htb]
\renewcommand{\arraystretch}{1.3}
\caption{\CP\ asymmetries 
$A^{}_{\CP}=[\Gamma(D^0)-\Gamma(\dbar)]/[\Gamma(D^0)+\Gamma(\dbar)]$
for two-body $D^0,\dbar$ decays. 
In each entry, the first uncertainty is statistical, and the second (if quoted) is systematic, 
unless explicitly stated that they have been combined.
The third uncertainty in the Belle and LHCb $A^{}_{\CP}(D^0 \to K_S K_S)$ measurements is due to $A^{}_{\CP}$ of the normalization channels $D^0 \to K_S \pi^0$ (Belle) and $D^0 \to K^+ K^-$ (LHCb).
\label{tab:cp_neutral}}
\footnotesize
\begin{center}

\end{center} 
\end{table}

\begin{table}[!htb]
\renewcommand{\arraystretch}{1.3}
\caption{\CP\ asymmetries 
$A^{}_{\CP}=[\Gamma(D^0)-\Gamma(\dbar)]/[\Gamma(D^0)+\Gamma(\dbar)]$
for three- and four-body $D^0,\dbar$ decays. 
In each entry, the first uncertainty is statistical, and the second (if quoted) is systematic, unless explicitly stated that they have been combined.
\label{tab:cp_neutral2}}
\footnotesize
\begin{center}
%
\end{center} 
\end{table}

\begin{table}[!htb]
\renewcommand{\arraystretch}{1.3}
\caption{\CP\ asymmetries 
$A^{}_{\CP}=[\Gamma(D^0)-\Gamma(\dbar)]/[\Gamma(D^0)+\Gamma(\dbar)]$
for rare $D^0,\dbar$ decays. 
In each entry, the first uncertainty is statistical, and the second is systematic.
\label{tab:cp_neutral_rare}}
\footnotesize
\begin{center}
%
\end{center} 
\end{table}

\begin{table}[!htb]
\renewcommand{\arraystretch}{1.3}
\caption{\CP\ asymmetries of $D^0, \bar{D}^0 \to h^+h^- \mu^+ \mu^-$ decays in different dimuon invariant mass ranges, measured by LHCb. 
In each entry, the first uncertainty is statistical, and the second is systematic. Measurements are not performed for mass intervals with
insignificant signal yields. 
\label{tab:cp_neutral_binned}}
\footnotesize
\begin{center}
%
\end{center} 
\end{table}

\begin{table}[!htb]
\renewcommand{\arraystretch}{1.4}
\caption{\CP\ asymmetries 
$A^{}_{\CP}=[\Gamma(D_s^+)-\Gamma(D_s^-)]/[\Gamma(D_s^+)+\Gamma(D_s^-)]$ for two-body $D_s^\pm$ decays. In each entry, the first
uncertainty is statistical and the second is systematic.  
\label{tab:cp_ds}}
\footnotesize
\begin{center}
%
\end{center} 
\end{table}

\begin{table}[!htb]
\renewcommand{\arraystretch}{1.4}
\caption{\CP\ asymmetries 
$A^{}_{\CP}=[\Gamma(D_s^+)-\Gamma(D_s^-)]/[\Gamma(D_s^+)+\Gamma(D_s^-)]$ for multi-body $D_s^\pm$ decays. In each entry, the first
uncertainty is statistical, and the second is systematic. 
\label{tab:cp_ds2}}
\footnotesize
\begin{center}
%
\end{center} 
\end{table}

\mysubsubsection{Measurements for charmed baryons}\label{sec:cp_asym:baryons}

In the charm baryon sector, there is no evidence of \CP\ violation.
Until recently, there were only two measurements 
for $\Lambda_c^+$; these were performed by CLEO~\cite{Hinson:2004pj} 
and FOCUS~\cite{Link:2005ft} and had limited sensitivity.
The CLEO measurement used the semileptonic decay 
$\Lambda_c^+ \to \Lambda e^+ \nu_{e}$, while the FOCUS measurement 
used the CF decay $\Lambda_c^+ \to \Lambda \pi^+$. Both searched 
for \cp\ violation through an angular analysis. 
Exploiting the $\Lambda$ helicity angle,   
\CP\ violation was probed by comparing the $P$ asymmetry 
in decays of $\Lambda_c^+$ and $\bar{\Lambda}_c^-$.
This was done by measuring the weak-asymmetry parameter $\alpha^{}_{\Lambda_c}$, defined for $\Lambda_c^+$ decays as 
\begin{equation}
\alpha^{}_{\Lambda_c} = \frac{\sum _{\lambda_2}|H_{\lambda_1=+\frac{1}{2}, \, \lambda_2}|^2-|H_{\lambda_1=-\frac{1}{2}, \, \lambda_2}|^2}{\sum _{\lambda_2} |H_{\lambda_1=+\frac{1}{2}, \, \lambda_2}|^2+|H_{\lambda_1=-\frac{1}{2}, \, \lambda_2}|^2},
\label{eq:alphaLc}
\end{equation}
and the corresponding parameter $\alpha^{}_{\bar{\Lambda}_c}$ defined for $\bar{\Lambda}_c^-$ decays.
$H_{\lambda_1, \, \lambda_2}$ in Eq.~\ref{eq:alphaLc} is the helicity amplitude with $\lambda_1$ denoting the helicity of $\Lambda$ baryon and $\lambda_2$ denoting the helicity of the accompanying particle, i.e., $\pi^+$ in $\Lambda_c^+ \to \Lambda \pi^+$ decays and $\W^+ \to e^+ \nu_e$ in the semileptonic decays. 
Thus $\alpha^{}_{\Lambda_c}$ ($\alpha^{}_{\bar{\Lambda}_c}$) describes a longitudinal polarisation of the $\Lambda$ ($\bar{\Lambda}$) baryon.

The  $\alpha^{}_{\Lambda_c}$ parameter is measured from the angular distribution for the decay chain $\Lambda_c^+ \to \Lambda \pi^+$ 
 followed by $\Lambda \to p \pi^-$,
\begin{equation}
\frac{d\Gamma}{d \cos{\theta_p}} \simeq  1+ \alpha^{}_{\Lambda_c} \alpha^{}_{\Lambda} \cos{\theta_p},
\label{eq:angularLc}
\end{equation}
where $\theta_p$ is the $\Lambda$ helicity angle, defined as the angle between the proton momentum and the vector opposite to the  $\Lambda_c^+$ momentum, both in the $\Lambda$ rest frame.  A weak-asymmetry parameter for 
$\Lambda \to p \pi^-$ decays, $\alpha^{}_{\Lambda}$, is defined similar to Eq.~\ref{eq:alphaLc} but with $\lambda_1$ denoting  the proton helicities of $\pm \frac{1}{2}$ and $\lambda_2$ being the $\pi^-$ helicity. 
The angular distribution in Eq.~\ref{eq:angularLc} for the charge conjugate process, $\bar{\Lambda}_c^- \to \bar{\Lambda} \pi^-$ with $\bar{\Lambda} \to \bar{p} \pi^+$, involves  
$\alpha^{}_{\bar{\Lambda}_c}$ and $\alpha^{}_{\bar{\Lambda}}$. The weak-asymmetry parameters for the $\Lambda$ and $\bar{\Lambda}$ decays are well measured~\cite{PDG_2020} and used as external parameters in the fits for $\alpha^{}_{\Lambda_c}$ and $\alpha^{}_{\bar{\Lambda}_c}$. The angular distribution of the semileptonic decays $\Lambda_c^+ \to \Lambda e^+ \nu_{e}$ is more complicated, owing to contribution from one more weakly-decaying system, $W^+ \to e^+ \nu_{e}$ (see Sec.~\ref{sec:charm:semileptonic}). Therefore, in addition to the $\Lambda$ helicity angle, it also depends on the $W^+$ helicity angle, the angle between the decay planes of the $W^+$ and $\Lambda$, as well as  $q^2\equiv m^2(e^+ \nu_{e})$, making the CLEO measurement a four-dimensional analysis~\cite{Hinson:2004pj}.

As $\alpha^{}_{\Lambda_c} = - \alpha^{}_{\bar{\Lambda}_c}$ in the case of 
$P$-parity conservation, the \CP-violating asymmetry is defined as
\begin{eqnarray}  
A_{\CP}^{\alpha} & = & 
\frac{ \alpha_{\Lambda_c} + \alpha_{\bar{\Lambda}_c}}
{\alpha_{\Lambda_c} - \alpha_{\bar{\Lambda}_c}} \, .
\end{eqnarray}

Recently, $A_{\CP}^{\alpha}$ has been measured by Belle~\cite{Belle:2022uod} for the SCS decays $\Lambda_c^+ \to \Lambda K^+$ and $\Lambda_c^+ \to \Sigma^0 K^+$, $\Sigma^0 \to \Lambda \gamma$. The method of accessing \cp\ violation occurring through $P$~violation has also been applied by Belle~\cite{Belle:2021crz} to the channel 
$\Xi_c^0 \to \Xi^- \pi^+,\,\Xi^- \to \Lambda \pi^-$. All the measurements of $\alpha$-induced \CP asymmetry are summarised in Table~\ref{tab:cp_alpha_baryons}.

\begin{table}[!htb]
\renewcommand{\arraystretch}{1.4}
\caption{\CP\ asymmetries 
$A^{\alpha}_{\CP}$ induced through the weak-asymmetry parameter $\alpha$, 
for decays of charmed baryons and anti-baryons. For each entry, the first uncertainty is statistical, 
and the second is systematic. The third uncertainty in the CLEO measurement is related to the uncertainty of the $\Lambda$ weak-asymmetry parameter.}
\label{tab:cp_alpha_baryons}
\footnotesize
\begin{center}
\begin{tabular}{|l|c|c|c|} 
\hline
{\bf Mode} & {\bf Year} & {\bf Collaboration} & {\boldmath $A^{\alpha}_{\CP}$} \\
\hline
{\boldmath $\Lambda_c^+ \to \Lambda e^+ \nu_{e}$} &
  2005 & CLEO~\cite{Hinson:2004pj} &  $ 0.00 \pm 0.03 \pm 0.01 \pm 0.02 $ \\
\hline
{\boldmath $\Lambda_c^+ \to \Lambda \pi^+$} &
  2006 & FOCUS~\cite{Hinson:2004pj} &  $ -0.07 \pm 0.19 \pm 0.12 $ \\
  \hline
  {\boldmath $\Lambda_c^+ \to \Lambda K^+$} &
  2022 & Belle~\cite{Belle:2022uod} &  $ -0.023 \pm 0.086 \pm 0.071 $ \\
  \hline
{\boldmath $\Lambda_c^+ \to \Sigma^0 K^+$} &
  2022 & Belle~\cite{Belle:2022uod} &  $ +0.08 \pm 0.35 \pm 0.14 $ \\
  \hline 
{\boldmath $\Xi_c^0 \to \Xi^- \pi^+$} &
  2021 & Belle~\cite{Belle:2021crz} &  $ +0.024 \pm 0.052 \pm 0.014 $ \\
  \hline 
\end{tabular}
\end{center} 
\end{table}

The first \cpv \ measurement with charm baryons and precision reaching the 1\% level  
was performed by LHCb, who measured $A_{\CP}$ difference for 
the  SCS 
decays $\Lambda_c^+\to p K^+K^-$ and $\Lambda_c^+\to p \pi^+\pi^-$~\cite{Aaij:2017xva}, obtaining the result 
\begin{equation*}
\Delta A_{\CP}(\Lambda_c^+ \to p h^+h^-) 
\equiv A_{\CP}(p K^+K^-) - A_{\CP}(p \pi^+\pi^-) 
= 0.003 \pm 0.009 \pm 0.006.
\end{equation*}
However, the $U$-spin argument which motivated measuring the $\Delta A_{\CP}$ for $D^0 \to K^+K^-$ and $D^0 \to \pi^+\pi^-$ decays does not exist for  $\Lambda_c^+\to p K^+K^-$ and $\Lambda_c^+\to p \pi^+\pi^-$ decays. The measurement, performed in a phase-space-integrated manner, 
has limited sensitivity and does not facilitate an interpretation. Nevertheless, the production asymmetry between $\Lambda_c^+$ and $\bar{\Lambda}_c^-$
baryons cancels in this difference.
Given the potentially rich dynamics of these decays in their 
five-dimensional phase space\footnote{For unpolarized $\Lambda_c$, 
the decay phase space reduces to a two-dimensional Dalitz distribution.}, 
$\Delta A_{\CP}$ measured in phase-space regions or a model-dependent
measurement of intermediate amplitude asymmetries would be very desirable. 
The first $A_{\CP}$ measurements for $\Lambda_c^+$ decays have been only recently performed by Belle; they are summarised in Table~\ref{tab:cp_baryons}.

\begin{table}[!htb]
\renewcommand{\arraystretch}{1.4}
\caption{\CP\ asymmetries 
$A^{}_{\CP}$ of decay widths 
for decays of charmed baryons and anti-baryons. For each entry, the first uncertainty is statistical, 
and the second is systematic.}
\label{tab:cp_baryons}
\footnotesize
\begin{center}
\begin{tabular}{|l|c|c|c|} 
\hline
{\bf Mode} & {\bf Year} & {\bf Collaboration} & {\boldmath $A^{}_{\CP}$} \\
\hline
  {\boldmath $\Lambda_c^+ \to \Lambda K^+$} &
  2022 & Belle~\cite{Belle:2022uod} &  $ +0.021 \pm 0.026 \pm 0.001 $ \\
  \hline
{\boldmath $\Lambda_c^+ \to \Sigma^0 K^+$} &
  2022 & Belle~\cite{Belle:2022uod} &  $ +0.025 \pm 0.054 \pm 0.004 $ \\
  \hline 
\end{tabular}
\end{center} 
\end{table}

\mysubsubsection{Sum rules}\label{sec:cp_asym:sumrules}

For charm decays, one can construct various SU(3)-based sum rules which, 
in addition to testing SU(3) symmetry itself, are also useful for 
performing model-independent tests of the SM. Particularly useful 
are sum rules exploiting SU(3) subgroups such as $U$-spin or isospin (I), 
as they involve fewer decays and offer more 
precise tests. While $U$-spin symmetry in charm decays is broken by 
a non-negligible amount due to the $s$-quark mass, isospin symmetry holds 
at the $(m_u-m_d)$ level and thus is very precise. 
Important for consideration are sum rules that relate 
individual \cp\ asymmetries of isospin-related processes; 
such rules allow for tests to be performed that have reduced 
uncertainty arising from strong interaction effects. 

Such a sum rule has been proposed for $D \to \pi \pi$ decays in 
Ref.~\cite{Grossman:2012eb}. Following the phase convention of
\cite{Grossman:2012eb}, the isospin decomposition of $D \to \pi \pi$
amplitudes gives
\begin{eqnarray}
A_{\pi^+\pi^-} & = & \sqrt{2} {\cal A}_3  + \sqrt{2} {\cal A}_1\,, \\
A_{\pi^0\pi^0} & = & 2 {\cal A}_3  - {\cal A}_1\,, \\
A_{\pi^+\pi^0} & = & 3 {\cal A}_3\,,  
\end{eqnarray}
where ${\cal A}_1$ and ${\cal A}_3$ are amplitudes corresponding to 
the $\Delta I = 1/2$ and $\Delta I = 3/2$ transitions, respectively
(\ie, transitions to $\pi\pi$ final states with $I=0$ and $I=2$).
From this, one can get an amplitude isospin sum rule
\begin{equation}
\frac{1}{\sqrt{2}}A_{\pi^+\pi^-} + A_{\pi^0\pi^0} - A_{\pi^+\pi^0} \ =\ 0.
\end{equation}
Probing such a sum requires knowledge of strong phases, which are 
accessible only at charm-threshold experiments. However, even without 
this knowledge, the sum of differences of decay rates for $D$ and 
$\bar{D}$ decays can be measured:
\begin{equation}
|A_{\pi^+\pi^-}|^2 - |\bar{A}_{\pi^+\pi^-}|^2 + |A_{\pi^0\pi^0}|^2 - |\bar{A}_{\pi^0\pi^0}|^2  
- \frac{2}{3}(|A_{\pi^+\pi^0}|^2 - |\bar{A}_{\pi^-\pi^0}|^2)  \ =\ 
3( |{\cal A}_1|^2 -|{\cal \bar{A} }_1|^2 )\,.
\label{eq:rate_sum_rule}
\end{equation}
This equation suggests several SM tests. %
As the penguin amplitude is, 
to excellent approximation within the SM, purely $\Delta I = 1/2$, it does not contribute to $D^+ \to \pi^+\pi^0$. This decay is therefore expected to be \CP symmetric within the SM. Any \CP asymmetry observed in $D^+ \to \pi^+\pi^0$ 
would be a sign of New Physics, which would be associated with the $\Delta I = 3/2$ amplitude.

If the sum in Eq.~(\ref{eq:rate_sum_rule}), 
depending only on ${\cal A}_1$, is found to be non-zero,
this would mean that \CP\ violation arises from the $\Delta I = 1/2$ 
transitions. Moreover, a scenario in which the sum in 
Eq.~(\ref{eq:rate_sum_rule}) is zero and individual asymmetries 
are non-zero would suggest New Physics contributing 
to the $\Delta I = 3/2$ amplitude. 

To facilitate an experimental test, one can exploit also the combination of decay rates:
\begin{equation}
|A_{\pi^+\pi^-}|^2 + |\bar{A}_{\pi^+\pi^-}|^2 + |A_{\pi^0\pi^0}|^2 + |\bar{A}_{\pi^0\pi^0}|^2  
- \frac{2}{3}(|A_{\pi^+\pi^0}|^2 + |\bar{A}_{\pi^-\pi^0}|^2)  \ =\ 
3( |{\cal A}_1|^2 +|{\cal \bar{A} }_1|^2 )\,.
\label{eq:rate_sum_rule1}
\end{equation}
Dividing Eq.~(\ref{eq:rate_sum_rule}) by Eq.~(\ref{eq:rate_sum_rule1})
gives
\begin{equation}
R\ \equiv\ \frac{|A_{\pi^+\pi^-}|^2 - |\bar{A}_{\pi^+\pi^-}|^2 
+ |A_{\pi^0\pi^0}|^2 - |\bar{A}_{\pi^0\pi^0}|^2  
- \frac{2}{3}(|A_{\pi^+\pi^0}|^2 - |\bar{A}_{\pi^-\pi^0}|^2)}
{|A_{\pi^+\pi^-}|^2 + |\bar{A}_{\pi^+\pi^-}|^2 
+ |A_{\pi^0\pi^0}|^2 + |\bar{A}_{\pi^0\pi^0}|^2  
- \frac{2}{3}(|A_{\pi^+\pi^0}|^2 + |\bar{A}_{\pi^-\pi^0}|^2)}\,.
\label{eq:rate_sum_ratio}
\end{equation}
Note that the last term of the denominator enters with the sign opposite compared to the one in the sum tested in Refs.~\cite{Babu:2017bjn} and~\cite{LHCb:2021rou}. An advantage of the ratio in Eq.~\ref{eq:rate_sum_ratio} is that, unlike the sum in Refs.~\cite{Babu:2017bjn} and~\cite{LHCb:2021rou}, it corresponds to the \CP asymmetry in the $\Delta I = 1/2$ process and has no dependence on the $\Delta I = 3/2$ amplitude, which facilitates an interpretation.

Relating the amplitude, the branching fraction, the lifetime, and the asymmetry with
$|A|^2 \propto {\cal B}/\tau^{}_D$ and
$|A|^2 - |\bar{A}|^2 = A_{CP}(|A|^2 + |\bar{A}|^2)$,
we rewrite Eq.~(\ref{eq:rate_sum_ratio}) as
\begin{equation}
R\ =\  
\frac{A_{CP}(D^0 \to \pi^+ \pi^-)}
{1+\frac{\tau_{D^0}}{{\cal B}_{+-}}
\large(\frac{{\cal B}_{00}}{\tau_{D^0}} 
-\frac{2}{3}\frac{{\cal B}_{+0}}{\tau_{D^+}}\large) } 
  + 
\frac{A_{CP}(D^0 \to \pi^0 \pi^0)}
{1+\frac{\tau_{D^0}}{{\cal B}_{00}}
\large(\frac{{\cal B}_{+-}}{\tau_{D^0}} 
-\frac{2}{3}\frac{{\cal B}_{+0}}{\tau_{D^+}}\large) } 
  + 
\frac{A_{CP}(D^+ \to \pi^+ \pi^0)}
{1-\frac{3}{2}\frac{\tau_{D^+}}{{\cal B}_{+0}}
\large(\frac{{\cal B}_{00}}{\tau_{D^0}} 
+\frac{{\cal B}_{+-}}{\tau_{D^0}}\large) }\,,
\label{eq:rate4test}
\end{equation}
where ${\cal B}_{+-}$, ${\cal B}_{00}$, and ${\cal B}_{+0}$
denote the branching fractions for $D^0 \to \pi^+ \pi^-$,
$D^0 \to \pi^0 \pi^0$, and $D^+ \to \pi^+ \pi^0$, respectively.
The sum $R$ is calculated using our averages for \cp\ asymmetries 
(Tables~\ref{tab:cp_charged} and \ref{tab:cp_neutral}) and
PDG averages~\cite{PDG_2020} for branching fractions and 
lifetimes. The result is
\begin{equation}
R\ =\ (+0.86 \pm 3.09) \times 10^{-3},
\end{equation}
which is consistent with zero. In addition, all the individual 
asymmetries contributing to $R$ are consistent with zero. The 
uncertainty on $R$ is dominated by the uncertainties on 
individual asymmetries.

The analogous sum rule for $D \to \bar{K} K$ decays involves full SU(3) 
considerations and thus is imprecise. 
Ref.~\cite{Grossman:2012eb} proposes a set of isospin sum rules 
for $D\to \rho \pi$ or $D\to \bar{K}^{(*)} K^{(*)}\pi$, but testing
these sum rules requires a number of experimental
measurements that have not yet been performed. A number of $U$-spin sum rules for three-body decays of charmed baryons is proposed in Ref.~\cite{Grossman:2018ptn}.   

\clearpage
\mysubsection{$T$-odd asymmetries}\label{sec:todd_asym}
                                               
Measuring $T$-odd asymmetries provides a complementary way 
to search for \CP\ violation in the charm sector, exploiting ${\CP}T$ invariance. 
$T$-odd asymmetries are measured using triple-products %
of the form $\vec{a}\cdot(\vec{b}\times\vec{c})$, where $a$, $b$, 
and $c$ are spins or momenta. This combination is odd under time 
reversal~($T$). Because triple products are $T$-odd, measuring a nonzero value for them was originally proposed as a way to search for $T$ violation~\cite{Datta:2003mj}~\cite{Gronau:2011cf}.

If a triple product is formed using {\it both\/} 
spin and momenta, \ie, 
\begin{equation*}
\vec{s_1} \cdot(\vec{p_2} \times \vec{p}_3),
\end{equation*}
it can be even under $P$-conjugation. 
However, if only momenta are used, \ie, 
\begin{equation*}
\vec{p_1} \cdot(\vec{p_2} \times \vec{p}_3),
\end{equation*}
it is odd under $P$-conjugation. 
Thus, in this case the $T$-odd method becomes $P$-odd and 
allows one to probe \CP\ violation occurring via $P$ violation.
This type of \cpv, arising in $P$-odd amplitudes, can be 
studied in decays of mesons into final states with at least 
four spinless particles. Two- and three-body hadronic decays 
of charm mesons to spinless particles do not provide $T$-odd or $P$-odd observables\footnote{They can be fully described with $P$-even variables, such as invariant 
masses or helicity angles.}, thus only \CP\ violation occurring through $C$-violation can be probed for them.

Taking as an example the decay mode $D^0 \to K^+K^-\pi^+\pi^-$, 
involving spinless particles only, one forms a triple-product 
correlation using momenta of the final-state particles in 
the $D^0$ center-of-mass frame.\footnote{For momentum-only 
triple products, at least four-daughter final states are 
required to give a nonzero correlation, as only three out 
of four momenta are independent. For three-body decays, the 
daughters are in a plane and the triple product is zero.}
Defining the $T$-odd (and $P$-odd) correlation for $D^0$
\begin{equation} 
C_T \equiv \vec{p}^{}_{K^+}\cdot(\vec{p}_{\pi^+}\times \vec{p}_{\pi^-}),
\end{equation}  
and the corresponding quantity for $\dbar$
\begin{equation}
\overline{C}_T \equiv 
      \vec{p}^{}_{K^-}\cdot(\vec{p}_{\pi^-}\times \vec{p}_{\pi^+}),
\label{eq:CT}
\end{equation}      
one can construct the asymmetry for the $D^0$ decays as
\begin{equation}
 A_{T} = 
    \frac{\Gamma(C_T>0)-\Gamma(C_T<0)}{\Gamma(C_T>0)+\Gamma(C_T<0)},
\end{equation}
and for their \CP-conjugate decays as 
\begin{equation}
\overline{A}_{T} = 
   \frac{\Gamma(-\overline{C}_T>0)-\Gamma(-\overline{C}_T<0)}
                        {\Gamma(-\overline{C}_T>0)+\Gamma(-\overline{C}_T<0)}.
\label{eq:CTbar}
\end{equation} 
In these expressions, $\Gamma$ represents a partial width, 
and the following applies: 
\begin{equation}
P(C_T)=-C_T, \quad C(C_T)=\overline{C_T}, \quad C\!P(A_T) = \overline{A}_T.
\label{eq:parities}
\end{equation}
The asymmetries $A_T$ and $\overline{A}_T$ depend on angular 
distributions of the daughter particles and may be nonzero due to 
final-state interactions or $P$ violation in weak decays. 
Given Eq.~(\ref{eq:parities}), one can construct the \CP-violating, 
\ie\ \CP-odd (and $P$-odd, $T$-odd) asymmetry 
\begin{equation}
{\cal A}^{}_{T} \equiv \frac{A_{T}-\overline{A}_{T}}{2};
\label{eqn:atodd}
\end{equation}
where a nonzero value indicates \CP\ violation
(see Refs.~\cite{Golowich:1988ig,Bigi:2001sg,Bensalem:2002ys,
Bensalem:2000hq,Bensalem:2002pz,Gronau:2011cf}).
This asymmetry is also referred to in the literature by 
other names: $A^{}_{T\,{\rm viol}}$, $a^P_{\CP}$, and $a^{T-{\rm odd}}_{\CP}$.

\vspace{0.2cm}
Values of ${\cal A}^{}_{T}$ for $D^+$, $D^+_s$, and
$D^0$ decay modes are listed in Table~\ref{tab:t_odd}. 
Despite relatively high precision ($<1\%$), 
there is no evidence for \CP\ violation. 
In order to increase sensitivity to \CP\ violation (see Sec.~\ref{sec:cp_asym}), some of the measurements in Table~\ref{tab:t_odd} are also performed locally in phase-space regions. Decay phase space is divided according to two- or three-body invariant mass (for $D^0 \to K_S \pi^+\pi^-\pi^0$ decays in Ref.~\cite{Prasanth:2017beu}, and $D^+_{(s)} \to K^+K^-\pi^+\pi^0$ and $D^+_{(s)} \to K^+\pi^-\pi^+\pi^0$ decays in Ref.~\cite{Belle:2023str}), helicity angles of the two-body systems ($D^0 \to K^+K^-\pi^+\pi^-$ in Ref.~\cite{Kim:2018mtf}), or using both mass and angular observables ($D^0 \to K^+K^-\pi^+\pi^-$ in Ref.~\cite{Aaij:2014qwa}).
None of the local ${\cal A}^{}_{T}$ asymmetries are found to be significant.
\begin{table}[htb]
\renewcommand{\arraystretch}{1.4}
\caption{Measurements of the $T$-odd \CP\ asymmetry 
${\cal A}^{}_{T} = (A_{T}-\overline{A}_{T})/2$.
\label{tab:t_odd}}
\footnotesize
\begin{center}

\end{center} 
\end{table}

All $P$-even contributions contributing to ${\cal A}^{}_{T}$ cancel out 
in the difference. Thus, ${\cal A}^{}_{T}$ is only sensitive to $P$-odd amplitudes or to
interference between $P$-odd and $P$-even ones. 
The cancellation typically applies also to detection asymmetries 
and, at the hadron-collider experiments, 
the production asymmetry, making
this is a significant advantage of the $T$-odd method.

Another way to probe $P$-odd amplitudes is through amplitude analysis 
using $P$-odd variables. One example is $\sin{\Phi}$, 
where $\Phi$ is the angle in the $D^0$ frame between 
the $K^+K^-$ decay plane and the $\pi^+\pi^-$ decay plane for the decay
$D^0 \to K^+K^-\pi^+\pi^-$~\cite{Aaij:2018nis}.
It can be shown that $\sin{\Phi}$ is proportional to the triple product as in Eqs.~\ref{eq:CT} or~\ref{eq:CTbar}~\cite{Gronau:2011cf}. 
However, $P$-odd amplitudes in four-body decays of charm mesons, for instance $D \to [VV]_{L=1}$, \ie \ final states involving 
two vector mesons in a $P$-wave state, are typically quite suppressed ($<10\%$)~\cite{Aaij:2018nis,Aaij:2017kbo}. This makes searches for \CP\ violation in these amplitudes challenging. 
As discussed in Sec.~\ref{sec:cp_asym}, no significant \CP\ violation was observed 
for any of the amplitudes contributing into the $D^0 \to K^+K^-\pi^+\pi^-$ decays studied by LHCb~\cite{Aaij:2018nis}. The most
significant asymmetry of $2.8\sigma$ was observed for the
phase of the $P$-odd amplitude $D^0 \to [\phi(1020) \rho(1450)^0]_{L=1}$. In another method, the model-independent technique used to search for \CP\ asymmetries in
$D^0 \to \pi^+\pi^-\pi^+\pi^-$decays (see Sec.~\ref{sec:cp_asym})
has been carried out separately for $P$-odd and $P$-even contributions, 
separated out using a triple product~\cite{Aaij:2013aa}. 
The $p$-value of 0.6\%, corresponding to a significance for 
\CP\ violation of $2.7\sigma$, was obtained for the $P$-odd test of $D^0 \to \pi^+\pi^-\pi^+\pi^-$ decays. 

Decays of charm baryons also offer access to $P$-odd amplitudes, \eg, 
$\Lambda_c^+$ decays with a weakly-decaying baryon in the final state,
such as $\Lambda_c^+ \to \Lambda\,\pi^+$.
Moreover, for polarized charm baryons, \eg, $\Lambda_c$ 
produced weakly in $\Lambda_b$ decays, one can build a triple 
product using the $\Lambda_c$ spin.
Recently, the topic of symmetries has been revisited (see 
Refs.~\cite{Bevan:2015xra,Durieux:2015zwa}), with the suggestion 
to exploit additional asymmetries constructed from triple products
in multi-body decays.

\clearpage
\subsection{Interplay between direct and indirect \cp\ violation}
\label{sec:charm:cpvdir}

In decays of $D^0$ mesons, \cp\ asymmetry measurements have contributions from 
both direct and indirect \cp\ violation, as discussed in Sec.~\ref{sec:charm:mixcpv}.
The contribution from indirect \cp\ violation depends on the decay-time distribution 
of the data sample~\cite{Kagan:2009gb}. This section describes a combination of 
measurements that allows the determination of the individual contributions of the 
two types of \cp\ violation.
At the same time, the level of agreement for a no-\cp-violation hypothesis is 
tested.

The first observable is
\begin{equation}
A_{\Gamma} \equiv \frac{\hat{\tau}(\dbar \ra f) - \hat{\tau}(D^0 \ra f )}
{\hat{\tau}(\dbar \ra f) + \hat{\tau}(D^0 \ra f )},
\end{equation}
where $f$ can be $K^+ K^-$ or $\pi^+\pi^-$ and $\hat{\tau}(D^0 \ra f )$ indicates the effective $D^0$ lifetime, measured with an exponential approximation in the decay to $f$. More recently, the related observable~\cite{Kagan:2020vri}
\begin{equation}
    \Delta Y\equiv\frac{\hat{\Gamma}(\overline{D}^0\rightarrow f)-\hat{\Gamma}(D^0\rightarrow f)}{2},
\end{equation}
has been measured~\cite{LHCb:2021vmn} instead of $A_\Gamma$, where $\hat{\Gamma}$ denotes the effective decay width and is the inverse of the effective lifetime.
They are related by 
\begin{equation}
    A_\Gamma = \frac{-\Delta Y}{1+y_{\CP}}.
\end{equation}

The second observable is
\begin{equation}
\Delta A_{\CP}   \equiv A_{\CP}(K^+K^-) - A_{\CP}(\pi^+\pi^-),
\end{equation}
where $A_{\CP}$ are time-integrated \cp\ asymmetries. The underlying 
theoretical parameters are 
\begin{eqnarray}
a_{\CP}^{\rm dir} & \equiv & 
\frac{|{\cal A}_{D^0\rightarrow f} |^2 - |{\cal A}_{\dbar\rightarrow f} |^2} 
{|{\cal A}_{D^0\rightarrow f} |^2 + |{\cal A}_{\dbar\rightarrow f} |^2} ,\nonumber\\ 
a_{\CP}^{\rm ind}  & \equiv & \frac{1}{2} 
\left[ \left(\left|\frac{q}{p}\right| + \left|\frac{p}{q}\right|\right) x \sin \phi - 
\left(\left|\frac{q}{p}\right| - \left|\frac{p}{q}\right|\right) y \cos \phi \right] ,
\end{eqnarray}
where ${\cal A}_{D\rightarrow f}$ is the amplitude for $D\ra f$~\cite{Grossman:2006jg}. 
The observables are related to the underlying parameters via~\cite{Gersabeck:2011xj}
\begin{eqnarray}
\Delta Y & = & -A_{\Gamma}(1+y_{\CP}) = a_{\CP}^{\rm ind} + a_{\CP}^{\rm dir} y_{\CP}\approx a_{\CP}^{\rm ind},\label{eqn:charm_MG_AGamma}\\ 
\Delta A_{\CP} & = &  \Delta a_{\CP}^{\rm dir} \left(1 + y_{\CP} 
\frac{\overline{\langle t\rangle}}{\tau} \right)   +   
   a_{\CP}^{\rm ind} \frac{\Delta\langle t\rangle}{\tau}   +   
  \overline{a_{\CP}^{\rm dir}} y_{\CP} \frac{\Delta\langle t\rangle}{\tau},\nonumber\\ 
& \approx & \Delta a_{\CP}^{\rm dir} \left(1 + y_{\CP} 
\frac{\overline{\langle t\rangle}}{\tau} \right)   +   a_{\CP}^{\rm ind} 
\frac{\Delta\langle t\rangle}{\tau}.\label{eqn:charm_MG_DACP}
\end{eqnarray}
Equation~(\ref{eqn:charm_MG_AGamma}) shows that $\Delta Y$ constrains mostly indirect \cp\ violation. 
The 
direct \cp\ violation contribution can differ for different final states. 
However, the $a_{\CP}^{\rm dir}$ term is neglected as it is suppressed by $y_{CP}$ and not directly accessible in the fit below.
In Eq.~(\ref{eqn:charm_MG_DACP}), $\langle t\rangle/\tau$ denotes the mean decay 
time in units of the $D^0$ lifetime; $\Delta X = X_{KK}-X_{\pi\pi}$ denotes the difference 
in quantity $X$ between the $K^+K^-$ and $\pi^+\pi^-$ final states; and $\overline{X} = (X_{KK}+X_{\pi\pi})/2$ 
denotes the average for quantity $X$. 
We neglect the term $\overline{a_{\CP}^{\rm dir}} y_{\CP} \frac{\Delta\langle t\rangle}{\tau}$ in Eq.~(\ref{eqn:charm_MG_DACP}), as the factors constituting it make it negligible 
with respect to the other two terms. 

We perform a $\chi^2$ fit to extract  $\Delta a_{\CP}^{\rm dir}$ 
and $a_{\CP}^{\rm ind}$.
The value for $y_{\CP}$ is taken to be $y_{\CP}=(0.655\pm0.028)\%$, obtained by taking $y_{\CP}-y_{\CP}(K\pi)=(0.697\pm0.028)\%$ (see Table~\ref{tab:observables1}) %
and subtracting $y_{\CP}(K\pi)=-4.19\times10^{-4}$~\cite{Schwartz:2022egt}.
The input measurements are listed in 
Table~\ref{tab:charm:dir_indir_comb}. 
For the \babar measurements of $A_{\CP}(K^+K^-)$ and $A_{\CP}(\pi^+\pi^-)$, we calculate $\Delta A_{\CP}$
adding all uncertainties in quadrature. 
This may overestimate the systematic uncertainty for the difference, 
as it neglects correlated uncertainties. However, the result is conservative, 
and the effect is small, as all measurements are statistically limited. 
For all measurements, statistical and systematic uncertainties are added 
in quadrature when calculating the $\chi^2$. 
In this fit, $A_\Gamma(KK)$ and $A_\Gamma(\pi\pi)$ are assumed to be identical but are plotted separately in Fig.~\ref{fig:charm:dir_indir_comb} to visualise their level of agreement.
This approximation, which holds in the SM, is supported by all measurements to date.
An alternative assessment that avoids relying on this approximation by analyzing only the $KK$ final state is presented below.
The inclusion of these measurements in the fit excludes a significant contribution to $\Delta A_{\CP}$ that is due to final-state-dependent $A_\Gamma$ values and different mean decay times, corresponding to a contribution from the term proportional to $a_{\CP}^{\rm ind}$ in Eq.~\ref{eqn:charm_MG_DACP}.
The latest LHCb measurement in the fit measures $\Delta{}Y$~\cite{LHCb:2021vmn}, which is approximately equal to $-A_\Gamma$ up to a fractional difference of $y_{\CP}$.

\begin{table}
\centering 
\caption{Inputs to the fit for direct and indirect \cp\ violation. 
The first uncertainty listed is statistical and the second is systematic. The uncertainties on $\Delta \langle t\rangle/\tau$ and $\langle t\rangle/\tau$ are $\le0.01$ and are not quoted here.}
\label{tab:charm:dir_indir_comb}
\vspace{3pt}
\begin{tabular}{ll|ccccc}
\hline \hline
Year & 	Experiment	& Results
& $\Delta \langle t\rangle/\tau$ & $\langle t\rangle/\tau$ & Reference\\
\hline
2012	& \babar	& $A_\Gamma = (+0.09 \pm 0.26 \pm 0.06 )\%$ &	-&	-&	 
\cite{BaBar:2012bho}\\
2014	& CDF & $A_\Gamma = (-0.12 \pm 0.12 )\%$ &	-&	-&	 
\cite{CDF:2014wyb}\\
2015	& Belle	& $A_\Gamma = (-0.03 \pm 0.20 \pm 0.07 )\%$ &	-&	-&	 
\cite{Belle:2015etc}\\
2021	& LHCb	 & $\Delta{}Y(KK) = (-0.003 \pm 0.013 \pm 0.003 )\%$ &	-&	-&	 
\cite{LHCb:2021vmn}\\
    	&     	& $\Delta{}Y(\pi\pi) = (-0.036 \pm 0.024 \pm 0.004 )\%$ &  -&	-&	 
                   \\
2008	& \babar	& $A_{\CP}(KK) = (+0.00 \pm 0.34 \pm 0.13 )\%$&&&\\ 
& & $A_{\CP}(\pi\pi) = (-0.24 \pm 0.52 \pm 0.22 )\%$ &	$0.00$ &	
$1.00$ &	 \cite{Aubert:2007if}\\
2012	& CDF	 & $\Delta A_{\CP} = (-0.62 \pm 0.21 \pm 0.10 )\%$ &	
$0.25$ &	$2.58$ &	 \cite{Collaboration:2012qw}\\
2014	& LHCb	SL & $\Delta A_{\CP} = (+0.14 \pm 0.16 \pm 0.08 )\%$ &	
$0.01$ &	$1.07$ &	 \cite{Aaij:2014gsa}\\
2016	& LHCb	prompt & $\Delta A_{\CP} = (-0.10 \pm 0.08 \pm 0.03 )\%$ &	
$0.12$ &	$2.10$ &	 \cite{Aaij:2016cfh}\\
2019	& LHCb	SL2 & $\Delta A_{\CP} = (-0.09 \pm 0.08 \pm 0.05 )\%$ &	
$0.00$ &	$1.21$ &	 \cite{LHCb:2019hro}\\
2019	& LHCb	prompt2 & $\Delta A_{\CP} = (-0.182 \pm 0.032 \pm 0.009 )\%$ &	
$0.13$ &	$1.74$ &	 \cite{LHCb:2019hro}\\
\hline
\end{tabular}
\end{table}

The fit results are shown in Fig.~\ref{fig:charm:dir_indir_comb}.
\begin{figure}
\begin{center}
\includegraphics[width=0.90\textwidth]{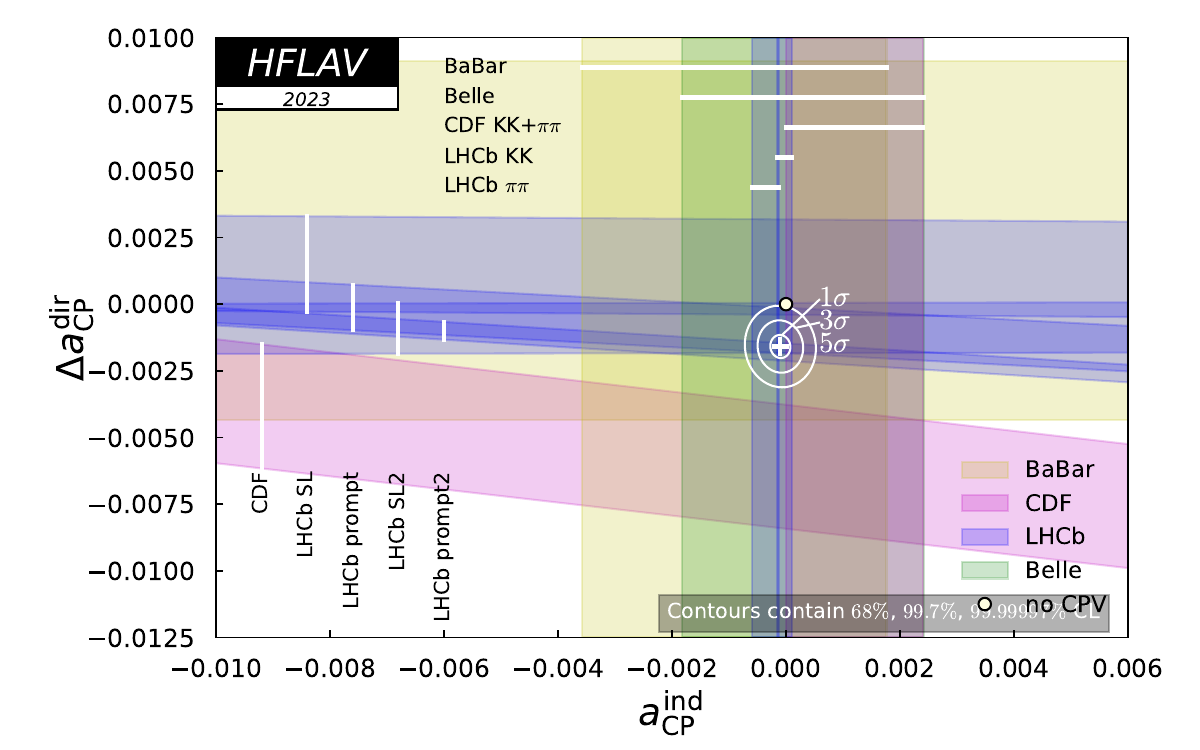}
\includegraphics[width=0.90\textwidth]{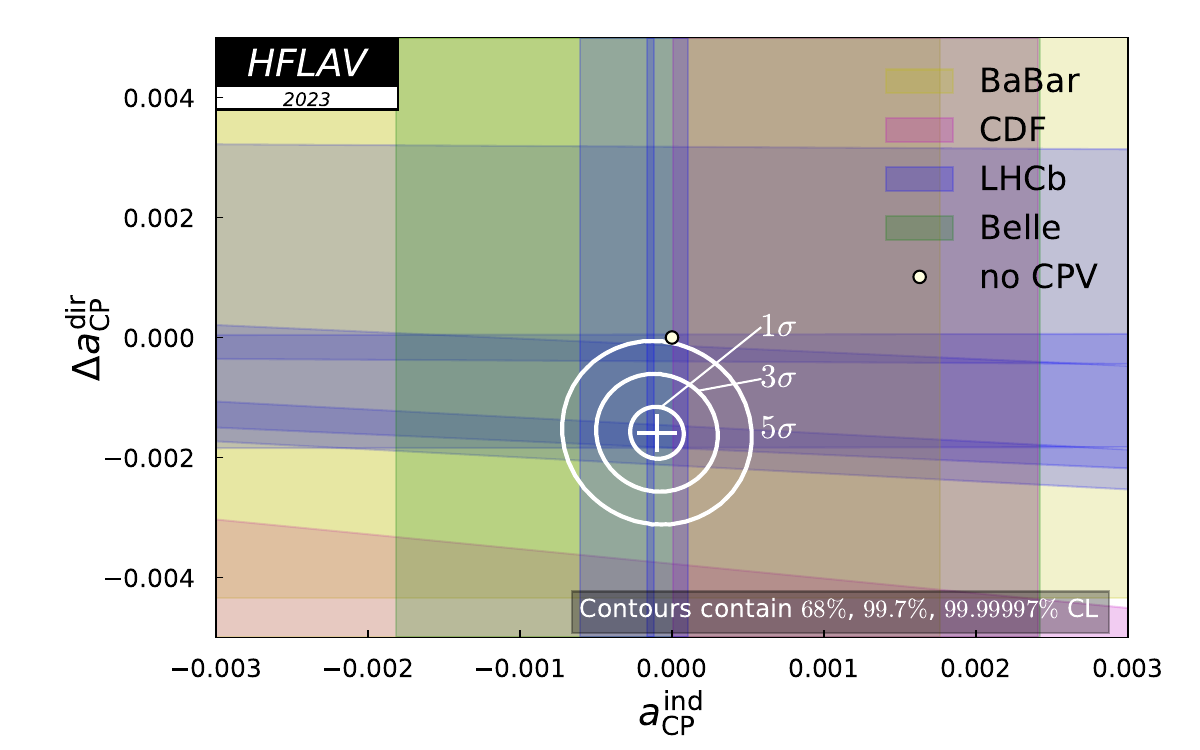}
\caption{(Top) Plot of all data and the fit result for $\Delta a^{\rm dir}_{CP}$ and $a^{\rm ind}_{CP}$. Individual 
measurements are plotted as bands showing their $\pm1\sigma$ range. 
The no-\cpv\ point (0,0) is shown as a filled circle, and the best 
fit value is indicated by a cross showing the one-dimensional uncertainties. 
Two-dimensional $1\sigma$ ($68.3\%$ C.L.), $3\sigma$ ($99.7\%$ C.L.), 
and $5\sigma$ ($99.99997\%$ C.L.) regions are plotted as ellipses.
(Bottom) Zoom of the same plot, highlighting the best fit region.}
\label{fig:charm:dir_indir_comb}
\end{center}
\end{figure}
From the fit, the change in $\chi^2$ from the minimum value for the no-\cpv\ 
point (0,0) is $33.0$, which corresponds to a C.L.\ of $6.7\times 10^{-8}$ for 
two degrees of freedom or $5.4$ standard deviations. The central
values and $\pm1\sigma$ uncertainties for the individual parameters are
\begin{eqnarray}
a_{\CP}^{\rm ind} & = & (-0.010 \pm 0.012 )\% \nonumber\\
\Delta a_{\CP}^{\rm dir} & = & (-0.159 \pm 0.029 )\%.
\label{eq:charm_dir_indir:fit_results}
\end{eqnarray}
Relative to the average reported in our previous report~\cite{HFLAV:2022esi}, the results remain approximately unchanged, with the main update being the value of $y_{\CP}$, which only plays a minor role.
The average clearly demonstrates direct \CP violation in the singly Cabibbo-suppressed decays to two charged hadrons.

A new combination with respect to the previous report is performed to extract the parameters for the $KK$ final state alone.
This fit avoids the assumption of universal indirect \CP violation.
As not all measurements were published with a separate result for the $KK$ final state, only the results in Table~\ref{tab:charm:dir_indir_comb_KK} are used.
Compared to the fits above, the second term in Eq.~\ref{eqn:charm_MG_AGamma} is no longer neglected, resulting in the time-dependent measurements being represented by sloped bands, although the smallness of $y_{\CP}$ suppresses this effect so they still appear as vertical bands in Fig.~\ref{fig:charm:dir_indir_comb_KK}.
The value for $y_{\CP}$ is also extracted from the $KK$ final state alone, for which the only available measurement is that of the LHCb collaboration, $y_{\CP}(KK)-y_{\CP}(K\pi)=(0.708\pm0.033)\%$~\cite{LHCb:2022gnc}, corrected for $y_{\CP}(K\pi)$ as described above to yield $y_{\CP}(KK)=(0.666\pm0.033)\%$.
The time integrated asymmetry enters the fit according to the equation
\begin{equation}
A_{\CP} \approx a_{\CP}^{\rm dir} + a_{\CP}^{\rm ind} 
\frac{\langle t\rangle}{\tau}.\label{eqn:charm_MG_ACP}
\end{equation}
The fit results are shown in Fig.~\ref{fig:charm:dir_indir_comb_KK}.
From the fit, the change in $\chi^2$ from the minimum value for the no-\cpv\ 
point (0,0) is $1.44$, which corresponds to a C.L.\ of $0.49$ for 
two degrees of freedom or $0.7$ standard deviations. The central
values and $\pm1\sigma$ uncertainties for the individual parameters are
\begin{eqnarray}
a_{\CP}^{\rm ind}(KK) & = & (-0.001 \pm 0.013 )\% \nonumber\\
a_{\CP}^{\rm dir}(KK) & = & \phantom{-}(0.054 \pm 0.060 )\%.
\end{eqnarray}
The results are in agreement with \CP symmetry in $D^0\rightarrow K^+K^-$ decays.
The result for $a_{\CP}^{\rm ind}(KK)$ is in agreement with that obtained with the fit above assuming universality of indirect \CP violation (see Eq.~\ref{eq:charm_dir_indir:fit_results}) and hence supports this assumption.
The result for $a_{\CP}^{\rm dir}(KK)$ indicates that the \CP violation observed in $\Delta a_{\CP}$ is driven by $D^0\rightarrow \pi^+\pi^-$ decays.

\begin{table}
\centering 
\caption{Inputs to the fit for direct and indirect \cp\ violation in the $KK$ final state. 
The first uncertainty listed is statistical and the second is systematic. The uncertainties on $\langle t\rangle/\tau$ are $\le0.01$ and are not quoted here.}
\label{tab:charm:dir_indir_comb_KK}
\vspace{3pt}
\begin{tabular}{ll|cccc}
\hline \hline
Year & 	Experiment	& Results
& $\langle t\rangle/\tau$ & Reference\\
\hline
2014	& CDF & $A_\Gamma(KK) = (-0.19 \pm 0.15 \pm 0.04 )\%$ &	-&	 
\cite{CDF:2014wyb}\\
2021	& LHCb	 & $\Delta{}Y(KK) = (-0.003 \pm 0.013 \pm 0.003 )\%$ &	-& \cite{LHCb:2021vmn}\\
2008	& \babar	& $A_{\CP}(KK) = (+0.00 \pm 0.34 \pm 0.13 )\%$ &	
$1.00$ &	 \cite{Aubert:2007if}\\
2008	& \belle	& $A_{\CP}(KK) = (-0.43 \pm 0.30 \pm 0.11 )\%$ &	
$1.00$ &	 \cite{Belle:2008ddg}\\
2023	& LHCb & $ A_{\CP}(KK) = (0.068 \pm 0.054 \pm 0.016 )\%$ &	
$1.75$ &	 \cite{LHCb:2022lry}\\
\hline
\end{tabular}
\end{table}

\begin{figure}
\begin{center}
\includegraphics[width=0.90\textwidth]{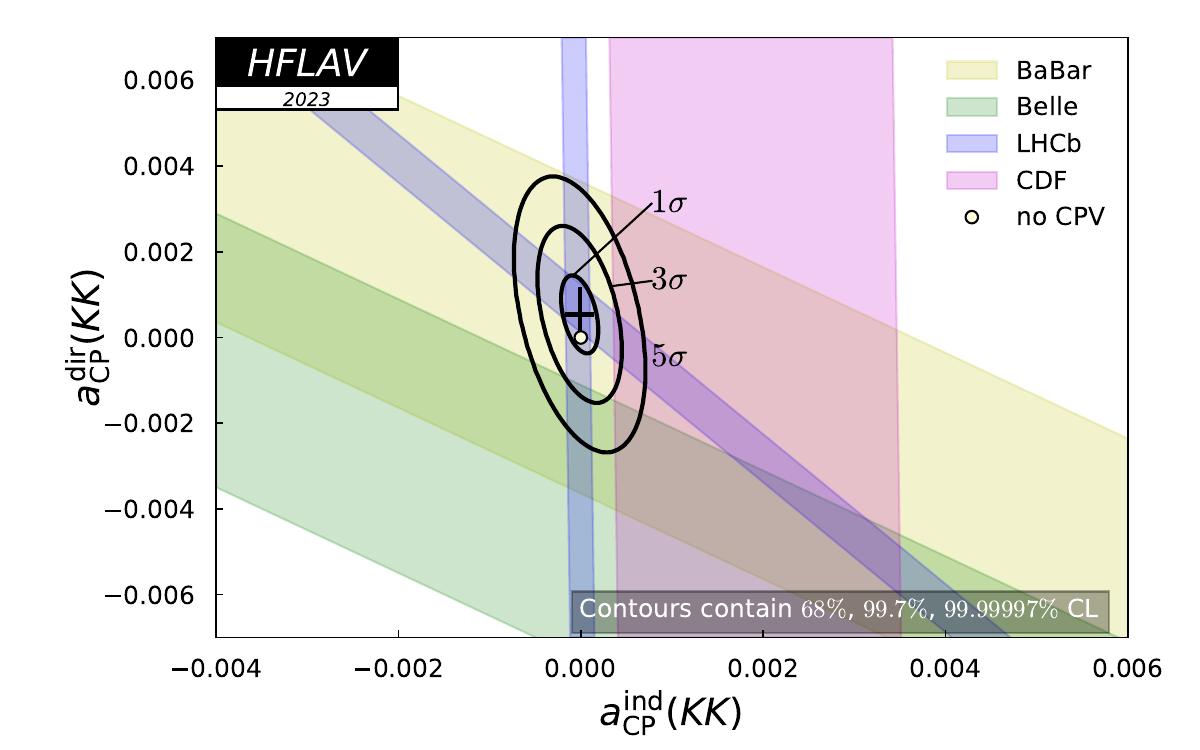}
\caption{Plot of data and the fit result for $a^{\rm dir}_{CP}(KK)$ and $a^{\rm ind}_{CP}(KK)$. Individual 
measurements are plotted as bands showing their $\pm1\sigma$ range. 
The no-\cpv\ point (0,0) is shown as a filled circle, and the best 
fit value is indicated by a cross showing the one-dimensional uncertainties. 
Two-dimensional $1\sigma$ ($68.3\%$ C.L.), $3\sigma$ ($99.7\%$ C.L.), 
and $5\sigma$ ($99.99997\%$ C.L.) regions are plotted as ellipses.}
\label{fig:charm:dir_indir_comb_KK}
\end{center}
\end{figure}

\clearpage

\section{Charm decays}
\label{sec:charm_decays}

\subsection{Semileptonic decays}
\label{sec:charm:semileptonic}

Semileptonic $D$ decays involve the interaction of a leptonic
current with a hadronic current. The latter is nonperturbative
and 
is usually
parameterized in terms of form factors. The transition matrix element 
is written as:
\begin{eqnarray}
  {\cal M} & = & -i\,\frac{G_F}{\sqrt{2}}\,V^{}_{cq}\,L^\mu H_\mu\,,
  \label{Melem}
\end{eqnarray}
where $G_F$ is the Fermi constant and $V^{}_{cq}$ is a CKM matrix element.
The leptonic current $L^\mu$ is evaluated directly from the lepton spinors 
and has a simple structure; this allows one to extract information about 
the form factors (in $H^{}_\mu$) from data on semileptonic decays~\cite{Becher:2005bg}.  
Conversely, because there are no strong final-state interactions between the
leptonic and hadronic systems, semileptonic decays for which the form 
factors can be calculated allow one to 
determine~$|V^{}_{cq}|$~\cite{Kobayashi:1973fv}.
In experiment, one determines the form factors after taking the $|V_{cq}|$ from the standard model global fit as input;
or determines the $|V_{cq}|$ after taking the form factors calculated by theory as input.

\mysubsubsection{$D\ra P \ell^+\nu_\ell$ decays and $|V_{cq}|$}

\subsubsubsection{Theoretical formula}

When the final state hadron is a pseudoscalar, the hadronic 
current is given by~\cite{Hill:2006ub}
\begin{eqnarray}
\hspace{-1cm}
H_\mu & = & \left< P(p) | \bar{q}\gamma_\mu c | D(p') \right> \ =\  
f_+(q^2)\left[ (p' + p)_\mu -\frac{m_D^2-m_P^2}{q^2}q_\mu\right] + 
 f_0(q^2)\frac{m_D^2-m_P^2}{q^2}q_\mu\,,
\label{eq:hadronic}
\end{eqnarray}
where $m_D$ and $p'$ are the mass and four momentum of the 
parent $D$ meson, $m_P$ and $p$ are those of the daughter meson, 
$f_+(q^2)$ and $f_0(q^2)$ are the form factors, and $q = p' - p$.  
Kinematics require that $f_+(0) = f_0(0)$.
The contraction $L^\mu q_\mu$ results in terms proportional 
to $m_\ell$\cite{Gilman:1989uy}, and thus for $\ell=e $
the terms proportional to $q_\mu$ in Eq.~(\ref{eq:hadronic}) are negligible and only the $f_+(q^2)$ vector form factor 
is relevant. The corresponding differential partial width is
\begin{eqnarray}
\frac{d\Gamma(D \to P e^+\nu_e)}{dq^2\, d\cos\theta_e} & = & 
   \frac{G_F^2|V_{cq}|^2}{32\pi^3} p^{*\,3}|f_{+}(q^2)|^2\sin\theta^2_e\,,
\label{eq:dGamma}
\end{eqnarray}
where ${p^*}=\frac{\left [\left (m_D^2-(m_P+q)\right )^2\left (m_D^2-(m_P-q)\right )\right ]^{1/2}}{2 m_D}$ is the magnitude of the momentum of the final state hadron
in the $D$ rest frame, and $\theta_e$ is the angle of the electron in the 
$e\nu$ rest frame with respect to the direction of the pseudoscalar meson 
in the $D$ rest frame.

The form factor is traditionally parameterized with an explicit pole 
and a sum of effective poles:
\begin{eqnarray}
f_+(q^2) & = & \frac{f_+(0)}{(1-\alpha)}\Bigg [
\left(\frac{1}{1- q^2/m^2_{\rm pole}}\right)\ +\ 
\sum_{k=1}^{N}\frac{\rho_k}{1- q^2/(\gamma_k\,m^2_{\rm pole})}\Bigg ]\,,
\label{eqn:expansion}
\end{eqnarray}
where $\rho_k$ and $\gamma_k$ are expansion parameters and $\alpha$ is 
a parameter that normalizes the form factor at $q^2=0$, $f_+(0)$. 
The parameter $m_{{\rm pole}}$ is the mass of the lowest-lying $c\bar{q}$ 
resonance with the vector quantum numbers; this is expected to 
provide the largest contribution to the form factor for the $c\ra q$ 
transition. The sum over $N$ gives the contribution of higher mass states.  
For example, for $D\to\pi$ transitions the dominant resonance is
expected to be the $D^*(2010)$, and thus $m^{}_{\rm pole}=m^{}_{D^*(2010)}$. 
For $D\to K$ transitions, the dominant resonance is expected to be the
$D^{*}_s(2112)$, and thus $m^{}_{\rm pole}=m^{}_{D^{*}_{s}(2112)}$.

Equation~(\ref{eqn:expansion}) can be simplified by neglecting the 
sum over effective poles, leaving only the explicit vector meson pole. 
This approximation is referred to as ``nearest pole dominance'' or 
``vector-meson dominance.''  The resulting parameterization is
\begin{eqnarray}
  f_+(q^2) & = & \frac{f_+(0)}{(1-q^2/m^2_{\rm pole})}\,. 
\label{SimplePole}
\end{eqnarray}
However, values of $m_{{\rm pole}}$ that give a good fit to the data 
do not agree with the expected vector meson masses~\cite{Hill:2006ub}. 
To address this problem, the ``modified pole'' or Becirevic-Kaidalov~(BK) 
parameterization~\cite{Becirevic:1999kt} was introduced.
In this parameterization $m_{\rm pole} /\sqrt{\alpha_{\rm BK}}$
is interpreted as the mass of an effective pole higher than 
$m_{\rm pole}$, i.e., it is expected that $\alpha_{\rm BK}<1$.
The parameterization takes the form
\begin{eqnarray}
f_+(q^2) & = & \frac{f_+(0)}{(1-q^2/m^2_{\rm pole})}
\frac{1}{\left(1-\alpha^{}_{\rm BK}\frac{q^2}{m^2_{\rm pole}}\right)}\,,
\end{eqnarray}
where $\alpha^{}_{\rm BK}$ is a free parameter
that takes into account contributions from higher states in the form of an effective pole.
This phenomenological parameterization is used by several experiments to 
determine form factor parameters.

Alternatively, a power series expansion around some value $q^2=t_0$ can be used to 
parameterize $f^{}_+(q^2)$~\cite{Boyd:1994tt,Boyd:1997qw,Arnesen:2005ez,Becher:2005bg}. 
This parameterization is model-independent and %
correctly incorporates the branch points associated with $m_D$ and $m_P$.
The expansion is given in terms of a complex parameter $z$, which is 
the analytic continuation of $q^2$ into the complex plane:
\begin{eqnarray}
z(q^2,t_0) & = & \frac{\sqrt{t_+ - q^2} - \sqrt{t_+ - t_0}}{\sqrt{t_+ - q^2}
	  + \sqrt{t_+ - t_0}}\,, 
\end{eqnarray}
where $t_{0}= t_{+} (1-\sqrt{1-t_{-}/t_{+}})$ and $t_\pm \equiv (m_D \pm m_P)^2$. 
In this parameterization, $q^2=t_0$ corresponds to $z=0$, and the physical 
region extends in either direction up to $\pm|z|_{\rm max} = \pm 0.051$ for 
$D\to K \ell^+\nu_\ell$ decays, and up to $\pm 0.17$ for $D\to \pi \ell^+\nu_\ell$ 
decays. 

The form factor is expressed as
\begin{eqnarray}
f_+(q^2) & = & \frac{1}{P(q^2)\,\phi(q^2,t_0)}\sum_{k=0}^\infty
a_k(t_0)[z(q^2,t_0)]^k\,,
\label{z_expansion}
\end{eqnarray}
where the Blaschke factor $P(q^2)$ is used to remove sub-threshold poles, for instance, $P(q^2)=1$ for $D\to \pi$ and $P(q^2)=z(q^2,M^2_{D^*_s})$ for $D\to K$.
The ``outer'' function $\phi(t,t_0)$ can be any analytic function, but a preferred 
choice (see, \eg, Refs.~\cite{Boyd:1994tt,Boyd:1997qw,Bourrely:1980gp}), obtained
from the Operator Product Expansion (OPE), is
\begin{eqnarray}
\phi(q^2,t_0) & =  & \alpha 
\left(\sqrt{t_+ - q^2}+\sqrt{t_+ - t_0}\right) \times  \nonumber \\
 & & \hskip0.20in \frac{t_+ - q^2}{(t_+ - t_0)^{1/4}}\  
\frac{(\sqrt{t_+ - q^2}\ +\ \sqrt{t_+ - t_-})^{3/2}}
     {(\sqrt{t_+ - q^2}+\sqrt{t_+})^5}\,,
\label{eqn:outer}
\end{eqnarray}
with $\alpha = \sqrt{\pi m_c^2/3}$.
The OPE analysis provides a constraint upon the 
expansion coefficients, $\sum_{k=0}^{N}a_k^2 \leq 1$.
These coefficients receive $1/M_D$ corrections, and thus
the constraint is only approximate. However, the
expansion is expected to converge rapidly since 
$|z|<0.051\ (0.17)$ for $D\ra K$ ($D\ra\pi$) over 
the entire physical $q^2$ range, and Eq.~(\ref{z_expansion}) 
remains a useful parameterization. 
As a major advantage, the z-expansion complies with the analytic structure of the form factors predicted by general principles of quantum field theory.
The main disadvantage as compared to 
phenomenological approaches is that there is no physical interpretation 
of the fitted coefficients~$a_K$.

\subsubsubsection{Experimental results of $D\to K\ell^+\nu_\ell$ and $D\to \pi\ell^+\nu_\ell$}

Various techniques have been used by several experiments to measure 
semileptonic $D$ decays with a pseudoscalar particle in the 
final state. The most recent results are provided by the \babar~\cite{Lees:2014ihu} 
and BESIII~\cite{Ablikim:2015ixa, BESIII:2015jmz} collaborations.
Belle~\cite{Widhalm:2006wz}, \babar~\cite{Aubert:2007wg}, and 
\mbox{CLEO-c}~\cite{Besson:2009uv,Dobbs:2007aa} have all  
previously reported results. Belle fully 
reconstructs $e^+e^- \to D \bar D X$ events from the continuum 
under the $\Upsilon(4S)$ resonance, achieving very good $q^2$ 
resolution (15${\rm~MeV}^2$) and a low background level but with 
a low efficiency. Using 282~$\fb^{-1}$ of data, about 
1300 $D\to K\ell^+\nu_\ell$ (Cabibbo-favored) and 115 $D\to\pi\ell^+\nu_\ell$ 
(Cabibbo-suppressed) decays are reconstructed, considering the electron 
and muon channels together. The \babar experiment uses a partial 
reconstruction technique in which the semileptonic decays 
are tagged via $ D^{\ast +}\to D^0\pi^+$ decays. 
The $D$ direction and neutrino energy are obtained 
using information from the rest of the event. 
With 75~$\fb^{-1}$ of data, 74000 signal events in the 
$D^0 \to {K}^- e^+ \nu_e$ mode are obtained. This technique 
provides a large signal yield but also a high background level 
and a poor $q^2$ resolution (ranging from 66 to 219 MeV$^2$). In this 
case, the measurement of the branching fraction is obtained by normalizing 
to the $D^0 \to K^- \pi^+$ decay channel; thus the measurement would
benefit from future improvements in the determination of the branching fraction for this
reference channel. The Cabibbo-suppressed mode has been 
measured using the same technique and 350~fb$^{-1}$ data. For
this measurement, 5000 $D^0 \to {\pi}^- e^+ \nu_e$ 
signal events were reconstructed~\cite{Lees:2014ihu}.  

The CLEO-c experiment uses two different methods to measure charm semileptonic 
decays. The {\it tagged\/} analyses~\cite{Besson:2009uv} rely on the full 
reconstruction of 
$\psi(3770)\to D {\overline D}$ events. One of the $D$ mesons is reconstructed 
in a hadronic decay mode, and the other in the semileptonic channel. The only missing 
particle is the neutrino, and thus the $q^2$ resolution is very good and the 
background level very low.   
With the entire CLEO-c data sample of 818 $\pb^{-1}$, 14123 and 1374 signal 
events are reconstructed for the $D^0 \to K^{-} e^+\nu_e$ and $D^0\to \pi^{-} e^+\nu_e$ 
channels, respectively, and 8467 and 838 are reconstructed for the 
$D^+\to {\overline K}^{0} e^+\nu_e$ and $D^+\to \pi^{0} e^+\nu_e$ decays, 
respectively. An alternative method that does not tag the $D$ decay in a 
hadronic mode (referred to as {\it untagged\/} analyses) has also been 
used by CLEO-c~\cite{Dobbs:2007aa}. In this method, the entire missing 
energy and momentum in an event are associated with the neutrino four 
momentum, with the penalty of larger backgrounds as compared to the 
tagged method. 

Using the tagged method, the BESIII experiment measures the 
$D^0 \to {K}^- e^+ \nu_e$ and $D^0 \to {\pi}^- e^+ \nu_e$ decay channels. 
With 2.93~fb$^{-1}$ of data, they fully reconstruct 70700 and 6300 signal 
events, respectively, for the two channels~\cite{Ablikim:2015ixa}. In 
a separate analysis, BESIII measures the semileptonic decay 
$D^+ \to K^{0}_{L} e^+ \nu_e$ \cite{BESIII:2015jmz}, with about 
20100 semileptonic candidates. 
Since 2016, BESIII has reported additional measurements of 
$D\to \bar K\ell^+\nu_\ell$ and $\pi\ell^+\nu_\ell$. The signal yields 
are 26008, 5013, 47100, 20714, 3402, 2265, and 1335 events for 
$D^+\to \bar K^0(\pi^+\pi^-)e^+\nu_e$, $D^+\to \bar K^0(\pi^0\pi^0)e^+\nu_e$, 
$D^0\to K^-\mu^+\nu_\mu$, $D^+\to \bar K^0(\pi\pi)\mu^+\nu_\mu$, 
$D^+\to \pi^0 e^+\nu_e$, $D^0\to \pi^-\mu^+\nu_\mu$, and 
$D^+\to \pi^0\mu^+\nu_\mu$~\cite{bes3:2017fay, bes3:2019zsf, 
bes3:2016hzl, bes3:2016wy, bes3:2018wy}, respectively. The branching fractions or hadronic form factors are determined with good precision. 

Results of the hadronic form factor parameters, $m_{\rm pole}$ and $\alpha_{\rm BK}$,
obtained from the measurements discussed above, are given in 
Tables~\ref{kPseudoPole} and~\ref{piPseudoPole}.
The $z$-expansion formalism has been used by \babar~\cite{Aubert:2007wg,Lees:2014ihu}, 
BESIII~\cite{Ablikim:2015ixa,BESIII:2015jmz,bes3:2017fay,bes3:2019zsf} and CLEO-c~\cite{Besson:2009uv,Dobbs:2007aa}.
Their fits use the first three terms of the expansion, %
and the results for the ratios $r_1\equiv a_1/a_0$ and $r_2\equiv a_2/a_0$ are 
listed in Tables~\ref{KPseudoZ} and~\ref{piPseudoZ}.

\begin{table}[tp]
\centering
\caption{Results for $m_{\rm pole}$ and $\alpha_{\rm BK}$ from various 
  experiments for $D^0\to K^-\ell^+\nu_\ell$ ($\ell=e$ or $\mu$) and $D^+\to \bar{K}^0\ell^+\nu_\ell$ decays.
  The last two rows list results for other $c\to s e^+\nu_e$ decays, for comparison, because some %
lattice calculations~\cite{Koponen:2012a,Koponen:2013b} %
find 
that the form factors are possibly insensitive to the spectator quarks.
\label{kPseudoPole}}
\resizebox{\textwidth}{!}{

}
\end{table}

\begin{table}[htbp]
\centering
\caption{Results for $m_{\rm pole}$ and $\alpha_{\rm BK}$ from various experiments for 
  $D^0\to \pi^-\ell^+\nu_\ell$ and $D^+\to \pi^0\ell^+\nu_\ell$ decays.
  The last two rows list results for other $c\to d e^+\nu_e$ decays, for comparison, because some %
lattice calculations~\cite{Koponen:2012a,Koponen:2013b} %
find 
that the form factors are possibly insensitive to the spectator quarks.
\label{piPseudoPole}}
\resizebox{\textwidth}{!}{
%
}
\end{table}

\begin{table}[tbp]
\caption{Results for $r_1$ and $r_2$ from various experiments for $D^0\to K^-\ell^+\nu_\ell$ and $D^+\to \bar{K}^0\ell^+\nu_\ell$  decays.
Some %
lattice calculations~\cite{Koponen:2012a,Koponen:2013b} %
find 
that the form factors are possibly insensitive to the spectator quarks.  For comparison, the last four rows list results for $c\to s e^+\nu_e$ decays in which
  only the first two terms of the $z$ expansion were used.} 
\label{KPseudoZ}
\begin{center}
%
\end{center}
\end{table}

\begin{table}[htbp]
  \caption{Results for $r_1$ and $r_2$ from various experiments for  $D^0\to \pi^-\ell^+\nu_\ell$ and $D^+\to \pi^0\ell^+\nu_\ell$
    decays.  Some %
lattice calculations~\cite{Koponen:2012a,Koponen:2013b} %
find 
that the form factors are possibly insensitive to the spectator quarks.
    For comparison, the last three rows list results for $c\to d e^+\nu_e$ decays in
    which only the first two terms of the $z$ expansion were used.} 
\label{piPseudoZ}
\begin{center}
%
\end{center}
\end{table}

\mysubsubsubsection{Determinations of $|V_{cs}|$ and $|V_{cd}|$}
 
Results and world averages for the products $f_+^{D\to K}(0)|V_{cs}|$ and 
$f_+^{D\to \pi}(0)|V_{cd}|$ as measured by CLEO-c, Belle, \babar, and BESIII 
are summarized in Tables~\ref{tab:aver_FF_D_K} and \ref{tab:aver_FF_D_pi}, and plotted in Fig.~\ref{fig:aver_FF_D}. 
When calculating these world averages, the systematic uncertainties of the
BESIII analyses are conservatively taken to be fully correlated.

Assuming unitarity of the CKM matrix, the values of the CKM matrix elements entering in charm semileptonic decays are evaluated as~\cite{PDG_2022}
\begin{equation}
\label{eq:charm:ckm}
\begin{aligned}
|V_{cs}| & = 0.97349\pm 0.00016 \, ,\\
|V_{cd}| & = 0.22486\pm 0.00067 \, .
\end{aligned}
\end {equation}
The unitarity assumption means that the measurement of charm decays themselves have insignificant influence on these averages, and they can thus be used to give a determination of the form factors under the unitarity assumption. Using the world average values of $f_+^K(0)|V_{cs}|$ and $f_+^{\pi}(0)|V_{cd}|$
from Tables~\ref{tab:aver_FF_D_K} and \ref{tab:aver_FF_D_pi} leads to the 
form factor values 
\begin{equation}
\begin{aligned}
 f_+^{D\to K}(0) & = 0.7376 \pm 0.0034  \, , \\ 
 f_+^{D\to \pi}(0) &= 0.6342 \pm 0.0082 \,. 
\end{aligned}\nonumber 
\end {equation}
Table \ref{table:FF_DKpi} summarizes $f^{D\to\pi}_+(0)$ and $f^{D\to K}_+(0)$ 
results based on $N_f=2+1+1$ flavour lattice QCD of the 
ETM~\cite{etm:2017lub}, HPQCD~\cite{Parrott:2022rgu}, and Fermilab Lattice and MILC~\cite{FermilabLattice:2022gku} Collaborations. The weighted averages 
are $f^{D\to K}_+(0)=0.7449\pm0.0025$ and
$f^{D\to\pi}_+(0)=0.6296\pm0.0050$. 
The experimental averages of $f_+^{D\to K}(0)$ and $f_+^{D\to \pi}(0)$ agree with the corresponding lattice QCD calculations within $2\sigma$.

Alternatively, if one assumes the lattice QCD form factor values,
the averages in Tables~\ref{tab:aver_FF_D_K} and \ref{tab:aver_FF_D_pi} give
\begin{equation}
\begin{aligned}
|V_{cs}| &= 0.9639 \pm 0.0044({\rm exp.})\pm 0.0032({\rm LQCD})  \, , \\ 
|V_{cd}| &= 0.2265 \pm 0.0029({\rm exp.})\pm 0.0018({\rm LQCD})\,.
\end{aligned}\nonumber 
\end {equation} 
Here, the uncertainties are dominated by the lattice QCD calculations.

\begin{table}[htbp]
\centering
\caption{Results for $f_+^{D\to K}(0)|V_{cs}|$ from various experiments.
 \babar~2007~\cite{Aubert:2007wg} and Belle 2006~\cite{Widhalm:2006wz} only reported $f_{+}^{D\to K}(0)$ values. The listed $f_+^{D\to K}(0)|V_{cs}|$ values of these two experiments are obtained by multiplying $f_{+}^{D\to K}(0)$ with their quoted $|V_{cs}|$.
}
\label{tab:aver_FF_D_K}

\end{table}

\begin{table}[htbp]
\centering
\caption{Results for $f_+^{D\to \pi}(0)|V_{cd}|$ from various experiments.}
\label{tab:aver_FF_D_pi}
%
 \\
 \hline 
{\bf World average}   &    & {\bf 0.1426(18)} & BESIII syst.~fully correlated \\
\hline
\end{tabular}
\end{table}

\begin{figure}[hbt!]
\centering
\includegraphics[width=0.49\textwidth]{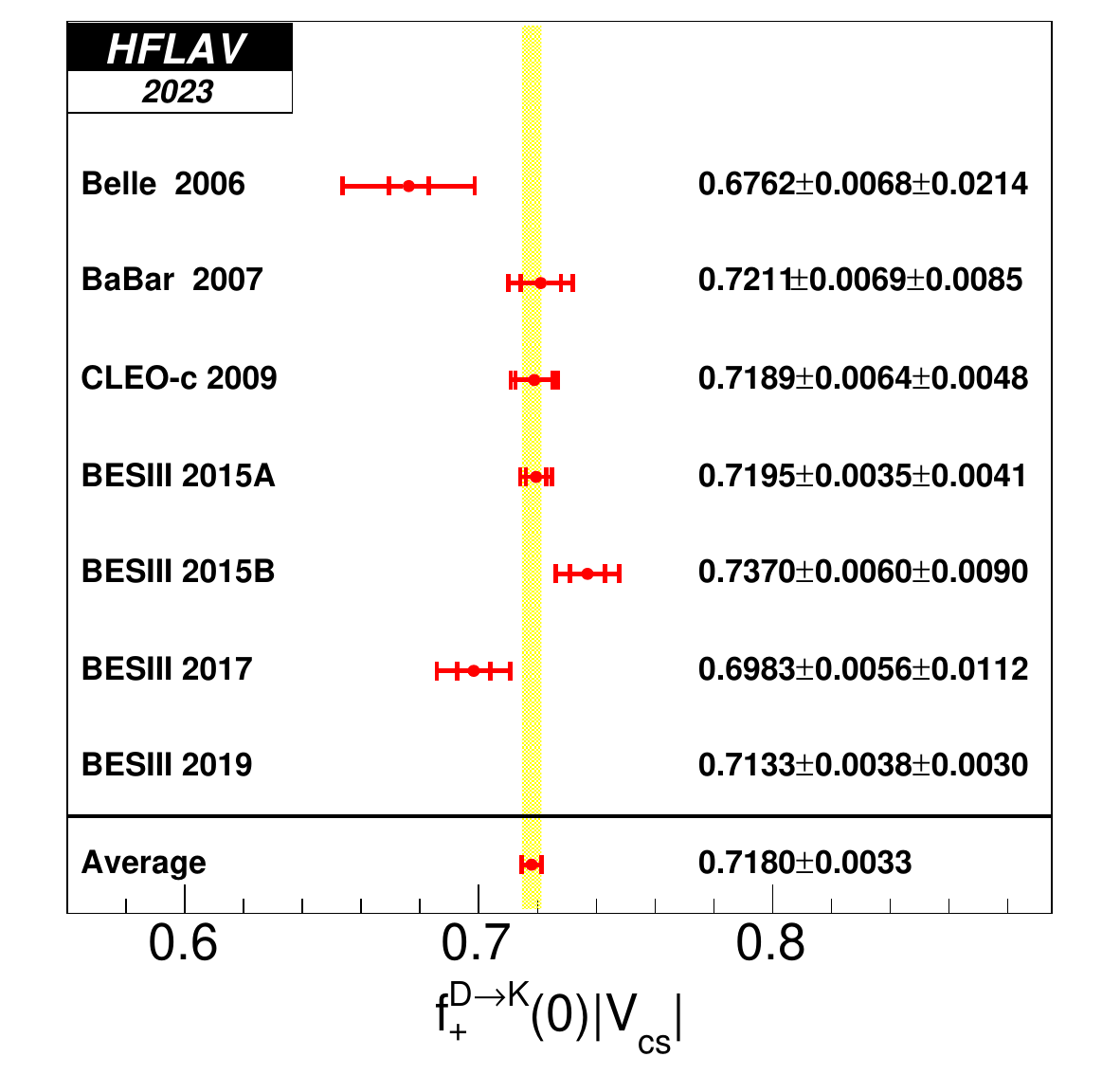}~
\includegraphics[width=0.49\textwidth]{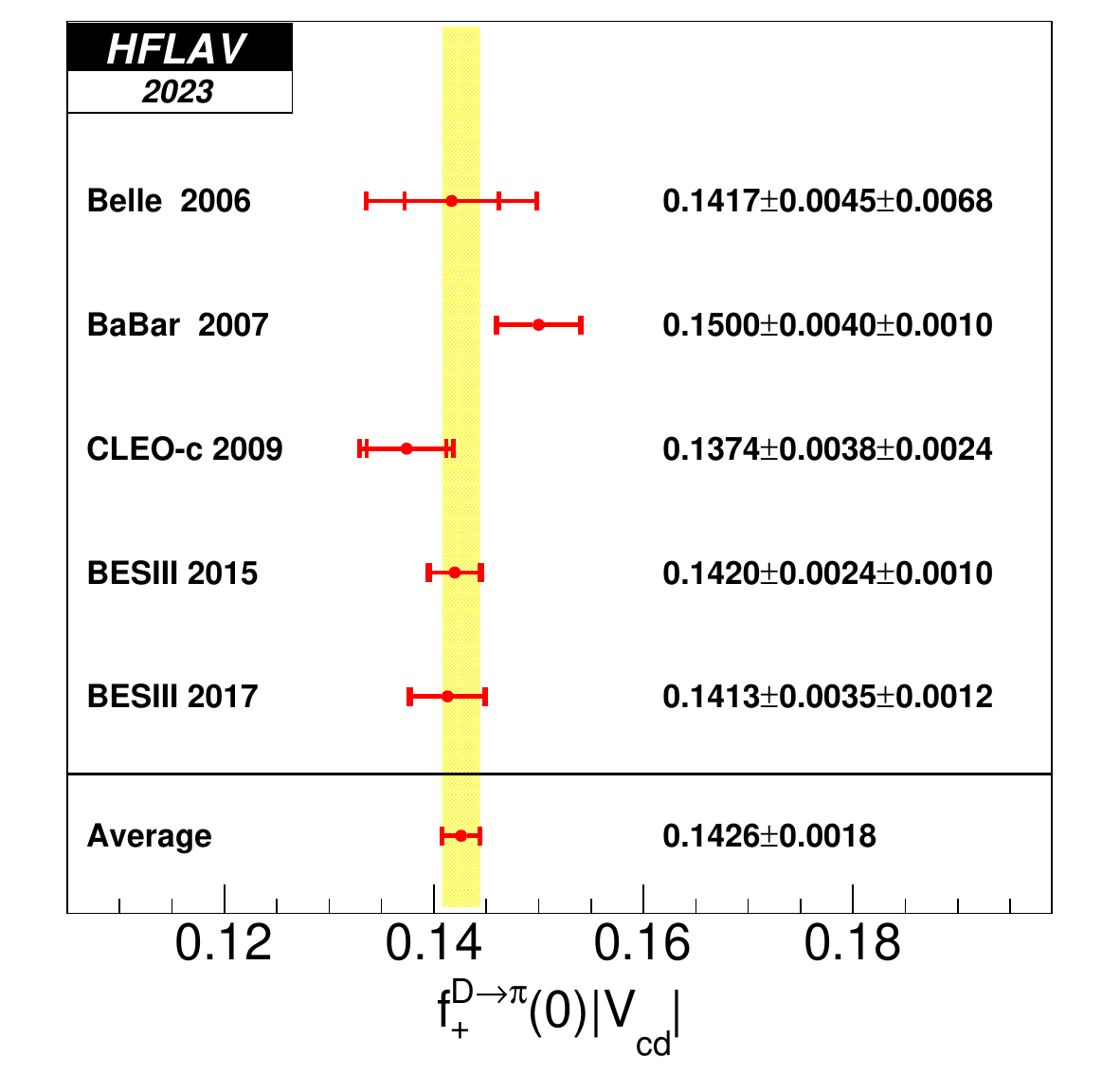}
\caption{
Comparison of the experimental results of $f_+^{D\to K}(0)|V_{cs}|$ and $f_+^{D\to \pi}(0)|V_{cd}|$.
\label{fig:aver_FF_D}
}
\end{figure}

\begin{table*}[htp]
\centering
\caption{\label{table:FF_DKpi}
\small Summary of the latest LQCD calculations of $f^{D\to\pi}_+(0)$ and 
$f^{D\to K}_+(0)$ from the Fermilab/MILC, ETM, and HPQCD collaborations.}
\begin{tabular}{lcc}
\hline
Collaboration    &$f^{D\to\pi}_+(0)$&$f^{D\to K}_+(0)$\\ \hline
Fermilab Lattice and MILC(2+1+1)~\cite{FermilabLattice:2022gku}&$0.6300\pm0.0051$&$0.7452\pm0.0031$\\
ETM(2+1+1)~\cite{etm:2017lub}&$0.612\pm0.035$&$0.765\pm0.031$\\
HPQCD(2+1+1)~\cite{Parrott:2022rgu}&--&$0.7441\pm0.0040$\\ \hline
Average&$0.6296\pm0.0050$&$0.7449\pm0.0025$\\ \hline
\end{tabular}
\end{table*}

\mysubsubsubsection{Experimental results of other $D_{(s)}\to P\ell^+\nu_\ell$ decays}

In the past two decades, rapid progress in lattice QCD calculations of 
$f_+^{D\to K\,(\pi)}(0)$ has been achieved, motivated by much improved 
experimental measurements of $D\to \bar K\ell^+\nu_\ell$ and $D\to \pi\ell^+\nu_\ell$. 
However, in contrast, progress in theoretical calculations of form 
factors in other $D_{(s)}\to P\ell^+\nu_\ell$ decays has been slow, 
and experimental measurements sparse. Before BESIII, only CLEO reported 
a measurement of $f_+^{D\to\eta}(0)$~\cite{cleo:2011jy}. For this 
analysis both tagged and untagged methods were used.
In 2018 and 2019, BESIII reported measurements of $f_+^{D\to\eta}(0)$, 
$f_+^{D_s\to\eta^{(\prime)}}(0)$, and $f_+^{D_s\to K}(0)$ with a tagged method. The former one was measured by using 2.93 fb$^{-1}$ of $e^+e^-$ collision data taken at $\sqrt s=3.773$ GeV~\cite{bes3:2018zhy,bes3:2020lius}, and the latter two were measured by analyzing  3.19 fb$^{-1}$ of $e^+e^-$ collision data taken at $\sqrt s=4.178$ GeV~\cite{bes3:2019yyh,bes3:2018sll}. In 2023, BESIII reported the updated measurements of $f_+^{D_s\to\eta^\prime}(0)$ with the same method, by analyzing the  $D^+_s\to \eta^{(\prime)} e^+\nu_e$~\cite{BESIII:2023ajr} and  $D^+_s\to \eta^{(\prime)} \mu^+\nu_\mu$~\cite{BESIII:2023gbn} decays from 7.33 fb$^{-1}$ of $e^+e^-$ collision data taken at $\sqrt s=4.128-4.226$ GeV. 
These 
measurements greatly expand experimental knowledge of hadronic form factors 
in $D\to P\ell^+\nu_\ell$ decays. To date, there is still no measurement 
of $f_+^{D\to\eta^\prime}(0)$ due to the small amount of data available.

The results and world averages of the products $f_+^{D^+_s\to\eta}(0)|V_{cs}|$, $f_+^{D^+_s\to\eta^\prime}(0)|V_{cs}|$, and $f_+^{D\to\eta}(0)|V_{cd}|$, which have been measured by BESIII and CLEO-c, are summarized in Tables \ref{tab:aver_FF_Ds_eta}, \ref{tab:aver_FF_Ds_eta}, and~\ref{tab:aver_FF_D_eta}  and plotted in Fig.~\ref{fig:aver_FF_others}.  In averaging, the systematic uncertainties of the two BESIII analyses are conservatively taken to be fully correlated.

\begin{table*}[htbp]
\centering
\caption{Results for $f_+^{D^+_s\to \eta}(0)|V_{cs}|$ from various experiments.}
\label{tab:aver_FF_Ds_eta}
\begin{tabular}{l|cc|l}
\hline
Experiment & Mode & $f_+^{D^+_s\to \eta}(0)|V_{cs}|$ & Comment \\
\hline
BESIII~\cite{BESIII:2023ajr} & $D^+_s\to \eta e^+\nu_e$           & 0.452(07)(07)  &
      $z$ expansion, 2 terms \\
BESIII~\cite{BESIII:2023gbn} & $D^+_s\to \eta\mu^+\nu_\mu$        & 0.452(10)(07)  &
      $z$ expansion, 2 terms \\ \hline
{\bf World average}   &    & {\bf 0.452(09)} & Syst.~fully correlated \\
\hline
\end{tabular}
\end{table*}

\begin{table*}[htbp]
\centering
\caption{Results for $f_+^{D^+_s\to \eta^\prime}(0)|V_{cs}|$ from various experiments.}
\label{tab:aver_FF_Ds_etap}
\begin{tabular}{l|cc|l}
\hline
Experiment & Mode & $f_+^{D^+_s\to \eta^\prime}(0)|V_{cs}|$ & Comment \\
\hline
BESIII~\cite{BESIII:2023ajr} & $D^+_s\to \eta^\prime e^+\nu_e$           & 0.525(24)(09)  &
      $z$ expansion, 2 terms \\
BESIII~\cite{BESIII:2023gbn} & $D^+_s\to \eta^\prime\mu^+\nu_\mu$        & 0.504(37)(12)  &
      $z$ expansion, 2 terms \\ \hline
{\bf World average}   &    & {\bf 0.519(10)} & Syst.~fully correlated \\
\hline
\end{tabular}
\end{table*}

\begin{table}[htbp]
\centering
\caption{Results for $f_+^{D\to\eta}(0)|V_{cd}|$ from various experiments.}
\label{tab:aver_FF_D_eta}
\begin{tabular}{l|cc|l}
\hline
 Experiment & Mode & $|V_{cd}| f_{+}^{D\to\eta}(0)$ & Comment \\
\hline
BESIII 2020~\cite{bes3:2020lius} & $D^+\to \eta \mu^+\nu_\mu$            & 0.087(8)(2)  &
      $z$ expansion, 2 terms \\
BESIII 2018~\cite{bes3:2018zhy} & $D^+\to \eta e^+\nu_e$        & 0.079(6)(2)  &
      $z$ expansion, 2 terms \\
CLEO-c 2009~\cite{cleo:2011jy} & $D^+\to \eta e^+\nu_e$     & 0.085(6)(1)  &
      $z$ expansion, 2 terms \\
{\bf World average}   &    & {\bf 0.083(4)} & BESIII syst.~fully correlated \\
\hline
\end{tabular}
\end{table}

\begin{figure}[hbt!]
\centering
\includegraphics[width=0.495\textwidth]{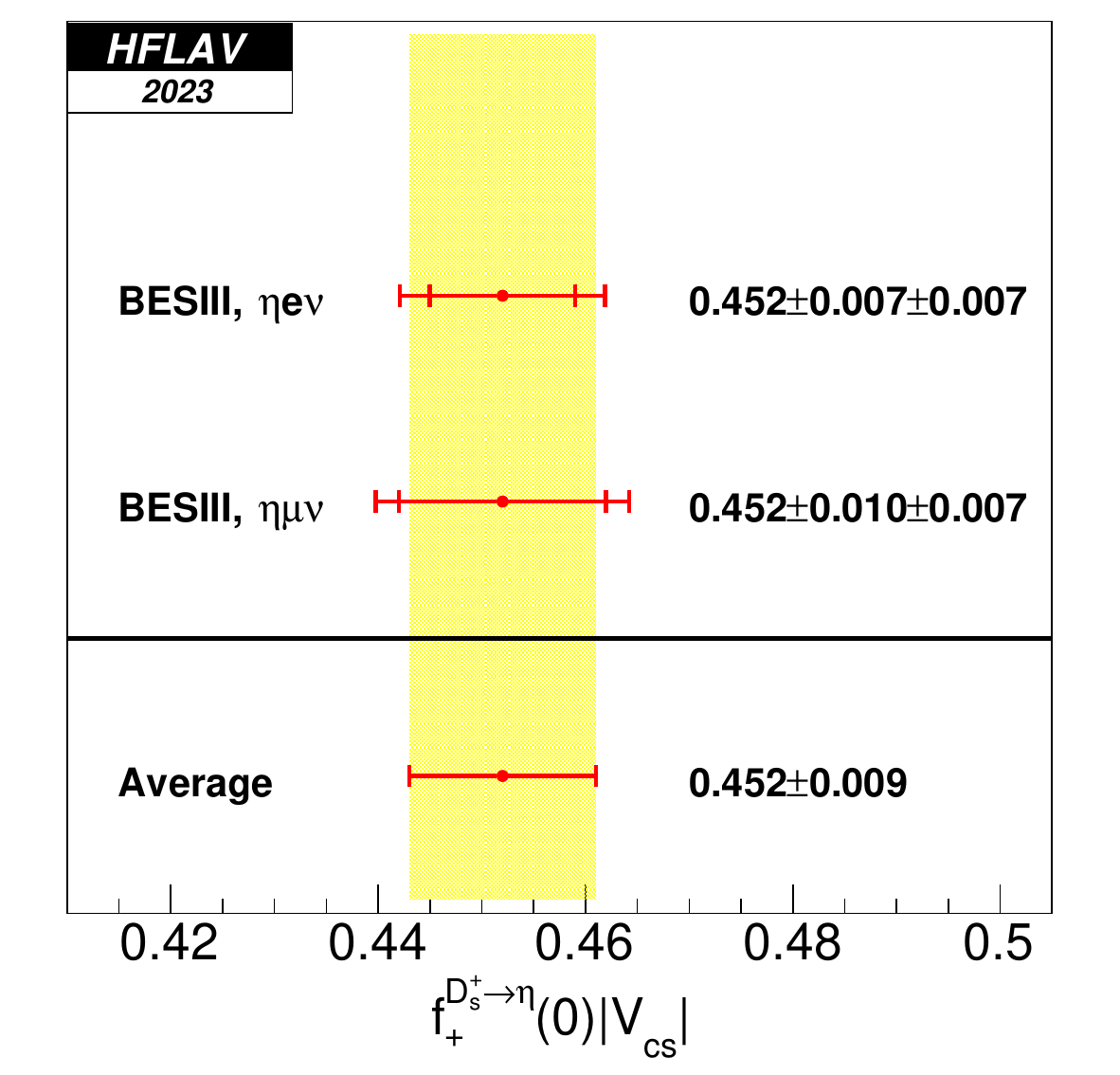}~
\includegraphics[width=0.495\textwidth]
{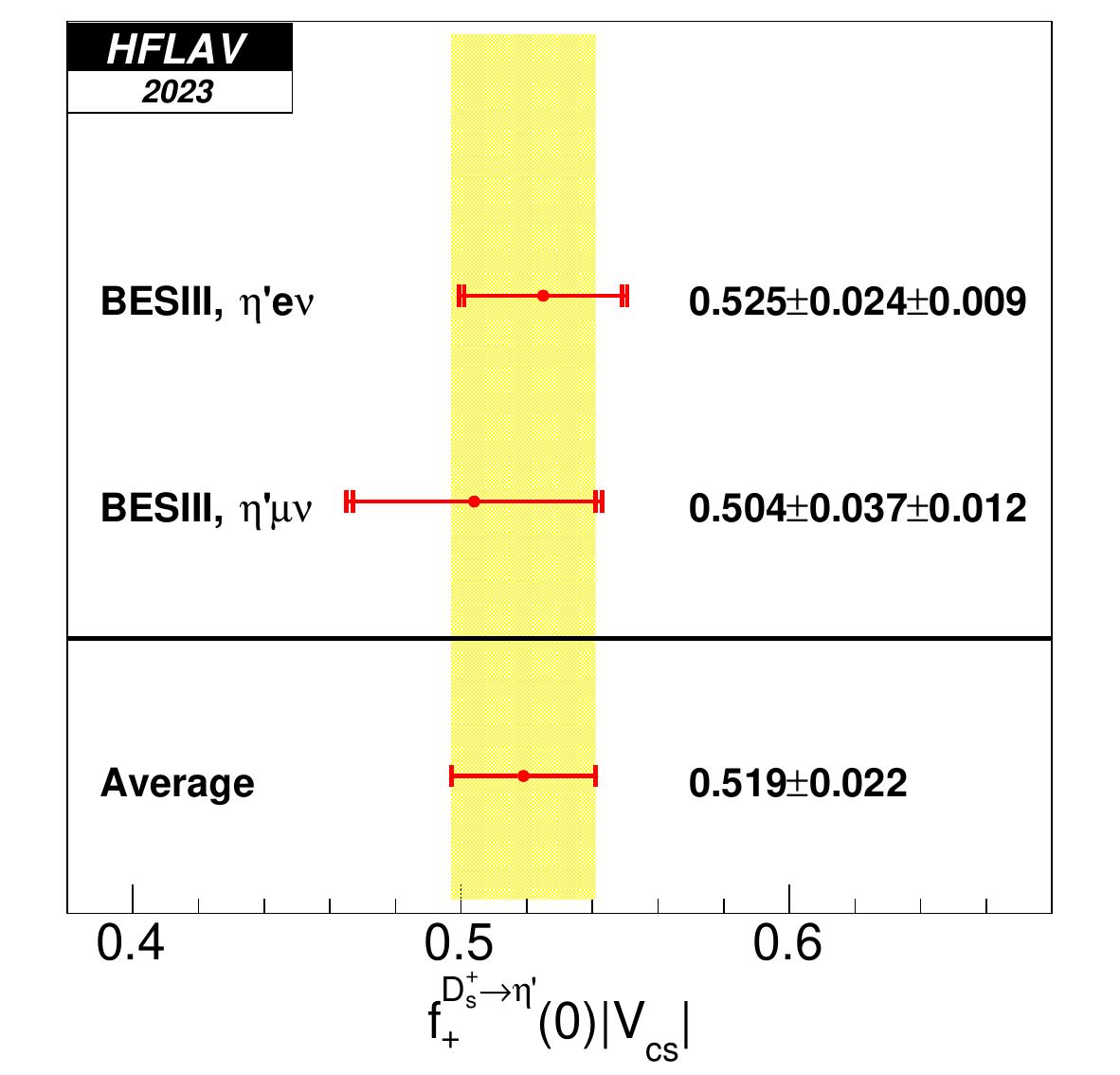}
\includegraphics[width=0.495\textwidth]{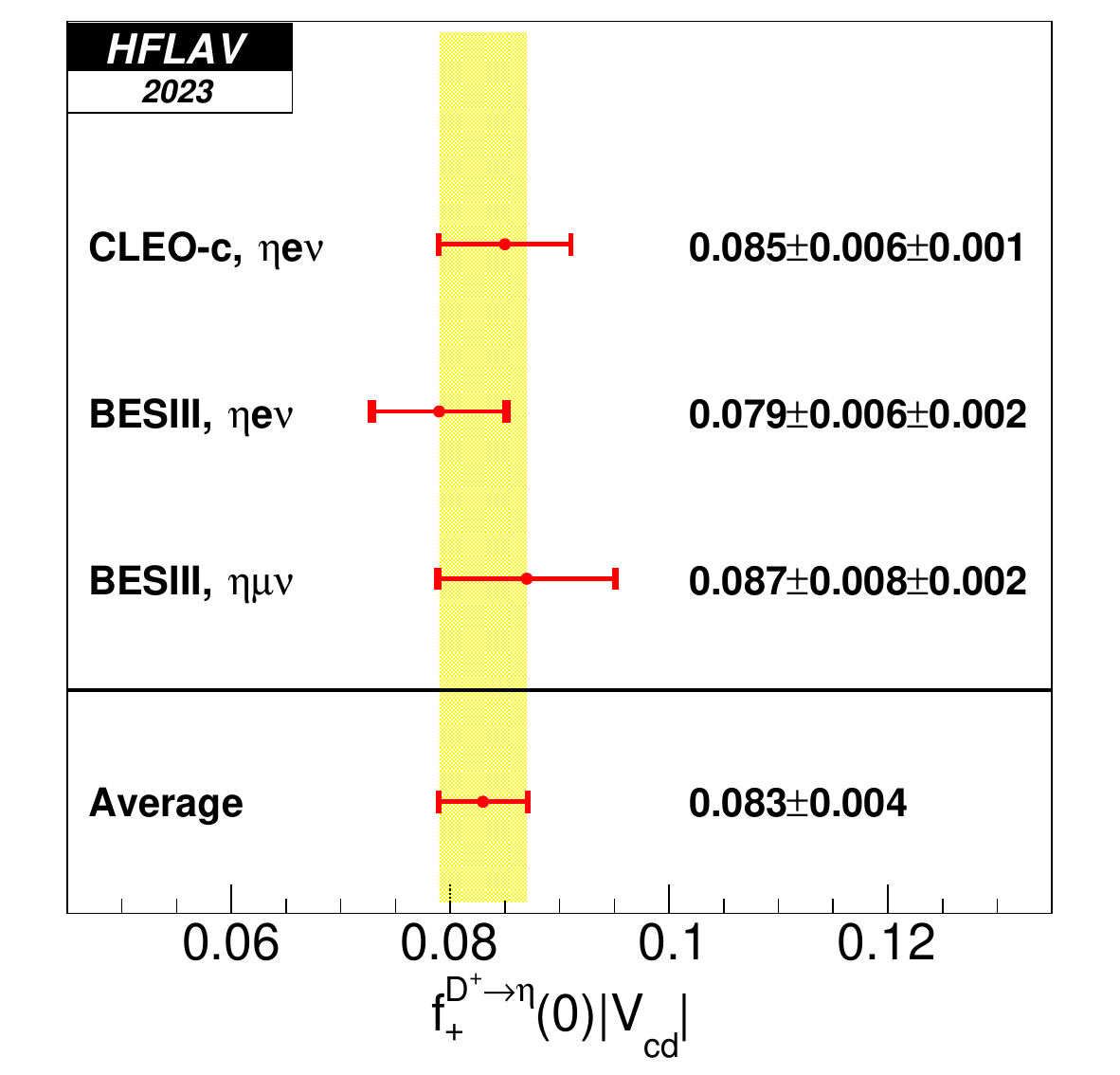}
\caption{
Comparison of the experimental results of $f_+^{D^+_s\to \eta}(0)|V_{cs}|$, $f_+^{D^+_s\to \eta^\prime}(0)|V_{cs}|$  and $f_+^{D^+\to \eta}(0)|V_{cd}|$.
\label{fig:aver_FF_others}
}
\end{figure}

\mysubsubsection{$D\ra V \ell^+\nu_\ell$ decays}

\subsubsubsection{Theoretical formula}

When the final state hadron is a vector meson, the decay can proceed through
both vector and axial vector currents, and four form factors are needed.
The hadronic current is $H^{}_\mu = V^{}_\mu + A^{}_\mu$, 
where~\cite{Gilman:1989uy} 
\begin{eqnarray}
V_\mu & = & \left< V(p,\varepsilon) | \bar{q}\gamma_\mu c | D(p') \right> \ =\  
\frac{2V(q^2)}{m_D+m_V} 
\varepsilon_{\mu\nu\rho\sigma}\varepsilon^{*\nu}p^{\prime\rho}p^\sigma, \\
 & & \nonumber\\
A_\mu & = & \left< V(p,\varepsilon) | -\bar{q}\gamma_\mu\gamma_5 c | D(p') \right> 
 \ =\  -i\,(m_D+m_V)A_1(q^2)\varepsilon^*_\mu \nonumber \\
 & & \hskip2.10in 
  +\ i \frac{A_2(q^2)}{m_D+m_V}(\varepsilon^*\cdot q)(p' + p)_\mu \\
 & & \hskip2.10in 
+\ i\,\frac{2m_V}{q^2}\left(A_3(q^2)-A_0(q^2)\right)[\varepsilon^*\cdot (p' +
p)] q_\mu\,. \nonumber 
\end{eqnarray}
In this expression, $m_V$ is the invariant mass of the daughter particles of the $V$ meson  and
\begin{equation}
  A_3(q^2) = \frac{m_D + m_V}{2m_V}A_1(q^2)\ -\ \frac{m_D - m_V}{2m_V}A_2(q^2)\,.
\end{equation}
To avoid divergence of the $i\,\frac{2m_V}{q^2} \left (A_3(q^2)-A_0(q^2) \right )$ item, kinematics require that $A_3(0) = A_0(0)$. 
Terms proportional to $q_\mu$ are important for the $\tau$ case. Thus, only the three form factors 
$A_1(q^2)$, $A_2(q^2)$ and $V(q^2)$ are relevant for charm decays.
Usually, the $q^2$ dependent hadronic form factors $A_{1/2}(q^2)$ and $V(q^2)$ take simple pole forms, $A_{1/2}(q^2)=A_{1/2}(0)/(1-q^2/M^2_A)$ and $V(q^2)=V(0)/(1-q^2/M^2_V)$, with pole masses of $M_V=M_{D^*(1^-)}\backsimeq 2.1$~GeV/$c^2$ and $M_A=M_{D^*(1^+)}\backsimeq 2.5$~GeV/$c^2$, respectively.

The differential decay rate is
\begin{eqnarray}
\frac{d\Gamma(D \to V \overline \ell \nu_\ell)}{dq^2\, d\cos\theta_\ell} & = & 
  \frac{G_F^2\,|V_{cq}|^2}{128\pi^3m_D^2}\,p^*\,q^2 \times \nonumber \\
 & &  
\left[\frac{(1-\cos\theta_\ell)^2}{2}|H_-|^2\ +\  
\frac{(1+\cos\theta_\ell)^2}{2}|H_+|^2\ +\ \sin^2\theta_\ell|H_0|^2\right]\,,
\end{eqnarray}
where $H^{}_\pm$ and $H^{}_0$ are helicity amplitudes, corresponding to 
helicities of the vector ($V$) meson. The helicity amplitudes
can be expressed in terms of the form factors as
\begin{eqnarray}
H_\pm & = & \frac{1}{m_D + m_V}\left[(m_D+m_V)^2A_1(q^2)\ \mp\ 
      2m^{}_D\,p^* V(q^2)\right], \\
 & & \nonumber \\
H_0 & = & \frac{1}{|q|}\frac{m_D^2}{2m_V(m_D + m_V)}\ \times\ \nonumber \\
 & & \hskip0.01in \left[
    \left(1- \frac{m_V^2 - q^2}{m_D^2}\right)(m_D + m_V)^2 A_1(q^2) 
    \ -\ 4{p^*}^2 A_2(q^2) \right]\,.
\label{HelDef}
\end{eqnarray}
Here 
$p^*$ is the magnitude of the three-momentum of the $V$ system as measured 
in the $D$ rest frame, and $\theta_\ell$ is the angle of the lepton momentum 
with respect to the direction opposite that of the $D$ in the $W$ rest frame 
(see Fig.~\ref{DecayAngles} for the electron case, $\theta_e$).
The left-handed nature of the quark current manifests itself as
$|H_-|>|H_+|$. The differential decay rate for $D\ra V\ell^+\nu_\ell$ 
followed by the vector meson decaying into two pseudoscalars is
\begin{eqnarray}
\frac{d\Gamma(D\ra V \overline \ell\nu, V\ra P_1P_2)}{dq^2 d\cos\theta_V d\cos\theta_\ell d\chi} 
 &  = & \frac{3G_F^2}{2048\pi^4}
       |V_{cq}|^2 \frac{p^*(q^2)q^2}{m_D^2} {\cal B}(V\to P_1P_2)\ \times \nonumber \\ 
 & &  \Big\{ (1 + \cos\theta_\ell)^2 \sin^2\theta_V |H_+(q^2)|^2 \nonumber \\
 & & +\ (1 - \cos\theta_\ell)^2 \sin^2\theta_V |H_-(q^2)|^2 \nonumber \\
 & &  +\ 4\sin^2\theta_\ell\cos^2\theta_V|H_0(q^2)|^2 \nonumber \\
 & &  -\ 4\sin\theta_\ell (1 + \cos\theta_\ell) 
             \sin\theta_V \cos\theta_V \cos\chi H_+(q^2) H_0(q^2) \nonumber \\
 & &  +\ 4\sin\theta_\ell (1 - \cos\theta_\ell) 
          \sin\theta_V \cos\theta_V \cos\chi H_-(q^2) H_0(q^2) \nonumber \\
 & &  -\ 2\sin^2\theta_\ell \sin^2\theta_V 
                \cos 2\chi H_+(q^2) H_-(q^2) \Big\}\,,
\label{eq:dGammaVector}
\end{eqnarray}
where the helicity angles $\theta^{}_\ell$, $\theta^{}_V$, and 
acoplanarity angle $\chi$ are defined as shown in Fig.~\ref{DecayAngles}. 
Inclusion of the $S$-wave $D\to K\pi \ell^+\nu_\ell$ contribution, denoted an s-wave form factor $h_0(q^2)$, leads to an interference term between them.
Usually, the ratios of the form factors at $q^2=0$ are defined as
\begin{eqnarray}
r^{}_V  & \equiv & \frac{V(0)}{A_1(0)}\,, \\
 & & \nonumber \\
r^{}_2 & \equiv & \frac{A_2(0)}{A_1(0)}\,. \label{rVr2_eq}
\end{eqnarray}
From the experimental point of view, these ratios can be obtained without any assumption
about the total decay rates or the CKM matrix elements.

\begin{figure}[htbp]
  \begin{center}
\includegraphics[width=2.5in, viewport=0 0 320 200]{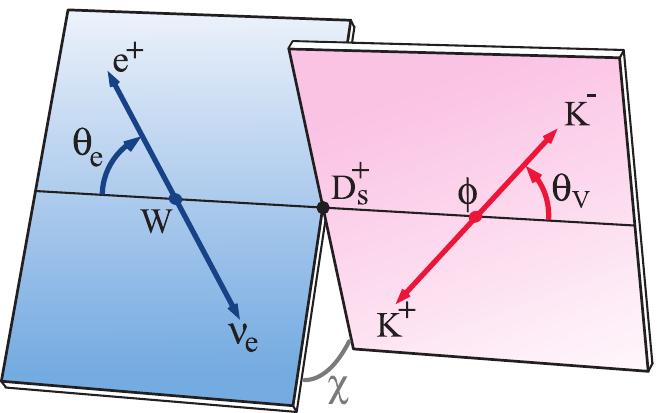}
  \end{center}
  \caption{
    Decay angles $\theta_V$, $\theta_\ell$ 
    and $\chi$. Note that the angle $\chi$ between the decay
    planes is defined in the $D$-meson reference frame, whereas
    the angles $\theta^{}_V$ and $\theta^{}_\ell$ are defined
    in the $V$ meson and $W$ reference frames, respectively.}
  \label{DecayAngles}
\end{figure}

\subsubsubsection{Experimental results}

In 2002 FOCUS reported an asymmetry in the observed $\cos(\theta_V)$ 
distribution in $D^+\to K^-\pi^+\mu^+\nu_\mu$ decays~\cite{Link:2002ev}. 
This was interpreted as
evidence for an $S$-wave $K^-\pi^+$ component in the decay amplitude. 
It should be noted that $H_0(q^2)$ is equal to zero at $q^2 = q^2_{\rm max}$ but dominated over a wide range of $q^2$~\cite{Ivanov:2019nqd}. 
The distribution given 
by Eq.~(\ref{eq:dGammaVector}) is, after integration over $\chi$,
roughly proportional to $\cos^2\theta_V$. 
Inclusion of a constant $S$-wave amplitude of the form $A\,e^{i\delta}$ 
leads to an interference term proportional to 
$|A H_0 \sin\theta_\ell \cos\theta_V|$ which then causes an asymmetry 
in $\cos(\theta_V)$.
When FOCUS fit their data including this $S$-wave amplitude, 
they obtained $A = 0.330 \pm 0.022 \pm 0.015~\gev^{-1}$ and 
$\delta = 0.68 \pm 0.07 \pm 0.05$~\cite{Link:2002wg}. 
Both \babar~\cite{Aubert:2008rs} and CLEO-c~\cite{Ecklund:2009fia} 
have also found evidence for an $f^{}_0 \to K^+ K^-$ component in semileptonic $D^{}_s$ decays.

The CLEO-c collaboration extracted the form factors $H_+(q^2)$, $H_-(q^2)$, 
and $H_0(q^2)$ from 11000 $D^+ \rightarrow K^- \pi^+ \ell^+ \nu_\ell$ events with purity greater than 96\%
in a model-independent fashion directly as functions of $q^2$~\cite{Briere:2010zc}. 
They also determined the $S$-wave form factor $h_0(q^2)$ via the interference term, despite the
fact that the $K\pi$ mass distribution appears dominated by the vector
$K^*(892)$ state. 
The transverse form factor,
\begin{eqnarray}
H_t(\qsq) &=&
   {M_D K\over m_{K\pi}\sqrt{\qsq}}
   \left[ (M_D+m_{K\pi})A_1(\qsq)
    -{(M^2_D -m^2_{K\pi}+\qsq) \over M_D+m_{K\pi}}A_2(\qsq) \right. \nonumber\\
          & & \mbox{} \hspace{2cm} \left.
    +{2\qsq\over M_D+m_{K\pi}}A_3(\qsq) \right] \,,
\end{eqnarray}
which can be related 
to $A_3(q^2)$, is small compared to LQCD calculations 
and suggests that the form factor ratio $r_3 \equiv A_3(0) / A_1(0)$ is large and negative.

The \babar collaboration selected a large sample of 
$244\times 10^3$ $D^+ \rightarrow K^- \pi^+ e^+ \nu_e$ candidates 
with a ratio $S/B\sim 2.3$ from an integrated luminosity of 
$347~\fb^{-1}$~\cite{delAmoSanchez:2010fd}. With four particles emitted 
in the final state, the differential decay rate depends on five variables.
In addition to the four variables defined in previous sections there is 
also $m^2$, the mass squared of the $K\pi$ system.
To analyze the $D^+ \rightarrow K^- \pi^+ e^+ \nu_e$ decay channel, 
it was assumed that all form factors have a $q^2$ variation given by 
the simple pole model, and an effective pole mass of 
$m_A=(2.63 \pm 0.10 \pm 0.13)~\gevcc$ is fitted. This value is compatible
with expectations when comparing to the mass of $J^P=1^+$ charm mesons. 
For the mass dependence of the form factors, a Breit-Wigner with a
mass-dependent width and a Blatt-Weisskopf damping factor is used. For the 
$S$-wave amplitude, a polynomial below the $\overline{K}^*_0(1430)$, and 
a Breit-Wigner distribution above, are used~\cite{delAmoSanchez:2010fd}. These are consistent with
measurements of $D^+ \rightarrow K^- \pi^+\pi^+$ decays.
For the polynomial part, a linear term is sufficient to fit the data.
It is verified that the variation of the $S$-wave phase is compatible 
with expectations from elastic $K\pi$ 
scattering~\cite{Estabrooks:1977xe,Aston:1987ir} (after correcting for
$\delta^{3/2}$) according to the Watson theorem~\cite{Watson:1954uc}.
As compared with elastic $K^-\pi^+$ scattering, there is an additional
negative sign between the $S$ and $P$ waves.
Contributions from other spin-1 and spin-2 resonances decaying into $K^-\pi^+$ 
are also considered.

From an $e^+e^-$ collision data sample corresponding to an integrated luminosity of $2.92$~fb$^{-1}$ taken at $\sqrt s=3.773$ GeV,
the BESIII Collaboration obtained an almost background free sample which contains 18262 $D^{+} \to K^{-} \pi^+ e^+ \nu_e$ events~\cite{bes3:2016aff}, benefiting from double-tag technique.
A partial wave analysis on this sample reveals that the $D^{+} \to K^{-} \pi^+ e^+ \nu_e$ decay is dominated by $D^{+} \to \bar K^{*}(892)^0 e^+ \nu_e$
with a $K\pi$ \emph S-wave component with a fraction of $(6.05\pm0.22\pm0.18)\%$,
and possible contributions from the $\bar K^{*}(1410)^{0}$ and $\bar K_{2}^{*}(1430)^{0}$ are negligible. The parameterizations of hadronic form factors follow those of the \babar collaboration~\cite{delAmoSanchez:2010fd}.
The hadronic transition form factor ratios ($r_V$ and $r_2$) for $D^{+} \to \bar K^{*}(892)^0 e^+ \nu_e$ are extracted with 
the vector pole mass $m_{V}$ floated or fixed at 2.0 GeV/$c^2$;
and the pole mass of the axial form factor is free. 
The obtained results in different cases are consistent with each other.
The \emph S-wave phase variation with $m_{K\pi}$  is measured in a
model-independent way, and it agrees with the partial wave analysis results based on the parameterization in the LASS scattering experiment~\cite{Aston:1987ir}. 
A model-independent measurement of the $q^{2}$ dependence of the helicity basis form factors is also reported
and the obtained results are consistent with the CLEO-c result.
From the same data sample, BESIII obtained 3112 $D^0 \to K^0_S \pi^- e^+ \nu_e$ events~\cite{bes3:2018dll}
with background fraction of 0.6\%. Based on the first amplitude analysis on the $D^0 \to K^0_S \pi^- e^+ \nu_e$ sample, 
the hadronic form factor ratios for $D^0 \to \bar K^{*}(892)^- e^+ \nu_e$ are extracted.

In 2013, the CLEO-c experiment reported a measurement of the $D^{0(+)}\to \rho^{-(0)} e^+\nu_e$ decay dynamics~\cite{Dobbs:2013sd}
from the analysis of $0.818$~fb$^{-1}$ of data taken at $\sqrt s=3.774$ GeV.
Unlike the other measurements, the $\pi\pi$ invariant mass is not used in their amplitude analysis. 
Based on several hundreds of signal events, the hadronic form factor ratios for $D^{0(+)}\to \rho^{-(0)} e^+\nu_e$ are determined for the first time.
In the measurement, the interference term between a possible $s$-wave $\pi\pi$ component and the $\rho$ amplitude has not been included,
and its absence is treated as a source of systematic uncertainty.

Later, the BESIII Collaboration reported the analysis of the decay dynamics of $D^0\to \pi^-\pi^0 e^+\nu_e$ and $D^+\to \pi^+\pi^-e^+\nu_e$~\cite{bes3:2018zhl}
by using $2.92$~fb$^{-1}$ of data taken at $\sqrt s=3.773$ GeV.
The signal yields for $D^0$ and $D^+$ decays are 1498 and 2017 with background fractions of 33\% and 24\%, respectively.
The hadronic system in $D^+\to \pi^+\pi^-e^+\nu_e$ is found to be dominated by the $P$ wave, which is mostly from $\rho^0$ contribution 
along with much smaller $\omega$ contribution.
Additionally, the $S$-wave component of $D^+\to f_0(500)e^+\nu_e$ is observed for the first time with a relative contribution of $(25.7\pm1.6\pm1.1)\%$.
Meanwhile, the $P$-wave component of $D^0\to \rho^- e^+\nu_e$ dominates in the $\pi^-\pi^0$ system of $D^0\to \pi^-\pi^0e^+\nu_e$,
but no significant $S$-wave component is observed.
Based on the same data sample, BESIII also presented the amplitude analysis of $D^+\to \omega e^+\nu_e$
with hundreds of signal events  after ignoring the non-$\omega$ component due to limited statistics. 
In these two measurements, the hadronic form factor ratios for $D^{0(+)}\to \rho^{-(0)} e^+\nu_e$ and $D^+\to \omega e^+\nu_e$ are measured.

Using $3.19$~fb$^{-1}$ of $e^+e^-$ collision data taken at $\sqrt s=4.178$ GeV,
the BESIII Collaboration presented the first amplitude analysis of $D^+_s \to K^+\pi^- e^+ \nu_e$
with only about one hundred of signal events~\cite{bes3:2018sll}.
In 2023, by analyzing $7.33$~fb$^{-1}$ of data sample taken at in the energy region of $\sqrt s=4.188-4.226$ GeV,
BESIII reported an amplitude analysis of $D^+_s \to K^+K^-e^+ \nu_e$~\cite{BESIII:2023opt}.
The obtained $D^+_s \to K^+K^-e^+ \nu_e$ sample contains about one thousand of signal events with background fraction of 10\%.
The former and latter channels are found to be dominated by $D^+_s \to K^*(892)^0 e^+ \nu_e$ and $D^+_s \to \phi e^+ \nu_e$,
while the \emph S-wave component is found to be negligible in both two measurements.
Based on these, the hadronic form factor ratios for $D^+_s \to K^*(892)^0 e^+ \nu_e$ and $D^+_s \to \phi e^+ \nu_e$ are determined.

Table \ref{Table1} lists the obtained results of $r_V$ and $r_2$ from different experiments. The measured form factors are important to test different theoretical calculations based on quark model, QCD sum rule, and lattice QCD. The list of references is huge and see a recent review of theoretical calculations \cite{Ivanov:2019nqd} for example.

\begin{table}[htbp]
\caption{Results for $r_V$ and $r_2$ from various experiments. Experiments marked with $^*$ did not consider a separate $S$-wave contribution.
\label{Table1}}
\begin{center}

\end{center}
\end{table}

\mysubsubsection{$D\to S \ell^+\nu_\ell$ decays}

In 2018, BESIII reported measurements of semileptonic $D$ decays into a 
scalar meson, $D\to S \ell^+\nu_\ell$. The experiment measured $D^{0,+}\to a_0(980) e^+\nu_e$, with 
$a_0(980)\to\eta\pi$. Signal yields of $25.7^{+6.4}_{-5.7}$ events for
$D^0\to a_0(980)^-e^+\nu_e$, and $10.2^{+5.0}_{-4.1}$ events for 
$D^+\to a_0(980)^0e^+\nu_e$, were obtained, resulting in statistical
significances of greater than 6.5$\sigma$ and 3.0$\sigma$, 
respectively~\cite{bes3:2018dzl}. As the branching fraction for
$a_0(980)\to\eta\pi$ is not well-measured, BESIII reports the
product branching fractions
\begin{eqnarray}
\br(D^0\to a_0(980)^-e^+\nu_e)\times \br(a_0(980)^-\to\eta\pi^-) 
   & = & (1.33^{+0.33}_{-0.29}\pm0.09)\times 10^{-4}\,, \\
\br(D^+\to a_0(980)^0e^+\nu_e)\times \br(a_0(980)^0\to\eta\pi^0)
   & = & (1.66^{+0.81}_{-0.66}\pm0.11)\times 10^{-4}\,.
\end{eqnarray}
The ratio of these values can be compared to a prediction based on
QCD light-cone sum rules~\cite{Cheng:2017fkw} 
Taking the lifetimes of the $D^0$ and $D^+$ into account, and assuming 
$\br(a_0(980)^-\to\eta\pi^-) = \br(a_0(980)^0\to\eta\pi^0)$, the 
ratio of the partial widths is
\begin{eqnarray}
\frac{\Gamma(D^0\to a_0(980)^-e^+\nu_e)}{\Gamma(D^+\to a_0(980)^0e^+\nu_e)}
 & = & 2.03\pm 0.95\pm 0.06 \,.
\end{eqnarray}
This value is consistent with the prediction based on  isospin symmetry.

In 2021, BESIII searched for the semileptonic decay of $D^{+}_s\to a_0(980) e^+\nu_e$, with 
$a_0(980)^0\to\eta\pi^0$. No significant signal is observed. The product branching fraction upper limit at  the 90\% confidence level is 
$\br(D^{+}_s\to a_0(980) e^+\nu_e)\times\br(a_0(980)^0\to\eta\pi^0)<1.2\times 10^{-4}$~\cite{bes3:2021kbc}.

In 2022, BESIII reported the studies of $D^+_s\to \pi^0\pi^0e^+\nu_e$ and $D^+_s\to K^0_SK^0_Se^+\nu_e$~\cite{BESIII:2021drk}, by analyzing 6.32 fb$^{-1}$ of data taken at 4.178-4.226 GeV. The $f_0(980)$ is clearly observed in the $\pi^0\pi^0$ mass spectrum and the product branching fraction is determined to be 
$\br(D^+_s\to f_0(980)e^+\nu_e)\times \br( f_0(980)\to\pi^0\pi^0)=(7.9\pm1.4\pm0.4)\times 10^{-4}$.
No significant signal of $D^+_s\to f_0(500)e^+\nu_e$ and $D^+_s\to K^0_SK^0_Se^+\nu_e$ is observed. 
The (product) branching fraction upper limits at the 90\% confidence level are 
$\br(D^+_s\to f_0(500)e^+\nu_e)\times \br( f_0(500)\to\pi^0\pi^0)<7.3\times 10^{-4}$ and
$\br(D^+_s\to K^0_SK^0_Se^+\nu_e)<3.8\times 10^{-4}$, respectively.

In 2023, BESIII reported the studies of $D^+_s\to \pi^+\pi^-e^+\nu_e$~\cite{BESIII:2023wgr}, by analyzing 7.33 fb$^{-1}$ of data taken at 4.126-4.226 GeV. The $f_0(980)$ is observed in the $\pi^+\pi^-$ mass spectrum and the product branching fraction is determined to be 
$\br(D^+_s\to f_0(980)e^+\nu_e)\times \br( f_0(980)\to\pi^0\pi^0)=(1.72\pm0.13\pm0.10)\times 10^{-3}$.
From the studies of the dynamics of the $D^+_s\to f_0(980)e^+\nu_e$ decay, the product of the
form factor $f^{D_s\to f_0}_+(0)$  and the $c\to s$ Cabibbo-Kobayashi-Maskawa matrix element $|V_{cs}|$ is determined
for the first time to be $f^{D_s\to f_0}_+(0)|V_{cs}|=0.504\pm0.017\pm0.035$.

\mysubsubsection{$D\to A \ell^+ \nu_\ell$ decays}

Experimental studies of semileptonic $D$ decays into an 
axial-vector ($A$) meson $D\to A \ell^+\nu_\ell$ are challenging due to low statistics and high backgrounds.
In 2007, CLEO-c reported first evidence 
for the Cabibbo-favored decay $D^0\to K_1(1270)^-e^+\nu_e$ with a statistical significance of 
$4\sigma$~\cite{cleo:2007Artuso}. The branching fraction was measured
to be $\br(D^0\to K_1(1270)^-e^+\nu_e) = 
(7.6^{+4.1}_{-3.0}\pm0.6\pm0.7)\times 10^{-4}$. In 2019, BESIII reported 
the first observation of $D^+\to \bar K_1(1270)^0e^+\nu_e$, with statistical 
significance greater than $10\sigma$~\cite{bes3:2019liuk}. The branching fraction was measured to be 
$\br(D^+\to \bar K_1(1270)^0e^+\nu_e) = (23.0\pm2.6^{+1.8}_{-2.1}\pm2.5)\times 10^{-4}$.
In 2021, the $D^0\to K_1(1270)^-e^+\nu_e$ decay was observed for the first time by BESIII with a statistical significance greater than $10\sigma$~\cite{bes3:2021fyl}. The reported branching fraction is 
$\br(D^+\to \bar K_1(1270)^0e^+\nu_e) = (10.9\pm1.3^{+0.9}_{-1.3}\pm1.2)\times 10^{-4}$.
Here, the third errors listed arise from 
the branching fraction for $K_1(1270)\to K\pi\pi$. The obtained branching fractions are consistent with the theoretical calculations with the $K_1$ mixing angle of $33^\circ$ or $57^\circ$.
Taking the lifetimes of $D^0$ and $D^+$ into account, the ratio of the partial widths is
\begin{eqnarray}
\frac{\Gamma(D^+\to \bar K_1(1270)^0e^+\nu_e)}{\Gamma(D^0\to K_1(1270)^-e^+\nu_e)}
  & = & 1.20\pm0.20\pm0.15.
\end{eqnarray}
This value agrees with unity as predicted by isospin symmetry.

In 2020, BESIII searched for the Cabibbo-suppressed semileptonic decays $D^+\to b_1(1235)^0e^+\nu_e$ and $D^0\to b_1(1235)^0e^+\nu_e$, by analyzing 2.93 fb$^{-1}$ of data taken at 3.773 GeV. No significant signal is observed. The product branching fraction upper limits at the 90\% confidence level are 
$\br(D^+\to b_1(1235)^0e^+\nu_e)\times\br( b_1(1235)^0\to\omega\pi^0)<1.12\times 10^{-4}$ and
$\br(D^0\to b_1(1235)^-e^+\nu_e)\times \br( b_1(1235)^0\to\omega\pi^0)<1.75\times 10^{-4}$,
respectively~\cite{bes3:2020wux}.

In 2023, BESIII searched for $D^+_s\to K_1(1270)^0e^+\nu_e$ and $D^+_s\to b_1(1235)^0e^+\nu_e$, by analyzing 7.33 fb$^{-1}$ of data taken at 4.128-4.226 GeV. Also, no significant signal is observed. The product branching fraction upper limits at the 90\% confidence level are 
$\br(D^+_s\to K_1(1270)^0e^+\nu_e)<4.1\times 10^{-4}$ and
$\br(D^+_s\to b_1(1235)^0e^+\nu_e)\times \br( b_1(1235)^0\to\omega\pi^0)<6.4\times 10^{-4}$,
respectively~\cite{BESIII:2023clm}.

\mysubsubsection{Test of $e$-$\mu$ lepton flavour universality}

In the Standard Model (SM), the couplings between the three families 
of leptons and gauge bosons are expected to be equal; this is known 
as lepton flavour universality (LFU). The semileptonic decays of 
pseudoscalar mesons are well understood in the SM and thus offer
a robust way to test LFU and search for new physics. Various tests 
of LFU with $B$ semileptonic decays have been reported by \babar, Belle, 
and LHCb. The average of the ratio of the branching fractions 
${\mathcal B}_{B\to \bar D^{(*)}\tau^+\nu_\tau}/
{\mathcal B}_{B\to \bar D^{(*)}\ell^+\nu_\ell}$~($\ell=\mu,\,e$) 
deviates from the SM prediction by $3.1\sigma$ 
(see section~\ref{slbdecays_b2dtaunu}).  Precision 
measurements of the semileptonic $D$ decays also test LFU, and in a manner complimentary to that of $B$ decays~\cite{th:2017svj}. 
Within the SM, the ratios
${\mathcal B}_{D\to \bar K\mu^+\nu_\mu}/{\mathcal B}_{D\to \bar K e^+\nu_e}$
and 
${\mathcal B}_{D\to \pi\mu^+\nu_\mu}/{\mathcal B}_{D\to \pi e^+\nu_e}$
are predicted to be $0.975\pm0.001$ and 
$0.985\pm0.002$, respectively~\cite{epjc:2018rig}.
The ratios are expected to be close to unity with negligible uncertainty mainly due to high correlation of the corresponding hadronic form factors~\cite{epjc:2018rig}.

In the SM, the semimuonic $D$ decays are expected to have lower branching fraction than their  semielectronic counterparts. Before BESIII, however, the information related to the semimuonic $D$ decays is relatively poor, mainly due to higher backgrounds caused due to difficulty of distinguishing muon and charged pions. In the charmed meson sector, only $D^0\to K^-\mu^+\nu_\mu$, $D^0\to K^{*-}\mu^+\nu_\mu$, $D^0\to \pi^-\mu^+\nu_\mu$, $D^+\to \bar K^0\mu^+\nu_\mu$, $D^+\to \rho^0\mu^+\nu_\mu$, and $D^+\to \bar K^{*0}\mu^+\nu_\mu$ have been investigated in experiments previously. Except for $D^+\to \bar K^{*0}\mu^+\nu_\mu$, all measurements of the other decays are dominated by FOCUS and Belle experiments and the existing measurements suffer large uncertainties.

Since 2016, BESIII performed a series of studies of semimuonic $D$ decays, including improved measurements of $D^+\to\bar K^0\mu^+\nu_\mu$~\cite{bes3:2016hzl}, 
$D^0\to \pi^-\mu^+\nu_\mu$~\cite{bes3:2018wy}, and
$D^0\to K^-\mu^+\nu_\mu$~\cite{bes3:2019zsf}, and the first observations of $D^+\to \pi^0\mu^+\nu_\mu$~\cite{bes3:2018wy}, 
$D^+\to \omega\mu^+\nu_\mu$~\cite{bes3:2020lufx},
$D^+\to \eta\mu^+\nu_\mu$~\cite{bes3:2020lius}. 
All these analyses used the tagged method and 2.93 fb$^{-1}$ of data taken at 3.773 GeV. The reported branching fractions are summarized in the second column of Table~\ref{compare:bfs}. 
Combining these results with previous BESIII 
measurements of their counterparts of the semielectronic decays using the same data sample or the world average values, the ratios of branching fractions are obtained, as shown in the last column of Table~\ref{compare:bfs}.

In 2018, using 0.482 fb$^{-1}$ of data taken at the center-of-mass energy 
of 4.009~GeV, BESIII reported measurements of the branching fractions 
for semileptonic decays $D^+_s\to \phi\,\mu^+\nu_\mu$, 
$D^+_s\to \eta \mu^+\nu_\mu$, and 
$D^+_s\to \eta^\prime \mu^+\nu_\mu$~\cite{bes3:2018grp}. In 2023, BESIII reported the precision measurements of the branching fractions of $D^+_s\to \phi\,\mu^+\nu_\mu$~\cite{BESIII:2023opt}, 
$D^+_s\to \eta \mu^+\nu_\mu$, and 
$D^+_s\to \eta^\prime \mu^+\nu_\mu$~\cite{BESIII:2023gbn}, from analyses of 7.33 fb$^{-1}$ of data taken at center-of-mass energies between 4.128 and 4.226~GeV,
Combining these results with the world average value of the branching fraction of  
$D^+_s\to \phi\,e^+\nu_e$~\cite{PDG_2022}, and those of
$D^+_s\to \eta e^+\nu_e$, and $D^+_s\to \eta^\prime e^+\nu_e$ reported in Ref.~\cite{BESIII:2023ajr}. 
The obtained branching fractions, the comparison with their counterparts of the semielectronic decays using the same data sample or the world average value are also given in Table~\ref{compare:bfs}.

These results indicate that any $e$-$\mu$ LFU violation in semileptonic $D$  decays 
has to be at the level of a few percent or less. BESIII also tested $e$-$\mu$ LFU
in separate $q^2$ intervals using 
$D^{0,+}\to \pi^{-(0)}\ell^+\nu_\ell$~\cite{bes3:2018wy},
$D^0\to K^-\ell^+\nu_\ell$~\cite{bes3:2019zsf}, and 
$D^+_s\to \eta^{(\prime)} \mu^+\nu_\mu$~\cite{BESIII:2023gbn} decays.
No indication of LFU above the $2\sigma$ level was found.

\begin{table*}[htbp]
	\centering
	\caption{Comparison of the branching fractions of the semileptonic $D$ decays.}
	\label{compare:bfs}
 \resizebox{1.0\textwidth}{!}{
	\begin{tabular}{c|c|c|l}
		\hline Decay mode & $\br_\mu$ (\%) & $\br_e$ (\%) & \multicolumn{1}{c}{$\br_\mu/\br_e$} \\	\hline
$D^0\to K^-\ell^+\nu_\ell$      &$3.413\pm0.019\pm0.035$\cite{bes3:2019zsf}  &$3.505\pm0.014\pm0.033$\cite{Ablikim:2015ixa}&$0.974\pm0.014$\cite{bes3:2019zsf}\\
$D^0\to \pi^-\ell^+\nu_\ell$    &$0.272\pm0.008\pm0.006$\cite{bes3:2018wy}   &$0.295\pm0.004\pm0.003$\cite{Ablikim:2015ixa}&$0.922\pm0.037$\cite{bes3:2018wy}\\
$D^+\to \bar K^0\ell^+\nu_\ell$ &$8.72\pm0.07\pm0.18$   \cite{bes3:2016hzl}  &$8.72\pm0.09$    \cite{PDG_2022}&$1.000\pm0.024$\\
$D^+\to \pi^0\ell^+\nu_\ell$    &$0.350\pm0.011\pm0.010$\cite{bes3:2018wy}   &$0.363\pm0.008\pm0.005$\cite{bes3:2017fay}&$0.964\pm0.045$\cite{bes3:2018wy}\\
$D^+\to \omega\ell^+\nu_\ell$   &$0.177\pm0.018\pm0.011$\cite{bes3:2020lufx} &$0.169\pm0.011$  \cite{PDG_2022} &$1.05\pm0.14$           \\
$D^+\to \eta \ell^+\nu_\ell$    &$0.104\pm0.010\pm0.005$\cite{bes3:2020lius} &$0.111\pm0.007$  \cite{PDG_2022} &$0.94\pm0.12$           \\
$D^+_s\to \phi\,\ell^+\nu_\ell$ &$2.25\pm0.09\pm0.07$   \cite{BESIII:2023opt}&$2.39\pm0.16$    \cite{PDG_2022} &$0.94\pm0.08$\cite{BESIII:2023opt}           \\
$D^+_s\to \eta\ell^+\nu_\ell$   &$2.235\pm0.051\pm0.052$\cite{BESIII:2023gbn}&$2.255\pm0.039\pm0.051$\cite{BESIII:2023ajr}& $0.984\pm0.032$\cite{BESIII:2023gbn} \\
$D^+_s\to \eta'\ell^+\nu_\ell$  &$0.801\pm0.055\pm0.028$\cite{BESIII:2023gbn}&$0.810\pm0.038\pm0.024$\cite{BESIII:2023ajr}& $0.989\pm0.089$\cite{BESIII:2023gbn} \\
		\hline
	\end{tabular}}
\end{table*}

\clearpage
\subsection{Leptonic decays}

Purely leptonic decays of $\Dp$ and $\dsp$ mesons are among the 
simplest and best understood probes of $c\to d$ and $c\to s$ 
quark flavour-changing transitions. The amplitude of purely leptonic 
decays consists of the annihilation of the initial quark-antiquark 
pair ($c\overline{d}$ or $c\overline{s}$) into a virtual $W^+$ that 
subsequently materializes as an antilepton-neutrino pair ($\ellnu$). 
The Standard Model (SM) branching fraction is given by
\begin{equation}
  \br(D_{q}^+\to \ell^+\nu_{\ell})=
  \frac{G_F^2}{8\pi}\tau^{}_{D_q} f_{D_{q}}^2 |V_{cq}|^2 m_{D_{q}}m_{\ell}^2
  \left(1-\frac{m_{\ell}^2}{m_{D_{q}}^2} \right)^2\,,
  \label{eq:brCharmLeptonicSM}
\end{equation}
where $m_{D_{q}}$ is the $D_{q}$ meson mass, 
$\tau^{}_{D_q}$ is its lifetime, 
$m_{\ell}$ is the charged lepton mass, 
$|V_{cq}|$ is the magnitude of the relevant CKM matrix element, and 
$G_F$ is the Fermi coupling constant. The parameter $f_{D_{q}}$ is the 
$D_q$ meson decay constant and parameterizes the overlap of the wave 
functions of the constituent quark and anti-quark. 
This formulae and the similar formulae later do not include QED corrections. In view of the small errors quoted later these corrections may be relevant. The decay constants 
have been calculated using several methods, the most accurate 
and robust being that of lattice QCD (LQCD). 
Using the $N_f=2+1+1$ flavour LQCD calculations of $f_{D^+}$ and 
$f_{D^+_s}$ from the ETM~\cite{etm:2015lqcd} and FNAL/MILC~\cite{Bazavov:2017lyh} 
Collaborations, the Flavour Lattice Averaging Group (FLAG) calculates world 
average values~\cite{flag:2019}
\begin{eqnarray}
f^{\rm FLAG}_{D^+} & = & 212.0\pm0.7~{\rm MeV}\,, \label{eqn:fDplus_FLAG}  \\
 & & \nonumber \\
f^{\rm FLAG}_{D^+_s} & = & 249.9\pm0.5~{\rm MeV}\,, \label{eqn:fDs_FLAG}  
\end{eqnarray}
and the ratio
\begin{eqnarray}
\left(\frac{f_{D^+_s}}{f_{D^+}}\right)^{\rm FLAG} & = & 1.1783\pm0.0016\,. \label{eqn:ratio_FLAG}  
\end{eqnarray}
These values are used within this section to determine the magnitudes 
$|V_{cd}|$ and $|V_{cs}|$ from the measured branching fractions of 
$D^+\to \ell^+\nu_{\ell}$ and $D_s^+\to \ell^+\nu_{\ell}$.

The leptonic decays of pseudoscalar mesons 
are helicity-suppressed, meaning their decay rates are 
proportional to the square of the charged lepton mass. 
Thus, decays to $\tau^+\nu^{}_\tau$ are favored over decays 
to $\mu^+\nu^{}_\mu$, and decays to $e^+\nu^{}_e$, with an 
expected $\br\lesssim 10^{-7}$, are not yet experimentally 
observable. The ratio of $\tau^+\nu^{}_\tau$ to $\mu^+\nu^{}_\tau$ 
decays is given by
\begin{eqnarray}
R^{D_q}_{\tau/\mu}\ \equiv\  
\frac{\br(D^+_q\to\tau^+\nu_{\tau})}{\br(D^+_q\to\mu^+\nu_{\mu})} &  = &
\left(\frac{m_{\tau}^2}{m_{\mu}^2}\right)
\frac{(m^2_{D_q}-m^2_{\tau})^2}{(m^2_{D_q}-m^2_{\mu})^2}\,,
\end{eqnarray}
and equals $9.75\pm0.01$ for $D_s^+$ decays and 
$2.67\pm0.01$ for $D^+$ decays, based on the well-measured 
values of $m^{}_\mu$, $m^{}_\tau$, and $m^{}_{D_{(s)}}$~\cite{PDG_2022}. 
A significant deviation from this expectation would be 
interpreted as LFU violation in charged currents~\cite{Filipuzzi:2012mg}.

In this section we present world average values for the product 
$f_{D_{q}} |V_{cq}|$, where $q=d,s$. For these averages, 
correlations between measurements and dependencies on 
input parameters are taken into account. 

\subsubsection{Recent experimental results}

In 2019, BESIII reported the first observation of $D^+\to\taunu$ with a statistical signficance of $5.1\sigma$ from analysis of 2.93 fb$^{-1}$ of $e^+e^-$ collision data taken at the center-of-mass energy of 3.773 GeV.~\cite{bes3:2019hajm}. 
In the same year, BESIII reported a measurement of $D^+_s\to\munu$~\cite{Ablikim:2018jun}, by analyzing 3.19 fb$^{-1}$ of $e^+e^-$ collision data taken at a center-of-mass energy of 4.178 GeV, in which this analysis used the muon counter to identify $\mu^+$ lepton, and thereby obtain low background. 

In 2021, BESIII reported an updated measurement of  $D^+_s\to\munu$~\cite{bes3:2021hajm} without using the muon counter,
the measurements of $D^+_s\to\taunu$ with $\tau^+\to\pi^+\overline{\nu}_{\tau}$~\cite{bes3:2021hajm},
$\tau^+\to\rho^+\bar \nu_\tau$~\cite{bes3:2021px} and 
$\tau^+\to e^+\nu_e\bar \nu_\tau$~\cite{bes3:2021lhj} decays,
by analyzing 6.32 fb$^{-1}$ of $e^+e^-$ collision data taken at center-of-mass energies of 4.178-4.226 GeV. Compared to Ref.~\cite{Ablikim:2018jun}, there are much higher backgrounds in the measurement of $D^+_s\to\munu$~\cite{bes3:2021hajm} due to not using the muon counter. The updated measurement of $D^+_s\to\munu$ in Ref.~\cite{bes3:2021hajm} supersedes that reported in Ref.~\cite{Ablikim:2018jun}.

In 2023, BESIII reported a further updated measurement of  $D^+_s\to\munu$~\cite{BESIII:2023cym} by using the muon counter, an updated measurement of $\tau^+\to\pi^+\overline{\nu}_{\tau}$~\cite{BESIII:2023fhe} with a multivariate analysis method, and a measurement of
$\tau^+\to \mu^+\nu_\mu\bar \nu_\tau$~\cite{BESIII:2023ukh}, based on the analysis of 7.33 fb$^{-1}$ of $e^+e^-$ collision data taken at center-of-mass energies of 4.128-4.226 GeV. The latest measurements of $D^+_s\to\munu$ and $D^+_s\to\taunu$ with $\tau^+\to\pi^+\overline{\nu}_{\tau}$ reported in Refs.~\cite{BESIII:2023cym,BESIII:2023fhe} supersede those reported in Ref.~\cite{bes3:2021hajm}.

\mysubsubsection{$D^+\to \ell^+\nu_{\ell}$ decays and $|V_{cd}|$}

The branching fraction $\br(D^+\to\munu)$ has been determined by  \mbox{CLEO-c}~\cite{Eisenstein:2008aa} and 
BESIII~\cite{Ablikim:2013uvu}. These lead to the 
world average (WA) value
\begin{equation}
 \br^{\rm WA}(D^+\to\munu) = (3.77\pm0.18)\times10^{-4}.
 \label{eq:Br:WA:DtoMuNu}
\end{equation}
For $D^+\to\taunu$, the recent BESIII measurement~\cite{bes3:2019hajm} gives
\begin{equation}
 \br^{\rm WA}(D^+\to\taunu) = (1.20\pm0.27)\times10^{-3}.
 \label{eq:Br:WA:DtoTauNu}
\end{equation}
Based on these two branching fractions and 
using the most recent values for $m^{}_\tau$, $m^{}_{D}$, 
and $\tau^{}_{D}$~\cite{PDG_2022}, we determine the products of the decay constant and $|V_{cd}|$ to be
\begin{equation}
 f_{D}|V_{cd}| = \left(46.2\pm1.0\pm0.3\pm0.1\right)~\mbox{MeV}
 \label{eq:expFDVCD1}
\end{equation}
and
\begin{equation}
 f_{D}|V_{cd}| = \left(50.5\pm5.1\pm2.5\pm0.1\right)~\mbox{MeV},
 \label{eq:expFDVCD2}
\end{equation}
respectively.
These give the weighted product of the decay constant and $|V_{cd}|$ to be
\begin{equation}
 f_{D}|V_{cd}| = \left(46.3\pm1.0\right)~\mbox{MeV}\,.
 \label{eq:expFDVCD}
\end{equation}
The uncertainty listed includes the uncertainty on $\br^{\rm WA}(D^+\to\munu)$, 
and also uncertainties on the external parameters $m^{}_\mu$, $m^{}_D$, and 
$\tau^{}_D$~\cite{PDG_2022} needed to extract $f_{D}|V_{cd}|$ 
from the branching fraction via Eq.~(\ref{eq:brCharmLeptonicSM}). 
Using the LQCD value for $f_D$ from FLAG [Eq.~(\ref{eqn:fDplus_FLAG})], 
we calculate the magnitude of the CKM matrix element $V_{cd}$ to be
\begin{equation}
 |V_{cd}| = 0.2186\pm0.0049\,(\rm exp.)\pm0.0007\,(\rm LQCD),
 \label{eq:Vcd:WA:Leptonic}
\end{equation}
where the first and second uncertainties are from experiment and from LQCD, 
respectively. All input values and the resulting world average are 
summarized in Table~\ref{tab:DExpLeptonic} 
and plotted in Fig.~\ref{fig:ExpDLeptonic} (left).

Using the WA values of the branching fractions $\br(D^+\to\munu)$ and
$\br(D^+\to\taunu)$ [Eq. \ref{eq:Br:WA:DtoMuNu} and Eq. \ref{eq:Br:WA:DstoTauNu}],
the ratio of these two branching fractions is determined to be
\begin{equation}
R_{\tau/\mu}^{D^+} = 3.18\pm0.73\,,
\label{eq:R:WA:DpLeptonic}
\end{equation}
which is consistent with the ratio expected in the SM.

\begin{table}[t!]
\caption{Experimental results and world averages for 
${\cal{B}}(D^+\to \ell^+\nu_{\ell})$ and $f_{D}|V_{cd}|$.
The first uncertainties are statistical and the second are experimental 
systematic. The third uncertainties in the case of $f_{D^+}|V_{cd}|$ are
due to external inputs (dominated by the uncertainty on $\tau_D$). 
Here, we take the unconstrained result from CLEO-c.
\label{tab:DExpLeptonic}}
\vskip0.15in
\begin{center}
\begin{tabular}{lccll}
\toprule
Mode 	& ${\cal{B}}$ ($10^{-4}$)	& $f_{D}|V_{cd}|$ (MeV)		& Reference & \\ 
\midrule
\multirow{2}{*}{$\munu$} & $3.95\pm0.35\pm 0.09$ 	& $47.3\pm2.1\pm0.5\pm0.1$	& CLEO-c & \cite{Eisenstein:2008aa}\\ 
			& $3.71\pm0.19\pm 0.06$ 	& $45.9\pm1.2\pm0.4\pm0.1$	& BESIII & \cite{Ablikim:2013uvu}\\
\midrule
  & {\boldmath $3.77\pm0.17\pm 0.05$}  & {\boldmath $46.2\pm1.0\pm0.3\pm0.1$} 
 & {\bf Average} & \\
 & & & & \\
\midrule
$\taunu$ 		& {$12.0\pm2.4\pm1.2$}		&{$50.5\pm5.1\pm2.5\pm0.1$}& BESIII & \cite{bes3:2019hajm}\\
\midrule
{\boldmath $\munu\,+\,\taunu$}  &  & {\boldmath $46.3\pm1.0\pm0.3\pm0.1$} & {\bf Average} & \\
 & & & & \\
\midrule
$\enu$	 		& {$<0.088$ at 90\% C.L.}	&& CLEO-c & \cite{Eisenstein:2008aa}\\
\\ \bottomrule
\end{tabular}
\end{center}
\end{table}

\mysubsubsection{$D_s^+\to \ell^+\nu_{\ell}$ decays and $|V_{cs}|$}

We use the measurements of the branching fraction $\br(D_s^+\to\munu)$ 
from CLEO-c~\cite{Alexander:2009ux}, \babar~\cite{delAmoSanchez:2010jg},
Belle~\cite{Zupanc:2013byn}, and BESIII~\cite{Ablikim:2016duz,BESIII:2023cym}
to obtain a WA value of
\begin{equation}
 \br^{\rm WA}(\dsmunu) = (5.36\pm0.12)\times10^{-3}.
 \label{eq:Br:WA:DstoMuNu}
\end{equation}
The WA value for $\br(D_s^+\to\taunu)$ is calculated from 
the CLEO-c, \babar, Belle, and BESIII measurements. 
For the CLEO-c experiment, the separate measurements of $\br(D_s^+\to\taunu)$ were performed with the
$\tau^+\to e^+\nu_e\overline{\nu}{}_{\tau}$~\cite{Naik:2009tk},
$\tau^+\to\pi^+\overline{\nu}{}_{\tau}$~\cite{Alexander:2009ux}, and $\tau^+\to\rho^+\overline{\nu}{}_{\tau}$ decays~\cite{Onyisi:2009th}.
The Belle measurement was based on the tau decay modes of
$\tau^+\to e^+\nu_e\overline{\nu}{}_{\tau}$, 
$\tau^+\to \mu^+\nu_{\mu}\overline{\nu}{}_{\tau}$, 
and $\tau^+\to\pi^+\overline{\nu}{}_{\tau}$~\cite{Zupanc:2013byn};
while the \babar measurement was only based on the
$\tau^+\to e^+\nu_e\overline{\nu}{}_{\tau}$ and 
$\tau^+\to \mu^+\nu_{\mu}\overline{\nu}{}_{\tau}$ decays were adopted~\cite{delAmoSanchez:2010jg};
The BESIII Collaboration reported individual measurements of  $\br(D_s^+\to\taunu)$ by using the $\tau^+\to\pi^+\overline{\nu}{}_{\tau}$~\cite{Ablikim:2016duz,BESIII:2023fhe},
$\tau^+\to\rho^+\overline{\nu}_\tau$~\cite{bes3:2021px}, 
$\tau^+\to e^+\nu_e\overline{\nu}_\tau$~\cite{bes3:2021lhj} and
 $\tau^+\to \mu^+\nu_\mu\overline{\nu}_\tau$~\cite{BESIII:2023ukh} decays.
Combining all these results and accounting for correlations, we obtain 
a WA value of
\begin{equation}
 \br^{\rm WA}(\dsp\to\taunu) = (5.33\pm0.12)\times10^{-2}.
 \label{eq:Br:WA:DstoTauNu}
\end{equation}
The ratio of branching fractions is found to be
\begin{equation}
R_{\tau/\mu}^{\ds} = 10.01\pm0.29\,,
\label{eq:R:WA:Leptonic}
\end{equation}
which is consistent with the ratio expected in the SM.

Taking the average of 
$\br^{\rm WA}(D^+_s\to\mu^+\nu_\mu)$ and $\br^{\rm WA}(D^+_s\to\tau^+\nu_\tau)$
[Eqs.~(\ref{eq:Br:WA:DstoMuNu}) and (\ref{eq:Br:WA:DstoTauNu})],
and using the most recent values for $m^{}_\tau$, $m^{}_{D_s}$, 
and $\tau^{}_{D_s}$~\cite{PDG_2022}, the products of 
the $D_s$ decay constant and $|V_{cs}|$ are determined to be
\begin{equation}
 \fds|V_{cs}|=\left(243.2\pm2.1\pm1.6\pm1.0\right)~\mbox{MeV}
 \label{eq:expFDSVCS1}
\end{equation}
and 
\begin{equation}
 \fds|V_{cs}|=\left(246.5\pm1.5\pm1.7\pm1.0\right)~\mbox{MeV},
 \label{eq:expFDSVCS2}
\end{equation}
respectively.
The average of these two results is
\begin{equation}
 \fds|V_{cs}|=\left(245.1\pm2.0\right)~\mbox{MeV},
 \label{eq:expFDSVCS}
\end{equation}
where the uncertainty is due to the uncertainties on 
$\br^{\rm WA}(D_s^+\to\munu)$, $\br^{\rm WA}(D_s^+\to\taunu)$,
and the external inputs. 
All input values and the resulting world average are 
summarized in Table~\ref{tab:DsLeptonic} and plotted in 
Fig.~\ref{fig:ExpDLeptonic} (right). To calculate this average,
we take into account correlations within each 
experiment\footnote{In the case of \babar, we use 
the covariance matrix from 
the Errata of~Ref.\cite{delAmoSanchez:2010jg}.} for 
uncertainties related to normalization, tracking, 
particle identification, signal and background 
parameterizations, and peaking background contributions.

Using the LQCD value for $\fds$ from FLAG [Eq.~(\ref{eqn:fDs_FLAG})], 
we calculate the magnitude of the CKM matrix element $V_{cs}$ to be
\begin{equation}
 |V_{cs}| = 0.9808\pm0.0079\,(\rm exp.)\pm0.0020\,(\rm LQCD),
 \label{eq:Vcs:WA:Leptonic}
\end{equation}
where the first and second uncertainties are from experiment and from lattice calculations, respectively.

\begin{table}[t!]
\caption{Experimental results and world averages for ${\cal{B}}(\dsellnu)$ 
and $f_{D_s}|V_{cs}|$. The first uncertainties are statistical and the second 
are experimental systematic. The third uncertainties in the case of 
$f_{D_s}|V_{cs}|$ are due to external inputs (dominated by the uncertainty 
on $\tau_{D_s}$). We have adjusted the $\br(\dsp\to\taunu)$ values quoted 
by CLEO-c and \babar to account for the most recent values of
$\br(\tau^+\to\pi^+\bar{\nu}^{}_\tau)$, $\br(\tau^+\to\mu^+\nu^{}_\mu\bar{\nu}^{}_\tau)$, 
and $\br(\tau^+\to e^+\nu^{}_e\bar{\nu}^{}_\tau)$~\cite{PDG_2022}.
CLEO-c and \babar include the uncertainty in the number of $\ds$ tags 
(denominator in the calculation of the
branching fraction) in the statistical uncertainty of $\br$; however,
we subtract this uncertainty from the statistical one and include it 
in the systematic uncertainty. When averaging the BESIII results of  $\br(\dsp\to\taunu)$, small correlations among various measurements have been taken into account. 
\label{tab:DsLeptonic}}
\vskip0.15in
\begin{center}
\resizebox{1.0\textwidth}{!}{

}
\end{center}
\end{table}

\begin{figure}[hbt!]
\centering
\includegraphics[width=0.65\textwidth]{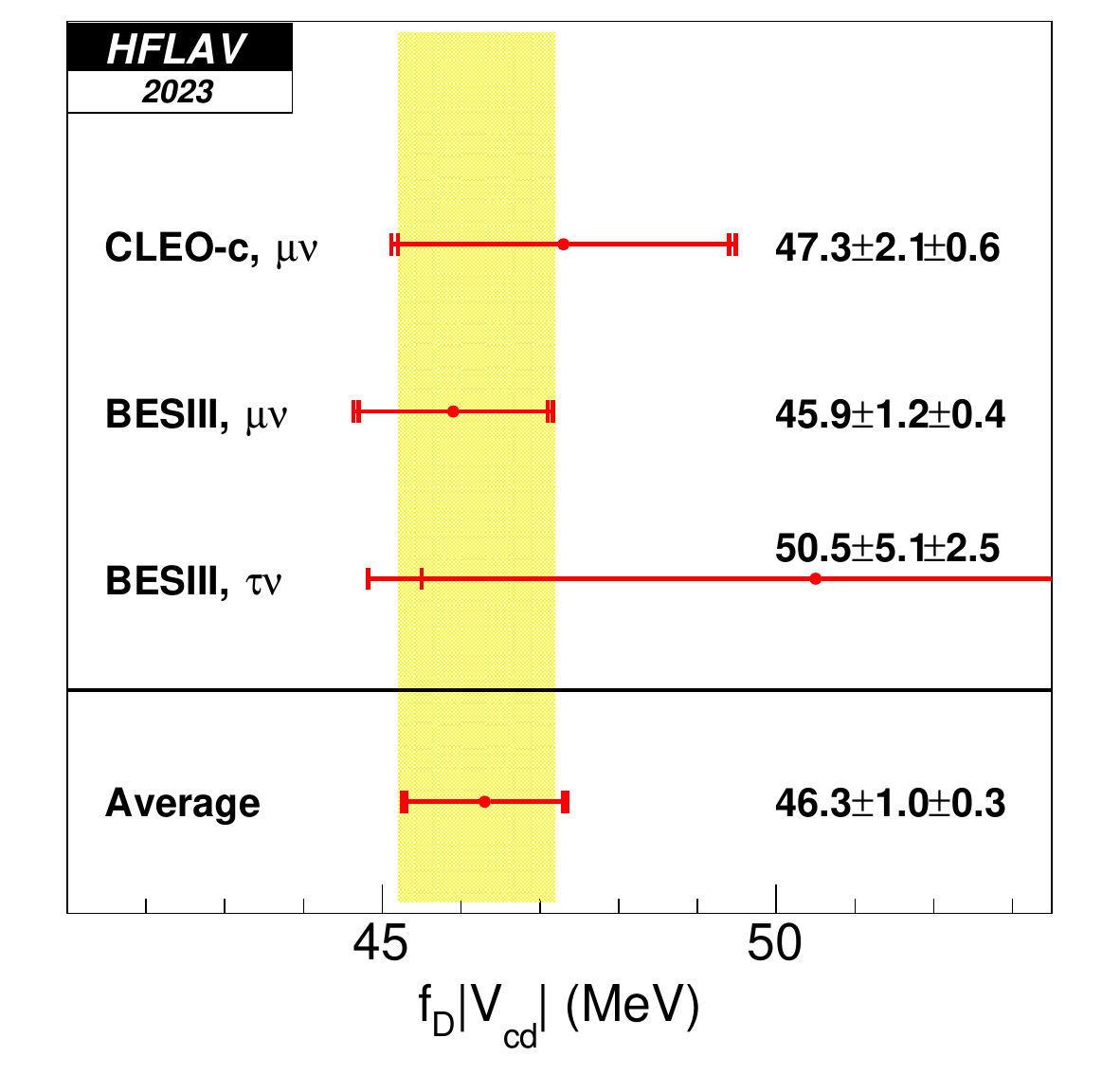}
\includegraphics[width=0.65\textwidth]{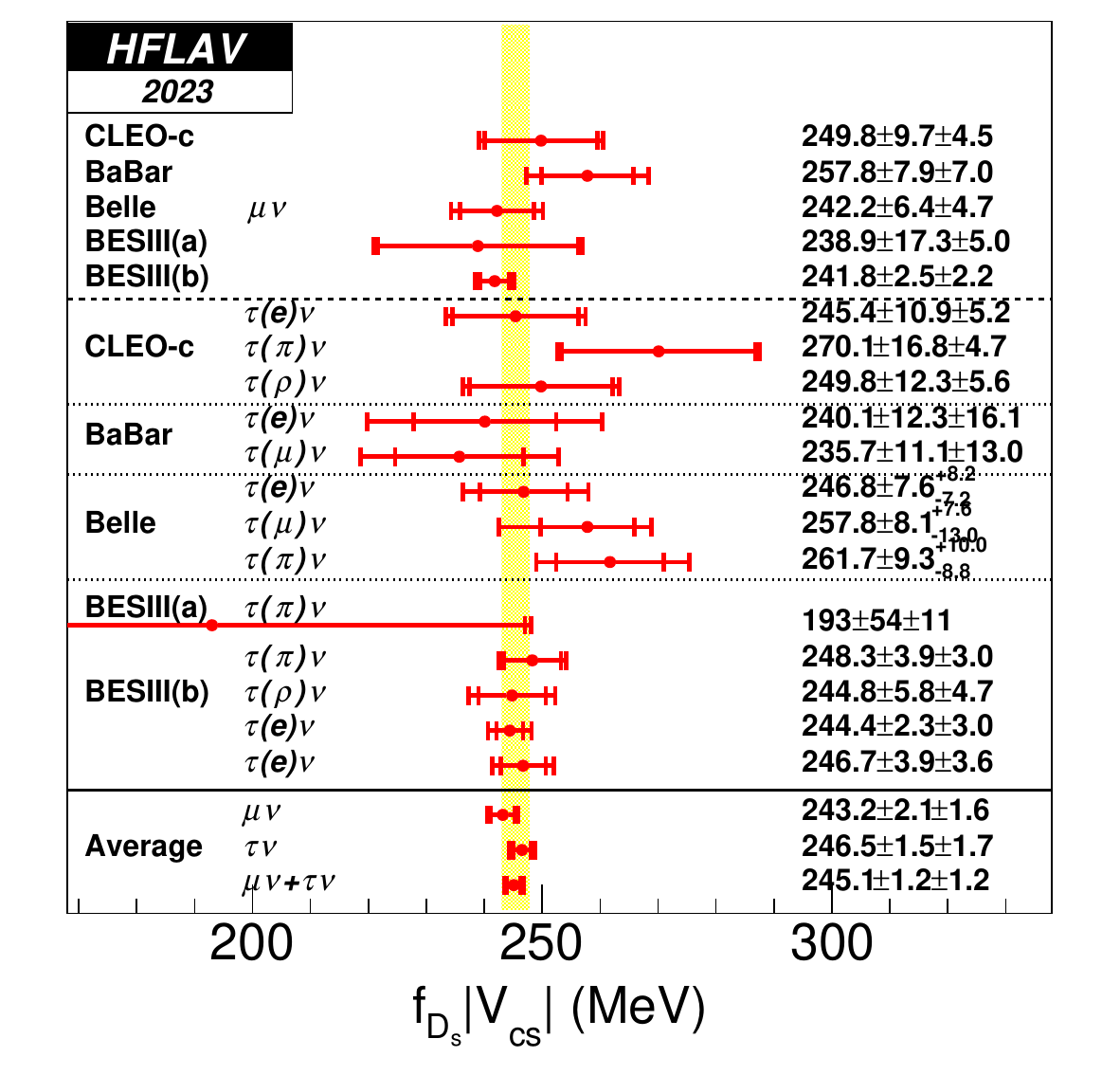}
\vskip-0.10in
\caption{
WA values for $f_{D}|V_{cd}|$  (top) and $f_{D_s}|V_{cs}|$ (bottom). 
For each point, the first errors are 
statistical and the second errors are systematic. 
BESIII(a) represents results based on 0.48~fb$^{-1}$ of 
data recorded at $\sqrt{s}=4.009$~GeV~\cite{Ablikim:2016duz},
and BESIII(b) represents results based on 7.33~(or 6.32)~fb$^{-1}$ of 
data recorded at $\sqrt{s}=4.128-4.226$~GeV~\cite{BESIII:2023fhe,bes3:2021px,bes3:2021lhj,BESIII:2023ukh}.
\label{fig:ExpDLeptonic}
}
\end{figure}

\mysubsubsection{Comparison with other determinations of $|V_{cd}|$ and $|V_{cs}|$}

Table~\ref{tab:CKMVcdVcs} summarizes, and Fig.~\ref{fig:VcdVcsComparions} 
displays, all determinations of the magnitudes $|V_{cd}|$ and $|V_{cs}|$. The
table and figure show that, currently, the most precise direct determinations 
are from leptonic $D^+$ and $D^+_s$ decays. The obtained $|V_{cd}|$ and $|V_{cs}|$ values are in agreement within uncertainties with that obtained from a global fit assuming CKM unitarity~\cite{Charles:2004jd}. 

\begin{table}[htb]
\centering
\caption{Averages of the magnitudes of CKM matrix elements $|V_{cd}|$ and 
$|V_{cs}|$, as determined from leptonic and semileptonic $D_{(s)}$ decays.
In calculating these averages, we conservatively assume that uncertainties 
due to LQCD are fully correlated. For comparison, values determined from 
neutrino scattering, from $W$ decays, and from a global fit to the 
CKM matrix assuming unitarity~\cite{Charles:2004jd} are also listed. 
\label{tab:CKMVcdVcs}}
\vskip0.15in
\begin{tabular}{lcc}
\toprule
Method & Reference & Value \\ 
\midrule
&&{$|V_{cd}|$}\\
\cline{3-3}
$D\to\ell\nu_{\ell}$ 	 & This section	                 & $0.2186\pm0.0049(\rm exp.)\pm0.0007(\rm LQCD)$\\
$D\to\pi\ell\nu_{\ell}$  & Section~\ref{sec:charm:semileptonic}	& $0.2265\pm0.0029(\rm exp.)\pm0.0018(\rm LQCD)$\\
\midrule
$D\to\ell\nu_{\ell}$ 	& \multirow{2}{*}{Average}	& \multirow{2}{*}{$0.2240\pm0.0028$}\\
$D\to\pi\ell\nu_{\ell}$ & \multirow{-2}{*}{Average}	& \multirow{-2}{*}{$0.2240\pm0.0028$}\\
\midrule
$\nu N$			& PDG~\cite{PDG_2022}	& $0.230\pm0.011$\\
Global CKM Fit		& CKMFitter~\cite{Charles:2004jd}	& $0.22486\pm0.00067$\\
\midrule
\midrule
&&{$|V_{cs}|$}\\
\cline{3-3}
$D_s\to\ell\nu_{\ell}$ 	 & This section            	& $0.9808\pm0.0079(\rm exp.)\pm0.0020(\rm LQCD)$\\
$D\to K\ell\nu_{\ell}$   & Section~\ref{sec:charm:semileptonic}	& $0.9639\pm0.0044(\rm exp.)\pm0.0032(\rm LQCD)$\\
\midrule
$D_s\to\ell\nu_{\ell}$ 	& \multirow{2}{*}{Average}	& \multirow{2}{*}{$0.9691\pm0.0045$}\\
$D\to K\ell\nu_{\ell}$ & \multirow{-2}{*}{Average}	& \multirow{-2}{*}{$0.9691\pm0.0045$}\\
\midrule
$W\to c\overline{s}$	& PDG~\cite{PDG_2022}	& $0.94^{+0.32}_{-0.26}\pm0.13$\\
Global CKM Fit		& CKMFitter~\cite{Charles:2004jd} & $0.97349\pm0.00016$\\
\bottomrule
\end{tabular}
\end{table}

\begin{figure}[hbt]
\centering
\includegraphics[width=0.45\textwidth]{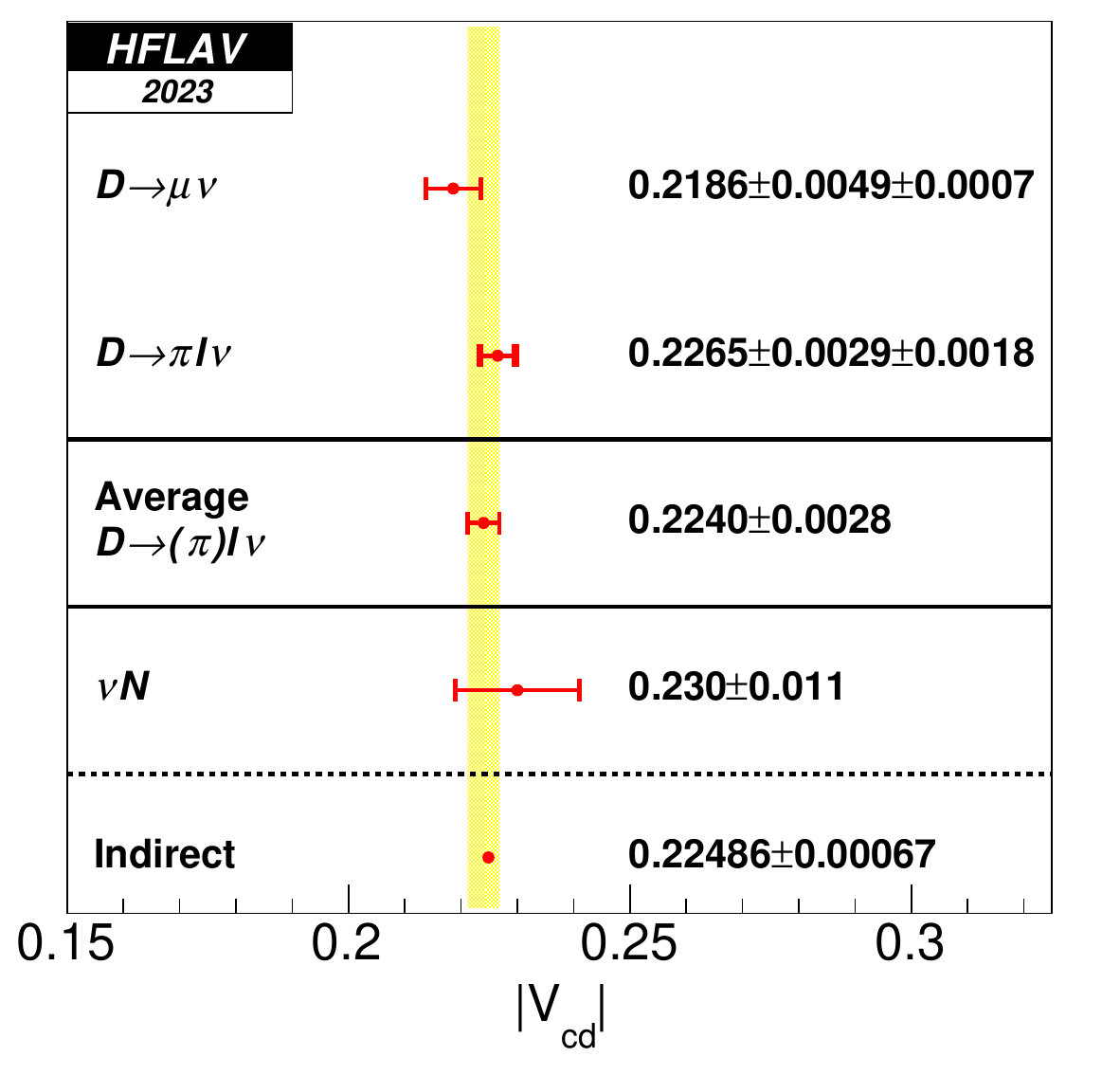}~
\includegraphics[width=0.45\textwidth]{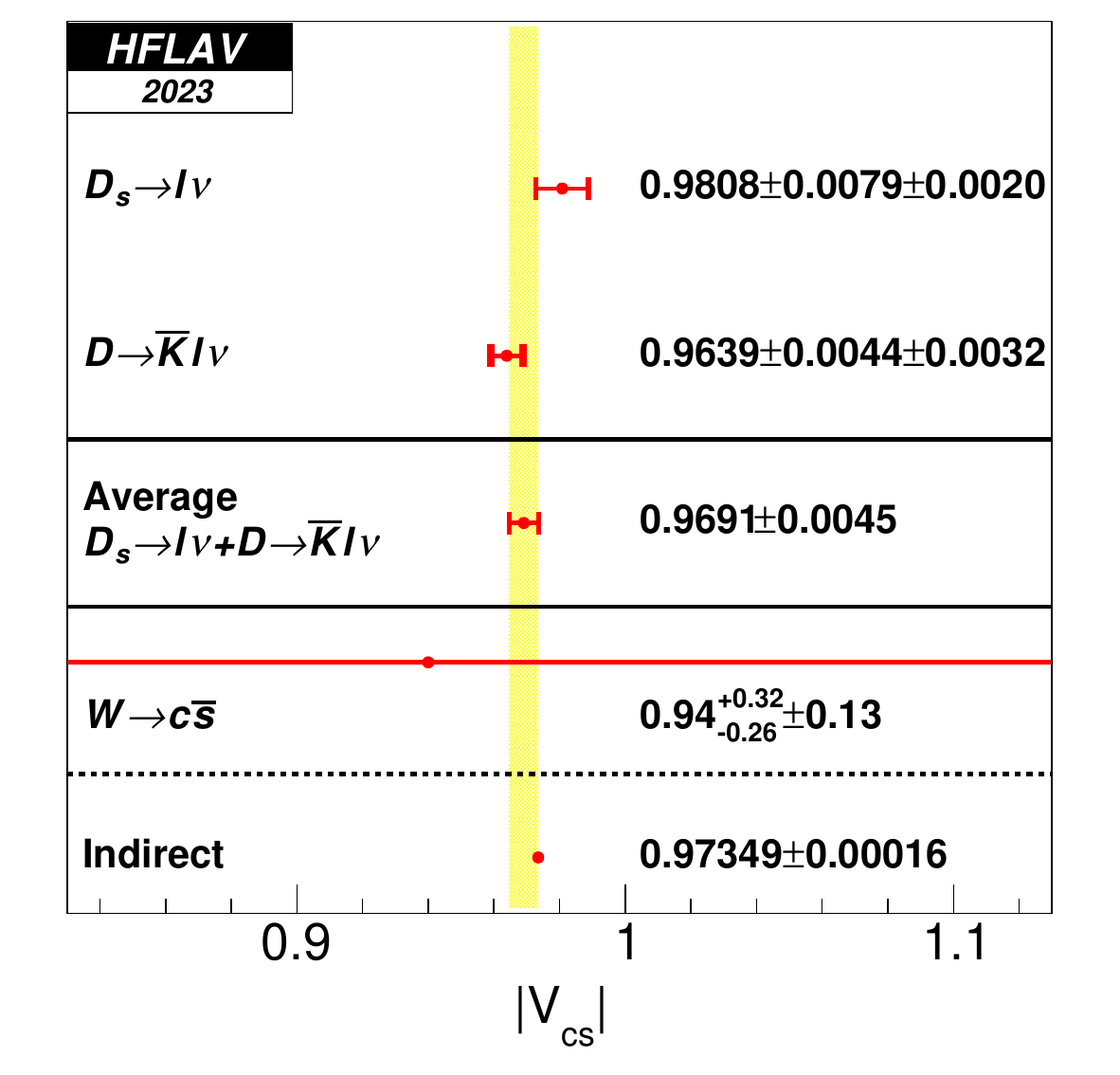}
\vskip-0.10in
\caption{
Comparison of magnitudes of CKM matrix elements $|V_{cd}|$ (left) and 
$|V_{cs}|$ (right), as determined from leptonic and semileptonic $D_{(s)}$ 
decays. Also listed are results from neutrino scattering, from $W$ decays,
and from a global fit of the CKM matrix assuming unitarity~\cite{Charles:2004jd}.
\label{fig:VcdVcsComparions}
}
\end{figure}

\mysubsubsection{Extraction of $D_{(s)}$ meson decay constants}

As listed in Table~\ref{tab:CKMVcdVcs} (and plotted in 
Fig.~\ref{fig:VcdVcsComparions}), the values of $|V^{}_{cs}|$ 
and $|V^{}_{cd}|$ can be determined from a global fit of the
CKM matrix assuming unitarity~\cite{Charles:2004jd}. 
These values can be used to extract the $D^+$ and $D^+_s$ decay constants 
from the world average values of $f_{D}|V_{cd}|$ and $f_{D_s}|V_{cs}|$ 
given in Eqs.~(\ref{eq:expFDVCD}) and~(\ref{eq:expFDSVCS}). 
The results are
\begin{eqnarray}
f_D^{\rm exp} & = & (206.0\pm4.7)~{\rm MeV,}\\ 
f_{D_s}^{\rm exp} & = & (251.8\pm2.0)~{\rm MeV,}
\end{eqnarray}
and the ratio of the decay constants is
\begin{equation}
\frac{f_{D_s}^{\rm exp}}{f_D^{\rm exp}} = 1.223\pm0.030\,.
\label{eq:fDsfDRatio:WA}
\end{equation}
These values are in agreement within their uncertainties 
with the LQCD values given by FLAG [Eqs.~(\ref{eqn:fDplus_FLAG})-(\ref{eqn:ratio_FLAG})].

\clearpage
\subsection{Hadronic $D^0$ decays and final state radiation}

Measurements of the branching fractions for the decays $D^0\to K^\mp\pi^\pm$,
$D^0\to \pi^+\pi^-$, and $D^0\to K^+ K^-$ have reached sufficient precision to
allow averages with ${\cal O}(1\%)$ relative uncertainties. 
At this precision, QED final 
state radiation (FSR) must be treated correctly and consistently across 
the input measurements for the accuracy of the averages to match the 
precision.  The sensitivity of measurements to FSR arises because of 
a tail in the distribution of radiated energy that extends to the 
kinematic limit.  The tail beyond $\sum{E_\gamma} \approx 30$ MeV causes 
typical selection variables like the hadronic invariant mass to 
shift outside the selection range dictated by experimental 
resolution, as shown in Fig.~\ref{fig:FSR_mass_shift}.  While the 
differential rate for the tail is small, the integrated rate 
amounts to several percent of the total $h^+ h^-(n\gamma)$ 
rate because of the tail's extent.  The tail therefore 
translates directly into a several percent loss in 
experimental efficiency.

All measurements that include an FSR correction 
have a correction based on the use of 
PHOTOS~\cite{Barberio:1990ms,Barberio:1993qi,Golonka:2005pn,Golonka:2003xt,Golonka:2006tw} 
within the experiment's Monte Carlo (MC) simulation.  
PHOTOS itself, however, has evolved over the period spanning the set of
measurements~\cite{Golonka:2003xt}.  
In particular, the incorporation of interference between
radiation from 
the two separate mesons has proceeded in stages: it was first
available for particle--antiparticle pairs in version 2.00 (1993), extended 
to any two-body, all-charged, final states in version 2.02 (1999), and 
further extended to multi-body final states in version 2.15 (2005).
The effects of interference are clearly visible, as shown in
Figure~\ref{fig:FSR_mass_shift}, and cause a 
roughly 30\% increase in the integrated rate into 
the high energy photon tail.  To evaluate the FSR 
correction incorporated into a given measurement, 
we must therefore note whether any correction was 
made, the version of PHOTOS used in correction, 
and whether the interference terms in PHOTOS were 
turned on. Also worth noting, an exponentiated multiple-photon mode was introduced in PHOTOS version 2.09, which allows PHOTOS to also simulate photons with low energies; this mode can be switched on or off. %
\begin{figure}[bh]
\begin{center}
\includegraphics[width=0.48\textwidth,angle=0.]{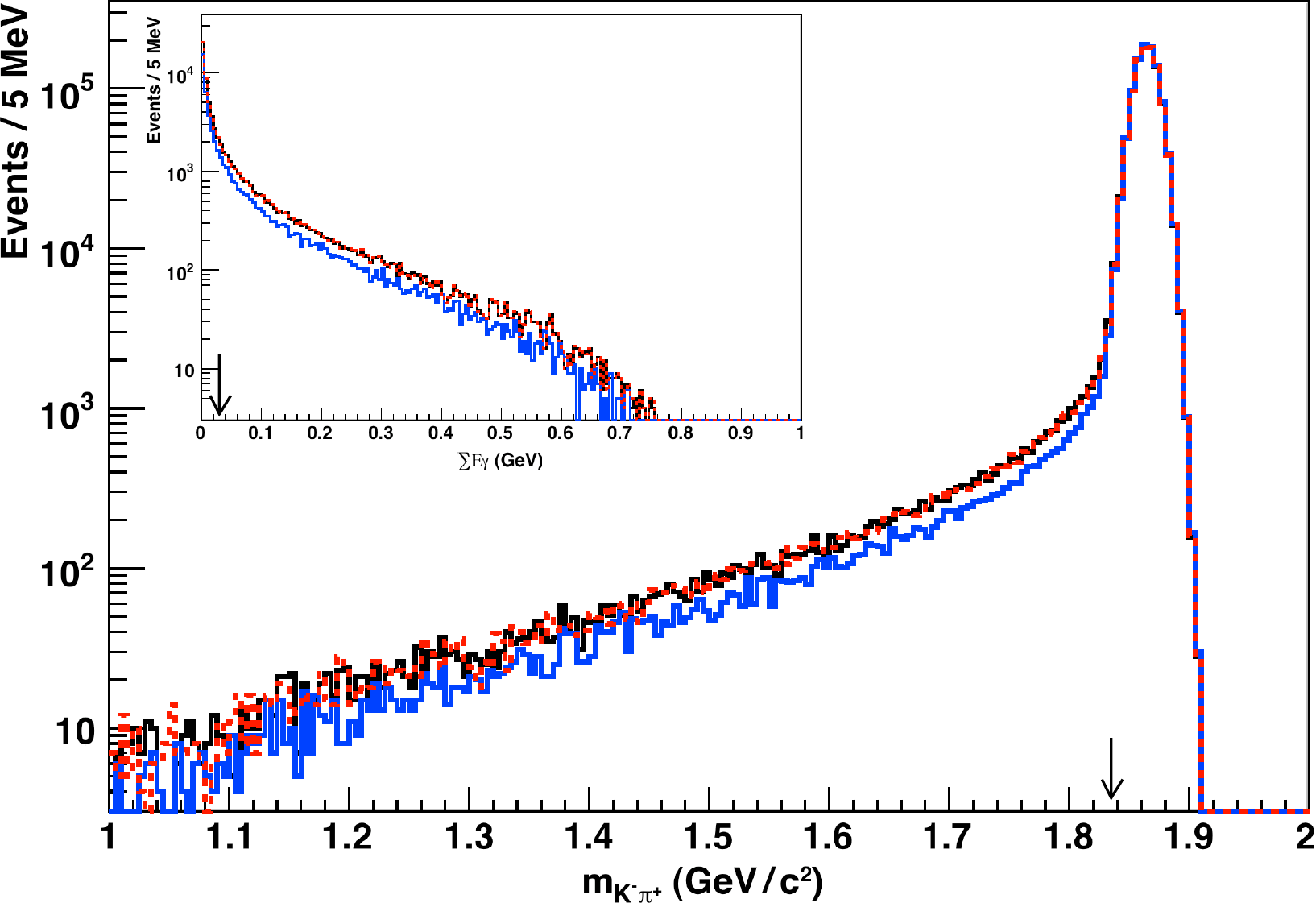}
\caption{The $K\pi$ invariant mass distribution for equally-sized samples of
$D^0\to K^-\pi^+ (n\gamma)$ decays. The three curves correspond 
to three different configurations of PHOTOS for modeling FSR: 
version 2.02 without interference (blue/grey), version 2.02 with 
interference (red dashed) and version 2.15 with interference (black).  
The true invariant mass has been smeared with a typical experimental 
resolution of 10 MeV${}/c^2$.  Inset: The corresponding spectrum of 
total energy radiated per event.  The arrow indicates the $\sum E_\gamma$ 
value that begins to shift kinematic quantities outside of the range 
typically accepted in a measurement.}
\label{fig:FSR_mass_shift}
\end{center}
\end{figure}

\subsubsection{Updates to the branching fractions} %

Before averaging the measured branching fractions, the published 
results are updated, as necessary, to the FSR prediction of 
PHOTOS~2.15 with interference included and exponentiated multiple-photon mode turned on.  The update %
will always shift a branching fraction to a higher value: with no 
FSR correction or an FSR correction suboptimally modeled, %
the experimental efficiency determination will be biased high, 
and therefore the branching fraction will be biased low. 

Most of the branching fraction analyses used the kinematic quantity 
sensitive to FSR in the candidate selection criteria. For the 
analyses at the $\psi(3770)$, this variable was $\Delta E$, the 
difference between the candidate $D^0$ energy and the beam energy 
(\eg, 
$E_K + E_\pi - E_{\rm beam}$ 
for $D^0\to K^-\pi^+$).  
In the remainder of the analyses, the relevant quantity was the 
reconstructed hadronic two-body mass $m_{h^+h^-}$. To make an FSR 
correction, 
we need to evaluate the fraction of decays that FSR moves 
outside of the range accepted for the analysis. 
The corrections were evaluated using an event generator ({\sc EvtGen} 
\cite{Ryd:2005zz, Lange:2001uf}) that incorporates PHOTOS to simulate the 
portions of the decay process most relevant to the correction. 

We compared corrections determined both with and without smearing 
to account for experimental resolution; for the analyses using $m_{h^+h^-}$ as the kinematic quantity sensitive to FSR, the differences were 
negligible, typically of ${\cal O}(1\%)$ of the correction itself. 
The immunity of the correction to resolution effects comes about because 
most of the long FSR-induced tail in %
the $m_{h^+h^-}$ 
distribution resides well away from the selection boundaries.  The 
smearing from resolution, on the other hand, mainly affects the 
distribution of events right at the boundary. 
For the analyses using $\Delta E$ however, events with low energy photons are found to substantially move events across the selection boundary; thus PHOTOS versions with exponentiated multiple-photon mode turned on and off, respectively, can give substantially different FSR corrections. In the case that this mode is on, smearing of the events with low energy photons increases the amount of the FSR correction by about 10\%. This is well within the uncertainty on the FSR correction, as discussed later in this section, and thus ignored.

For measurements incorporating an FSR correction that did not 
include interference and/or use exponentiated multiple-photon mode, 
we update by assessing the FSR-induced 
efficiency loss for both the PHOTOS version and configuration 
used in the analysis and our nominal version 2.15 (with interference included and exponentiated multiple-photon mode turned on).  
For measurements that published their sensitivity to FSR, our 
generator-level predictions for the original efficiency loss 
agreed to within a few percent of the correction. 
This agreement lends additional credence to the procedure.

Once the event loss from FSR in the most sensitive kinematic 
quantity is accounted for, the event loss in other quantities 
is typically very small. For example, analyses using $D^{*+}$ tags show 
very little sensitivity to FSR in the reconstructed $D^{*+}\!-D^0$ 
mass difference, \ie, in $m^{}_{h^+h^-\pi^+}\!-m^{}_{h^+h^-}$. 
In this case, the effect of FSR tends to cancel in the difference 
of reconstructed masses. 
In the $\psi(3770)$ analyses, the 
beam-constrained mass distributions 
(\eg $\sqrt{E_{\rm beam}^2 - |\vec{p}_K + \vec{p}_\pi|^2}$)  
have some sensitivity, but provide negligible independent sensitivity after the  %
$\Delta E$ selection. %
\begin{figure}[t]
\begin{center}
\adjincludegraphics[trim={0 0 {0.65\width} 0},clip,height=0.25\textheight]{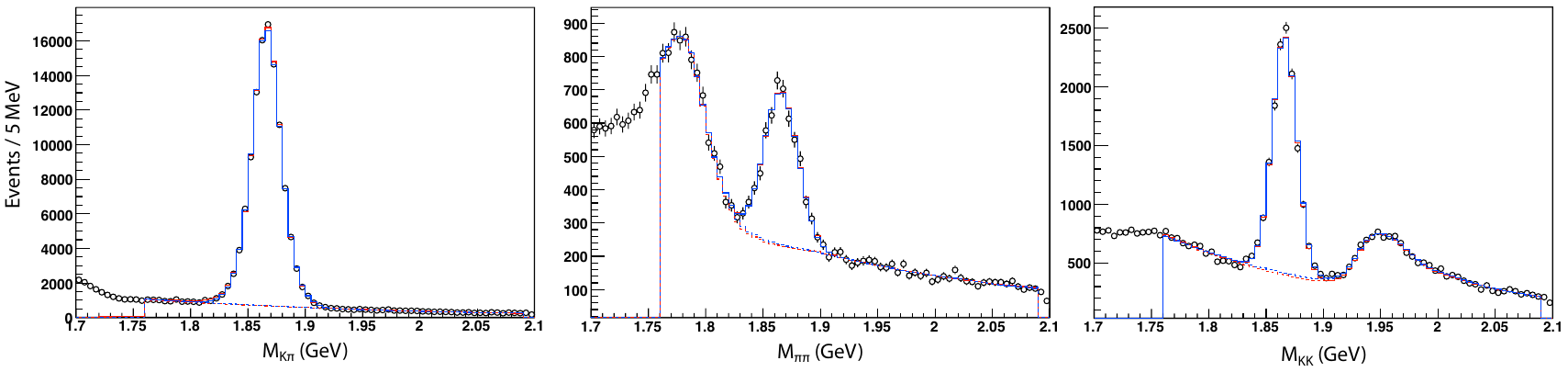}\\
\adjincludegraphics[trim={0 0 {0.9825\width} 0},clip,height=0.25\textheight]{figures/charm/FocusFits.pdf}
\adjincludegraphics[trim={{0.36\width} 0 0 0},clip,height=0.25\textheight]{figures/charm/FocusFits.pdf}
\caption{FOCUS data (dots), original fits (blue) and 
toy MC parameterization (red) for $D^0\to K^-\pi^+$ (top), 
$D^0\to \pi^+\pi^-$ (bottom left), and $D^0\to \pi^+\pi^-$ (bottom right).}\label{fig:FocusFits}
\end{center}
\end{figure}

The FOCUS~\cite{Link:2002hi} analysis of the branching fraction
ratios ${\cal B}(D^0 \to \pi^+\pi^-)/{\cal B}(D^0\to K^- \pi^+)$ and 
${\cal B}(D^0 \to K^+ K^-)/{\cal B}(D^0 \to K^- \pi^+)$ obtained 
yields using fits to the two-body mass distributions.  FSR will 
both distort the low end of the signal mass peak, and will 
contribute a signal component to the low side tail used to 
estimate the background.  The fitting procedure is not sensitive 
to signal events out in the FSR tail, which would be counted as 
part of the background.

A more complex toy MC procedure was required to analyze 
the effect of FSR on the fitted yields, which were published with 
no FSR corrections applied.  
Determining the update %
involved an iterative 
procedure in which samples of similar size to the FOCUS sample were 
generated and then fit using the FOCUS signal and background 
parameterizations. The MC parameterizations were tuned based 
on differences between the fits to the toy MC data and the FOCUS 
fits, and the procedure was repeated. These steps were iterated until 
the fit parameters matched the original FOCUS parameters.  
\begin{table}[!htb]
  \centering 
  \caption{The experimental measurements relating to ${\cal B}(D^0\to K^-\pi^+)$, ${\cal B}(D^0\to \pi^+\pi^-)$, and ${\cal B}(D^0\to K^+ K^-)$ after updating %
  them to the common version and configuration of PHOTOS.  The uncertainties are statistical and total systematic, with the FSR-related systematic estimated in this procedure shown in parentheses.  Also listed are the percent shifts in the results from those with the original correction (if any), in the case an update is applied here, as well as the original PHOTOS and interference configuration for each publication. %
  }
  \label{tab:FSR_corrections}

\end{table}

The toy MC samples for the first iteration were based on the generator-level 
distributions of $m_{K^-\pi^+}$, $m_{\pi^+\pi^-}$, and $m_{K^+K^-}$, including 
the effects of FSR, smeared according to the original FOCUS resolution 
function,  and on backgrounds 
generated 
using the parameterization from the final
FOCUS fits.  For each iteration, 400 to 1600 individual 
data-sized samples were 
generated 
and fit. The central values %
of the parameters from these fits determined the 
corrections to the generator parameters for the following iteration.  The 
ratio between the number of signal events generated and the final signal 
yield provides the required FSR correction in the final iteration.  Only a 
few iterations were required in each mode.  Figure~\ref{fig:FocusFits} 
shows the FOCUS data, the published FOCUS fits, and the final toy MC 
parameterizations.  The toy MC provides an excellent description of the 
data.

The corrections obtained to the individual FOCUS yields were 
$1.0298 \pm 0.0001$ for $K^-\pi^+$, $1.062 \pm 0.001$ for $\pi^+\pi^-$, 
and $1.0183 \pm 0.0003$ for $K^+K^-$.  These corrections tend to 
cancel in the branching ratios, leading to corrections (update shifts)  
of 
$1.031 \pm 0.001$ 
($3.10\%$)
for 
${\cal B}(D^0\to \pi^+\pi^-)/{\cal B}(D^0\to K^-\pi^+)$, and 
$0.9888 \pm 0.0003$ 
($-1.12\%$)
for 
${\cal B}(D^0\to K^+ K^-)/{\cal B}(D^0\to K^-\pi^+)$.

Table~\ref{tab:FSR_corrections} summarizes the updated %
branching fractions. 
The published FSR-related modeling uncertainties have been replaced with a
new, common estimate; this estimate is based on the assumption that the dominant
uncertainty in the FSR corrections comes from the fact that the mesons are treated 
as structureless particles. No contributions from structure-dependent terms in 
the decay process (\eg, radiation from individual quarks) are included in PHOTOS. 
Internal studies performed by various experiments (reconstructing decays including FSR photons) have indicated that in $K\pi$
decays, the PHOTOS corrections agree with data at the 20-30\% level. We therefore
attribute a 25\% uncertainty to the (updated) FSR correction from potential 
structure-dependent contributions. For the other two modes, the only difference 
in structure is the final state valence quark content. While radiative corrections 
typically enter with a $1/M$ dependence, the additional contribution from the
structure terms enters on a time scale shorter than the hadronization time scale.
Thus, this contribution corresponds to $M\!\sim\!\Lambda_{\rm QCD}$ rather than that of
the quark masses and would be the same for all three modes. We make this
assumption when treating the correlations among measurements. We also assume
that the PHOTOS amplitudes and any missing structure amplitudes interfere
constructively.
The uncertainties largely cancel 
in the branching fraction ratios. For the final average branching 
fractions, the FSR uncertainty on $K\pi$ %
is as large as the uncertainty due to other systematic effects. 
Note that because 
of the relative sizes of FSR in the different modes, the $\pi\pi/K\pi$ 
branching ratio uncertainty from FSR is 
positively correlated with that 
for the $K\pi$ branching fraction, while the $KK/K\pi$ branching ratio FSR
uncertainty is negatively correlated.

The ${\cal B}(D^0\to K^-\pi^+)$ measurement of reference~\cite{Coan:1997ye} (CLEO II), the  
${\cal B}(D^0\to \pi^+\pi^-)/{\cal B}(D^0\to K^-\pi^+)$ measurements of 
references~\cite{Aitala:1997ff} (E791) 
and~\cite{CLEO:2001lgl} (CLEO II.V), and the 
${\cal B}(D^0\to K^+ K^-)/{\cal B}(D^0\to K^-\pi^+)$ measurement
of reference~\cite{CLEO:2001lgl} are excluded from the branching 
fraction averages presented here.
These measurements appear not to have incorporated any FSR corrections, 
and insufficient information
is available to determine the 2-3\% update shifts %
that would be required.
\begin{sidewaystable}[p]
  \centering 
  \caption{The correlation matrix corresponding to the full covariance matrix. 
  Subscripts $h \in \{\pi, K\}$ denote which of the $D^0 \to h^+ h^-$ decay results from a single experiment
  is represented in that row or column.}\label{tab:correlations}
  \scriptsize

\end{sidewaystable}

\subsubsection{Average branching fractions for
\texorpdfstring{$D^0\to K^-\pi^+$, $D^0\to \pi^+\pi^-$ and $D^0\to K^+ K^-$}{D0 to K-pi+, D0 to pi+pi-, D0 to K+K-}} %
\begin{figure}[b]
\begin{center}
\includegraphics[width=0.6\textwidth,angle=0.]{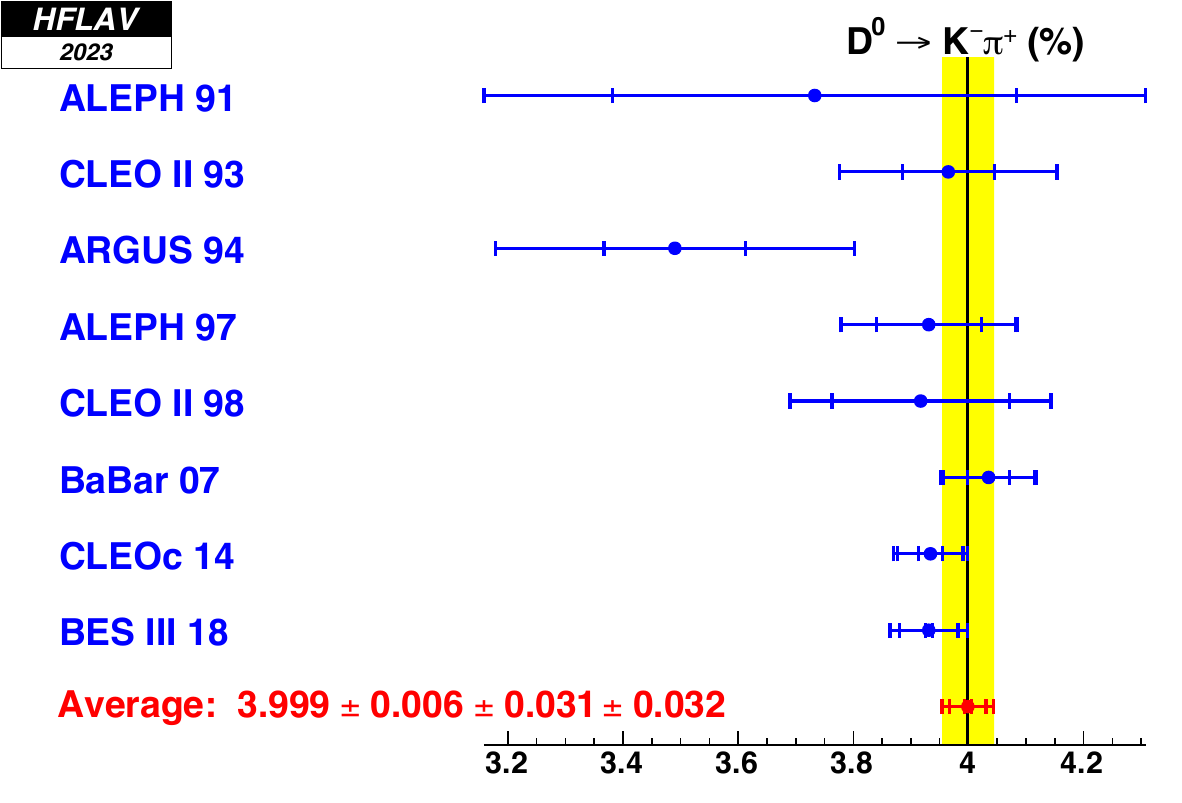}
\caption{Comparison of measurements of 
${\cal B}(D^0\to K^-\pi^+)$ (blue) with the average 
branching fraction obtained here (red, and yellow band).  For these measurements only, the partial $\chi^2$ is 4.9 in the final fit.}
\label{D0bfs}
\end{center}
\end{figure}

The average branching fractions for 
$D^0\to K^-\pi^+$, $D^0\to \pi^+\pi^-$ and $D^0\to K^+ K^-$ decays
are obtained from a single $\chi^2$ minimization procedure, 
in which the three branching fractions are floating parameters and the measurements listed in Table~\ref{tab:FSR_corrections} are the inputs. 
The central values are obtained from a fit in which the full covariance matrix, accounting for all statistical, systematic (excluding FSR), and FSR 
measurement uncertainties, is used.  
Table~\ref{tab:correlations} presents the correlation matrix for 
this nominal fit. %
We then obtain the three reported uncertainties on those central values as follows:
The statistical uncertainties are obtained from a fit using only the statistical 
covariance matrix.  
The systematic uncertainties are obtained by subtracting (in quadrature) the 
statistical uncertainties 
from the uncertainties determined via a fit using a covariance matrix that 
accounts for both statistical and systematic measurement uncertainties.  
The FSR uncertainties are obtained by subtracting (in quadrature)
the uncertainties determined via a fit using a covariance matrix that accounts 
for both statistical and systematic measurement uncertainties
from the uncertainties determined via the fit using the full covariance matrix.

In forming the full covariance matrix, the FSR
uncertainties are treated as fully correlated (or anti-correlated) as 
described above.  %
For the covariance matrices involving systematic measurement uncertainties, ALEPH's systematic %
uncertainties in the $\theta_{D^*}$ parameter are treated
as fully correlated between the ALEPH 97 and ALEPH 91 measurements.  Similarly,
the tracking efficiency uncertainties in the CLEO II 98 and the
CLEO II 93 measurements are treated as fully correlated. For the three BESIII 18 results, both tracking and particle identification efficiencies for any particles shared between decay modes are treated as fully correlated. Finally, the BESIII 18 results also have a fully correlated statistical dependence on the number %
of ${D^0\bar{D}^0}$ pairs produced.

The averaging procedure results in a 
final $\chi^2$ of $36.0$ for $13$ ($16-3$) degrees 
of freedom ($p\mathrm{-value} = 5.9\times10^{-4}$).  The branching
fractions obtained are
\begin{eqnarray}
\label{DHad_results}
  {\cal B}(D^0\to K^-\pi^+)   & = & ( 3.999~\, \pm 0.006~\, \pm 0.031~\, \pm 0.032~\, )\,\%, \\
  {\cal B}(D^0\to \pi^+\pi^-) & = & ( 0.1490 \pm 0.0012 \pm 0.0015 \pm 0.0019 )\,\%, \\
  {\cal B}(D^0\to K^+ K^-)    & = & ( 0.4113 \pm 0.0017 \pm 0.0041 \pm 0.0025 )\,\%\,. 
\end{eqnarray}The uncertainties, estimated as described above, are statistical, 
systematic (excluding FSR), and
FSR modeling.  The correlation coefficients from the fit using the 
total uncertainties are 
\begin{center}
\begin{tabular}{llll}
               & $K^-\pi^+$ & $\pi^+\pi^-$ & $K^+ K^-$ \\
$K^-\pi^+$     &  1.00 & 0.77 & 0.76  \\
$\pi^+\pi^-$   &  0.77 & 1.00 & 0.58  \\
$K^+ K^-$      &  0.76 & 0.58 & 1.00  \\
\end{tabular}
\end{center}

\begin{table}[!htb]
  \centering 
  \caption{Evolution of the $D^0\to K^-\pi^+$ branching fraction from a fit with
  no FSR updates %
  or correlations (similar to the average in the 
  PDG 2023 update~\cite{PDG_2022}) to the nominal fit presented
here.} \label{tab:fit_evolution}
\resizebox{\textwidth}{!}{
\begin{tabular}{cccll}
\hline\hline
Modes &  Description & ${\cal B}(D^0\to K^-\pi^+)$ (\%)  & $\chi^2$/(deg.~of freedom) \\
fit   &              &                                   &     \\ \hline
$K^-\pi^+$ & PDG 2023\,\cite{PDG_2022} equivalent    
     & $3.910 \pm 0.006 \pm 0.033$ & $~5.1/(9-1)=0.64$ \\
$K^-\pi^+$ & drop Ref.~\cite{Coan:1997ye}  & $3.913 \pm 0.006 \pm 0.033$ & $~5.1/(8-1)=0.73$ \\
$K^-\pi^+$ & add FSR updates           & $3.948 \pm 0.006 \pm 0.032 \pm 0.019$ & $~3.5/(8-1)=0.50$  \\ %
$K^-\pi^+$ & add FSR correlations          & $3.949 \pm 0.006 \pm 0.032 \pm 0.033$ & $~3.7/(8-1)=0.53$  \\
all        & add CLEO-c, CDF, and FOCUS $h^+ h^-$   & $3.956 \pm 0.006 \pm 0.032 \pm 0.033$ &   $11.1/(14-3)=1.01$  \\
all        & add BESIII $h^+ h^-$  & $3.999 \pm 0.006 \pm 0.031 \pm 0.032$ &   $36.0/(16-3)=2.77$  \\
\hline
\end{tabular}  
}
\end{table}
These results are explained in detail as follows. 
As %
Fig.~\ref{D0bfs} shows, the average 
value for ${\cal B}(D^0\to K^-\pi^+)$ and
the input branching fractions agree very well. 
For the ${\cal B}(D^0\to K^-\pi^+)$ measurements only, the partial $\chi^2$ is 4.9 in the final fit.
With the estimated 
uncertainty in the FSR modeling used here,
the FSR uncertainty dominates the statistical uncertainty 
in the average, suggesting that experimental
work in the near future should focus on verification of FSR with 
$\sum E_\gamma \simge 100$ MeV.  
Note that the systematic uncertainty excluding FSR
has now approached the level of 
the FSR uncertainty; in the most 
precise measurements of these branching fractions, the 
competing
uncertainty is the uncertainty on the tracking efficiency. 

The ${\cal B}(D^0\to
K^+K^-)$ and ${\cal B}(D^0\to \pi^+\pi^-)$ measurements inferred
from the branching ratio measurements %
do not agree as well
(Fig.~\ref{fig:kkpipi}). There is some tension among the results 
when all measurements related to ${\cal B}(D^0 \to K^+K^-)$
and ${\cal B}(D^0\to \pi^+\pi^-)$ are included in the average together.
For the measurements related to ${\cal B}(D^0\to K^+ K^-)$ [${\cal B}(D^0\to \pi^+\pi^-)$] only, the partial $\chi^2$ is 15.7 [6.0] in the final fit.

The ${\cal B}(D^0\to K^-\pi^+)$ average obtained here 
is 
approximately 
two standard deviations %
higher than 
the PDG 2023 update average~\cite{PDG_2022}.
Table~\ref{tab:fit_evolution} shows the evolution from a
fit similar to the PDG fit (no FSR updates %
or correlations, 
reference~\cite{Coan:1997ye} 
included)
 to the average presented here.
There are three main contributions to the difference. 
The
branching fraction in reference~\cite{Coan:1997ye} is
low, and its exclusion shifts the result upwards. 
A subsequently larger shift
($+0.035\%$) is due to the FSR updates, %
which as
expected shift the result upwards.
The largest shift ($+0.050\%$) occurs as all of the measurements related to ${\cal B}(D^0 \to K^+K^-)$
and ${\cal B}(D^0 \to \pi^+\pi^-)$ are included in the average together with the ${\cal B}(D^0\to K^-\pi^+)$ measurements. 
\begin{figure}
\begin{center}
\includegraphics[width=0.47\textwidth,angle=0.]{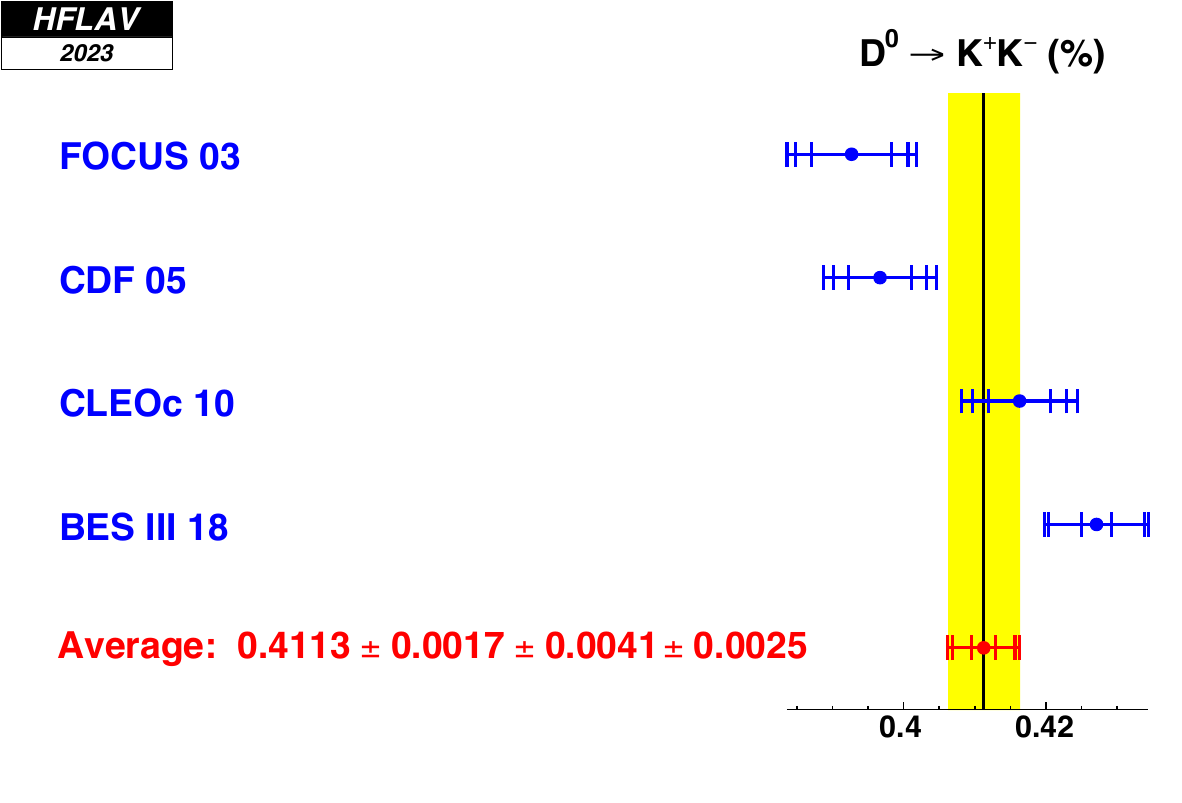}\hfill
\includegraphics[width=0.47\textwidth,angle=0.]{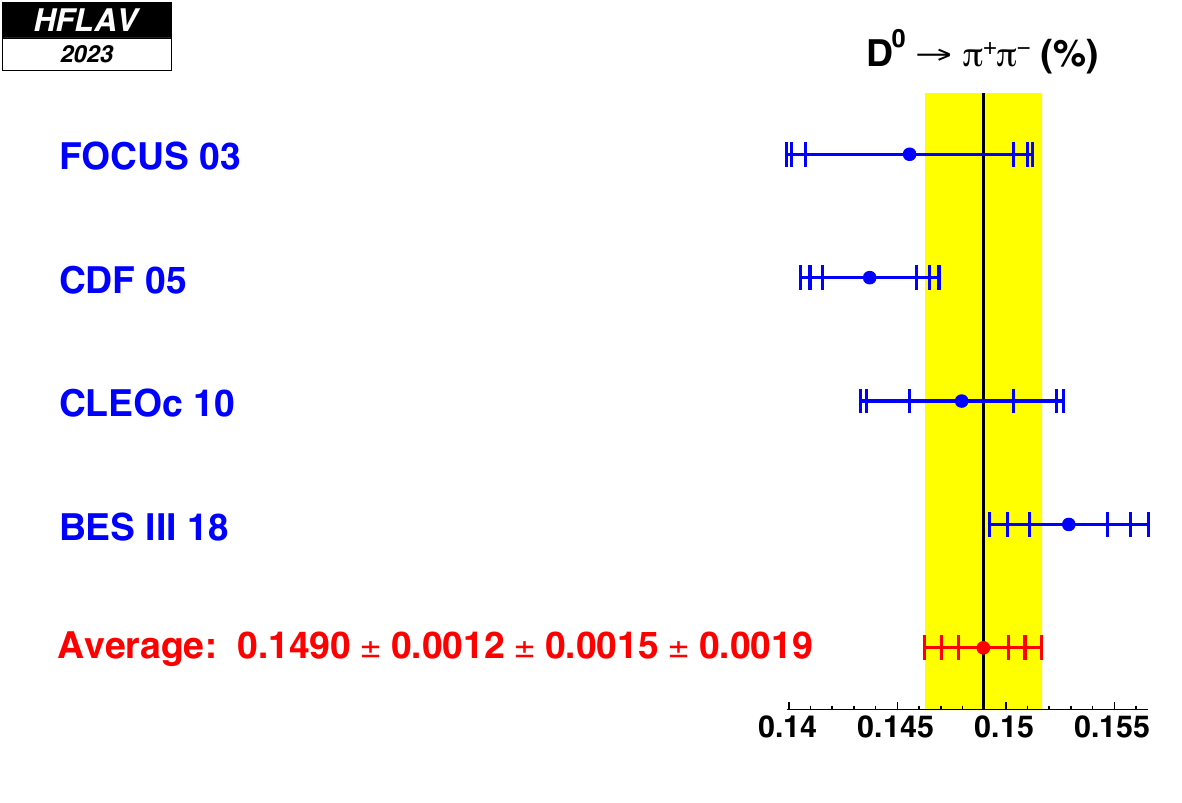}
\caption{The ${\cal B}(D^0\to K^+K^-)$ (left) and ${\cal B}(D^0\to \pi^+\pi^-)$ (right) 
values obtained either from absolute measurements or by scaling the measured branching ratios with the ${\cal B}(D^0\to K^-\pi^+)$ branching fraction
average obtained here.  For the measurements (blue points), the error bars correspond to the statistical, systematic
and either the $K\pi$ normalization uncertainties or, in case of an absolute measurement, the FSR modeling uncertainty.  The average obtained here (red point, yellow band) lists the statistical,
systematics excluding FSR, and the FSR systematic. 
 For the measurements related to ${\cal B}(D^0\to K^+K^-)$ [${\cal B}(D^0 \to \pi^+\pi^-)$] only, the partial $\chi^2$ is 15.7 [6.0] in the final fit.
\label{fig:kkpipi}}
\end{center}
\end{figure}

\subsubsection{Average branching fraction for 
\texorpdfstring{$D^0\to K^+\pi^-$}{D0 to K+ pi-}} 

There is no reason to presume that the effects of FSR should be
different in $D^0\to K^+\pi^-$ and $D^0\to K^-\pi^+$ decays, as both decay to
one charged kaon and one charged pion; indeed, for the same version of PHOTOS
the FSR simulations of these decays are identical. Measurements of the relative
branching fraction ratio between the doubly Cabibbo-suppressed decay
$D^0\to K^+\pi^-$ and the Cabibbo-favored decay $D^0\to K^-\pi^+$
($R_D$, determined in Section~\ref{sec:charm:mixcpv})
have now approached ${\cal O}(1\%)$ relative uncertainties. 
This makes it worthwhile to combine our $R_D$ average with the
${\cal B}(D^0\to K^-\pi^+)$ average obtained in Eq.~(\ref{DHad_results}),
to provide a measurement of the branching fraction:
\begin{eqnarray}
{\cal B}(D^0\to K^+\pi^-)   & = & ( 1.376 \pm 0.017 ) \times 10^{-4}.%
\end{eqnarray} 
Note that, by definition of $R_D$, these branching fractions do not include
any contribution from Cabibbo-favored $\Dzb \to K^+\pi^-$ decays. 
Our result is more precise than the PDG 2023 value of 
$(1.363 \pm 0.025 ) \times 10^{-4}$~\cite{PDG_2022} due to our using
a more precise value for the ratio $R^{}_D$ (obtained from a global fit
to a range of mixing data, see Section~\ref{sec:charm:mixcpv}).

\subsubsection{Consideration of PHOTOS++}

The versions of PHOTOS that existing measurements were performed with are now well
over a decade %
out of date. The newest version, PHOTOS++ 3.61~\cite{Davidson:2010ew},
is now fully based on C++ instead of the original FORTRAN. 
None of the measurements used in our branching fraction averages use PHOTOS++, so we have
not yet undertaken an effort to update all results to this newest version. However, at this
time it is worth continuing our procedure to evaluate whether there is any continued low
bias in the the branching fractions, due to sub-optimal modeling of FSR.

We find that the FSR spectra for PHOTOS 2.15, with interference included and exponentiated
multiple-photon mode turned on, and PHOTOS++ (in its default mode) are compatible. The
distributions of $m_{K\pi}$ for simulated $D$ mesons from $B \to D^* X$ decays produced
at $\Upsilon(4S)$ threshold are nearly identical. As an example, the \babar 07 selection
criteria were applied to decays simulated with PHOTOS++ and our nominal version of PHOTOS~2.15;
both produce identical FSR corrections to within 0.01\%. 

The distributions of $\Delta E$ for simulated $D$ mesons produced at $\psi(3770)$ threshold
also are nearly identical. As an example, for the BESIII 18
$D^0 \to K^-\pi^+$, $D^0 \to \pi^+\pi^-$, and $D^0 \to K^+K^-$ branching fraction results,
the additional update shifts required to correct from our nominal version of PHOTOS 2.15
to PHOTOS++ are less than or equal to 0.02\%. However, if smearing is applied with the
BESIII 18 $\Delta E$ resolution, while the update for $D^0 \to K^-\pi^+$ remains negligible,
the update shifts for $D^0 \to \pi^+\pi^-$ and $D^0 \to K^+K^-$ are modest at $-0.25\%$ and 0.19\%,
respectively; this level of shifts are well within the systematic uncertainty of our averages.

\clearpage
\mysubsection{Excited $D_{(s)}$ mesons}

Excited ``open'' charm mesons have received increased attention
since the first observation of low mass, narrow $D_{sJ}$ states that were inconsistent with
QCD predictions~\cite{Aubert:2003fg,Besson:2003cp,Abe:2003jk,Aubert:2003pe}.
Their properties can be measured in both prompt analyses as well as in
amplitude analyses of multi-body $B$ decays.
Tables~\ref{table:charm:spect:D1}, \ref{table:charm:spect:D2}, and 
\ref{table:charm:spect:Ds} summarize the measurements of masses and
widths of excited $D$ and $D_{s}$ states.
If a preferred assignment of spin and parity was measured, 
it is listed in the column $J^{P}$, where the label ``natural'' denotes
$P = (-1)^J$ ($J^{P}=0^{+},1^{-},2^{+}\ldots$) and ``unnatural'' denotes
$P = (-1)^{J+1}$ ($J^{P}=0^{-},1^{+},2^{-}\ldots$). 
In some studies, it was possible to identify only whether the state has
natural or unnatural spin-parity, but not the values of the quantum numbers. 

\begin{table}[htb!]
\caption{\label{table:charm:spect:D1} Measurements of masses 
and widths for excited $D$ mesons. The column $J^{P}$ lists 
the assignment of spin and parity. If possible,
an average mass or width is calculated. Table 1 of 2.}
\begin{adjustbox}{width=\textwidth,center} 
  {\setlength\tabcolsep{0pt} 
 
  } 
\end{adjustbox} 
\end{table} 

\begin{table}[htb!]
\caption{\label{table:charm:spect:D2} Measurements of masses 
and widths for excited $D$ mesons. The column $J^{P}$ lists 
the assignment of spin and parity. If possible,
an average mass or width is calculated. Table 2 of 2.}
\begin{adjustbox}{width=\textwidth,center} 
  {\setlength\tabcolsep{0pt} 
    %
 
  } 
\end{adjustbox} 
\end{table} 

\begin{table}[htb!]
\caption{\label{table:charm:spect:Ds} Measurements of masses 
and widths for excited $D^{}_s$ mesons. The column $J^{P}$ lists 
the of spin and parity. If possible,
an average mass or width is calculated.}
\begin{adjustbox}{width=\textwidth,center} 
  {\setlength\tabcolsep{0pt} 
    %
 
  } 
\end{adjustbox} 
\end{table}

For states in which multiple measurements are available, an average mass and width
are calculated; these are listed in the grey shaded rows. For simplicity, when calculating
averages we neglect possible correlations among individual measurements. All averaged masses
and widths are summarized in Fig.~\ref{fig:charm:spect:1}. The resonances listed in the
tables and figures are as they appear in the respective publications. In some cases, it
is unclear whether separately listed states are in fact distinct or are the same resonance.
An example is the $D^{*}_{1}(2680)^{0}$ state~\cite{Aaij:2016fma}, which
has parameters close to those of the $D^{*}(2650)^{0}$. Further measurements are needed to resolve these ambiguities. Additionally, subsequent measurements can change the average value such as to change the relative masses of states, which may be in contradiction with the naming. An example is the $D_{1}(2430)^{0}$ state, whose average mass has been lowered by the latest LHCb measurement~\cite{Aaij:2019sqk} to become smaller than the average mass of the $D_{1}(2420)$ states. 

\begin{figure}[htb!]
\begin{centering}
\includegraphics[width=0.49\textwidth]{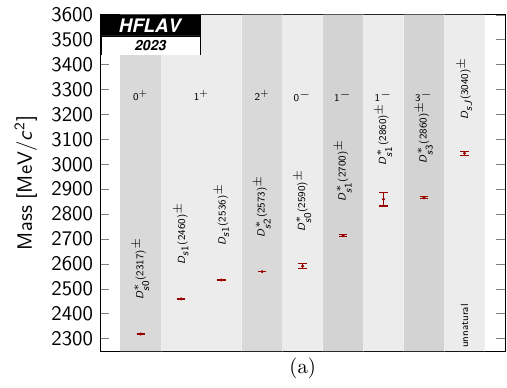}
\includegraphics[width=0.49\textwidth]{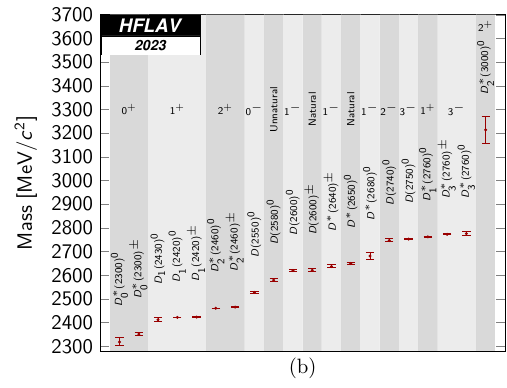}\\
\includegraphics[width=0.49\textwidth]{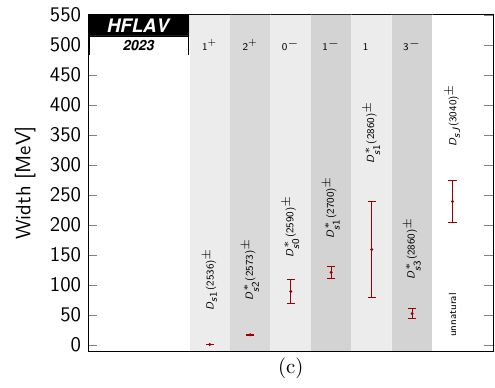}
\includegraphics[width=0.49\textwidth]{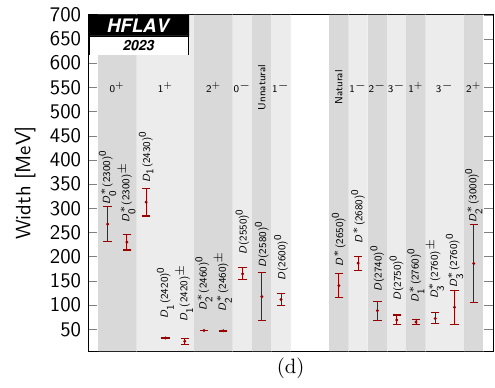}
\caption{\label{fig:charm:spect:1}
(a) Average masses for excited $D_{s}$ mesons;
(b) average masses for excited $D$ mesons;
(c) average widths for excited $D_{s}$ mesons;
(d) average widths for excited $D$ mesons.
The vertical shaded regions distinguish between different spin-parity states.}
\end{centering}
\end{figure}

The masses and widths of narrow ($\Gamma<50$~MeV) orbitally excited 
$D$ mesons (1P states), both neutral and charged, are 
well-established. Measurements of broad states ($\Gamma\sim$ 200--400~MeV) 
are less abundant, as identifying the signal is more challenging. The measured masses and widths, as well as the $J^P$ 
values, are in agreement with theoretical predictions based on potential 
models~\cite{Godfrey:1985xj, Godfrey:1986wj, Isgur:1991wq, Schweitzer:2002nm}. Following precise measurements from LHCb, recent studies based on unitarized chiral perturbation theory and lattice QCD propose a two-pole structure of the $D^{*}(2300)^{0}$ and predict the existence of a lighter state $D^{*}(2100)^{0}$~\cite{Albaladejo:2016lbb, Du:2020pui}.

The spectroscopic assignment of heavier states remains less clear. Further
theoretical studies taking into account the experimentally observed mesons suggest the identity of some 2S and 1D states~\cite{Chen:2015lpa,Godfrey:2015dva}
and tentatively discuss possible 1F, 3S and 2P states. Possible new states to
be found in the future were suggested in Refs.~\cite{Godfrey:2015dva,Badalian:2020ngz}. In the region around 3000 MeV, the state $D_{2}^{*}(3000)^{0}$ has been observed by LHCb~\cite{Aaij:2016fma}  and is explained as potentially a 2P or 1F state~\cite{Song:2015fha}. A pseudoscalar state $D(2900)$ is predicted in this region along with its decay channels~\cite{Malabarba:2021gyq}, laying out the path for further experimental investigations.

Tables~\ref{table:charm:spect:3a}, \ref{table:charm:spect:3b} and \ref{table:charm:spect:4} summarize the
branching fractions of $B$ meson decays to excited $D$ and $D_{s}$ 
states, respectively. The measurements listed are the products of the
$B$ meson branching fraction and the daughter $D$ meson branching fraction.
It is notable that the branching fractions for 
$B$ mesons decaying to a narrow $D^{\ast}$ state and a pion are similar 
for charged and neutral $B$ initial states, while the branching fractions 
to a broad $D^{\ast}$ state and $\pi^+$ are much larger for $B^+$ than 
for $B^0$. This may be due to the fact that color-suppressed amplitudes 
contribute only to the $B^+$ decay and not to the $B^0$ decay (for a 
theoretical discussion, see Refs.~\cite{Jugeau:2005yr,Colangelo:2004vu}). 
Values for the branching fractions of the $D$ mesons are difficult
to extract due to 
the unknown (and difficult to calculate) $B \to D^{\ast} X$ branching fractions. Recently, BESIII produced the first absolute branching fraction measurement of $D_{s0}^\ast(2317)^{-} \to D_{s}^{-}\pi^{0}$~\cite{Ablikim:2017rrr}.

The discoveries of the $D_{s0}^\ast(2317)^{\pm}$ and $D_{s1}(2460)^{\pm}$ 
have triggered increased interest in properties of, and searches for, 
excited $D_s$ mesons. While 
the masses and widths of the $D_{s1}(2536)^{\pm}$ and $D_{s2}^{\ast}(2573)^{\pm}$ 
states are in relatively good agreement with potential model predictions, 
the masses of the $D_{s0}^\ast(2317)^{\pm}$ and $D_{s1}(2460)^{\pm}$ states are 
significantly lower than expected (see Ref.~\cite{Cahn:2003cw} for a 
discussion of $c\bar{s}$ models). Moreover, the mass splitting between 
these two states greatly exceeds that between the $D_{s1}(2536)^{\pm}$ 
and $D_{s2}(2573)^{\pm}$. The internal structure of the $D_{s0}^\ast(2317)^{\pm}$ and 
$D_{s1}(2460)^{\pm}$ remains a subject of intense study to this date.  The unexpected properties have led to 
interpretations
as exotic four-quark states~\cite{Barnes:2003dj,Lipkin:2003zk}. A molecule-like ($DK$) interpretation of these states ~\cite{Barnes:2003dj,Lipkin:2003zk} that can account 
for their low masses and isospin-breaking decay modes, or a mixture or molecular and $c\overline{s}$ states, are among the most prominent models in recent works~\cite{Cleven:2014oka, Liu:2012zya, Guo:2017jvc, Fu:2021wde, Tang:2023yls, Gil-Dominguez:2023huq}. A study of the states in an unquenched quark model shows that the best match is a model with a DK component around 53\%~\cite{Tan:2021bvl}. 
Searches for charged and neutral isospin partners of these states have so far yielded negative results. A crucial tool to test the molecular picture is also measuring the partial decay widths, particularly isospin-breaking hadronic decays which would have a much larger width in a hadronic molecule approach. Another prominent test would be measuring
rates for the radiative processes 
$D_{s0}^\ast(2317)^{\pm}/D_{s1}(2460)^{\pm}\to D_s^{(\ast)}\gamma$ and 
comparing to theoretical predictions. The predicted rates, however, are below the sensitivity of current
experiments. A step toward this goal recent is the branching fraction measurement of $D_{s0}^\ast(2317)^{-} \to D_{s}^{-}\pi^{0}$ by BESIII~\cite{Ablikim:2017rrr}. A study of the transition $D_{s1}(2460)^{+} \to D_{s}^{+}\pi^{+}\pi^{-}$ indicates that in a molecular scenario, the $\pi^{+}\pi^{-}$ invariant mass would show a double bump structure which is not present in a compact state scenario~\cite{Tang:2023yls}, thus offering another potentially discriminating variable for experimental studies. Future measurements with more statistics are expected to shed further light on the topic of a molecular interpretation.

Another model successful in explaining the total widths and the 
$D_{s0}^\ast(2317)^{\pm}$ -- $D_{s1}(2460)^{\pm}$ mass splitting is based 
on the assumption that these states are chiral partners of the ground 
states $D_{s}^{+}$ and~$D_{s}^{*}$~\cite{Bardeen:2003kt}. While some 
measured branching fraction ratios agree with predicted values, further 
experimental tests with better sensitivity are needed to confirm or 
refute this scenario. A summary of the mass difference measurements 
is given in Table~\ref{table:charm:spect:5}. 

Measurements by \babar{}~\cite{Aubert:2009ah} and LHCb~\cite{Aaij:2012pc} 
first indicated the existence of the $D^{*}_{sJ}(2860)^{\pm}$ meson. An LHCb study of $B_{s}^{0}\to \overline{D}{}^{0}K^{-}\pi^{+}$ decays, 
in which they searched for excited $D_{s}$ mesons~\cite{Aaij:2014xza}, showed 
with $10\sigma$ significance that this state comprises two different 
particles,  one of spin 1 and one of spin 3. This represents the first 
measurement of a heavy flavoured spin-3 particle, and the first observation 
of $B$ meson decays to spin-3 particles. A subsequent study of $D^{(*)}_{sJ}$ mesons 
by the LHCb collaboration~\cite{Aaij:2016utb} supports the natural parity 
assignment for these states. This study also shows weak evidence 
for a further structure at a mass around 3040 MeV$/c^2$ with unnatural 
parity, which was first hinted at by a \babar\ analysis~\cite{Aubert:2009ah}.
The second observation of a spin-3 charm meson was a subsequent LHCb analysis
of $B^0 \to \overline{D}{}^0\pi^+\pi^-$ decays, which measured the spin-parity
assignment of the state $D^*_3(2760)^{\pm}$ to be $J^P=3^-$~\cite{Aaij:2015sqa}. This resonance was
in fact observed previously by \babar{}~\cite{delAmoSanchez:2010vq} and
LHCb~\cite{Aaij:2013sza}. The measurement suggests a spectroscopic assignment
of ${}^3D_3$. Recently, also the corresponding neutral state was observed by
LHCb, the $D^{*}_{3}(2760)^{0}$~\cite{Aaij:2016fma}.

Other observed excited $D_s$ states include $D_{s1}^{\ast}(2700)^{\pm}$  and 
$D_{s2}^{\ast}(2573)^{\pm}$. The properties of both (mass, width, $J^P$) have 
been measured and determined in several analyses. A theoretical 
discussion~\cite{Matsuki:2006rz} investigates the possibility that 
the $D_{s1}(2700)^{\pm}$ could represent radial excitations of the 
$D_s^{\ast\pm}$. Similarly, the  $D_{s1}^{\ast}(2860)^{\pm}$ and $D_{sJ}(3040)^{\pm}$ 
could be excitations of $D_{s0}^\ast(2317)^{\pm}$ and $D_{s1}(2460)^{\pm}$ or 
$D_{s1}(2536)^{\pm}$, respectively. The most recently discovered state is denoted as $D_{s0}(2590)^{\pm}$, and is a strong candidate to be the missing $2^{1}S_0$ state, the radial excitation of the pseudoscalar ground-state $D_s^{+}$ meson~\cite{Aaij:2020voz}. Matching newly observed states to theoretical predictions is not always trivial, as the predicted widths often heavily depend on the assumed mass. In this particular case, however, a re-calculation of the width at the experimentally measured mass nevertheless remains inconclusive as to the interpretation of the $D_{s0}(2590)^{\pm}$ as the $2^{1}S_0$ state~\cite{Wang:2021orp}.

Table~\ref{table:charm:spect:6} summarizes measurements of the helicity parameter
 $A_{D}$ (also referred to as the polarization parameter). In decays of orbitally excited charm mesons ($D^{\ast\ast}$) to $D^{\ast}\pi$, $D^{\ast} \to D \pi$, the helicity
distribution varies like $1 + A_{D}\cos^{2}\theta_{H}$, where $\theta_{H}$
is the angle in the $D^{\ast}$ rest frame between the two pions emitted
by the decays $D^{\ast\ast} \to D^{\ast}\pi$ and $D^{\ast} \to D \pi$. The
parameter is sensitive to possible $S$-wave contributions in the decay. 
In the case of a $D$ meson decaying purely via $D$-wave, 
the helicity parameter is predicted to give $A_{D}=3$. Studies of the 
$D_{1}(2420)^{0}$ meson by the ZEUS and \babar{} collaborations suggest 
that there is an S-wave admixture in the decay, which is  contrary to 
the expectation based on Heavy Quark Effective Theory~\cite{Isgur:1989vq,Neubert:1993mb}.

\begin{table}[htb!]
\begin{center}
\caption{\label{table:charm:spect:3a} 
Product of the $B$ meson branching fraction and the
daughter (excited) $D$ meson branching fraction. Table 1 of 2.}

\begin{adjustbox}{width=0.84\textwidth,center}
{\setlength\tabcolsep{0pt}

}
\end{adjustbox}

\end{center}
\end{table} 

\begin{table}[htb!]
\begin{center}
\caption{\label{table:charm:spect:3b} 
Product of the $B$ meson branching fraction and the
daughter (excited) $D$ meson branching fraction. Table 2 of 2.}
\begin{adjustbox}{width=0.95\textwidth,center}
{\setlength\tabcolsep{0pt}
%
}
\end{adjustbox}

\end{center}
\end{table} 

\begin{table}[htb!]
\begin{center}
\caption{\label{table:charm:spect:4} 
Product of the $B$ meson branching fraction and the
daughter (excited) $D^{}_s$ meson branching fraction.}
\begin{adjustbox}{width=\textwidth,center}
{\setlength\tabcolsep{0pt}
 %
}
\end{adjustbox}

\end{center}
\end{table}

\begin{table}[htb!]
\begin{center}
\caption{\label{table:charm:spect:5} 
Measurements of mass differences for excited $D$ mesons.}
{\setlength\tabcolsep{0pt}
 %
}

\end{center}
\end{table} 

\begin{table}[htb!]
\begin{center}
\caption{\label{table:charm:spect:6}
Measurements of polarization amplitudes for excited $D$ mesons.}
{\setlength\tabcolsep{0pt}
 %
}

\end{center}
\end{table}

\clearpage
\subsection{Excited charmed baryons}

In this section we summarize the present status of excited charmed
baryons decaying strongly or electromagnetically. We list their
masses (or the mass difference between the excited baryon and the
corresponding ground state), natural widths, decay modes, and 
assigned quantum numbers. 
The present ground-state measurements are: $M(\Lambda_c^+)=(2286.46\pm0.14)$~MeV/$c^2$
measured by \babar~\cite{Aubert:2005gt},
$M(\Xi_c^0)=(2470.90^{+0.22}_{-0.29}$)~MeV/$c^2$ and
$M(\Xi_c^+)=(2467.94^{+0.17}_{-0.20}$)~MeV/$c^2$, both 
dominated by CDF~\cite{Aaltonen:2014wfa}, and
$M(\Omega_c^0)=(2695.2\pm1.7$)~MeV/$c^2$, dominated 
by Belle~\cite{Solovieva:2008fw}. Should these values 
change, so will many of the values for the masses of the excited states.

Table~\ref{sumtable1} summarizes the excited $\Lambda_c^+$ baryons.  
The first two states listed, namely the $\Lambda_c(2595)^+$ and $\Lambda_c(2625)^+$,
are well-established. 
The measured masses and decay patterns suggest that 
they are orbitally 
excited $\Lambda_c^+$ baryons with total angular momentum of the
light quarks $L=1$. Thus their quantum numbers are assigned to be 
$J^P=\frac{1}{2}^-$ and $J^P=\frac{3}{2}^-$, respectively. 
Their mass measurements are  
dominated by CDF~\cite{Aaltonen:2011sf}: 
$M(\Lambda_c(2595)^+)=(2592.25\pm 0.28$)~MeV/$c^2$ and
$M(\Lambda_c(2625)^+)=(2628.11\pm 0.19$)~MeV/$c^2$. 
Earlier measurements did not fully take into account the restricted
phase-space of the $\Lambda_c(2595)^+$ decays.

The next two states, $\Lambda_c(2765)^+$ and $\Lambda_c(2880)^+$, 
were discovered by CLEO~\cite{Artuso:2000xy} in the $\Lambda_c^+\pi^+\pi^-$ 
final state. CLEO found that a significant fraction of the $\Lambda_c(2880)^+$ decays 
proceeds via an intermediate $\Sigma_c(2445)^{++,0}\pi^{-,+}$.  
Later, \babar~\cite{Aubert:2006sp} 
observed that this state has also a $D^0 p$ decay mode. This was the 
first example of an excited charmed baryon decaying into a charm meson 
plus a baryon; previously all excited charmed baryons were found via 
hadronic transitions into lower lying charmed 
baryons. In the same analysis, \babar observed for the
first time an additional state, $\Lambda_c(2940)^+$, 
decaying into $D^0 p$. Studying the $D^+ p$ final state,
\babar found no signal and this implies that the $\Lambda_c(2880)^+$ 
and $\Lambda_c(2940)^+$ are $\Lambda_c^+$ excited states
rather than $\Sigma_c$ excitations. 
Belle reported the result of an angular analysis that favors
$5/2$ for the $\Lambda_c(2880)^+$ spin hypothesis~\cite{Belle:2006xni}. 
Moreover, the measured ratio of branching fractions 
${\cal B}(\Lambda_c(2880)^+\rightarrow \Sigma_c(2520)\pi^{\pm})/{\cal B}(\Lambda_c(2880)^+\rightarrow \Sigma_c(2455)\pi^{\pm})
=0.225\pm 0.062\pm 0.025$, combined 
with theoretical predictions based on HQS~\cite{Isgur:1991wq,Cheng:2006dk}, 
favor even parity. However this prediction is only valid if the P-wave 
portion of $\Sigma_c(2520)\pi$ is suppressed. 
LHCb~\cite{Aaij:2017vbw} have analysed the $D^0p$ system in the resonant substructure of $\Lambda_b$
decays. They confirm the $\frac{5}{2}^+$ identification of the $\Lambda_c(2880)^+$. In addition they find 
evidence for a further, wider, state they name the $\Lambda_c(2860)^+$, with $J^P=\frac{3}{2}^+$ (the
parity is measured with respect to that of the $\Lambda_c(2880)^+$). The explanation for these states
in the heavy quark-light diquark model is that they are a pair of orbital D-wave excitations.
Furthermore, LHCb~\cite{Aaij:2017vbw} find evidence for the spin-parity of the $\Lambda_c(2940)^+$
to be $\frac{3}{2}^-$, and improve the world average measurements of both the mass and width of this particle.

The Belle analysis that measured the $\Lambda_c(2880)$ spin~\cite{Belle:2006xni}
also showed a statistically significant excess in the $\Sigma_c\pi$ mass distribution 
between the $\Lambda_c(2880)$ and $\Lambda_c(2940)$ states. 
Further information was lacking until 2022 when Belle analysed the decay 
$B\to\Sigma_c\pi p$~\cite{Belle:2022hnm}. Here in the $\Sigma_c\pi$ mass spectrum, no signals for the 
$\Lambda_c(2880)$ or $\Lambda_c(2940)$~resonances were observed, 
but a large, wide peak spanning the region and centered
at a mass of 2910~MeV/$c^2$.
They conjecture that this peak could be due to a $J^P\ =\ \frac{1}{2}^-,\ 2P$ state. However, we must also be aware that this could be the overlap of two or more particles as has been seen in $\Xi_c$ production from $B$ decays, and it is clearly an area that requires more investigation.

A current open question concerns the nature of the $\Lambda_c(2765)^+$ state.
However, it has now been experimentally shown by Belle~\cite{Abdesselam:2019bfp} to be a $\Lambda_c$ rather than a
$\Sigma_c$, and implicit in this analysis is that the data can be explained by 
one resonance.
The state has also been observed, but not measured, by LHCb~\cite{Aaij:2017svr}.

\begin{table}[htb]
\caption{Summary of excited $\Lambda_c^+$ baryons. The uncertainties are the total of the statistical and systematic uncertainties.} 
\vskip0.15in
\begin{center}
\renewcommand{\arraystretch}{1.2}
\begin{tabular}{c|c|c|c|c}
\hline
Charmed baryon   & Mode  & Mass & Natural width  & $J^P$  \\
excited state &  &  (MeV/$c^2$) & (MeV)  \\
\hline
$\Lambda_c(2595)^+$ & $\Lambda_c^+\pi^+\pi^-$, $\Sigma_c(2455)\pi$ &  $2592.25\pm 0.28$ &
$2.59 \pm 0.57$  & $\frac{1}{2}^-$  \\
\hline
$\Lambda_c(2625)^+$ & $\Lambda_c^+\pi^+\pi^-$   & $2628.11\pm 0.19$ & $<0.97$ & $\frac{3}{2}^-$  \\
\hline
$\Lambda_c(2765)^+$ & $\Lambda_c^+\pi^+\pi^-$, $\Sigma_c(2455)\pi$ & $2766.6\pm 2.4$ & $50$ & ?  \\
\hline
$\Lambda_c(2860)^+$ & $D^0p$ & $2856.1\,^{+2.3}_{-5.6}$ &
$67.6\,^{+11.8}_{-21.6}$ & $\frac{3}{2}^+$ \\
\hline
$\Lambda_c(2880)^+$ & $\Lambda_c^+\pi^+\pi^-$, $\Sigma_c(2455)\pi$,  &$2881.63\pm 0.24$ &
$5.6\,^{+11.3}_{-10.0}$ & $\frac{5}{2}^+$ \\
 &  $\Sigma_c(2520)\pi$, $D^0p$     & & &  \\
\hline
$\Lambda_c(2910)^+$ & $\Sigma_c(2455)\pi$,  &$2913.8\pm 5.6 \pm 3.8$ &
$51.8\pm {20.3}\pm 18.8$ & ? \\
\hline

$\Lambda_c(2940)^+$ & $D^0p$, $\Sigma_c(2455)\pi$ & $2939.6\,^{+1.3}_{-1.5}$ & $20\,^{+6}_{-5}$  & ?  \\
\hline 
\end{tabular}
\end{center}
\label{sumtable1} 
\end{table}

Table~\ref{sumtable2} summarizes the excited $\Sigma_c^{++,+,0}$ baryons.
The ground state iso-triplets of $\Sigma_c(2455)^{++,+,0}$ and
$\Sigma_c(2520)^{++,+,0}$ baryons are well-established. 
Belle~\cite{Lee:2014htd} 
precisely measured the mass differences  
and widths of the doubly charged and neutral members of this triplet.
The short list of excited $\Sigma_c$ baryons is completed by the triplet 
of $\Sigma_c(2800)$ states observed by Belle~\cite{Mizuk:2004yu}. Based 
on the measured masses and theoretical predictions~\cite{Copley:1979wj,Pirjol:1997nh}, 
these states are thought by some to be members of the predicted $\Sigma_{c2}$ $\frac{3}{2}^-$
triplet, where the subscript 2 refers to the total spin of the light quark degrees of freedom. From a study of resonant substructure 
in $B^-\rightarrow \Lambda_c^+\bar{p}\pi^-$ decays, \babar found 
a significant signal in the $\Lambda_c^+\pi^-$ final state with a mean value 
higher than measured for the $\Sigma_c(2800)$ by Belle by about $3\sigma$
(Table~\ref{sumtable2}). The decay widths measured by
Belle and \babar are consistent, but it is an open question if the 
observed state is the same as the Belle state. It is possible that the present excesses will prove to be due to two or more
overlapping states. Circumstantial evidence for this can be found by comparing with the $\Xi_c$ and $\Omega_c$ states of similar excitation energies.

\begin{table}[!htb]
\caption{Summary of the excited $\Sigma_c^{++,+,0}$ baryon family. The mass difference is given with respect to the $\Lambda_c^+$} 
\vskip0.15in
\begin{center}
\renewcommand{\arraystretch}{1.2}
\begin{tabular}{c|c|c|c|c}
\hline
Charmed baryon   & Mode  & Mass Difference & Natural width  & $J^P$  \\
excited state &  &  (MeV/$c^2$) & (MeV)  \\
\hline
$\Sigma_c(2455)^{++}$ &$\Lambda_c^+\pi^+$  & $167.510 \pm 0.17$ & $1.89\,^{+0.09}_{-0.18}$ & $\frac{1}{2}^+$   \\
$\Sigma_c(2455)^{+}$ &$\Lambda_c^+\pi^0$  & $166.4\pm 0.4$ & $<4.6$~@~90$\%$~C.L. & $\frac{1}{2}^+$ \\
$\Sigma_c(2455)^{0}$ &$\Lambda_c^+\pi^-$  & $167.29\pm 0.17$ & $1.83\,^{+0.11}_{-0.19}$ & $\frac{1}{2}^+$    \\
\hline
$\Sigma_c(2520)^{++}$ &$\Lambda_c^+\pi^+$  & $231.95\,^{+0.17}_{-0.12}$ & $14.78\,^{+0.30}_{-0.40}$ & $\frac{3}{2}^+$   \\
$\Sigma_c(2520)^{+}$ &$\Lambda_c^+\pi^0$  & $231.0\pm 2.3$ & $<17$~@~90$\%$~C.L. & $\frac{3}{2}^+$ \\
$\Sigma_c(2520)^{0}$ &$\Lambda_c^+\pi^-$  & $232.02\,^{+0.15}_{-0.14}$ & $15.3\,^{+0.4}_{-0.5}$ & $\frac{3}{2}^+$    \\
\hline
$\Sigma_c(2800)^{++}$ & $\Lambda_c^+\pi^{+}$ & $514\,^{+4}_{-6}$ & $75\,^{+22}_{-17}$ & $\frac{3}{2}^-$?     \\
$\Sigma_c(2800)^{+}$ & $\Lambda_c^+\pi^{0}$&$505\,^{+15}_{-5}$ &$62\,^{+64}_{-44}$ &   \\
$\Sigma_c(2800)^{0}$ & $\Lambda_c^+\pi^{-}$&$519\,^{+5}_{-7}$ & $72\,^{+22}_{-15}$ &   \\
 & $\Lambda_c^+\pi^{-}$ & $560\pm 13$ & $86\,^{+33}_{-22}$  \\

\hline 
\end{tabular}
\end{center}
\label{sumtable2} 
\end{table}

Table~\ref{sumtable3} summarises the excited $\Xi_c^{+,0}$. The list of excited $\Xi_c$
baryons has many states, of unknown quantum numbers, having masses 
above 2900~MeV/$c^2$ and decaying through three different types of modes:
$\Lambda_c K n\pi$ or $\Sigma_c K n\pi$, $\Xi_c n\pi$, and $\Lambda D$.  
Some of these states ($\Xi_c(2970)^+$, $\Xi_c(3055)$ and $\Xi_c(3080)^{+,0}$) have been
observed by
both Belle~\cite{Chistov:2006zj,YKato:2014,YKato:2016} 
and \babar~\cite{Aubert:2007eb}, are produced in the charm continuum, 
and are considered well-established.
Recently LHCb~\cite{Aaij:2020yyt} reported three narrow states in $\Lambda_c^+K^-$. The masses
of two of these states, named the $\Xi_c(2923)$ and $\Xi_c(2939)$ bookend the already
discovered, wide
$\Xi_c(2930)^0$. This latter state, also decaying into $\Lambda_c^+ K^-$, was found in $B$ 
decays
by \babar~\cite{Aubert:2007bd}. It was also observed by Belle~\cite{Li:2017uvv}, who in addition observed a similar charged state. Table~\ref{sumtable3} shows the parameters measured by
Belle, who allow for interference and other possible resonances in their analysis.
However, given the low statistics of these observations in $e^+e^-$ annihilation, 
it seems likely that the
LHCb data explains the peaks found by Belle and \babar as the manifestation of the two states seen by LHCb 
overlapping, and thus there are no distinct
$\Xi_c(2930)$ states. They are thus not included in ~\ref{sumtable3}.

LHCb~\cite{Aaij:2020yyt} note that there may well be a fourth resonance at a lower mass of $\approx$ 2880~MeV/$c^2$. This was then confirmed, albeit still with low statistical precision, in a subsequent LHCb paper~\cite{LHCb:2022vns} using low statistics decays from $B$ decays. The authors note that the pattern of masses of these states bears a remarkable similarity to that of the excited $\Omega_c$ masses,
implying the same underlying spin-structure of the quarks. The highest mass of the LHCb $\Omega_c$ quintuplet does not appear to have an analogous state here. This state had the lowest signal of the five in the LHCb data and was not confirmed by Belle, so this may imply that it is of a completely different nature to the other four. Alternatively there may exist an equivalent state in the excited $\Lambda_c^+K^-$ data but it cannot be seen because of its large width and alternative decay modes. 
The highest-mass of the new $\Xi_c$ states reported by LHCb is very close in mass
to the $\Xi_c(2970)$; formerly named by the PDG as the $\Xi_c(2980)$. A recent analysis by Belle~\cite{Moon:2020gsg} reveals that the favoured spin-parity of this state is $J^P=\frac{1}{2}^+$, which
corresponds to it being a (2S) radial excitation. The $\Xi_c(2765)$ and $\Xi_c(2970)$ properties seem to differ by enough to be able to 
say with reasonable confidence that they are completely different states, though the confusion due to their similar masses might help explain the poor consistency of the historical 
measurements of the $\Xi_c(2970)$. The properties of the $\Xi_c(2970)$ 
can be seen to have similarities to the $\Lambda_c(2765)$, not only in its mass difference with respect to the ground state, but also its decay pattern and large production cross-section in $e^+e^-$ annihilation data. This strengthens the confidence in their identification as "Roper-like" resonances~\cite{Roper:1964zza}, that is, as the charmed equivalents of the light quark $P_{11}(1440)$ which is often 
identified as a radial excitation.

In addition to the $\Xi_c(2970)$, Belle~\cite{Yelton:2016fqw} has analysed large samples of 
$\Xi_c^\prime$, $\Xi_c(2645)$, $\Xi_c(2790)$, and $\Xi_c(2815)$ decays. From this analysis they obtain the most 
precise mass measurements of all five iso-doublets, and the first
significant width measurements of the $\Xi_c(2645)$, $\Xi_c(2790)$ and $\Xi_c(2815)$.
Though the spin-parity of these particles have not been directly measured, there seems little controversy that the simple quark-diquark model can explain the data. We note 
that the precision of the mass measurements allows for added information from the isospin mass-splitting to be used in identifying the underlying structure. In addition, the recent observation by Belle~\cite{Yelton:2020awh} of photon 
transitions to the ground state from the neutral (but not charged) $\Xi_c(2815)$, and probably also the $\Xi_c(2790)$,
can be interpreted as confirmation of the standard quark interpretation.
 
Several of the width and mass measurements for the $\Xi_c(3055)$ and $\Xi_c(3080)$ 
isodoublets are only in marginal agreement between experiments and 
decay modes. However, there seems little doubt that the differing 
measurements are of the same particle. The masses do indicate that their spin-parity might
match those of the $\Lambda_c(2860)$ and $\Lambda_c(2880)$.

The $\Xi_c(3123)^+$ reported by \babar~\cite{Aubert:2007eb}
in the $\Sigma_c(2520)^{++}\pi^-$ final state has not been
confirmed by Belle~\cite{YKato:2014} with twice the statistics; 
thus its existence is in doubt and it is omitted from Tab.~\ref{sumtable3} which summarises
the present situation in the excited $\Xi_c$ sector.

\begin{table}[b]
\caption{Summary of excited $\Xi_c^{+,0}$ states. 
For the first four isodoublets, it is the mass difference with respect to the 
ground state to which they decay that is quoted as this avoids uncertainties in the ground state masses. For the remaining cases, the uncertainty on the 
measurement of the excited state itself dominates.} 
\vskip0.15in
\resizebox{\textwidth}{!}{
\renewcommand{\arraystretch}{1.2}
}
\label{sumtable3} 
\end{table}
 The $\Omega_c^{*0}$ doubly-strange charmed baryon has been seen by both 
\babar~\cite{Aubert:2006je} and Belle~\cite{Solovieva:2008fw}.
The mass differences $\Delta M=M(\Omega_c^{*0})-M(\Omega_c^0)$ 
measured by the experiments are in good agreement
and are also consistent with most theoretical 
predictions~\cite{Rosner:1995yu,Glozman:1995xy,Jenkins:1996de,
Burakovsky:1997vm}. 
LHCb~\cite{Aaij:2017nav} has found a family of five excited $\Omega_c^{0}$ baryons
decaying into $\Xi_c^+K^-$. A natural explanation is that they are the five states with $L=1$
between the heavy quark and the light ($ss$) diquark; however, there is no consensus as to
which state is which, and this overall interpretation is controversial. 
Four of the five states
were confirmed by Belle~\cite{Yelton:2017qxg}. These same four states were then confirmed by LHCb in the decays of the $\Omega_b$~\cite{LHCb:2021ptx}. The lack of confirmation of the $\Omega_c(3120)^0$ in these latter two production mechanisms indicate that it might well belong to a different multiplet - for instance a $\rho$ excitation where the $L=1$ is between the two lighter quarks. 
Further data from LHCb~\cite{LHCb:2023sxp} added two more clear observations. One, the $\Omega_c(3185)$ is copiously produced and wide, and the other, of width $\approx 20$~MeV at a considerably higher mass.

\begin{table}[b]
\caption{Summary of excited $\Omega_c^{0}$ baryons. 
For the $\Omega_c(2770)^0$, the mass difference with respect to the 
ground state is given, as the uncertainty is dominated by the uncertainty
in the ground state mass. In the remaining cases the total mass is shown,
though the uncertainty in the $\Xi_c^+$ mass (the last uncertainty quoted) makes an important contribution
to the total uncertainty. The latest LHCb results supercede their initial ones and the Belle result can no longer contribute significantly to the average values. } 
\vskip0.15in
\renewcommand{\arraystretch}{1.2}
\begin{center}
    
\begin{tabular}{c|c|c|c|c}
\hline
Charmed baryon   & Mode  & Mass difference & Natural width  & $J^P$  \\
excited state &  & (MeV/$c^2$)  & (MeV)  \\
\hline
$\Omega_c(2770)^0$ & $\Omega_c^0\gamma$ & $70.7^{+0.8}_{-0.9}$ &      &$\frac{3}{2}^+$  \\
\hline 
\hline
Charmed baryon   & Mode  & Mass  & Natural width  & $J^P$  \\
excited state &  & (MeV/$c^2)$  & (MeV) & \\
\hline
$\Omega_c(3000)^0$  & $\Xi_c^+K^-$  & $3000.44\pm0.07^{+0.07}_{-0.13}\pm0.23$   &   $3.83\pm0.23^{+1.50}_{-0.29} $ &? \\
\hline
$\Omega_c(3050)^0$  & $\Xi_c^+K^-$  & $3050.18\pm0.04^{+0.06}_{-0.06}\pm0.23$   &   $0.67\pm0.17^{+0.64}_{-0.72}$ & ?\\
\hline
$\Omega_c(3065)^0$  & $\Xi_c^+K^-$  & $3065.63\pm0.06\pm0.06\pm0.23$      &   $3.79\pm0.20^{+0.38}_{-0.47}$ & ? \\
\hline
$\Omega_c(3090)^0$  & $\Xi_c^+K^-$  & $3090.16\pm0.11^{+0.06}_{-0.10}\pm0.23$       &   $8.48\pm0.44^{+0.61}_{-1.62}$ & ? \\
\hline
$\Omega_c(3120)^0$  & $\Xi_c^+K^-$  &  $3118.98\pm0.12^{+0.09}_{-0.23}\pm0.23$     &   $0.60\pm0.63^{+0.90}_{-1.05}$ & ? \\
\hline
$\Omega_c(3185)^0$  & $\Xi_c^+K^-$  &  $3185.1\pm1.7^{+7.4}_{-0.9}\pm0.2$     &   $50\pm7^{+10}_{-20}$ & ? \\
\hline
$\Omega_c(3327)^0$  & $\Xi_c^+K^-$  &  $3327.1\pm1.2^{+0.1}_{-1.3}\pm0.2$     &   $20\pm5^{+13}_{-1}$ & ? \\
\hline
\end{tabular}
\label{sumtable4} 
\end{center}
\end{table}

Figure~\ref{charm:leveldiagram2023} shows the levels of excited charm
baryons along with corresponding transitions between them, and
also transitions to the ground states.
\begin{figure}[!htb]
\includegraphics[width=1.0\textwidth]{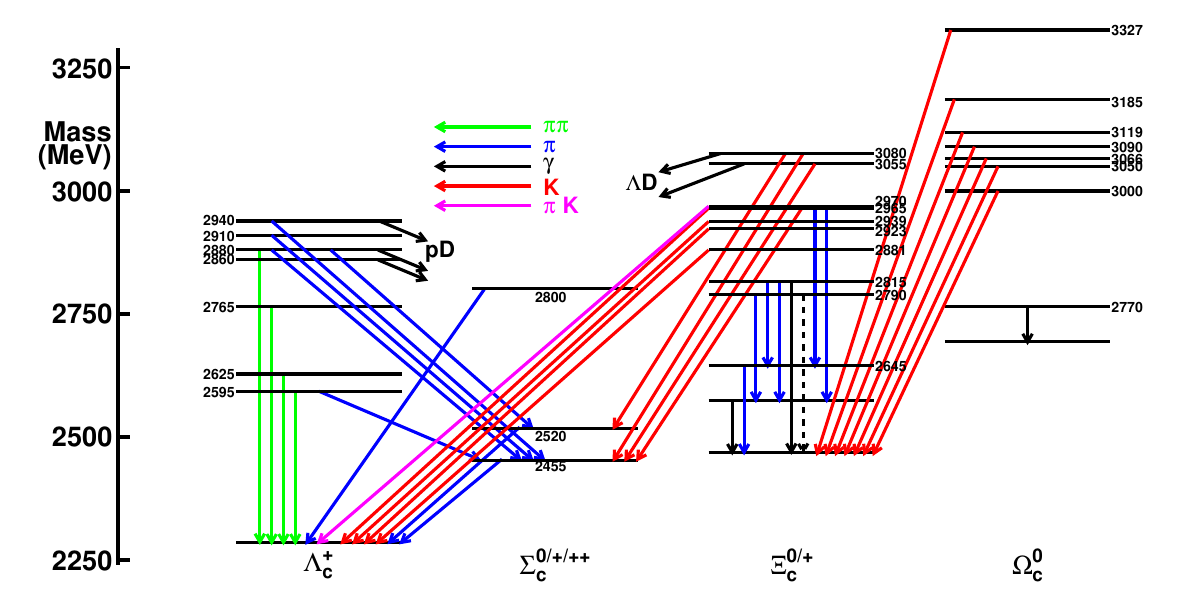}
\caption{Level diagram for multiplets and transitions for excited charm baryons.}
\label{charm:leveldiagram2023}
\end{figure} 
We note that Belle and \babar\ discovered
that transitions between "families" of baryons are possible, \ie, between 
 the charmed-strange ($\Xi_c$) and charmed-nonstrange ($\Lambda_c^+$ and $\Sigma_c$) families of excited charmed 
baryons~\cite{Chistov:2006zj,Aubert:2007eb}, and that 
highly excited states are found to decay into a
non-charmed baryon and a $D$ meson\cite{Aubert:2006sp,YKato:2016}. 
Transitions of the ground state $\Xi_c^0$ to the $\Lambda_c^+$, corresponding to the weak decay of the stange quark, were observed first by Belle\cite{Lee:2014htd}, and then studied and measured by LHCb\cite{Aaij:2020wtg}

\clearpage
\subsection{Rare and forbidden decays}
\label{sec:charm:rare}

This section provides a summary of searches for rare and forbidden charm decays
in tabular form. The decay modes can be categorized as 
flavour-changing neutral currents, including decays with and without hadrons in the final state, and radiative, lepton-flavour-violating, 
lepton-number-violating, and both baryon- and lepton-number-violating decays.
Figures~\ref{fig:charm:rare_d0}-\ref{fig:charm:lambdac} plot the 
upper limits for $D^0$, $D^+$, $D_s^+$, $D^{*0}$, $D_s^{*+}$, and $\Lambda_c^+$ decay branching fractions. 
Tables~\ref{tab:charm:rare_d0}-\ref{tab:charm:rare_lambdac} give the 
corresponding numerical results. Some theoretical predictions are given in 
Refs.~\cite{Burdman:2001tf,Fajfer:2002bu,Fajfer:2007dy,Golowich:2009ii,Paul:2010pq,Borisov:2011aa,Wang:2014dba,deBoer:2015boa,Fajfer:2015mia,Sahoo:2017lzi,deBoer:2017que,deBoer:2018buv,Bause:2019vpr,Bause:2020xzj,Bhattacharya:2018msv,Fajfer:2021woc}.

Some $D^0$ decay modes have been observed and are quoted as branching fractions with uncertainties in the tables and shown as a symbol with a line representing the $68\%$ C.L. interval in the plots.

In several cases the rare-decay final states have been observed with 
the di-lepton pair being the decay product of a vector meson.
For these measurements the quoted limits are those expected for the 
non-resonant di-lepton spectrum.
For the extrapolation to the full spectrum a phase-space distribution 
of the non-resonant component has been assumed.
This applies to the CLEO measurement of the decays 
$D_{(s)}^+\to(K^+,\pi^+)e^+e^-$~\cite{Rubin:2010cq}, to the D0 measurements 
of the decays $D_{(s)}^+\to\pi^+\mu^+\mu^-$~\cite{Abazov:2007aj}, and to 
the \babar measurements of the decays $D_{(s)}^+\to(K^+,\pi^+)e^+e^-$ and 
$D_{(s)}^+\to(K^+,\pi^+)\mu^+\mu^-$, where the contribution from 
$\phi\to l^+l^-$ ($l=e,\mu$) has been excluded.
In the case of the LHCb measurements of the decays 
$D^0\to\pi^+\pi^-\mu^+\mu^-$~\cite{Aaij:2013uoa} as 
well as the decays $D_{(s)}^+\to\pi^+\mu^+\mu^-$~\cite{Aaij:2013sua} 
the contributions from $\phi\to l^+l^-$ as well as from 
$\rho,\omega\to l^+l^-$ ($l=e,\mu$) have been excluded. 

For the $D^{*0}$ meson, the limit for the $e^+e^-$ final state was obtained by searching for the meson's on-shell production in $e^+e^-$ collisions by the CMD-3 experiment.
The $D_{\rm s}^{*+}$ limit for the final state $e^+\nu_e$ seemingly refers to a Cabibbo-favoured weak decay.
However, the limit on its partial width is comparable to that of other rare decays.
This is due to its total width being among the smallest of any excited meson state.

\begin{figure}
\begin{center}
\includegraphics[width=1.0\textwidth]{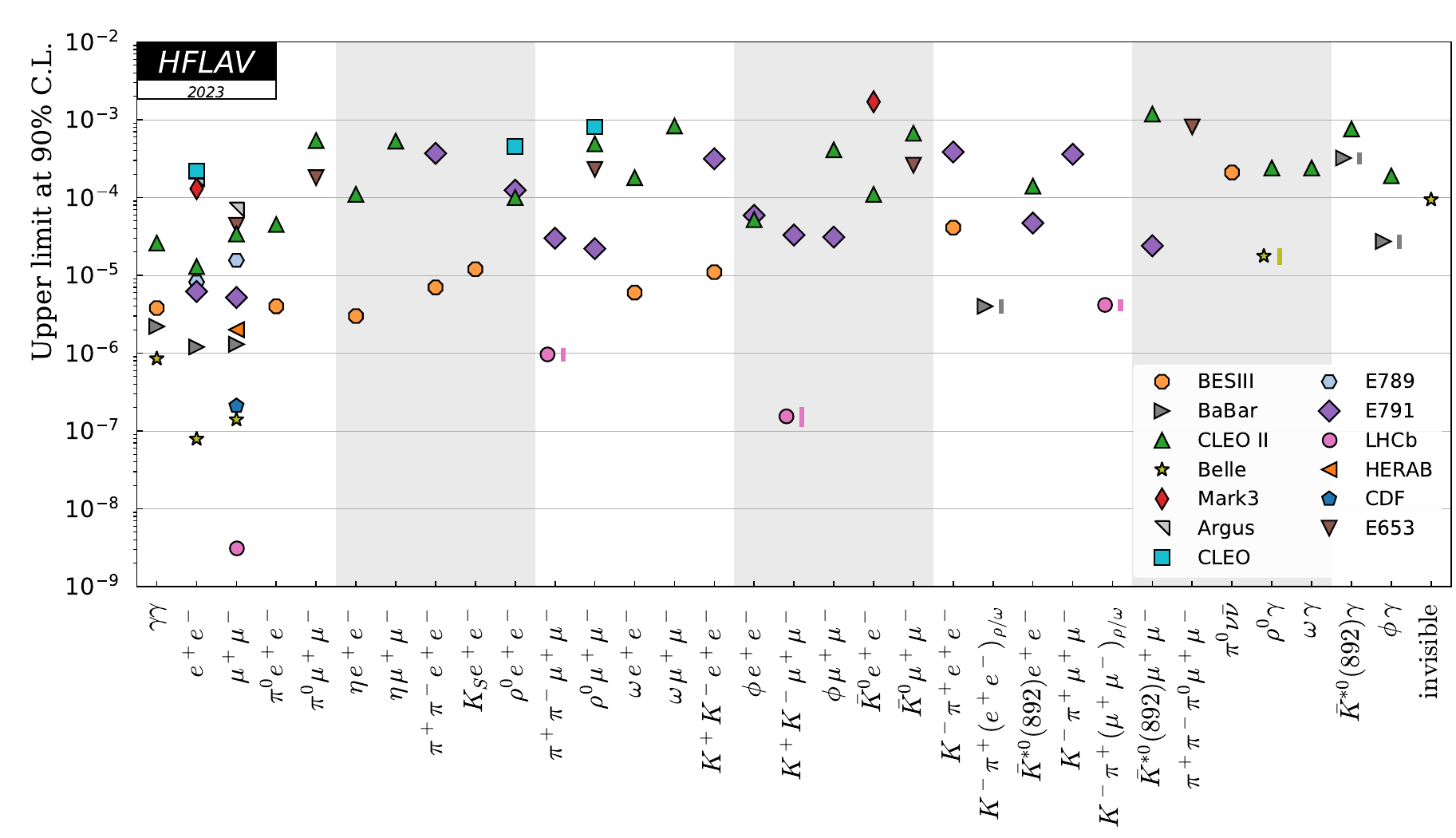}
\vskip0.10in
\includegraphics[width=5.0in]{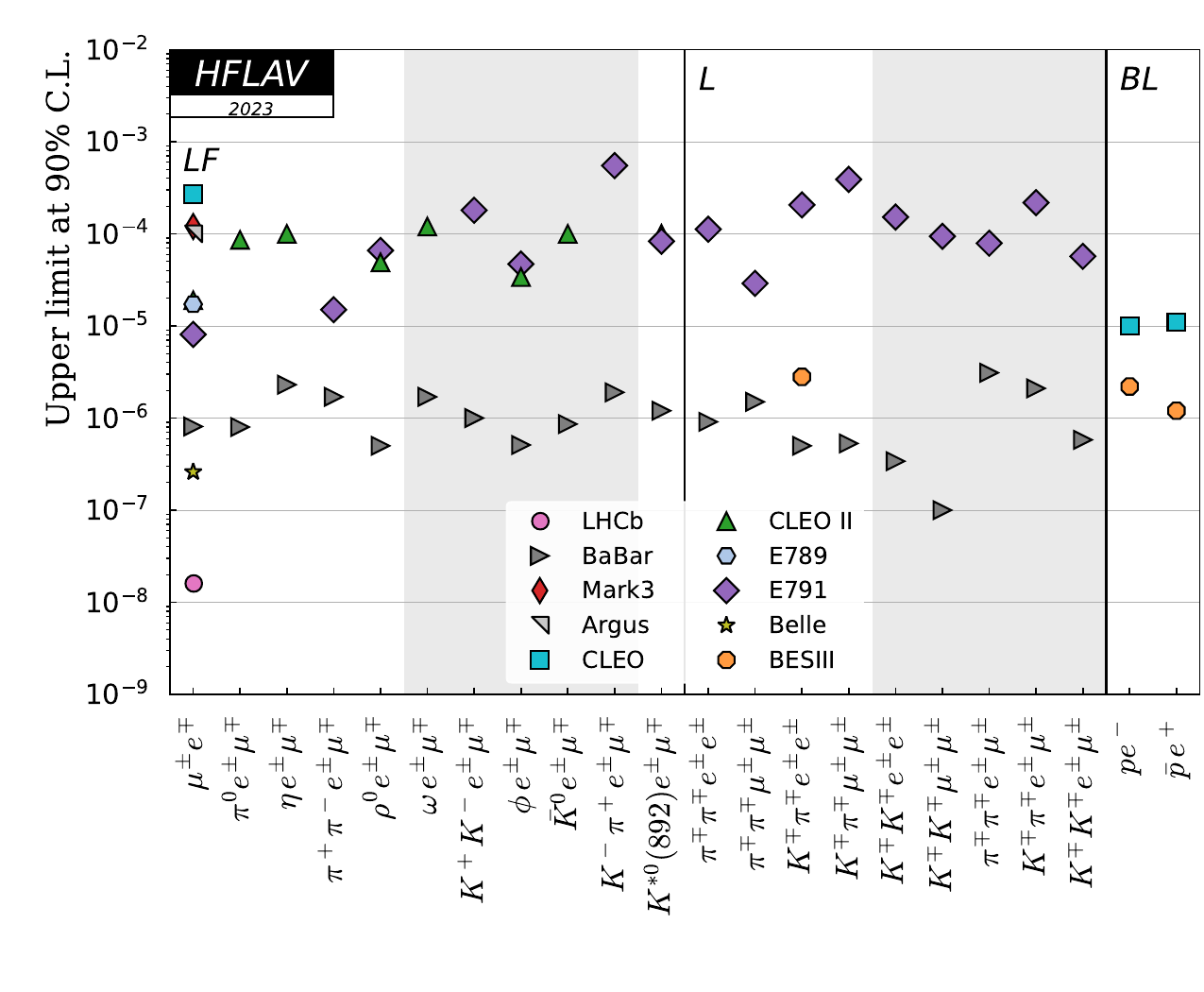}
\caption{Upper limits at $90\%$ C.L.\ for $D^0$ decay branching fractions. The top plot
shows flavour-changing neutral current and radiative decays, and the bottom plot
shows lepton-flavour-changing (LF), lepton-number-changing (L), and 
both baryon- and lepton-number-changing (BL) decays.
}
\label{fig:charm:rare_d0}
\end{center}
\end{figure}

\begin{figure}
\begin{center}
\includegraphics[width=5.0in]{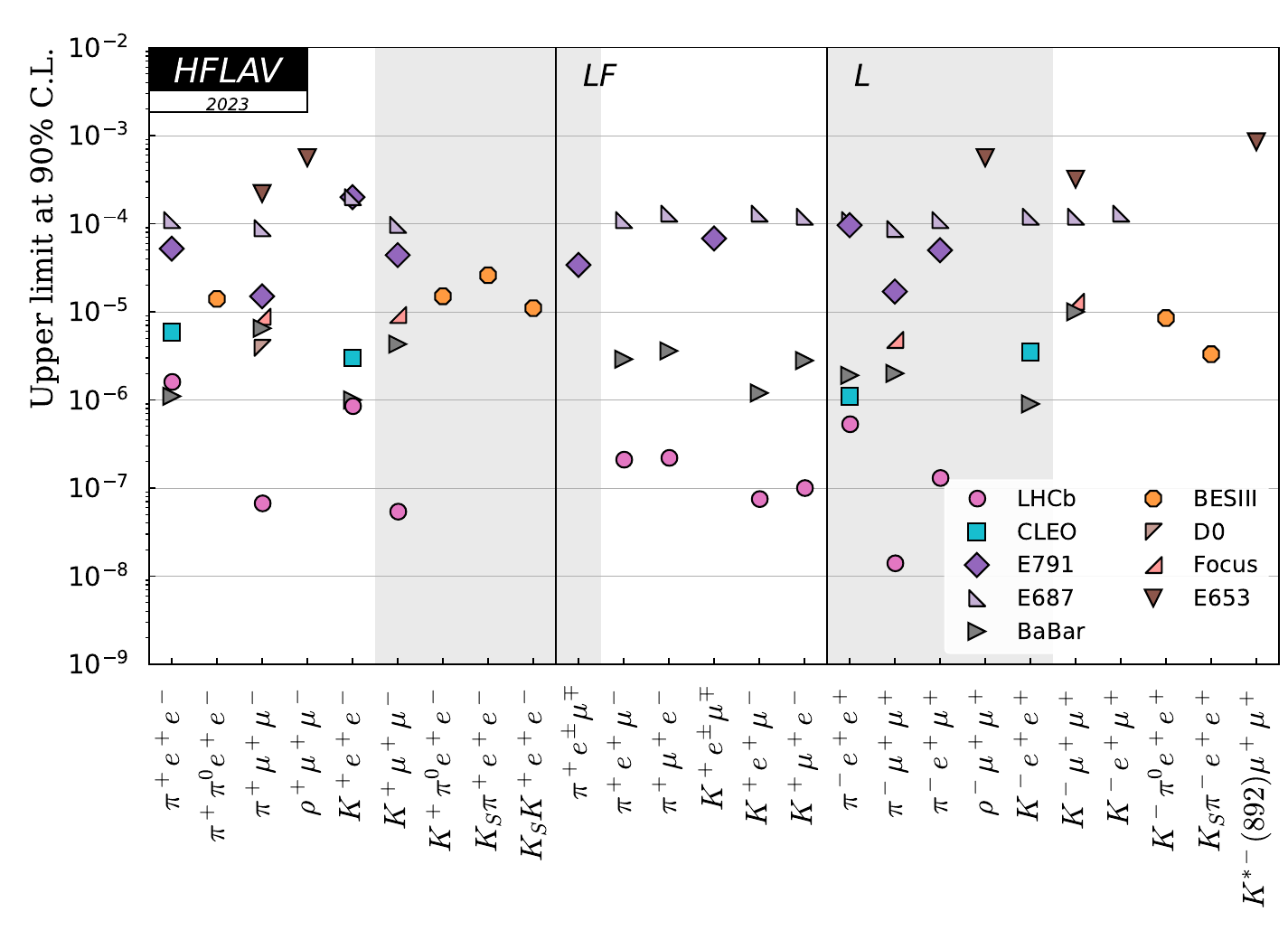}
\vskip0.10in
\includegraphics[width=5.0in]{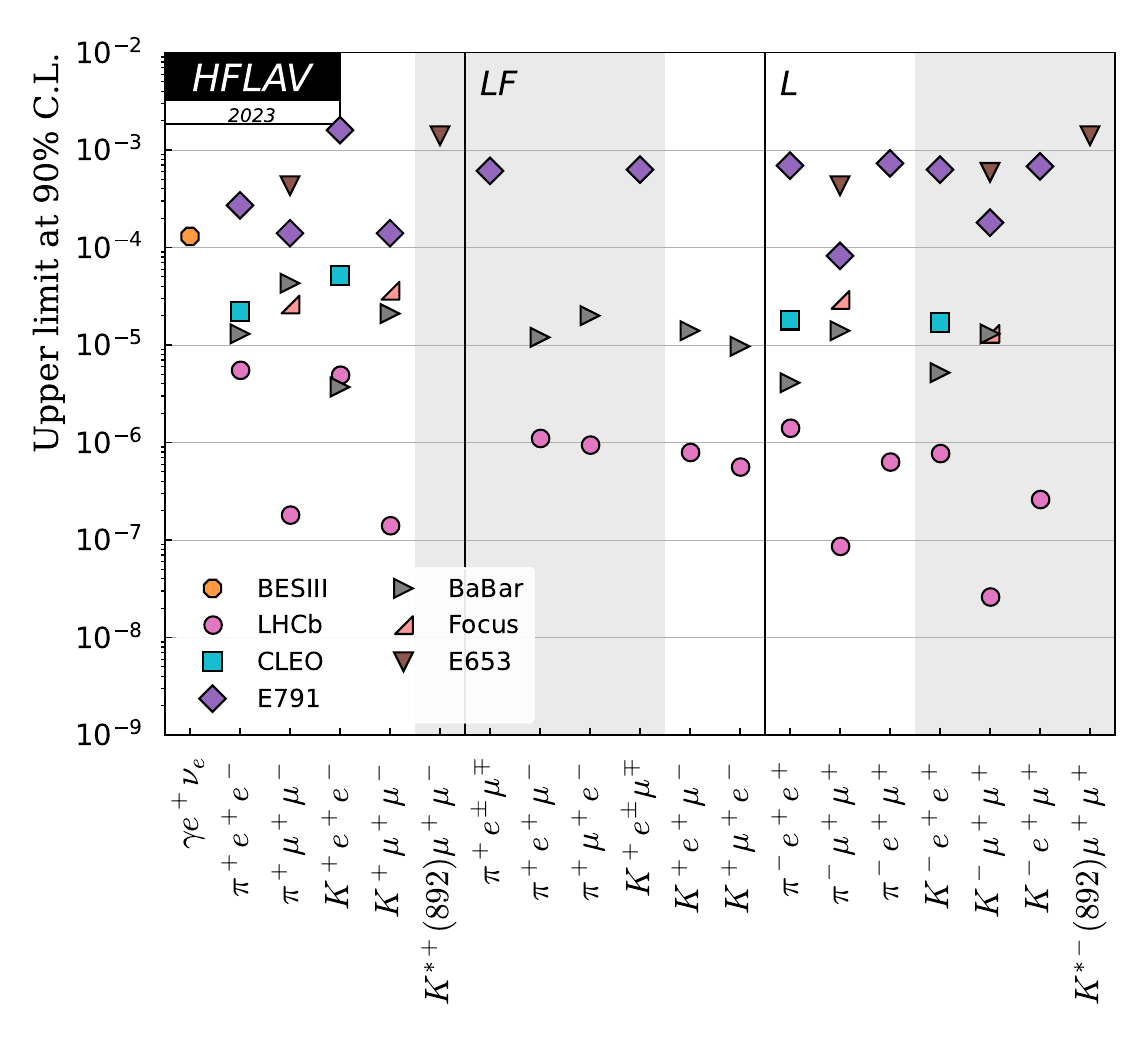}
\caption{Upper limits at $90\%$ C.L.\ for $D^+$ (top) and $D_s^+$ (bottom) 
decay branching fractions. Each plot shows flavour-changing neutral current and rare decays, 
lepton-flavour-changing decays (LF), and lepton-number-changing (L) decays. 
}
\label{fig:charm:rare_charged}
\end{center}
\end{figure}

\begin{figure}
\begin{center}
\includegraphics[width=3.0in]{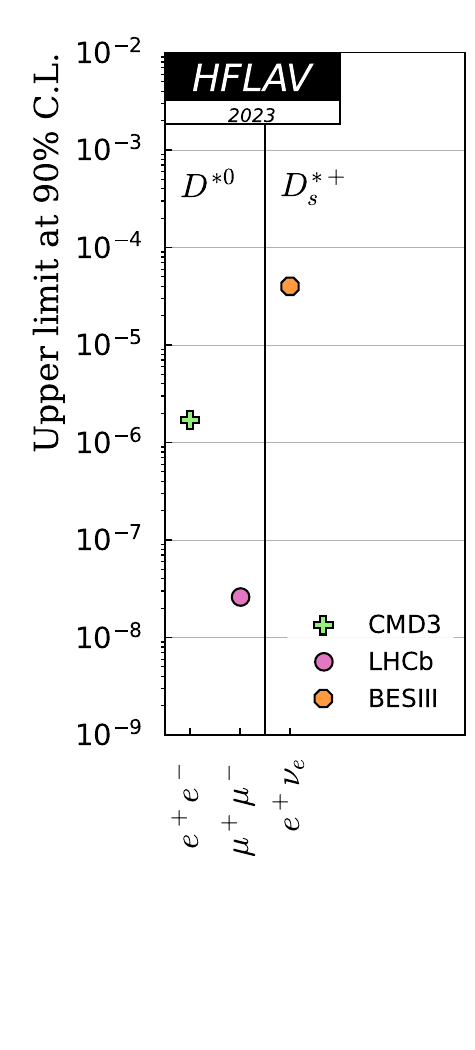}
\caption{Upper limits at $90\%$ C.L.\ for $D^{*0}$ and $D_{\rm s}^{*+}$ decay branching fractions. }
\label{fig:charm:Dstar}
\end{center}
\end{figure}

\begin{figure}
\begin{center}
\includegraphics[width=3.0in]{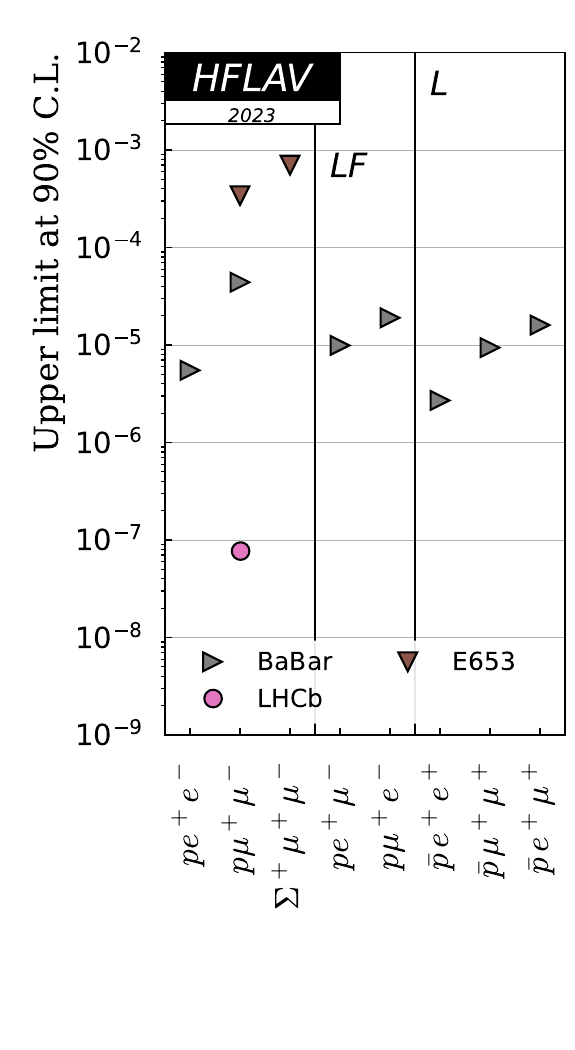}
\caption{Upper limits at $90\%$ C.L.\ for $\Lambda_c^+$ decay branching fractions. Shown are 
flavour-changing neutral current decays, lepton-flavour-changing (LF) 
decays, and lepton-number-changing (L) decays. }
\label{fig:charm:lambdac}
\end{center}
\end{figure}



\begin{table}
\centering
\caption{Upper limits at $90\%$ C.L.\ for $D^{*0}$ and $D_{\rm s}^{*+}$ decay
branching fractions.\label{tab:charm:rare_dstar}} 
%
\end{table}

\begin{table}
\centering
\caption{Upper limits at $90\%$ C.L.\ for $\Lambda_c^+$ decay
branching fractions.\label{tab:charm:rare_lambdac}}
%
\end{table}

\clearpage
\section{Tau lepton properties}
\label{sec:tau}
\input{tau/tau_all_common}

\section{Acknowledgments}

We thank 
Lu Cao,
Mirco Dorigo,
Sevda Esen,
Alessandro Gaz,
Carla G\"obel,
Yuval Grossman,
Judd Harrison,
Tobias Huber,
Gianluca Inguglia,
Martin Jung,
Alexander Lenz,
Ryan Mitchell,
Ulrich Nierste,
Emilie Passemar,
and Keri Vos
for their careful review of the
text in preparation of this paper for publication.
We also thank all past members of HFLAV and co-authors previous HFLAV reports.
We are grateful for the strong support of the
\atlas, \babar, \belle, \belletwo, \besthree, CLEO, \cdf, CMS, \dzero\ and \lhcb collaborations, without whom this
compilation of results and world averages would not have
been possible. 
The success of these experiments was made possible by the excellent performance and operation of the
BEPC-II, CESR, KEKB, LHC, PEP-II, SuperKEKB, and Tevatron colliders.
We recognize the interplay between theoretical and experimental communities that has provided a stimulus for many of the measurements in this document.
We thank CERN for providing computing resources to HFLAV.

Members of HFLAV are supported by the following funding agencies:
Australian Research Council (Australia);
Ministry of Science and Technology, National Natural Science Foundation, Key Research Program of Frontier Sciences of the Chinese Academy of Sciences (China);
European Research Council;
CNRS/IN2P3 (France);
Deutsche Forschungsgemeinschaft, Bundesministerium für Forschung, Technologie und Raumfahrt (Germany);
Israel Science Foundation, US-Israel Binational Science Fund (Israel);
Istituto Nazionale di Fisica Nucleare (Italy);
Ministry of Education, Culture, Sports, Science and Technology, Japan Society for the Promotion of Science (Japan);
National Agency for Academic Exchange, Ministry of Science and Higher Education, National Science Centre (Poland);
MCINN and XuntaGal (Spain);
Swiss National Science Foundation (Switzerland);
Science and Technology Facilities Council (UK);
and Department of Energy (USA).

\clearpage
\bibliographystyle{tex/LHCb}
\raggedright
\setlength{\parskip}{0pt}
\setlength{\itemsep}{0pt plus 0.3ex}
\begin{small}
\bibliography{main,prodfrac/prodfractions,life_mix/life_mix,slbdecays/slb_ref,rare/RareDecaysBib,rare/notincluded/RareDecaysNotIncludedBib,b2charm/b2charm,charm/charm_refs,tau/tau-refs}
\end{small}

\end{document}